\documentclass[
dsingle,
]{Dissertate}
\usepackage{graphicx}
\usepackage{xcolor}
\usepackage{dcolumn}
\usepackage{mathtools}
\usepackage{amsthm}
\usepackage{amssymb}
\usepackage{bm}
\usepackage{geometry}
\usepackage{hyperref}
\usepackage{lmodern}
\usepackage[T1]{fontenc}
\usepackage[utf8]{inputenc}
\usepackage[english]{babel}

\usepackage[capitalise]{cleveref}
\usepackage{booktabs}
\usepackage[version=4]{mhchem}
\usepackage{algorithm2e}
\RestyleAlgo{ruled}

\usepackage{epigraph}
\epigraphnoindent

\makeatletter
\newtheorem*{rep@theorem}{\rep@title}
\newcommand{\newreptheorem}[2]{%
\newenvironment{rep#1}[1]{%
 \def\rep@title{#2 \ref{##1}}%
 \begin{rep@theorem}}%
 {\end{rep@theorem}}}
\makeatother

\theoremstyle{plain}
\newtheorem{theorem}{Theorem}[chapter]
\newreptheorem{theorem}{Theorem}
\newtheorem{lemma}[theorem]{Lemma}
\newtheorem{corollary}[theorem]{Corollary}
\newtheorem{proposition}[theorem]{Proposition}

\theoremstyle{remark}
\newtheorem{remark}[theorem]{Remark}

\theoremstyle{definition}
\newtheorem{definition}[theorem]{Definition}
\newreptheorem{definition}{Definition}
\newtheorem{assumptions}{Assumptions}[chapter]

\renewcommand{\O}{\mathcal{O}}
\newcommand{\Ord}{\O}
\newcommand{\Ot}{\tilde{\mathcal{O}}}
\newcommand{\Omegat}{\tilde{\Omega}}
\DeclareMathOperator{\poly}{poly}
\newcommand{\R}{\mathbb{R}}
\newcommand{\C}{\mathbb{C}}

\newcommand{\Z}{\mathbb{Z}}

\newcommand{\I}{\mathbb{I}}
\newcommand{\openone}{\I}
\newcommand{\Had}{\mathsf{H}}
\newcommand{\Ph}{S}
\newcommand{\G}{\mathsf{G}}
\newcommand{\HS}{\mathcal{H}}
\renewcommand{\Re}{\operatorname{Re}}
\renewcommand{\Im}{\operatorname{Im}}
\renewcommand{\i}{\mathrm{i}}
\newcommand{\MA}[2]{\mathcal{A}_{#2}^{(#1)}}
\newcommand{\Spin}{\mathrm{Spin}}
\newcommand{\Pin}{\mathrm{Pin}}
\newcommand{\Orth}{\mathrm{O}}
\newcommand{\SO}{\mathrm{SO}}
\newcommand{\SU}{\mathrm{SU}}
\newcommand{\U}{\mathrm{U}}
\newcommand{\FGU}{\mathrm{FGU}}
\newcommand{\NC}{\mathrm{NC}}
\newcommand{\Cl}{\mathrm{Cl}}
\newcommand{\Pauli}{\mathcal{P}}
\newcommand{\Sym}{\mathrm{Sym}}
\newcommand{\Alt}{\mathrm{Alt}}
\newcommand{\B}{\mathrm{B}}
\newcommand{\SymCl}[1]{\Cl(1)^{\otimes #1}_\Sym}
\newcommand{\Stab}{\mathrm{Stab}}
\newcommand{\spec}{\mathrm{spec}}
\DeclareMathOperator{\diag}{diag}
\DeclareMathOperator{\spn}{span}
\DeclareMathOperator*{\E}{\mathbb{E}}
\DeclareMathOperator{\V}{Var}

\DeclareMathOperator{\sgn}{sign}
\DeclareMathOperator{\sign}{sign}
\DeclareMathOperator{\tr}{tr}

\newcommand{\comb}[2]{\mathcal{C}_{#1, #2}}
\newcommand{\diags}[2]{\mathcal{D}_{#1, #2}}
\newcommand{\op}[2]{\ket{#1}\!\bra{#2}}
\newcommand{\ip}[2]{\langle #1 | #2 \rangle}
\newcommand{\ev}[3]{\langle #1 | #2 | #3 \rangle}
\newcommand{\ket}[1]{| #1 \rangle}
\newcommand{\bra}[1]{\langle #1 |}

\newcommand{\rrangle}{\rangle \! \rangle}
\newcommand{\llangle}{\langle \! \langle}
\newcommand{\kket}[1]{| #1 \rrangle}
\newcommand{\bbra}[1]{\llangle #1 |}
\newcommand{\vop}[2]{\kket{#1} \! \bbra{#2}}
\newcommand{\vip}[2]{\llangle #1 | #2 \rrangle}
\newcommand{\vev}[3]{\llangle #1 | #2 | #3 \rrangle}

\newcommand{\T}{\mathsf{T}}
\newcommand{\sn}[1]{\| #1 \|_{\mathrm{shadow}}}
\newcommand{\sns}[2]{\| #1 \|_{#2}}
\renewcommand{\l}[1]{\mathopen{}\left#1}
\renewcommand{\r}[1]{\right#1\mathclose{}}
\renewcommand{\mod}{\:\mathrm{mod}\:}
\newcommand{\loc}{\mathrm{loc}}

\newcommand{\fs}{f_{\mathrm{samp}}}
\newcommand{\tl}{t_{\mathrm{load}}}

\newcommand{\SWAP}{\mathrm{SWAP}}
\newcommand{\iSWAP}{\i\mathrm{SWAP}}
\newcommand{\fSim}{\mathrm{fSim}}
\newcommand{\PhXZ}{\mathrm{PhXZ}}

\DeclareMathOperator{\conv}{conv}
\DeclareMathOperator*{\argmin}{arg\,min}
\DeclareMathOperator*{\argmax}{arg\,max}

\begin{document}

\title{Learning, Optimizing, and Simulating Fermions with Quantum Computers}
\author{Andrew Zhao}
\previousdegrees{
	B.S., Physics, University of Maryland, College Park, 2018\\
	M.S., Physics, The University of New Mexico, 2021
}

\advisor{Akimasa Miyake}

\committeeInternalOne{Ivan H. Deutsch}
\committeeInternalTwo{Milad Marvian}


\committeeExternal{Andrew J. Landahl}

\degree{Doctor of Philosophy}
\degreeabbrv{Ph.D.}
\field{Physics}
\degreeyear{2024}
\degreeterm{Spring}
\degreemonth{May}
\department{Physics and Astronomy}
\defensedate{November 30\textsuperscript{th}, 2023}

\frontmatter
\setstretch{\dnormalspacing}

\setcounter{chapter}{0}  
\chapter{Introduction}\label{chap:introduction}

The subject of this dissertation is centered around the study of many-fermion systems within the model of quantum computation. Fermions make up essentially all of the matter that we encounter in everyday life, so it is evidently a worthwhile endeavor to study them. We will do so through a variety of lenses, each of which is inherited from a different perspective of what it means to study, or understand, a quantum system more broadly.

\section{Simulating}

For a physicist, ``understanding'' typically lies in the ability to make predictions. Given some system in Nature and a theoretical model of it in one's mind, how accurately does that model describe the behavior of the actual system? To make that assessment, we require the ability to query the model and receive meaningful information back. In the broadest sense possible, we will define this task as \emph{simulation}. That is, we only care about the correct input--output pairs, and not necessarily whether the underlying calculation actually ``emulated'' the physical system in any sense.\footnote{After all, what we think the system ``actually does'' is based on yet another model, for instance nonrelativistic quantum theory. It's models all the way down.}

Models of quantum systems are in an interesting position, in that they can be relatively simple to write down, yet exceedingly complex to simulate. Indeed, Dirac provocatively remarked as early as 1929 that~\cite{dirac1929quantum},
\begin{quote}
	The underlying physical laws necessary for the mathematical theory of a large part of physics and the whole of chemistry are thus completely known, and the difficulty is only that the exact application of these laws leads to equations much too complicated to be soluble. 
\end{quote}
At one end of this spectrum are analytical solutions, for instance those arising from one-dimensional models of interacting electrons~\cite{bethe1931theorie,lieb1962theory,lieb1968absence}. Such instances, however, are special and rare. For the vast swath of models that do not admit nice, exact solutions, one resorts to numerical calculations, be they with slide rules or supercomputing clusters. Such tools fall under the paradigm of \emph{classical computation}, as they operate under the familiar laws of classical physics. While computational methods have revolutionized our ability to simulate Nature~\cite{thijssen2007computational,dykstra2011theory}, there still remain many important examples in the landscape of quantum systems which continue to elude accurate solutions.\footnote{It should be pointed out that there exists many important classical problems that cannot be efficiently solved either.}

In light of this challenge, Feynman in 1981 proffered the use of a controllable quantum system to emulate other quantum systems~\cite{feynman1982simulating}. Just as modern computers trick classical ensembles of electrons (bits) into performing mathematical operations, a \emph{quantum computer} manipulates individual quantum constituents (qubits) into behaving like an arbitrary quantum system of one's choosing. Because the quantum computer operates by the same physical laws that the simulated system obeys, Feynman conjectured that the hardness of quantum simulation could be circumvented in this manner. Indeed, in 1996, Lloyd rigorously confirmed this intuition, under the paradigm that simulation strictly means emulating quantum dynamics~\cite{lloyd1996universal}. Assuming quantum theory to be the most fundamental description of reality, such a machine would be \emph{universal}~\cite{deutsch1985quantum}.

However, the story does not end there. Beyond dynamical simulation, physicists are also interested in static properties of quantum systems. The canonical example is the ground state, or minimal-energy configuration. Ground states represent systems as they approach the zero-temperature limit, wherein genuinely quantum effects begin to dominate the physics. Alas, computing the ground-state energy is intractable even with a universal quantum machine. This was shown in 1999 by Kitaev for quantum systems involving five-body interactions~\cite{kitaev2002classical}, which was subsequently improved to a more physically relevant three-body instance in 2003 by Kempe and Regev~\cite{kempe20033}, and finally brought down to two-body interactions in 2004 by all three researchers~\cite{kempe2006complexity}. These results are couched in the language of complexity theory, a reflection of the broader melding of computer science and physics in order to study this emerging paradigm of computation.

Rather than interpret these results pessimistically, it is better to view them as guardrails to keep us in check as we develop algorithms for simulating quantum systems. Indeed, this dissertation is not about complexity theory, but rather the more practical considerations for quantum computation. Nonetheless, we will sometimes comment on complexity and hardness when relevant, because it is important to know where the boundary between the possible and the impossible lies.

\section{Optimizing}

In retrospect, the hardness of the ground-state problem is perhaps not so surprising. This is because finding ground states is the quantum analogue of the classical Boolean satisfiability problem, whose intractability is given by the celebrated Cook--Levin theorem~\cite{cook1971complexity,levin1973universal}.\footnote{In a technical sense, this ``classical intractability'' is weaker than ``quantum intractability'';~regardless, it is believed that quantum computers cannot overcome either notion.} This classical problem asks for a global assignment of Boolean variables which satisfies as many competing clauses as possible. Similarly, in the quantum setting we seek a global configuration of the particles which minimizes as many competing interactions as possible. In both cases, the problem can be reformulated as one of \emph{optimization}:~there is an energy landscape over all possible configurations, over which we want navigate toward its extremal points.

Despite the underlying hardness, framing such problems in terms of optimization inspires the development of practical algorithms and heuristics. Even if the globally optimal solution cannot be found efficiently, variational principles allow one to reach an approximately satisfying solution, which may be sufficient for the given purpose. For classical algorithms, the theory of approximation is rigorously understood, in part because of the existence of the so-called PCP theorem~\cite{hochba1997approximation}. In the absence of a quantum analogue~\cite{aharonov2013guest}, it is more difficult to make similar claims about the theory of quantum approximations.

Nonetheless, quantum heuristics for optimization have grown to occupy a considerable amount of space within the landscape of quantum algorithms. Some, like quantum annealing~\cite{kadowaki1998quantum,farhi2001quantum}, use natural physical mechanisms to adiabatically close in on a good solution. Others, like the variational quantum eigensolver~\cite{peruzzo2014variational}, have universal gate-based devices in mind, so they opt to perform the optimization with the help of a classical processor on the side.

Such approaches have become particularly popular in recent times because of the current state of quantum technology, coined by Preskill in 2017 as the noisy intermediate-scale quantum (NISQ) era~\cite{preskill2018quantum}. Already, devices have on the order of 50--100 qubits, potentially placing them beyond the limits of classical simulation despite their modest size. At the same time, they are heavily limited in how long they can maintain their quantum coherence, thus limiting them to run relatively shallow circuits. The heuristic nature of variational algorithms therefore makes them appealing for NISQ machines, as short-depth circuits can be tailored to compactly represent fairly complex quantum states through variational parametrization. Improving the performance of NISQ algorithms, to eventually (hopefully) exhibit a quantum advantage over classical competitors, is one of the high-level motivations for the work presented in this dissertation. In \cref{chap:nisq}, we overview such ideas within the NISQ paradigm.

\section{Learning}

Recall that our definition for simulation involves receiving \emph{meaningful} information from the model. When that model is encoded into a quantum device, this aspect is complicated by the unique mechanics of quantum measurement:~the process of extracting information from the quantum domain into the classical is probabilistic and destructive. We therefore require multiple, identically prepared copies of a quantum state in order to statistically \emph{learn} its properties. This terminology is derived from connections with computational learning theory, which broadly concerns the complexity of learning a concept from just a few examples of it~\cite{kearns1994introduction}.

In our setting, the concept is a classical description of a quantum state, and the examples that one learns from are the random measurement outcomes. This task is known as \emph{quantum state tomography} (QST), which was considered as early as 1957 by Fano~\cite{fano1957description} based on the density-matrix representation of quantum states. It would not be until 1993, however, for the first experimental demonstration by Smithey \emph{et al.}~\cite{smithey1993measurement}, performed on a single mode of squeezed light. One shortcoming of their results is not related to the experimental setup itself, but rather concerns how they reconstructed the density matrix from the raw data. Their method employed an inverse Radon transform,\footnote{This is a standard technique in computed tomography, hence their use of the term ``tomography'' which has since become universal throughout quantum information.} which involves a heuristic smoothing hyperparameter. D'Ariano, Macchiavello, and Paris~\cite{dariano1994detection} pointed out that this could lead to uncontrolled errors in the reconstruction, inspiring them to develop a more theoretically sound alternative (which they then demonstrated on that same experimental data).

The quest for efficient and accurate reconstruction algorithms has since become a central theme in the development of QST techniques~\cite{gebhart2023learning}. By efficient, we mean both in the number of copies that need to be measured and the computational complexity of the reconstruction algorithm. By accurate, we mean that the protocol has good theoretical guarantees for how well it can estimate the density matrix. For these reasons, the tools of learning theory (as well as of signal processing and matrix analysis) have become a natural fit for the theory of quantum tomography.

The notion of learning quantum states has also since broadened beyond this original formulation. Indeed, generic QST protocols are necessarily efficient only in the dimension of the state space~\cite{haah2017sample,yuen2023improved}, which for a many-body system grows exponentially in the number of individual constituents. This has motivated the search for efficient schemes for \emph{partial state learning},\footnote{Fano was remarkably prescient and had already envisioned this reduced notion of quantum learning~\cite[Section~6]{fano1957description}, although he did not pursue the idea in further detail.} whereby some reduced description of the state suffices to make accurate predictions of the properties of interest (but not all properties of the state). A major portion of this dissertation is devoted to partial state learning in theory and in practice, which largely draws influence from the recently introduced paradigm of \emph{shadow tomography}~\cite{aaronson2020shadow,huang2020predicting,paini2021estimating}. We review these ideas in detail in \cref{chap:learning}.

Note that the concept of quantum learning extends beyond that of states, for example to learning unknown quantum processes~\cite{eisert2020quantum}. Quantum machine learning is also a particularly burgeoning field~\cite{biamonte2017quantum,cerezo2022challenges}. Universal themes from learning theory run throughout all of these subjects, however, this dissertation will not cover such topics.

\section{Outline of this dissertation}

I will now briefly outline the remaining chapters. This dissertation assumes a background in quantum computation at a level comparable to Nielsen and Chuang's textbook~\cite{nielsen2010quantum}.

The remaining chapters are roughly divided into two parts. The first part, \cref{chap:fermions,chap:learning,chap:nisq}, consists of background material which frames the context for the main contributions of this dissertation. \cref{chap:fermions} reviews the physics and chemistry of fermions, with special attention to a solvable class of models called \emph{free fermions}. \cref{chap:learning} covers the subject of learning quantum states, with an emphasis on the paradigm of partial state learning. Within this context, I highlight \emph{classical shadows}, an efficient and experimentally friendly protocol for learning partial descriptions of quantum states. \cref{chap:nisq} gives an overview of the NISQ era, and provides a survey of prominent ideas in \emph{quantum error mitigation}, a near-term-feasible method for combating the ``noise'' obstacle within NISQ.

The second part, \cref{chap:MR,chap:FCS,chap:QR,chap:EM}, is derived from a collection of works completed throughout my PhD (see \cref{sec:pub_list}). These chapters constitute the main contributions of this dissertation. \cref{chap:FCS} introduces \emph{fermionic classical shadows},\footnote{Also coined later as ``matchgate shadows'' in the literature.} an extension of classical shadows for optimally learning all local properties of a many-fermion system~\cite{zhao2021fermionic}. \cref{chap:EM} introduces an economical error-mitigation scheme called \emph{symmetry-adjusted classical shadows}, which derives its effectiveness from unifying the theory of classical shadows with symmetries present in quantum systems~\cite{zhao2023group}. \cref{chap:MR} proposes an alternative approach to learning local observables called \emph{unitary partitioning}, which is shown to be particularly effective for addressing highly complex observables such as the electronic energy in chemical and molecular systems~\cite{zhao2020measurement}. \cref{chap:QR} then pivots directions by initiating the study of a hard classical optimization problem on quantum computers, based on a surprisingly natural formulation of the classical problem in terms of fermions~\cite{zhao2023expanding}.

Finally, \cref{chap:conclusion} concludes the dissertation with an outlook for the future of the ideas presented within.

\section{List of publications}\label{sec:pub_list}

Below is a chronological list of the papers that I coauthored during my PhD. Not all works listed here appear as chapters in this dissertation.
\begin{itemize}
	\item \cite{zhao2020measurement} A.~Zhao, A.~Tranter, W.~M.~Kirby, S.~F.~Ung, A.~Miyake, and P.~J.~Love. Measurement reduction in variational quantum algorithms. \href{http://dx.doi.org/10.1103/PhysRevA.101.062322}{\emph{Physical Review A}, 101(6):062322, 2020}. (Appears in \cref{chap:MR})
	
	\item \cite{zhao2021fermionic} A.~Zhao, N.~C.~Rubin, and A.~Miyake. Fermionic partial tomography via classical shadows. \href{http://dx.doi.org/10.1103/PhysRevLett.127.110504}{\emph{Physical Review Letters}, 127(11):110504, 2021}. (Appears in \cref{chap:FCS})
	
	\item \cite{daniel2022quantum} A.~K.~Daniel, Y.~Zhu, C.~H.~Alderete, V.~Buchemmavari, A.~M.~Green, N.~H.~Nguyen, T.~G.~Thurtell, A.~Zhao, N.~M.~Linke, and A.~Miyake. Quantum computational advantage attested by nonlocal games with the cyclic cluster state. \href{http://dx.doi.org/10.1103/PhysRevResearch.4.033068}{\emph{Physical Review Research}, 4:033068, 2022}.
	
	\item \cite{babbush2023quantum} R. Babbush, W.~J.~Huggins, D.~W.~Berry, S.~F.~Ung, A.~Zhao, D.~R.~Reichman, H.~Neven, A.~D.~Baczewski, and J.~Lee. Quantum simulation of exact electron dynamics can be more efficient than classical mean-field methods. \href{http://dx.doi.org/10.1038/s41467-023-39024-0}{\emph{Nature Communications}, 14:4058, 2023}.
	
	\item \cite{zhao2023expanding} A.~Zhao and N.~C.~Rubin. Expanding the reach of quantum optimization with fermionic embeddings. \href{https://arxiv.org/abs/2301.01778}{\emph{arXiv:2301.01778}, 2023}. (Appears in \cref{chap:QR})
	
	\item \cite{zhao2023group} A.~Zhao and A.~Miyake. Group-theoretic error mitigation enabled by classical shadows and symmetries. \href{https://arxiv.org/abs/2310.03071}{\emph{arXiv:2310.03071}, 2023}. (Appears in \cref{chap:EM})
\end{itemize}
\chapter{Theory of Fermions}\label{chap:fermions}


Fermions are the elementary building blocks of matter. The canonical example to keep in the back of one's mind is the humble electron, whose central role to our world can hardly be understated. The goal of this chapter is to provide the mathematical background of nonrelativistic many-fermion systems necessary to understand how one might simulate them on a computer (be it classical or quantum).

\section{From first to second quantization}

In this dissertation, we will largely work with fermions in second quantization. However, to gain some physical intuition it is useful to introduce concepts in first quantization, and then show how they translate over to second quantization. Let $\HS_n$ be a single-particle Hilbert space of dimension $n$ over $\C$. We will assume $n < \infty$ unless otherwise stated. The Hilbert space of a composite system of $\eta$ particles is then the tensor product $\HS_n^{\otimes \eta}$. Because fermions are identical particles obeying the Pauli exclusion principle, any two fermions are forbidden from occupying the same state in $\HS_n$.

Mathematically, this implies that many-fermion systems obey \emph{antisymmetrization}:~let $S_{(i,j)}$ be the operator which swaps the states of particles $i$ and $j$. An antisymmetric state $\ket{\psi} \in \HS_n^{\otimes \eta}$ obeys
\begin{equation}\label{eq:antisymmetry_swap}
	S_{(i,j)} \ket{\psi} = -\ket{\psi}.
\end{equation}
Such states lie in the antisymmetric subspace of $\HS_n^{\otimes \eta}$, which we denote by $\wedge^\eta \HS_n$. For notation, let $[n] \coloneqq \{1, \ldots, n\}$ be the set of integers from $1$ to $n$. Choosing some arbitrary orthonormal basis $\{\ket{p} \mid p \in [n] \}$ for $\HS_n$, we equip $\wedge^\eta \HS_n$ with the convenient basis $\{ \ket{\bm{p}} \mid \bm{p} \in \binom{[n]}{\eta} \}$,
where $\binom{[n]}{\eta}$ is the set of all subsets of $[n]$ of cardinality $\eta$:
\begin{equation}
    \binom{[n]}{\eta} \coloneqq \{ \bm{p} =(p_1, \ldots p_\eta) \mid 1 \leq p_1 < \cdots < p_\eta \leq n \}.
\end{equation}
Note that we abide by the convention that $\bm{p}$ is in ascending order. The antisymmetrized basis states are then defined as
\begin{equation}
	\ket{\bm{p}} \coloneqq \frac{1}{\sqrt{\eta!}} \sum_{\pi \in \Sym(\eta)} (-1)^\pi \ket{p_{\pi(1)}} \otimes \cdots \otimes \ket{p_{\pi(\eta)}},
\end{equation}
where the symmetric group $\Sym(\eta)$ is the set of all bijections $\pi : [\eta] \to [\eta]$, and $(-1)^\pi$ is the permutational parity of $\pi$. By construction, these states obey \cref{eq:antisymmetry_swap}.

The idea of second quantization is to promote the antisymmetry from the description of the states to that of the operators. This results in a more compact mathematical representation of the fermions, particularly within a computational context. The formalism also enables us to describe fermionic systems wherein the particle number is not a conserved quantity, such as is the case in the Bardeen--Cooper--Schrieffer (BCS) theory of superconductivity~\cite{bardeen1957theory}. To this end, we define a \emph{Fock space} as the direct sum of all antisymmetric subspaces of all possible particle numbers:
\begin{equation}
	\mathcal{F}_n \coloneqq \bigoplus_{\eta=0}^n \wedge^\eta \HS_n.
\end{equation}
Note that because each fermion must be in a different state, there can only be at most $\eta \leq n$ antisymmetrized particles. The dimension of this space is  $\sum_{\eta=0}^n \binom{n}{\eta} = 2^n$.

Rather than concatenate antisymmetrized basis states to form a basis for $\mathcal{F}_n$, it is more convenient to use the \emph{occupation-number basis}:~for each $n$-bit string $b \in \{0, 1\}^n$, the Fock state $\ket{b}$ describes a fermion occupying the energy level $p \in [n]$ whenever $b_p = 1$, and all other levels $q$ (for which $b_q = 0$) are unoccupied. Explicitly, the first-quantized representation $\bm{p} \subseteq [n]$ is identified with the bit string $b \in \{0,1\}^n$ satisfying
\begin{equation}
	b_p = \begin{cases}
	1 & \text{ if } p \in \bm{p},\\
	0 & \text{ else}.
	\end{cases}
\end{equation}
We will write $\ket{b} \simeq \ket{\bm{p}}$ to indicate this equivalence between first- and second-quantized objects. The state $\ket{0^n} \simeq \ket{\varnothing}$ is called the \emph{vacuum}. As a point of language, we would like to abstract the single-particle levels (basis states of $\HS_n$) away from any notion of energy, so we typically refer to them as \emph{fermionic modes}. In contexts such as quantum chemistry, they are also referred to as \emph{orbitals}, as they are the many-atom (molecular) generalizations of the familiar hydrogen-like electron orbitals.

\section{Fermion ladder algebra}

The second-quantized representation of states no longer explicitly features antisymmetry;~instead, this information is encoded into the operators on $\mathcal{F}_n$. For each $p \in [n]$, define $a_p^\dagger$ and $a_p$ as the creation and annihilation (ladder) operators which either create or annihilate a fermion in the $p$th mode. We equip these operators with the canonical anticommutation relations:
\begin{align}
	\{a_p^\dagger, a_q^\dagger\} &= \{a_p, a_q\} = 0, \label{eq:car_same}\\
	\{a_p, a_q^\dagger\} &= \delta_{pq} \I_{2^n}, \label{eq:car_diff}
\end{align}
where $\I_d$ is the $d$-dimensional identity operator and $\{A, B\} \coloneqq AB + BA$ is the anticommutator (cf.~the commutator $[A, B] \coloneqq AB - BA$). These relations precisely enforce the antisymmetry:~let $a_{p_1}^\dagger \cdots a_{p_\eta}^\dagger \ket{0^n}$ be an $\eta$-fermion Fock state (all $p_i$ are different unless otherwise specified). Consider swapping any two particles $i$ and $j$;~by \cref{eq:car_same}, this incurs a minus sign, as desired:
\begin{equation}
	a_{p_1}^\dagger \cdots a_{p_i}^\dagger \cdots a_{p_j}^\dagger \cdots a_{p_\eta}^\dagger \ket{0^n} = -a_{p_1}^\dagger \cdots a_{p_j}^\dagger \cdots a_{p_i}^\dagger \cdots a_{p_\eta}^\dagger \ket{0^n}.
\end{equation}
The anticommutation relations also imply $a_p^2 = (a_p^\dagger)^2 = 0$, which is the content of the Pauli exclusion principle (each fermionic mode carry at most one particle). Concretely, we can fully specify the ladder operators by their action on any basis state:
\begin{align}
	a_p^\dagger \ket{b} &= \delta_{b_p,0} (-1)^{\sum_{q < p} b_q} \ket{b_1 \cdots b_p + 1 \cdots b_n},\\
	a_p \ket{b} &= \delta_{b_p,1} (-1)^{\sum_{q < p} b_q} \ket{b_1 \cdots b_p - 1 \cdots b_n}.
\end{align}

\subsection{Generating a complete operator basis}

The creation and annihilation operators generate a complete basis for the space of linear operators on $\mathcal{F}_n$, denoted by $\mathcal{L}(\mathcal{F}_n)$. One way to show this is by demonstrating that any $A \in \mathcal{L}(\mathcal{F}_n)$ can be written as a linear combination of arbitrary products of the ladder operators. Here, we will instead show how to construct a complete basis of operators.  We do so by invoking the Jordan--Wigner transformation~\cite{jordanwigner}, which provides a concrete matrix representation of the isomorphism between Fock space and a more familiar $n$-qubit space, $\mathcal{F}_n \cong (\C^2)^{\otimes n}$.

Let $X, Y, Z$ be the $2 \times 2$ Pauli matrices. It is well known that the set of all $4^n$ Pauli operators $\Pauli_n \coloneqq \{\I_2, X, Y, Z\}^{\otimes n}$ is a complete basis for $\mathcal{L}((\C^2)^{\otimes n})$, satisfying orthogonality with respect to the trace inner product:
\begin{equation}
	\langle P, Q \rangle \equiv \tr(P^\dagger Q) = 2^n \delta_{PQ}
\end{equation}
for all $P, Q \in \Pauli_n$. The Jordan--Wigner isomorphism identifies ladder operators with $n$-qubit operators as
\begin{equation}
	a_p^\dagger \mapsto Z^{\otimes (p - 1)} \otimes \l( \frac{X - \i Y}{2} \r) \otimes \I_2^{\otimes (n - p)}.
\end{equation}
It is straightforward to verify that these operators on the qubit space obey the canonical anticommutation relations, \cref{eq:car_same,eq:car_diff}. Thus they completely specify the algebra.

To generate $\Pauli_n$, it will be convenient to introduce Majorana operators:
\begin{equation}
	\gamma_{2p-1} \coloneqq a_p + a_p^\dagger, \quad \gamma_{2p} \coloneqq -\i(a_p - a_p^\dagger),
\end{equation}
for each $p \in [n]$. The canonical anticommutation relations translate to the Majorana operators as a single Clifford algebra identity,
\begin{equation}\label{eq:majorana_clifford_background}
	\{\gamma_\mu, \gamma_\nu\} = 2\delta_{\mu\nu} \I_{2^n}.
\end{equation}
Applying the Jordan--Wigner transformation reveals that these Majorana operators are just representations of Pauli operators:
\begin{align}
	\gamma_{2p-1} &\mapsto Z^{\otimes (p - 1)} \otimes X \otimes \I_2^{\otimes (n - p)},\\
	\gamma_{2p} &\mapsto Z^{\otimes (p - 1)} \otimes Y \otimes \I_2^{\otimes (n - p)}.
\end{align}
Then, to generate all $4^n$ unique Pauli operators, we take all possible products of Majorana operators. The unique products correspond to distinct indices, due to the anticommutation relation of Majorana operators (as any colliding indices yield $\gamma_\mu^2 = \I_{2^n}$). For each $\bm{\mu} \subseteq [2n]$, where we order $1 \leq \mu_1 < \cdots < \mu_{|\bm{\mu}|} \leq 2n$, we define a $|\bm{\mu}|$-degree Majorana operator as
\begin{equation}\label{eq:majorana_op_def}
	\Gamma_{\bm{\mu}} \coloneqq (-\i)^{\binom{|\bm{\mu}|}{2}} \gamma_{\mu_1} \cdots \gamma_{\mu_{|\bm{\mu}|}}.
\end{equation}
The phase factor is chosen such that $\Gamma_{\bm{\mu}}$ is Hermitian. From the anticommutation relation \cref{eq:majorana_clifford_background}, one can show that these Majorana operators are also trace orthogonal, $\langle \Gamma_{\bm{\mu}}, \Gamma_{\bm{\nu}} \rangle = 2^n \delta_{\bm{\mu} \bm{\nu}}$. Because \cref{eq:majorana_op_def} is a product of $\gamma_{\mu_j} \in \Pauli_n$ and products of Pauli operators are closed within $\{\pm 1, \pm \i\} \times \Pauli_n$, it follows that $\Gamma_{\bm{\mu}}$ is also a Pauli operator (potentially with an overall sign). From counting, there are $|\{\bm{\mu} \subseteq [2n]\}| = 4^n$ such linearly independent (indeed, trace-orthogonal) operators. Hence $\{\Gamma_{\bm{\mu}} \mid \bm{\mu} \subseteq [2n]\} \cong \Pauli_n$ and so we have generated a complete basis for $\mathcal{L}(\mathcal{F}_n) \cong \mathcal{L}((\C^2)^{\otimes n})$ from the fermion ladder operators.

\section{Noninteracting fermions}

Let us return to a fixed-particle sector in the first-quantized picture. A noninteracting unitary transformation on a composite system of $\eta$ particles, each with local dimension $n$, factorizes as the product operator $\bigotimes_{i \in [\eta]} u_i$, where each $u_i \in \U(n)$. For many-fermion systems, the class of noninteracting unitaries must be restricted to respect antisymmetry. Specifically, because the particles are identical we cannot apply a different $u_i$ to each fermion;~the same single-particle transformation $u \in \U(n)$ must act on all $\eta$ fermions. Supposing $\ket{\psi} \in \mathcal{H}_n^{\otimes \eta}$ is antisymmetrized, then this transformation is described simply as $\ket{\psi} \mapsto u^{\otimes \eta} \ket{\psi}$.\footnote{Alternatively, directly writing $\ket{\psi} \in \wedge^\eta \mathcal{H}_n$ demands one to define the appropriate antisymmetric representation, $\wedge^\eta \U(n)$.}

Because $\U(n)$ is a connected and compact Lie group, it can be described as $\U(n) = e^{\mathfrak{u}(n)}$, where the Lie algebra $\mathfrak{u}(n)$ consists of all anti-Hermitian $n \times n$ matrices, i.e., all $\kappa \in \C^{n \times n}$ obeying $\kappa = -\kappa^\dagger$. Thus every unitary $u = e^\kappa \in \U(n)$ corresponds to time evolution generated by a single-particle Hamiltonian $h = \i\kappa$. This Hamiltonian can be lifted to act on the Fock space as the second-quantized operator
\begin{equation}\label{eq:noninteracting_hamiltonian}
	H = \sum_{p,q \in [n]} h_{pq} a_p^\dagger a_q.
\end{equation}
We understand this lifting intuitively by recognizing that the matrix elements $h_{pq} = \ev{p}{h}{q}$ are transition amplitudes between modes $p$ and $q$, corresponding to the action of the hopping terms $a_p^\dagger a_q$. The anticommutation relations in the ladder operators guarantee that these hoppings preserve antisymmetry.

\subsection{Single-particle basis rotations}\label{sec:basis_rotation_bkgd}

One approach to demonstrating \cref{eq:noninteracting_hamiltonian} is the correct lifting from first- to second-quantization is to show that $e^{-\i H}$ on $\mathcal{F}_n$ enacts the appropriate transformation by $u = e^{-\i h}$ on $\HS_n$ (up to global phase). This is particularly interesting for two reasons;~first, it describes the unitary dynamics of any system of noninteracting fermions, which are classically solvable. Second, such unitary transformations are \emph{basis rotations} of the single-particle Hilbert space, which are invaluable tools in the study of interacting-fermion models.

The action of $u$ is a local unitary transformation on each particle, which seems almost trivial in the first-quantized picture:
\begin{equation}
	\ket{p} \mapsto u \ket{p} = \sum_{q \in [n]} u_{qp} \ket{q} \eqqcolon \ket{\tilde{p}},
\end{equation}
where $\ket{\tilde{p}}$ denotes the rotated basis states. Just as $a_p^\dagger$ places a fermion in mode $\ket{p}$, it follows that the linear combination $\tilde{a}_{p}^\dagger \coloneqq \sum_{q \in [n]} u_{qp} a_q^\dagger$ places a fermion in the rotated mode $\ket{\tilde{p}}$. Indeed, $u \in \U(n)$ is necessary and sufficient for the rotated ladder operators to maintain the canonical anticommutation relations:
\begin{equation}
	\begin{split}
	\{\tilde{a}_p, \tilde{a}_q^\dagger\} &= \sum_{p',q' \in [n]} u_{p'p}^* u_{q'q} \{a_{p'}, a_{q'}^\dagger\}\\
	&= \sum_{p',q' \in [n]} [u^\dagger]_{pp'} \delta_{p'q'} u_{q'q}\\
	&= [u^\dagger u]_{pq} = \delta_{pq}.
	\end{split}
\end{equation}
Furthermore, because unitary conjugation preserves (anti)commutation relations, we can describe the rotated operator system as $\tilde{a}_p = U a_p U^\dagger$ for some $U \in \U(2^n)$. Our goal is then to show that $U = e^{-\i H}$ is the correct choice. Specifically, we will show the following.

\begin{proposition}\label{prop:basis_rotation_proof}
The adjoint action by $U = e^{-\i H}$ for any one-body Hamiltonian $H = \sum_{p,q \in [n]} h_{pq} a_p^\dagger a_q$ obeys
\begin{equation}\label{eq:basis_rotation_adj}
	U^\dagger a_p U = \sum_{q \in [n]} u_{pq} a_q,
\end{equation}
where $u = e^{-\i h} \in \U(n)$.
\end{proposition}

While the convention of \cref{eq:basis_rotation_adj} is that of Heisenberg evolution for operators, recall that $\tilde{a}_p = U a_p U^\dagger$ defined prior corresponded to a Schr\"{o}dinger-type picture. This is because when we had defined $\tilde{a}_p$, we had in mind the action of the unitary on states, i.e., $\ket{p} \mapsto \ket{\tilde{p}}$. Of course, the two pictures can be related through the inverse of $u$. Nonetheless, it will be useful to maintain this definition for the symbol $\tilde{a}_p$, while we we aim to prove \cref{eq:basis_rotation_adj}.

\begin{proof}
The standard approach for showing \cref{eq:basis_rotation_adj} is by studying the commutators arising from the Heisenberg equation. Instead, here we will offer an alternative proof through the lens of $n \times n$ matrix operations that does not require invoking differential equations. The key idea is that we can diagonalize the single-particle Hamiltonian efficiently because it is an $n \times n$ matrix:
\begin{equation}
	h = v \varepsilon v^\dagger,
\end{equation}
where $v \in \U(n)$ and $\varepsilon = \diag(\varepsilon_1, \ldots, \varepsilon_n) \in \R^{n \times n}$. Lifted to the Fock space, this corresponds to a diagonalization of $H$ in the basis of modes $\tilde{a}_p^\dagger \coloneqq \sum_{q \in [n]} v_{qp} a_q^\dagger$:
\begin{equation}\label{eq:noninteracting_diag}
	\begin{split}
	H &= \sum_{p,q \in [n]} h_{pq} a_p^\dagger a_q\\
	&= \sum_{p,q \in [n]} \sum_{r \in [n]} v_{pr} \varepsilon_{r} v_{qr}^* a_p^\dagger a_q\\
	&= \sum_{r \in [n]} \varepsilon_r \tilde{a}_r^\dagger \tilde{a}_r.
	\end{split}
\end{equation}
We emphasize that the rotated modes $\tilde{a}_r^\dagger$ (here, transformed by $v$) are, at this point, defined merely as linear combinations obeying the canonical anticommutation relations. As discussed above, because $\tilde{a}_r^\dagger$ preserves the commutation relations of $a_r^\dagger$ we know that there exists some unitary $V$ on the Fock space such that $\tilde{a}_r^\dagger = V a_r^\dagger V^\dagger$. We do not assume any further properties about $V$ beyond unitarity (as otherwise we would be using a circular argument). Instead, it suffices to write
\begin{equation}
	H = V \l( \sum_{p \in [n]} \varepsilon_r a_r^\dagger a_r \r) V^\dagger.
\end{equation}
We can therefore exponentiate $H$, obtaining
\begin{equation}
	\begin{split}
	U &= e^{-\i H} = V \exp\l( -\i \sum_{r \in [n]} \varepsilon_r a_r^\dagger a_r \r) V^\dagger\\
	&= V \l( \prod_{r \in [n]} e^{-\i \varepsilon_r n_r} \r) V^\dagger,
	\end{split}
\end{equation}
where the final equality follows because the occupation-number operators $n_r \coloneqq a_r^\dagger a_r$ all mutually commute.

We are now in a position to compute the adjoint action
\begin{equation}\label{eq:UaU_expanded}
	U^\dagger a_p U = V \l( \prod_{r \in [n]} e^{\i \varepsilon_r n_r} \r) V^\dagger a_p V \l( \prod_{s \in [n]} e^{-\i \varepsilon_s n_s} \r) V^\dagger.
\end{equation}
To faciliate a smoother calculation, we will employ the Jordan--Wigner transformation here:
\begin{align}
	a_p &= Z^{\otimes (p - 1)} \otimes \l( \frac{X + \i Y}{2} \r) \otimes \I_2^{\otimes (n - p)}\\
	n_p &= \I_2^{\otimes (p - 1)} \otimes \l( \frac{\I_2 - Z}{2} \r) \otimes \I_2^{\otimes (n - p)}.
\end{align}
We treat this merely as a useful matrix representation of $\mathcal{F}_n$ with which we can perform concrete calculations (i.e., not an actual system of $n$ qubits). First, write $V^\dagger a_p V = \sum_{q \in [n]} v_{pq} a_q$ and
\begin{equation}
	\prod_{r \in [n]} e^{\i \varepsilon_r n_r} = e^{\i\tr\varepsilon/2} \bigotimes_{r \in [n]} e^{-\i \varepsilon_r Z / 2}.
\end{equation}
Then each Pauli-$Z$ rotation $R_Z(\theta) \coloneqq e^{-\i \theta Z / 2}$ commutes with the tensor factors in $a_q$ except at the $r = q = s$ site:
\begin{align}
	\l( \prod_{r \in [n]} e^{\i \varepsilon_r n_r} \r) V^\dagger a_p V &\l( \prod_{s \in [n]} e^{-\i \varepsilon_s n_s} \r) = \sum_{q \in [n]} v_{pq} \l( \bigotimes_{r \in [n]} e^{-\i \varepsilon_r Z/2} \r) a_q \l( \bigotimes_{s \in [n]} e^{\i \varepsilon_s Z/2} \r) \label{eq:UaU_jw}\\
	&= \sum_{q \in [n]} v_{pq} Z^{\otimes (q - 1)} \otimes R_Z(\varepsilon_q) \l( \frac{X + \i Y}{2} \r) R_Z^\dagger(\varepsilon_q) \otimes \I_2^{\otimes (n - q)} \notag.
\end{align}
Thus, in this matrix representation we can import well-known identities for Pauli rotations:
\begin{equation}\label{eq:pauli_matrix_adjoint}
	R_Z(\theta) \begin{pmatrix}
	X\\
	Y
	\end{pmatrix} R_Z^\dagger(\theta) = \begin{pmatrix}
	\cos\theta & \sin\theta\\
	-\sin\theta & \cos\theta
	\end{pmatrix} \begin{pmatrix}
	X\\
	Y
	\end{pmatrix}.
\end{equation}
This implies the relation
\begin{equation}
	R_Z(\theta) (X + \i Y) R_Z^\dagger(\theta) = e^{-\i\theta} (X + \i Y),
\end{equation}
and so the creation and annihilation operators are transformed as simply $a_p^\dagger \mapsto e^{\i\theta} a_p^\dagger$ and $a_p \mapsto e^{-\i\theta} a_p$. Applying these results to \cref{eq:UaU_jw} implies
\begin{equation}\label{eq:UaU_final}
	\begin{split}
	U^\dagger a_p U &= V \l( \sum_{q \in [n]} v_{pq} e^{-\i \varepsilon_q} a_q \r) V^\dagger\\
	&= \sum_{q,r \in [n]} v_{pq} e^{-\i \varepsilon_q} v_{rq}^* a_r\\
	&= \sum_{r \in [n]} [v e^{-\i \varepsilon} v^\dagger]_{pr} a_r.
	\end{split}
\end{equation}
Finally, using the fact that $v e^{-\i \varepsilon} v^\dagger = e^{-\i h} = u$, we see that \cref{eq:UaU_final} is identical to \cref{eq:basis_rotation_adj}, which is what we had set out to show.
\end{proof}

A corollary of this result is that we can recognize that the unitary $V$ diagonalizing $H$ is itself of the form
\begin{equation}\label{eq:basis_rotation_bkgd}
	V = \exp\l( \sum_{p,q \in [n]} [\log v]_{pq} a_p^\dagger a_q \r),
\end{equation}
where $\log v$ is the (principal branch) matrix logarithm of $v \in \U(n)$. Additionally, it can be shown that $V : \U(n) \to \U(2^n)$ is a group homomorphism, i.e., satisfies the property $UV = \exp\l( \sum_{p,q} [\log(uv)]_{pq} a_p^\dagger a_q \r)$ for all $u, v \in \U(n)$. This was shown in \cite[Supplemental Material]{kivlichan2018quantum} using the tools of Grassmann algebras. Below we demonstrate a slightly weaker property but with much simpler techniques. Namely, that the adjoint action $\mathcal{V}(\cdot) = V (\cdot) V^\dagger$ is a homomorphism. This implies that $V$ is at least a projective representation.\footnote{Indeed, this is sufficient for quantum mechanics because (pure) states are elements of \emph{projective} Hilbert space.}

First, use the conjugation identity for one-body rotations:
\begin{equation}
	\begin{split}
	UV a_p^\dagger V^\dagger U^\dagger &= U \l( \sum_{q \in [n]} v_{qp} a_q^\dagger \r) U^\dagger\\
	&= \sum_{q \in [n]} \sum_{r \in [n]} v_{qp} u_{rq} a_r^\dagger\\
	&= \sum_{r \in [n]} [uv]_{rp} a_r^\dagger.
	\end{split}
\end{equation}
The result for the annihilation operators is analogous. To extend this to the full operator space, we can compute the action of $\mathcal{V}$ on higher-order fermionic operators. For example, for a $k$-fold product of creation operators, we have
\begin{equation}\label{eq:basis_rotation_product}
	\begin{split}
	\mathcal{V}(a_{p_1}^\dagger \cdots a_{p_k}^\dagger) &= V a_{p_1}^\dagger V^\dagger \cdots V a_{p_k}^\dagger V^\dagger\\
	&= \sum_{q_1, \ldots, q_k \in [n]} v_{q_1 p_1} \cdots v_{q_k p_k} a_{q_1}^\dagger \cdots a_{q_k}^\dagger\\
	&= \sum_{1 \leq q_1 < \cdots < q_k \leq n} \sum_{\pi \in \Sym(k)} v_{q_{\pi(1)} p_1} \cdots v_{q_{\pi(k)} p_k} (-1)^\pi a_{q_1}^\dagger \cdots a_{q_k}^\dagger\\
	&= \sum_{\bm{q} \in \binom{[n]}{k}} \det(v_{\bm{q}, \bm{p}}) a_{q_1}^\dagger \cdots a_{q_k}^\dagger,
	\end{split}
\end{equation}
where $v_{\bm{q}, \bm{p}}$ is the $k \times k$ submatrix of $v$ obtained by extracting the rows and columns corresponding to the indices in $\bm{q}$ and $\bm{p}$, respectively. Note that the third equality follows from the fact that $(a_p^\dagger)^2 = 0$ and $a_{p_i}^\dagger a_{p_j}^\dagger = -a_{p_j}^\dagger a_{p_i}^\dagger$. These unitaries therefore preserve the locality of fermionic operators, as must be the case of any noninteracting transformation;~this is merely a second-quantized picture of this fact. Furthermore, we can see that the coefficients in the linear combination can be efficiently computed via determinants, a signature of fermions that we see again later.

To see that the homomorphism property persists for these higher-order operators, we use the Cauchy--Binet formula which states that the determinant of a product of matrices $uv$ is the sum of products of subdeterminants of $u$ and $v$ individually:
\begin{equation}
	\det([uv]_{\bm{q}, \bm{p}}) = \sum_{\bm{r} \in \binom{[n]}{k}} \det(u_{\bm{q}, \bm{r}}) \det(v_{\bm{r}, \bm{p}}).
\end{equation}
These are the coefficients in the expansion of $\mathcal{U}(\mathcal{V}(a_{p_1}^\dagger \cdots a_{p_k}^\dagger))$.

\subsection{Noninteracting systems and Slater determinants}

Recall that any one-body (noninteracting) Hamiltonian can be diagonalized as
\begin{equation}
	\begin{split}
	H &= \sum_{p, q \in [n]} h_{pq} a_p^\dagger a_q\\
	&= V \l( \sum_{p \in [n]} \varepsilon_p n_p \r) V^\dagger.
	\end{split}
\end{equation}
The eigenvalues $\varepsilon_p$ of $h$ are called the single-particle energies. The occupation-number operators $n_p$ have spectrum $\{0, 1\}$, corresponding to whether that mode is occupied or not. Because they commute, the spectrum of $H$ is
\begin{equation}\label{eq:noninteracting_spectrum}
	\spec(H) = \l\{ \sum_{p \in [n]} b_p \varepsilon_p~\middle\vert~b \in \{0, 1\}^n \r\}.
\end{equation}
This is an example of a \emph{free-fermion} spectrum, because the total energy of a given eigenstate is simply the sum of independent energies $\varepsilon_p$ from each occupied level $p \in [n]$. Indeed, each $b \in \{0, 1\}^n$ maps to an eigenstate of $H$ in terms of the Fock basis as $\ket{\psi_b} \coloneqq V \ket{b}$.

Noninteracting-fermion states with a fixed number particles are called \emph{Slater determinants}, and they always take the form of $\ket{\psi_b}$ for some $V \in \U(n)$ and $b \in \{0, 1\}^n$. Let $\ket{b}$ have $|b| = \eta$ particles occupying the modes $q_1 < \cdots < q_\eta$. We can expand $\ket{\psi_b}$ as
\begin{equation}
	\begin{split}
	\ket{\psi_b} &= V a_{q_1}^\dagger \cdots a_{q_\eta}^\dagger \ket{0^n}\\
	&= V a_{q_1}^\dagger \cdots a_{q_\eta}^\dagger V^\dagger V \ket{0^n}\\
	&= \sum_{p_1, \ldots, p_\eta \in [n]} v_{p_1,q_1} \cdots v_{p_\eta,q_\eta} a_{p_1}^\dagger \cdots a_{p_\eta}^\dagger V^\dagger \ket{0^n}.
	\end{split}
\end{equation}
To simplify this expression, we make two observations. First, because $V$ (and its inverse $V^\dagger$) preserves particle number, it acts as a one-dimensional representation on the vacuum. In fact, it can checked that under our convention the global phase is trivial:~$V\ket{0^n} = V^\dagger \ket{0^n} = \ket{0^n}$. Second, by the anticommutation relations we know that swapping the order of any $a_{p_i}^\dagger$ and $a_{p_j}^\dagger$ incurs a minus sign, while $(a_p^\dagger)^2 = 0$. Thus the sum over $p_1, \ldots, p_\eta \in [n]$ only has $\binom{n}{\eta}$ unique terms, say $p_1 < \cdots < p_\eta$, by recognizing repeated operators in the expansion which are related by permutations of those $\eta$ indices. Note that the parity of that permutation potentially incurs a sign in order to sort the creation operators into the same convention. This is coincides precisely with the definition of a determinant, whence their nomenclature:
\begin{equation}\label{eq:slater_derive}
	\begin{split}
	\ket{\psi_b} &= \sum_{\bm{p} \in \binom{[n]}{\eta}} \sum_{\pi \in \Sym(\eta)} (-1)^\pi v_{p_{\pi(1)}, q_1} \cdots v_{p_{\pi(\eta)}, q_\eta} a_{p_1}^\dagger \cdots a_{p_\eta}^\dagger \ket{0^n}\\
	&\simeq \sum_{\bm{p} \subseteq \binom{[n]}{\eta}} \det[v_{\bm{p}, \bm{q}}] \ket{\bm{p}}.
	\end{split}
\end{equation}
Above, we have used the first-quantized representation $\ket{\bm{p}}$ for notational simplicity. As before, the notation $v_{\bm{p}, \bm{q}}$ denotes the $\eta \times \eta$ submatrix of $v$ formed by extracting its rows and columns indexed by $\bm{p}$ and $\bm{q}$, respectively.

Thus we see that a Slater determinant of $\eta$ particles is specified by the columns labeled by $q_1, \ldots, q_\eta$ of an $n \times n$ unitary matrix. From here on, we will always assume that $\bm{q} = [\eta]$, which can be achieved by simply reordering the relevant columns of $v$ to the first $\eta$ spots. The set of all such $n \times \eta$ matrices is called a complex Stiefel manifold, each of which specifies an $\eta$-dimensional subspace of $\C^n$ (recall that the columns are orthonormal). Identifying equivalent subspaces (i.e., two Stiefel elements which are connected by a $\U(\eta)$ transformation) yields a so-called Grassmann manifold, which is equivalent to all $\eta$-particle Slater determinants over $n$ modes (the $\U(\eta)$ freedom manifests as a mere global phase). Hence Slater determinants have sometimes been referred to as subspace states in contexts outside of fermionic physics~\cite{kerenidis2022quantum}.

We have just seen that the amplitudes of $\ket{\psi_b}$ in the basis of $\ket{\bm{p}}$ are given by determinants, $\ip{\bm{p}}{\psi_b} = \det[v_{\bm{p}, [\eta]}]$. Indeed, this was recognized by Slater in 1929~\cite{slater1929theory},\footnote{Heisenberg~\cite{heisenberg1926mehrkorperproblem} and Dirac~\cite{dirac1926theory} also independently formulated the idea three years prior.} who developed these states by enforcing antisymmetry by fiat on simple product states. In contexts such as quantum chemistry, Slater determinants are often described in more traditional language as an $\eta$-body wavefunction
\begin{equation}
	\psi(\bm{x}_1, \ldots, \bm{x}_\eta) = \frac{1}{\sqrt{\eta!}} \det \begin{pmatrix}
	\chi_1(\bm{x}_1) & \cdots & \chi_\eta(\bm{x}_1)\\
	\vdots & \ddots & \vdots\\
	\chi_1(\bm{x}_\eta) & \cdots & \chi_\eta(\bm{x}_\eta)
	\end{pmatrix},
\end{equation}
where $\chi_i$ are single-particle wavefunctions (cf.~the columns of $v$) and $\bm{x}_j$ are the coordinates of the $j$th electron (cf.~the modes $\ket{p}$). Note the normalization factor of $(\eta!)^{-1/2}$ appearing due to the ``basis'' coordinates $(\bm{x}_1, \ldots, \bm{x}_\eta)$ not being antisymmetrized, whereas the basis $\ket{\bm{p}}$ in the formulation that we presented are already antisymmetrized. From a dynamical perspective, Slater determinants are all states which can be reached from a simple initial product state by means of a one-body rotation. This is what we showed in Section~\ref{sec:basis_rotation_bkgd}, and first proven (by a different argument) in 1960 by Thouless~\cite{thouless1960stability}.

\section{Reduced density matrices}

As uncorrelated states, Slater determinants have significantly fewer independent parameters than arbitrary $\eta$-particle states. We have already seen one realization of this, through their description by an $n \times \eta$ complex matrix with orthonormal columns. An equivalent representation (which removes the $\U(\eta)$ freedom) is the one-body \emph{reduced density matrix} (1-RDM) ${}^1 D$ of a Slater determinant $\ket{\psi}$.

This representation is equivalent because Slater determinants are noninteracting, so their one-body information is sufficient to uniquely specify the state. In general, the higher-order information stored in the $k$-RDMs ($1 \leq k \leq \eta$) of a correlated many-body quantum system is required. The $k$-RDM, introduced in 1940 by Husimi\footnote{Whose name would more commonly be romanized as ``Fushimi'' today.}~\cite{husimi1940some}, is the mixed state obtained by tracing out all but a $k$-particle subsystem. This section is devoted to understanding the important properties of this object.

We first aim to unify the presentation of this object through two different but equivalent formulations:~the partial trace ${}^1 D = \eta \tr_{2, \ldots, \eta} \op{\psi}{\psi}$ in first quantization, and the matrix of expectation values ${}^1 D_{pq} = \ev{\psi}{a_q^\dagger a_p}{\psi}$ in second quantization. For Slater determinants there is another convenient description of the 1-RDM as ${}^1 D = A A^\dagger$, where $A \coloneqq v_{[n], [\eta]}$ is the first $\eta$ columns of the unitary $v \in \U(n)$ that specifies the Slater determinant. Then we will see how the $k$-RDM arises from the 1-RDM of these special, noninteracting states.

\subsection{RDMs in first and second quantization}

Arguably, the first-quantized definition of RDMs is the fundamental one. In this section, we will consider $k$-RDMs, $1 \leq k \leq \eta$, for \textit{arbitrary} antisymmetric states $\rho$ (i.e., not just Slater determinants):
\begin{equation}
	{}^k D \coloneqq \binom{\eta}{k} \tr_{k+1, \ldots, \eta}(\rho).
\end{equation}
Due to (anti)symmetry, it does not matter which of the $\eta - k$ particles we trace out---the $k$-particle reduced states are all the same. For consistency with the literature, e.g., in chemistry, we have chosen to define ${}^k D$ as the sum over all $k$-particle subsystems,
\begin{equation}
	{}^k D = \sum_{\bm{i} \in \binom{[\eta]}{k}} \tr_{[\eta] \setminus \bm{i}}(\rho) = \binom{\eta}{k} \tr_{k+1, \ldots, \eta}(\rho),
\end{equation}
hence the normalization of $\tr({}^k D) = \binom{\eta}{k}$.\footnote{Frustratingly, the conventions are further split between summing only over unique $k$-particle subsystems, or including a sum over all $k!$ orderings, in which case the trace is $\eta!/(\eta - k)!$. The latter convention arises from viewing ${}^k D$ as an $n \times \cdots \times n$ tensor of rank $2k$.} Recall that the partial trace is defined as
\begin{equation}
	\tr_{k+1, \ldots, \eta}(\rho) = \sum_{r_{k+1}, \ldots, r_\eta \in [n]} (\I_{n}^{\otimes k} \otimes \bra{r_{k+1} \cdots r_\eta}) \rho (\I_{n}^{\otimes k} \otimes \ket{r_{k+1} \cdots r_\eta}).
\end{equation}
We thus seek an expression for the matrix element
\begin{equation}
	[\tr_{k+1, \ldots, \eta}(\rho)]^{p_1 \cdots p_k}_{q_1 \cdots q_k} = \ev{p_1 \cdots p_k}{\tr_{k+1, \ldots, \eta}(\rho)}{q_1 \cdots q_k},
\end{equation}
where without loss of generality we only consider $p_1 < \cdots < p_k$ and $q_1 < \cdots < q_k$ (due to antisymmetry, swapping the order of any two $p_i, p_j$ or $q_i, q_j$ indices merely incurs a minus sign, so those matrix elements are not unique).

To do so, let us introduce the antisymmetrizer
\begin{equation}
	\mathcal{A}_\eta \coloneqq \frac{1}{\eta!} \sum_{\pi \in \Sym(\eta)} (-1)^\pi S_\pi,
\end{equation}
where $S_\pi$ permutes the $\eta$ particles as $S_\pi \ket{i_1 \cdots i_\eta} = \ket{i_{\pi^{-1}(1)} \cdots i_{\pi^{-1}(\eta)}}$. By construction, $\mathcal{A}_\eta$ is an orthogonal projector onto $\wedge^\eta \mathcal{H}_n$, so $\mathcal{A}_\eta^2 = \mathcal{A}_\eta = \mathcal{A}_\eta^\dagger$. More generally, we can define an antisymmetrizer which acts nontrivially only some subset of $\eta' \leq \eta$ particles. We will particularly make use of $\mathcal{A}_{\eta - k}$, which is defined to antisymmetrize the last $\eta - k$ particles and act trivially on the first $k$ particles. Notably, this partial antisymmetrizer obeys
\begin{equation}
	\mathcal{A}_\eta \mathcal{A}_{\eta - k} = \mathcal{A}_{\eta - k} \mathcal{A}_\eta = \mathcal{A}_\eta,
\end{equation}
which can be seen by the fact that $\Sym(\eta - k)$ is a subgroup of $\Sym(\eta)$. Because $\rho$ is already antisymmetrized, we use the fact that $\rho = \mathcal{A}_{\eta - k}^\dagger \rho \mathcal{A}_{\eta - k}$ to write
\begin{equation}
	\begin{split}
	[\tr_{k+1, \ldots, \eta}(\rho)]^{p_1 \cdots p_k}_{q_1 \cdots q_k} &= \sum_{r_{k+1}, \ldots, r_\eta \in [n]} \bra{p_1 \cdots p_k \, r_{k+1} \cdots r_\eta} \mathcal{A}_{\eta - k} \rho \mathcal{A}_{\eta - k} \ket{q_1 \cdots q_k \, r_{k+1} \cdots r_\eta}\\
	&= \sum_{\bm{r} \in \binom{[n]}{\eta - k}} (\bra{p_1 \cdots p_k} \bra{\bm{r}}) \rho (\ket{q_1 \cdots q_k} \ket{\bm{r}}).
	\end{split}
\end{equation}
Then we further use $\rho = \mathcal{A}_\eta^\dagger \rho \mathcal{A}_\eta$ and sum over all $k$-particle subsystems to find
\begin{equation}\label{eq:kRDM_from_FQ}
	\begin{split}
	{}^k D^{p_1 \cdots p_k}_{q_1 \cdots q_k} &= \sum_{\bm{i} \in \binom{[\eta]}{k}} [\tr_{[\eta] \setminus \bm{i}}(\rho)]^{p_{1} \cdots p_{k}}_{q_{1} \cdots q_{k}}\\
	&= \sum_{\bm{r} \in \binom{[n]}{\eta - k}} \sum_{\bm{i} \in \binom{[\eta]}{k}} (\bra{p_{i_1} \cdots p_{i_k}} \bra{\bm{r}}) \mathcal{A}_\eta^\dagger \rho \mathcal{A}_\eta (\ket{q_{i_1} \cdots q_{i_k}} \ket{\bm{r}})\\
	&= \sum_{\substack{\bm{r} \in \binom{[n]}{\eta - k} \\ \bm{p}, \bm{q} \not\subseteq \bm{r}}} \ev{\bm{p} \cup \bm{r}}{\rho}{\bm{q} \cup \bm{r}} (-1)^{\mathrm{ord}(\bm{p}, \bm{r})} (-1)^{\mathrm{ord}(\bm{q}, \bm{r})},
	\end{split}
\end{equation}
where $\mathrm{ord}(\bm{x}, \bm{r})$ is the permutation that reorders $\bm{x}$ such that $\bm{x} \cup \bm{r}$ is brought into ascending order ($\bm{r}$ being kept fixed). 

Now we show that \cref{eq:kRDM_from_FQ} is equivalent to the expectation value of the $2k$-product of ladder operators, $\langle a_{p_1}^\dagger \cdots a_{p_k}^\dagger a_{q_k} \cdots a_{q_1}, \rho \rangle = \tr(a_{q_1}^\dagger \cdots a_{q_k}^\dagger a_{p_k} \cdots a_{p_1} \rho)$. Indeed, using the fact that the product inside the trace is an operator of $\eta - k$ particles, we can evaluate the trace by considering only the $(\eta - k)$-particle subspace. Using the definition of how creation operators act on occupation-number basis states, we find that
\begin{equation}
	\begin{split}
	\tr(a_{q_1}^\dagger \cdots a_{q_k}^\dagger a_{p_k} \cdots a_{p_1} \rho) &= \sum_{\bm{r} \in \binom{[n]}{\eta - k}} \ev{\bm{r}}{a_{p_k} \cdots a_{p_1} \rho a_{q_1}^\dagger \cdots a_{q_k}^\dagger}{\bm{r}}\\
	&= \sum_{\substack{\bm{r} \in \binom{[n]}{\eta - k} \\ \bm{p}, \bm{q} \not\subseteq \bm{r}}} \ev{\bm{p} \cup \bm{r}}{\rho}{\bm{q} \cup \bm{r}} (-1)^{\mathrm{ord}(\bm{p}, \bm{r})} (-1)^{\mathrm{ord}(\bm{q}, \bm{r})}.
	\end{split}
\end{equation}
For the 1-RDM, this reduces to $\tr(a_q^\dagger a_p \rho) = {}^1 D_q^p \equiv {}^1 D_{pq}$.

\subsection{RDMs of Slater determinants and Wick's theorem}

Here we return to the case where $\rho = \op{\psi}{\psi}$ is a Slater determinant and show how ${}^1 D$ is related to $V$. Using the fact that $\ket{\psi} = V\ket{[\eta]}$, we have
\begin{equation}
	\begin{split}
	{}^1 D_{pq} &= \ev{\psi}{a_q^\dagger a_p}{\psi}\\
	&= \ev{[\eta]}{V^\dagger a_q^\dagger a_p V}{[\eta]}\\
	&= \ev{[\eta]}{\sum_{q',p' \in [n]} v_{qq'}^* v_{pp'} a_{q'}^\dagger a_{p'}}{[\eta]}\\
	&= \sum_{p',q' \in [n]} v_{pp'} [\Pi_\eta]_{p'q'} [v^\dagger]_{q'q}\\
	&= [v \Pi_\eta v^\dagger]_{pq},
	\end{split}
\end{equation}
where we define $\Pi_\eta \coloneqq \diag(\underbrace{1, \ldots, 1}_{\eta}, 0, \ldots, 0)$ as the projector onto the first $\eta$ rows. Note that $\Pi_\eta$ is the 1-RDM of $\ket{[\eta]}$. Since $v \Pi_\eta v^\dagger = AA^\dagger$, where $A \in \C^{n \times \eta}$ is formed from the first $\eta$ columns of $v$, this implies that ${}^1 D = AA^\dagger$ for Slater determinants. Furthermore, observe that there is a $\U(\eta)$ gauge freedom in the definition of $A$, as $A' = AQ$ for any $Q \in \U(\eta)$ yields the same 1-RDM, $A'A^{\prime\dagger} = AA^\dagger$.

Mixed states of Slater determinants are convex combinations $\rho = \sum_{j} c_j \op{\psi_j}{\psi_j}$, hence by linearity their 1-RDMs are simply the same convex combination of 1-RDMs of each $\ket{\psi_j}$. Note that if the Slater determinants in this mixture have different particle numbers $\eta_j$, then the 1-RDM of $\rho$ has trace $\sum_j c_j \eta_j$, so it may not diagonalize into the form of an orthogonal projector. Nonetheless, by the Pauli exclusion principle (which applies to mixed states equally well) we can constrain the eigenvalues of any 1-RDM as $0 \preceq {}^1 D \preceq \I_n$. This is necessary and sufficient for ${}^1 D$ to be a valid 1-RDM of some antisymmetric, possibly mixed state (not just Slater determinants), and if $\tr({}^1 D) = \eta$ then the underlying state is guaranteed to contain $\eta$ particles~\cite{coleman1963structure}.\footnote{The necessary and sufficient conditions on the 1-RDM of a \emph{pure} state are significantly more complicated~\cite{klyachko2006quantum,altunbulak2008pauli}.}

For Slater determinants, all higher-order $k$-RDMs can be determined from the 1-RDM alone. Since the $\eta$-RDM is equivalent to the global state itself, this implies that the 1-RDM contains all the information about such states. This is a consequence of Wick's theorem~\cite{wick1950evaluation,bach1994generalized}, which implies that for Slater determinants $\ket{\psi}$ with 1-RDM ${}^1 D$,
\begin{equation}\label{eq:wicks_slater}
	\ev{\psi}{a_{q_1}^\dagger \cdots a_{q_k}^\dagger a_{p_k} \cdots a_{p_1}}{\psi} = \det[ {}^1 D_{\bm{p}, \bm{q}} ],
\end{equation}
where recall that ${}^1 D_{\bm{p}, \bm{q}}$ is the $k \times k$ submatrix of ${}^1 D$ corresponding to the rows and columns labeled by $\bm{p}$ and $\bm{q}$ respectively. For example, 2-RDM elements of Slater determinants obey
\begin{equation}\label{eq:wicks_thm_2rdm}
	{}^2 D^{sr}_{pq} = {}^1 D^s_p \, {}^1 D^r_q - {}^1 D^r_p \, {}^1 D^s_q.
\end{equation}
More generally, a mixed state $\rho$ of Slater determinants obeys this identity as well, wherein we make the replacement $\ev{\psi}{a_q^\dagger a_p}{\psi} \to \tr(a_q^\dagger a_p \rho)$.

\cref{eq:wicks_slater} implies that the higher-order moments of the state are completely determined by its first ($\ev{\psi}{a_p}{\psi} = 0$) and second ($\ev{\psi}{a_p^\dagger a_q}{\psi} = {}^1 D_p^q, \ev{\psi}{a_p^\dagger a_q^\dagger}{\psi} =  0$) moments. In analogy with how classical Gaussian probability distributions are completely determined by their first (mean) and second (variance) moments, Slater determinants are an example of so-called \emph{fermionic Gaussian states}. In Section~\ref{sec:gaussian_fermions_bkgd}, we will expand on this notion of fermionic Gaussianity and extend it to its natural conclusion, exhibiting a rather broad class of systems which fall under this category.

\subsection{Calculating properties with $k$-RDMs}

Here we show explicitly how RDMs are used to compute local observables. That is, any $k$-body observable can be calculated from knowledge of the $k$-RDM~\cite{coleman1980reduced}. Recall from \cref{eq:noninteracting_hamiltonian} that a one-body Hamiltonian ${}^1 h \in \C^{n \times n}$ lifts to the Fock space as $H = \sum_{p,q \in [n]} {}^1 h_{pq} a_p^\dagger a_q$. For a fermionic state $\rho$ with 1-RDM ${}^1 D$, its energy is
\begin{equation}
	\begin{split}
	\tr(H \rho) &= \sum_{p,q \in [n]} {}^1 h_{pq} \tr(a_p^\dagger a_q \rho)\\
	&= \sum_{p,q \in [n]} {}^1 h_{pq} \, {}^1 D _{qp}\\
	&= \tr({}^1 h \, {}^1 D),
	\end{split}
\end{equation}
so we only require computing a matrix product in the $n$-dimensional space. Indeed, this is true for any one-body observable.

More generally, a $k$-body observable ${}^k h \in \C^{n^k \times n^k}$ lifts to the Fock space as
\begin{equation}\label{eq:k-body_fermion_op}
	O = \sum_{\bm{p}, \bm{q} \in \binom{[n]}{k}} {}^k h_{q_1 \cdots q_k}^{p_1 \cdots p_k} a_{p_1}^\dagger \cdots a_{p_k}^\dagger a_{q_k} \cdots a_{q_1},
\end{equation}
where ${}^k h_{q_1 \cdots q_k}^{p_1 \cdots p_k} \coloneqq \ev{p_1 \cdots p_k}{h}{q_1 \cdots q_k}$.\footnote{Note that the literature often writes these objects with repeated indices, rather than our unique indexing by $\bm{p} \in \binom{[n]}{k}$, hence we do not have the factors of $1/k!$ that typically appear there.} Just as with the one-body observables, we can compute the expectation value of this $k$-body observable as
\begin{equation}\label{eq:klocal_fermion_value}
	\begin{split}
	\tr(O \rho) &= \sum_{\bm{p}, \bm{q} \in \binom{[n]}{k}} {}^k h_{q_1 \cdots q_k}^{p_1 \cdots p_k} \, {}^k D_{p_1 \cdots p_k}^{q_1 \cdots q_k}\\
	&= \tr({}^k h \, {}^k D).
	\end{split}
\end{equation}
As long as $k = \O(1)$, i.e., the observable has bounded locality, the complexity of this calculation is polynomial in the system size.

It is worth pointing out that this is completely analogous to the unsymmetrized case. Let $(\C^{n})^{\otimes \eta}$ be the Hilbert space of $\eta$ distinguishable $n$-level particles. Let $O_I$ be a $k$-local operator on subsystem $I \in \binom{[\eta]}{k}$, i.e., $O_I = h_I \bigotimes_{i \notin I} \I_n$ for Hermitian $h_I \in \C^{n^k \times n^k}$. Then the expectation value of $O_I$ with respect to a state $\rho$ is
\begin{equation}
	\tr(O_I \rho) = \tr(h_I \rho_I),
\end{equation}
where $\rho_I \coloneqq \tr_{[\eta] \setminus I} \rho$ is the marginal state on subsystem $I$. More generally, any $k$-body observable can be written as
\begin{equation}
	O = \sum_{I \in \binom{[\eta]}{k}} O_I,
\end{equation}
whose expectation value is
\begin{equation}\label{eq:klocal_qudit_value}
	\tr(O \rho) = \sum_{I \in \binom{[\eta]}{k}} \tr(h_I \rho_I).
\end{equation}
We can recover \cref{eq:klocal_fermion_value} by imposing (anti)symmetry on the system. Suppose all $h_I = {}^k h$ for some fixed $k$-body observable ${}^k h$, and furthermore the state is symmetric about all $k$-body marginals, $\rho_I = {}^k \rho$. Then clearly \cref{eq:klocal_qudit_value} reduces to
\begin{equation}
	\tr(O \rho) = \binom{\eta}{k} \tr({}^k h \, {}^k \rho),
\end{equation}
where we note that the $\binom{\eta}{k}$ factor above was absorbed into the normalizaion of ${}^k D$.

The use of second-quantization is therefore, in some sense, a double-edged sword. It very concisely wraps up the many-body information along with the desired symmetry properties. At the same time, this highly compact representation can be opaque when relating back to the physical picture of multiple particles in a composite tensor-product space.

\subsection{Representability}

Let us make a brief comment about the complexity of using RDMs to solve quantum many-body problems. It appears enticing to say that, because any $\O(1)$-local observable can be computed in time $n^{\O(1)}$ using RDMs, they might provide an avenue to construct efficient algorithms for calculating ground-state energies. Indeed, this idea was explored as early as the 1950's, when Coleman worked on a reduction of the many-electron problem to a mere two-electron problem. However, as he reminisces in \cite{coleman2007representability},
\begin{quote}
I did \emph{too well}, obtaining a level about 10\% BELOW the observed ground-state energy!

Impossible!
\end{quote}
The problem was recognized soon afterward:~there are a number of important constraints required of ${}^2 D$, beyond positivity and fixed trace, for it to correspond to any global $N$-electron wavefunction.\footnote{In this subsection only, we make the notational change $\eta \to N$ for the number of electrons, because of the well-established nomenclature ``$N$-representability.''} Without these additional constraints, the calculated 2-RDM is unphysical, therefore violating the variational principle of quantum mechanics. Thus birthed the \emph{$N$-representability problem} in quantum chemistry~\cite{coleman1963structure}, which simply asks:~what are the necessary and sufficient conditions on ${}^2 D$ such that there exists an $N$-particle antisymmetric state $\rho$ for which ${}^2 D = \tr_{3, \ldots, N}(\rho)$?\footnote{Pure-state $N$-representability further requires that $\tr(\rho^2) = 1$.}

At a very high level, the only missing property of $\rho$ from the naive description of ${}^2 D$ is the positivity of the \emph{global} state, $\rho \succeq 0$. Alas, this property is incredibly difficult to characterize at with only a two-body description. A large body of work has been devoted to understanding the problem, from the perspectives of chemistry, physics, and mathematics~\cite{mazziotti2012two}. The tools of complexity theory have been used to formalize the computational hardness of the problem, showing that $N$-representability is $\mathsf{QMA}$-complete~\cite{liu2007quantum,liu2007complexity}.\footnote{As is its unsymmetrized version, the quantum marginal-consistency problem~\cite{liu2006consistency}.} This implies that, even with a large, perfect quantum computer, the problem will remain intractable in general.

The reason for this complexity is intuitive to understand if we start from the celebrated result that determining ground-state energies of local Hamiltonians is $\mathsf{QMA}$-complete~\cite{kempe2006complexity}. Intuitively, if one had access to an oracle which can determine whether or not a given matrix is a $N$-representable, then they could use a polynomial-sized semidefinite program (SDP) to variationally optimize a trial 2-RDM. Guided in the ``$N$-representable directions'' by this oracle, \cite{liu2007quantum,liu2007complexity} showed that this SDP converges (in polynomial time) to the ground-state 2-RDM. But then we have solved the ground-state problem efficiently, which should not be possible because of its hardness. Therefore $N$-representability is at least as hard as the ground-state problem itself. $\mathsf{QMA}$-completeness follows from the fact that $N$-representability is also in $\mathsf{QMA}$:~if ${}^2 D$ is indeed $N$-representable, then we can verify this fact by simply measuring the 2-RDM of its parent $N$-electron state. This only requires measuring a polynomial number of expectation values $\langle a_p^\dagger a_q^\dagger a_s a_r \rangle$ up to inverse-polynomial precision, all of which can be done in polynomial time with a polynomial number of copies of the parent quantum state. Otherwise, if ${}^2 D$ is not representable, it is unphysical and so there does not exist any $N$-electron state whose measured 2-RDM will be consistent with ${}^2 D$.\footnote{One must also show that the prover (Merlin) cannot use entangled copies of the wrong quantum state to trick the verifier (Arthur) in this non-representable instance.}

Despite this ultimate complexity, chemists and mathematicians continue to chip away at the challenges imposed by $N$-representability. For chemists, it provides a practical algorithm to approximate the ground-state 2-RDM through variational minimization, using only a subset of the $N$-representability conditions~\cite{garrod1964reduction,garrod1975variational}. One hopes that the conditions employed are the most important in some sense, although in general this method is always bounded from below. For mathematicians and theorists, the $N$-representability problem is one of spectral analysis, representation theory, and convex geometry, and therefore is an enticingly rich but difficult subject to tackle~\cite{kummer1967n,borland1972conditions,coleman1972necessary,klyachko2006quantum,altunbulak2008pauli}.

\section{Canonical example of interacting fermions:~electronic structure}

As an example of a two-body Hamiltonian, ubiquitous within quantum chemistry~\cite{szabo1989modern,helgaker2000molecular}, let us consider the electronic-structure problem. The model is usually framed within the Born--Oppenheimer approximation~\cite{bornoppenheimer}, wherein a molecule of $L$ nuclei and $\eta$ electrons is treated semiclassically. Because the nuclei are large and heavy relative to the electrons, they are treated as classical point particles emitting a background electrostatic potential for the electrons to move through. Thus, the electron--nuclei interactions are semiclassical and easy to handle. The challenge, then, lies in simulating the electron--electron interactions, which are treated fully quantumly. The electronic-structure problem is to resolve these electron--electron interactions, usually by finding the ground state of the electrons in some fixed geometry of the nuclei. Varying this molecular geometry as a classical parameter generates different electronic structures, and solving them enables access to the study of phenomena such as bond dissociation and chemical reactions at a quantum level~\cite{szabo1989modern,helgaker2000molecular}.

The electronic-structure Hamiltonian is constructed from \emph{ab initio} principles, wherein the potential is nothing more than the Coulomb interaction between charged particles:
\begin{equation}
\begin{split}
	H &= -\sum_{i=1}^\eta \frac{\nabla_i^2}{2} - \sum_{i=1}^\eta \sum_{\ell=1}^L \frac{\zeta_\ell}{\| R_\ell - r_i \|} + \sum_{1 \leq i < j \leq \eta} \frac{1}{\|r_i - r_j\|}\\
	&\quad\, - \sum_{\ell=1}^L \frac{\nabla_\ell^2}{2 m_\ell} + \sum_{1 \leq \ell < k \leq L} \frac{\zeta_\ell \zeta_k}{\|R_\ell - R_k\|},
\end{split}
\end{equation}
where $\zeta_\ell$ and $m_\ell$ are the atomic charges and masses of the nuclei, respectively, and $r_i$ and $R_\ell$ are the position (operators) of the electrons and nuclei, respectively. Note that we are using atomic units, so $\hbar$ and the charge and mass of electrons are all unity. Because we take the nuclei to be classical, the last two terms reduce to a constant energy shift, so for simplicity we will drop them henceforth.

This first-quantized representation assumes that each electron's Hilbert space is $L^2(\R^3)$, i.e., infinite-dimensional. In order to study this model computationally, we discretize it using some finite basis set. Let $\{\ket{p} \mid p \in [n]\}$ be such an orthonormal basis, with associated square-integrable basis functions $\phi_p \in L^2(\R^3)$. The error of this discretization scales as $\O(1/n)$ for any choice of basis functions~\cite{babbush2018low}, so we can converge toward the continuum limit of the model by systematically growing the basis set. The discretized Hamiltonian (suppressing the purely nuclear contributions) is
\begin{equation}
	H = \sum_{i \in [\eta]} \sum_{p, q \in [n]} T_{pq} \op{p}{q}_i + \sum_{i \in [\eta]} \sum_{p, q} U_{pq} \op{p}{q}_i + \frac{1}{2} \sum_{i \neq j \in [\eta]} \sum_{p,q,r,s \in [n]} V_{pqrs} \op{p}{s}_i \op{q}{r}_j,
\end{equation}
where the notation $\op{p}{q}_i$ indicates $\op{p}{q}$ acting on the $i$th electron and $\I_n$ on the remaining electrons, and the coefficients $T_{pq}, U_{pq}, V_{pqrs}$ are the following integrals involving the electronic kinetic energy, electron--nuclei potential, and electron--electron interaction, respectively:
\begin{align}
	T_{pq} &= \ev{p}{\l(-\frac{\nabla^2}{2} \r)}{q} \label{eq:T_int}\\
	&= \int dr \, \phi_p^*(r) \l( -\frac{\nabla^2}{2} \r) \phi_q(r), \notag\\
	U_{pq} &= \ev{p}{\sum_{\ell \in [L]} \l( -\frac{\zeta_\ell}{\| R_\ell - r \|} \r)}{q} \label{eq:U_int}\\
	&= \sum_{\ell \in [L]} \int dr \, \phi_p^*(r) \l( -\frac{\zeta_\ell}{\| R_\ell - r \|} \r) \phi_q(r), \notag\\
	V_{pqrs} &= \ev{pq}{\l( \frac{1}{\| r_1 - r_2 \|} \r)}{sr} \label{eq:V_int}\\
	&= \int dr_1 \, dr_2 \, \phi_p^*(r_1) \phi_q^*(r_2) \l( \frac{1}{\| r_1 - r_2 \|} \r) \phi_r(r_2) \phi_s(r_1). \notag
\end{align}
Observe that the first two terms are both one-body operators, while the third term is two-body (interacting). Then following \cref{eq:k-body_fermion_op}, the second-quantized electronic-structure Hamiltonian is
\begin{equation}\label{eq:elec_struct_bkgd}
	H = \sum_{p,q \in [n]} h_{pq} a_p^\dagger a_q + \frac{1}{2} \sum_{p,q,r,s \in [n]} V_{pqrs} a_p^\dagger a_q^\dagger a_r a_s,
\end{equation}
where $h = T + U$. Note that the sums over $i \in [\eta]$ have disappeared, because lifting to the Fock space gives a representation of the Hamiltonian in \emph{every} particle-number sector. In other words, the information about $\eta$ is absent from this representation;~one must specify the desired electron number when working in second quantization, for example by restricting to Fock-basis states with the correct particle number.

\subsection{Computational complexity of electronic structure}

Although this Hamiltonian's ground-state energy is $\mathsf{QMA}$-hard to compute~\cite{ogorman2022intractability}, its central importance to quantum chemistry motivates the development of a wide range of both classical and quantum algorithms to study it~\cite{foulkes2001quantum,geerlings2003conceptual,mcardle2020quantum,bauer2020quantum}. One of the most important algorithms is the Hartree--Fock method, a mean-field approximation to the ground state. At its core, the goal of Hartree--Fock is to find the lowest-energy noninteracting state (Slater determinant). Hartree--Fock theory is central to quantum chemistry, as some formulation of it almost always serves as the starting point for a more sophisticated approximation by correlated states.

The Hartree--Fock problem can be expressed as follows. By Wick's theorem, \cref{eq:wicks_thm_2rdm}, we can write the energy of any $\eta$-electron Slater determinant $\ket{\psi}$ in terms of its $1$-RDM ${}^1 D_{qp} = \ev{\psi}{a_p^\dagger a_q}{\psi}$,
\begin{equation}\label{eq:HF_math_form}
\begin{split}
	\ev{\psi}{H}{\psi} &= \sum_{p,q \in [n]} h_{pq} \, {}^1 D_{qp} + \frac{1}{2} \sum_{p,q,r,s \in [n]} V_{pqrs} \l({}^1 D_{sp} \, {}^1 D_{rq} - {}^1 D_{rp} \, {}^1 D_{sq}\r)\\
	&= \sum_{p,q \in [n]} h_{pq} \, {}^1 D_{qp} + \frac{1}{2} \sum_{p,q,r,s \in [n]} \l(V_{pqrs} - V_{pqsr}\r) \, {}^1 D_{sp} \, {}^1 D_{rq},
\end{split}
\end{equation}
where the second equality is a mere index relabeling for clarity. The minimization problem is therefore a quadratic program over $n \times n$ complex matrices ${}^1 D = AA^\dagger$, where $A \in \C^{n \times \eta}$ has orthonormal columns. Alternatively, this constraint can be phrased as $0 \preceq {}^1 D \preceq \I_n$ having trace and rank $\eta$.

In practice, Hartree--Fock algorithms do not directly work with the form of \cref{eq:HF_math_form}, but rather leverage additional structure to heuristically solve the optimization problem. In part, this is because Hartree--Fock is computationally hard in the worst case:~it is $\mathsf{NP}$-hard. This was shown originally in \cite{schuch2009computational} for arbitrary coefficients $h_{pq}, V_{pqrs}$, wherein they reduced the ground-state problem of classical Ising spin glasses to \cref{eq:HF_math_form}. Later, \cite{ogorman2022intractability} showed that this hardness persists when coefficients are restricted to obey the form of the integrals as in \cref{eq:T_int,eq:U_int,eq:V_int}. Similar to the approach of \cite{schuch2009computational}, this was shown by a reduction from a classical Hamiltonian capable of encoding $\mathsf{NP}$-hard problems, for instance the independent-set problem.

\section{Gaussian fermions}\label{sec:gaussian_fermions_bkgd}

The class of fermionic Gaussian states can be defined as antisymmetric states obeying the Gaussian-distribution property:~its first and second moments completely characterize all other moments. As such, the systems they describe are integrable, allowing for their efficient classical simulation~\cite{terhal2002classical,bravyi2004lagrangian}. This turns out to be equivalent to a restricted class of quantum circuits in 1D known as matchgates~\cite{valiant2001quantum,knill2001fermionic,jozsa2008matchgates} under the Jordan--Wigner transformation. Gaussian fermions go by a variety of other names, such as free fermions,\footnote{The seminal paper on the subject~\cite{bach1994generalized} referred to this class as \emph{quasi}-free fermions;~here we do away with the qualifier.} generalized Hartree--Fock~\cite{bach1994generalized}, or fermionic linear optics, the latter being due to their mathematical similarity to bosonic linear optics. Slater determinants lie in the number-conserving manifold of fermionic Gaussian states, and in this section we will see how this generalization naturally arises.

\subsection{Quadratic Hamiltonians}

Let us start from the physical perspective. Define again the Majorana operators,
\begin{equation}\label{eq:majorana_def_gauss_section}
	\gamma_{2p-1} \coloneqq a_p + a_p^\dagger, \quad \gamma_{2p} \coloneqq -\i(a_p - a_p^\dagger),
\end{equation}
which express the $n$ creation and annihilation operators into $2n$ Hermitian operators. The anticommutation relations of the ladder operators translate into a single Clifford-algebraic identity on the Majorana operators,
\begin{equation}
	\{\gamma_\mu, \gamma_\nu\} = 2 \delta_{\mu\nu} \I_{2^n},
\end{equation}
for all $\mu, \nu \in [2n]$. Majorana operators can also be understood as real and imaginary operator components of $a_p, a_p^\dagger$. Because each Majorana operator is a sum of both a creation and an annihilation operator, they inherently do not preserve fermion number. Thus there is no $u \in \U(n)$ which transforms the ladder operators into Majorana operators. Instead, a broader class of transformations, called \emph{Bogoliubov transformations}, is required. Indeed, one can see that \cref{eq:majorana_def_gauss_section} can be rewritten as
\begin{equation}
	\frac{1}{\sqrt{2}} \begin{pmatrix}
	\gamma_{2p-1}\\
	\gamma_{2p}
	\end{pmatrix} = \frac{1}{\sqrt{2}} \begin{pmatrix}
	1 & 1\\
	\i & -\i
	\end{pmatrix}
	\begin{pmatrix}
	a_p^\dagger\\
	a_p
	\end{pmatrix},
\end{equation}
where the normalization of $1/\sqrt{2}$ makes the transformation unitary. Collecting the vectors of operators $\bm{\gamma} \coloneqq (\gamma_1, \ldots, \gamma_{2n})^\T$ and $\bm{a} \coloneqq (a_1, \ldots, a_n)^\T$, we have
\begin{equation}
	\frac{1}{\sqrt{2}} \bm{\gamma} = \Omega \begin{pmatrix}
	\bm{a}^\dagger\\
	\bm{a}
	\end{pmatrix},
\end{equation}
where
\begin{equation}
	\Omega \coloneqq \frac{1}{\sqrt{2}} \begin{pmatrix}
	\I_n & \I_n\\
	\i \I_n & -\i \I_n
	\end{pmatrix} \in \U(2n).
\end{equation}
Because particle-number symmetry is broken, the Fock-space representation is necessary here.

Consider a Hamiltonian which is quadratic in the creation and annihilation operators;~this generalizes beyond the previously defined noninteracting Hamiltonians by including non-number-preserving terms:\footnote{Such terms are sometimes referred to as ``superconducting terms,'' due to their historical connection to mean-field models of superconductivity (BCS theory)~\cite{bardeen1957theory}.}
\begin{equation}\label{eq:quadratic_fermionic_hamiltonian}
	\begin{split}
	H &= \sum_{p,q \in [n]} h_{pq} a_p^\dagger a_q + \frac{1}{2} \sum_{p,q \in [n]} \l( \Delta_{pq} a_p^\dagger a_q^\dagger - \Delta_{pq}^* a_p a_q \r)\\
	&= \frac{1}{2} \begin{pmatrix}
	(\bm{a}^\dagger)^\T & \bm{a}^\T
	\end{pmatrix}
	\begin{pmatrix}
	\Delta & h\\
	-h^* & -\Delta^*
	\end{pmatrix}
	\begin{pmatrix}
	\bm{a}^\dagger\\ \bm{a}
	\end{pmatrix} + \frac{1}{2} \tr(h) \I_{2^n}.
	\end{split}
\end{equation}
The matrices $h, \Delta \in \C^{n \times n}$ obey $h = h^\dagger$ and $\Delta = -\Delta^\T$ to ensure that $H$ is Hermitian. Denote the composite $2n \times 2n$ block matrix above as $M$. Using the inverse transformation, $\Omega^{-1} = \Omega^\dagger$, we can write $H$ in terms of quadratic Majorana operators as
\begin{equation}
	H = \frac{1}{4} \bm{\gamma}^\T \Omega^* M \Omega^\dagger \bm{\gamma}.
\end{equation}
Note that we use complex conjugtion on the left because $((\bm{a}^\dagger)^\T \, \bm{a}^\T) = \frac{1}{\sqrt{2}} (\Omega^\dagger \bm{\gamma})^\T = \frac{1}{\sqrt{2}} \bm{\gamma}^\T \Omega^*$.

Let us examine the properties of the matrix $\Omega^* M \Omega^\dagger$. Defining the elementwise real and imaginary components of a complex matrix $X$ as
\begin{equation}
	\Re X \coloneqq \frac{X + X^*}{2}, \quad \Im X \coloneqq \frac{X - X^*}{2\i},
\end{equation}
a straightforward calculation reveals
\begin{equation}
	\begin{split}
	\Omega^* M \Omega^\dagger &= \frac{1}{2} \begin{pmatrix}
	\I_n & \I_n\\
	-\i \I_n & \i \I_n
	\end{pmatrix}
	\begin{pmatrix}
	\Delta & h\\
	-h^* & -\Delta^*
	\end{pmatrix}
	\begin{pmatrix}
	\I_n & -\i \I_n\\
	\I_n & \i \I_n
	\end{pmatrix}\\
	&= \i \begin{pmatrix}
	\Im h + \Im \Delta & \Re h - \Re \Delta\\
	-\Re h - \Re \Delta & \Im h - \Im \Delta
	\end{pmatrix}.
	\end{split}
\end{equation}
Therefore the matrix $A \coloneqq -\i \Omega^* M \Omega^\dagger$ is purely real in this basis. In fact, $A$ is antisymmetric, which can be seen by applying the following identities,
\begin{align}
	(\Re h)^\T = \Re h, &\quad (\Im h)^\T = -\Im h\\
	(\Re \Delta)^\T = -\Re \Delta, &\quad (\Im \Delta)^\T = -\Im \Delta,
\end{align}
to obtain
\begin{equation}
	A^\T = \begin{pmatrix}
	-\Im h - \Im \Delta & -\Re h + \Re \Delta\\
	\Re h + \Re \Delta & -\Im h + \Im \Delta
	\end{pmatrix} = -A,
\end{equation}
as claimed. We can further see that all $2n \times 2n$ real antisymmetric matrices correspond to all quadratic Hamiltonians, \cref{eq:quadratic_fermionic_hamiltonian}, up to an arbitrary energy shift $\tr h/2 \in \R$. To see that this association is complete, consider a simple parameter-counting argument:~the space of $2n \times 2n$ real antisymmetric matrices has $n(2n - 1)$ real parameters. Meanwhile, the space of $n \times n$ complex Hermitian matrices $h = h^\dagger$ has $n(n-1) + n$ real parameters, and the space of $n \times n$ complex antisymmetric matrices $\Delta = -\Delta^\T$ has $n(n-1)$ real parameters, for a total of $n(2n - 1)$ real parameters.

Henceforth, we will write all quadratic Hamiltonians in the Majorana basis, and without loss of generality we set the energy shift such that $\tr H = 0$:
\begin{equation}
	H = -\frac{\i}{4} \sum_{\mu, \nu \in [2n]} A_{\mu\nu} \gamma_\mu \gamma_\nu = -\frac{\i}{4} \bm{\gamma}^\T A \bm{\gamma}
\end{equation}
for any $A = -A^\T \in \R^{2n \times 2n}$. Just as number-preserving one-body Hamiltonians could be diagonalized within the single-particle basis, we can diagonalize $H$ within the smaller $2n$-dimensional space. We use the fact that any antisymmetric matrix can be brought into canonical form
\begin{equation}
	A = Q \Lambda Q^\T,
\end{equation}
where
\begin{equation}
	\Lambda \coloneqq \bigoplus_{p=1}^n \begin{pmatrix}
	0 & \varepsilon_p\\
	-\varepsilon_p & 0
	\end{pmatrix}
\end{equation}
with $\varepsilon_p \in \R$ and $Q \in \Orth(2n)$. Note that the eigenvalues of $A$ are $\pm\i\varepsilon_p$. This leads to the expression
\begin{equation}
	\begin{split}
	H &= -\frac{\i}{4} \sum_{\mu,\nu \in [2n]} \sum_{\mu',\nu' \in [2n]} Q_{\mu\mu'} \Lambda_{\mu'\nu'} Q_{\nu\nu'} \gamma_{\mu} \gamma_{\nu}\\
	&= -\frac{\i}{4} \sum_{\mu',\nu' \in [2n]} \Lambda_{\mu'\nu'} \tilde{\gamma}_{\mu'} \tilde{\gamma}_{\nu'}\\
	&= -\frac{\i}{2} \sum_{p \in [n]} \varepsilon_p \tilde{\gamma}_{2p-1} \tilde{\gamma}_{2p},
	\end{split}
\end{equation}
where $\tilde{\gamma}_\mu \coloneqq \sum_{\nu \in [2n]} Q_{\nu\mu} \gamma_\nu$. This defines a basis rotation of the Majorana modes by a $2n \times 2n$ orthogonal matrix, in the same way that we saw the single-particle modes can be rotated by an $n \times n$ unitary matrix.

Indeed, the rotated Majorana operators satisfy the anticommutation relations for all $Q \in \Orth(2n)$:
\begin{equation}
	\begin{split}
	\{\tilde{\gamma}_\mu, \tilde{\gamma}_\nu\} &= \sum_{\mu',\nu' \in [2n]} Q_{\mu'\mu} Q_{\nu'\nu} \{\gamma_{\mu'}, \gamma_{\nu'}\}\\
	&= \sum_{\mu',\nu' \in [2n]} Q_{\mu'\mu} Q_{\nu'\nu} 2 \delta_{\mu'\nu'} \I_{2^n}\\
	&= [Q^\T Q]_{\mu\nu} 2 \I_{2^n} = 2 \delta_{\mu\nu} \I_{2^n}.
	\end{split}
\end{equation}
This classifies all possible unitary Bogoliubov transformations of fermions. As a consequence of preserving commutation relations, this implies that there exists some unitary $U_Q \in \U(2^n)$ such that $\tilde{\gamma}_\mu = U_Q \gamma_\mu U_Q^\dagger$. Hence
\begin{equation}
	H = U_Q \l( \frac{1}{2}  \sum_{p \in [n]} \varepsilon_p (-\i \gamma_{2p-1} \gamma_{2p} ) \r) U_Q^\dagger
\end{equation}
is the canonical diagonalization of $H$, as seen by the fact that $-i \gamma_{2p-1} \gamma_{2p} = \I_{2^n} - 2a_p^\dagger a_p$. Alternatively, we can appeal to the Jordan--Wigner transformation which reveals $-\i \gamma_{2p-1} \gamma_{2p} = Z_p$, where $Z_p$ is the Pauli-$Z$ matrix acting on the $p$th qubit and $\I_2$ elsewhere. The eigensystem of $H$ is therefore
\begin{equation}
	H \ket{\phi_b} = E_b \ket{\phi_b},
\end{equation}
where $\ket{\phi_b} = U_Q \ket{b}$ and $E_b = \sum_{p \in [n]} (-1)^{b_p} \varepsilon_p$ for each $b_1 \cdots b_n \in \{0,1\}^n$. Such a spectrum is called \emph{free} because it is a sum of $n$ independent energies:
\begin{equation}
	E = \pm \varepsilon_1 \pm \cdots \pm \varepsilon_n,
\end{equation}
where each sign corresponds to whether or not a particular mode (called the natural orbitals in the number-preserving setting) is occupied. Indeed, we saw this defining property in \cref{eq:noninteracting_spectrum}, wherein by convention we kept the global energy constant of $\tr h/2$. Thus while the eigenstates of Gaussian fermions may not have particle-number symmetry, their spectrum is still identical to that of a noninteracting system. That is, each eigenenergy is built up by simply placing either an excitation or hole into each eigenmode.

\subsubsection{Free-fermion solvability}

Beyond BCS theory~\cite{bardeen1957theory}, free-fermion (quadratic) Hamiltonians have become an indispensable tool for studying many-body physics. These are usually known as free-fermion solutions for non-fermionic models. For example, inverting the Jordan--Wigner transformation allows us to write Hamiltonians of spin-$1/2$ particles in terms of second-quantized fermions;~if that transformed Hamiltonian is quadratic in fermion operators, then the techniques discussed above can be used to solve the model, even if it involves interactions in the original description. The canonical example is the 1D transverse-field Ising chain~\cite{schultz1964two}. Even before that, however, the \emph{classical} Ising model on a 2D square lattice was shown to be solvable by the mathematics of free fermions (albeit before the connection to fermions was explicitly realized)~\cite{onsager1944crystal,kaufman1949crystal}. More involved quantum models include the antiferromagnetic XY Heisenberg and Heisenberg--Ising chains, wherein the free-fermion solvability enables a rigorous understanding of these systems in the thermodynamic limit ($n \to \infty$)~\cite{lieb1961two}. More recently, extending beyond the Jordan--Wigner transformation has resulted in a rich research program for identifying spin models that are free fermions ``in disguise''~\cite{fendley2014free,fendley2019free,chapman2020characterization,elman2021free,chapman2023unified}.

\subsection{Fermionic Gaussian unitaries}\label{subsec:FGU_bkgd}

We now turn to study the properties of the diagonalizing unitaries $U_Q$ which generate the free-fermion eigenstates $\ket{\phi_b}$. In analogy with the particle-conserving case, it will not be surprising to see that such unitaries are generated by a quadratic Hamiltonian themselves. We technically work backwards, by exponentiating the Hamiltonian $H$ and demonstrating its adjoint action on the Majorana operators. Parallel to \cref{prop:basis_rotation_proof}, we will demonstrate this through the perspective of a matrix decompositions, rather than appealing to the Heisenberg equation.

\begin{proposition}\label{prop:gaussian_adjoint_result}
	The adjoint action by $e^{-\i H}$ for any quadratic Hamiltonian $H = (-\i/4) \sum_{\mu,\nu \in [2n]} A_{\mu\nu} \gamma_{\mu} \gamma_{\nu}$ obeys
	\begin{equation}\label{eq:majorana_basis_rotation_adj}
	e^{\i H} \gamma_\mu e^{-\i H} = \sum_{\mu \in [2n]} R_{\mu\nu} \gamma_\nu,
	\end{equation}
	where $R \coloneqq e^{A} \in \SO(2n)$.
\end{proposition}

\begin{proof}
Using our diagonal form for $H$, we can write
\begin{equation}
	e^{-\i H} = U_Q \l( \bigotimes_{p \in [n]} e^{-\i \varepsilon_p Z / 2} \r) U_Q^\dagger.
\end{equation}
For ease of notation, we stack the Majorana operators into an operator vector $\bm{\gamma}$, and let the adjoint action by operators act elementwise, e.g., $U_Q \bm{\gamma} U_Q^\dagger = Q^\T \bm{\gamma}$. Then
\begin{equation}
	\begin{split}
	e^{\i H} \bm{\gamma} e^{-\i H} &= U_Q \l( \bigotimes_{p \in [n]} e^{\i \varepsilon_p Z / 2} \r) U_Q^\dagger \bm{\gamma} U_Q \l( \bigotimes_{p' \in [n]} e^{-\i \varepsilon_{p'} Z / 2} \r) U_Q^\dagger\\
	&= U_Q \l( \bigotimes_{p \in [n]} e^{\i \varepsilon_p Z / 2} \r) (Q\bm{\gamma}) \l( \bigotimes_{p' \in [n]} e^{-\i \varepsilon_{p'} Z / 2} \r) U_Q^\dagger.
	\end{split}
\end{equation}
Using the fact that
\begin{equation}
	\gamma_{\mu} = Z_1 \cdots Z_{\lceil \mu / 2 \rceil - 1} \begin{cases}
	X_p & \text{if } \mu = 2p-1,\\
	Y_p & \text{if } \mu = 2p,
	\end{cases}
\end{equation}
we see that the Pauli-$Z$ rotations only act nontrivially on sites $p = p'$ when $\mu \in \{2p-1, 2p\}$. The Pauli-rotation identity from \cref{eq:pauli_matrix_adjoint} thus yields
\begin{align}
	e^{\i \varepsilon_p Z_p / 2} \gamma_{\nu} e^{-\i \varepsilon_p Z_p / 2} &= \begin{cases}
	\cos(\varepsilon_p) \gamma_{2p-1} - \sin(\varepsilon_p) \gamma_{2p} & \text{if } \nu = 2p-1,\\
	\sin(\varepsilon_p) \gamma_{2p-1} + \cos(\varepsilon_p) \gamma_{2p} & \text{if } \nu = 2p,\\
	\gamma_\nu & \text{else}.
	\end{cases}
\end{align}
In matrix notation, we can define the block-diagonal matrix of Givens rotations,
\begin{equation}
	G \coloneqq \bigoplus_{p \in [n]} \begin{pmatrix}
	\cos\varepsilon_p & -\sin\varepsilon_p\\
	\sin\varepsilon_p & \cos\varepsilon_p
	\end{pmatrix},
\end{equation}
to write
\begin{equation}
	\begin{split}
	e^{\i H} \bm{\gamma} e^{-\i H} &= U_Q \l( \bigotimes_{p \in [n]} e^{\i \varepsilon_p Z / 2} \r) (Q\bm{\gamma}) \l( \bigotimes_{p' \in [n]} e^{-\i \varepsilon_{p'} Z / 2} \r) U_Q^\dagger\\
	&= U_Q (Q G \bm{\gamma}) U_Q^\dagger\\
	&= Q G Q^\T \bm{\gamma}.
	\end{split}
\end{equation}
Finally, we observe that $G = e^\Lambda$, so $Q G Q^\T = e^{Q \Lambda Q^\T} = e^A$. Thus $e^{-iH} \in \U(2^n)$ is the Fock-space representation of $e^A \in \SO(2n)$, in the sense that $e^{-\i H} = U_{e^A}$.
\end{proof}

The astute reader may recognize that this only covers the special orthogonal group. Indeed, because $\Orth(2n)$ has two connected components, the exponential map can only reach the component containing the identity. This is physically consequential:~it is a reflection of fermionic parity superselection rules~\cite{streater2000pct}. While particle-number symmetry may be violated, the parity (even or odd number of particles) cannot be changed. Indeed, one can check that $e^{-\i H}$ commutes with fermionic parity operator $P \coloneqq (-\i)^n \prod_{\mu=1}^{2n} \gamma_\mu \stackrel{\text{JW}}{=} Z_1 \cdots Z_n$. More generally, an operator commutes with $P$ if and only if its degree is even.

On the other hand, the transformation $U_Q$ when $\det Q = -1$ will flip particle-number parity, thereby violating parity superselection. Nonetheless, it is still a unitary transformation that can be implemented, for example when using qubits to simulate the fermions. It is also mathematically straightforward to reach the negative-determinant component:~any single reflection, say $R = \diag(1, \ldots, 1, -1)$, maps between the two connected components of $\Orth(2n)$. The transformation by $R$ on the Fock space is represented by $X_n$, which (under the Jordan--Wigner transformation) commutes with all Majorana operators except $\gamma_{2n}$, for which it acts as a reflection, $X_n^\dagger \gamma_{2n} X_n = -\gamma_{2n}$. Then $U_{e^A R} = U_{e^A} X_n$ allows us to capture the other component of $\Orth(2n)$, thereby extending \cref{prop:gaussian_adjoint_result} to all orthogonal matrices.

Observe that, as a fermionic operator, $X_n = (-\i)^{n-1} \gamma_{1} \cdots \gamma_{2n - 1}$ has odd degree, as expected. This broader class of non-parity-preserving unitaries are referred to as \emph{generalized matchgates}~\cite{helsen2022matchgate}, as they generalize the class of matchgate circuits introduced by Valiant~\cite{valiant2001quantum} which were demonstrated to be equivalent to $\SO(2n)$-Gaussian unitaries~\cite{knill2001fermionic,terhal2002classical,jozsa2008matchgates}. In Chapter~\ref{chap:EM}, \cref{sec:FGU_circuit_design}, we introduce an algorithm for compiling the quantum circuits for $U_Q$ for any $Q \in \Orth(2n)$ under the Jordan--Wigner transformation~\cite{zhao2023group}. Compared to the prior state of the art, our circuit design has substantially improved gate count and depth (1/2 and 1/3 reduction, respectively).

While odd-dimensional representations, $\SO(2n + 1)$ or $\Orth(2n + 1)$, are also possible, these merely reduce to the even-dimensional case with $2n + 2$. We can show this following the argument of \cite{knill2001fermionic} (considering only the $\SO(2n + 1)$ case for now). Let $\mathcal{L}_2^\circ = \spn\{\gamma_\mu, \gamma_p \gamma_q \mid \mu \in [2n], 1 \leq p < q \leq 2n\}$ be the (complex) linear span of one- and two-degree Majorana operators (excluding the zero-degree identity). This space has dimension $ 2n^2 + n = \binom{2n+1}{2}$, which we observe is the dimension of the complexified Lie algebra $\mathfrak{so}_{\C}(2n+1)$. Indeed, by the commutation relations of the Majorana operators, we can identify the two algebras. Thus the fermionic Gaussian representation of $\SO(2n + 1)$ is given by the invertible elements of $e^{\mathcal{L}_2^\circ}$, which is generated by Hamiltonians both quadratic \emph{and linear} in the Majorana operators.

To see how this reduces to the even-dimensional case (with purely quadratic Hamiltonians), we merely append a zeroth qubit and multiply each linear term in $\mathcal{L}_2^\circ$ by $\gamma_0 = X_0$. This makes all elements of the algebra strictly quadratic, hence a subalgebra of $\mathcal{L}_2' = \spn\{\gamma_p \gamma_q \mid 0 \leq p < q \leq 2n\}$. Note that the invertible elements of $e^{\mathcal{L}_2'}$ is precisely the $\SO(2n + 2)$ group of fermionic Gaussian unitaries described in this section. This enables working with $\SO(2n+1)$ as a subgroup of the strictly quadratic $\SO(2n + 2)$ representation.

We now comment on the comparison to the particle-conserving basis rotations. First, the fact that fermionic Gaussian unitaries are a superset can be seen by the embedding $\U(n) \to \Orth(2n)$,
\begin{equation}
	u \mapsto \widetilde{Q} = \begin{pmatrix}
	R_{11} & \cdots & R_{1n}\\
	\vdots & \ddots & \vdots\\
	R_{n1} & \cdots & R_{nn}
	\end{pmatrix}, \quad \text{where } R_{ij} \coloneqq \begin{pmatrix}
	\Re u_{ij} & -\Im u_{ij}\\
	\Im u_{ij} & \Re u_{ij}
	\end{pmatrix}.
\end{equation}
It can be verified that $\mathcal{U}_{\widetilde{Q}} = \mathcal{U}_u$. Clearly, these do not have any issues with superselection rules, since particle conservation is a special instance of parity symmetry. Indeed, the representation revealed this in a rather deep manner:~$\U(n)$ is a connected and compact Lie group, thus its exponential map is surjective, whereas $\Orth(2n)$ is compact but has two separate connected components.

Finally, just as the particle-conserving unitaries were (projective) homomorphisms of $\U(n)$, so too are the Gaussian unitaries for $\Orth(2n)$:
\begin{equation}
	\begin{split}
	U_Q U_{Q'} \gamma_\mu U_{Q'}^\dagger U_Q^\dagger &= \sum_{\sigma \in [2n]} \sum_{\nu \in [2n]} \gamma_\sigma Q_{\sigma\nu} Q'_{\nu\mu}\\
	&= \sum_{\sigma \in [2n]} [QQ']_{\sigma\mu} \gamma_\sigma = U_{QQ'} \gamma_\mu U_{QQ'}^\dagger
	\end{split}
\end{equation}
for any $Q, Q' \in \Orth(2n)$. Also analogously to \cref{eq:basis_rotation_product}, the adjoint action on $k$-fold products of Majorana operators (hence any operator by linear extension) is
\begin{equation}
	\begin{split}
	U_Q \gamma_{\mu_1} \cdots \gamma_{\mu_k} U_Q^\dagger &= \sum_{\nu_1, \ldots, \nu_k \in [2n]} Q_{\nu_1 \mu_1} \cdots Q_{\nu_k \mu_k} \gamma_{\nu_1} \cdots \gamma_{\nu_k}\\
	&= \sum_{1 \leq \nu_1 < \ldots < \nu_k \leq 2n} \sum_{\pi \in \Sym(k)} Q_{\nu_{\pi(1)} \mu_1} \cdots Q_{\nu_{\pi(k)} \mu_k} (-1)^\pi \gamma_{\nu_1} \cdots \gamma_{\nu_k}\\
	&= \sum_{\bm{\nu} \in \binom{[2n]}{k}} \det[Q_{\bm{\nu}, \bm{\mu}}] \gamma_{\nu_1} \cdots \gamma_{\nu_k}.
	\end{split}
\end{equation}
Note that the second equality follows because, whenever any $\nu_i = \nu_j$, then using the fact that $\gamma_{\nu_i}^2 = \I_{2^n}$ we have the following term within the sum:
\begin{equation}
	\sum_{\nu_i \in [2n]} Q_{\nu_i \mu_i} Q_{\nu_i \mu_j} = [Q^\T Q]_{\mu_i \mu_j} = \delta_{\mu_i \mu_j}.
\end{equation}
But because we assume all $\mu_1 < \cdots < \mu_k$ (because otherwise it reduces to, say, a $(k - 2)$-degree operator), such terms always vanish. Thus we are only left with the summands wherein all $\nu_i$ are different. The Cauchy--Binet formula also applies to show how the homomorphism acts on $k$-degree operators:
\begin{equation}
	\det[QQ']_{\bm{\mu},\bm{\nu}} = \sum_{\bm{\sigma} \in \binom{[2n]}{k}} \det[Q_{\bm{\mu},\bm{\sigma}}] \det[Q'_{\bm{\sigma},\bm{\nu}}].
\end{equation}

\subsection{Fermionic Gaussian states}

We now turn to the study of the \emph{fermionic Gaussian states} directly. Just as the 1-RDM served as a complete description for Slater determinants, every fermionic Gaussian state is fully characterized by a one-body object called its \emph{covariance matrix}. This essentially essentially embeds the 1-RDM along with additional information pertaining to the non-number-preserving operators $\langle a_p^\dagger a_q^\dagger \rangle$. However, the Majorana operators will be a significantly more convenient representation to use.

Define the covariance matrix $\Gamma \in \R^{2n \times 2n}$ of any quantum state $\rho$ by
\begin{equation}
	\Gamma_{\mu\nu} \coloneqq -\frac{\i}{2} \tr([\gamma_\mu, \gamma_\nu] \rho).
\end{equation}
Observe that $\Gamma = -\Gamma^\T$ is antisymmetric:~when $\mu = \nu$, the commutator is $0$, and otherwise $(-\i/2)[\gamma_\mu, \gamma_\nu] = -\i \gamma_\mu \gamma_\nu = \i\gamma_\nu \gamma_\mu$. Just as with the 1-RDM, rotating $\rho$ by a free-fermion unitary $U_Q$ is represented compactly on the covariance matrix:
\begin{equation}\label{eq:cov_mat_transformation}
	\begin{split}
	-\frac{\i}{2} \tr([\gamma_\mu, \gamma_\nu] U_Q \rho U_Q^\dagger) &= \sum_{\mu',\nu' \in [2n]} Q_{\mu\mu'} Q_{\nu\nu'} \Gamma_{\mu'\nu'}\\
	&= [Q \Gamma Q^\T]_{\mu\nu}.
	\end{split}
\end{equation}
Since any pure free-fermion state can be written as $U_Q \ket{0^n}$ for some $Q \in \Orth(2n)$, it suffices to know the covariance matrix of $\op{0^n}{0^n}$,
\begin{equation}
	\Gamma^{(0)} = \bigoplus_{p \in [n]} \begin{pmatrix}
	0 & 1\\
	-1 & 0
	\end{pmatrix},
\end{equation}
from which the covariance matrix of $U_Q \ket{0^n}$ can be calculated via \cref{eq:cov_mat_transformation}. More generally, any fermionic Gaussian state (pure or mixed) can be described as a rotation of a mixed product state in the standard basis:
\begin{equation}\label{eq:gaussian_density_matrix}
	\rho = U_Q \l( \prod_{p \in [n]} \frac{1}{2} (\I_{2^n} - \lambda_p \i \gamma_{2p-1} \gamma_{2p}) \r) U_Q^\dagger,
\end{equation}
whose covariance matrix is therefore
\begin{equation}\label{eq:cov_matrix_normal_form}
	\Gamma = Q \bigoplus_{p \in [n]} \begin{pmatrix}
	0 & \lambda_p\\
	-\lambda_p & 0
	\end{pmatrix} Q^\T.
\end{equation}
Because $\pm \i \lambda_p$ are the eigenvalues of $\Gamma$, and $|\lambda_p| \leq 1$ is necessary for \cref{eq:gaussian_density_matrix} to be a valid quantum state (saturated by pure states), we have that any valid covariance matrix must obey $-\I_{2n} \preceq \i\Gamma \preceq \I_{2n}$, or equivalently $0 \preceq -\Gamma^2 \preceq \I_{2n}$. This further implies that pure Gaussian states obey $\Gamma\Gamma^\T = \I_{2n}$, i.e., antisymmetric and orthogonal.

Just as pure Gaussian states are ground states of some one-body Hamiltonian, mixed Gaussian states are their thermal states. We can see this by computing the Gibbs state of $H = (-\i/4) \bm{\gamma}^\T A \bm{\gamma} = (-\i/2) U_Q \sum_{p \in [n]} \varepsilon_p \gamma_{2p-1} \gamma_{2p} U_Q^\dagger$ at a temperature $1/\beta$:
\begin{equation}
	\begin{split}
	\frac{e^{-\beta H}}{\tr(e^{-\beta H})} &= U_Q \l( \bigotimes_{p \in [n]} \frac{e^{-\beta \varepsilon_p Z / 2}}{\tr(e^{-\beta \varepsilon_p Z / 2})} \r) U_Q^\dagger\\
	&= U_Q \l( \bigotimes_{p \in [n]} \frac{1}{2} (\I_2 + \tanh(-\beta \varepsilon_p / 2) Z) \r) U_Q^\dagger.
	\end{split}
\end{equation}
Hence any mixed Gaussian state with covariance matrix as in \cref{eq:cov_matrix_normal_form} corresponds to the thermal state of such a Hamiltonian, with single-mode occupancies $\lambda_p = \tanh(-\beta \varepsilon_p / 2)$. Taking the limit $\beta \to \infty$ corresponds to the ground state, revealing $|\lambda_p| = 1$ as discussed above.

\subsection{Wick's theorem and mean-field approximations}

Analogous to the 1-RDM, the covariance matrix of a fermionic Gaussian state also determines the entire state by Wick's theorem~\cite{bach1994generalized}:
\begin{equation}
	(-\i)^k\tr(\gamma_{\mu_1} \cdots \gamma_{\mu_{2k}} \rho) = \mathrm{Pf}(\Gamma_{\bm{\mu}, \bm{\mu}}),
\end{equation}
where $\mathrm{Pf}$ is the Pfaffian, an antisymmetric matrix polynomial related to the determinant via $\mathrm{Pf}(A)^2 = \det(A)$. Recall that $\Gamma_{\bm{\mu}, \bm{\mu}}$ is the $2k \times 2k$ submatrix of $\Gamma$ indexed by $\bm{\mu} \subseteq [2n]$, and that fermionic states have no support on odd-product Majorana operators (seen here by the fact that Pfaffians of odd-dimensional matrices always vanish).

For example, any two-body fermionic correlation of a Gaussian state can be computed from the one-body expectations via
\begin{equation}
	-\tr(\gamma_p \gamma_q \gamma_r \gamma_s \rho) = \Gamma_{pq} \Gamma_{rs} - \Gamma_{pr} \Gamma_{qs} + \Gamma_{ps} \Gamma_{qr}.
\end{equation}
This property allows for a mean-field approximation of ground states of interacting-fermion Hamiltonians without particle-number symmetry, in terms of the lowest-energy Gaussian state~\cite{bravyi2017complexity,bravyi2019approximation,herasymenko2023optimizing}. Up to arbitrary energy shift, any two-body Hamiltonian can be written as
\begin{equation}
	H = -\i \sum_{p,q \in [2n]} V_{pq}  \gamma_p \gamma_q - \sum_{p,q,r,s \in [2n]} W_{pqrs} \gamma_p \gamma_q \gamma_r \gamma_s,
\end{equation}
where $V \in \R^{2n \times 2n}$ and $W \in \R^{2n \times 2n \times 2n \times 2n}$ are totally antisymmetric to ensure Hermiticity of $H$. Then any Gaussian state $\rho$ has energy
\begin{equation}\label{eq:HFB_energy}
	\tr(H \rho) = \sum_{p,q \in [2n]} V_{pq} \Gamma_{pq} + 3 \sum_{p,q,r,s \in [2n]} W_{pqrs} \Gamma_{pq} \Gamma_{rs},
\end{equation}
and searching for the optimal Gaussian state is equivalent to minimizing this functional over all antisymmetric $\Gamma \in \R^{2n \times 2n}$ with the quadratic constraint $\Gamma \Gamma^\T \preceq \I_{2n}$.

This is the underlying optimization problem behind Hartree--Fock--Bogoliubov theory, a generalization of Hartree--Fock theory to systems which do not preserve particle number~\cite{bach1994generalized}. Unfortunately, just as Hartree--Fock theory is $\mathsf{NP}$-complete~\cite{schuch2009computational,ogorman2022intractability}, so too is this generalization. Trivially, since Hartree--Fock is a special instance, the more general problem is also at least as hard. More interestingly, it was shown in \cite{bravyi2019approximation} that \cref{eq:HFB_energy} with the appropriate constraints is an instance of quadratic programming with orthogonality constraints, a known $\mathsf{NP}$-hard optimization problem~\cite{luo2010semidefinite}. In Chapter~\ref{chap:QR}, we exhibit a subclass called the \emph{little noncommutative Grothendieck (LNCG) problem}, which can be naturally addressed on a quantum computer~\cite{zhao2023expanding}. By using this connection between quadratic optimization and fermionic physics, we show how to produce high-quality approximations to the LNCG problem using quantum-simulation techniques. Analogous to how quantum annealers are envisioned to find good classical solutions by searching over a larger space of entangled states, so too does our proposed scheme use a quantum computer to explore the landscape of interacting states, eventually projecting onto a near-optimal classical solution (Gaussian state).
\chapter{Learning from Quantum Systems}\label{chap:learning}


Learning properties about quantum systems is at the heart of quantum information processing. A quantum experiment or computation is only meaningful if one can extract actionable (classical) information from the system. Unfortunately, quantum measurements abide by strange rules:~they are random (quantum states describe generalized probability distributions), destructive (the measured state ``collapses'' and is rendered useless for subsequent measurements), and reveal relatively little information (the measurement only yields samples from the distribution in a particular basis). In this chapter, we review ideas and techniques for handling these challenges.

\section{Quantum state tomography}

The standard approach to extracting quantum information is to repeatedly run the experiment multiple times and perform measurements on each fresh copy of the quantum state. Assuming that the experiment consistently produces identical copies, we can learn about the quantum state in a Monte Carlo fashion as follows. The quantum state, represented as a density matrix $\rho \in \C^{d \times d}$ with unit trace ($\tr\rho = 1$) and nonnegative eigenvalues ($\rho \succeq 0$), can be thought of as a collection of probability distributions. Each such distribution is specified by a positive-operator-valued measure (POVM), a set of operators $\{E_j\}_j$ obeying $E_j \succeq 0$ and $\sum_j E_j = \I$, wherein the probability of seeing an outcome that we label as $j$ is given by $p_j = \tr(E_j \rho)$. These POVMs correspond to the physical measurements that are performed in experiment;~it is typically assumed that the experiment has native access to projective measurements in the standard basis, corresponding to the POVM $\{\op{b}{b} \mid b \in \{0, 1\}^n\}$.

By repeatedly preparing and measuring copies of $\rho$ according to some POVM, we are effectively sampling outcomes from the distribution $\{p_j\}_j$ and storing them as classical data. Statistical techniques then allow us to construct estimates for various properties of $\rho$. As in classical statistics, we would like to estimate with high accuracy using as few copies of $\rho$ as possible---this is called the copy or sample complexity.

In the standard task of quantum state tomography, we aim to learn the entire density matrix. This can be achieved by taking an informationally complete POVM (i.e., whose elements span the operator space, $\spn(\{E_j\}_j) = \C^{d \times d}$) and reconstructing the density matrix from classically postprocessing the measurement outcomes. Examples of such postprocessing algorithms include linear inversion~\cite{sugiyama2013precision}, matrix completion (compressed sensing)~\cite{gross2010quantum,flammia2012quantum}, and least squares~\cite{opatrny1997least,guta2020fast}. Equipped with an estimate of the density matrix $\rho$, any property of the state can be predicted.

This is however very expensive:~$N = \Theta(rd)$ copies is both necessary and sufficient to learn $\rho$ with constant accuracy, where $r$ is the rank of $\rho$~\cite{odonnell2016efficient,wright2016learn,haah2017sample}.\footnote{Tight bounds when the accuracy metric (usually either the trace distance $\delta$ or infidelity $\epsilon$) is not constant is a subtly open problem, however it is commonly conjectured that $\Theta(rd/\delta^2)$ and $\Theta(rd/\epsilon)$ are the optimal scalings~\cite{yuen2023improved}.} This is true even with the power of entangled measurements, wherein we are allowed to perform one large collective measurement on the state $\rho^{\otimes N}$. Thus even when the state is pure ($r = 1$), a many-body system of $n$ qubits (hence dimension $d = 2^n$) requires exponentially many copies to be completely learned. Furthermore, the computational complexity for reconstructing the $d \times d$ matrix is inherently polynomial in $d$, hence also exponential in $n$.

\section{Observable estimation}

In many practical applications, one rarely seeks a complete description of a large many-body system. Instead, it suffices to learn only a collection of target properties of the system, such as its energy or many-body correlations. Indeed, this is standard fare for experimentalists:~suppose we have an observable $O$ which admits a spectral decomposition
\begin{equation}
	O = \sum_{j=1}^{d} \lambda_j \op{\psi_j}{\psi_j}.
\end{equation}
The ability to measure in the eigenbasis of this observable means one has access to the POVM elements $E_j = \op{\psi_j}{\psi_j}$. Recall that the expectation value of $O$ for the state $\rho$ is
\begin{equation}
	\tr(O \rho) = \sum_{j=1}^d p_j \lambda_j,
\end{equation}
where $p_j = \tr(E_j \rho)$ defines a probability distribution. Sampling copies of $\rho$ in this POVM therefore yields statistical estimates for $\tr(O \rho)$. A straightforward application of Hoeffding's inequality implies that a sample complexity of $N = \O(\|O\|_\infty^2/\epsilon^2)$, where $\|O\|_\infty$ is the spectral norm of $O$, suffices to learn the value of $\tr(O \rho)$ to within additive error $\epsilon$, with high probability.

\subsection{Operator decompositions}

Many hard computational problems correspond to knowing the eigenvalues and eigenvectors of a many-body observable, so we cannot generally expect prior knowledge of $\lambda_j$, nor the ability to measure in the basis of $\{\ket{\psi_j}\}_j$. Instead, we typically have access to the observable only in some sparse representation. For instance, if $O$ is a $k$-local observable (for $k = \O(1)$ a small constant), then there always exists a decomposition of $O$ into only polynomially many $k$-local, ``simple'' terms. The canonical decomposition is via the Pauli basis, which indeed satisfies the simple criterion (the eigenvalues and eigenvectors of Pauli operators are completely known). In such a basis, $O$ decomposes into $R \leq \sum_{j=0}^k 3^j \binom{n}{j} = \O(n^k)$ many terms:
\begin{equation}
	O = \sum_{i=1}^R h_i P_i, \quad h_i = \frac{1}{2^n} \tr(P_i O),
\end{equation}
where each $P_i \in \{\I_2, X, Y, Z\}^{\otimes n}$ acts nontrivially on at most $k$ qubits. Then we can instead estimate $\tr(O \rho)$ by, for each $i = 1, \ldots, R$, measuring in the much simpler POVM for $P_i$ (whose eigenvectors are simple product states and eigenvalues are just $\pm 1$) and constructing an estimate for each $\tr(P_j \rho)$. By linearity this enables an estimate for $\tr(O \rho) = \sum_{i=1}^R h_i \tr(P_i \rho)$. More generally, this is always an efficient decomposition as long as $R = \poly(n)$ (regardless of locality).

The sample complexity for this approach can be made to be $N = \O(\|\bm{h}\|_1^2 / \epsilon^2)$, where $\|\bm{h}\|_1 = \sum_{j=i}^R |h_i|$ is the 1-norm of the coefficients, by allocating the measurements according to the magnitudes $|h_j|$ of the terms~\cite{wecker2015progress,rubin2018application}. That is, if we set a total copy budget to some fixed $N$, then $\lfloor (|h_i| / \|\bm{h}\|_1) N \rfloor$ rounds of the experiment should be spent measuring in the POVM for $P_i$.

This idea of decomposing a highly complex observable into simple-to-measure terms is at the heart of low-depth quantum heuristics, such as variational quantum algorithms, which are designed to be suitable for the near-term, noisy quantum hardware~\cite{peruzzo2014variational,mcclean2016theory}. We will discuss such algorithms in Chapter~\ref{chap:nisq}.

\subsection{Measurement reduction strategies}

While this is a sufficient paradigm for estimating observables in polynomial complexity, there still leaves much room for practical improvement. Shortly after the introduction of the variational quantum eigensolver for approximating many-body ground states~\cite{peruzzo2014variational}, Wecker \emph{et al.}~\cite{wecker2015progress} quickly recognized that the total number of measurements (number of POVMs times the number of shots per POVM) would be ``astronomically large for quantum chemistry applications to molecules.'' They particularly addressed the quantum-chemistry problem, due to its high scientific and industrial relevance. However, the measurement of such electronic-structure Hamiltonians is highly challenging due to the fact that it decomposes into $\O(n^4)$ terms, as seen by \cref{eq:elec_struct_bkgd}.

In light of this bottleneck, a large body of work soon emerged to address it. One immediate approach is to use the fact that any two $P_j, P_{j'}$ which commute share a joint POVM, so they can both be estimated from the same sample. This seemingly simple observation is in fact highly challenging to optimally take advantage of;~for example, because commutativity is not transitive, $[P_j, P_{j'}] = 0$ and $[P_{j'}, P_{j''}] = 0$ for some third term $P_{j''}$ does not imply that $[P_j, P_{j''}] = 0$. Thus if we wish to minimize the number of unique POVMs required (hence finding the largest groups of mutually commuting Pauli operators), this turns out to reduce to the $\mathsf{NP}$-hard problem of finding the minimal clique cover of a graph~\cite{verteletskyi2020measurement,yen2020measuring,jena2019pauli,gokhale2019on3}. Nonetheless, these and other works have proposed heuristics to obtain approximate solutions which are sufficient, reducing the number of unique POVMs from $\O(n^4)$ to $\O(n^3)$~\cite{yen2020measuring,gokhale2019on3}. However, experimental feasibility must also be considered, such as whether we allow only local transformations to implement the POVM (referred to as ``qubitwise'' commutativity)~\cite{verteletskyi2020measurement}, or whether more costly entangling operations are allowed (e.g., Bell-basis-type measurements)~\cite{yen2020measuring}.

Alongside commutativity, it turns out that anticommutativity between Pauli terms can also be leveraged to effect measurement reduction. This technique, known as \emph{unitary partitioning}~\cite{izmaylov2019unitary,bonet2020nearly,zhao2020measurement}, is described in detail in Chapter~\ref{chap:MR}. The key idea is that linear combinations of mutually anticommuting operators possess a nice structure that allows them to be unitarily rotated into a single basis direction in operator space. We show that, by doing so, one can achieve a linear (in $n$) term reduction for the electronic-structure Hamiltonian~\cite{zhao2020measurement}.

A third direction is to decompose the observable into a more compact representation, for example by a method called basis-rotation grouping~\cite{huggins2021efficient}. This idea is based on the fact that the two-body coefficient tensor of the electronic-structure Hamiltonian is positive semidefinite and low rank, hence admits a low-rank Cholesky decomposition~\cite{peng2017highly}.\footnote{This property also enables highly efficient Trotter circuits~\cite{motta2021low} and block encodings~\cite{berry2019qubitization} for these Hamiltonians.} This in turn allows one to write the underlying Hamiltonian as $H = H_0 + \sum_{\ell=1}^L H_\ell$, where $H_0 = U_0 \l( \sum_{p \in [n]} g_p^{(0)} n_p \r) U_0^\dagger$ is single-body and the $L = \O(n)$ two-body terms are of the form
\begin{equation}
    H_\ell = U_\ell \l( \sum_{p,q \in [n]} g_p^{(\ell)} g_q^{(\ell)} n_p n_q \r) U_\ell^\dagger.
\end{equation}
Each $U_\ell$ is a single-particle basis rotation, as described in \cref{eq:basis_rotation_bkgd}. Although each $H_\ell$ furthermore has $\O(n^2)$ terms, they are number operators which all mutually commute, hence can all be estimated simultaneously from the same measurement outcome. These basis rotations precisely define the $L + 1 = \O(n)$ unique POVMs to estimate the energy, as opposed to the naive decomposition with $\O(n^4)$ terms. Even further reduction with this idea by numerical optimizations has been subsequently developed~\cite{yen2020cartan}.

With regards to the sample complexity, covariances between joint outcomes of different operators also should be considered. Indeed, the total number of samples required to achieve some desired estimation error is ultimately controlled by the variance of the estimator for $\tr(O \rho)$, for which the number of unique POVMs is only one component. The problem is further complicated if one is additionally interested in a collection of many observables $O_1, \ldots, O_L$, each of which potentially decomposes into partially intersecting sets of Pauli operators~\cite{cotler2020quantum,bonet2020nearly}.

\subsection{Quadratically more precise learning}

The dependence of $1/\epsilon^2$ in the sample complexity is a generic feature of learning with classical data (i.e., the independently drawn measurement outcomes). Sometimes referred to the standard quantum limit\footnote{Some would argue that this should be referred to as the standard \emph{classical} limit, although the established jargon is rigid at this point.} in contexts of quantum metrology, its universality is essentially a consequence of the central-limit theorem:~in the $N \to \infty$ limit, all distributions of the mean converge to a normal distribution with width $\sigma/\sqrt{N} \sim \epsilon$ (where $\sigma^2$ is the variance of a single sample).

Quantum coherence can be harnessed to surpass the standard limit, with the ultimate rate of $1/\epsilon$ called the \emph{Heisenberg limit}. This is typically encountered in metrological contexts~\cite{giovannetti2006quantum}, using techniques such as phase estimation~\cite{kitaev1995quantum,nielsen2010quantum}. Beyond that context, it is also possible to learn expectation values at the Heisenberg limit, so long as one can afford the cost of the additional quantum coherence. (Unfortunately, near-term quantum processors do not have such a budget, so these ideas are mostly aimed at a fault-tolerant machine sometime in the future.)

Early ideas were based on quantum amplitude amplification~\cite{brassard1997exact}, a generalization of Grover's celebrated search algorithm~\cite{Grover1996,grover1998quantum}. In particular, \cite{brassard2002quantum} showed how to estimate the amplitude $\ip{x}{\psi}$ for some basis state $\ket{x}$ with precision $\O(1/N)$ in a single shot, given access to the unitary $U$, and its inverse $U^\dagger$, which prepares $\ket{\psi} = U\ket{0^n}$. They achieve this by a multi-controlled variant of the Grover iterate, using an ancilla register of $N$ qubits and $\O(N)$ calls to $U$.

Based on this procedure, \cite{knill2007optimal} constructed an algorithm to estimate $\ev{\psi}{O}{\psi}$ with similar precision. The key idea is to make the replacement $\ket{x} \to e^{-\i O t} \ket{\psi}$ and use the fact that, for small $t$,
\begin{equation}
    \Im\ev{\psi}{e^{-\i O t}}{\psi} = -\ev{\psi}{O}{\psi} t + \O(t^3).
\end{equation}
Thus the imaginary part of the amplitude can be converted into an approximation for the desired expectation value. Under mild assumptions about the spectral distribution of $A$ with respect to $\ket{\psi}$, the algorithm returns an estimate of $\ev{\psi}{O}{\psi}$ with error $\epsilon$, using $\epsilon^{-1 + o(1)}$ applications of $U$ and $e^{-\i O t}$. Furthermore, the total evolution time over all applications of $e^{-\i O t}$ is also $\O(\epsilon^{-1})$. Beyond this method, other Heisenberg-limited schemes have been developed, for example using maximum-likelihood estimation to reduce the circuit complexity~\cite{suzuki2020amplitude} or block-encoding techniques to access observables which cannot be easily exponentiated, e.g., time-correlation functions~\cite{rall2020quantum}.

One drawback of these approaches is that they are not (easily) parallelizable for estimating multiple expectation values. Naively, learning $L$ separate observables by the method above would involve a total of $\O(L/\epsilon)$ queries to the state-preparation circuit $U$. To avoid this unfavorable scaling in $L$, \cite{huggins2021nearly} adapted a fast quantum gradient-estimation algorithm~\cite{gilyen2019optimizing} to solve this problem. Their algorithm requires only $\Ot\l(\sqrt{\sum_{j=1}^L \|O_j\|_\infty^2}/\epsilon\r)$ queries to either $U$ or $U^\dagger$, where $\|O_j\|_\infty$ is the spectral norm of $O_j$. In the setting where all operators have unit-bounded spectra, $\|O_j\|_\infty \leq 1$, this reduces to a query complexity of $\Ot(\sqrt{L} / \epsilon)$, a quadratic improvement in $L$ over the naive approach.

To describe their measurement algorithm, we will suppose all $\|O_j\|_\infty \leq 1$ for simplicity. Computationally, the algorithm additionally requires $\Ot(\sqrt{L}/\epsilon)$ doubly controlled gates of the form c-c-$e^{-\i t O_j}$, where each $t$ is at most $\O(1/\sqrt{L})$ in magnitude and the ancilla register holds a total of $\O(L \log\epsilon^{-1})$ qubits. The large size of this register is one of the main catches of this algorithm, because if $L = \omega(n)$ then the number of ancilla qubits will dominate the space complexity over the computational register itself. For example, the number of operators required to characterize all $k$-RDMs of the $n$ qubits is $L = \O(n^k)$, while the fermionic $k$-RDM features $L = \O(n^{2k})$.

At a high level, the algorithm works as follows. Defining the parametrized circuit
\begin{equation}
    V(\bm{t}) \coloneqq \prod_{j=1}^L e^{-2\i t_j O_j},
\end{equation}
and subsequently the function
\begin{equation}
    f(\bm{t}) \coloneqq -\frac{1}{2} \Im \ev{\psi}{V(\bm{t})}{\psi} + \frac{1}{2},
\end{equation}
a straightforwad calculation reveals that the $L$-dimensional gradient of $f$ evaluated at $\bm{t} = \bm{0}$ is a vector of the desired expectation values:
\begin{equation}
    \nabla f(\bm{0}) = \begin{pmatrix}
    \ev{\psi}{O_1}{\psi}\\
    \vdots\\
    \ev{\psi}{O_L}{\psi}
    \end{pmatrix}.
\end{equation}
Thus the algorithm of~\cite{gilyen2019optimizing} can be used to estimate this gradient, given oracular access to $f$ as a unitary. \cite{huggins2021nearly} give an explicit quantum circuit for such an oracle:~first, the Hadamard-test circuit gives a way to evaluate the imaginary part of $\ev{\psi}{V(\bm{t})}{\psi}$. One can encode this information, for each $j$th vector component, with $\epsilon$ precision into a register of $\O(\log\epsilon^{-1})$ qubits (i.e., in binary) by applying c-$e^{-\i t_j O_j}$ for appropriate values of $t_j$. Hence we require $\O(L \log\epsilon^{-1})$ total ancilla to store all $j \in [L]$ components. A second control on the $e^{-\i t_j O_j}$ gates is then required to actually implement the Hadamard test. This furnishes the unitary oracle for $f$;~note that only one application of $U$ was required, to simply prepare $\ket{\psi}$ at the start of the circuit. Then by the analysis of~\cite{gilyen2019optimizing}, the gradient of $f$ can be estimated to within $\infty$-norm error $\epsilon$ with only $\Ot(\sqrt{L}/\epsilon)$ queries to the oracle, hence $U$.

In the setting where the spectral norms are not uniformly bounded, \cite{van2023quantum} achieved an improvement to $\Ot\l(\sqrt{\|\sum_{j=1}^L O_j^2\|_\infty} / \epsilon\r)$ queries to $U$. Their algorithm does not use the gradient-estimation algorithm, but rather improves techniques from shadow tomography~\cite{aaronson2020shadow} when one has access to $U$ and the ability to perform block encodings~\cite{low16a,gilyen2019quantum}. Unfortunately, they still require an $\O(L \log\epsilon^{-1})$-sized ancilla register, in addition to the space overhead to block encode. Below, we will explore the ideas of this so-called shadow tomography, and related ideas, in detail.

\section{Learning partial descriptions of quantum states}

Here we examine a contemporaneous line of thinking in quantum learning, one based on the theory of computational learning~\cite{arunachalam2017survey}. We will see how this framework naturally leads to effective and succinct descriptions of quantum systems.

\subsection{Probably approximately correct learning}

Based on Valiant's \emph{probably approximately correct (PAC) learning} framework~\cite{valiant1984theory}, Aaronson introduced an operational meaning to the notion of learning quantum states~\cite{aaronson2007learnability}, not dissimilar from the observable estimation paradigm described above. Let $\mathcal{D}$ be a distribution of two-outcome POVMs $\{E, \I - E\}$ where $0 \preceq E \preceq \I$.\footnote{Pauli measurements are an example, by setting $E = (\I + P)/2$ for any Pauli observable $P$.} The goal is to determine a good ``hypothesis'' state $\sigma$ such that, with probability at least $1 - \delta$ over the draw of $E \sim \mathcal{D}$, the error $|{\tr(E\sigma) - \tr(E \rho)}| \leq \gamma$ is small.

To do so, the learner is given access to a training set of $N$ measurements (i.e., independently drawn POVMs $E_1, \ldots, E_N \sim \mathcal{D}$). Aaronson showed that this is possible, with high probability, as long as $N \geq \Omegat(n \gamma^{-4} \delta^{-4})$~\cite{aaronson2007learnability}. Thus one can learn the quantum state with a sample complexity only linear in the system size, where ``learning'' here is a restricted notion that suffices for the types of measurements in $\mathcal{D}$ that the learner cares about, not all possible observables. This information-theoretic bound is established by bootstrapping the classical learning theory of distinguishing between different hypotheses~\cite{anthony2000function,bartlett1998prediction} to the quantum setting~\cite{ambainis2002dense}.

In practice, an experimenter can repeatedly implement the training POVMs such that they learn each $\tr(E_i \rho)$ up to additive error at most $\O(\gamma\delta)$. Then the satisfying hypothesis state $\sigma$ can be found by solving a semidefinite program (SDP) minimizing the mean-square error of the training data~\cite{rocchetto2019experimental}. As SDPs run in time polynomial in the dimension of the matrix $d = 2^n$, this classical computation is exponentially expensive. Indeed, Aaronson's original result only comments on the sample, not computational, complexity.

\subsection{Shadow tomography}

The goal of PAC learning for quantum states is to output good approximations, with high probability over a distribution $\mathcal{D}$ of POVMs. One can instead pose a stronger task:~given a concrete, fixed set of two-outcome POVMs $E_1, \ldots, E_L$, output numbers $b_1, \ldots, b_L \in [0, 1]$ such that $|b_i - \tr(E_i \rho)| \leq \epsilon$ for all $i \in [L]$, with success probability at least $1 - \delta$. This is precisely the problem of \emph{shadow tomography}, also introduced by Aaronson~\cite{aaronson2020shadow}.\footnote{The qualifier ``shadow'' was suggested to Aaronson by Steve Flammia, although Flammia preferred the terminology shadow \emph{estimation}, rather than \emph{tomography}, for this particular task~\cite{classical_shadow_scirate}.} In the original formulation, the learner is allowed to accomplish this task by performing any POVM on $\rho^{\otimes N}$, and the goal is to find the smallest sufficient $N$. A highly entangled measurement is therefore allowed.

Under this scenario, Aaronson showed that $N = \Ot(n \epsilon^{-4} \log^4 L \log\delta^{-1})$ copies suffice, where the $\Ot(\cdot)$ notation hides a factor of $\poly(\log\log L, \log n, \log\epsilon^{-1})$. Thus the copy complexity is nearly linear in system size for this problem, as it was for PAC learning, and only polylogarithmic in the number of POVM outcomes to learn. Unfortunately, the dependence on $1/\epsilon$ remains quartic, which although polynomial is impractically large.

Along with the conceptual departure from quantum PAC learning, shadow tomography also allows the use of quantum resources to address its computational hardness. Aaronson's initial procedure required $G = \O(4^n)$ quantum gates to implement each POVM, and $2^{\O(\epsilon^{-2} \, n \log n)}$ time and space to store and update a classical description of the hypothesis state. However, the quantum SDP solver developed by Brand{\~ a}o \emph{et al.}~\cite{brandao2019quantum} was shown to greatly improve these costs. Primarily, the ability to solve SDPs on a quantum computer bypasses the need to manipulate an exponentially large classical representation. That being said, while this makes the space complexity polynomial, the total number of quantum gates (hence time complexity) remains exponential, as $\Ot(\sqrt{L} G) + \poly(2^n)$. These costs can be somewhat improved by placing further restrictions, for instance by only considering POVMs that can be implemented in $G = \poly(n)$ gates.

\section{Classical shadows}

While a breakthrough in sample complexity for learning quantum states, it is clear that the computational cost of shadow tomography is still impractical. Even ignoring gate complexity, the problem involves coherently manipulating the state $\rho^{\otimes N}$, which requires a massive register of $\Ot(n^2 \epsilon^{-4} \log^4 L \log\delta^{-1})$ qubits. A pressing question is therefore:~can the philosophy of shadow tomography be adapted into a practical, near-term-friendly protocol?

In their seminal 2020 paper, Huang, Kueng, and Preskill~\cite{huang2020predicting} affirmed this question, introducing the method of \emph{classical shadows}.\footnote{Paini \emph{et al.}~\cite{paini2019approximate_superseded,paini2021estimating} also introduced essentially the same idea independently, albeit specializing to $\mathcal{D} = \SU(2)^{\otimes n}$ rather than providing a completely general formulation.} Importing ideas from quantum state tomography, in particular linear-inversion estimation~\cite{guta2020fast}, into the ``shadow'' framework, they developed a protocol which converts single copies of a quantum state into a minimal classical sketch, called its classical shadow. This efficient description is designed such that different properties statistically converge to the ground truth at different rates;~for example, a classical shadow built out of $\Ot(\epsilon^{-2})$ samples may already predict local observables to within $\epsilon$ error, while nonetheless being exponentially far from the true density matrix. Importantly, the required gates can be as simple as single-qubit rotations.

The central idea is to craft a distribution of POVMs such that the desired properties are the ones converged quickly. Thus the task is somewhat of an inverse to the PAC learning problem:~instead of being given $\mathcal{D}$, we seek such a distribution which yields nice prediction features with respect to the target set of observables. The setting can be framed as follows. Given single-copy measurement access to the $n$-qubit state $\rho$, we wish to predict the values of $\tr(O_j \rho)$ for a collection of observables $O_1, \ldots, O_L$. Specifically, for accuracy and confidence parameters $\epsilon, \delta \in (0, 1)$, output estimates $\hat{o}_1, \ldots, \hat{o}_L \in \R$ such that
\begin{equation}\label{eq:classical_shadow_error_target}
	|\hat{o}_j - \tr(O_j \rho)| \leq \epsilon \quad \forall j \in [L],
\end{equation}
with success probability at least $1 - \delta$. The difference from shadow tomography is that, rather than demanding the protocol be efficient over any possible collection of observables, we accept that it may only be efficient over certain classes of observables (for example, local observables or low-rank observables).

\subsection{Warm-up example:~random Pauli measurements}\label{sec:warm-up}

To build some intuition for classical shadows, we first examine a specific but highly relevant protocol of \emph{random Pauli measurements}. Such measurements are important because they are arguably the simplest informationally complete set of POVMs, requiring only local, single-qubit control to implement. Note that random Pauli measurements have been considerably studied prior to the advent of classical shadows, especially in the context of partial state learning~\cite{yu2020sample,cotler2020quantum,paini2019approximate_superseded,evans2019scalable}. The intuition we develop here draws from those works as well.

\subsubsection{A single qubit}

We begin with a single qubit. A Pauli measurement is achieved by mapping the computational basis, defined by $Z = \op{0}{0} - \op{1}{1}$, to one of either $X = \Had^\dagger Z \Had$, $Y = \G^\dagger Z \G$, or itself, $Z = \I^\dagger Z \I$. These transformations are achieved by the unitaries
\begin{equation}
	\Had \coloneqq \frac{1}{\sqrt{2}} \begin{pmatrix}
	1 & 1\\
	1 & -1
	\end{pmatrix}, \quad \G \coloneqq \Had \Ph^\dagger \text{ where } \Ph \coloneqq \begin{pmatrix}
	1 & 0\\
	0 & \i
	\end{pmatrix}.
\end{equation}
Let us first review how Pauli measurements are typically performed. As our native POVM is limited to the $Z$ basis, predicting expectation values of an arbitrary Pauli observable $W \in \{X, Y, Z\}$ makes use of the identity
\begin{equation}
	\tr(W \rho) = \tr(U^\dagger Z U \rho)  = \tr(Z U \rho U^\dagger)
\end{equation}
for an appropriate choice of $U \in \{\I, \Had, \G\}$. Thus, by applying $U$ to the state immediately before measurement in the native $Z$-basis, we obtain statistics of the original state in the $W$ basis.

By the postulates of quantum mechanics, each measurement outcome $x \in \{0, 1\}$ is obtained with probability $p_x = \ev{x}{U \rho U^\dagger}{x}$. Using $Z\ket{x} = (-1)^x\ket{x}$, we can construct an estimator for $\tr(W \rho)$ from
\begin{align}
	\tr(W \rho) = \tr(Z U \rho U^\dagger) &= \sum_{x \in \{0, 1\}} p_x (-1)^x\\
	&\approx \frac{1}{N} \sum_{\ell=1}^N (-1)^{x_\ell},
\end{align}
where $x_1, \ldots, x_N \in \{0, 1\}$ are the measurement outcomes of $N$ rounds of this experiment. We want the absolute error of this estimate to be no larger than $\epsilon$:
\begin{equation}
	\l| \tr(W \rho) - \frac{1}{N} \sum_{\ell=1}^N (-1)^{x_\ell} \r| \leq \epsilon.
\end{equation}

To show how many samples $N$ we need for this bound to hold, one can appeal to Chebyshev's inequality. For a random variable $\mathcal{X}$ with finite first and second moments, $|{\E[\mathcal{X}]}| < \infty$ and $\V[\mathcal{X}] < \infty$, Chebyshev's inequality states that the probability that $\mathcal{X}$ deviates from its mean value $\E[\mathcal{X}]$ by more than $\epsilon$ obeys
\begin{equation}
	\Pr[|\mathcal{X} - \E[\mathcal{X}]| \geq \epsilon] \leq \frac{\V[\mathcal{X}]}{\epsilon^2}.
\end{equation}
This is the probability that our experiment has failed to produce $\epsilon$-accurate estimates, so we want it to be small. In our setting, the random variable is simply the average of the measurement outcomes:
\begin{equation}
	\mathcal{X} = \frac{1}{N} \sum_{\ell=1}^N (-1)^{x_\ell},
\end{equation}
which obeys $\E[\mathcal{X}] = \tr(W \rho)$. Because the samples are independent and identically distributed (i.i.d.), we can compute its variance as
\begin{equation}
\begin{split}
	\V[\mathcal{X}] = \V\l[\frac{1}{N} \sum_{\ell=1}^N (-1)^{x_\ell}\r] = \frac{1}{N^2} \sum_{\ell=1}^N \V\l[(-1)^{x_\ell}\r] &= \frac{1}{N} \V\l[(-1)^x\r]\\
	&= \frac{1}{N} \l[ 1 - \tr(W \rho)^2 \r].
\end{split}
\end{equation}
Assuming no prior information about $\rho$, we can apply the trivial bound $\tr(W \rho)^2 \geq 0$ to bound the variance as $\V[\mathcal{X}] \leq 1/N$. Chebyshev's inequality thus implies that the failure probability is at most $\delta = 1/(N\epsilon^2)$. Rearranging this statement informs us that taking $N = 1/(\delta \epsilon^2)$ samples suffices\footnote{More sophisticated tail bounds, such as Bernstein's inequality, can exponentially improve the dependence on $\delta$ to the optimal scaling of $\log(1/\delta)$. We stick with Chebyshev's inequality here for simplicity, and besides we take $\delta$ to be constant.} to learn $\tr(W \rho)$ to error $\leq \epsilon$, with success probability $\geq 1 - \delta$. Doing this for all three Pauli matrices means we need to a total of $N \to 3/(\delta \epsilon^2)$ measurements. Taking $\delta$ to be some small constant, say $0.01$, implies that the statistical error scales as $\epsilon \sim \sqrt{3/N}$.

Now we invert the approach:~rather than choosing the Pauli matrix $W \in \{X, Y, Z\}$ that we want to measure and selecting the appropriate $U \in \{\I, \Had, \G\}$, we \emph{first} select the unitary and then ask which Pauli basis it rotates $Z$ into. This perspective is perfectly valid if we are interested in learning all three Pauli observables, as we will eventually need to choose all three unitaries anyways. Since we need to perform many rounds of the experiment to collect ample statistics in the first place, we can allocate the three rotations uniformly among the $N$ rounds.

Suppose instead that we have decision anxiety about which unitary to implement at any given round of the experiment.\footnote{For example, perhaps we can only run the experiment for an unknown amount of time before funding runs out and power to the lab gets cut off.} Fortunately, we can appeal to classical randomness to make the decision for us:~for each $\ell \in [N]$, flip a fair three-sided coin\footnote{We leave the construction of such an object as an exercise to the reader.} to decide which $U$ to apply. This promotes $U$ to a random variable $U_\ell$, $\ell = 1, \ldots, N$, with uniform probabilities $\Pr[U_\ell = \I] = \Pr[U_\ell = \Had] = \Pr[U_\ell = \G] = 1/3$. This defines our distribution $\mathcal{D}$ of POVMs.

For clarity of exposition, suppose we want to estimate $\tr(X \rho)$ from the $N$ samples of this \emph{randomized} measurement scheme. (By uniformity, the following argument will apply to $Y$ and $Z$ as well.) On average, only $N/3$ samples will contain any information about $X$, which we assign a value of $(-1)^{x_\ell}$ as usual. The other $2N/3$ experiments do not measure in the $X$ basis (on average), and so are assigned a value of $0$. Thus in expectation,
\begin{equation}
	\E_{U_\ell \sim \mathcal{D}}[\tr(U_\ell^\dagger Z U_\ell \rho)] = \frac{1}{3} \tr(\Had^\dagger Z \Had \rho) + \frac{2}{3} \cdot 0 = \frac{1}{3} \tr(X \rho).
\end{equation}
To correct for this factor of $1/3$, we merely need to redefine the estimator as
\begin{equation}\label{eq:1q_classical_shadow_est}
	\hat{o}_\ell = \begin{cases}
	3 (-1)^{x_\ell} & \text{if } U_\ell = \Had,\\
	0 & \text{else},
	\end{cases}
\end{equation}
which obeys $\E[\hat{o}_\ell] = \tr(X \rho)$. To get an accurate estimate, average over all $N$ samples,
\begin{equation}
	\hat{o}(N) \coloneqq \frac{1}{N} \sum_{\ell=1}^N \hat{o}_\ell,
\end{equation}
which by linearity still obeys $\E[\hat{o}(N)] = \tr(X \rho)$. The convergence of this estimator also scales as $\epsilon \sim \sqrt{3/N}$;~we can show this using the same Chebyshev argument from before. The only modification we need to make is to compute the variance of $\hat{o}(N)$. As the samples are i.i.d.~(including the random draws of $U_\ell$), we have
\begin{equation}
\begin{split}
	\V[\hat{o}(N)] &= \frac{1}{N} \V[\hat{o}(1)]\\
	&= \frac{1}{N} \l( \frac{1}{3} \cdot 3^2 + \frac{2}{3} \cdot 0^2 \r) - \frac{1}{N} \tr(X \rho)^2\\
	&\leq \frac{3}{N}.
\end{split}
\end{equation}
Note that because all $N$ samples are used to estimate all three Pauli matrix expectation values, this is the final variance bound. Hence, we have sidestepped our decision anxiety through the use of randomization, while enjoying the same statistical guarantees as the deterministic approach.

\subsubsection{Many qubits}

It turns out that this sidestepping of decision making has fundamental consequences for constructing an efficient protocol. To see this, we must generalize to $n \geq 1$ qubits. Fortunately, it is simple to generalize the distribution $\mathcal{D}$ from the single-qubit setting:~simply perform random Pauli measurements on each qubit independently. Estimating any single-qubit Pauli observable, for example $X \otimes \I^{\otimes (n - 1)}$, proceeds as described above;~we merely have to keep track of qubit labels for the random unitaries $U = U^{(1)} \otimes \cdots \otimes U^{(n)}$ and measurement outcomes $b = b^{(1)} \cdots b^{(n)} \in \{0, 1\}^n$ that we acquire from each experimental round. We can estimate such observables in parallel over the $n$ qubits.

Since we have a composite system, we would like to learn not only single-body, but also many-body properties. For example, consider the set of two-local operators, $\I^{\otimes (i-1)} \otimes W^{(i)} \otimes \I^{\otimes (j - i - 1)} \otimes W^{(j)} \otimes \I^{\otimes (n-j)}$, where $1 \leq i < j \leq n$ and $W^{(i)}, W^{(j)} \in \{X, Y, Z\}$. In this scenario, a randomly selected $n$-qubit Pauli measurement ``hits'' $W^{(i)}$ and $W^{(j)}$ with probability $1/9$. Then, the appropriate estimator that we should define will be analogous to \cref{eq:1q_classical_shadow_est}, but with a factor of $9$ instead of $3$:
\begin{equation}\label{eq:two-qubit_shadow_est}
	\hat{o}_\ell = \begin{cases}
	9 (-1)^{b_\ell^{(i)} + b_\ell^{(j)}} & \text{if } (U_\ell^{(i)})^\dagger Z U_\ell^{(i)} = W^{(i)} \text{ and } (U_\ell^{(j)})^\dagger Z U_\ell^{(j)} = W^{(j)},\\
	0 & \text{else}.
	\end{cases}
\end{equation}
By the same analysis, the error of this estimate scales as $\epsilon \sim 3/\sqrt{N}$.

Now we make the key observation that reveals the power of randomization. For each $n$-qubit Pauli basis, there are $\binom{n}{2}$ two-local Pauli observables that are ``hit,'' i.e., acquire a nonzero estimate according to \cref{eq:two-qubit_shadow_est}. On the other hand, there are $9 \binom{n}{2}$ two-local Pauli operators in total. Thus we can conceptually lift our probabilistic interpretation of hitting individual operators to the hitting of collections of operators:~with probability $1/9$, a whole collection of $\binom{n}{2}$ operators are learned by a single choice of random measurement basis. By a coupon-collector argument, one may expect that on the order of ${\sim} (9 \log 9)$ random draws of measurement bases will result in every two-local Pauli operator being hit at least once.\footnote{This argument is not entirely valid, because commutation relations only allow certain collections of operators to be grouped together. Nonetheless, the Pauli matrices are sufficiently uniform such that this intuition at least morally holds.} Thus, up to some logarithmic corrections, taking $N \sim 9 / \epsilon^2$ random Pauli measurements suffices to obtain $\epsilon$-accurate estimates of all $9 \binom{n}{2}$ two-local Pauli observables. Very crucially, this sample complexity is independent of $n$, despite being able to learn quadratically many observables!

In the absence of randomization, if one were to learn each operator individually (say, by allocating each experimental round to only estimating a single two-local Pauli observable), this would require taking $N \sim 9\binom{n}{2}/\epsilon^2$ measurements. But this is far too naive:~as we have already pointed out, each measurement basis hits many operators. The question then is, can one construct a deterministic set of Pauli measurements which covers all two-local operators with only around $9$ Pauli POVMs? This is certainly possible in principle;~however, we need to construct this cover efficiently for any $n$. What about covering the set of three-local Paulis instead---do we need an entirely new construction algorithm? Four-local? What if we had performed the experiment with our measurement scheme optimized for two-local operators, but then in retrospect we now want to extract some three-body information from that data without running more experiments?

The elegance of randomized measurement schemes, and classical shadows in particular, is that all these challenges and concerns are sidestepped. Instead, an appropriately chosen distribution of random measurements does much of the heavy lifting for us, and we are left in the comfortable position of ``measure first, ask questions later''~\cite{elben2023randomized}. Furthermore, this approach is fully rigorous:~using statistical analysis, we can control the estimation accuracy and bound our confidence intervals.

\subsection{Description of the general theory}

This section provides an overview of full theory of classical shadows, generalizing beyond Pauli measurements~\cite{huang2020predicting}. The randomized measurement primitive can described as follows:~fix an ensemble of $n$-qubit unitaries $\mathcal{D} \subseteq \U(2^n)$ equipped with a probability distribution. For each copy of $\rho$, draw an independently random $U \sim \mathcal{D}$ and apply it to $\rho$. Then measure the state $U \rho U^\dagger$, obtaining a bit string $b \in \{0, 1\}^n$ with probability $\ev{b}{U \rho U^\dagger}{b}$. Finally, store a classical description of the postmeasurement state with the inverse unitary applied, $U^\dagger \op{b}{b} U$.

This measurement primitive implements the POVM $\{p(U) U^\dagger \op{b}{b} U \mid U \in \mathcal{D}, b \in \{0, 1\}^n\}$, where $p(U)$ is the probability measure on $\mathcal{D}$. In expectation, the quantum process is a channel (completely positive trace-preserving map):
\begin{equation}\label{eq:shadow_channel_background}
	\mathcal{M}(\rho) = \E_{U \sim \mathcal{D}} \sum_{b \in \{0, 1\}^n} \ev{b}{U \rho U^\dagger}{b} U^\dagger \op{b}{b} U.
\end{equation}
One can also view this as the partial trace of a two-fold twirl by $\mathcal{D}$:
\begin{equation}
	\mathcal{M}(\rho) = \tr_1\l[ \sum_{b \in \{0, 1\}^n} \mathcal{T}_{2,\mathcal{D}}(\op{b}{b}^{\otimes 2}) (\rho \otimes \I) \r],
\end{equation}
where the $t$-fold twirl is defined as
\begin{equation}
	\mathcal{T}_{t,\mathcal{D}}(A) \coloneqq \E_{U \sim \mathcal{D}} (U^\dagger)^{\otimes t} A U^{\otimes t}
\end{equation}
for any $A \in (\C^{2^n \times 2^n})^{\otimes t}$. Alternatively, from a representation of superoperator composition, one can write
\begin{equation}
	\mathcal{M} = \E_{U \sim \mathcal{D}} \mathcal{U}^\dagger \mathcal{M}_Z \mathcal{U},
\end{equation}
where $\mathcal{U}(\cdot) \coloneqq U (\cdot) U^\dagger$ and $\mathcal{M}_Z(\cdot) \coloneqq \sum_{b \in \{0, 1\}^n} \ev{b}{(\cdot)}{b} \op{b}{b}$.

If the effective POVM is informationally complete, then it is possible to invert $\mathcal{M}$,\footnote{If not, then one can still define a pseudoinverse which inverts over the span of the POVM elements and annihilates the kernel.} which leads to the seemingly trivial statement
\begin{equation}
	\rho = \mathcal{M}^{-1} \l(\mathcal{M}(\rho)\r).
\end{equation}
However, from the sampling perspective of $\mathcal{M}$ according to \cref{eq:shadow_channel_background}, we can use linearity to write
\begin{equation}
	\rho = \E_{U \sim \mathcal{D}} \E_{b \sim U \rho U^\dagger} \mathcal{M}^{-1}( U^\dagger \op{b}{b} U ),
\end{equation}
where $b \sim U \rho U^\dagger$ denotes draws of $\ket{b}$ according to the Born rule. Thus each sample $(U, b)$ of the measurement primitive is converted to a \emph{classical shadow snapshot} $\hat{\rho}_{U,b} \coloneqq \mathcal{M}^{-1}( U^\dagger \op{b}{b} U )$, called such because it is an unbiased estimator for $\rho$. That is, $\hat{\rho}_{U,b}$ obeys $\E_{U,b}[\hat{\rho}_{U,b}] = \rho$. While $\mathcal{M}^{-1}$ is not a physical process (it is not completely positive), the postmeasurement state $U^\dagger \op{b}{b} U$ is already stored as some classical description, so applying $\mathcal{M}^{-1}$ amounts to further classical computation. This can be achieved, for example, by computing an analytic expression for $\mathcal{M}$ using tools from representation theory whenever $\mathcal{D}$ is a compact group with the Haar measure. See Chapter~\ref{chap:EM} for further details.

Because classical shadows reproduce the density matrix in expectation, by linearity they can be used to predict expectation values:
\begin{equation}
	\tr(O_j \rho) = \E_{U,b} \tr(O_j \hat{\rho}_{U, b}).
\end{equation}
To study how quickly these estimates converge to the ground truth, we can analyze their variance. Let $T = KN$ be the total number of samples, $\hat{\rho}_1, \ldots, \hat{\rho}_T$, partitioned into $K$ groups of $N$ snapshots each. Huang \emph{et al.}~\cite{huang2020predicting} employ the median-of-means estimator
\begin{equation}
	\hat{o}_j(T, K) \coloneqq \mathrm{median}\l\{ \tr(O_j \hat{\rho}_{(1)}), \ldots, \tr(O_j \hat{\rho}_{(K)}) \r\},
\end{equation}
where each $\hat{\rho}_{(k)}$ is defined as the average classical shadow of the $k$th batch,
\begin{equation}
	\hat{\rho}_{(k)} \coloneqq \frac{1}{N} \sum_{\ell=(k-1)N + 1}^{kN} \hat{\rho}_\ell.
\end{equation}
An application of the Chebyshev and Hoeffding bounds implies that a sample complexity of
\begin{equation}
	T = \O\l( \epsilon^{-2} \log(L/\delta) \max_{j \in [L]} \V[\hat{o}_j] \r)
\end{equation}
suffices to solve the problem stated in \cref{eq:classical_shadow_error_target}. The variance appearing here is the single-shot variance of $\hat{o}_j \equiv \hat{o}_j(1, 1)$, defined in the standard way:
\begin{equation}
	\begin{split}
	\V[\hat{o}_j] &= \E_{U,b}\l[ \tr(O_j \hat{\rho}_{U,b})^2 \r] - \tr(O_j \rho)^2\\
	&= \E_{U \sim \mathcal{D}} \sum_{b \in \{0, 1\}^n} \ev{b}{U \rho U^\dagger}{b} \ev{b}{U \mathcal{M}^{-1}(O_j) U^\dagger}{b}^2 - \tr(O_j \rho)^2\\
	&= \tr\l[ \sum_{b \in \{0,1\}^n} \mathcal{T}_{3,\mathcal{D}}(\op{b}{b}^{\otimes 3}) \l( \rho \otimes \mathcal{M}^{-1}(O_j)^{\otimes 2} \r) \r] - \tr(O_j \rho)^2.
	\end{split}
\end{equation}
Above, we used the fact that $\mathcal{M}$ and its inverse are self-adjoint, i.e., $\tr(A \mathcal{M}^{-1}(B)) = \tr(\mathcal{M}^{-1}(A) B)$. Observe that the variance depends on the third moment of the distribution $\mathcal{D}$, as captured by the three-fold twirl $\mathcal{T}_{3,\mathcal{D}}$.

Using the universal bound $\tr(O_j \rho)^2 \geq 0$ and taking the supremum over all quantum states yields a state-independent variance bound called the \emph{shadow norm}:
\begin{equation}
	\begin{split}
	\sn{O_j} &\coloneqq \max_{\text{states } \sigma} \sqrt{ \E_{U \sim \mathcal{D}} \sum_{b \in \{0, 1\}^n} \ev{b}{U \rho U^\dagger}{b} \ev{b}{U \mathcal{M}^{-1}(O_j) U^\dagger}{b}^2 }\\
	&\geq \V[\hat{o}_j].
	\end{split}
\end{equation}
Indeed, the shadow norm obeys all the properties of a norm on the space of linear operators. This quantity is particularly useful for evaluating the efficiency of the classical shadows protocol as it is valid for all quantum states, and it can be (in principle) computed before running the experiment, given only knowledge of the user-specified inputs to the protocol (namely, the unitary ensemble $\mathcal{D}$ and the observables $O_j$ of interest).

The two examples considered by \cite{huang2020predicting} were the local Clifford group $\Cl(1)^{\otimes n}$ and the global $n$-qubit Clifford group $\Cl(n)$. For the former, they showed that $k$-local observables $O_j$ exhibit a shadow norm $\sn{O_j}^2 \leq 4^k \|O_j\|_\infty$, which simplifies to $\sn{O_j}^2 = 3^k$ whenever $O_j$ is a single $k$-local Pauli operator. Thus this protocol, essentially equivalent to the random Pauli measurements highlighted in \cref{sec:warm-up}, is highly efficient for estimating local qubit observables. For the latter distribution, the variance instead scales with the Frobenius norm of the operator, $\sn{O_j}^2 \leq 3 \tr(O_j^2)$, which is favorable for low-rank observables such as $\op{\psi}{\psi}$ (e.g., as in fidelity estimation).

In Chapter~\ref{chap:FCS} we will demonstrate another distribution, essentially of fermionic Gaussian unitaries, which achieves optimal shadow norm $\sn{O_j}^2 = \binom{2n}{2k} / \binom{n}{k}$ for $k$-body fermionic (Majorana) observables~\cite{zhao2021fermionic}.

Altogether we see that, unlike in shadow tomography, the scaling with respect to $1/\epsilon$ is only quadratic, which is much more manageable (and is in fact optimal without additional quantum resources or special assumptions). The scaling with the number of target observables $L$ is also only logarithmic, as is with $1/\delta$, both of which are optimal.

It is a somewhat subtle point that median-of-means estimation was necessary in order to achieve this optimal scaling with $L/\delta$, as otherwise (say, with a standard mean estimator) a union bound over all failure probabilities would result in a sampling bound linear in $L/\delta$, not logarithmic. This would be unacceptably loose, as there are many applications for which we would like to take $L = \poly(n)$ (for instance, local observable estimation). In Chapter~\ref{chap:FCS}, \cref{sec:hoeffding}, we touch on this subject, showing that under certain scenarios one can in fact apply a Bernstein inequality to the mean estimator to nonetheless achieve the desired $\log(L/\delta)$ scaling. \cite{helsen2022thrifty} also examined this point in some more detail, showing that median-of-means estimation can also be successfully replaced with the mean estimator when $\mathcal{D} = \U(2^n)$, the full unitary group, but not when $\mathcal{D} = \Cl(n)$, the $n$-qubit Clifford group. This is a particularly surprising result because the Clifford group is a unitary $3$-design, and as such has identical estimation formulas and variances to $\U(2^n)$ itself. This discrepancy is due to the difference in the higher moments ($t > 3$) of the distributions, which causes the tails of $\hat{o}_j$ to be exponentially bounded under $\U(2^n)$ but heavy under $\Cl(n)$. Unfortunately, while Clifford circuits can be efficiently constructed from $\O(n^2)$ gates~\cite{aaronson2004improved,bravyi2021hadamard}, Haar-random unitaries require exponentially many gates~\cite{knill1995approximation}. Thus it seems that the nicer tail properties of $\U(2^n)$ cannot be efficiently taken advantage of.

\subsection{Learning nonlinear functions}

As an aside, it turns out that classical shadows yield more than what was bargained for:~they can also be used to predict nonlinear functions of $\rho$. This is accomplished by manipulating the classical shadow $\hat{\rho}$ appropriately. For example, any quadratic function, such as $\tr(O \rho^2)$, can be expressed as the expectation value of some $2n$-qubit operator $O'$ on two copies of the state:~$\tr(O \rho^2) = \tr(O' \rho^{\otimes 2})$. Under this equivalence, the $T$ classical shadows can be used to construct estimates for $\rho^{\otimes 2}$, and hence the quadratic function, as
\begin{equation}
	\frac{1}{T(T-1)} \sum_{\ell \neq \ell' \in [T]} \tr(O' (\hat{\rho}_\ell \otimes \hat{\rho}_{\ell'})) \approx \tr(O \rho^2).
\end{equation}
The estimator is unbiased due to linearity and independence of each $\hat{\rho}_\ell$.

This aspect of classical shadows further broadens its appeal, as quantities such as the $\alpha$-R{\'e}nyi entanglement entropy [$\log(\tr(\rho^\alpha)) / (1 - \alpha)$] are both of high importance to many-body physics and also experimentally difficult to learn with standard techniques~\cite{elben2020mixed}. Although we will only be interested in predicting linear functions within this dissertation, further analysis of the nonlinear scenario can be found in the original paper~\cite{huang2020predicting}.
\chapter{Noisy Quantum Computation}\label{chap:nisq}


One of the greatest challenges facing the development and deployment of large-scale quantum processors is their susceptibility to noise. By virtue, quantum computers manipulate delicate states of Nature which are fragile to the effects of their harsh environment. Thus while quantum computation is theoretically believed to be more powerful than classical computation (expressed as the complexity-theoretic conjecture that the inclusion $\mathsf{BPP} \subset \mathsf{BQP}$ is strict), the task of building a quantum machine that is fully robust to its environment is both a necessary yet daunting one.

In light of such challenges, the clearest path forward is a paradigm known as quantum error correction~\cite{lidar2013quantum}. 
Shor first codified the idea in 1995~\cite{shor1995scheme}, which was soon followed by his celebrated threshold theorem~\cite{shor1996fault}. These results established the ability of a quantum computer, with access only to noisy components, to achieve \emph{fault tolerance}---the ability to correct errors faster than the rate at which they occur. His theorem set an upper bound on the required physical error rate, known as the \emph{threshold}, for fault tolerance;~above this threshold, errors would occur too frequently for a quantum error-correction code to keep up with. Initially, Shor's threshold required that each noisy gate have error rate at most $\O(1/\poly(\log t))$ to fault-tolerantly run a quantum circuit of $t$ gates. Within about a year, others improved this threshold to $\O(1)$, i.e., an error rate per gate independent of the size of the circuit~\cite{aharonov1997fault,knill1998resilient_rspa,kitaev2003fault}. This strengthened threshold solidified the promise of quantum error correction to enable universal, scalable quantum computation~\cite{gottesman1998theory}.

Unfortunately, current quantum technology is not yet below the required threshold for promising error-correction codes, such as the surface code~\cite{fowler2012surface}. At the time of this writing, substantial experimental progress has been made in demonstrating the essential components for a fault-tolerant, logically encoded qubit~\cite{egan2021fault,postler2022demonstration,zhao2022realization,sundaresan2023demonstrating,google2023suppressing,sivak2023real,ni2023beating}. However, device imperfections and limited qubit counts ultimately impede the realization of fault-tolerant quantum computation in the near term. In the face of this chasm, researchers began to investigate the capabilities of noisy devices in the absence of quantum error correction, a time frame coined by Preskill as the noisy intermediate-scale quantum (NISQ) era~\cite{preskill2018quantum}. The broad motivation is that, already at a modest number of qubits, say $n = 100$, the corresponding Hilbert space of complex dimension $2^n \approx 10^{30}$ is well beyond the reach of being simulated by the world's most powerful supercomputers. For example, strongly correlated quantum systems can be studied using classical algorithms on up to $50 \sim 70$ qubits, beyond which one must typically resort to approximate methods which may fail to provide sufficient accuracy~\cite{wecker2014gate}. The hope then is that quantum devices of such size, despite the imperfections and errors, might be able to outperform the best-known classical methods.

\section{Quantum supremacy and advantage}

There are two tall asks associated with demonstrating such a quantum advantage. First, because we can always compare against existing classical approximations, even if the the quantum algorithm could solve the problem exactly, we must assess how much accuracy is retained in the presence of decoherence and errors. Second, we are ultimately seeking a meaningful application of NISQ computers, i.e., the target problem should be one of potential scientific or industrial relevance.

To date, there have been few experimental demonstrations of so-called \emph{quantum supremacy}---a computational task that would take up to multiple millennia for the fastest classical supercomputers to solve, but at most a few hours for the quantum machine. The most salient example is \emph{random circuit sampling} (RCS), first experimentally demonstrated in 2019 by Google~\cite{arute2019quantum} and then in 2021 by USTC~\cite{wu2021strong}. Both experiments ran circuits of up to 20 cycles (layers of random single- and two-qubit gates) with 53 and 56 qubits, respectively. Both groups also recently reran their RCS experiments at larger sizes with improved devices (24 cycles with 70 and 60 qubits, respectively)~\cite{morvan2023phase,zhu2022quantum}.

The idea of RCS is that its hardness is derived from the hardness of simulating the quantum computer itself~\cite{aaronson2017complexity,boixo2018characterizing,bouland2019complexity}. The goal is essentially to output samples consistent with a distribution of random quantum circuits. Thus the task is designed such that the quantum computer can address it natively. On the other hand, the distribution is crafted so that the best-known classical algorithm is simply to simulate it with exponential cost. Even with such an inherent head start, the largest and most recent iteration of this experiment in 2023 (70 qubits at 24 cycles)~\cite{morvan2023phase} exhibited a fidelity of only $F \sim 0.0017$. That paper also further improved the best-known classical algorithms for simulating such low-fidelity RCS experiments~\cite{pan2022solving}, from which they estimated that it would take roughly $4.4 \times 10^{27}$ floating-point operations to classically simulate RCS with the same $F$. On the Frontier supercomputer at full capacity (ignoring issues of insufficient memory), they translated this to a wall-clock time of 3.3 millennia.

Regardless of its ultimate fate as a demonstration of quantum supremacy, RCS is is widely believed to be too artificially constructed to be relevant to meaningful problems. With that in consideration, researchers have instead looked toward problems such as combinatorial optimization, machine learning, and quantum simulation as potential avenues~\cite{bharti2022noisy}. The overarching strategy to tackle such problems is the use of short, parametrized quantum circuits to variationally optimize a cost function. This heuristic is broadly classified under variational quantum algorithms, which we discuss below.

\section{Variational quantum algorithms}

Noisy-device fidelities rapidly degrade with the length of the quantum circuit. Thus, there is a strong restriction on the quantum complexity on a noisy machine before its output becomes entirely meaningless. At the same time, many challenging computational tasks can be framed as optimization problems, e.g., searching for the minimum eigenvalue of some Hermitian operator $H$:
\begin{equation}
	E^\star = \min_{\text{states }\ket{\psi}} \ev{\psi}{H}{\psi}.
\end{equation}
For example, when $H$ is the Hamiltonian of a quantum system, then $E^\star$ is its ground-state energy. If $H$ is a classical (i.e., diagonal) Hamiltonian, then the problem corresponds to finding an optimal combinatorial solution.

The space of all quantum states is exponentially large. \emph{Variational quantum algorithms} (VQAs) therefore restrict this search space via a parametrized quantum circuit (PQC) $U(\bm{\theta})$, where $\bm{\theta} \in \Theta \subseteq \R^m$ is a vector of $m$ real parameters. For instance, the parameters are typically encoded as the angles of the gates making up the PQC. This restricts the problem to finding
\begin{equation}
	E(\bm{\theta}^\star) = \min_{\bm{\theta} \in \Theta} \ev{\psi(\bm{\theta})}{H}{\psi(\bm{\theta})},
\end{equation}
where, without loss of generality, $\ket{\psi(\bm{\theta})} \coloneqq U(\bm{\theta}) \ket{0^n}$.

To offload arithmetic complexity from the quantum device, VQAs perform the minimization via a quantum--classical feedback loop. This involves calls to the quantum computer only to evaluate (measure) the cost function $E(\bm{\theta})$ by executing the PQC. Meanwhile, the navigation through parameter space via estimated energy gradients is performed on a classical computer (e.g., by the Nelder--Mead simplex method). The philosophy behind VQAs is to identify the essential ``quantum hardness'' of a problem, which the PQC handles;~the rest of the algorithm is handed to a reliable classical processor which we assume can do arbitrary amounts of polynomial-time computation. By the variational principle, we always have that $E(\bm{\theta}^\star) \geq E^\star$, with equality only when the PQC is expressible enough to capture a ground state within its parameter space.

While it is known that the ground-state problem is $\mathsf{QMA}$-hard in general~\cite{kempe2006complexity} (hence intractable even for a quantum computer), the hope is that either some Hamiltonian of interest has enough structure to avoid that complexity, or otherwise PQCs can achieve higher-quality approximations than competing classical algorithms in practice. Indeed, this was the motivation behind the first proposed VQA, the \emph{variational quantum eigensolver} (VQE) for approximating ground-state energies of quantum-chemical systems~\cite{peruzzo2014variational}. To date, a variety of experiments have been performed to investigate the performance of VQE in state-of-the-art quantum processors~\cite{omalley2016scalable,kandala2017hardware,colless2018computation,dumitrescu2018cloud,hempel2018quantum,kandala2019error,kokail2019self,nam2020ground,arute2020hartree,stanisic2022observing,motta2023quantum,obrien2023purification};~however, neither the system sizes nor hardware performance are yet sufficient to demonstrate a quantum advantage in this domain.

The framework of VQAs has been imported into a variety of other domains, such as the quantum approximate optimization algorithm (QAOA\footnote{This acronym was later co-opted and reworked into an generalized framework called the ``quantum alternating operator ansatz''~\cite{hadfield2019quantum}.})~\cite{farhi2014quantum} for combinatorial optimization (which is $\mathsf{NP}$-hard in general) and quantum neural networks for machine-learning tasks~\cite{mitarai2018quantum}. While differing in scope and precise methodology, all such variational approaches face a number of universal challenges:
\begin{enumerate}
	\item \textbf{Ansatz expressibility.} Clearly, we desire PQCs whose parameter space actually contains a good approximation to the problem solution. This is essentially an educated guess, or \emph{ansatz}, based on perhaps some physical intuition or symmetry. Expressibility, however, is hindered by the fact that we desire compact PQCs to avoid the overwhelming accumulation of device errors. Broadly speaking, $\poly(\log n)$-depth circuits may already be sufficient to demonstrate quantum advantage, since current classical algorithms for simulating arbitrary quantum circuits are only efficient at constant depth~\cite{bravyi2021classical}). This is however a subtle issue to consider, because certain classes of quantum circuits may have $\poly(n)$ gates yet be simulable by special techniques (e.g., Clifford~\cite{gottesman1998heisenberg} or matchgate~\cite{valiant2001quantum} circuits).
	
	\item \textbf{Barren plateaus.} It was first realized in \cite{mcclean2018barren} that the parameter landscape for common cost functions (e.g., local Hamiltonians) exhibits \emph{barren plateaus}:~large regions of parameter space wherein the gradients in any direction are exponentially small. This effectively renders the search for the optimal parameters intractable. The problem of barren plateaus has been investigated in a wide number of scenarios, revealing close relations to a number of other PQC properties such as circuit depth and cost-function locality~\cite{cerezo2021cost}, the presence of noise~\cite{wang2021noise}, and trade-offs with ansatz expressibility~\cite{holmes2022connecting}. On the other hand, the barren-plateaus phenomenon is a statement made on average over the parameter landscape, and it has been proposed that good initial guesses might circumvent the problem~\cite{zhang2022escaping}. Alternatively, diagnostics can be probed to detect and potentially avoid the encounter of a barren plateau~\cite{sack2022avoiding}.
	
	\item \textbf{Number of measurements.} The actual value of any $E(\bm{\theta})$ cannot be extracted exactly from a quantum computer. Instead, it must be learned from sampling the quantum circuit, as discussed in Chapter~\ref{chap:learning}. This can be ``astronomically large''~\cite{wecker2015progress};~for example, the electronic-structure Hamiltonian naively decomposes into $\O(n^4)$ simple-to-measure terms. Estimating gradients of the cost function also has this issue;~the analytic-gradient method~\cite{schuld2019evaluating} is a numerically stable approach (in contrast to finite difference), but requires measuring even more properties from the quantum state. Note that these costs are further magnified by the fact that we require this large number of measurements at each step of the variational minimization.
	
	\item \textbf{Noisy quantum hardware.} Even ignoring all the above challenges, the effects of noise in non-error-corrected devices impose a floor to the ultimate accuracy of the VQA. While one can always hope to build hardware with less imperfections and more precise control, one question to ask is what kinds of \emph{algorithmic} protocols can be applied, to bootstrap more accurate quantum calculations from an inherently noisy machine.
\end{enumerate}
Quantum error mitigation was developed to address this final point, and below we discuss this topic in closer detail.

\section{Quantum error mitigation}

Quantum error correction aims to detect and, as the name suggests, correct errors throughout the course of the computation in a coherent fashion. In contrast, \emph{quantum error mitigation} (QEM) works by running the noisy device at or above its inherent error rate, over an ensemble of various circuits. The errors are not coherently corrected, but rather the effects of the errors (say, in the estimation of observables) are approximately countered by appropriate classical postprocessing of the outcomes from this ensemble of circuit runs.

Different QEM strategies have been developed, although they share a common penalty:~the variance of error-mitigated estimation increases, thereby requiring more samples compared to a noiseless protocol. Generally speaking, the goal is to decrease the bias of estimation at the cost of higher variance, so that although it takes longer to converge to the mean, this mean is closer to the output of an ideal quantum computer than what the noisy device can produce on its own. How this is accomplished, however, varies from strategy to strategy. The best way to get a sense of what QEM looks like is by examining a few of the most popular ideas. This is by no means a comprehensive survey, and we recommend the interested reader to see \cite{cai2022quantum} for an extensive review article on QEM.

In all the examples below, we take the task to be the estimation of the expectation value $\tr(O \rho)$ of some observable $O$ with respect to an ideally prepared state $\rho$ (although we do not have the ability to prepare $\rho$ noise-free). The sampling overhead factor will be denoted by $C$.

\subsection{Zero-noise extrapolation}

Zero-noise extrapolation (ZNE)~\cite{li2017efficient,temme2017error} operates under the assumption that the accumulation of errors is sufficiently well-behaved. Suppose that the noise can be effectively modeled in terms of a ``circuit fault rate'' $\lambda \geq 0$, such that the noisy quantum circuit prepares the state $\rho(\lambda)$. The ideal, noiseless state is $\rho \equiv \rho(0)$, although at best we only have access to $\rho(\lambda_{\min})$ for some $\lambda_{\min} > 0$. Then, rather than reduce the fault rate, one can consider \emph{boosting} the error rate within a sequence $\lambda_{\min} \leq \lambda_1 < \lambda_2 < \cdots$. This might be achieved, for example, by inserting ``noisy identities'' within the circuit (e.g., $G^\dagger G = \I$ for any gate $G$)~\cite{giurgica2020digital}.

The expectation value $\tr(O \rho(\lambda))$ can then be modeled as a function $f(\lambda; \bm{p})$ with some free model parameters $\bm{p} \in \R^M$ to be fit. If $\lambda$ is small, one can make the following approximation:
\begin{equation}
	\tr(O \rho(\lambda)) \approx f(\lambda; \bm{p}) = \sum_{\ell=0}^{M-1} p_\ell \frac{\lambda^\ell}{\ell!}.
\end{equation}
Supposing we evaluate the noisy device at $M$ different fault rates $\lambda_1, \ldots, \lambda_M$, we can find the optimal parameters $\bm{p}^\star$ such that the model best fits the measured expectation values:~$\tr(O \rho(\lambda_m)) \approx f(\lambda_m; \bm{p})$. Then we extrapolate to the zero-noise value by evaluating the model $f(\lambda; \bm{p}^\star)$ at $\lambda = 0$. For example, a popular option is Richardson extrapolation, for which the ZNE prediction is equal to the zeroth component of the optimized parameters~\cite{giurgica2020digital}:
\begin{equation}
	\tr(O \rho(0)) \approx p_0^\star = \sum_{m=1}^M \tr(O \rho(\lambda_m)) \prod_{k \neq m} \frac{\lambda_k}{\lambda_k - \lambda_m}.
\end{equation}
Assuming that the noise behaves under this circuit-fault-rate model, the bias in this estimate is of order $\lambda^M$. The sampling overhead scales by a factor of
\begin{equation}
	C \sim \l( \sum_{m=1}^M \prod_{k \neq m} \frac{\lambda_k}{\lambda_k - \lambda_m} \r{)^2},
\end{equation}
which can be shown from a variance analysis of $p_0^\star$.

\subsection{Probabilistic error cancellation}

Probabilistic error cancellation (PEC)~\cite{temme2017error} is an approach using Monte Carlo sampling techniques to, on average, cancel the effects of noisy gates. The general theory can be described as follows. Let $U = U_L \cdots U_1$ be the ideal quantum circuit desired, decomposed into a sequence of elementary gates $U_j$. The channel representation of a unitary $U$ will be denoted by $\mathcal{U}$. In the experiment, we only have access to noisy physical gate operations $\{\widetilde{\mathcal{G}}_k\}_k$. Suppose each ideal gate $\mathcal{U}_j$ can be decomposed into a linear combination over the noisy gate set,
\begin{equation}
	\mathcal{U}_j = \sum_k \gamma_k^{(j)} \widetilde{\mathcal{G}}_k, \quad \gamma_k^{(j)} \in \R.
\end{equation}
This is called a \emph{quasiprobability decomposition} of $\mathcal{U}_j$, since it can be implemented in expectation by treating the coefficients $\gamma_k$ as a quasiprobability distribution. That is, for each gate $j \in [L]$, we sample $k$ from the probability distribution $|\gamma_k^{(j)}| / \|\bm{\gamma}^{(j)}\|_1$, implement the corresponding $\widetilde{\mathcal{G}}_k$, and store the sign of $\gamma_k^{(j)}$. Then after performing all the experiments, when averaging over the measurement data we reweight this particular sample by a factor of $\|\bm{\gamma}^{(j)}\|_1 \sgn(\gamma_k^{(j)})$.

As a simple demonstrative example of how this works, suppose the noisy operation $\widetilde{\mathcal{U}}$ is the ideal unitary $\mathcal{U}$, followed by a depolarizing channel of strength $\lambda$. The form of this channel can be written as $\widetilde{\mathcal{U}} = (1 - \lambda)\mathcal{U} + \lambda \mathcal{E}$, where $\mathcal{E}(\rho) \coloneqq \tr(\rho) \I / 2^n$ sends every input to the maximally mixed state. Rewriting this expression reveals
\begin{equation}
	\mathcal{U} = \frac{1}{1 - \lambda} \widetilde{\mathcal{U}} - \frac{\lambda}{1 - \lambda} \mathcal{E}.
\end{equation}
The channel $\mathcal{E}$ must be further written in terms of the physical gate set. For example, we can use its Kraus decomposition in terms of all Pauli operators:
\begin{equation}
	\mathcal{E} = \frac{1}{4^n} \sum_{P \in \{\I, X, Y, Z\}^{\otimes n}} \mathcal{P}.
\end{equation}
Assuming we can perform perfect Pauli gates,\footnote{This is typically a reasonable assumption, because in most platforms single-qubit gates have infidelities at least an order of magnitude lower than those of two-qubit gates.} this gives a physically implementable quasiprobability decomposition as desired. With probability $1/(1 + \lambda)$, we implement the noisy unitary $\widetilde{\mathcal{U}}$ as usual. But with probability $\lambda/(1 + \lambda)$, we implement a random Pauli matrix on each qubit instead. Then after a sufficient number of rounds of this randomized protocol, each measurement outcome wherein we performed $\widetilde{\mathcal{U}}$ is scaled by $(1 + \lambda)/(1 - \lambda)$, while the outcomes associated with the random Pauli gates is scaled by $-(1 + \lambda)/(1 - \lambda)$ instead.

Returning to the general framework with $L$ gates, the overall rescaling factor is a product of each gate's quasiprobability $1$-norm, and in a straightforward manner the variance of estimation is augmented by the square of that factor:
\begin{equation}
	C \sim \prod_{j=1}^L \| \bm{\gamma}^{(j)} \|_1^2.
\end{equation}

An alternative approach to PEC is to assume that each noisy gate takes the form $\widetilde{\mathcal{U}}_j = \mathcal{E}_j \mathcal{U}_j$. After each gate, we would like to apply the channel inverse $\mathcal{E}_j^{-1}$. However, this inverse is typically not a physically implementable map (usually failing to be completely positive). Instead, we find a quasiprobability decomposition for $\mathcal{E}_j^{-1} = \sum_k \gamma_k^{(j)} \widetilde{\mathcal{G}}_k$ and effectively implement it using the Monte Carlo technique outlined in the previous approach. The sampling overhead for this approach is of the same form, although the $\gamma_k^{(j)}$'s will be different and therefore might be advantageous depending on the details of the noise.

Essential to either of these PEC methods is an accurate characterization of the error channels, as otherwise one does not know how to define the correct quasiprobability decomposition. While in principle this can be achieved by careful calibration of the quantum device, in practice this is both costly and might not accurately capture error effects if they do not appear in calibration (e.g., crosstalk or time-dependent phenomena). Therefore, a number of learning-based approaches have been developed, wherein effective models for the quasiprobability decomposition are fit to via training data~\cite{strikis2021learning,van2023probabilistic}. For example, the training data might consist of circuits which are structurally similar to the desired circuit of interest, but can be efficiently validated by classical simulation.

\subsection{Training error-mitigation models}

Ideas from machine learning can be imported into QEM more broadly as well. Here we will describe two similar methods that train a simple regression model using noisy experimental data from classically simulable quantum circuits:~Clifford data regression (CDR)~\cite{czarnik2021error} and training with fermionic linear optics (TFLO)~\cite{montanaro2021error}. As their names imply, they train models from (nearly) Clifford and free-fermion circuits, respectively, both of which can be simulated efficiently with classical computation. The structure of the circuits, however, are chosen to be close to that of the target circuits (which are ideally not classically simulable), such that we expect the generalization error of this model to be small.

Both ideas use a linear regression model, based on the fact that many forms of decohering errors corrupt the quantum state as $\rho \mapsto \widetilde{\rho} = (1 - \lambda) \rho + \lambda \sigma$, where $\sigma$ is some other state and $0 \leq \lambda \leq 1$. The quantum state itself is produced by some PQC with parameters $\bm{\theta} \in \Theta$, so we write the ideal states as $\rho(\bm{\theta})$. Assuming that neither the overall fault rate $\lambda$ nor the erroneous state $\sigma$ change when the parameters are varied, the noisy expectation value can be modeled as
\begin{equation}
	\tr(O \widetilde{\rho}(\bm{\theta})) = (1 - \lambda) \tr(O \rho(\bm{\theta})) + \lambda \tr(O \sigma).
\end{equation}
The noisy values $\tr(O \widetilde{\rho}(\bm{\theta}))$ are measured from the quantum computer, while the noise-free values $\tr(O \rho(\bm{\theta}))$ can be classically computed at special values of $\bm{\theta} \in \mathcal{T} \subset \Theta$, corresponding to the set of training circuits. Then $a = 1/(1 - \lambda)$ and $b = -\lambda \tr(O \sigma) / (1 - \lambda)$ are the unknown model parameters that we optimize via least-squares minimization:
\begin{equation}
	(a^\star, b^\star) = \argmin_{a, b} \sum_{\bm{\theta} \in \mathcal{T}} \l| a\tr(O \widetilde{\rho}(\bm{\theta})) + b - \tr(O \rho(\bm{\theta})) \r{|^2}.
\end{equation}
The linear-regression model is therefore
\begin{equation}
	f(\bm{\theta}; a^\star, b^\star) = a^\star \tr(O \widetilde{\rho}(\bm{\theta})) + b^\star,
\end{equation}
which we hope will be close to the noise-free expectation value $\tr(O \rho(\bm{\theta}))$ at any $\bm{\theta} \in \Theta$, not just within the training set.

In the case of CDR, one sets the training parameters such that most of the gates are Clifford and only at most $\O(\log n)$ gates are non-Clifford;~such nearly Clifford circuits can be classically simulated in polynomial time~\cite{bravyi2016improved}. For example, the parametrized gate $e^{-\i \theta P/2}$, where $P$ is any Pauli operator, is Clifford whenever $\theta \in \{-\pi/2, 0, \pi/2, \pi\}$.

For TFLO, the circuits are usually based on elementary one- and two-body fermionic unitaries. For example, $e^{-\i \theta (a_p^\dagger a_q + a_q^\dagger a_p) / 2}$ is one-body, while $e^{-\i \theta a_p^\dagger a_p a_q^\dagger a_q}$ is two-body. Then, given some circuit structure of such gates, we simply remove all the two-body gates\footnote{Or set their angles to be a full $2\pi$ rotation such that they implement a noisy identity.} such that the resultant circuit is purely one-body, or a fermionic linear-optical network. Such circuits can be efficiently classically simulated~\cite{terhal2002classical} (see also Chapter~\ref{chap:fermions}).

Unfortunately, the sampling overhead of these approaches is difficult to analytically study. In part, this is because the assumption that the noise obeys the linear regression model is virtually never observed in practice, although it is a reasonable approximation. It is also difficult to rigorously ascertain that the training data is representative of the full parameter space, which may require, e.g., out-of-distribution techniques~\cite{caro2023out} to properly analyze. However, numerical and experimental evidence suggests that the resources required are reasonable to achieve substantial error mitigation in practice~\cite{czarnik2021error,montanaro2021error,stanisic2022observing}.

\subsection{Symmetry verification}

Symmetry-based QEM is a conceptually simple approach that uses the ubiquitous presence of symmetries in physical systems. If we know that the ideal state $\rho$ should lie in a symmetry sector (eigenspace) of some symmetry operator $S$, then we can detect and potentially mitigate against errors which leak the state into the wrong sectors by measuring that symmetry. Note that this is the core idea behind much of quantum error correction~\cite{lidar2013quantum};~for example in stabilizer codes, the joint $+1$-eigenspace of a set of Pauli operators defines a subspace of quantum states called a codespace, in which the encoded, error-free quantum information should lie. Correctable errors then take the state outside of that codespace, which we detect and correct through syndrome measurements of the code.

In the case of QEM, we do not aim to correct the errors coherently, but rather use classical postprocessing (and perhaps low-overhead detection circuitry) to counteract the effects of symmetry-violating noise. Besides, it is too costly to build a quantum code which can detect a large variety of errors;~instead, we typically restrict ourselves to whichever symmetries are inherently provided by the physics of the system. The simplest method is postselection on measurement outcomes, wherein we discard all the data that do not obey the symmetry. This is broadly known as symmetry verification (SV)~\cite{mcardle2019error,bonet2018low}. However, this is only possible, without additional circuitry, when the symmetry operator commutes with the observable $O$ (or more generally, its expansion terms). For example, suppose the system obeys a parity symmetry, corresponding to the operator $S = Z^{\otimes n}$. This means the ideal state only has support on either all even- or odd-Hamming-weight basis states. Then only observables which are also diagonal in the $Z$ basis can be estimated from the data that measured $S$.

If we allow for additional circuitry, then it is possible to develop SV techniques with broader scope. For example, one can introduce an ancilla qubit and apply $n$ $\mathrm{CNOT}$ gates to it, where each $\mathrm{CNOT}$ is controlled on a qubit from the system register. If the system did not experience a parity-violating error, then the measurement outcome of the ancilla is predictable (either $\ket{0}$ or $\ket{1}$, depending on the correct parity sector). Otherwise, if such an error did occur, then the ancilla will reveal this fact by returning the wrong output. The system itself, meanwhile, is free to be measured in whichever basis one desires. The sampling overhead of performing postselection in this manner is inversely proportional to the ``pass rate,'' or the percentage of outcomes that pass the symmetry check. Specifically, let $\widetilde{\rho}$ be the noisy state and $\Pi$ the projector onto the correct symmetry sector. That is,
\begin{equation}
	C \sim \tr(\Pi \widetilde{\rho})^{-1}
\end{equation}
for this form of coherent SV. While this overhead is more favorable compared to some other approaches (for example, the incoherent SV that we will describe below), parity verification already requires an additional $n$ two-qubit ($\mathrm{CNOT}$) gates. The verification circuit for more complex symmetries, for example particle number $\eta$, requires $n \lceil \log \eta \rceil$ controlled-phase gates~\cite{mcardle2019error}.

An alternative to the coherent detection of errors is to perform a virtual projection via classical postprocessing on the measurement record~\cite{bonet2018low}.\footnote{This idea is closely related to techniques called quantum subspace expansion~\cite{mcclean2017hybrid,mcclean2020decoding}.} This is an incoherent approach to SV. Let $\Pi$ be the projector as before;~for example, in our simple parity example, $\Pi = (\I + S) / 2$ projects onto the even-parity subspace. Using the fact that the postselected state is $\rho' = \Pi \widetilde{\rho} \Pi / \tr(\Pi \widetilde{\rho})$, we can write expectation values for $\rho'$ as
\begin{equation}\label{eq:SV_postprocessing}
	\tr(O \rho') = \frac{\tr(O' \widetilde{\rho})}{\tr(\Pi \widetilde{\rho})},
\end{equation}
where $O' = \Pi O \Pi = (O + SO + OS + SOS) / 4$. Then rather than actually postselect the measurements, we instead estimate the noisy expectation values these four operators and $S$. Using \cref{eq:SV_postprocessing}, we can construct estimates for $\tr(O \rho')$ without needing the additional controlled operations and ancilla. The cost of this virtual postselection is that the sampling overhead now scales as
\begin{equation}
	C \sim \tr(\Pi \widetilde{\rho})^{-2},
\end{equation}
because the appearance of the pass rate in the denominator of \cref{eq:SV_postprocessing} translates to an inverse-quadratically scaling factor in the variance~\cite{huggins2021efficient,cai2021quantum}.

This idea of postprocessing to effect postselection has been extended in a number of ways, for example, recombining the expansion operators with different weights to reduce bias~\cite{cai2021quantum} or using classical shadows to efficiently estimate the multiple expansion operators with a single protocol~\cite{jnane2023quantum}. However, these SV approaches suffer from only being blind to symmetry-respecting errors. Thus projecting the noisy state into the correct symmetry sector has no guarantee for the closeness of the projected state to the ideal state. In Chapter~\ref{chap:EM}, we introduce a technique called \emph{symmetry-adjusted classical shadows}~\cite{zhao2023group}, which combines classical shadows with symmetry information in a paradigm beyond that of SV. Under simplifying noise assumptions, one can guarantee the correctness of the error-mitigated estimates with this protocol for a wide range of noise models. The key idea is that the twirling aspect of classical shadows scrambles the noise uniformly throughout the quantum state, so that (1) a much wider class of errors become scrambled into symmetry-violating form, and (2) detectable errors from just the symmetry become valid diagnostics for how the errors affect the rest of the state.

\subsection{Virtual distillation}

Virtual distillation (VD)~\cite{huggins2021virtual}, also introduced contemporaneously as error suppression by derangement (ESD)~\cite{koczor2021exponential}, takes advantage of the fact that the ideal state is typically pure, $\rho = \op{\psi}{\psi}$. It furthermore assumes that the dominant errors in real hardware are mostly incoherent, or can be made as such~\cite{wallman2016noise}. Consider the spectral decomposition of a noisy density matrix,
\begin{equation}
    \widetilde{\rho} = \sum_{i=0}^{2^n - 1} p_i \op{\chi_i}{\chi_i},
\end{equation}
where the labels are such that $p_0 \geq \cdots \geq p_{2^n-1}$. Assuming that $p_0 > p_1$ holds strictly, then taking powers of the density matrix
\begin{equation}
    \rho^{(M)} = \frac{\widetilde{\rho}^M}{\tr(\widetilde{\rho}^M)}
\end{equation}
converges to $\lim_{M\to\infty} \rho^{(M)} = \op{\chi_0}{\chi_0}$ exponentially quickly in $M$.

In order to access the state $\rho^{(M)}$, VD/ESD proposes to ``virtually distill'' the dominant eigenvector by preparing $M$ copies of $\widetilde{\rho}$ in parallel and performing entangled measurements across the copies. Rather than coherently purify the state, it estimates properties of $\rho^{(M)}$ by employing the identity
\begin{equation}
    \tr(O \rho^M) = \tr(\overline{O} S_M \widetilde{\rho}^{\otimes M}),
\end{equation}
where $S_M$ is the cyclic shift operator on the $M$ systems and
\begin{equation}
    \overline{O} = \frac{1}{M} \sum_{j=1}^M \I \otimes \cdots \otimes \I \otimes \underbrace{O}_{j\text{th system}} \otimes \I \otimes \cdots \otimes \I 
\end{equation}
is the target observable, symmetrized over the $M$ systems. Then the mitigated expectation value can be calculated as
\begin{equation}
    \tr(O \rho^{(M)}) = \frac{\tr(O \rho^M)}{\tr(\rho^M)} = \frac{\tr(\overline{O} S_M \widetilde{\rho}^{\otimes M})}{\tr(S_M \widetilde{\rho}^{\otimes M})},
\end{equation}
which involves measuring the observables $\overline{O} S_M$ and $S_M$ on the $M$-system state. Is it worth noting that this technique can be viewed as another use of symmetry to perform QEM;~in this case, the symmetry is artificially introduced through the preparation of $M$ copies of the same $n$-qubit state.

While a powerful technique for suppressing device errors, VD/ESD involves a number of important considerations. First, note that the pure state (the dominant eigenvector) that is being distilled is $\ket{\chi_0}$, which has no guarantee to be the ideal state $\ket{\psi}$. As a trivial counterexample, a purely coherent error will unitarily rotate $\ket{\psi}$ to some other pure state, against which VD/ESD has no hope of mitigating. In general, this ``coherent mismatch,'' defined as the infidelity $c = 1 - |\ip{\psi}{\chi_0}|^2$, was studied in depth in \cite{koczor2021dominant}. One important result from that work is that, while the ultimate noise floor is bounded as $\Delta = |\ev{\psi}{O}{\psi} - \ev{\chi_0}{O}{\chi_0}| \leq 2 \sqrt{c} \|O\|_\infty$ for arbitrary states, a quadratically smaller bound of $\Delta \leq 2 c \|O\|_\infty$ holds whenever $\ket{\psi}$ is an eigenstate of $O$, which is the setting of most VQAs.

Another challenge is the measurement of the highly global observables $\overline{O} S_M$ and $S_M$. The presence of the cyclic shift operator particularly complicates matters. For $M = 2$ and $O$ being a single-qubit observable, circuits were given in \cite{huggins2021virtual} that require only a single layer of transversal two-qubit gates between identically labeled qubits across the copies. However, already for $M = 3$ the measurement circuit needed to be numerically approximated and optimized by \cite{huggins2021virtual}. For larger $M$, ancilla qubits are needed to perform Hadamard tests for the measurement~\cite{huggins2021virtual,koczor2021exponential}. Furthermore, if $O$ is not a single-qubit observable then even more circuitry and ancilla are required. The use of classical shadows for VD/ESD has been considered to circumvent these challenges~\cite{seif2023shadow,hu2022logical}, requiring only a single copy and using the shadows to predict the nonlinear functions of the state. However, such approaches feature an exponentially large sample complexity upfront.

For $M = 2$, the sampling overhead of the standard VD/ESD technique vanishes quadratically with the impurity of the noisy state:
\begin{equation}
    C \sim \tr(\widetilde{\rho}^2)^{-2}.
\end{equation}
One may observe that this behavior is similar to that of SV via postprocessing, as a consequence of the ``virtual'' (incoherent) aspect. For general $M$, $\tr(\widetilde{\rho}^M)$ is exponentially small in $M$ (unless $\widetilde{\rho}$ is pure) and so the sampling overhead will grow exponentially in $M$.

\subsection{Verified phase estimation}

The last example we review here is verified phase estimation (VPE)~\cite{obrien2021error}, which falls broadly under a category of techniques known as echo verification~\cite{cai2021resource,huo2022dual}. In a sense, VPE is similar to VD/ESD in that it uses two copies of a(n ideally pure) state $\rho = \op{\psi}{\psi}$ to perform error mitigation;~however, whereas in VD those copies are spatially separated, in VPE they are \emph{temporally} separated. The basic premise is that, after preparing $\ket{\psi} = U\ket{0^n}$ from some state-preparation circuit $U$, applying the inverse $U^\dagger$ (hence the ``echo'') and measuring the system should always return a deterministic output, $\ket{0^n}$, in the absence of noise. Of course, in the presence of errors corrupting both $U$ and $U^\dagger$, the final noisy state will support some nontrivial distribution in the computational basis. Thus we can use this measurement as a verification of whether or not any errors have occurred---if any output other than $\ket{0^n}$ is seen, we throw that run out (technically, we do not postselect but rather assign such unverified runs an estimate of $0$ while retaining that sample in our statistics). It can be shown that this is equivalent to measuring the state $(\widetilde{\rho} \rho + \rho \widetilde{\rho}) / (2 \tr(\widetilde{\rho} \rho))$~\cite{huo2022dual}, hence the similarity to second-order VD/ESD.

Because we are measuring the system qubits to perform the echo verification, we need an alternative way to extract the information about observables from the state. This is achieved using ideas from phase estimation~\cite{somma2019quantum}. First, decompose the target observable $O$ into a linear combination of terms $H_s$,
\begin{equation}
	O = \sum_{s=1}^L h_s H_s.
\end{equation}
such that the unitary $V_s(t) = e^{-\i H_s t}$ can be implemented for any $t$ in a NISQ-friendly manner. For example, a local Pauli decomposition satisfies this condition. Then for each $s \in [L]$, we perform a single-ancilla phase estimation experiment with $V_s(t)$:~apply the unitary c-$V_s(t)$ on $\ket{\psi}$, controlled on an ancilla qubit prepared in $\ket{+} = (\ket{0} + \ket{1}) / \sqrt{2}$. This yields the $(n + 1)$-qubit state
\begin{equation}
	\ket{\Psi(t)} = \sum_j a_{s,j} \l(\frac{\ket{0} + e^{-\i E_{s,j} t} \ket{1}}{\sqrt{2}}\r{)_{\mathrm{anc}}} \otimes \ket{E_{s,j}}_{\mathrm{sys}},
\end{equation}
where $H_s = \sum_j E_{s,j} \op{E_{s,j}}{E_{s,j}}$ is the spectral decomposition of the term and $a_{s,j} = \ip{E_{s,j}}{\psi}$. The state of the control qubit is then
\begin{equation}
	\rho_{\mathrm{anc}}(t) = \tr_{\mathrm{sys}}(\op{\Psi(t)}{\Psi(t)}) = \frac{1}{2} \begin{pmatrix}
	1 & g_s(t)\\
	g_s^*(t) & 1
	\end{pmatrix},
\end{equation}
where
\begin{equation}
	g_s(t) = \sum_j |a_{s,j}|^2 e^{-\i E_{s,j} t}
\end{equation}
is the phase function containing information about the observable $H_s$ within $\ket{\psi}$. Note that performing the inverse circuit $U^\dagger$ on the system does not change the state of the ancilla qubit. Thus we can both perform verification and learn properties from the system by probing the ancilla.

Specifically, one can learn this function by implementing c-$V_s(t_k)$ at various time points $t_1, t_2, \ldots$ and measuring the control qubit in the $X$ and $Y$ bases. This reveals the signal because we can express $g_s(t) = \tr(X \rho_{\mathrm{anc}}(t)) + \i \tr(Y \rho_{\mathrm{anc}}(t))$. Classical signal-analysis methods can then be employed to extract the amplitudes $|a_{s,j}|^2$. Finally, the expectation value can be estimated from these amplitudes using $\ev{\psi}{H_s}{\psi} = \sum_j |a_{s,j}|^2 E_{s,j}$. This requires knowledge of the eigenvalues of $H_s$, which is generally available because they are our basis operators for the decomposition of $O$;~for example, the spectrum of Pauli operators is simply $\{\pm 1\}$. For such bounded operators, the number of time points required to resolve the signal is also bounded, so the number of shots required is the usual $\O(\epsilon^{-2})$ incurred from sampling the quantum circuits.

Repeating this for each $s = 1, \ldots, L$ yields $\epsilon$-accurate estimates for each $\ev{\psi}{H_s}{\psi}$, and so by linearity we have an $\O(\|\bm{h}\|_1 \epsilon)$-accurate estimate for $\ev{\psi}{O}{\psi}$. Furthermore, a variance analysis reveals that the overhead due to noise is
\begin{equation}
	C \sim p^{-1}_{\mathrm{NE}},
\end{equation}
where $p_{\mathrm{NE}}$ is the probability of no error occurring in the entire quantum circuit (recall that this is $U^\dagger (\text{c-}V_s(t)) U$, which has essentially double the size of $U$ alone). Thus to obtain an $\epsilon$-accurate estimate for $\ev{\psi}{O}{\psi}$, VPE naively requires taking $\O(L \|\bm{h}\|_1^2 \epsilon^{-2} p_{\mathrm{NE}}^{-1})$ samples, although this can be ameliorated, e.g., by parallelizing over groups of commuting terms (although this requires a separate control qubit for each such term). After paying these costs, however, VPE is notable in that it has the potential to mitigate errors down to order of $(1 - p_{\mathrm{NE}})^2$ or even $(1 - p_{\mathrm{NE}})^3$, whereas other QEM techniques typically only suppress errors to first order. (For comparison, VD/ESD suppresses errors to $M$th order.)

\subsection{Comment on sampling overhead}

As we have seen from these examples, there is always some overhead $C$ in the number of samples required to resolve the same accuracy in error-mitigated estimates. Indeed, this is a necessary feature of QEM~\cite{takagi2022fundamental,takagi2022universal,tsubouchi2022universal,quek2022exponentially}. Furthermore, the accumulation of noise generally implies that this factor will scale exponentially in the circuit size. As a simple example, let $D$ be the depth of the circuit and consider the VPE method. Assuming a simple depolarizing noise model wherein each qubit has error probability $p = \O(1)$ at each gate, then $p_{\mathrm{NE}} = (1 - p)^{nD}$, hence generically~\cite{obrien2021error}
\begin{equation}
    C \sim \l( \frac{1}{1 - p} \r{)^{nD}} = 2^{\poly(n)}.
\end{equation}
Qualitatively speaking, the overheads for other QEM methods behave similarly~\cite{cai2022quantum}.
\chapter{Fermionic Partial Tomography via Classical Shadows}\label{chap:FCS}

\section*{Preface}

This chapter is based on \cite{zhao2021fermionic}, coauthored by the author of this dissertation, Nicholas C.~Rubin, and Akimasa Miyake. New material has been added to this chapter in \cref{thm:informal_lower_bound}, proven in \cref{sec:FCS_lower_bound}, which establishes an information-theoretic lower bound matching the performance of the protocol constructed here.

\section{Introduction}

One of the most promising applications of quantum computation is the study of strongly correlated systems such as interacting fermions. While quantum algorithms such as phase estimation~\cite{nielsen2010quantum,kitaev2002classical} allow for directly computing important quantities such as ground-state energies with quantum speedup~\cite{abrams1999quantum,somma2002simulating,aspuru2005simulated}, current hardware limitations~\cite{preskill2018quantum} have directed much attention toward variational methods. Of note is the variational quantum eigensolver (VQE)~\cite{peruzzo2014variational,mcclean2016theory}, where short-depth quantum circuits are repeatedly executed in order to estimate observable expectation values.

Initial bounds on the number of these circuit repetitions associated with fermionic two-body Hamiltonians were prohibitively high~\cite{wecker2015progress}, spurring on much recent work addressing this problem. We roughly classify these strategies into two categories:~those that specifically target energy estimates~\cite{mcclean2014exploiting,mcclean2016theory,kandala2017hardware,babbush2018low,rubin2018application,izmaylov2019revising,izmaylov2019unitary,huggins2021efficient,crawford2019efficient,zhao2020measurement,torlai2020precise,arrasmith2020operator,paini2019approximate_superseded,hadfield2022measurements,yen2020cartan,gonthier2022measurements,huang2021efficient,garcia2021learning,hillmich2021decision,hadfield2021adaptive,wu2023overlapped}, referred to as Hamiltonian averaging, and more general techniques that can learn the $ k $-body reduced density matrices ($ k $-RDMs) of a quantum state~\cite{aaronson2020shadow,aaronson2018online,aaronson2019gentle,yu2019quantum,yu2020sample,verteletskyi2020measurement,jena2019pauli,yen2020measuring,gokhale2019on3,cotler2020quantum,bonet2020nearly,hamamura2020efficient,garcia2020pairwise,jiang2020optimal,evans2019scalable,huang2020predicting,smart2020lowering,tilly2021reduced}. (Not all works fit neatly into this dichotomy, e.g., Refs.~\cite{harrow2019low,wang2019accelerated,kubler2020adaptive,sweke2020stochastic,van2020measurement,wang2021minimizing}.) Hamiltonian averaging is ultimately interested in a single observable, allowing for heavy exploitation in its structure. In contrast, reconstructing an RDM requires estimating all the observables that parametrize it.

Though generally more expensive than Hamiltonian averaging, calculating the $ k $-RDM allows one to determine the expectation value of any $ k $-body observable~\cite{coleman1980reduced}. For example, the electronic energy of chemical systems is a linear functional of the 2-RDM, while in condensed-matter systems, effective models for electrons can require knowledge of the 3-RDM~\cite{tsuneyuki2008transcorrelated,PhysRevB.87.245129}. Beyond the energy, other important physical properties include pair-correlation functions and various order parameters~\cite{mazziotti2012two,jensen2017introduction}. The 2-RDM is also required for a host of error-mitigation techniques for near-term quantum algorithms~\cite{mcclean2017hybrid,rubin2018application,takeshita2020increasing}, which have been experimentally demonstrated to be crucial in obtaining accurate results~\cite{colless2018computation,sagastizabal2019experimental,mccaskey2019quantum,arute2020hartree}. Additionally, promising extensions to VQE such as adaptive ansatz construction~\cite{grimsley2019adaptive,ryabinkin2020iterative,tang2019qubit,wang2020resource} and multireference- and excited-state calculations~\cite{mcclean2017hybrid,parrish2019quantum,takeshita2020increasing,huggins2020non,stair2020multireference,urbanek2020chemistry} can require up to the 4-RDM.

Motivated by these considerations, in this work we focus on partial tomography for fermionic RDMs. While numerous works have demonstrated essentially optimal sample complexity for estimating qubit RDMs~\cite{cotler2020quantum,bonet2020nearly,jiang2020optimal,evans2019scalable,huang2020predicting}, such approaches necessarily underperform in the fermionic setting. Recognizing this fundamental distinction, Bonet-Monroig \emph{et al.}~\cite{bonet2020nearly} developed a measurement scheme that achieves optimal scaling for fermions.\footnote{Ref.~\cite{jiang2020optimal} also claims an optimally scaling protocol (which requires a specific fermion-to-qubit mapping and doubling the number of qubits);~unfortunately, the claim does not hold due to an error in their conversion of a derived ``attenuation factor'' to sample complexity.} However, the construction is not readily generalizable for $ k > 2 $, in part due to its highly complicated design.


In this chapter, we propose a randomized scheme that is free from these obstacles. It is based on the theory of classical shadows~\cite{huang2020predicting}:~a protocol of randomly distributed measurements from which one acquires a partial classical representation of an unknown quantum state (its ``shadow''). Classical shadows are sufficient for learning a limited collection of observables, making this framework ideal for partial state tomography. Our key results identify efficient choices for the ensemble of random measurements, suitable for the structure of fermionic RDMs.

\section{Fermionic RDMs}

Consider a fixed-particle state $ \rho $ represented in second quantization on $ n $ fermion modes. The $ k $-RDM of $ \rho $, obtained by tracing out all but $ k $ particles, is typically represented as a $ 2k $-index tensor,
\begin{equation}
{}^k D_{q_1 \cdots q_k}^{p_1 \cdots p_k} \coloneqq \tr(a_{p_1}^\dagger \cdots a_{p_k}^\dagger a_{q_k} \cdots a_{q_1} \rho),
\end{equation}
where $ a_p^\dagger, a_p $ are fermionic creation and annihilation operators, $ p \in \{0, \ldots, n-1\} $.
By linearity, these matrix elements may be equivalently expressed using Majorana operators, starting with the definitions
\begin{equation}
\gamma_{2p} \coloneqq a_p + a_p^\dagger, \quad \gamma_{2p+1} \coloneqq -\i (a_p - a_p^\dagger).
\end{equation}
Then for each $ 2k $-combination $ \bm{\mu} \equiv (\mu_1, \ldots, \mu_{2k}) $, where $ 0 \leq \mu_1 < \cdots < \mu_{2k} \leq 2n-1 $, we define a $ 2k $-degree Majorana operator
\begin{equation}\label{eq:majorana_defn}
\Gamma_{\bm{\mu}} \coloneqq (-\i)^{k} \, \gamma_{\mu_1} \cdots \gamma_{\mu_{2k}}.
\end{equation}
All unique $ 2k $-degree Majorana operators are indexed by the set of all $ 2k $-combinations of $ \{ 0, \ldots, 2n-1 \} $, which we shall denote by $ \comb{2n}{2k} $. Because Majorana operators possess the same algebraic properties as Pauli operators (Hermitian, self-inverse, and Hilbert--Schmidt orthogonal), any fermion-to-qubit encoding maps between the two in a one-to-one correspondence.

The commutativity structure inherited onto $ \comb{2n}{2k} $ constrains the maximum number of mutually commuting (hence simultaneously measurable) operators to be $ \O(n^k) $~\cite{bonet2020nearly}. As there are $ \O(n^{2k}) $ independent $ k $-RDM elements, this implies an optimal scaling of $ \O(n^k) $ measurement settings to account for all matrix elements.

\section{Classical shadows and randomized measurements}

We briefly review the framework of classical shadows introduced by Huang \emph{et al.}~\cite{huang2020predicting}, upon which we build our fermionic extension and prove sampling bounds. Let $ \rho $ be an $ n $-qubit state and $ \{ O_1, \ldots, O_L \} $ a set of $L$ traceless observables for which we wish to learn $ \tr(O_1 \rho), \ldots, \tr(O_L \rho)$. Classical shadows require a simple measurement primitive:~for each preparation of $ \rho $, apply the unitary map $ \rho \mapsto U \rho U^\dagger $, where $ U $ is randomly drawn from some ensemble $ \mathcal{U} $;~then perform a projective measurement in the computational basis, $ \{ \ket{z} \mid z \in \{ 0,1 \}^n \} $.

Suppose we have an efficient classical representation for inverting the unitary map on postmeasurement states, yielding $ U^\dagger \op{z}{z} U $. Then the process of repeatedly applying the measurement primitive and classically inverting the unitary may be viewed, in expectation, as the quantum channel
\begin{equation}
\mathcal{M}_{\mathcal{U}}(\rho) \coloneqq \E_{U\sim\mathcal{U}, \ket{z} \sim U \rho U^\dagger} \l[ U^\dagger \op{z}{z} U \r],
\end{equation}
where $ \ket{z} \sim U \rho U^\dagger $ is defined by the usual probability distribution from Born's rule, $ \Pr[\ket{z} \mid U \rho U^\dagger] = \ev{z}{U \rho U^\dagger }{z} $. Informational completeness of $\mathcal{U}$  ensures that this channel is invertible, which allows us to define the classical shadow
\begin{equation}
\hat{\rho}_{U,z} \coloneqq \mathcal{M}_{\mathcal{U}}^{-1}\l( U^\dagger \op{z}{z} U \r)
\end{equation}
associated with the particular copy of $ \rho $ for which $ U $ was applied and $ \ket{z} $ was obtained. Classical shadows form an unbiased estimator for $ \rho $, and so they can be used to estimate the expectation value of any observable $ O $:
\begin{equation}
\E_{U\sim\mathcal{U}, \ket{z} \sim U \rho U^\dagger} \l[ \tr(O \hat{\rho}_{U,z}) \r] = \tr(O \rho).
\end{equation}

The number of repetitions $ M $ required to obtain an accurate estimate for each $ \tr(O_j \rho) $ is controlled by the estimator's variance, which may be upper bounded by
\begin{equation}\label{eq:var_shadows}
\max_{\text{states } \sigma} \mathbb{E}_{\substack{U \sim \mathcal{U} \\ \ket{z} \sim U \sigma U^\dagger}} \l[ \ev{z}{U \mathcal{M}_{\mathcal{U}}^{-1}(O_j) U^\dagger}{z}^2 \r] \eqqcolon \sns{O_j}{\mathcal{U}}^2.
\end{equation}
This quantity is referred to as the (squared) shadow norm. Then by median-of-means estimation, one may show that
\begin{equation}\label{eq:cs_M}
M = \O\l( \frac{\log L}{\varepsilon^2} \max_{1 \leq j \leq L} \sns{O_j}{\mathcal{U}}^2 \r)
\end{equation}
samples suffice to estimate all expectation values to within additive error $ \varepsilon $. To minimize Eq.~\eqref{eq:cs_M} for a fixed collection of observables, the only available freedom is in $ \mathcal{U} $. One must therefore properly choose the ensemble of unitaries, with respect to the target observables.

\section{Naive application to fermionic observables}

A natural ensemble for near-term considerations is the group of single-qubit Clifford gates, $ \Cl(1)^{\otimes n} $ (i.e., Pauli measurements). For an $ \ell $-local Pauli observable $ P $, Huang \emph{et al.}~\cite{huang2020predicting} showed that $ \sns{P}{\Cl(1)^{\otimes n}}^2 = 3^\ell $, similar to the results of prior approaches also based on Pauli measurements~\cite{cotler2020quantum,bonet2020nearly,jiang2020optimal,evans2019scalable}. Although optimal for qubit $ \ell $-RDMs, such strategies cannot achieve the desired $ \O(n^k) $ scaling in the fermionic setting due to the inherent nonlocality of fermion-to-qubit mappings. Indeed, assuming that the $ n $ fermion modes are encoded into $ n $ qubits, the 1-degree Majorana operators necessarily possess an average qubit locality of at least $ \log_3(2n) $~\cite{jiang2020optimal}. This implies that, under random Pauli measurements, the squared shadow norm maximized over all $ 2k $-degree Majorana operators cannot do better than $ 3^{2k \log_3 (2n)} = 4^k n^{2k} $. In fact, for commonly used mappings such as the Jordan--Wigner~\cite{jordanwigner} or Bravyi--Kitaev~\cite{bravyi2002fermionic,seeley2012bravyi,tranter2015bravyi,havlivcek2017operator} transformations, the scalings are poorer ($ 3^n $ and $ {\sim} \, 9^k n^{3.2k} $, respectively).

\section{Randomized measurements with fermionic Gaussian unitaries}

To obtain optimal scaling in the shadow norm for fermionic observables, we propose randomizing over a different ensemble:~the group of fermionic Gaussian Clifford unitaries. First, the group of fermionic Gaussian unitaries $ \FGU(n) $ comprises all unitaries of the form
\begin{equation}
U(e^{A}) \coloneqq \exp\l( \frac{1}{4} \sum_{\mu,\nu=0}^{2n-1} A_{\mu\nu} \gamma_\mu \gamma_\nu \r),
\end{equation}
where $ A = -A^\T \in \R^{2n \times 2n} $. This condition implies that $ \FGU(n) $ is fully characterized by the Lie group $ \SO(2n) $~\cite{sattinger1986lie}. In particular, the adjoint action
\begin{equation}\label{eq:gaussian_adjoint_action}
U(Q)^\dagger \gamma_\mu U(Q) = \sum_{\nu=0}^{2n-1} Q_{\mu\nu} \gamma_\nu \quad \forall Q \in \SO(2n)
\end{equation}
allows for efficient classical simulation of this group~\cite{knill2001fermionic,terhal2002classical,bravyi2004lagrangian,divincenzo2005fermionic,jozsa2008matchgates}. Second, the Clifford group $ \Cl(n) $ is the set of all unitary transformations that permute $ n $-qubit Pauli operators among themselves. It also admits an efficient classical representation~\cite{gottesman1998heisenberg,aaronson2004improved}. 

Because Majorana operators are equivalent to Pauli operators, we may deduce from Eq.~\eqref{eq:gaussian_adjoint_action} that a unitary that is both Gaussian and Clifford corresponds to $ Q $ being a signed permutation matrix. Note that this defines the full group of Majorana swap circuits~\cite{bonet2020nearly}. As the signs are irrelevant for our purpose, we simply consider the group of $ 2n \times 2n $ permutation matrices with determinant 1, known as (the faithful matrix representation of) the alternating group, $ \Alt(2n) $.

Concretely, we set
\begin{equation}\label{eq:FGU_ensemble}
\mathcal{U}_{\FGU} \coloneqq \{ U(Q) \in \FGU(n) \mid Q \in \Alt(2n) \}.
\end{equation}
Given the context of fermionic tomography, the motivation for studying $ \FGU(n) $ is clear, as it preserves the degree of Majorana operators. On the other hand, the restriction to the discrete Clifford elements is valuable for practical considerations. As we show in Section~\ref{sec:fgu_appendix} of the Supplemental Material (SM), the permutational property of Clifford transformations necessarily implies that $ \mathcal{M}_{\FGU} $, as a linear map on the algebra of fermionic observables, is diagonal in the Majorana-operator basis,
\begin{equation}
\mathcal{M}_{\FGU}(\Gamma_{\bm{\mu}}) = \lambda_{\bm{\mu}} \Gamma_{\bm{\mu}} \quad \forall \bm{\mu} \in \comb{2n}{2k},
\end{equation}
with eigenvalues
\begin{equation}
\lambda_{\bm{\mu}} = \l. \binom{n}{k} \middle/ \binom{2n}{2k} \r. \equiv \lambda_{n,k}.
\end{equation}
In this diagonal form, the channel is readily invertible. Thus one may obtain closed-form expressions for the classical shadows $ \hat{\rho}_{Q,z} $, and, importantly, their corresponding estimators for $ \tr(\Gamma_{\bm{\mu}} \rho) $:
\begin{equation}\label{eq:ev_estimator}
\tr(\Gamma_{\bm{\mu}} \hat{\rho}_{Q,z}) = \lambda_{n,k}^{-1} \sum_{\bm{\nu} \in \comb{2n}{2k}} \ev{z}{\Gamma_{\bm{\nu}}}{z} \det[ Q_{\bm{\nu},\bm{\mu}} ].
\end{equation}
Here, $ Q_{\bm{\nu},\bm{\mu}} $ denotes the submatrix of $ Q $ formed from its rows and columns indexed by $ \bm{\nu} $ and $ \bm{\mu} $, respectively~\cite{chapman2018classical}. Because $ Q $ is a permutation matrix, for each $ \bm{\mu} $ there is exactly one $ \bm{\nu}' $ such that $ \det[ Q_{\bm{\nu}',\bm{\mu}} ] \neq 0 $. Thus Eq.~\eqref{eq:ev_estimator} is nonzero if and only if that $ \Gamma_{\bm{\nu}'} $ is diagonal (i.e., maps to a Pauli-$ Z $ operator under a fermion-to-qubit transformation). In other words, the Clifford operation $ U(Q) $ sends $ \Gamma_{\bm{\mu}} $ to $ \pm\Gamma_{\bm{\nu}'} $, which can be estimated only if it is diagonal in the computational basis.

From Eq.~\eqref{eq:var_shadows}, the eigenvalues $ \lambda_{n,k}^{-1} $ of the inverse channel $ \mathcal{M}^{-1}_{\FGU} $ determine the shadow norm. The sample complexity of our approach then follows from Eq.~\eqref{eq:cs_M}. We summarize this first key result with the following theorem.

\begin{theorem}\label{thm:fgu_performance}
	Consider all $2k$-degree Majorana operators $\Gamma_{\bm{\mu}}$ on $n$ fermionic modes, labeled by $ \bm{\mu} \in \comb{2n}{2k} $. Under the ensemble $ \mathcal{U}_{\FGU} $ defined in Eq.~\eqref{eq:FGU_ensemble}, the shadow norm satisfies
	\begin{equation}\label{eq:norm_fgu}
	\sns{\Gamma_{\bm{\mu}}}{\FGU}^2 = \l. \binom{2n}{2k} \middle/ \binom{n}{k} \r. \approx \binom{n}{k}\sqrt{\pi k}
	\end{equation}
	for all $ \bm{\mu} \in \comb{2n}{2k} $. Thus the method of classical shadows estimates the fermionic $k$-RDM of any state $\rho$, i.e., $ \tr(\Gamma_{\bm{\mu}} \rho) \  \forall \bm{\mu} \in \bigcup_{j=1}^k \comb{2n}{2j} $, to additive error $ \varepsilon $, given
	\begin{equation}
	M = \O\l[ \binom{n}{k} \frac{k^{3/2} \log n}{\varepsilon^2} \r]
	\end{equation}
	copies of $ \rho $. Additionally, there is no subgroup $ G \subset \FGU(n) \cap \Cl(n) $ for which $ \sns{\Gamma_{\bm{\mu}}}{G} < \sns{\Gamma_{\bm{\mu}}}{\FGU} $.
\end{theorem}

The proof is presented in the SM, Section~\ref{sec:fgu_appendix}. Furthermore, noting from Eq.~\eqref{eq:ev_estimator} that $ \l|\tr(\Gamma_{\bm{\mu}} \hat{\rho}_{Q,z})\r| \leq \lambda_{n,k}^{-1} $, we also show in the SM that Bernstein's inequality~\cite{boucheron2013concentration} guarantees the above sample complexity via standard sample-mean estimation, rather than requiring the median-of-means technique proposed in the original work on classical shadows~\cite{huang2020predicting}.

This result has an intuitive conceptual interpretation. In the computational basis, there are precisely $ \binom{n}{k} $ diagonal Majorana operators within $ \comb{2n}{2k} $, corresponding to the unique $ k $-fold products of occupation-number operators (e.g., $ \prod_{j=1}^k a_{p_j}^\dagger a_{p_j} $) on $ n $ modes. As a permutation on $ \comb{2n}{2k} $, each element of $ \mathcal{U}_{\FGU} $ defines a different basis in which some other subset of $ \binom{n}{k} $ operators are diagonal. Then, one may expect to account for all $ |\comb{2n}{2k}| = \binom{2n}{2k} $ Majorana operators by randomly selecting on the order of $ \binom{2n}{2k}/\binom{n}{k} $ such bases;~Theorem~\ref{thm:fgu_performance} makes this claim rigorous.

We also establish a matching lower bound on the sample complexity for the task of estimating multiple $2k$-degree Majorana-operator expectation values with single-copy, local fermionic measurements. Such measurements encompass any POVM that can be implemented with fermionic Gaussian operations, such as our Gaussian Clifford ensemble. This result implies that, under single-copy measurements, any protocol that can uniformly learn all $k$-body fermionic observables necessarily requires the same number of samples as our constructive protocol described above, up to constant factors.

\begin{theorem}[Informal]\label{thm:informal_lower_bound}
    Any learning protocol based on single-copy, local-fermionic measurements that can estimate the expectation values of any collection of $L \leq \binom{2n}{2k}$ $k$-body fermionic operators to additive error $\varepsilon$ requires at least
    \begin{equation}
    M \geq \Omega\l( \frac{\lambda_{n,k}^{-1} \log L}{\varepsilon^2} \r)
    \end{equation}
    copies of $\rho$.
\end{theorem}

The formal statement of the theorem and its proof are provided in the SM, \cref{sec:FCS_lower_bound}. The proof technique is a fermionic generalization of the local-qubit argument presented in Huang \emph{et al.}~\cite{huang2020predicting}, which is based on information-theoretic limits of communicating classical information through quantum states. Our result is complementary to the lower bound proven by Bonet-Monroig \emph{et al.}~\cite{bonet2020nearly}, which addresses nearly the same task. Their bound differs from ours by a logarithmic factor, and their proof method is also drastically different from ours. Also, they considered the scenario of general Clifford measurements, whereas our result concerns local fermionic (Gaussian) measurements. It interesting that both our and their constructive protocols employ measurements that are at the intersection of Gaussian and Clifford transformations, hence the two lower bounds nicely complement each other. Finally, we note that our result applies to arbitrary states $\rho$, so it does not apply to, e.g., the protocol of Low~\cite{low2022classical} which targets only fixed-particle-number states.

Fermionic Gaussian circuits have a well-studied compilation scheme based on a Givens-rotation decomposition~\cite{wecker2015solving,kivlichan2018quantum,jiang2018quantum}. For a general element of $ \mathcal{U}_{\FGU} $, we require a circuit depth of at most $ 2n $ with respect to this decomposition~\cite{jiang2018quantum}. Additionally, as pointed out in Ref.~\cite{bonet2020nearly}, Gaussian unitaries commute with the global parity operator $ \Gamma_{(0,\ldots,2n-1)} $, allowing for error mitigation via symmetry verification~\cite{bonet2018low,mcardle2019error}.

Such compilation schemes make use of a group homomorphism property, $ U(Q_1) U(Q_2) = U(Q_1 Q_2) $. Therefore, if the circuit preparing $ \rho $ itself features fermionic Gaussian operations at the end, then we may further compile the measurement unitary into the state-preparation circuit~\cite{takeshita2020increasing}. In the case of indefinite particle number, this concatenation is essentially free. However, rotations with particle-number symmetry have depth at most $ n $~\cite{kivlichan2018quantum,jiang2018quantum}, so they must be embedded into the larger Gaussian unitary of depth $ 2n $. This observation motivates us to explore classical shadows over the number-conserving (NC) subgroup of $ \FGU(n) $.

\section{Modification based on particle-number symmetry}

Fermionic Gaussian unitaries that preserve particle number are naturally parametrized by $ \U(n) $. We express an element of this NC subgroup as
\begin{equation}
U(e^\kappa) \coloneqq \exp\l( \sum_{p,q=0}^{n-1} \kappa_{pq} a_p^\dagger a_q \r),
\end{equation}
where $ \kappa = -\kappa^\dagger \in \C^{n \times n} $, hence $ e^{\kappa} \in \U(n) $. Because the particle-number symmetry manifests as a global phase factor $ e^{\tr\kappa/2} \in \U(1) $, without loss of generality we may consider $ \tr\kappa = 0 $, or equivalently, $ e^{\kappa} \in \SU(n) $. Such unitaries are also called orbital-basis rotations, owing to their adjoint action,
\begin{equation}
U(u)^\dagger a_p U(u) = \sum_{q=0}^{n-1} u_{pq} a_{q} \quad \forall u \in \SU(n).
\end{equation}
This action on Majorana operators follows by linear extension.

Taking the intersection with the Clifford group requires that $ u $ be an $ n \times n $ generalized permutation matrix, with nonzero elements taking values in $ \{ \pm 1, \pm i \} $. This corresponds to the group of fermionic swap circuits~\cite{bravyi2002fermionic,kivlichan2018quantum}. Again, the phase factors on the matrix elements are irrelevant, so we shall restrict to $ u \in \Alt(n) $. By itself, this ensemble is insufficient to perform tomography. To see this, consider an arbitrary reduced density operator $ A_{\bm{p}}^\dagger A_{\bm{q}} \coloneqq a_{p_1}^\dagger \cdots a_{p_k}^\dagger a_{q_k} \cdots a_{q_1} $, where $ \bm{p}, \bm{q} \in \comb{n}{k} $. Such operators are diagonal in the computational basis only if $ \bm{p} = \bm{q} $. Informational completeness thus requires that there exists some $ U(u) $ that maps $ A_{\bm{p}}^\dagger A_{\bm{q}} $ to $ A_{\bm{r}}^\dagger A_{\bm{r}} $, for some $ \bm{r} \in \comb{n}{k} $. Because $ u \in \Alt(n) $, conjugation by $ U(u) $ simply permutes $ \bm{p} $ and $ \bm{q} $ independently. However, as permutations are bijective, it is not possible to permute both $ \bm{p} $ and $ \bm{q} $ to the same $ \bm{r} $ if $ \bm{p} \neq \bm{q} $.

Therefore, this ensemble will necessarily require operations beyond either the NC or Gaussian constraints. The simplest option for maintaining the low-depth structure of the basis rotations is to append Pauli measurements at the end of the circuit. Although the resulting circuit no longer preserves particle number, this addition incurs only a single layer of single-qubit gates. Specifically, we define the ensemble
\begin{equation}\label{eq:ncu_ensemble}
\mathcal{U}_{\NC} \coloneqq \{ V \circ U(u) \mid V \in \Cl(1)^{\otimes n}, \,  u \in \Alt(n) \}.
\end{equation}
By virtue of introducing the notion of ``single-qubit'' gates, this method is dependent on the choice of fermion-to-qubit mapping. Let $ \loc(\Gamma_{\bm{\mu}}) $ denote the qubit locality of $ \Gamma_{\bm{\mu}} $ under some chosen mapping. While Pauli measurements incur a factor of $ 3^{\loc(\Gamma_{\bm{\mu}})} $ in the variance, the randomization over fermionic swap circuits effectively averages this quantity over all same-degree Majorana operators (rather than depending solely on the most nonlocal operator). Formally, we find that the shadow norm here is
\begin{equation}
\sns{\Gamma_{\bm{\mu}}}{\NC}^2 = \E_{u \sim \Alt(n)} \l[ 3^{- \loc[ U(u)^\dagger \Gamma_{\bm{\mu}} U(u) ]} \r{]^{-1}}.
\end{equation}
Although this expression does not possess a closed form, the following theorem provides a universal upper bound.

\begin{theorem}\label{thm:NC_bounds}
	Under the ensemble $ \mathcal{U}_{\NC} $ defined in Eq.~\eqref{eq:ncu_ensemble}, the shadow norm obeys
	\begin{equation}
	\max_{\bm{\mu} \in \comb{2n}{2k}} \sns{\Gamma_{\bm{\mu}}}{\NC}^2 \leq \l. 9^k \binom{n}{2k} \middle/ \binom{n-k}{k} \r. = \O(n^k)
	\end{equation}
	for a fixed integer $ k $ and for all fermion-to-qubit mappings. Thus the method of classical shadows with $ \mathcal{U}_{\NC} $ estimates the $ k $-RDM to additive error $ \varepsilon $ with sample complexity
	\begin{equation}
	M = \O\l( \frac{n^k \log n}{\varepsilon^2} \r).
	\end{equation}
\end{theorem}

We provide derivations for the above results in the SM, Section~\ref{sec:ncu_appendix}.
Note that we have fixed $ k $ as a constant here, so the asymptotic notation may hide potentially large prefactors depending on $ k $. To understand such details, we turn to numerical studies.

\section{Numerical calculations}

Instead of drawing a new circuit for each repetition, here we employ a simplification more amenable to practical implementation. Fixing some integer $ r \geq 1 $, we generate a random collection $ \{ U^{(j)} \sim \mathcal{U} \}_{j=1}^{K_r} $ of $ K_r $ unitaries such that all target observables are covered at least $ r $ times. We say a Majorana operator $ \Gamma_{\bm{\mu}} $ is covered by the measurement unitary $ U $ if $ U \Gamma_{\bm{\mu}} U^\dagger $ is diagonal in the computational basis. Because the ensembles considered here consist of Gaussian and Clifford unitaries, we can determine all covered operators efficiently. Additionally, for the $ \mathcal{U}_{\NC} $ calculations, the qubit mappings were automated through OpenFermion~\cite{openfermion}.

To achieve precision corresponding to $ S = \O(1/\varepsilon^2) $ samples per observable, one repeats each circuit $ \lceil S/r \rceil $ times. The total number of circuit repetitions for our randomized protocols is then $ \lceil S/r \rceil K_r $. For practical purposes, we fix $ r = 50 $ in this work (see Section~\ref{sec:hyperparameter} of the SM for further details). To compare against prior deterministic strategies, we compute $S \times C$ for each such strategy, where $C$ is the number of sets of commuting observables constructed by a given strategy.

\begin{figure}
	\centering
	\includegraphics[width=\textwidth]{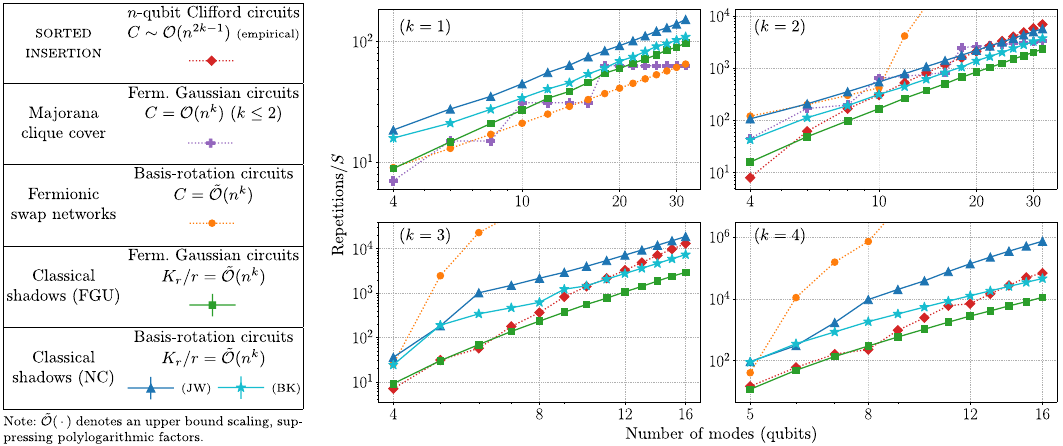}
	\caption[Comparison of methods to estimate the fermionic $k$-RDM.]{(Left)~Summary of the methods compared here, cataloging their required circuit types and scalings in the number of measurement settings. Because graph-based methods~\cite{jena2019pauli,yen2020measuring,gokhale2019on3,hamamura2020efficient} are resource intensive, we employ \textsc{sorted insertion}~\cite{crawford2019efficient} as a more tractable alternative. The Majorana clique cover~\cite{bonet2020nearly}, which employs the same class of fermionic Gaussian Clifford circuits as our classical shadows (FGU) unitaries, possesses optimal asymptotic scaling;~however, it exhibits jumps at powers of 2 due to a divide-and-conquer approach. Furthermore, the construction exists only for $ k \leq 2 $. The measurement strategy using fermionic swap networks is a generalization of the optimal 1-RDM strategy introduced in Ref.~\cite{arute2020hartree}, which we describe in Section~\ref{sec:rdmswap_appendix} of the SM. (Right)~Numerical performances  (log--log scale). Note that \textsc{sorted insertion} and the Majorana clique cover are equivalent for $ k = 1 $. Because our scheme uses randomization, we include error bars of 1 standard deviation, averaged over 10 instances. However, they are not visible at the scale of the plots, indicating the consistency of our method.}
	\label{fig:mainfig}
\end{figure}

For the comparisons presented in Fig.~\ref{fig:mainfig}, we focus on the most competitive prior strategies applicable to fermionic RDM tomography. Because the 1-RDM has a relatively simple structure, optimal strategies are known~\cite{bonet2020nearly,arute2020hartree}, and so randomization underperforms for $ k = 1 $. However, the advantage of our $ \mathcal{U}_{\FGU} $-based method becomes clear for $ k \geq 2 $. When comparing against the Majorana clique cover, which features asymptotically optimal $ \O(n^2) $ scaling for the 2-RDM~\cite{bonet2020nearly}, we find a roughly twofold factor improvement by our approach.

For the $ \mathcal{U}_{\NC} $ case, we observe a trade-off between circuit size and measurement efficiency. As expected, the choice of fermion-to-qubit mapping matters here;~the Jordan--Wigner (JW) mapping performs worse than Bravyi--Kitaev (BK), as the former possesses more qubit nonlocality. Although more measurement settings are required compared to the $ \mathcal{U}_{\FGU} $ ensemble (e.g., a factor of $ {\sim} \, \text{2--5} $ under BK, depending on $ k $), each circuit itself requires only half the depth of general fermionic Gaussian circuits. Notably, however, $ \mathcal{U}_{\NC} $ classical shadows for the 2-RDM under the BK mapping is closely comparable to the Majorana clique cover.

\section{Conclusions}

We have adapted the framework of classical shadows to the efficient tomography of fermionic $k$-RDMs, applicable for all $k$. Numerical calculations demonstrate that our approach consistently outperforms prior strategies using measurement circuits of comparable sizes when $k \geq 2$, despite the logarithmic factor in the sample complexity (a consequence of rigorously bounding the worst-case probabilistic instances). The power of randomization here lies in avoiding the $\mathsf{NP}$-hard problem of partitioning observables into commuting cliques~\cite{verteletskyi2020measurement,yen2020measuring,jena2019pauli,gokhale2019on3}. Instead, we show that a highly overlapping cover of the observables suffices to perform partial tomography efficiently, as a factor of $ \O(1/\varepsilon^2) $ repetitions is already required for this task.

An outlook for further applications is to adapt these ensembles, e.g., for Hamiltonian averaging. As expected, our method is less efficient in this context than those tailored for the task (see Section~\ref{sec:hamiltonian_appendix} of the SM for preliminary numerical calculations). Possible modifications may include biasing the distribution of unitaries~\cite{hadfield2022measurements,hillmich2021decision,hadfield2021adaptive,wu2023overlapped}, or derandomization techniques~\cite{huang2021efficient}.


\newpage
\section*{Supplemental Material}

\section{Additional notation}

First, we define some notation and preliminary concepts not discussed in the main text. For completeness, we generalize the definition of Majorana operators to include odd-degrees:
\begin{equation}
	\Gamma_{\bm{\mu}} \coloneqq (-\i)^{\binom{k}{2}} \gamma_{\mu_1} \cdots \gamma_{\mu_k},
\end{equation}
where $ \bm{\mu} \in \comb{2n}{k} $. Where required, we may define the empty product ($ k = 0 $) as the identity, i.e., $ \Gamma_{\varnothing} \coloneqq \openone $. Then for each $ 0 \leq k \leq 2n $, we define the $ \R $-linear span of $ k $-degree Majorana operators on $ n $ fermion modes,
\begin{equation}
	\MA{n}{k} \coloneqq \spn\{\Gamma_{\bm{\mu}} \mid \bm{\mu} \in \comb{2n}{k} \},
\end{equation}
which is isomorphic (as a vector space) to $ \R^{\binom{2n}{k}} $. The algebra of all even-degree fermionic observables shall be denoted by
\begin{equation}
	\MA{n}{\mathrm{even}} \coloneqq \bigoplus_{k=0}^{n} \MA{n}{2k}.
\end{equation}
Due to the parity superselection rule~\cite{streater2000pct}, physical fermionic operators lie in $ \MA{n}{\mathrm{even}} $;~thus the notions of informational completeness for fermionic tomography are understood with respect to this algebra. To see this, suppose we interpret $ \ket{z} $ as a Fock basis state. Only products of occupation-number operators, $ \prod_p a_p^\dagger a_p $ (modulo anticommutation relations), have nonzero diagonal matrix elements in the Fock basis. Thus if $ U $ is a unitary generated by a fermionic Hamiltonian (hence respecting parity supersymmetry), only those $ O \in \MA{n}{\mathrm{even}} $ are able to satisfy $ \ev{z}{U O U^\dagger}{z} \neq 0 $. This argument also holds if one insists on viewing $ \ket{z} $ as a qubit computational basis state, since they are necessarily mapped to Fock basis states under a fermion-to-qubit encoding~\cite{steudtner2018fermion}.

The group of fermionic Gaussian unitaries $ \FGU(n) $ is generated by $ \i\MA{n}{2} $, and its adjoint action on any $ k $-degree Majorana operator straightforwardly generalizes as
\begin{equation}\label{eq:gaussian_adjoint_action_k}
	U(Q)^\dagger \Gamma_{\bm{\mu}} U(Q) = \sum_{\bm{\nu}\in\comb{2n}{k}} \det[ Q_{\bm{\mu},\bm{\nu}} ] \Gamma_{\bm{\nu}},
\end{equation}
where $ Q_{\bm{\mu},\bm{\nu}} $ denotes the submatrix formed by taking the rows and columns of $ Q $ indexed by $ \bm{\mu} $ and $ \bm{\nu} $, respectively~\cite[Appendix A]{chapman2018classical}. 
This defines an adjoint representation $ \FGU(n) \to \SO[\mathfrak{su}(2^n)] \cong \SO(4^n - 1) $, understood in the sense that $ \FGU(n) \subseteq \SU(2^n) $. We will rather be interested in the orthogonal representation $ \Phi $ of $ \SO(2n) $ induced by this adjoint representation, through the group homomorphism $ U \colon \SO(2n) \to \FGU(n) $. That is, we define $ \Phi \colon \SO(2n) \to \SO(4^n - 1) $ by the matrix elements
\begin{equation}\label{eq:big_rep}
	[ \Phi(Q) ]_{\bm{\mu}\bm{\nu}} \coloneqq \det[ Q_{\bm{\mu},\bm{\nu}} ].
\end{equation}
By the Cauchy--Binet formula one may verify that this indeed defines an orthogonal matrix~\cite{chapman2018classical}. Additionally, $ \Phi $ inherits the homomorphism property from $ U $;~hence it is an orthogonal representation.

Since determinants are defined only for square matrices, $ \Phi $ possesses a natural decomposition as $ \Phi = \bigoplus_{k=0}^{2n} \phi_k $, where each $ \phi_k \colon \SO(2n) \to \SO(\MA{n}{k}) \cong \SO\l[ \binom{2n}{k} \r] $ is defined just as in \cref{eq:big_rep}, restricted a particular $ k $. These subrepresentations will be the main focus of our analysis.

Finally, because of their relation to the Clifford group, we make frequent use of permutations and their generalizations. We establish the relevant notation here. Let $ d,m > 0 $ be integers. We denote the symmetric group on $ d $ symbols by $ \Sym(d) $, which is faithfully represented by the group of $ d \times d $ permutation matrices. The alternating group $ \Alt(d) $ is the subgroup of all even parity permutations. The generalized symmetric group of cyclic order $ m $ over $ d $ symbols is defined via the wreath product, $ \Sym(m,d) \coloneqq \Z_m \wr \Sym(d) \equiv \Z_m^d \rtimes \Sym(d) $. Its faithful matrix representation is the group of $ d \times d $ generalized permutation matrices, wherein each nonzero matrix element can take on values from the $ m $th roots of unity. The determinant of such matrices is the sign of the underlying permutation, multiplied by all $ d $ nonzero elements. In particular, we shall refer to $ \Sym(2,d) $ as the group of signed permutation matrices, and denote its $ ({+1}) $-determinant subgroup by $ \Sym^{+}(2,d) $.

\section{\label{sec:fgu_appendix}Computations with the fermionic Gaussian Clifford ensemble}

We now derive the main results leading to \cref{thm:fgu_performance} of the main text. In \cref{sec:fgu_channel} we find an expression for the channel $ \mathcal{M}_{\FGU} $, showing that the permutational property of the Clifford group necessarily implies that the channel is diagonalized by the basis of Majorana operators [\cref{eq:fgu_sum_z}]. Then in \cref{sec:fgu_norm} we compute the corresponding eigenvalues of $ \mathcal{M}_{\FGU} $, which are directly related to the ensemble's shadow norm (\cref{lem:shadow_norm_bound}). In particular, \cref{lem:shadow_norm_bound} provides necessary and sufficient conditions for saturating the minimum value of the shadow norm under Gaussian Clifford measurements. In \cref{thm:tight_frame_bound} we explicitly calculate this minimum value, and finally with \cref{thm:G_irrep} we prove that both $ \FGU(n) \cap \Cl(n) $ and its $ \Alt(2n) $-generated subgroup $ \mathcal{U}_\FGU $ (the one presented in the main text) satisfy the necessary and sufficient conditions.

After proving these main results, in \cref{sec:class_shadow_appendix} we explicitly derive the expression for the classical shadows $ \hat{\rho}_{Q,z} $. For completeness, we show how to compute the shadow norm for an arbitrary observable in \cref{sec:arb_observable_appendix}. Finally, in \cref{sec:hoeffding} we show how the boundedness of classical shadow estimators and Bernstein's inequality allow one to avoid requiring median-of-means estimation, as mentioned in the main text.

\subsection{\label{sec:fgu_channel}The classical shadows channel}

Our goal is to find an analytic expression for the channel~\cite{huang2020predicting} $ \mathcal{M}_\FGU \colon \MA{n}{\mathrm{even}} \to \MA{n}{\mathrm{even}} $,
\begin{equation}
	\begin{split}
	\mathcal{M}_{\FGU}(O) &= \mathbb{E}_{U \sim \FGU(n) \cap \Cl(n)} \l[ \sum_{z\in\{0,1\}^n} \ev{z}{U O U^\dagger}{z} U^\dagger \op{z}{z} U \r]\\
	&= \mathbb{E}_{Q \sim \Sym^{+}(2,2n)} \l[ \sum_{z\in\{0,1\}^n} \ev{z}{U(Q) O U(Q)^\dagger}{z} U(Q)^\dagger \op{z}{z} U(Q) \r].
	\end{split}
\end{equation}
Since this is a linear map, we need only to evaluate it on a basis of $ \MA{n}{\mathrm{even}} $, the most natural choice being the Majorana operators. Distinguishing the Majorana operators which are diagonal (with respect to the computational basis) is highly important. For each $ 1 \leq k \leq n, $ we shall define the subset $ \diags{2n}{2k} \subseteq \comb{2n}{2k} $ of $ 2k $-combinations corresponding to the diagonal $ 2k $-degree Majorana operators. Formally,
\begin{equation}
	\diags{2n}{2k} \coloneqq \{ \bm{\mu} \in \comb{2n}{2k} \mid \ev{z}{\Gamma_{\bm{\mu}}}{z} \neq 0 \  \forall z \in \{0,1\}^n \}.
\end{equation}
Since there are $ \binom{n}{k} $ independent $ k $-fold products of occupation-number operators, each set has cardinality $ |\diags{2n}{2k}| = \binom{n}{k} $, so that
\begin{equation}
	\bigg| \bigcup_{k=1}^{n} \diags{2n}{2k} \bigg| = \sum_{k=1}^n \binom{n}{k} = 2^n - 1,
\end{equation}
which indeed matches the maximal number of simultaneously commuting Pauli operators~\cite{lawrence2002mutually} (e.g., all Pauli-$ Z $ operators of locality $ 1 $ to $ n $). For instance, under the Jordan--Wigner transformation, the corresponding sets are
\begin{equation}
\begin{split}
\diags{2n}{2} &\coloneqq \{ (p,p+1) \mid 0 \leq p \leq 2n-2 : p \text{ even} \},\\
\diags{2n}{4} &\coloneqq \{ (p,p+1,q,q+1) \mid 0 \leq p < q \leq 2n-2 : p,q \text{ even} \},\\
\diags{2n}{6} &\coloneqq \{ (p,p+1,q,q+1,r,r+1) \mid 0 \leq p < q < r \leq 2n-2 : p,q,r \text{ even} \},
\end{split}
\end{equation}
and so forth up to $ \diags{2n}{2n} $.

With this formalism, we can express basis states as
\begin{equation}
	\op{z}{z} = \frac{1}{2^n} \l( \openone + \sum_{j=1}^{n} \sum_{\bm{\mu} \in \diags{2n}{2j}} \ev{z}{\Gamma_{\bm{\mu}}}{z} \Gamma_{\bm{\mu}} \r),
\end{equation}
which undergo fermionic Gaussian evolution as
\begin{equation}\label{eq:UzzU}
	\begin{split}
	U(Q)^\dagger \op{z}{z} U(Q) &= \frac{1}{2^n} \l( \openone + \sum_{j=1}^{n} \sum_{\bm{\mu} \in \diags{2n}{2j}} \ev{z}{\Gamma_{\bm{\mu}}}{z} U(Q)^\dagger \Gamma_{\bm{\mu}} U(Q) \r)\\
	&= \frac{1}{2^n} \l( \openone + \sum_{j=1}^{n} \sum_{\bm{\mu} \in \diags{2n}{2j}} \ev{z}{\Gamma_{\bm{\mu}}}{z} \sum_{\bm{\nu} \in \comb{2n}{2j}} \det\l[ Q_{\bm{\mu},\bm{\nu}} \r] \Gamma_{\bm{\nu}} \r).
	\end{split}
\end{equation}
For the corresponding Born-rule probability factor, we use the fact that, since $ U $ is a group homomorphism, inverses are preserved [$ U(Q)^\dagger = U(Q^\T) $], and so for any $ \bm{\tau} \in \comb{2n}{2k} $ we have
\begin{equation}\label{eq:zUGUz}
	\begin{split}
	\ev{z}{U(Q) \Gamma_{\bm{\tau}} U(Q)^\dagger}{z} &= \sum_{\bm{\sigma} \in \comb{2n}{2k}} \det\l[ \l( Q^\T \r)_{\bm{\tau},\bm{\sigma}} \r] \ev{z}{\Gamma_{\bm{\sigma}}}{z}\\
	&= \sum_{\bm{\sigma} \in \diags{2n}{2k}} \det\l[ \l( Q^\T \r)_{\bm{\tau},\bm{\sigma}} \r] \ev{z}{\Gamma_{\bm{\sigma}}}{z}.
	\end{split}
\end{equation}
Now recall some basic properties of Majorana (equiv.~Pauli) operators. They are traceless,
\begin{equation}
	\sum_{z \in \{0,1\}^n} \ev{z}{\Gamma_{\bm{\sigma}}}{z} = \tr \Gamma_{\bm{\sigma}} = 0,
\end{equation}
and more generally they are Hilbert--Schmidt (trace) orthogonal,
\begin{equation}
	\tr\l( \Gamma_{\bm{\sigma}} \Gamma_{\bm{\mu}} \r) = 2^n \delta_{\bm{\sigma} \bm{\mu}}.
\end{equation}
In the case that $ \Gamma_{\bm{\sigma}} $ and $ \Gamma_{\bm{\mu}} $ are both diagonal, we have that
\begin{equation}
	\begin{split}
	\sum_{z \in \{0,1\}^n} \ev{z}{\Gamma_{\bm{\sigma}}}{z} \ev{z}{\Gamma_{\bm{\mu}}}{z} &= \sum_{z \in \{0,1\}^n} \ev{z}{\Gamma_{\bm{\sigma}} \Gamma_{\bm{\mu}}}{z}\\
	&= \tr\l( \Gamma_{\bm{\sigma}} \Gamma_{\bm{\mu}} \r) = 2^n \delta_{\bm{\sigma} \bm{\mu}}.
	\end{split}
\end{equation}
With these relations in hand, we multiply \cref{eq:UzzU,eq:zUGUz} and sum over all $ z\in\{0,1\}^n $ to obtain
\begin{equation}\label{eq:fgu_sum_z}
	\begin{split}
	\sum_{z\in\{0,1\}^n} \ev{z}{U(Q) \Gamma_{\bm{\tau}} U(Q)^\dagger}{z} U(Q)^\dagger \op{z}{z} U(Q) &= \frac{1}{2^n} \sum_{\bm{\sigma} \in \diags{2n}{2k}} \det\l[ \l( Q^\T \r)_{\bm{\tau},\bm{\sigma}} \r] \Bigg( \tr (\Gamma_{\bm{\sigma}}) \openone\\
	&\quad + \sum_{j=1}^{n} \sum_{\bm{\mu} \in \diags{2n}{2j}} 2^n \delta_{\bm{\sigma} \bm{\mu}} \sum_{\bm{\nu} \in \comb{2n}{2j}} \det\l[ Q_{\bm{\mu},\bm{\nu}} \r] \Gamma_{\bm{\nu}} \Bigg)\\
	&= \sum_{\substack{\bm{\sigma} \in \diags{2n}{2k} \\ \bm{\nu} \in \comb{2n}{2k}}} \det\l[ Q_{\bm{\sigma},\bm{\tau}} \r] \det\l[ Q_{\bm{\sigma},\bm{\nu}} \r] \Gamma_{\bm{\nu}}.
	\end{split}
\end{equation}
One may be tempted to use the Cauchy--Binet formula to evaluate the sum over $ \bm{\sigma} $;~crucially, however, the sum is restricted to $ \diags{2n}{2k} $, so the identity does not apply here. Instead, we will first evaluate the sum over $ \bm{\nu} $ by formalizing our notion of Gaussian Clifford transformations as degree-preserving permutations of Majorana operators. For generality, we state the following lemma with regards to any generalized permutation matrix.

\begin{lemma}\label{lem:unique_submatrix}
	Let $ Q \in \Sym(m,d) $ and fix $ \bm{\tau} \in \comb{d}{j} $, $ 1 \leq j \leq d $. Then there exists exactly one $ \bm{\sigma} \in \comb{d}{j} $ for which $ |{\det[ Q_{\bm{\sigma},\bm{\tau}} ]}| = 1 $;~otherwise, $ \det[ Q_{\bm{\sigma},\bm{\tau}} ] = 0 $. In particular, we have
	\begin{equation}\label{eq:subdet_lemma}
		\det[ ( Q^\dagger )_{\bm{\tau},\bm{\sigma}} ] \det[ Q_{\bm{\sigma},\bm{\nu}} ] = | {\det[ Q_{\bm{\sigma},\bm{\tau}} ]} | \, \delta_{\bm{\tau} \bm{\nu}}
	\end{equation}
	for all $ \bm{\nu} \in \comb{d}{j} $.
\end{lemma}

\begin{proof}
	By definition of permutation matrices, for each column $ q $ of $ Q $ there is exactly one row $ p $ for which $ Q_{pq} \neq 0 $. Furthermore, this row is unique to each column. This property generalizes from matrix elements to subdeterminants:~for each set of columns indexed by $ \bm{\tau} $, there is exactly one unique set of rows $ \bm{\sigma} $ for which $ Q_{\bm{\sigma},\bm{\tau}} $ has a nonzero element in each row. In other words, $ Q_{\bm{\sigma},\bm{\tau}} \in \Sym(m,j) $ and hence $ |{\det[ Q_{\bm{\sigma},\bm{\tau}} ]}| = 1 $ (recall that the determinant of these matrices is the sign of the underlying permutation multipled by $ m $th roots of unity). Otherwise, for all $ \bm{\sigma}' \neq \bm{\sigma} $, $ Q_{\bm{\sigma}',\bm{\tau}} $ possesses at least one row or column of all zeros, and hence has determinant $ 0 $.
	
	Next, we show that submatrices behave under conjugate transposition as $ ( Q_{\bm{\sigma},\bm{\tau}} )^\dagger = ( Q^\dagger )_{\bm{\tau},\bm{\sigma}} $, which can be seen by examining their matrix elements:
	\begin{equation}
	\begin{split}
		[ Q_{\bm{\sigma},\bm{\tau}} ]_{pq} &= Q_{\sigma_p \tau_q} = [ Q^\dagger ]_{\tau_q \sigma_p}^* = [ ( Q^\dagger )_{\bm{\tau},\bm{\sigma}} ]_{qp}^* = [ ( [ Q^\dagger ]_{\bm{\tau},\bm{\sigma}} )^\dagger ]_{pq}.
	\end{split}
	\end{equation}
	Since the determinant is invariant under transposition and preserves complex conjugation, $ \det[ ( Q^\dagger )_{\bm{\tau},\bm{\sigma}} ] = \det[ ( Q_{\bm{\sigma},\bm{\tau}} ) ]^* $. This gives us
	\begin{equation}
		\det[ ( Q^\dagger )_{\bm{\tau},\bm{\sigma}} ] \det[ Q_{\bm{\sigma},\bm{\nu}} ] = \det[ ( Q_{\bm{\sigma},\bm{\tau}} ) ]^* \det[ Q_{\bm{\sigma},\bm{\nu}} ].
	\end{equation}
	But since $ \bm{\tau} $ is the unique $ j $-combination for which $ \det[ ( Q_{\bm{\sigma},\bm{\tau}} ) ] \neq 0 $ for a fixed $ \bm{\sigma} $, the above expression can only be nonzero when $ \bm{\nu} = \bm{\tau} $. Equation~\eqref{eq:subdet_lemma} thus follows.
\end{proof}

\begin{remark}\label{rem:unique_submatrix}
    This uniqueness property of nonzero subdeterminants implies that the image of the orthogonal representation $ \Phi \colon \SO(2n) \to \SO(4^n - 1) $, when restricted to a subgroup $ G \subseteq \Sym^{+}(2,2n) \subset \SO(2n) $, lies in $ \Sym^{+}(2,4^n - 1) $. Its subrepresentations also satisfy $ \phi_k(G) \subseteq \Sym\l[2,\binom{2n}{k}\r] $. From a physical perspective, these are all straightforward consequences of the Clifford property imposed on our unitary ensemble.
\end{remark}

Applying \cref{lem:unique_submatrix} to \cref{eq:fgu_sum_z} reveals that the Majorana operators are in fact the eigenbasis of $ \mathcal{M}_\FGU $:
\begin{equation}\label{eq:M_eigenvalues_eq}
	\begin{split}
	\mathcal{M}_\FGU(\Gamma_{\bm{\tau}}) &= \E_{Q \sim \Sym^{+}(2,2n)} \l[ \sum_{z\in\{0,1\}^n} \ev{z}{U(Q) \Gamma_{\bm{\tau}} U(Q)^\dagger}{z} U(Q)^\dagger \op{z}{z} U(Q) \r]\\
	&= \E_{Q \sim \Sym^{+}(2,2n)} \l[ \sum_{\bm{\sigma} \in \diags{2n}{2k}} | {\det[ Q_{\bm{\sigma},\bm{\tau}} ]} | \r] \Gamma_{\bm{\tau}}.
	\end{split}
\end{equation}

\subsection{\label{sec:fgu_norm}The shadow norm}

To evaluate the eigenvalues of \cref{eq:M_eigenvalues_eq}, we invoke the theory of finite frames~\cite{han2000frames,waldron2018introduction}. We begin with the definition of a frame.

\begin{definition}
	Let $ V $ be a Hilbert space with inner product $ \langle \cdot, \cdot \rangle $. A \emph{frame} is a sequence $ \{ x_j \}_j \subset V $ which satisfies
	\begin{equation}
	\alpha \| v \|_V^2 \leq \sum_{j} | \langle x_j, v \rangle |^2 \leq \beta \| v \|_V^2 \quad \forall v \in V,
	\end{equation}
	for some real constants $ \alpha,\beta > 0 $ (called the \emph{frame bounds}). Here, $ \| v \|_V \coloneqq \sqrt{\langle v, v \rangle} $. A frame is called \emph{tight} if $ \alpha=\beta $.
\end{definition}

Recall that $ \det[ Q_{\bm{\sigma},\bm{\tau}} ] $ defines the matrix elements of the orthogonal representation $ \phi_{2k} \colon \SO(2n) \to \SO\l[ \binom{2n}{2k} \r] $. In order to demonstrate the optimality of our ensemble, we shall generalize \cref{eq:M_eigenvalues_eq} to take the average over any subgroup $ G \subseteq \Sym^{+}(2,2n) $, in which case we consider the restricted representations $ \phi_{2k}\big|_G \colon G \to \SO\l[ \binom{2n}{2k} \r] $. (Whenever the context is clear, we shall simply write $ \phi_{2k} $.) To simplify notation, we take $ \R^{\binom{2n}{2k}} $ as our representation space, spanned by the standard basis $ \{ e_{\bm{\mu}} \mid \bm{\mu} \in \comb{2n}{2k} \} $.

Consider the group orbit
\begin{equation}
	\phi_{2k}(G) e_{\bm{\sigma}} \coloneqq \{ \phi_{2k}(Q) e_{\bm{\sigma}} \mid Q \in G \}.
\end{equation}
A sufficient condition for the eigenvalues of $ \mathcal{M}_G $ to be nonzero (hence guaranteeing the existence of $ \mathcal{M}_G^{-1} $) is that $ \phi_{2k}(G) e_{\bm{\sigma}} $ be a frame for some $ \bm{\sigma} \in \diags{2n}{2k} $. Indeed, assuming the frame condition, we have
\begin{equation}
	\begin{split}
	\E_{Q \sim G} | {\det[ Q_{\bm{\sigma},\bm{\tau}} ]} | &= \frac{1}{|G|} \sum_{Q \in G} | {\det[ Q_{\bm{\sigma},\bm{\tau}} ]} |^2\\
	&= \frac{1}{|G|} \sum_{Q \in G} | \langle \phi_{2k}(Q)^\T e_{\bm{\sigma}} , e_{\bm{\tau}} \rangle |_2^2\\
	&\geq \frac{\alpha}{|G|} \| e_{\bm{\tau}} \|_2^2 > 0
	\end{split}
\end{equation}
for all $ \bm{\tau} \in \comb{2n}{2k} $. Note that the transpose is irrelevant since $ G $ is a group and $ \phi_{2k} $ an orthogonal representation. While $ \mathcal{M}_G $ may be made positive even without taking $ G $ to be a group, the group structure has desirable implications for the measurement complexity (shadow norm) of our scheme. (It is also generally easier to sample from a well-established group like the symmetric group, rather than some ad hoc subset of $ \Sym^{+}(2,2n) $.) To see this, we shall focus attention exclusively to tight frames, which is motivated by the following observation.

\begin{lemma}\label{lem:shadow_norm_bound}
	Let $ G \subseteq \Sym^{+}(2,2n) $ be a subset such that $ \phi_{2k}(G) e_{\bm{\sigma}} $ is a frame for all $ \bm{\sigma} \in \diags{2n}{2k} $, $ 1 \leq k \leq n $. Let $ \alpha_k,\beta_k > 0 $ be the cumulative frame bounds over all $ \bm{\sigma} $ for each $ k $, i.e.,
	\begin{equation}\label{eq:general_frame_for_eigenvalues}
		\alpha_k \leq \frac{1}{|G|} \sum_{\bm{\sigma} \in \diags{2n}{2k}} \sum_{Q \in G} | {\det[ Q_{\bm{\sigma},\bm{\tau}} ]} |^2 \leq \beta_k
	\end{equation}
	for all $ \bm{\tau} \in \comb{2n}{2k} $. Then the shadow norm associated with the ensemble $ G $ satisfies
	\begin{equation}
		\alpha_k / \beta_k^2 \leq \sns{\Gamma_{\bm{\tau}}}{G}^2 \leq \beta_k/\alpha_k^{2}.
	\end{equation}
	These bounds are saturated if and only if the cumulative frame is tight, i.e., $ \sns{\Gamma_{\bm{\tau}}}{G}^2 = \alpha_k^{-1} $.
\end{lemma}

\begin{remark}
	In \cref{eq:general_frame_for_eigenvalues}, and hence in \cref{eq:M_eigenvalues_eq}, the actual frame being used is $ \phi_{2k}(G) (e_{\bm{\sigma}} |G|^{-1/2}) $, i.e., the ``orbit'' of the basis vector scaled by a factor of $ |G|^{-1/2} $. (We use scare quotes to indicate that, since we do not require $ G $ to be a group here, the corresponding set of vectors may not be an orbit proper.)
\end{remark}

\begin{proof}
	First, we recall the definition of the shadow norm:
	\begin{equation}
			\sns{\Gamma_{\bm{\tau}}}{G}^2 = \max_{\text{states } \rho} \l( \mathbb{E}_{Q \sim G} \l[ \sum_{z \in \{0,1\}^n} \ev{z}{U(Q) \rho U(Q)^\dagger}{z} \ev{z}{U(Q) \mathcal{M}^{-1}(\Gamma_{\bm{\tau}}) U(Q)^\dagger}{z}^2 \r] \r).
	\end{equation}
	Adapting \cref{eq:M_eigenvalues_eq} to the language of frames yields the inequalities
	\begin{equation}
		\beta_k^{-1} \leq \| \mathcal{M}_G^{-1}(\Gamma_{\bm{\tau}}) \| \leq \alpha_k^{-1},
	\end{equation}
	where we have used the fact that the spectral norm $ \| \cdot \| $ of the Majorana operators is $ 1 $. Thus
	\begin{equation}
		\begin{split}
		\sns{\Gamma_{\bm{\tau}}}{G}^2 &\leq \alpha_k^{-2} \max_{\text{states } \rho} \l( \mathbb{E}_{Q \sim G} \l[ \sum_{z \in \{0,1\}^n} \ev{z}{U(Q) \rho U(Q)^\dagger}{z} \ev{z}{U(Q) \Gamma_{\bm{\tau}} U(Q)^\dagger}{z}^2 \r] \r)\\
		&= \alpha_k^{-2} \max_{\text{states } \rho} \l( \mathbb{E}_{Q \sim G} \l[ \sum_{z \in \{0,1\}^n} \ev{z}{U(Q) \rho U(Q)^\dagger}{z} \l( \sum_{\bm{\mu} \in \comb{2n}{2k}} \det[ ( Q^\T )_{\bm{\tau},\bm{\mu}} ] \ev{z}{\Gamma_{\bm{\mu}}}{z} \r{)^2} \r] \r).
		\end{split}
	\end{equation}
	Let us examine the innermost bracketed term. Since $ \ev{z}{\Gamma_{\bm{\mu}}}{z} = \pm 1 $ when $ \bm{\mu} \in \diags{2n}{2k} $ and $ 0 $ otherwise, we can restrict the sum to run over $ \diags{2n}{2k} $. Furthermore, by \cref{lem:unique_submatrix}, we know that $ \det[ ( Q^\T )_{\bm{\tau},\bm{\mu}} ] $ is nonzero only for one particular $ \bm{\mu} $, which may or may not lie in $ \diags{2n}{2k} $. This means that the sum contains at most one nonzero term, which takes the value $ \pm 1 $, allowing us to ``transform'' the square into an absolute value:
	\begin{equation}\label{eq:norm_var_term}
		\l( \sum_{\bm{\mu} \in \comb{2n}{2k}} \det[ ( Q^\T )_{\bm{\tau},\bm{\mu}} ] \ev{z}{\Gamma_{\bm{\mu}}}{z} \r{)^2} = \sum_{\bm{\mu} \in \diags{2n}{2k}} | {\det[ Q_{\bm{\mu},\bm{\tau}} ]} |.
	\end{equation}
	Importantly, this quantity does not depend on $ z $, which makes the sum over $ z $ trivial (as well as the maximum over all states):
	\begin{equation}
		\sum_{z \in \{0,1\}^n} \ev{z}{U(Q) \rho U(Q)^\dagger}{z} = \tr \rho = 1.
	\end{equation}
	We are then left with taking the average over $ Q \sim G $ in \cref{eq:norm_var_term}, which is simply the original quantity of interest obeying the frame bounds [\cref{eq:general_frame_for_eigenvalues}]. Therefore
	\begin{equation}
		\begin{split}
			\sns{\Gamma_{\bm{\tau}}}{G}^2 &\leq \alpha_k^{-2} \E_{Q \sim G} \l[ \sum_{\bm{\mu} \in \diags{2n}{2k}} | {\det[ Q_{\bm{\mu},\bm{\tau}} ]} | \r]\\
			&\leq \alpha_k^{-2} \beta_k.
		\end{split}
	\end{equation}
	To obtain the lower bound, simply reverse the roles of $ \alpha_k $ and $ \beta_k $. Since the only inequalities invoked were those of the frame bounds, it follows that equality holds in both directions if and only if the frame is tight.
\end{proof}

\cref{lem:shadow_norm_bound} is useful in two ways:~first, it gives an estimate on the shadow norm for any valid subset of fermionic Gaussian Clifford unitaries, assuming one has bounds on the eigenvalues of $ \mathcal{M}_G $. Second, it tells us that those subsets which give rise to \emph{tight} frames exhibit optimal sample complexity in the sense of the shadow norm. Tight frames generated by the action of finite groups have been completely characterized through representation theory~\cite{vale2004tight,waldron2018introduction}. Below we restate the primary results relevant to our context, which will motivate us to restrict $ G $ to a group.

\begin{proposition}[{\cite[Theorems 6.3 and 6.5]{vale2004tight}}]
	\label{prop:tight_frames}
	Let $ H $ be a finite group, $ V $ a Hilbert space, and $ \varphi \colon H \to \U(V) $ a unitary representation. Then:
	\begin{enumerate}
		\item Every orbit $ \varphi(H)v $, $ v \in V \setminus \{0\} $, is a tight frame if and only if every orbit spans $ V $ [i.e., $ \varphi $ is an irreducible representation (irrep)].
		\item There exists $ v \in V \setminus \{0\} $ for which $ \varphi(H)v $ is a tight frame if and only if there exists $ w \in V \setminus \{0\} $ such that $ \spn(\varphi(H)w) = V $.
	\end{enumerate}
\end{proposition}

In our context, this means that if $ \phi_{2k}\big|_G $ is irreducible, then every orbit is tight, hence saturating the bound of \cref{lem:shadow_norm_bound}. Alternatively, if $ \phi_{2k}\big|_G $ is not an irrep, there may be only specific orbits which form tight frames. Fortunately, this complication does not arise in our setting due to the fact that we are dealing exclusively with (signed) permutation matrices. We formalize this notion with the following lemma. (Although we are only interested in the even-degree representations, we formulate the statement to apply to all $ 1 \leq k \leq 2n $ for completeness.)

\begin{lemma}\label{lem:all_orbits_span}
	Let $ G\subseteq\Sym^{+}(2,2n) $ be a group such that $ \spn(\phi_{k}(G) v_0) = \R^{\binom{2n}{k}} $
	for some nonzero $ v_0\in\R^{\binom{2n}{k}} $. Then $ \spn(\phi_{k}(G) v) = \R^{\binom{2n}{k}} $ for all nonzero $ v\in\R^{\binom{2n}{k}} $.
\end{lemma}
\begin{proof}
	Without loss of generality, we can consider $ v_0 $ an arbitrary element of the standard basis. From \cref{rem:unique_submatrix}, we have $ \phi_k(G) \subseteq \Sym\l[2,\binom{2n}{k}\r] $;~therefore, $ \spn(\phi_k(G)v_0) = \R^{\binom{2n}{k}} $ if and only if the orbit is the entire basis (modulo signs):
	\begin{equation}
	\phi_k(G) v_0 = \{ e_{\bm{\mu}} \text{ and/or } {-}e_{\bm{\mu}} \mid \bm{\mu} \in \comb{2n}{k} \}.
	\end{equation}
	Now consider some other basis vector $ w_0 \neq v_0 $. Since there exists some $ g\in G $ such that $ \phi_k(g)v_0 = \pm w_0 $, it follows that
	\begin{equation}
	\begin{split}
		\phi_k(G)v_0 &= \phi_k(G) [\phi_k(g^{-1}) (\pm w_0)]\\
		&= \phi_k(G g^{-1}) (\pm w_0) = \phi_k(G) (\pm w_0).
	\end{split}
	\end{equation}
	Since the sign is irrelevant when taking the span, we see that the orbit of any basis vector spans the space. Hence the orbit of every nonzero vector does as well.
\end{proof}

This result allows us to ignore the second part of \cref{prop:tight_frames}, so that we only have to consider the irreducibility of $ \phi_{2k}\big|_G $. Furthermore, we do not have to worry about choosing some particular $ \bm{\sigma} \in \diags{2n}{2k} $ to generate our cumulative frame;~although the specific form of these tuples changes depending on the choice of fermion-to-qubit mapping, the above results tell us that the behavior of such tight frames is uniform across all of $ \comb{2n}{2k} $.

Enumerating all possible subgroups to search for the one which gives the largest frame bound (hence smallest shadow norm) is a highly impractical task. Fortunately, this is not necessary, as it turns out that \emph{all} irreps $ \phi_{2k}\big|_G $ yield the same frame bound. To see this, we introduce an alternative, equivalent formulation of frames based on the \emph{frame operator} $ T \colon V \to V $,
\begin{equation}
	T \coloneqq \sum_{j} \langle x_j, \cdot \, \rangle x_j,
\end{equation}
where $ \{x_j\}_j $ is a frame for $ V $.

\begin{proposition}[{\cite[Theorem 3]{cotfas2010finite}}]\label{prop:frame_op}
	Let $ H $ be a finite group, $ V $ a Hilbert space, and $ \varphi \colon H \to \U(V) $ an irreducible unitary representation. For any nonzero $ v \in V $, the frame operator for the orbit $ \varphi(H)v $ is
	\begin{equation}
		T = \frac{|\varphi(H)v|}{\dim V} \| v \|_V^2 \, \openone_{V},
	\end{equation}
	where $ \openone_{V} $ is the identity operator on $ V $.
\end{proposition}

In general, $ T = \alpha\openone_V $ if and only if the frame is tight (with frame bound $ \alpha $). In the context of tight frames generated by irreps, this result is essentially a variant on Schur's lemma. The utility of \cref{prop:frame_op} in particular is that the frame bound is given explicitly. Our claim that all irreducible subgroups yield the same shadow norm then follows.

\begin{theorem}\label{thm:tight_frame_bound}
	Let $ G \subseteq \Sym^{+}(2,2n) $ be irreducible with respect to $ \phi_{2k} $. Then
	\begin{equation}\label{eq:frame_eigenvalue}
	\begin{split}
		\E_{Q \sim G} \l[ \sum_{\bm{\sigma} \in \diags{2n}{2k}} | {\det[ Q_{\bm{\sigma},\bm{\tau}} ]} | \r] &\equiv \frac{1}{|G|} \sum_{Q \in G} \sum_{\bm{\sigma} \in \diags{2n}{2k}} | [ \phi_{2k}(Q) ]_{\bm{\sigma} \bm{\tau}} |^2\\
		&= \l. \binom{n}{k} \middle/ \binom{2n}{2k} \r.
	\end{split}
	\end{equation}
	for all $ \bm{\tau} \in \comb{2n}{2k} $.
\end{theorem}

\begin{proof}
	Consider the orbit $ \phi_{2k}(G) (e_{\bm{\sigma}} |G|^{-1/2}) $. Let $ \Stab_{G}(e_{\bm{\sigma}}) \coloneqq \{ Q \in G \mid \phi_{2k}(Q) e_{\bm{\sigma}} = e_{\bm{\sigma}} \} $ be its stabilizer subgroup. By the orbit--stabilizer theorem and Lagrange's theorem~\cite{grillet2007abstract},
	\begin{equation}
		| \phi_{2k}(G) (e_{\bm{\sigma}} |G|^{-1/2}) | = \frac{|G|}{| \Stab_{G}(e_{\bm{\sigma}}) |}.
	\end{equation}
	Using \cref{prop:frame_op} and recalling that the Hilbert space $ V $ of our frame is $ \R^{\binom{2n}{2k}} $, we see that the corresponding frame bound is
	\begin{equation}
	\begin{split}
		\alpha_{\bm{\sigma}} = \| T_{\bm{\sigma}} \| &= \frac{|G|}{| \Stab_{G}(e_{\bm{\sigma}}) |} \binom{2n}{2k}^{-1} \l\| \frac{e_{\bm{\sigma}}}{|G|^{1/2}} \r{\|_2^2}\\
		&= \frac{1}{| \Stab_{G}(e_{\bm{\sigma}}) |} \binom{2n}{2k}^{-1}.
	\end{split}
	\end{equation}
	One must be careful when handling this $ | \Stab_{G}(e_{\bm{\sigma}}) | $ term, since, when constructing the frame operator, we sum over all elements of the orbit, rather than of the group. However, the particular sum that we are interested in, namely \cref{eq:frame_eigenvalue}, \emph{is} over the entire group, and thus precisely double counts the elements of the stabilizer subgroup:
	\begin{equation}
	\begin{split}
	\frac{1}{|G|} \sum_{Q \in G} | [ \phi_{2k}(Q) ]_{\bm{\sigma} \bm{\tau}} |^2 &= \sum_{Q \in G} | \langle \phi_{2k}(Q)^\T e_{\bm{\sigma}} |G|^{-1/2} , e_{\bm{\tau}} \rangle |^2\\
	&= | \Stab_{G}(e_{\bm{\sigma}}) | \sum_{w \in \phi_{2k}(G) (e_{\bm{\sigma}} |G|^{-1/2})} | \langle w, e_{\bm{\tau}} \rangle |^2\\
	&= | \Stab_{G}(e_{\bm{\sigma}}) | \alpha_{\bm{\sigma}} = \binom{2n}{2k}^{-1}.
	\end{split}
	\end{equation}
	Since this quantity does not depend on $ \bm{\sigma} $, the remaining sum over $ \diags{2n}{2k} $ simply incurs a factor of $ |\diags{2n}{2k}| = \binom{n}{k} $.
\end{proof}

\begin{corollary}
	In conjunction with \cref{lem:shadow_norm_bound}, it immediately follows that
	\begin{equation}
		\sns{\Gamma_{\bm{\tau}}}{G}^2 = \l. \binom{2n}{2k} \middle/ \binom{n}{k} \r.
	\end{equation}
	for all irreducible $ G \subseteq \Sym^{+}(2,2n) $.
\end{corollary}

Finally, in order to obtain a concrete example of such tight frames, we show that $ \Sym^{+}(2,2n) $ is irreducible with respect to $ \phi_{k} $ for all $ 1 \leq k \leq 2n $. Again, though we are only interested in the even-degree case, we prove the statement for general $ k $ for completeness. It is then straightforward to show that $ \Alt(2n) \subset \Sym^{+}(2,2n) $ is also irreducible.

\begin{theorem}\label{thm:G_irrep}
	Let $ G = \Sym^{+}(2,2n) $ or $ \Alt(2n) $. For all $ 1 \leq k \leq 2n $, $ \phi_k\big|_G $ is irreducible.
\end{theorem}

\begin{proof}
	We begin with the case $ G = \Sym^{+}(2,2n) $;~it will be apparent that the proof methods adapt fully to the $ G = \Alt(2n) $ case. We show irreducibility via a standard result of character theory~\cite{fulton2004representation}:~$ \phi_k\big|_G $ is an irrep if and only if
	\begin{equation}
		\frac{1}{|G|} \sum_{Q \in G} \tr[ \phi_k(Q) ]^2 = 1.
	\end{equation}
	Here, the trace of our representation is simply
	\begin{equation}
		\tr[ \phi_k(Q) ] = \sum_{\bm{\mu} \in \comb{2n}{k}} \det[ Q_{\bm{\mu},\bm{\mu}} ].
	\end{equation}
	Expanding the square yields
	\begin{equation}\label{eq:expanded_char}
		\frac{1}{|G|} \sum_{Q \in G} \tr[ \phi_k(Q) ]^2 = \frac{1}{|G|} \sum_{Q \in G} \l( \sum_{\bm{\mu} \in \comb{2n}{k}} \det[ Q_{\bm{\mu},\bm{\mu}} ]^2 + \sum_{\substack{\bm{\mu},\bm{\nu} \in \comb{2n}{k} \\ \bm{\mu} \neq \bm{\nu}}} \det[ Q_{\bm{\mu},\bm{\mu}} ] \det[ Q_{\bm{\nu},\bm{\nu}} ] \r).
	\end{equation}
	We will calculate the average of the each term separately.
	
	Since $ | { \det[ Q_{\bm{\mu},\bm{\mu}} ] } | \in \{0,1\} $, the diagonal sum is
	\begin{equation}
		\frac{1}{|G|} \sum_{Q \in G} \sum_{\bm{\mu} \in \comb{2n}{k}} \det[ Q_{\bm{\mu},\bm{\mu}} ]^2 = \sum_{\bm{\mu} \in \comb{2n}{k}} \E_{Q \sim G} | { \det[ Q_{\bm{\mu},\bm{\mu}} ] } |.
	\end{equation}
	Intuitively, $ \E_{Q \sim G} | { \det[ Q_{\bm{\mu},\bm{\mu}} ] } | $ is the ``density'' of the signed permutation matrices whose $ (\bm{\mu}, \bm{\mu}) $ submatrix has nonzero subdeterminant. To compute this average, we proceed by a visual argument using the structure of the matrix. Recall that $ \det[ Q_{\bm{\mu},\bm{\mu}} ] \neq 0 $ if and only if $ Q_{\bm{\mu},\bm{\mu}} \in \Sym(2,k) $. When writing down a matrix as an array, we can choose any ordering of the row and column indices, so long as this choice is consistent. Therefore, we shall order the indices such that $ \bm{\mu} $ lies in the first $ k $ rows/columns:
	\begin{equation}
		Q = \l( \begin{array}{c | c}
		Q_{\bm{\mu},\bm{\mu}} & * \\
		\hline
		* & *
		\end{array} \r).
	\end{equation}
	Requiring that $ Q_{\bm{\mu},\bm{\mu}} \in \Sym(2,k) $ immediately sets the off-diagonal blocks to be all zeros, hence such a $ Q $ is block diagonal in this ordering of rows and columns. Furthermore, since $ \det Q = 1 $, the remaining $ (2n-k) \times (2n-k) $ block must have the same determinant as the $ Q_{\bm{\mu},\bm{\mu}} $ block. In other words, a $ Q $ which satisfies $ \det[ Q_{\bm{\mu},\bm{\mu}} ] \neq 0 $ must take the form
	\begin{equation}
		Q \in \l( \begin{array}{c | c}
		\Sym^{+}(2,k) & 0 \\
		\hline
		0 & \Sym^{+}(2,2n-k)
		\end{array} \r)
		\cup
		\l( \begin{array}{c | c}
		\Sym^{-}(2,k) & 0 \\
		\hline
		0 & \Sym^{-}(2,2n-k)
		\end{array} \r),
	\end{equation}
	where $ \Sym^{-}(2,d) \coloneqq \{ R \in \Sym(2,d) \mid \det R = -1 \} $. Although $ \Sym^{-}(2,d) $ is not a group, it has the same number of elements as $ \Sym^{+}(2,d) $:~$ |\Sym^{-}(2,d)| = |\Sym^{+}(2,d)| = 2^d d!/2 $. The density is thus
	\begin{equation}\label{eq:avg_density_calc}
	\begin{split}
		\E_{Q \sim G} | { \det[ Q_{\bm{\mu},\bm{\mu}} ] } | &= \frac{ |\Sym^{+}(2,k) \oplus \Sym^{+}(2,2n-k)| + |\Sym^{-}(2,k) \oplus \Sym^{-}(2,2n-k)| }{ |\Sym^{+}(2,2n)| }\\
		&= \frac{ 2 \l( 2^k k! \, 2^{2n-k} (2n-k)! \r) / 4 }{ 2^{2n} (2n)! / 2 } = \binom{2n}{k}^{-1},
	\end{split}
	\end{equation}
	and so
	\begin{equation}
		\sum_{\bm{\mu} \in \comb{2n}{k}} \E_{Q \sim G} | { \det[ Q_{\bm{\mu},\bm{\mu}} ] } | = 1.
	\end{equation}
	
	Next we show that the off-diagonal sum of \cref{eq:expanded_char} vanishes. The argument follows by generalizing the above calculation. For $ \bm{\mu} \neq \bm{\nu} $, the tuples (thought of as sets) may overlap $ 0 \leq j \leq k-1 $ times. We order the matrix representation such that $ \bm{\mu} $ makes up the first $ k $ rows/columns as before, but additionally the overlapping indices $ \bm{\mu} \cap \bm{\nu} $ are placed at the last $ j $ spots of this $ k \times k $ block. We then place the remaining part of $ \bm{\nu} $ in the following $ (k-j) \times (k-j) $ block. Visually, we have
	\begin{equation}
		Q = \l( \begin{array}{c | c | c | c}
		Q_{\bm{\mu}\setminus\bm{\nu},\bm{\mu}\setminus\bm{\nu}} & 0 & 0 & 0 \\
		\hline
		0 & Q_{\bm{\mu}\cap\bm{\nu},\bm{\mu}\cap\bm{\nu}} & 0 & 0 \\
		\hline
		0 & 0 & Q_{\bm{\nu}\setminus\bm{\mu},\bm{\nu}\setminus\bm{\mu}} & 0 \\
		\hline
		0 & 0 & 0 & *
		\end{array} \r),
	\end{equation}
	where the linear sizes of the four blocks are $ k-j $, $ j $, $ k-j $, and $ 2(n-k)+j $, respectively. If either $ \det[ Q_{\bm{\mu},\bm{\mu}} ] $ or $ \det[ Q_{\bm{\nu},\bm{\nu}} ] $ are $ 0 $, then that term in the sum trivially vanishes. Thus consider the case in which they are both nonzero. Again, since we have the constraint $ \det Q = 1 $, the product of the determinants of all four blocks must be $ 1 $. There are eight cases in which $ \det[ Q_{\bm{\mu},\bm{\mu}} ] \det[ Q_{\bm{\nu},\bm{\nu}} ] \neq 0 $:~$ Q \in \Sym^a(2,k-j) \oplus \Sym^b(2,j) \oplus \Sym^c(2,k-j) \oplus \Sym^d(2,2(n-k)+j) $, where
	\begin{equation}\label{eq:det_signs}
		(a,b,c,d) \in
		\l\{ \begin{array}{c c c c}
			({+},{+},{+},{+}), &
			({+},{+},{-},{-}), &
			({+},{-},{+},{-}), &
			({+},{-},{-},{+}), \\
			({-},{+},{+},{-}), &
			({-},{+},{-},{+}), &
			({-},{-},{+},{+}), &
			({-},{-},{-},{-}) 
		\end{array} \r\}.
	\end{equation}
	With this formalism, we can read off the terms in the off-diagonal sum as $ \det[ Q_{\bm{\mu},\bm{\mu}} ] \det[ Q_{\bm{\nu},\bm{\nu}} ] = (ab)(bc) = ac $ (where our notation means $ a = \pm \equiv \pm 1 $, etc.). By examining \cref{eq:det_signs}, we see that four of the possibilities give $ ac = +1 $, while the other four give $ ac = -1 $. Since the number of elements in each of the eight subsets are all the same, exactly half the terms in the sum will cancel with the other half, hence
	\begin{equation}
		\sum_{Q \in G} \det[ Q_{\bm{\mu},\bm{\mu}} ] \det[ Q_{\bm{\nu},\bm{\nu}} ] = 0 \quad \forall \bm{\mu} \neq \bm{\nu}
	\end{equation}
	as desired.
	
	Thus,
	\begin{equation}
		\frac{1}{|G|} \sum_{Q \in G} \tr[ \phi_k(Q) ]^2 = 1,
	\end{equation}
	and so $ \phi_k\big|_G $ is irreducible.
	
	The proof readily adapts for $ G = \Alt(2n) $. As we saw in \cref{eq:avg_density_calc}, the fact that the matrix elements have signs is irrelevant, as the factors arising due to the wreath product (i.e., $ 2^k $, $ 2^{2n-k} $, and $ 2^{2n} $) exactly cancel out. One may then simply replace every appearance of $ \Sym^{+}(2,d) $ with $ \Sym^{+}(d) \equiv \Alt(d) $ and $ \Sym^{-}(2,d) $ with $ \Sym^{-}(d) $ without consequence.
\end{proof}

In conjunction with the rigorous guarantees of classical shadows, \cref{thm:fgu_performance} of the main text follows. To simplify the shadow norm expression, we use Stirling's approximation, yielding
\begin{equation}
    \left. \binom{2n}{2k} \middle/ \binom{n}{k} \right. \approx \binom{n}{k} \sqrt{\pi k}.
\end{equation}
Additionally, the $ \log L $ factor from using classical shadows ($ L $ being the number of Majorana operators here) is
\begin{equation}
	\log\l( \sum_{j=1}^k |\comb{2n}{2j}| \r) = \O(k \log n).
\end{equation}

\subsection{\label{sec:class_shadow_appendix}The classical shadow estimator}

Here we provide a derivation for the formal expressions of the linear-inversion estimator used in the classical shadows methodology. Recalling \cref{eq:UzzU}, and using the linearity of $ \mathcal{M}^{-1}_{\FGU} $, the classical shadow is simply
\begin{equation}
    \begin{split}
    \hat{\rho}_{Q,z} &\equiv \mathcal{M}^{-1}_{\FGU} \l( U(Q)^\dagger \op{z}{z} U(Q) \r)\\
    &= \frac{1}{2^n} \l( \openone + \sum_{j=1}^n \lambda_{n,j}^{-1} \sum_{\bm{\mu} \in \diags{2n}{2j}} \ev{z}{\Gamma_{\bm{\mu}}}{z} \sum_{\bm{\nu} \in \comb{2n}{2j}} \det[Q_{\bm{\mu},\bm{\nu}}] \Gamma_{\bm{\nu}} \r),
    \end{split}
\end{equation}
where $ \lambda_{n,j}^{-1} = \binom{2n}{2j}/\binom{n}{j} $. Passing this expression into the expectation value estimator, we then obtain
\begin{equation}\label{eq:ev_estimator_app}
    \begin{split}
	\tr(\Gamma_{\bm{\tau}} \hat{\rho}_{Q,z}) &= \frac{1}{2^n} \l( \tr(\Gamma_{\bm{\tau}}) + \sum_{j=1}^n \lambda_{n,j}^{-1} \sum_{\bm{\mu} \in \diags{2n}{2j}} \ev{z}{\Gamma_{\bm{\mu}}}{z} \sum_{\bm{\nu} \in \comb{2n}{2j}} \det[Q_{\bm{\mu},\bm{\nu}}] \tr(\Gamma_{\bm{\tau}} \Gamma_{\bm{\nu}}) \r)\\
	&= \lambda_{n,k}^{-1} \sum_{\bm{\mu} \in \diags{2n}{2k}} \ev{z}{\Gamma_{\bm{\mu}}}{z} \det[Q_{\bm{\mu},\bm{\tau}}]
	\end{split}
\end{equation}
for all $ \bm{\tau} \in \comb{2n}{2k} $.

\subsection{\label{sec:arb_observable_appendix}Variance bounds for arbitrary observables}

For completeness, we provide a bound on the shadow norm (and hence the estimator variance) of an arbitrary fermionic observable. This result is particularly useful in the context of Hamiltonian averaging (e.g., \cref{sec:hamiltonian_appendix}).

Any element of $ \MA{n}{\mathrm{even}} $ can be written as
\begin{equation}
	O = h_{\varnothing} \openone + \sum_{j=1}^n \sum_{\bm{\mu} \in \comb{2n}{2j}} h_{\bm{\mu}} \Gamma_{\bm{\mu}},
\end{equation}
where $ h_{\bm{\mu}} \in \R $. Without loss of generality, we shall take $ \tr O = 0 $, since the identity component is irrelevant for variance calculations. Furthermore, to simplify the following exposition we shall suppose that $ O $ is at most a $ k $-body operator, so that $ h_{\bm{\mu}} = 0 $ for all $ |\bm{\mu}| > 2k $:
\begin{equation}
	O = \sum_{j=1}^k \sum_{\bm{\mu} \in \comb{2n}{2j}} h_{\bm{\mu}} \Gamma_{\bm{\mu}}.
\end{equation}
While we do not impose any restriction on $ k $, it is worth noting that most physical observables of interest obey $ k \leq 4 $, with $ k = 2 $ being particularly important (for instance, in describing electron--electron interactions).

Because the shadow norm is indeed a norm, it obeys the triangle inequality~\cite{huang2020predicting}. This property allows us to place a bound on
\begin{equation}
    \begin{split}
    \sns{O}{\FGU} &= \l\| \sum_{j=1}^k \sum_{\bm{\mu} \in \comb{2n}{2j}} h_{\bm{\mu}} \Gamma_{\bm{\mu}} \r{\|_{\FGU}}\\
    &\leq \sum_{j=1}^k \sum_{\bm{\mu} \in \comb{2n}{2j}} |h_{\bm{\mu}}| \sns{\Gamma_{\bm{\mu}}}{\FGU}\\
    &= \sum_{j=1}^k \lambda_{n,j}^{-1/2} \sum_{\bm{\mu} \in \comb{2n}{2j}} |h_{\bm{\mu}}|.
    \end{split}
\end{equation}
We therefore obtain an upper bound on the variance of our classical shadow estimator for $ \tr(O \rho) $ as
\begin{equation}
    \begin{split}
    \V_{Q,z} \l[ \tr(O \hat{\rho}_{Q,z}) \r] &\leq \sns{O}{\FGU}^2 - \tr(O \rho)^2\\
    &\leq \l( \sum_{j=1}^k \lambda_{n,j}^{-1/2} \sum_{\bm{\mu} \in \comb{2n}{2j}} |h_{\bm{\mu}}| \r{)^2} - \tr(O \rho)^2.
    \end{split}
\end{equation}
To get a sense for the asymptotic scaling of this expression, one may further loosen the estimate to obtain
\begin{equation}
    \V_{Q,z} \l[ \tr(O \hat{\rho}_{Q,z}) \r] \leq \l( \max_{1 \leq \ell \leq k} \lambda_{n,\ell}^{-1} \r) \l( \sum_{j=1}^k \sum_{\bm{\mu} \in \comb{2n}{2j}} |h_{\bm{\mu}}| \r{)^2},
\end{equation}
where $ \max_{1 \leq \ell \leq k} \lambda_{n,\ell}^{-1} = \O(n^k) $ when $ k = \O(1) $.

In follow-up works to this work, exact expressions for the variance and tighter bounds were derived by Wan \emph{et al.}~\cite{wan2023matchgate} and O'Gorman~\cite{ogorman2022fermionic}.

\subsection{\label{sec:hoeffding}Performance guarantees without median-of-means estimation}

As remarked in the main text, we do not require the median-of-means technique proposed in the original work~\cite{huang2020predicting} to obtain the same rigorous sampling bounds. Instead, one may simply use the typical sample mean:~given $ M $ independently obtained classical shadows $ \hat{\rho}_1, \ldots, \hat{\rho}_M $, define
\begin{equation}
    \omega_j(M) \coloneqq \frac{1}{M} \sum_{i=1}^M \tr(O_j \hat{\rho}_i)
\end{equation}
for each $ j \in \{1,\ldots,L\} $. Below, we state a general condition for which this estimator yields sample complexity equivalent to that of the median-of-means estimator.

\begin{theorem}\label{thm:sample_mean_estimator}
    Suppose the classical shadow estimators $ \hat{\rho}_{U,z} $ satisfy
    \begin{equation}\label{eq:rv_bound}
        -\sns{O_j}{\mathcal{U}}^2 \leq \tr(O_j \hat{\rho}_{U,z}) \leq \sns{O_j}{\mathcal{U}}^2
    \end{equation}
    for all $ U \in \mathcal{U} $, $ z \in \{0,1\}^n $, and $ j \in \{1,\ldots,L\} $. Let $ \varepsilon, \delta \in (0,1) $. Then by setting
    \begin{equation}\label{eq:sample_mean_M}
        M = \l(1 + \frac{\varepsilon}{3}\r) \frac{2 \log(2L/\delta)}{\varepsilon^2} \max_{1 \leq j \leq L} \sns{O_j}{\mathcal{U}}^2,
    \end{equation}
    we ensure that all sample-mean estimators $ \omega_1(M), \ldots, \omega_L(M) $ satisfy
    \begin{equation}
        | \omega_j(M) - \tr(O_j \rho) | \leq \varepsilon,
    \end{equation}
    with probability at least $ 1 - \delta $.
\end{theorem}

\begin{proof}
    The claim follows straightforwardly from Bernstein's inequality~\cite[Eq.~(2.10)]{boucheron2013concentration}: for a collection of independent random variables $ X_1, \ldots, X_M $ satisfying $ |X_i| \leq b $ for all $ i \in \{1,\ldots,M\} $, the probability that their empirical mean $ \bar{X} \coloneqq \frac{1}{M}\sum_{i=1}^M X_i $ deviates from the true mean $ \E[\bar{X}] $ by more than $ \varepsilon $ is bounded as
    \begin{equation}
        \Pr\l[ | \bar{X} - \E[\bar{X}] | \geq \varepsilon \r] \leq 2 \exp\l( - \frac{M^2 \varepsilon^2 / 2}{v + bM\varepsilon/3} \r),
    \end{equation}
    where $ v \coloneqq \sum_{i=1}^M \E[X_i^2] $.
    
    In our setting, for each $ j \in \{1,\ldots,L\} $, we have $ \bar{X} = \omega_j(M) $, $ b = \sns{O_j}{\mathcal{U}}^2 $, and $ v = M \sns{O_j}{\mathcal{U}}^2 $ (recall that the shadow norm squared is precisely $ \E[X_i^2] $~\cite{huang2020predicting}). The concentration inequality then reads
    \begin{equation}
        \Pr\l[ | \omega_j(M) - \tr(O_j \rho) | \geq \varepsilon \r] \leq 2 \exp\l[ - \frac{M \varepsilon^2 / 2}{\sns{O_j}{\mathcal{U}}^2 (1 + \varepsilon/3)} \r].
    \end{equation}
    If we require that each probability of failure be no more than $ \delta/L $, then from a union bound over all $ L $ events, we can succeed with probability at least $ 1 - \delta $ by setting
    \begin{equation}
        2 \exp\l( - \frac{M \varepsilon^2 / 2}{\max_{1 \leq j \leq L} \sns{O_j}{\mathcal{U}}^2 (1 + \varepsilon/3)} \r) = \frac{\delta}{L}.
    \end{equation}
    Solving for $ M $ yields \cref{eq:sample_mean_M}.
\end{proof}

For practical purposes, one typically desires that $ \varepsilon $ be small, thus $ (1 + \varepsilon/3) \approx 1 $. Importantly, \cref{thm:sample_mean_estimator} guarantees optimal scaling with the failure probability $ \delta $ and the number $ L $ of observables, using the sample mean rather than median-of-means estimation. The key detail which enables this observation is the boundedness of the classical shadows estimators, \cref{eq:rv_bound}. This condition holds for the ensembles presented in this work, $ \mathcal{U} = \mathcal{U}_{\FGU} $ and $ \mathcal{U}_{\NC} $ (see \cref{sec:ncu_appendix}), when taking the observables $ O_j $ as Majorana operators. It is also satisfied for estimating Pauli observables using the $ \Cl(1)^{\otimes n} $ ensemble of the original work~\cite{huang2020predicting}, indicating that median-of-means is redundant for estimating qubit RDMs as well.

Interestingly, \cref{eq:sample_mean_M} features significantly smaller numerical factors than what is obtained from the median-of-means approach, although we recognize that the proof techniques presented in Ref.~\cite{huang2020predicting} were not particularly optimized in this regard.

\section{Information-theoretic lower bounds on predicting local fermionic observables}\label{sec:FCS_lower_bound}

Our main result, \cref{thm:fgu_performance}, establishes that, with high probability,
\begin{equation}
    M = \O\l(\frac{\lambda_{n,k}^{-1} \log L}{\varepsilon^2}\r), \quad \lambda_{n,k}^{-1} = \l. \binom{2n}{2k} \middle/ \binom{n}{k} \r. = \O(n^k),
\end{equation}
samples from our fermionic classical shadows protocol suffice to estimate all $L = \O(n^{2k})$ $k$-body Majorana operators to accuracy $\varepsilon$. In this section, we prove a matching lower bound showing that, asymptotically, that many samples are also \emph{necessary} for any prediction algorithm to accomplish this task, given only the ability to perform local fermionic measurements. The claim is based on information-theoretic arguments~\cite{flammia2012quantum,haah2017sample}, which follows by adapting an analogous result of Huang \emph{et al.}~\cite{huang2020predicting} about local qubit observables to the fermion setting. We refer the reader to their paper for a lucid description of the intuition behind such arguments. We clarify that such lower bounds (including the one that we establish here) assumes only single-copy measurements, and does not apply to the task of predicting observables that are specified in advance. In other words, our result only applies to protocols which are tasked to learn all (local fermionic) observables equally well.

To state the formal theorem, we first need a general definition for a local fermionic measurement.

\begin{definition}
    Let $\{\ket{\varphi_i}\}_i$ be a collection of pure fermionic Gaussian states and $\{w_i\}_i \subset \R_{\geq 0}$ some weights. We say that $E = \{ w_i 2^n \op{\varphi_i}{\varphi_i} \}_i$ is a \emph{local fermionic measurement} if it is a POVM, i.e., if $2^n \sum_i w_i \op{\varphi_i}{\varphi_i} = \openone$.
\end{definition}

This definition encompasses any type of measurement in a (potentially overcomplete) basis of free-fermion states. Physically, this corresponds to any POVM that can be implemented with strictly noninteracting fermionic operations. A prime example is our $\FGU(n)$ ensemble. Furthermore, because we do not place any restrictions on the parity of the Gaussian states $\ket{\varphi_i}$, it also applies to the generalized matchgate ensembles of \cite{wan2023matchgate}, which are extensions of our $\FGU(n)$ to all of $\Orth(2n)$ [as well as its corresponding Clifford intersection, $\Sym(2, 2n)$] that retain the same performance guarantees for local-observable prediction.

We can now state the lower bound for learning from local fermionic measurements. We phrase the theorem parallel to the language of \cite[Theorem~S6]{huang2020predicting}, to underscore the similarity in transferring the concept to the fermionic setting.

\begin{theorem}[Formal version of \cref{thm:informal_lower_bound}]\label{thm:lower_bound}
	Fix a sequence of local fermionic measurements $ E_1,\ldots,E_M $ on an $ n $-mode system. Suppose that, given a collection of $L$ observables \emph{$ -\openone\preceq O_1,\ldots,O_L\preceq\openone $}, with Majorana degree exactly $2k$, there exists a machine (with arbitrary runtime, as long as it always terminates) that can use the outcomes of $ E_1,\ldots,E_M $ on $ M $ copies of an unknown quantum state $ \rho $ to predict $ \tr(O_1\rho),\ldots,\tr(O_L\rho) $, each up to additive error $ \varepsilon $, with high probability. Assuming $ L \leq \binom{2n}{2k} $, then necessarily
	\begin{equation}\label{eq:lower_bound_matchgate_result}
	M \geq \Omega\l( \frac{\lambda_{n,k}^{-1} \log L}{\varepsilon^2} \r),
	\end{equation}
	where $ \lambda_{n,k}^{-1}=\binom{2n}{2k}/\binom{n}{k} $.
\end{theorem}

The idea behind the proof of Huang \emph{et al.}~\cite{huang2020predicting} is a quantum communication protocol between two parties, Alice and Bob. They have agreed on a codebook of quantum states $\{\rho_1, \ldots, \rho_L\}$ that encodes classical information in the following way:~there exists a collection of observables $O_1, \ldots, O_L$ such that learning the $i$th property uniquely identifies the $i$th state. Suppose that this procedure can tolerate error at most $\varepsilon$ in the learned property to correctly identify the sent state. Then, Alice can use the codebook to pass some message $i$ to Bob by sending him $M$ copies of $\rho_i$. By performing the appropriate single-copy measurements, Bob can learn $\tr(O_j \rho_i)$ for each $j \in \{1, \ldots, L\}$ with sufficient accuracy, thereby identifying which state Alice sent and decoding her message.

However, this protocol does not fully embody the task we are interested in. Namely, Bob (the learner) does not know in advance which observables $O_j$ correspond to which states $\rho_i$, and only after measuring all the copies is that knowledge revealed to him. The communication protocol can be modified to reflect this aspect by introducing an interfering agent Loki, who intercepts each state sent by Alice and applies a random unitary transformation $\rho_i \mapsto U \rho_i U^\dagger$ before sending it along to Bob. Note that Loki applies the same $U$ to all $M$ copies. If Bob knew which $U$ Loki performed, then he could simply learn $\{U O_j U^\dagger\}_{j=1}^L$ instead. However, without that information, the task is more challenging---Bob needs to devise a strategy which succeeds for all possible $U$ from Loki's distribution. Only after Bob measures all $M$ copies does Loki eventually reveal the $U$ that he performed (as otherwise Bob would be hopelessly lost). The question is then to determine the minimum number of copies that Alice must send in order for Bob to decode her message in this scenario (with high probability).

Because the task we are interested in only considers learning $k$-local fermionic properties, we make the following restrictions to this communication protocol. First, the random unitary that Loki chooses is restricted to be Gaussian, because otherwise he could transform a local observable into a nonlocal one, and Bob's prediction machine has no guarantee to work for nonlocal properties. Second, because the $2k$-degree Majorana operators $\Gamma_{\bm{\mu}}$, $\bm{\mu} \in \comb{2n}{2k}$, form a complete basis for the space of all $2k$-degree operators, without loss of generality we can take them to be the observables $O_j$. This requires us to assume $L \leq |\comb{2n}{2k}| = \binom{2n}{2k}$, so that the codebook uniquely encodes each message. To lighten notation, we will simply write $\Gamma_{\bm{\mu}} \equiv \Gamma_i$ for some ordering of $\bm{\mu}$ into $i = 1, \ldots, L$.

With the problem now set up, we can prove \cref{thm:lower_bound} using this modified quantum communication protocol.

\begin{proof}[Proof (of \cref{thm:lower_bound}).]
Define the codebook via
\begin{equation}
    \rho_i \coloneqq \frac{1}{2^n} \l( \openone + \varepsilon \Gamma_i \r).
\end{equation}
This construction guarantees that the states are uniquely identifiable, but $ \varepsilon $-hard to do so:
\begin{equation}
    \tr(\Gamma_j \rho_i) = \varepsilon\delta_{ij}.
\end{equation}
The protocol then proceeds as follows. Alice randomly selects an integer $ X \sim \{1, \ldots, L\} $, prepares $M$ copies of $\rho_X$, and sends them to Bob. In between, Loki intercepts the states, samples a random $Q \sim \Orth(2n)$, and applies the corresponding fermionic Gaussian unitary, mapping $\rho_X \mapsto U(Q) \rho_X U(Q)^\dagger$. Bob then performs his fixed sequence of local fermionic measurements $E_j$, $j = 1, \ldots, M$, to each copy of $U(Q) \rho_X U(Q)^\dagger$. Denote his measurement outcomes as $Y = (Y_1, \ldots, Y_M)$.

Loki then reveals the randomly drawn $Q$ to Bob. Because $U(Q)$ preserves fermionic locality, the observables $U(Q) \Gamma_j U(Q)^\dagger$ remain $2k$-degree. Thus Bob's machine can predict expectation values of $U(Q) \Gamma_j U(Q)^\dagger$ from $Y$, and so in principle he can recover the original properties, since $\tr(\Gamma_j \rho_X) = \tr[(U(Q) \Gamma_j U(Q)^\dagger) (U(Q) \rho_X U(Q)^\dagger)]$. His decryption $\overline{X}$ of Alice's message $X$ is simply the choice of $\Gamma_{\overline{X}}$ with the largest estimated value for $\tr(\Gamma_{\overline{X}} \rho_X)$ returned by his prediction machine.

To determine how many copies of $ \rho_X $ that Bob needs for this procedure, one can examine the information content of the measurement outcomes $Y$. This analysis was already performed in \cite{huang2020predicting} with fairly high generality, so we simply quote the pertinent results here [adapted from their Eqs.~(S108), (S109), and (S112)]:
\begin{align}
    \Omega(\log L) \leq I(X : Y | Q) &\leq \sum_{j=1}^M I(X : E_j \text{ on } U(Q) \rho_X U(Q)^\dagger), \label{eq:mutual_info_bound}\\
    I(X : E_j \text{ on } U(Q) \rho_X U(Q)^\dagger) &\leq \sum_i \frac{\E_{X,Q}[p_{j,i}^2] - \E_{X,Q}[ p_{j,i} ]^2}{\E_{X,Q}[ p_{j,i} ]}, \label{eq:POVM_info_bound}
\end{align}
where $I$ is the (conditional) mutual information and
\begin{equation}
    p_{j,i} \coloneqq w_{j,i} 2^n \ev{\varphi_{j,i}}{U(Q) \rho_X U(Q)^\dagger}{\varphi_{j,i}}
\end{equation}
are the probabilities associated with the local fermionic POVMs $E_j = \{ w_{j,i} 2^n \op{\varphi_{j,i}}{\varphi_{j,i}} \}_i$. Intuitively, \cref{eq:mutual_info_bound} establishes a connection between the amount of information that Alice can encode ($\log L$ bits, using a codebook of length $L$) and the maximum amount of information that the measurements $E_j$ on the disrupted state $U(Q) \rho_X U(Q)^\dagger$ can reveal about her message $X$. It does so through $I(X : Y | Q)$, the mutual information between Alice's message $X$ and Bob's measurement data $Y$, conditioned on Loki's random unitary $U(Q)$. Meanwhile, \cref{eq:POVM_info_bound} establishes a bound on the information content of each individual measurement $E_j$ in terms of their outcome distributions.

Note that \cref{eq:POVM_info_bound} requires evaluating Haar averages over $Q \sim \Orth(2n)$ with respect to ``twirling'' by $U(Q)$. Fortunately, recent work by Wan \emph{et al.}~\cite{wan2023matchgate} did precisely this,\footnote{For comparison, in \cref{sec:fgu_appendix} we had established specific variants of these results, considering subgroups $\Alt(2n) \subset \Sym^{+}(2, 2n) \subset \SO(2n) \subset \Orth(2n)$ and averaging only over operators of the form $A_t = \sum_{z \in \{0,1\}^n} \op{z}{z}^{\otimes t}$. See \cref{chap:EM}, \cref{subsec:matchgate_bkgd} for further exposition.} deriving expressions for the following averages:
\begin{align}
    \mathcal{E}^{(1)}(A_1) &\coloneqq \E_{Q \sim \Orth(2n)}\l[ U(Q)^\dagger (A_1) U(Q) \r] = \frac{\tr(A_1)}{2^n} \openone, \label{eq:matchgate_twirl_1}\\
    \mathcal{E}^{(2)}(A_2) &\coloneqq \E_{Q \sim \Orth(2n)}\l[ (U(Q)^{\otimes 2})^\dagger (A_2) U(Q)^{\otimes 2} \r] = \frac{1}{4^n} \sum_{\ell=0}^{2n} \tr(\Upsilon_\ell^\dagger A_2) \Upsilon_\ell \label{eq:matchgate_twirl_2}
\end{align}
where $A_t \in (\C^{2^n \times 2^n})^{\otimes t}$ are arbitrary, and
\begin{equation}
    \Upsilon_\ell \coloneqq \frac{(-1)^{\binom{\ell}{2}}}{\sqrt{\binom{2n}{\ell}}} \sum_{\bm{\mu} \in \comb{2n}{\ell}} \Gamma_{\bm{\mu}}^{\otimes 2}.
\end{equation}
The average over $p_{j,i}$ can be computed using \cref{eq:matchgate_twirl_1}:
\begin{equation}
\begin{split}
    \E_{X,Q}[p_{j,i}] &= w_{j,i} 2^n \tr\l( \E_{X}\l[ \mathcal{E}^{(1)}(\op{\varphi_{j,i}}{\varphi_{j,i}}) \rho_X \r] \r)\\
    &= w_{j,i} \tr\l( \E_X \rho_X \r) = w_{j,i}.
\end{split}
\end{equation}

Similarly, the average over $p_{j,i}^2$ can be computed using \cref{eq:matchgate_twirl_2}. We will use an additional property of the twirl, namely that it is $\Orth(2n)$-invariant:
\begin{equation}
    \mathcal{E}^{(t)}\l( U(R)^{\otimes t} (\cdot) (U(R)^{\otimes t})^\dagger \r) = \mathcal{E}^{(t)}(\cdot)
\end{equation}
for all $t$ and any $R \in \Orth(2n)$. Thus, in conjunction with the fact that every Gaussian state can be written as $\ket{\varphi} = U(R) \ket{0}$ for some $R \in \Orth(2n)$, it follows that
\begin{equation}
    \E_{Q \sim \Orth(2n)}\l[ (U(Q)^{\otimes t})^\dagger (\op{\varphi}{\varphi})^{\otimes t} U(Q)^{\otimes t} \r] = \E_{Q \sim \Orth(2n)}\l[ (U(Q)^{\otimes t})^\dagger (\op{0}{0})^{\otimes t} U(Q)^{\otimes t} \r]
\end{equation}
for all Gaussian states $\ket{\varphi}$. This property greatly simplifies the calculation of $\E_{X,Q}[p_{j,i}^2]$:
\begin{equation}
\begin{split}
    \E_{X,Q}[p_{j,i}^2] &= w_{j,i}^2 4^n \tr\l( \E_{X}\l[ \mathcal{E}^{(2)}(\op{\varphi_{j,i}}{\varphi_{j,i}}^{\otimes 2}) \rho_X^{\otimes 2} \r] \r)\\
    &= w_{j,i}^2 4^n \tr\l( \mathcal{E}^{(2)}(\op{0}{0}^{\otimes 2}) \E_{X}[\rho_X^{\otimes 2}] \r).
\end{split}
\end{equation}
Wan \emph{et al.}~\cite{wan2023matchgate} have evaluated $\mathcal{E}^{(2)}(\op{0}{0}^{\otimes 2})$ as well, finding that
\begin{equation}
    \mathcal{E}^{(2)}(\op{0}{0}^{\otimes 2}) = \frac{1}{4^n} \sum_{\ell=0}^{n} \frac{\binom{n}{\ell}}{\binom{2n}{2\ell}} \sum_{\bm{\mu} \in \comb{2n}{2\ell}} \Gamma_{\bm{\mu}}^{\otimes 2}.
\end{equation}
Tracing this with $\rho_X^{\otimes 2}$ then yields
\begin{equation}
\begin{split}
    \tr\l( \mathcal{E}^{(2)}(\op{0}{0}^{\otimes 2}) \rho_X^{\otimes 2} \r) &= \frac{1}{4^n} \sum_{\ell=0}^{n} \frac{\binom{n}{\ell}}{\binom{2n}{2\ell}} \sum_{\bm{\mu} \in \comb{2n}{2\ell}} \tr( \Gamma_{\bm{\mu}} \rho_X )^2\\
    &= \frac{1}{4^n} \l( 1 + \frac{\binom{n}{k}}{\binom{2n}{2k}} \varepsilon^2 \r),
\end{split}
\end{equation}
which implies that
\begin{equation}
    \E_{X,Q}[p_{j,i}^2] = w_{j,i}^2 (1 + \lambda_{n,k} \varepsilon^2).
\end{equation}

Inserting these results into the mutual information bounds \cref{eq:POVM_info_bound} gives
\begin{equation}
\begin{split}
    I(X : E_j \text{ on } U(Q) \rho_X U(Q)^\dagger) &\leq \sum_i \frac{\E_{X,Q}[p_{j,i}^2] - \E_{X,Q}[ p_{j,i} ]^2}{\E_{X,Q}[ p_{j,i} ]}\\
    &= \sum_i \frac{w_{j,i}^2 (1 + \lambda_{n,k} \varepsilon^2) - w_{j,i}^2}{w_{j,i}}\\
    &= \lambda_{n,k} \varepsilon^2 \sum_i w_{j,i} = \lambda_{n,k} \varepsilon^2.
\end{split}
\end{equation}
Finally, we apply the inequality of \cref{eq:mutual_info_bound},
\begin{equation}
\begin{split}
    \Omega(\log L) &\leq \sum_{j=1}^M I(X : E_j \text{ on } U(Q) \rho_X U(Q)^\dagger)\\
    &\leq M \lambda_{n,k} \varepsilon^2,
\end{split}
\end{equation}
which implies \cref{eq:lower_bound_matchgate_result} as claimed.
\end{proof}

\section{\label{sec:ncu_appendix}Computations with the number-conserving modification}

We now prove \cref{thm:NC_bounds} of the main text. Many of the techniques used here follow straightforwardly from the fermionic formalism developed in \cref{sec:fgu_appendix}, along with the tools used to study the single-qubit Clifford ensemble in the original work on classical shadows~\cite{huang2020predicting}. In \cref{sec:ncu_calcs}, we first evaluate the channel $ \mathcal{M}_\NC $ [which again is diagonalized by the Majorana operators;~\cref{eq:NC_eig}] and provide an expression for its eigenvalues/shadow norm [\cref{eq:nc_norm}]. Then, recognizing that generically do not possess a closed-form expression, in \cref{sec:ncu_ub} we obtain an upper bound on the shadow norm for this ensemble, which is asymptotically optimal [\cref{eq:nc_norm_bound}].

For ease of notation, with $ u \in \Sym(n) $, we shall write $ U(u)^\dagger a_p U(u) = a_{u(p)} $, where $ u(p) $ is understood in the sense of the action of the permutation $ u $ on the mode indices $ \{ 0,\ldots,n-1 \} $. We further generalize this notation to act on $ \{ 0,\ldots,2n-1 \} $, in accordance with the definition of Majorana operators. Consider $ \bm{\mu} \in \comb{2n}{k} $, where each index takes the form $ \mu_j = 2q_j + x_j $ for $ q_j \in \{ 0,2,\ldots,2n-4,2n-2 \} $ and $ x_j \in \{ 0,1 \} $. We define $ \tilde{u}(\bm{\mu}) \coloneqq (2u(q_1)+x_1, \ldots, 2u(q_{k})+x_{k} ) $, so that $ U(u)^\dagger \Gamma_{\bm{\mu}} U(u) = \Gamma_{\tilde{u}(\bm{\mu})} $. Note that $ \tilde{u}(\bm{\mu}) $ is not necessarily ordered monotonically, so $ \Gamma_{\tilde{u}(\bm{\mu})} $ may differ from our standard definition of the Majorana operators by a minus sign. This detail is irrelevant to our present analysis, so we shall ignore it.

\subsection{\label{sec:ncu_calcs}The shadow norm}

The ensemble we consider here is
\begin{equation}
    \mathcal{U}_{\mathrm{NC}} = \{ V \circ U(u) \mid V \in \Cl(1)^{\otimes n}, \,  u \in \Alt(n) \}.
\end{equation}
We wish to evaluate the following expression for the classical shadows channel:
\begin{equation}
    \begin{split}
	\mathcal{M}_{\mathrm{NC}}(\Gamma_{\bm{\mu}}) &= \E_{\substack{V \sim \Cl(1)^{\otimes n} \\ u \sim \Alt(n)}} \l[ \sum_{z\in\{0,1\}^n} \ev{z}{V U(u) \Gamma_{\bm{\mu}} U(u)^\dagger V^\dagger}{z} U(u)^\dagger V^\dagger \op{z}{z} V U(u) \r]\\
	&= \E_{u \sim \Alt(n)} \l[ U(u)^\dagger \sum_{z\in\{0,1\}^n} \E_{V \sim \Cl(1)^{\otimes n}} \Big[ \ev{z}{V \Gamma_{\tilde{u}^{-1}(\bm{\mu})} V^\dagger}{z} V^\dagger \op{z}{z} V \Big] U(u) \r].
	\end{split}
\end{equation}
The average over $ \Cl(1)^{\otimes n} $ is precisely the same quantity evaluated in Ref.~\cite{huang2020predicting}. For convenience, we restate their results here:~let $ \ket{x} = \ket{x_1} \otimes \cdots \otimes \ket{x_n} $ be a product state over $ n $ qubits and $ A,B,C $ be Hermitian matrices which decompose into the same tensor product structure, $ A = A_1 \otimes \cdots \otimes A_n $, etc. Then
\begin{equation}\label{eq:cl_avg_1}
    \E_{V \sim \Cl(1)^{\otimes n}} \l[ V^\dagger \op{x}{x} V \ev{x}{V A V^\dagger}{x} \r] = \bigotimes_{j=1}^n \l( \frac{A_j + \tr(A_j) I}{6} \r),
\end{equation}
and for each $ j $ such that $ \tr B_j = \tr C_j = 0 $,
\begin{equation}\label{eq:cl_avg_2}
    \E_{V_j \sim \Cl(1)} \l[ V_j^\dagger \op{x_j}{x_j} V_j \ev{x_j}{V_j B_j V_j^\dagger}{x_j} \ev{x_j}{V_j C_j V_j^\dagger}{x_j} \r] = \frac{B_j C_j + C_j B_j + \tr(B_j C_j) I}{24}.
\end{equation}
To understand the tensor product structure of Majorana operators, we must fix some qubit mapping. Understanding $ \Gamma_{\tilde{u}^{-1}(\bm{\mu})} $ as a Pauli operator under such a mapping, we use \cref{eq:cl_avg_1} to obtain
\begin{equation}\label{eq:NC_eig}
    \begin{split}
    \mathcal{M}_{\mathrm{NC}}(\Gamma_{\bm{\mu}}) &= \E_{u \sim \Alt(n)} \l[ \frac{1}{3^{\loc(\Gamma_{\tilde{u}^{-1}(\bm{\mu})})}} U(u)^\dagger \Gamma_{\tilde{u}^{-1}(\bm{\mu})} U(u) \r]\\
    &= \E_{u \sim \Alt(n)} \l[ \frac{1}{3^{\loc(\Gamma_{\tilde{u}^{-1}(\bm{\mu})})}} \r] \Gamma_{\bm{\mu}},
    \end{split}
\end{equation}
where $ \loc(\Gamma_{\tilde{u}^{-1}(\bm{\mu})}) $ is the qubit locality of $ \Gamma_{\tilde{u}^{-1}(\bm{\mu})} $. Since $ u \mapsto u^{-1} $ is a bijection, we may equivalently express the average over $ \Alt(n) $ using $ \tilde{u}(\bm{\mu}) $ rather than its inverse.

Let $ \lambda_{\bm{\mu}} $ be the eigenvalues of $ \mathcal{M}_{\mathrm{NC}} $, given above in \cref{eq:NC_eig}. For the calculation of the shadow norm, we have
\begin{align}
	\sns{\Gamma_{\bm{\mu}}}{\mathrm{NC}}^2 &= \max_{\text{states } \rho} \l( \E_{\substack{V \sim \Cl(1)^{\otimes n} \\ u \sim \Alt(n)}} \l[ \sum_{z\in\{0,1\}^n} \ev{z}{V U(u) \rho U(u)^\dagger V^\dagger}{z} \ev{z}{V U(u) \mathcal{M}^{-1}(\Gamma_{\bm{\mu}}) U(u)^\dagger V^\dagger}{z}^2 \r] \r),\\
	&= \lambda_{\bm{\mu}}^{-2} \max_{\text{states } \rho} \l( \E_{u \sim \Alt(n)} \l[ \tr\l( U(u)^\dagger \rho U(u) \sum_{z\in\{0,1\}^n} \E_{V \sim \Cl(1)^{\otimes n}} \Big[ V^\dagger \op{z}{z} V \ev{z}{V \Gamma_{\tilde{u}^{-1}(\bm{\mu})} V^\dagger}{z}^2 \Big] \r) \r] \r). \notag
\end{align}
We then use \cref{eq:cl_avg_1} to evaluate the Clifford average over the identity factors of $ \Gamma_{\tilde{u}^{-1}(\bm{\mu})} $, and \cref{eq:cl_avg_2} for the nontrivial factors:
\begin{equation}\label{eq:nc_norm}
    \begin{split}
    \sns{\Gamma_{\bm{\mu}}}{\mathrm{NC}}^2 &= \lambda_{\bm{\mu}}^{-2} \max_{\text{states } \rho} \l( \E_{u \sim \Alt(n)} \l[ \tr\l( U(u)^\dagger \rho U(u) \frac{1}{3^{\loc(\Gamma_{\tilde{u}^{-1}(\bm{\mu})})}} \r) \r] \r)\\
    &= \lambda_{\bm{\mu}}^{-1}.
    \end{split}
\end{equation}

\subsection{\label{sec:ncu_ub}Universal upper bounds on the shadow norm}

Although there is no closed-form expression for $ \lambda_{\bm{\mu}}^{-1} $, we can still obtain a nontrivial estimate for it. To do so, we will evaluate the qubit locality with respect to the Jordan--Wigner transformation. This serves as a universal upper bound for all encodings, since the Jordan--Wigner mapping is maximally nonlocal. We formalize this notion with Jensen's inequality:~since $ x \mapsto 3^{-x} $ is concave, we have that
\begin{equation}
	\E_{u \sim \Alt(n)} \l[ 3^{-\loc(\Gamma_{\tilde{u}(\bm{\mu})})} \r] \geq 3^{-\E_{u \sim \Alt(n)}\l[ \loc(\Gamma_{\tilde{u}(\bm{\mu})}) \r]},
\end{equation}
where $ \loc(\,\cdot\,) $ is with respect to any arbitrary encoding. Then let $ \loc_{\mathrm{JW}}(\,\cdot\,) $ be the qubit locality specifically under the Jordan--Wigner transformation. Since it is maximally nonlocal, so is its average locality (over a fixed fermionic degree of $ 2k $), and hence
\begin{equation}
	\max_{\bm{\mu} \in \comb{2n}{2k}} \lambda_{\bm{\mu}}^{-1} \leq \max_{\bm{\mu} \in \comb{2n}{2k}} 3^{\E_{u \sim \Alt(n)}\l[ \loc(\Gamma_{\tilde{u}(\bm{\mu})}) \r]} \leq \max_{\bm{\mu} \in \comb{2n}{2k}} 3^{\E_{u \sim \Alt(n)}\l[ \loc_{\mathrm{JW}}(\Gamma_{\tilde{u}(\bm{\mu})}) \r]}.
\end{equation}
Thus for the rest of this section, all notions of qubit locality will be understood with respect to the Jordan--Wigner transformation exclusively.

Since the qubit locality of Majorana operators varies within a given $ \comb{2n}{2k} $, we obtain an upper bound by considering the most nonlocal $ 2k $-degree operators. Fortunately, under Jordan--Wigner it is simple to identify such operators. Let $ W \in \{ X,Y \} $ and $ \bm{q} \in \comb{n}{2k} $;~maximum locality is achieved by operators of the form
\begin{align}
	P_{\bm{q}} &\coloneqq \prod_{i=1}^k W_{q_{2i-1}} Z_{q_{2i-1}+1} \cdots Z_{q_{2i}-1} W_{q_{2i}},\\
	\loc(P_{\bm{q}}) &= \sum_{i=1}^k \l( q_{2i} - q_{2i-1} + 1 \r).
\end{align}
Note that we require $ n \geq 2k $ for such a Pauli operator to exist. (For the fringe cases in which $ n < 2k $, since we are evaluating an upper bound, the final result still holds.)

By applying a permutation $ u $ on $ \bm{q} $, one changes this locality by reducing or lengthening the various ``Jordan--Wigner strings'' of Pauli-$ Z $ operators in between each pair $ (q_{2i-1}, q_{2i}) $. Observe that $ \loc(P_{\bm{q}}) \in \{ 2k, \ldots, n \} $. Therefore, if we can calculate the number of permutations which correspond to each level of locality, we may compute
\begin{equation}\label{eq:ncsq_norm_comb}
	\E_{u \sim \Alt(n)} \l[ 3^{-\loc(P_{u(\bm{q})})} \r] = \frac{2}{n!} \sum_{\ell=2k}^n 3^{-\ell} \, | \{ u \in \Alt(n) \mid \loc(P_{u(\bm{q})}) = \ell \} |.
\end{equation}
We determine the size of this set in three steps. First, for a given configuration $ u(\bm{q}) $, we count the number of equivalent permutations which give the same Pauli operator, modulo signs. These permutations merely reorder the $ 2k $ relevant indices and the $ n-2k $ remaining indices independently. Accounting for the fact that the entire permutation must be even parity, we have
\begin{equation}
	\begin{split}
	|\Alt(2k) \oplus \Alt(n - 2k)| + |\Sym^{-}(2k) \oplus \Sym^{-}(n - 2k)| &= \frac{(2k)!}{2} \frac{(n-2k)!}{2} + \frac{(2k)!}{2} \frac{(n-2k)!}{2}\\
	&= \frac{(2k)!(n-2k)!}{2}
	\end{split}
\end{equation}
such permutations. With this factor at hand, we now only need to consider unique combinations of indices---that is, we assume $ u(q_1) < \cdots < u(q_{2k}) $ in the sequel.

Next, we calculate how many configurations of Jordan--Wigner strings give rise to a qubit locality of exactly $ \ell $. Let $ l_i \coloneqq u(q_{2i}) - u(q_{2i-1}) > 0 $. This problem is equivalent to finding all $ k $-tuples $ (l_1, \ldots, l_k) $ of positive integers such that
\begin{equation}\label{eq:snb_constraint}
	\sum_{i=1}^k (l_i + 1) = \ell.
\end{equation}
This is an instance of the classic ``stars-and-bars'' combinatorial problem, wherein we wish to fit $ \ell - k $ objects into $ k $ bins such that each bin has at least $ 1 $ object. There are $ \binom{\ell-k-1}{k-1} $ ways to do so.

Finally, we need to account for the $ n-\ell $ remaining indices which were not fixed by \cref{eq:snb_constraint}. These indices correspond to the qubits \emph{outside} of the Jordan--Wigner strings, i.e., on which $ P_{u(\bm{q})} $ acts trivially. By a similar combinatorial argument, there are a total of $ k+1 $ such spaces to place these trivial indices, of which we may select $ j \in \{ 1,\ldots,k+1 \} $. Again we use the stars-and-bars argument:~there are $ n-\ell $ objects we wish to place into $ j $ bins, which can be accomplished in $ \binom{n-\ell-1}{j-1} $ unique ways. Summing over all possible values of $ j $ gives us
\begin{equation}
	\begin{split}
	\sum_{j=1}^{k+1} \binom{k+1}{j} \binom{n-\ell-1}{j-1} &= \sum_{j=1}^{k+1} \binom{k+1}{k+1-j} \binom{n-\ell-1}{j-1}\\
	&= \sum_{j=0}^{k} \binom{k+1}{k-j} \binom{n-\ell-1}{j} = \binom{n+k-\ell}{k}
	\end{split}
\end{equation}
different combinations of the $ n - \ell $ trivial indices. Note that the Chu--Vandermonde identity was applied to evaluate the sum in the final line.

Reconciling the three steps of our calculation, we obtain
\begin{equation}
	| \{ u \in \Sym(n) \mid \loc(P_{u(\bm{q})}) = \ell \} | = (2k)!(n-2k)! \binom{\ell-k-1}{k-1} \binom{n+k-\ell}{k},
\end{equation}
and so \cref{eq:ncsq_norm_comb} can be expressed as
\begin{equation}
	\begin{split}
	\E_{u \sim \Sym(n)} \l[ 3^{-\loc(P_{u(\bm{q})})} \r] &= \frac{1}{n!} \sum_{\ell=2k}^n 3^{-\ell} \, | \{ u \in \Sym(n) \mid \loc(P_{u(\bm{q})}) = \ell \} |\\
	&= \frac{(2k)!(n-2k)!}{n!} \sum_{\ell=2k}^n 3^{-\ell} \binom{\ell-k-1}{k-1} \binom{n+k-\ell}{k}.
	\end{split}
\end{equation}
To evaluate this sum, first we relabel the index $ \ell $ to run from $ 0 $ to $ n-2k $, so that the summand becomes $ 3^{-(\ell + 2k)} \binom{\ell+k-1}{k-1} \binom{n-k-\ell}{k} $. Using the combinatorial identity
\begin{equation}
    \binom{n-k}{\ell} \binom{n-k-\ell}{k} = \binom{n-k}{k} \binom{n-2k}{\ell},
\end{equation}
we have
\begin{equation}
    \begin{split}
    \binom{n-k-\ell}{k} &= \binom{n-k}{k} \frac{(n-2k)!}{(n-2k-\ell)! \ell!} \frac{(n-k-\ell)! \ell!}{ (n-k)! }\\
    &= \binom{n-k}{k} \frac{(2k-n)_\ell}{(k-n)_\ell},
    \end{split}
\end{equation}
where $ (m)_\ell \coloneqq \prod_{j=0}^{n-1} (m+j) $ is the rising Pochhammer symbol. We recognize that the other binomial coefficient can also be expressed using these Pochhammer symbols:
\begin{equation}
    \binom{\ell+k-1}{k-1} = \frac{(k)_\ell}{\ell!}.
\end{equation}
The sum can therefore be understood in terms of the Gauss hypergeometric function $ {}_2 F_1 $:
\begin{equation}\label{eq:avg_f}
    \begin{split}
    \E_{u \sim \Sym(n)} \l[ 3^{-\loc(P_{u(\bm{q})})} \r] &= \binom{n}{2k}^{-1} \binom{n-k}{k} 9^{-k} \sum_{\ell=0}^{n-2k} \frac{3^{-\ell}}{\ell!} \frac{(k)_\ell (2k-n)_\ell}{(k-n)_\ell}\\
    &= \binom{n}{2k}^{-1} \binom{n-k}{k} 9^{-k} \, {}_2 F_1(k, 2k-n; k-n; 1/3).
    \end{split}
\end{equation}
Using standard properties of the hypergeometric function, for $ n $ and $ k $ being positive integers and $ k $ a constant, we have the bounds $ 1 \leq {}_2 F_1(k, 2k-n; k-n; 1/3) \leq (3/2)^k $. In particular, this factor is independent of $ n $.

We are ultimately interested in the lower bound of \cref{eq:avg_f}, since the shadow norm is its reciprocal. Then for $ k = \O(1) $, we obtain
\begin{equation}\label{eq:nc_norm_bound}
	\sns{P_{\bm{q}}}{\mathrm{NC}}^2 \leq \binom{n}{2k} \binom{n-k}{k}^{-1} 9^k = \O(n^k).
\end{equation}

\section{\label{sec:rdmswap_appendix}Fermionic swap network bounds}

In this section we describe another strategy to measure the $ k $-RDM with almost optimal scaling in $ n $. Generalizing from the measurement scheme introduced in Ref.~\cite{arute2020hartree} for 1-RDMs, this scheme employs fermionic swap gates to relabel which qubits correspond to which orbitals such that each $ k $-RDM observable becomes a $ 2k $-local qubit observable. These local qubit observables may then be measured in a parallel fashion via Pauli measurements. As the circuit structure is equivalent, our $ \mathcal{U}_{\NC} $-based scheme may be viewed as a randomized version of this strategy.

For simplicity, we employ the Jordan--Wigner encoding throughout this section. The use of fermionic swap networks to minimize the qubit locality of fermionic operators was first utilized in the context of Hamiltonian simulation~\cite{kivlichan2018quantum}.

\subsection{The 1-RDM method}

We briefly describe the methods of Ref.~\cite{arute2020hartree} here. Consider estimating the 1-RDM elements $ \tr( a_p^\dagger a_q \rho ) $. The observables here are $ \frac{1}{2}(a_p^\dagger a_q + \mathrm{h.c.}) $ and $ \frac{1}{2\i}(a_p^\dagger a_q - \mathrm{h.c.}) $, corresponding to the real and imaginary parts of the RDM element. In the experiment of Ref.~\cite{arute2020hartree}, they implement unitaries such that the imaginary part vanishes;~for full generality, we will keep the imaginary parts. Diagonal elements are trivial to measure, since
\begin{equation}
	a_p^\dagger a_p = \frac{I - Z_p}{2}.
\end{equation}
The one-off-diagonal terms are precisely the local qubit operators we are interested in:
\begin{align}
	\frac{a_{p}^\dagger a_{p+1} + a_{p+1}^\dagger a_{p}}{2} &= \frac{X_{p} X_{p+1} + Y_{p} Y_{p+1}}{4},\\
	\frac{a_{p}^\dagger a_{p+1} - a_{p+1}^\dagger a_{p}}{2\i} &= \frac{X_{p} Y_{p+1} - Y_{p} X_{p+1}}{4}.
\end{align}
Consider even and odd pairs of orbitals---even pairs being those starting with even indices, and analogously for the odd pairs. For the expectation values $ \tr( a_p^\dagger a_{p+1} \rho ) $ on even pairs, we measure in 4 different bases:~$ X $ on all qubits, $ Y $ on all qubits, $ X $ on every even qubit and $ Y $ on every odd qubit, and vice versa. Formally, the observables we measure are
\begin{equation}
    \begin{split}
	O_1 = \prod_{p=0}^{n-1} X_p, \quad & O_2 = \prod_{p=0}^{n-1} Y_p\\
	O_3 = \prod_{\substack{p=0\\\text{even}}}^{n-2} X_p Y_{p+1}, \quad & O_4 = \prod_{\substack{p=0\\\text{even}}}^{n-2} Y_p X_{p+1}.
	\end{split}
\end{equation}
If $ n $ is odd, then we simply ignore the final $ Y $ (resp.~$ X $) in $ O_3 $ (resp.~$ O_4 $).

The remaining off-diagonal elements will incur Jordan--Wigner strings, making the Pauli operators highly nonlocal. To circumvent this, we perform fermionic swaps to relabel the indices such that we retain qubit locality. The fermionic swap gate between orbitals $ p $ and $ q $ is
\begin{equation}
	\mathcal{F}_{pq} = \exp\l[ -\i \frac{\pi}{2} ( a_p^\dagger a_q + a_q^\dagger a_p - a_p^\dagger a_p - a_q^\dagger a_q ) \r].
\end{equation}
This unitary is Gaussian and number-preserving, hence one may use the group homomorphism property to consolidate an arbitrarily large product of fermionic swaps into a single circuit of depth $ n $~\cite{kivlichan2018quantum,jiang2018quantum}.

As described in Ref.~\cite{kivlichan2018quantum}, a total of $ \lceil n/2 \rceil $ different swap circuits are required to move orbitals such that every pair is nearest-neighbor at least once. This quantity can be understood from a simple counting argument:~there are $ \binom{n}{2} $ off-diagonal 1-RDM elements (orbital pairs) to account for. Each ordering creates $ n-1 $ nearest-neighbor pairs---therefore, we require $ \binom{n}{2}/(n-1) = n/2 $ different orderings (hence unique swap circuits) to match all pairs of orbitals. In practice, this is achieved using a parallelized odd--even transposition sort~\cite{habermann1972parallel}.

Rounding up in the case that $ n $ is odd, and accounting for the four Pauli bases per permutation, the total number of different measurement circuits is $ 4 \lceil n/2 \rceil + 1 $.

\subsection{The 2-RDM method}

We now generalize the use of such swap networks for measuring $ k $-RDMs. We will build up intuition with the 2-RDM. Diagonal terms are trivial as usual, since
\begin{equation}
	a_p^\dagger a_q^\dagger a_q a_p = (1 - \delta_{pq}) \frac{I - Z_p - Z_q + Z_p Z_q}{4}.
\end{equation}
The terms with a single occupation-number operator, restricted to 3-qubit locality, are
\begin{align}
	\frac{a_p^\dagger a_{q}^\dagger a_{q} a_{p+1} + \mathrm{h.c.}}{2} &= \frac{ X_p X_{p+1} + Y_p Y_{p+1} - X_p X_{p+1} Z_{q} - Y_p Y_{p+1} Z_{q} }{8},\\
	\frac{a_p^\dagger a_{q}^\dagger a_{q} a_{p+1} - \mathrm{h.c.}}{2\i} &= \frac{ X_p Y_{p+1} - Y_p X_{p+1} - X_p Y_{p+1} Z_{q} + Y_p X_{p+1} Z_{q} }{8},
\end{align}
where $ q \notin \{ p,p+1 \} $. Lastly, we have the most general case:
\begin{align}\label{eq:general_tpdm}
	\frac{ a_{p}^\dagger a_{p+1}^\dagger a_{q} a_{q+1} + \mathrm{h.c.} }{2} &= \frac{1}{16} ( - X_{p} X_{p+1} X_{q} X_{q+1} + X_{p} X_{p+1} Y_{q} Y_{q+1} \\
	&\qquad\ \ \, - X_{p} Y_{p+1} X_{q} Y_{q+1} - X_{p} Y_{p+1} Y_{q} X_{q+1} \notag\\
	&\qquad\ \ \, - Y_{p} X_{p+1} X_{q} Y_{q+1} - Y_{p} X_{p+1} Y_{q} X_{q+1} \notag\\
	&\qquad\ \ \, + Y_{p} Y_{p+1} X_{q} X_{q+1} - Y_{p} Y_{p+1} Y_{q} Y_{q+1} ), \notag\\
	\frac{ a_{p}^\dagger a_{p+1}^\dagger a_{q} a_{q+1} - \mathrm{h.c.} }{2\i} &= \frac{1}{16} ( - X_{p} X_{p+1} X_{q} Y_{q+1} - X_{p} X_{p+1} Y_{q} X_{q+1} \\
	&\qquad\ \ \, + X_{p} Y_{p+1} X_{q} X_{q+1} - X_{p} Y_{p+1} Y_{q} Y_{q+1} \notag\\
	&\qquad\ \ \, + Y_{p} X_{p+1} X_{q} X_{q+1} - Y_{p} X_{p+1} Y_{q} Y_{q+1} \notag\\
	&\qquad\ \ \, + Y_{p} Y_{p+1} X_{q} Y_{q+1} + Y_{p} Y_{p+1} Y_{q} X_{q+1} ). \notag
\end{align}
The $ a_{p}^\dagger a_{q}^\dagger a_{q+1} a_{p+1} $ terms feature the same Pauli operators, but with a different sign pattern in the linear combination.

The 3- and 4-local terms are best handled separately. For the 3-local terms, the $ q $th index is free to take any value different from $ p $ and $ p+1 $. We measure these qubits in the computational basis, and the $ p $ and $ (p+1) $th qubits in the same fashion as in the 1-RDM case. Then to obtain all combinations (triples) $ (p,p',q) $, we have to swap all pairs $ (p,p') $ into, say, the qubit ordering $ (0,1) $. This allows a single ordering to account for $ (n-2) $ triples. There are $ \binom{n}{2}(n-2) $ total triples to permute into, and so we require $ \binom{n}{2} $ different swap circuits. To measure all 8 terms, we require 4 different Pauli bases, thus $ 4\binom{n}{2} $ different measurement circuits.  In practice we anticipate using the parallel transposition sort of the $1$-RDM measurement with an additional $n-1$ measurements at each swap circuit to account for each $(p,p', q)$ triple as $n-2$ $(p, p', q)$ triples can be acquired simultaneously by measuring $p, p'$ in either $X$ or $Y$ and all other qubits in $Z$.

The 4-local terms require us to swap the orbital orderings into 2-combinations of pairs in order to measure all general $ a_{p}^\dagger a_{p'}^\dagger a_{q} a_{q'} $ terms. First, the number of nearest-neighbor pairs in $ \{0,\ldots,n-1\} $ is $ n-1 $. Then we count how many 2-combinations of these pairs we can construct. Note that locality between 2-combinations is not a constraint, since $ p $ and $ q $ do not have to be local. However, they must be disjoint, otherwise we are double counting the 3-local terms. Each swap circuit can account for $ \binom{n-2}{2} $ different pairs of pairs;~thus, since there are $ \binom{n}{2}\binom{n-2}{2} $ unique $ a_{p}^\dagger a_{p'}^\dagger a_{q} a_{q'} $ terms, exactly $ \binom{n}{2} $ swap circuits are required. This statement is proved by a simple combinatorial argument, which we defer to the generalization in \cref{sec:k-generalization}.

The 16 Pauli operators cannot be measured in 16 bases whenever $n > 7$. In particular, from Eq.~\eqref{eq:general_tpdm} we observe that only the $XXXX$ and $YYYY$ geminals can be measured simultaneously for $n > 7$. Therefore, in order to read off the 4-local Pauli observables, we rely on quantum overlap tomography (QOT)~\cite{cotler2020quantum} to provide asymptotic bounds.  Each $(p,p',q,q')$ term requires measuring almost all $4$-qubit marginals.  More precisely, we need only the marginal elements corresponding to expectation values of Pauli operators composed of $X$ and $Y$ operators.  QOT provides a bound for $4$-qubit bound that has $\mathcal{O}(\log(n))$ scaling whenever the perfect hash family is known.  Though large sets of the $(n,4)$-perfect hash families are documented, there remain significant gaps in $n$.  

A more general procedure that does not achieve the same asymptotic bound of Ref.~\cite{cotler2020quantum} is the construction of a suboptimal perfect hash family by bootstrapping from the binary partitioning scheme described in Ref.~\cite{bonet2020nearly}.  Using this approach, the leading order complexity, which upper bounds the true scaling, in the number of measurement settings is defined as the function EQOT (explicit quantum overlap tomography):
\begin{equation}\label{eq:qot_extende_scaling}
\mathrm{EQOT}(k, n) = 3^{k}(k - 1)\sum_{m = 0}^{\lceil \log(n) \rceil - 1} m^{k - 2}.
\end{equation}
This gives the number of partitions to measure all $k$-qubit marginals, multiplied by all $3^k$ bases for each partition. In the case of the 4-local 2-RDM terms, however, we do not need to measure in any Pauli basis containing a $Z$, so we can straightforwardly reduce the base of this prefactor to $2$.  Altogether, we can upper bound the number of circuit configurations as
\begin{equation}
M_{T}(\text{2-RDM}) = 1 + 4\binom{n}{2} + \binom{n}{2} \mathrm{EQOT}(4, n) \l(\frac{2}{3}\r{)^4}
\end{equation}
circuits to measure the full fermionic 2-RDM with this approach.  Again we emphasize that, though this $ \mathrm{EQOT}(4, n) $ scaling is not optimal, it will work for any $n$.  In the following section we prove the partition scaling for the component of the $k$-RDM with $2k$ distinct indices, and thus the complexity, given an optimal construction of the swap circuits.

\subsection{\label{sec:k-generalization}A $ k $-RDM generalization}

Each $ k $-RDM observable decomposes into up to $ 4^k $ Pauli operators, thus is it impractical to write out such terms for general $ k $ by hand. Nonetheless, we can still obtain a scaling estimate on how many measurement circuits are required to reach all $ k $-RDM elements. The following argument also applies for the 4-local terms of the 2-RDM in the previous section. Consider the asymptotically dominant terms, $ a_{p_1}^\dagger \cdots a_{p_k}^\dagger a_{p_k+1} \cdots a_{p_{2k}} $ where all $ p_i\neq p_j $ for $ i\neq j $. There are $ \binom{n}{k}\binom{n-k}{k} $ such index combinations. The number of $ k $-combinations of disjoint nearest-neighbor pairs taken from $ \{(0,1),(1,2),\ldots,(n-2,n-1)\} $ is equivalent to counting how many unique sets of $ k $ nonconsecutive integers from $ \{0,\ldots,n-2\} $ exist. This is a classic ``stars and bars'' combinatorial problem and has solution $ \binom{(n-1)-k+1}{k}=\binom{n-k}{k} $. We can see this by a visual argument:~write down $ (n-1)-k $ spaces where the unchosen numbers will be placed in order. Then there are $ (n-1)-k-1 $ gaps in between the spaces, plus the $ 2 $ endpoints, where the chosen numbers can be placed, of which we choose $ k $. Thus we require $ \binom{n}{k} $ different swap circuits, hence $ \Omega\l[\binom{n}{k} 4^k\r] $ unique measurement circuits, to reach all $ k $-RDM elements. Similar to the 2-RDM case, this lower bound is in general not achievable, as we require (E)QOT to measure the Pauli operators in parallel, thus incurring polylogarithmic factors. We can upper bound this scaling by counting circuit repetitions for each $k$-RDM element partitioned by the number of unique indices.

\subsubsection{Upper bounds for $k = 3, 4$}
To derive an upper bound  for arbitrary $n$ in terms of measurement configurations for the 3-RDM and 4-RDM, we can follow the same procedure as the 2-RDM:~count circuits for measuring RDM terms after partitioning based on the number of unique indices in each RDM element.  The 3-RDM is partitioned into terms with 3, 4, 5, 6 different indices.  This can be checked by building a basis for the unique 3-RDM elements indexed by a tuple $(p,q,r)$ with $p < q < r$.  The terms with only 3 unique indices are analogous to the 2-index case for the 2-RDM---e.g., the Jordan--Wigner transformation of the three index term corresponds to diagonal 3-RDM elements and thus involves only Pauli-$Z$ operators, as follows. All three index terms are of the form
\begin{equation}
a_{p}^{\dagger}a_{q}^{\dagger}a_{r}^{\dagger}a_{r}a_{q}a_{p} = \frac{-Z_{p} - Z_{q} - Z_{r} + Z_{p}Z_{q} + Z_{p}Z_{r} + Z_{q}Z_{r} - Z_{p}Z_{q}Z_{r}}{8}
\end{equation}
and can be measured in one permutation of qubits (the identity permutation).

The terms with four unique indices are similar to the 3-index 2-RDM case.  For example, consider the real component of a 4-index 3-RDM term,
\begin{equation}
a_{p}^{\dagger}a_{q}^{\dagger}a_{r}^{\dagger}a_{r}a_{q}a_{p + 1} + \mathrm{h.c.} = \frac{1}{8}[X_{p}X_{p + 1} + Y_{p}Y_{p+1} + (X_{p}X_{p+1} + Y_{p}Y_{p+1})( Z_{q}Z_{r} -Z_{q} - Z_{r} )].
\end{equation}
These terms can be measured in $n/2$ circuits for the $XX$ and $YY$ parts, followed by $\binom{n}{2}$ circuits for the $XXZ$ and $YYZ$ terms, using the fact that $XX + YY$ commutes with $ZZ$, allowing us to use the measurement circuit from Ref.~\cite{arute2020hartree}.  At each of the $n/2$ circuit configurations, all $n-1$ pairs must account for all other $Z$ operators.  Thus in total, counting real and imaginary parts, we have a total of $4 \binom{n}{2}$ qubit permutations to measure all 4-index terms.

The 5-index terms contain one $Z$ term and can be measured with swap circuits analogous to the 4-index case in the 2-RDM.  Consider the example of the real-component of the 5-index 3-RDM term,
\begin{equation}
a_{p}^{\dagger}a_{q}^{\dagger}a_{r}^{\dagger}a_{r}a_{q+1}a_{p+1} + a_{p+1}^{\dagger}a_{q+1}^{\dagger}a_{r}^{\dagger}a_{r}a_{q}a_{p} = \frac{1}{16}\left( A_{pq} + Z_{r}A_{pq} \right),
\end{equation}
where we define
\begin{equation}
\begin{split}
A_{pq} &\equiv (X_{p}X_{p+1}X_{q}X_{q+1} + X_{p}X_{p+1}Y_{q}Y_{q+1} + X_{p}Y_{p+1}X_{q}Y_{q+1} - X_{p}Y_{p+1}Y_{q}X_{q+1} \\
&\quad + Y_{p}X_{p+1}Y_{q}X_{q+1} - Y_{p}X_{p+1}X_{q}Y_{q+1}  + Y_{p}Y_{p+1}X_{q}X_{q+1} + Y_{p}Y_{p+1}Y_{q}Y_{q+1} ),
\end{split}
\end{equation}
with $r \notin \{p, p+1, q, q+1\}$.  We can obtain an upper bound for the number of unique measurements settings as $\binom{n}{k}M$, where $M$ is the complexity of measuring $5$-qubit marginals, again using the technique of Ref.~\cite{bonet2020nearly}. The scaling for $M$ is given in Eq.~\eqref{eq:qot_extende_scaling} for $k=5$ and has a prefactor of $3^{5}$ to measure all $X$, $Y$, and $Z$ terms.  Finally, the 6-index term involves 64 separate terms, including the imaginary terms, which is measured by constructing the $\binom{n}{k}$ swap circuits and using EQOT for each permutation on the $6$-qubit marginal terms which make up the 3-RDM element.

\begin{figure}
    \centering
    \includegraphics[width=0.75\textwidth]{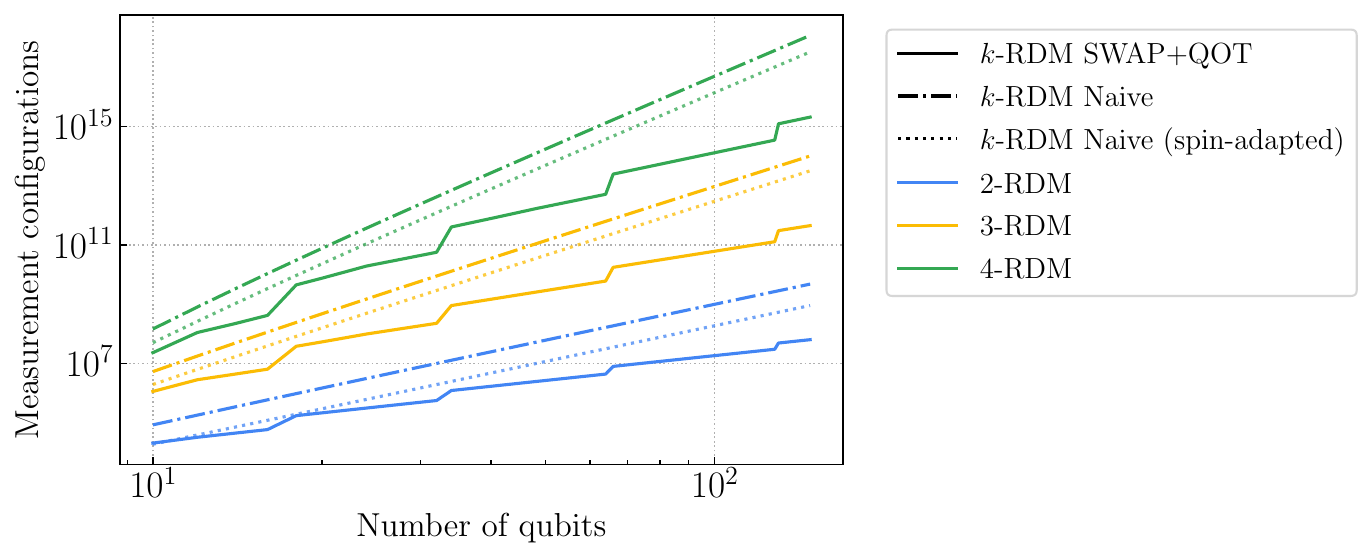}
    \caption[Scaling of the number of measurement configurations needed to measure the $k$-RDM via the swap network protocol.]{Scaling of the number of measurement configurations needed while measuring the $k$-RDM, for $k=2, 3, 4$ (blue, yellow, green), via the swap network protocol combined with quantum overlap tomography (solid lines) compared against the protocol where the unique upper triangle of the supermatrix representing the $k$-RDM is measured.  For the naive measurement strategy we scale the unique number  of terms in  the $k$-RDM by  $3^{2k}$ which is  obtained by counting products of $\sigma^{+},\sigma^{-},Z$. A constant-factor improvement is possible by taking advantage of $ S_z $-spin symmetry and only considering the spin-adapted blocks of the RDM supermatrix~\cite{rubin2018application}. The swap protocol upper bound is a quadratic improvement over the naive scaling. The solid curves corresponding to the SWAP+EQOT protocol are equivalent to the orange points in Fig.~\ref{fig:mainfig} of the main text.} 
    \label{fig:eqot_vs_naive}
\end{figure}

Overall, for the 3-RDM we can loosely bound the number of measurements as
\begin{equation}
M_{T}(\text{3-RDM}) = 1 + 4\binom{n}{2}  + \binom{n}{2}\mathrm{EQOT}(5)+ \binom{n}{3}\mathrm{EQOT}(6, n) \left(\frac{2}{3}\right)^{6},
\end{equation}
where each term corresponds to measuring the 3-, 4-, 5-, and 6-index terms of the 3-RDM, respectively. We note that this is an overestimate since Eq.~\eqref{eq:qot_extende_scaling} provides an upper bound to the $k$-qubit marginal measurement.  It is further loosened by the fact that we clearly do not need to measure \emph{all} $k$-qubit marginal terms.  

A similar accounting can be performed for the 4-RDM by breaking the unique 4-RDM elements into sets consisting of terms with 4, 5, 6, 7, and 8 unique indices.  This gives the upper bound on measurement configurations as
\begin{equation}
M_{T}(\text{4-RDM}) = 1 + 4\binom{n}{2} + \binom{n}{2}\mathrm{EQOT}(6, n) + \binom{n}{3}\mathrm{EQOT}(7, n) + \binom{n}{4}\mathrm{EQOT}(8, n) \left(\frac{2}{3}\right)^{8}.
\end{equation}

In \cref{fig:eqot_vs_naive} we plot the scaling of the swap network EQOT protocol against naively measuring the upper triangle of unique $k$-RDM elements in a supermatrix representation, as a function of the number of fermionic modes.

\section{\label{sec:numerics_appendix}Supplementary numerical calculations}

Here we provide some additional findings with our numerical studies. These are not essential to the primary results of the main text, but rather serve to explore some of the more subtler points of our partial tomography scheme.

\subsection{\label{sec:hyperparameter}Hyperparameter tuning}

\begin{figure}
\centering
	\includegraphics[width=0.6\textwidth]{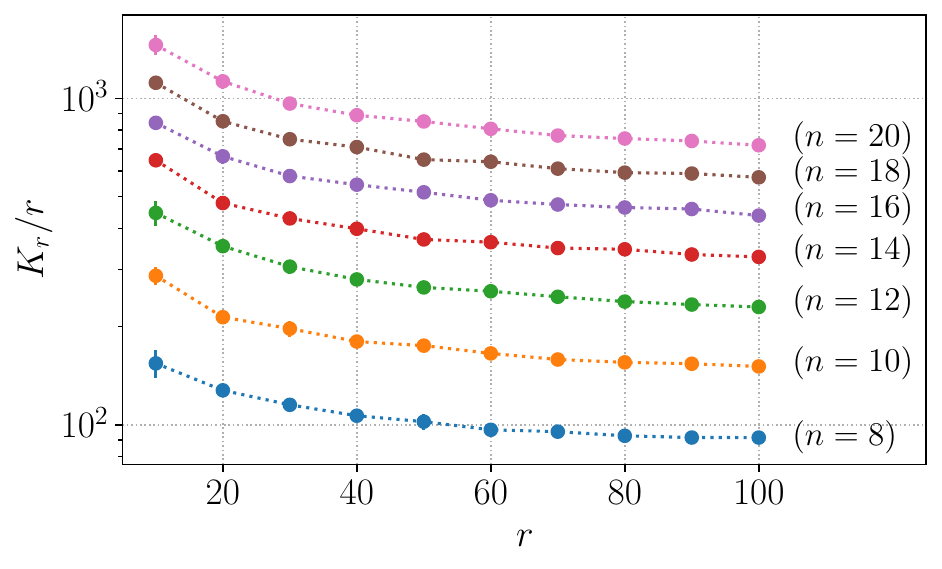}
	\caption[The relation between $ K_r/r $ and $ r $.]{The relation between $ K_r/r $ and $ r $ for various numbers of modes under the $ \mathcal{U}_{\FGU} $ ensemble, with respect to covering $ 2 $-RDM observables. Since $ K_r $ is a random variable, we average over 10 randomly generated circuit collections for each $ r $ and indicate 1 standard deviation with uncertainty bars.}
	\label{fig:K_vs_r}
\end{figure}

As mentioned in the main text, we may control how many circuits to randomly generate by setting a hyperparameter $ r $ such that all $ K_r $ unitaries correspond to measurements of all observables at least $ r $ times each. Since increasing the sample size decreases the frequency of outlier events (i.e., some subset of observables being accounted for more often than the rest), we then expect that the value of $ K_r/r $ decrease as a function of $ r $. Indeed, this is a generic feature of randomization, as observed in the numerical results of the original work on classical shadows~\cite{huang2020predicting}. This effect is demonstrated, with the 2-RDM as an example, in \cref{fig:K_vs_r}. The behavior is consistent as expected.

\subsection{Realistic time estimates}

The main experimental difference between deterministic and randomized measurement schemes is the number of unique circuits one must run. In general, the number of random unitaries ($ K_r $) will be larger than the deterministic clique cover size ($ C $). Thus depending on the architecture, reprogramming the quantum device for each new circuit may incur a nontrivial overhead in the actual wall-clock time of the algorithm. Here, we show that for realistic experimental parameters, this consideration does not affect the results presented in the main text.

\begin{figure}
	\centering
	\includegraphics[width=\textwidth]{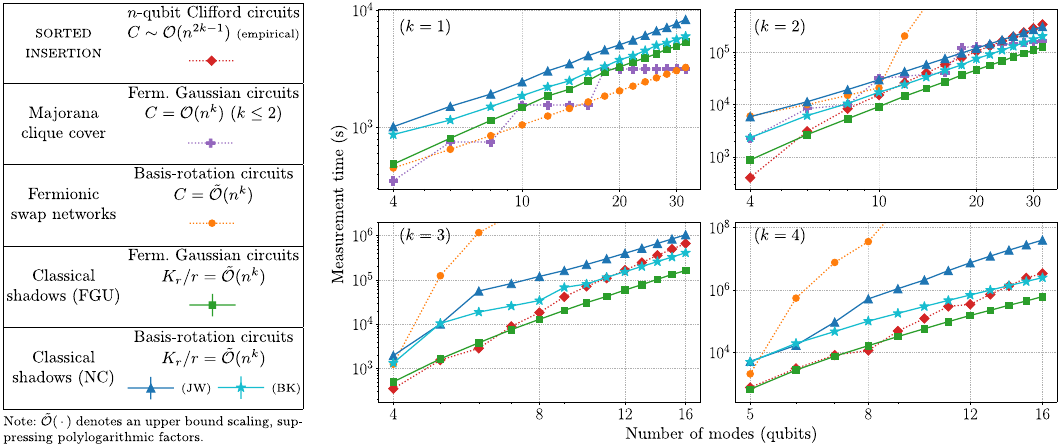}
	\caption[Measurement times for estimating $ k $-RDMs under the reported device parameters of the Google Sycamore chip.]{Measurement times for estimating $ k $-RDMs, calculated with \cref{eq:T_deter,eq:T_rand}, under the reported device parameters of the Google Sycamore chip~\cite{sung2020exploration} and with $ S = 2.5 \times 10^5 $. The underlying data is that of \cref{fig:mainfig} in the main text. For convenience, we reproduce the legend of the main text here.}
	\label{fig:google_meas_times}
\end{figure}

Suppose the quantum device can repeatedly sample a fixed circuit at a rate $ \fs $, but requires time $ \tl $ to load a new circuit. Then the total measurement time under the two paradigms are
\begin{align}
	T_{\mathrm{deter}} &= C \l( \frac{S}{\fs} + \tl \r), \label{eq:T_deter}\\
	T_{\mathrm{rand}} &= K_r \l( \frac{\lceil S/r \rceil}{\fs} + \tl \r), \label{eq:T_rand}
\end{align}
where we recall that $ S = \O(1/\varepsilon^2) $. Note that in the regime where $ S \gg \fs\tl $, we may directly compare $ C $ to $ K_r/r $, as in the main text. For hardware-dependent estimates, we take the specifications of the Google Sycamore chip as an example~\cite{arute2019quantum,arute2020hartree,harrigan2021quantum,arute2020observation}. The reported parameter values are $ \fs = 5 \times 10^3 $~Hz and $ \tl = 0.1 $~s~\cite{sung2020exploration}, and for \cref{fig:google_meas_times}, we set $ S = 2.5 \times 10^{5} $, in line with the number of shots taken to estimate 1-RDM elements in a recent Hartree--Fock experiment~\cite{arute2020hartree}. We observe no qualitative differences from the results of the main text.

\begin{figure}
	\centering
    \includegraphics[width=\textwidth]{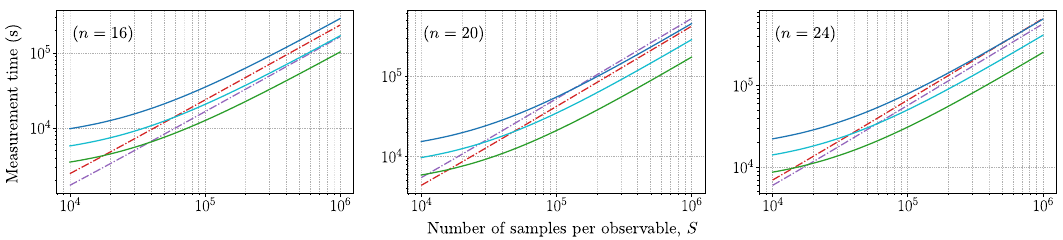}
    \caption[Examples of how the measurement times for our randomized schemes scale with the level of precision.]{Examples of how the measurement times for our randomized schemes scale with the level of precision, as given by \cref{eq:T_rand}. We use the same device parameters here as in \cref{fig:google_meas_times}. For comparison, we also plot the linear scaling of the prior deterministic strategies (excluding the swap network method), which gives an indication for the values of $ S $ beyond which we obtain an advantage with our methods under this time-cost model. The colors correspond to those in the legend of \cref{fig:google_meas_times}.}
    \label{fig:time_scaling_with_S}
\end{figure}

One may also study the performance of the different methods as a function of the target accuracy. In \cref{fig:time_scaling_with_S}, we show how $ T_{\mathrm{rand}} $ scales with $ S $ for the 2-RDM, using a few values of $ n $ as illustrative examples. Though the threshold at which randomization begins to outperform the other methods varies depending on $ n $ and $ k $, it typically lies below $ {\sim} \, 10^5 $, which is well under typical sampling requirements. Note that the $ S \gg \fs\tl $ regime corresponds to when $ T_{\mathrm{rand}} $ scales linearly with $ S $.

\subsection{\label{sec:hamiltonian_appendix}Hamiltonian averaging}

In the context of estimating a single observable, whose expectation value is a linear combination of RDM elements, the number of circuit repetitions required is more properly determined by taking into account the coefficients of the terms and covariances between simultaneously measured terms~\cite{wecker2015progress,mcclean2016theory,rubin2018application}. This can be directly calculated from the (single-shot) variance of the corresponding estimator. Here we provide some preliminary numerical calculations in this context, with respect to $ \mathcal{U}_{\FGU} $. Although our uniformly distributed ensemble is not tailored for Hamiltonian averaging, these calculations provide a benchmark for potential improvement.

Without loss of generality, consider a traceless fermionic $ k $-body Hamiltonian
\begin{equation}
	H = \sum_{j=1}^k \sum_{\bm{\mu} \in \comb{2n}{2j}} h_{\bm{\mu}} \Gamma_{\bm{\mu}}, \quad  h_{\bm{\mu}} \in \R.
\end{equation}
For the numerical studies presented here, we consider a sample of molecular Hamiltonians (hence $ k = 2 $) at equilibrium nuclear geometry, obtained through OpenFermion~\cite{openfermion} interfaced with the Psi4 electronic structure package~\cite{psi4}. We used a minimal STO-3G orbital basis set to generate these Hamiltonians, except for the $ \mathrm{H}_2 $ molecule, which was represented in the 6-31G basis.

In Table~\ref{tab:H_variances} we compare our classical shadows (CS) $ \mathcal{U}_{\FGU} $ ensemble against two prominent measurement schemes for electronic-structure Hamiltonians:~a strategy based on a low-rank factorization of the coefficient tensor, termed basis-rotation grouping (BRG)~\cite{huggins2021efficient}, and a locally biased adaptation on classical shadows (LBCS)~\cite{hadfield2022measurements}. For reference, we also report the variance under standard classical shadows with Pauli measurements~\cite{huang2020predicting}. The expressions for the variances in terms of the Hamiltonian terms and a reference state $ \rho $ (taken here to be the ground state) are provided in their respective references. While we provide a state-independent upper bound in \cref{sec:arb_observable_appendix}, the exact variance expression for our $ \mathcal{U}_\FGU $ ensemble on arbitrary observables has been derived in Refs.~\cite{wan2023matchgate,ogorman2022fermionic}.

Note that, in order to compare fairly between the deterministic (BRG) and randomized methods (CS, LBCS), we reframe the deterministic measurement scheduling such that an equivalent variance quantity may be computed. This principle was also used in the numerical comparisons of Ref.~\cite{hadfield2022measurements}, which we generalize here. We decompose the target Hamiltonian as
\begin{equation}
    H = \sum_{\ell=1}^L O_\ell,
\end{equation}
where each $ \tr(O_\ell \rho) $ may be estimated by a single measurement setting (as defined by the given strategy). The optimal distribution of measurements then allocates a fraction
\begin{equation}\label{eq:optimal_fraction}
    p_\ell \coloneqq \frac{\sqrt{\V_{\rho}[O_\ell]}}{\sum_{j=1}^L \sqrt{\V_{\rho}[O_j]}}
\end{equation}
of the total measurement budget to the $ \ell $th setting~\cite{rubin2018application}. The variance here is simply the quantum-mechanical operator variance,
\begin{equation}
    \V_{\rho}[O_\ell] = \tr(O_\ell^2 \rho) - \tr(O_\ell \rho)^2.
\end{equation}
Recognizing $ \{p_\ell\}_\ell $ as a collection of positive numbers which sum to unity, we may recast the deterministic strategy into the language of randomization, where the unbiased estimator is given by
\begin{equation}
    \E_{\ell, \rho}\l[ \frac{1}{p_\ell} O_\ell \r] = \E_{\ell}\l[ \frac{1}{p_\ell} \tr(O_\ell \rho) \r] = \tr(H \rho).
\end{equation}
The variance of this estimator is therefore
\begin{equation}\label{eq:variance_general}
    \begin{split}
    \V_{\ell, \rho}\l[ \frac{1}{p_\ell} O_\ell \r] &= \E_{\ell, \rho}\l[ \frac{1}{p_\ell^2} O_\ell^2 \r] - \E_{\ell, \rho}\l[ \frac{1}{p_\ell} O_\ell \r{]^2}\\
    &= \sum_{\ell=1}^L \frac{1}{p_\ell} \tr(O_\ell^2 \rho) - \tr(H \rho)^2.
    \end{split}
\end{equation}
Note that this analysis does not actually require a randomization the deterministic strategy, but merely normalizes the measurement allocations so as to produce an equivalent figure of merit.

\begin{table}
\centering
    \caption[Variances of Hamiltonian averaging estimators under various strategies.]{Variances of Hamiltonian averaging estimators under various strategies, reported in units of $ \mathrm{Ha}^2 $. The expressions for the CS (Pauli) and LBCS variances may be found in their original references, although the fundamental formula for all table entries is given by \cref{eq:variance_general}, where the decomposition into measurable terms $ O_\ell $ and probabilities $ p_\ell $ are determined by the particular method. For CS (FGU), the explicit expression is provided in Refs.~\cite{wan2023matchgate,ogorman2022fermionic}. The reference state used here is the exact ground state.}
    \begin{tabular}{c c c c c}
    \toprule
    {} & \multicolumn{4}{c}{Methods}\\
    \cmidrule{2-5}
    Molecule (qubits) & CS (Pauli)~\cite{huang2020predicting} & LBCS~\cite{hadfield2022measurements} & BRG~\cite{huggins2021efficient} & CS (FGU)\\
    \midrule
    $ \text{H}_2 $ (8) & 51.4 & 17.5 & 22.6 & 69.6\\
    $ \text{LiH} $ (12) & 266 & 14.8 & 7.0 & 155\\
    $ \text{BeH}_2 $ (14) & 1670 & 67.6 & 68.3 & 586\\
    $ \text{H}_2\text{O} $ (14) & 2840 & 257 & 6559 & 8440\\
    $ \text{NH}_3 $ (16) & 14400 & 353 & 3288 & 5846\\
    \bottomrule
    \end{tabular}
	\label{tab:H_variances}
\end{table}

While \cref{eq:optimal_fraction} provides the optimal distribution of measurements, one may use any distribution in its place. In particular, because $ \rho $ is unknown, a state-independent approximation to $ \{p_\ell\}_\ell $ is often a more practical option;~tighter bounds may be obtained by a classically tractable approximation to the true, unknown state. For simplicity, in \cref{tab:H_variances} we take $ \rho $ as the exact ground state within the model chemistry. We note that Ref.~\cite{huggins2021efficient} showed evidence that the discrepancy between using the exact ground state and the state obtained from a configuration interaction with single and double excitations (CISD) calculation is negligible in this context.

Our main takeaway from \cref{tab:H_variances} is that, similar to how Pauli measurements can be dramatically improved via solving an optimization problem targeted specifically at minimizing this variance~\cite{hadfield2022measurements}, our classical shadows method similarly has a large room for improvement. Since only LBCS performs such an optimization, it is perhaps not too surprising that it is the most efficient approach here, despite only employing Pauli measurements. Encouragingly, such a biasing (or similar optimization-based extensions) may also be applied to our classical shadows ensemble in principle.
It should be noted that these results assume the noiseless case;~for instance, the BRG strategy additionally offers resilience to device errors and the ability to postselect on particle number, which have the effect of reducing noise-induced contributions to the variance~\cite{huggins2021efficient}.
\chapter{Group-theoretic Error Mitigation Enabled by Classical Shadows and Symmetries}\label{chap:EM}

\section*{Preface}

This chapter is based on \cite{zhao2023group}, coauthored by the author of this dissertation and Akimasa Miyake.

\section{Introduction}

Quantum computers are highly susceptible to errors at the hardware level, posing a considerable challenge to realize meaningful applications in the so-called noisy intermediate-scale quantum (NISQ) era~\cite{preskill2018quantum,bharti2022noisy}. One particularly promising and natural candidate for NISQ applications is the simulation of quantum many-body physics and chemistry~\cite{feynman1982simulating,georgescu2014quantum,mcardle2020quantum,bauer2020quantum}. In order to minimize the accumulation of errors, such algorithms prioritize low-depth circuits, for instance, variational quantum circuits~\cite{peruzzo2014variational,mcclean2016theory,yuan2019theory,cerezo2021variational}. However, in order to exhibit quantum advantage, these circuits must also be beyond the capabilities of classical simulation~\cite{osborne2006efficient,bravyi2021classical,napp2022efficient,wild2023classical}, resulting in noise levels that nonetheless corrupt the calculations.

While quantum error correction is the long-term solution, current state-of-the-art hardware is still a few orders of magnitude from achieving scalable, fault-tolerant quantum computation~\cite{fowler2012surface,kelly2015state,egan2021fault,postler2022demonstration,zhao2022realization,sundaresan2023demonstrating,google2023suppressing,sivak2023real,ni2023beating}. In the meantime, there have been considerable theoretical and experimental efforts probing the beyond-classical potential of NISQ computers~\cite{omalley2016scalable,kandala2017hardware,colless2018computation,dumitrescu2018cloud,hempel2018quantum,kandala2019error,kokail2019self,nam2020ground,arute2019quantum,arute2020hartree,harrigan2021quantum,arute2020observation,zhong2020quantum,huggins2022unbiasing,kim2023scalable,huang2022quantum,stanisic2022observing,tazhigulov2022simulating,madsen2022quantum,motta2023quantum,obrien2023purification,morvan2023phase,kim2023evidence}. Should such an application be demonstrated, quantum error mitigation (QEM) is expected to play a crucial role. Broadly speaking, QEM aims to approximately recover the output of an ideal quantum computation, given only access to noisy quantum devices and offline classical resources. We refer the reader to Refs.~\cite{endo2021hybrid,cai2022quantum} for a review of prominent concepts and strategies in QEM.

A related but separate challenge for NISQ algorithms is the need to learn many observables in a rudimentary fashion, i.e., by repeatedly running and sampling from quantum circuits. The number of repetitions required can be immense, both to suppress shot noise and to handle the measurement of noncommuting observables~\cite{wecker2015progress,gonthier2022measurements}. While a variety of strategies have been proposed to address this bottleneck~\cite{cerezo2021variational,tilly2022variational}, one particularly promising approach is that of classical shadows~\cite{huang2020predicting,paini2021estimating}.

Classical shadows were developed primarily from the union of two themes in quantum learning theory:~linear-inversion estimators for state tomography~\cite{sugiyama2013precision,guta2020fast} (closed-form solutions that admit fast postprocessing and rigorous guarantees) and the framework of shadow tomography~\cite{aaronson2020shadow,aaronson2019gentle} (predict only a subset of observables, not the entire density matrix). The result is a simple but powerful protocol that accurately estimates a large collection of observables from relatively few samples. In terms of quantum resources, classical shadows only require the ability to measure in randomly selected bases, making the protocol particularly amenable to NISQ constraints. These desirable features have inspired a wide range of extensions and applications, 
for example:~entanglement detection~\cite{elben2020mixed}, quantum Fisher information bounds~\cite{rath2021quantum,vitale2023estimation}, learning quantum processes~\cite{levy2021classical,kunjummen2023shadow}, navigating variational landscapes~\cite{sack2022avoiding,boyd2022training}, energy-gap estimation~\cite{chan2022algorithmic}, and applications to fermions~\cite{zhao2021fermionic,wan2023matchgate,ogorman2022fermionic,low2022classical,babbush2023quantum,denzler2023learning} and bosons~\cite{gu2023efficient,becker2022classical}. For an overview of classical shadows and randomized measurement strategies, see Ref.~\cite{elben2023randomized}.

Due to their experimental friendliness and versatile prediction power, classical shadows naturally have been considered for QEM as well. For example, Refs.~\cite{seif2023shadow,hu2022logical} used classical shadows to approximately project a noisy quantum state toward a target subspace via classical postprocessing, the subspaces being either the logical subspace of an error-correcting code~\cite{mcclean2020decoding} and/or the dominant eigenvector (purification) of the noisy mixed state~\cite{koczor2021exponential,huggins2021virtual}. These shadow-based ideas circumvent some of the difficulties of performing subspace projection, at the cost of an exponential sample complexity. Meanwhile, Ref.~\cite{jnane2023quantum} intertwined classical shadows with other popular QEM strategies, with a particular focus on probabilistic error cancellation~\cite{temme2017error}. They establish rigorous estimators and performance guarantees, assuming an accurate characterization of the noisy quantum device. Finally, Refs.~\cite{chen2021robust,koh2022classical} described modifications to the classical linear-inversion step in order to mitigate errors in the randomized measurements. In particular, robust shadow estimation~\cite{chen2021robust} assumes no prior knowledge of the noise, instead implementing a separate calibration experiment that learns the necessary noise features.

In this work, we take this latter perspective~\cite{chen2021robust,koh2022classical}, with an eye on a more comprehensive mitigation of errors beyond readout errors. We introduce a QEM protocol, which we refer to as \emph{symmetry-adjusted classical shadows}, that takes advantage of known symmetries in the quantum system of interest. For example, in simulations of chemistry, the number of electrons is typically fixed. The corruption of such symmetries by noise informs us how to undo the effects of that noise. Crucially, because randomized measurements scramble the information, the other properties of the quantum system are corrupted (and therefore can be mitigated) in the same manner. Using these insights, symmetry-adjusted classical shadows appropriately modifies the linear-inversion based on the symmetry information alone.

A notable advantage of our protocol is that we do not run any extraneous calibration experiments. This has the added benefit of inherently accounting for errors that occur throughout the full quantum circuit, rather than the randomized measurements in isolation~\cite{karalekas2020quantum,chen2021robust,koh2022classical,van2022model,arrasmith2023development}. Also, the simplicity of the protocol allows for additional QEM techniques to be straightforwardly applied in tandem. Finally, in contrast to other symmetry-based ideas~\cite{bonet2018low,mcardle2019error,cai2021quantum,jnane2023quantum}, our approach goes beyond the concept of symmetry projection, instead utilizing a unified group-theoretic understanding of classical shadows in conjunction with symmetries.

This chapter is structured as follows. In Section~\ref{sec:background}, we establish preliminaries and background material. In Section~\ref{sec:summary_of_results} we provide a self-contained summary of results, describing symmetry-adjusted classical shadows and highlighting additional technical results that may be of independent interest. In Section~\ref{sec:main_theory}, we illustrate the theory of symmetry-adjusted classical shadows in further detail, including applications to fermion and qubit systems with global $\U(1)$ symmetry in Sections~\ref{sec:applications_fermions} and \ref{sec:applications_qubits}. We then turn to numerical experiments in Section~\ref{sec:numerics}, which include simulations of a noise model based on existing superconducting-qubit platforms to investigate the performance of our protocol in realistic scenarios. Finally, we summarize our findings and discuss future prospects in Section~\ref{sec:discussion}. Details regarding the mathematical proofs and numerical simulations are provided in the Appendix, and code for the latter is available at our open-source repository ({\small\url{https://github.com/zhao-andrew/symmetry-adjusted-classical-shadows}})~\cite{gitrepo}.

\section{Background}\label{sec:background}

Here, we provide a review of classical shadows~\cite{huang2020predicting,paini2021estimating} and robust shadow estimation~\cite{chen2021robust}. Readers familiar with this background material can skip to the summary of our results in Section~\ref{sec:summary_of_results}, after familiarizing themselves with the notation that we establish below.

\subsection{Notation and preliminaries}

For any integer $N > 1$, we define $[N] \coloneqq \{0, \ldots, N-1\}$ (note that we index starting from $0$). We use $\i \equiv \sqrt{-1}$ for the imaginary unit.

Throughout this chapter, we consider an $ n $-qubit system with Hilbert space $ \mathcal{H} \coloneqq (\C^{2})^{\otimes n} $. Its dimension is denoted by $ d \equiv 2^n $ unless otherwise specified. We often work with the space of linear operators $ \mathcal{L}(\mathcal{H}) \cong \C^{d \times d} $ as a vector space, so it will be convenient to employ the Liouville representation:~for any operator $ A \in \mathcal{L}(\mathcal{H}) $, its vectorization $\kket{A} \in \C^{d^2}$ in some orthonormal operator basis $\{B_1, \ldots, B_{d^2} \mid \tr(B_i^\dagger B_j) = \delta_{ij}\}$ is defined by the components $\vip{B_i}{A} \coloneqq \tr(B_i^\dagger A)$. Under this representation, superoperators are mapped to $ d^2 \times d^2 $ matrices:~any $\mathcal{E} \in \mathcal{L}(\mathcal{L}(\mathcal{H}))$ can be specified by its matrix elements $ \mathcal{E}_{ij} \coloneqq \vev{B_i}{\mathcal{E}}{B_j} = \tr(B_i^\dagger \mathcal{E}(B_j)) $. We let $\mathcal{E}$ denote both the superoperator and its matrix representation, and in a similar fashion we sometimes write $\kket{A} = A$.

For systems of qubits, the normalized Pauli operators $\Pauli(n) / \sqrt{d}$ are a convenient basis for $\mathcal{L}(\mathcal{H})$, where
\begin{equation}
    \Pauli(n) \coloneqq \{\I, X, Y, Z\}^{\otimes n}.
\end{equation}
This choice is called the Pauli transfer matrix (PTM) representation. The weight, or locality, of a Pauli operator $P \in \Pauli(n)$ is the number of its nontrivial tensor factors, denoted by $|P|$. For each $i \in [n]$, we define $W_i \in \Pauli(n)$ which acts as $W \in \{X, Y, Z\}$ on the $i$th qubit and trivially on the rest of the system.

For fermions in second quantization, a natural choice of basis is the set of Majorana operators, defined as $\{ \Gamma_{\bm{\mu}} / \sqrt{d} \mid \bm{\mu} \subseteq [2n] \}$ where
\begin{equation}\label{eq:majorana_def}
    \Gamma_{\bm{\mu}} \coloneqq (-\i)^{\binom{|\bm{\mu}|}{2}} \prod_{\mu \in \bm{\mu}} \gamma_{\mu}.
\end{equation}
The Hermitian generators $\{ \gamma_\mu \mid \mu \in [2n] \} \subset \mathcal{L}(\mathcal{H})$ obey the anticommutation relation $\gamma_\mu \gamma_\nu + \gamma_\nu \gamma_\mu = 2\delta_{\mu\nu} \I$ (we will use $\I$ to denote any identity operator whose dimension is clear from context). They are related to the fermionic creation and annihilation operators $a_p^\dagger, a_p$ via
\begin{equation}
    \gamma_{2p} = a_p + a_p^\dagger, \quad \gamma_{2p+1} = -\i(a_p - a_p^\dagger).
\end{equation}
By convention, the elements of $\bm{\mu}$ and the product in Eq.~\eqref{eq:majorana_def} are in strictly ascending order. We call $|\bm{\mu}|$ the degree of $\Gamma_{\bm{\mu}}$, or equivalently refer to them as ($|\bm{\mu}|/2$)-body operators whenever the degree is even. It is straightforward to check that Majorana operators are isomorphic to Pauli operators, in particular satisfying the orthogonality relation $\vip{\Gamma_{\bm{\mu}}}{\Gamma_{\bm{\nu}}} = d \delta_{\bm{\mu} \bm{\nu}}$.

For any unitary $U \in \U(d)$, its corresponding channel is denoted by $\mathcal{U}(\cdot) \coloneqq U (\cdot) U^\dagger$. For any $\ket{\varphi} \in \mathcal{H}$, $\kket{\varphi}$ is the vectorization of $\op{\varphi}{\varphi}$. We use tildes to indicate objects affected by quantum noise, e.g., $\widetilde{\mathcal{U}}$ denotes a noisy implementation of the $\mathcal{U}$. Hats indicate statistical estimators, e.g., $\hat{o}$ denotes an estimate for $o = \tr(O \rho)$. Asymptotic upper and lower bounds are denoted by $\O(\cdot)$ and $\Omega(\cdot)$ respectively, and $f(x) = \Theta(g(x))$ means that $f(x)$ is both $\O(g(x))$ and $\Omega(g(x))$.

\subsection{Classical shadows}

We summarize the method of classical shadows as formalized by Huang \emph{et al.}~\cite{huang2020predicting}, borrowing the PTM language of Chen \emph{et al.}~\cite{chen2021robust} which will make the robust extension clear later. Our task is to estimate the expectation values $\tr(O_j \rho) = \vip{O_j}{\rho}$ of a collection of $L$ observables $O_1, \ldots, O_L \in \mathcal{L}(\mathcal{H})$, ideally using as few copies of $\rho$ as possible. Classical shadows is based on a simple measurement primitive:~for each copy of $ \rho $, apply a unitary $U$ randomly drawn from a distribution of unitaries and measure in the computational basis. This produces a sample $ b \in \{0,1\}^n $ with probability $ \vev{b}{\mathcal{U}}{\rho} $. One then inverts the unitary on the outcome $\ket{b}$ in postprocessing, which amounts to storing a classical representation of $ U^\dagger \ket{b} $.

The unitary distribution determines the efficiency of this protocol with respect to the properties of interest. Throughout this chapter, we assume that the distribution is a finite group equipped with the uniform probability distribution.\footnote{It is straightforward to generalize to compact groups, using their Haar measures.} Specifically, let $ U : G \to \U(\mathcal{H}) $ be a unitary representation of a group $G$. The measurement primitives averaged over all random unitaries and measurement outcomes implement the quantum channel
\begin{equation}\label{eq:M_channel}
    \mathcal{M} \coloneqq \E_{g \sim G} \mathcal{U}_g^\dagger \mathcal{M}_Z \mathcal{U}_g \equiv \frac{1}{|G|} \sum_{g \in G} \mathcal{U}_g^\dagger \mathcal{M}_Z \mathcal{U}_g,
\end{equation}
where
\begin{equation}
    \mathcal{M}_Z = \sum_{b \in \{0,1\}^n} \vop{b}{b}
\end{equation}
describes the effective process of computational-basis measurements. The channel $ \mathcal{U}_g $ is the random unitary acting on the target state $ \rho $, while $ \mathcal{U}_g^\dagger $ is its classically computed inversion on the measurement outcomes $ \kket{b} $. Thus in expectation we produce the state
\begin{equation}\label{eq:M_rho_def}
    \mathcal{M}\kket{\rho} = \E_{g \sim G, b \sim \mathcal{U}_g \kket{\rho}} \mathcal{U}_g^\dagger \kket{b}.
\end{equation}
If $\mathcal{M}$ is invertible (corresponding to informational completeness of the measurement primitive), then applying $ \mathcal{M}^{-1} $ to Eq.~\eqref{eq:M_rho_def} recovers the state:
\begin{equation}
    \kket{\rho} = \mathcal{M}^{-1} \mathcal{M}\kket{\rho} = \E_{g \sim G, b \sim \mathcal{U}_g \kket{\rho}} \mathcal{M}^{-1} \mathcal{U}_g^\dagger \kket{b}.
\end{equation}
The objects $ \kket{\hat{\rho}_{g,b}} \coloneqq \mathcal{M}^{-1} \mathcal{U}_g^\dagger \kket{b} $ are called the classical shadows of $ \kket{\rho} $, for which they serve as unbiased estimators. Hence by construction they can predict expectation values,
\begin{equation}
    \E_{g \sim G, b \sim \mathcal{U}_g \kket{\rho}} \vip{O_j}{\hat{\rho}_{g,b}} = \vip{O_j}{\rho},
\end{equation}
as well as nonlinear functions of $\rho$~\cite{huang2020predicting}. While $ \mathcal{M}^{-1} $ is not a physical map (it is not completely positive), it only appears as classical postprocessing. Such a computation can be accomplished, for instance, by first deriving a closed-form expression for $ \mathcal{M} $.

One systematic approach to deriving such an expression is through the representation theory of $ G $. First, note that the $ d $-dimensional unitary $ U $ is promoted to a $ d^2 $-dimensional representation $ \mathcal{U} $. Equation~\eqref{eq:M_channel} reveals that $ \mathcal{M} $ is a twirl of $ \mathcal{M}_Z $ by the group $ G $ under the action of $ \mathcal{U} $. Such objects are well studied:~assuming that the irreducible components of $ \mathcal{U} $ have no multiplicities,\footnote{The general expression with multiplicities can be found in Ref.~\cite[Eq.~(A6)]{chen2021robust}.} an application of Schur's lemma implies that~\cite{fulton2004representation}
\begin{equation}\label{eq:channel_diag}
    \mathcal{M} = \sum_{\lambda \in R_G} f_\lambda \Pi_\lambda.
\end{equation}
Here, $ R_G $ is the set of labels $ \lambda $ for the irreducible representations (irreps) of $ G $. The superoperators $ \Pi_\lambda $ are orthogonal projectors onto the irreducible subspaces $V_\lambda \subseteq \mathcal{L}(\mathcal{H}) $. Choosing an orthonormal basis $ \{\kket{B_\lambda^j} \mid j = 1,\ldots,\dim V_\lambda\} $ for each subspace, we can write the projectors as
\begin{equation}\label{eq:proj_lambda}
    \Pi_\lambda = \sum_{j=1}^{\dim V_\lambda} \vop{B_\lambda^j}{B_\lambda^j}.
\end{equation}
The eigenvalues $ f_\lambda $ of $ \mathcal{M} $ can be computed using the orthogonality of projectors:
\begin{equation}\label{eq:channel_eigval}
    f_\lambda = \frac{\tr(\mathcal{M}_Z \Pi_\lambda)}{\tr(\Pi_\lambda)}.
\end{equation}
Note that $\tr(\Pi_\lambda) = \dim V_\lambda$. From this diagonalization, we immediately acquire an expression for the desired inverse:
\begin{equation}\label{eq:inverse_channel}
    \mathcal{M}^{-1} = \sum_{\lambda \in R_G} f_\lambda^{-1} \Pi_\lambda.
\end{equation}
If some $f_\lambda = 0$, then we may instead define 
$\mathcal{M}^{-1}$ as the pseudoinverse on the subspaces where $ f_\lambda $ is nonvanishing. This implies that the measurement primitive is informationally complete only within those subspaces.

To analyze the sample efficiency of this protocol, suppose we have performed $T$ experiments, yielding a collection of independent classical shadows $\hat{\rho}_1, \ldots, \hat{\rho}_T$ where each $\kket{\hat{\rho}_\ell} = \mathcal{M}^{-1} \mathcal{U}_{g_\ell}^\dagger \kket{b_\ell}$. From this data we can construct estimates
\begin{equation}\label{eq:noiseless_estimator}
    \hat{o}_j(T) = \frac{1}{T} \sum_{\ell=1}^T \vip{O_j}{\hat{\rho}_{\ell}},
\end{equation}
which by linearity converge to $\tr(O_j \rho)$. The single-shot variance of $\hat{o}_j$ can be bounded in terms of the so-called shadow norm:
\begin{equation}
\begin{split}
    \V[\hat{o}_j] &\leq \max_{\text{states } \sigma} \E_{g \sim G, b \sim \mathcal{U}_g \kket{\sigma}} \vev{O_j}{\mathcal{M}^{-1} \mathcal{U}_g^\dagger}{b}^2\\
    &\eqqcolon \sns{O_j}{\mathrm{shadow}}^2.
\end{split}
\end{equation}
This variance controls the prediction error, rigorously established via probability tail bounds.\footnote{For simplicity we have use the mean estimator throughout this chapter, which suffices whenever the ensemble is either local Cliffords or matchgates and the observables are Pauli or Majorana operators, repsectively~\cite[Supplemental Material, Theorem~12]{zhao2021fermionic}. In general, a median-of-means estimator can guarantee the advertised sample complexity regardless of ensemble.} In particular, taking a number of samples
\begin{equation}
    T = \O\l( \frac{\log(L/\delta)}{\epsilon^2} \max_{1 \leq j \leq L} \sns{O_j}{\mathrm{shadow}}^2 \r)
\end{equation}
ensures that, with probability at least $1 - \delta$, each estimate exhibits at most $\epsilon$ additive error:
\begin{equation}
    |\hat{o}_j(T) - \vip{O_j}{\rho}| \leq \epsilon.
\end{equation}

Finally, we comment on the classical computation of $\hat{o}_j$. In order to evaluate Eq.~\eqref{eq:noiseless_estimator}, one may use Eqs.~\eqref{eq:proj_lambda} and \eqref{eq:inverse_channel} to express the $\ell$th-sample estimate as
\begin{equation}\label{eq:shadow_estimate_formula}
    \vip{O_j}{\hat{\rho}_\ell} = \sum_{\lambda \in R_G} f_\lambda^{-1} \sum_{k=1}^{\dim V_\lambda} \vip{O_j}{B_\lambda^k} \vev{B_\lambda^k}{\mathcal{U}_{g_\ell}^\dagger}{b_\ell}.
\end{equation}
Thus it suffices to be able to efficiently compute the expansion coefficients $\vip{O_j}{B_\lambda^k} = \tr(O_j B_\lambda^k)$ of the observable $O_j$ in a basis of $V_\lambda$, as well as the matrix elements $\vev{B_\lambda^k}{\mathcal{U}_g^\dagger}{b} = \ev{b}{U_g (B_\lambda^k)^\dagger U_g^\dagger}{b}$. Note that this does not require explicitly representing the classical shadow $\mathcal{M}^{-1} \mathcal{U}_g^\dagger \kket{b}$;~we only need to determine the diagonal entry of the rotated operator $U_g (B_\lambda^k)^\dagger U_g^\dagger$ for a given basis state $\ket{b}$.

\subsection{Robust shadow estimation}\label{subsec:rse_background}

We now summarize the robust shadow estimation protocol by Chen \emph{et al.}~\cite{chen2021robust};~we note that Refs.~\cite{karalekas2020quantum,van2022model,arrasmith2023development} describe analogous ideas in the case of random single-qubit measurements. The basic premise is the fact that Schur's lemma applies to the twirl of any channel, not just $\mathcal{M}_Z$. Suppose that instead of $ \mathcal{U}_g $, the quantum computer implements a noisy channel $ \widetilde{\mathcal{U}}_g $ which obeys the following assumptions:
\begin{assumptions}[{\cite[Simplifying noise assumption \textbf{A1}]{chen2021robust}}]
\label{assumption_1}
The noise in $ \widetilde{\mathcal{U}}_g $ is gate independent, time stationary, and Markovian. Hence there exists the decomposition $ \widetilde{\mathcal{U}}_g = \mathcal{E} \mathcal{U}_g $, where $ \mathcal{E} $ is a completely positive, trace-preserving map, independent of both the ideal unitary and the experimental time.
\end{assumptions}
They also assume the ability to prepare the state $\ket{0^n}$ with sufficiently high fidelity. Given these conditions, the noisy version of the shadow channel implemented in experiment becomes
\begin{equation}
    \widetilde{\mathcal{M}} \coloneqq \E_{g \sim G} \mathcal{U}_g^\dagger \mathcal{M}_Z \widetilde{\mathcal{U}}_g = \frac{1}{|G|} \sum_{g \in G} \mathcal{U}_g^\dagger \mathcal{M}_Z \mathcal{E} \mathcal{U}_g,
\end{equation}
which is now a twirl over the composite channel $ \mathcal{M}_Z \mathcal{E} $. Although $\mathcal{E}$ is unknown, Schur's lemma implies that the eigenbasis is preserved, as we now have
\begin{equation}\label{eq:noisy_channel_diag}
    \widetilde{\mathcal{M}} = \sum_{\lambda \in R_G} \widetilde{f}_\lambda \Pi_\lambda,
\end{equation}
where the eigenvalues depend on $\mathcal{E}$,
\begin{equation}\label{eq:noisy_channel_eigval}
    \widetilde{f}_\lambda = \frac{\tr(\mathcal{M}_Z \mathcal{E} \Pi_\lambda)}{\tr(\Pi_\lambda)}.
\end{equation}
Therefore if one knows $\widetilde{f}_\lambda$, then one can perform the correct linear inversion in the presence of noise, i.e., by replacing $f_\lambda^{-1}$ with $\widetilde{f}_\lambda^{-1}$ in Eq.~\eqref{eq:shadow_estimate_formula}.

Because $ \mathcal{E} $ depends on the details of the quantum hardware, it is not possible to determine $ \widetilde{f}_\lambda $ without an \emph{a priori} accurate characterization of the noise. Absent such information, a calibration protocol is proposed to experimentally estimate the value of $ \widetilde{f}_\lambda $. This proceeds by performing the classical shadows protocol on a fiducial state $\ket{0^n}$, rather than the unknown target state $\rho$. This enables the study of errors in the random circuits $U_g$. Because $\ket{0^n}$ is known exactly, one can compare its noiseless properties against the noisy experimental data to determine a calibration factor.

Specifically, Chen \emph{et al.}~\cite{chen2021robust} construct an estimator $\mathrm{NoiseEst}_G(\lambda, g, b)$ for each sample $(U_g, b)$ of the calibration experiment, which converges to $\widetilde{f}_\lambda$ in expectation over $g$ and $b$. Although they do not prescribe a generic expression for $\mathrm{NoiseEst}_G$ (instead considering particular choices of $G$), it is straightforward to derive one following their ideas. Let $D_\lambda \in V_\lambda$ be an observable supported exclusively by a single irrep such that $\ev{0^n}{D_\lambda}{0^n} \neq 0$. Then we have
\begin{equation}
    \vev{D_\lambda}{\widetilde{\mathcal{M}}}{0^n} = \widetilde{f}_\lambda \ev{0^n}{D_\lambda}{0^n}.
\end{equation}
On the other hand, using the fact that
\begin{equation}
\begin{split}
    \vev{D_\lambda}{\widetilde{\mathcal{M}}}{0^n} &= \bbra{D_\lambda} \E_{g \sim G, b \sim \mathcal{U}_g \kket{0^n}} \mathcal{U}_g^\dagger \kket{b}\\
    &= \E_{g \sim G, b \sim \mathcal{U}_g \kket{0^n}} \ev{b}{U_g D_\lambda U_g^\dagger}{b},
\end{split}
\end{equation}
it follows that the random variable
\begin{equation}\label{eq:noiseest}
    \mathrm{NoiseEst}_G(\lambda, g, b) = \frac{\ev{b}{U_g D_\lambda U_g^\dagger}{b}}{\ev{0^n}{D_\lambda}{0^n}}
\end{equation}
obeys $\E_{g,b}\l[\mathrm{NoiseEst}_G(\lambda, g, b)\r] = \widetilde{f}_\lambda$.

One can recover the definitions for $\mathrm{NoiseEst}_G$ introduced by Chen \emph{et al.}~\cite{chen2021robust} as follows. The global Clifford group $\Cl(n)$ has two irreps:~the span of the identity operator, $V_0 = \spn\{\I\}$ (which is trivial), and its orthogonal complement $V_1 = V_0^\perp$ (the set of all traceless operators). Choosing $D_1 = d \op{0^n}{0^n} - \I$ gives
\begin{equation}
    \mathrm{NoiseEst}_{\Cl(n)}(1, U, b) = \frac{d |\ev{b}{U}{0^n}|^2 - 1}{d - 1},
\end{equation}
where $U \in \Cl(n)$.

On the other hand, the local Clifford group $\Cl(1)^{\otimes n}$ has $2^n$ irreps, labeled by all subsets $I \subseteq [n]$. Each $I$ indexes a subsystem of qubits, and each subspace $V_I$ is the span of all $n$-qubit Pauli operators which act nontrivially on exactly that subsystem. Defining
\begin{equation}
    D_I \coloneqq \prod_{i \in I} Z_i,
\end{equation}
one obtains
\begin{equation}
\begin{split}
    \mathrm{NoiseEst}_{\Cl(1)^{\otimes n}}(I, U, b) &= \frac{\ev{b}{U D_I U^\dagger}{b}}{\ev{0^n}{D_I}{0^n}}\\
    &= \prod_{i \in I} \ev{b_i}{C_i Z C_i^\dagger}{b_i}
\end{split}
\end{equation}
where now $U = \bigotimes_{i \in [n]} C_i \in \Cl(1)^{\otimes n}$.

Any QEM strategy necessarily incurs a sampling overhead dependent on the amount of noise~\cite{takagi2022fundamental,takagi2022universal,tsubouchi2022universal,quek2022exponentially}. For global Clifford shadows, Chen \emph{et al.}~\cite{chen2021robust} show that the sample complexity is augmented by a factor of $\O(F_Z(\mathcal{E})^{-2})$ for estimating observables with constant Hilbert--Schmidt norm, where $F_Z(\mathcal{E}) = 2^{-n} \sum_{b \in \{0,1\}^n} \vev{b}{\mathcal{E}}{b}$ is the average $Z$-basis fidelity of $\mathcal{E}$. Meanwhile for local Clifford shadows, they prove that product noise of the form $\mathcal{E} = \bigotimes_{i\in[n]} \mathcal{E}_i$, satisfying $\min_{i \in [n]} F_Z(\mathcal{E}_i) \geq 1 - \xi$, exhibits an overhead factor of $e^{\O(k\xi)}$ for estimating $k$-local qubit observables.

\section{Summary of results}\label{sec:summary_of_results}

\begin{figure}
\centering
\includegraphics[width=\textwidth]{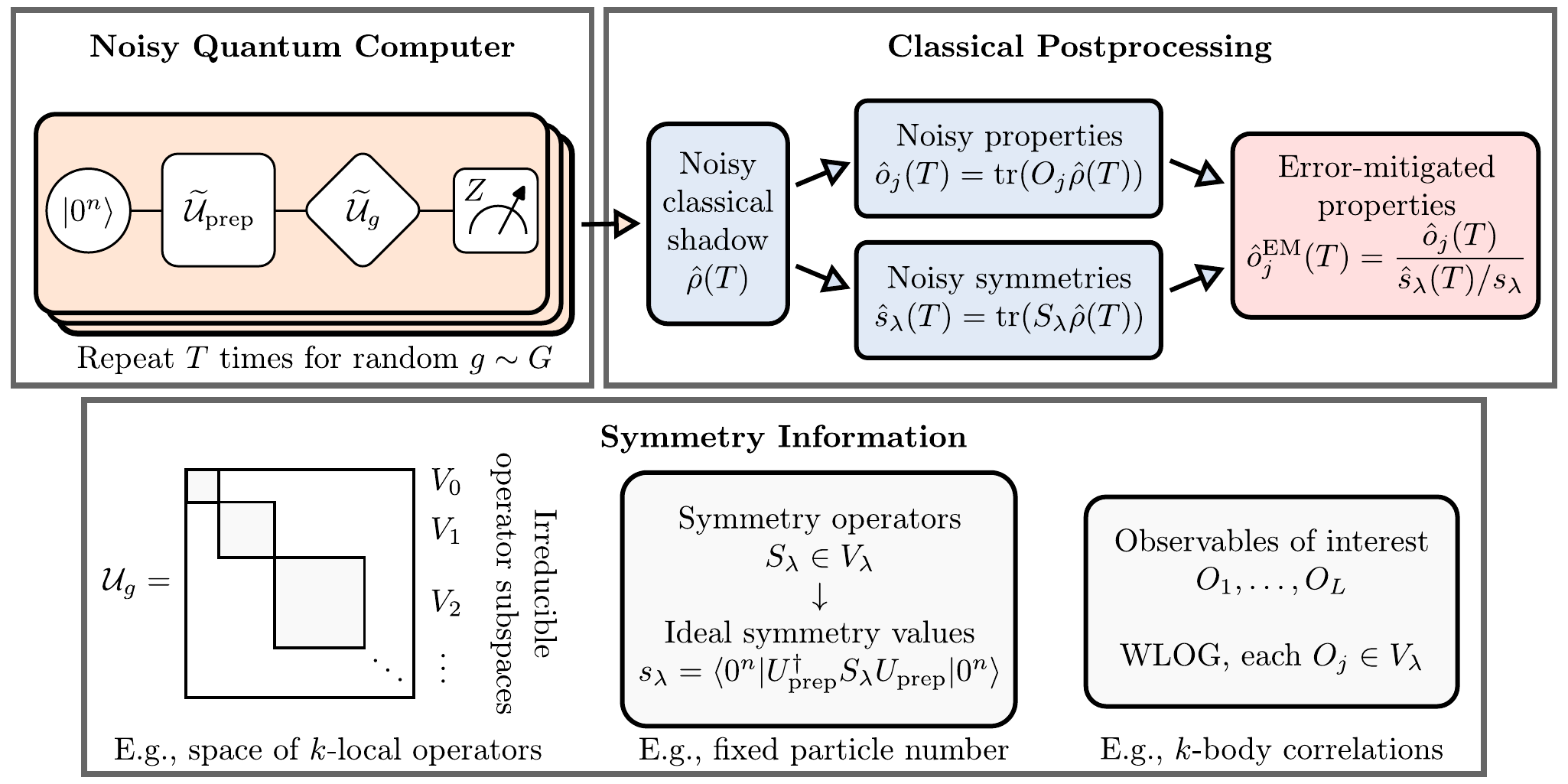}
\caption[Schematic of the \emph{symmetry-adjusted classical shadows} protocol.]{Schematic of the \emph{symmetry-adjusted classical shadows} protocol. Given an ideal unitary $\mathcal{U}(\cdot) = U(\cdot)U^\dagger$, its noisy implementation is denoted by $\widetilde{\mathcal{U}}$. Assuming the target state $\rho = \mathcal{U}_{\mathrm{prep}}(\op{0^n}{0^n})$ obeys certain symmetries $S_\lambda$, we can construct error-mitigated estimates using classical shadows produced by the noisy quantum computer. (While we depict the preparation of a pure state here, our formalism is equally valid if the target state is mixed.) In contrast to prior approaches that only address the noise in $\widetilde{\mathcal{U}}_g$, our protocol additionally incorporates the errors within $\widetilde{\mathcal{U}}_{\mathrm{prep}}$. Our method is applicable whenever the symmetry is compatible with the irreps $V_\lambda$ of the group $G$ describing the classical shadows protocol.}
\label{fig:cartoon_flowchart}
\end{figure}

The primary contribution of this paper, symmetry-adjusted classical shadows, is visualized in Figure~\ref{fig:cartoon_flowchart}. We summarize the main idea and results of this protocol in Section~\ref{subsec:EM_shadows_summary}. We follow by highlighting other notable technical innovations:~in Section~\ref{subsec:sym_pauli_summary}, we describe a modification to random Pauli measurements required to tailor its irreps for use with common symmetries;~in Section~\ref{subsec:improved_circuit_summary}, we discuss an improved design for compiling fermionic Gaussian unitaries with lower circuit depth and fewer gates than prior art;~and in Section~\ref{subsec:spin-adapt_summary}, we summarize a symmetry adaptation to fermionic classical shadows which reduces the quantum resources required, applicable to systems with spin symmetry.

\subsection{Symmetry-adjusted classical shadows}\label{subsec:EM_shadows_summary}

Consider a classical shadows protocol over $G$ with target observables $O_1, \ldots, O_L$. Without loss of generality, let each $O_j \in V_\lambda$ for some subset of irreps $\lambda \in R' \subseteq R_G$. Suppose the experiment experiences an unknown noise channel $\mathcal{E}$ obeying Assumptions~\ref{assumption_1}.

We show that, if $\rho$ obeys symmetries which are ``compatible'' with the irreps in $R'$, then it is possible to construct an estimator which accurately predicts the ideal, noiseless observables. By compatible, we mean that there exist symmetry operators $S_\lambda \in V_\lambda$ for each $\lambda \in R'$ for which their ideal expectation values
\begin{equation}
    s_\lambda \coloneqq \tr(S_\lambda \rho)
\end{equation}
are known \emph{a priori}. Using $T$ (noisy) classical shadows, we construct error-mitigated estimates as
\begin{equation}
    \hat{o}_j^{\mathrm{EM}}(T) \coloneqq \frac{\tr(O_j \hat{\rho}(T))}{\tr(S_\lambda \hat{\rho}(T)) / s_\lambda}.
\end{equation}
We find that the relevant noise characterization in this scenario is
\begin{equation}
    F_{Z, R'}(\mathcal{E}) \coloneqq \min_{\lambda \in R'} \frac{\tr(\mathcal{E} \mathcal{M}_Z \Pi_\lambda)}{\tr(\mathcal{M}_Z \Pi_\lambda)},
\end{equation}
which can be seen as a generalization of the noise fidelity $F_Z(\mathcal{E})$ described in Section~\ref{subsec:rse_background}. Here, $F_{Z, R'}(\mathcal{E})$ only considers how the noise channel acts within the irreducible subspaces of interest.

As two key applications, we study how symmetry-adjusted classical shadows perform in simulations of fermionic and qubit systems. For fermions, we consider $G$ corresponding to fermionic Gaussian unitaries~\cite{zhao2021fermionic} (also known as matchgate shadows~\cite{wan2023matchgate}). We establish the following performance bound for fermionic systems with particle-number symmetry, $N = \sum_{p \in [n]} a_p^\dagger a_p$.

\begin{theorem}[Fermions with particle-number symmetry, informal]\label{thm:informal_fermion_result}
    Let $\rho$ be an $n$-mode state with $\tr(N \rho) = \eta$ fermions. Under the noise model $\mathcal{E}$ satisfying Assumptions~\ref{assumption_1} and assuming $\eta = \O(n)$, matchgate shadows of size
    \begin{equation}
    T = \O( n^2 \log(n) \epsilon^{-2} F_{Z,\{2,4\}}(\mathcal{E})^{-2} )
    \end{equation}
    suffice to achieve prediction error
    \begin{equation}
    |\hat{o}_j(T) - \tr(O_j \rho)| \leq \epsilon + \O(\epsilon^2)
    \end{equation}
    with high probability, where the observables $O_j$ can be taken as all one- and two-body Majorana operators.
\end{theorem}
The dependence on system size $n$ and prediction error $\epsilon$ matches noiseless estimation with matchgate shadows~\cite{zhao2021fermionic,wan2023matchgate}. Meanwhile, the overhead of error mitigation is $\O(F_{Z,R'}(\mathcal{E})^{-2})$, analogous to prior related results~\cite{chen2021robust,koh2022classical}. The irreps $R' = \{2, 4\}$ correspond to the Majorana degree of the $k$-body observables.

For qubit systems, we consider $G$ essentially corresponding to the local Clifford group (i.e., random Pauli measurements)~\cite{huang2020predicting,paini2021estimating}. In order to make the irreducible structure compatible with commonly encountered symmetries, we introduce a technical modification that we call subsystem-symmetrized Pauli shadows (see Section~\ref{subsec:sym_pauli_summary} for a summary). The symmetry we consider here is generated by the total longitudinal magnetization, $M = \sum_{i \in [n]} Z_i$. For error-mitigated prediction of local qubit observables, we have the following result.

\begin{theorem}[Qubits with total magnetization symmetry, informal]\label{thm:informal_qubit_result}
    Let $\rho$ be an $n$-qubit state with a fixed magnetization, $\tr(M \rho) = m$. Under the noise model $\mathcal{E}$ satisfying Assumptions~\ref{assumption_1} and assuming $m = \Theta(1)$, subsystem-symmetrized Pauli shadows of size
    \begin{equation}
    T = \O( n \log(n) \epsilon^{-2} F_{Z,\{1,2\}}(\mathcal{E})^{-2} )
    \end{equation}
    suffices to achieve prediction error
    \begin{equation}
    |\hat{o}_j(T) - \tr(O_j \rho)| \leq \epsilon + \O(\epsilon^2)
    \end{equation}
    with high probability, where the observables $O_j$ can be taken as all one- and two-local Pauli operators.
\end{theorem}

Note that the irreps of subsystem-symmetrized Pauli shadows are labeled by Pauli weight. The variance bound we advertise here is linear in $n$, resulting from the extensive nature of the symmetry $M$. Specifically, we show that when $m = \Theta(1)$, $\sn{M}^2 = \O(n)$ dominates the asymptotic complexity over the $k$-local Pauli observables (for which our protocol exhibits the usual $\sn{O_j}^2 = 3^k$). This is consistent with standard Pauli shadows, wherein the shadow norm of arbitrary $k$-local observables scales at most linearly with spectral norm and exponentially in $k$~\cite{huang2020predicting,paini2021estimating}.

Besides these two examples, we describe symmetry-adjusted classical shadows for a more general class of groups $G$, and we establish accompanying bounds in Theorem~\ref{thm:main_theorem}. This allows for applications to other systems and unitary distributions. See Section~\ref{sec:main_theory} for the general theory, and Sections~\ref{sec:applications_fermions} and \ref{sec:applications_qubits} for the applications to fermion and qubit systems, respectively.

Because our protocol always runs the full noisy quantum circuit, it has the potential to mitigate a wider range of errors than those covered by Assumptions~\ref{assumption_1}, albeit without the rigorous theoretical guarantees. This is a significant feature of the method, as the preparation of $\rho$ often dominates the total circuit complexity (i.e., $U_{\mathrm{prep}}$ in Figure~\ref{fig:cartoon_flowchart}).

We explore this broader mitigation potential with a series of numerical experiments in Sections~\ref{subsec:fermion_qvm} and \ref{subsec:qubit_qvm}, wherein we simulate noisy Trotter circuits for systems of interacting fermions and spin-$1/2$ particles, respectively. The gate-level noise model is based on a superconducting architecture, with error rates derived from publicly available data of an existing Google Sycamore processor~\cite{arute2020hartree,arute2020observation,isakov2021simulations,cirq}. Overall, we assess that in this more realistic scenario, symmetry-adjusted classical shadows successfully mitigates errors, but with diminishing effectiveness as the circuit grows deeper. We observe an error floor to our approach, beyond which more samples does not improve prediction accuracy due to violations of Assumptions~\ref{assumption_1}. However, even in this regime we see substantially improved qualitative agreement of the mitigated results to the true dynamics.

\subsection{Subsystem-symmetrized Pauli shadows}\label{subsec:sym_pauli_summary}

While random Pauli measurements are efficient for predicting local qubit observables, the irreducible structure of the local Clifford group $\Cl(1)^{\otimes n}$ is difficult to reconcile with common symmetries under symmetry adjustment, such as the $\U(1)$ symmetry generated by $M = \sum_{i \in [n]} Z_i$. To remedy this issue, we modify the protocol by what we call \emph{subsystem symmetrization}:~define the group
\begin{equation}
    \SymCl{n} \coloneqq \Sym(n) \times \Cl(1)^{\otimes n},
\end{equation}
which has the unitary representation $U_{(\pi, C)} = S_\pi C$ where $S_\pi$ permutes the qubits according to $\pi \in \Sym(n)$ and $C \in \Cl(1)^{\otimes n}$. The circuit for $S_\pi$ can be obtained as a sequence of $\O(n^2)$ nearest-neighbor $\SWAP$ gates in $\O(n)$ depth via an odd--even decomposition of $\pi$~\cite{habermann1972parallel}. The following theorem summarizes its group-theoretic properties relevant to classical shadows.

\begin{theorem}[Irreducible representations of the subsystem-symmetrized local Clifford group]
    The representation $\mathcal{U} : \SymCl{n} \to \U(\mathcal{L}(\mathcal{H}))$, defined by $\mathcal{U}_{(\pi, C)}(\rho) = S_\pi C \rho C^\dagger S_\pi^\dagger$, decomposes into the irreps
    \begin{equation}
    V_k = \spn\{ P \in \Pauli(n) : |P| = k \}, \quad 0 \leq k \leq n.
    \end{equation}
    Under this group, the (noiseless) expressions for $\mathcal{M}$ and $\V[\hat{o}]$ coincide with those of standard Pauli shadows.
\end{theorem}

This modification therefore reduces the number of irreps from $2^n$ to $n + 1$, achieved by symmetrizing over all $k$-qubit subsystems. Meanwhile, the desirable estimation properties from standard Pauli shadows are retained:~for instance, the shadow norm obeys $\sn{P}^2 = 3^k$ for $k$-local Pauli operators $P$.

The upshot is that the symmetry $M$ is now compatible with this group, thereby enabling results such as Theorem~\ref{thm:informal_qubit_result}. We describe this construction in Section~\ref{sec:applications_qubits}, with technical details in \cref{sec:pauli_shadows_appendix}.

\subsection{Improved circuit design for fermionic Gaussian unitaries}\label{subsec:improved_circuit_summary}

Fermionic Gaussian unitaries are a broad class of free-fermion rotations, and they are ubiquitous primitives in algorithms for simulating (interacting) fermions. In the context of classical shadows, they form the basis for randomized measurements in matchgate shadows~\cite{zhao2021fermionic,wan2023matchgate,ogorman2022fermionic}. Such unitaries can be described by an orthogonal transformation $Q \in \Orth(2n)$ of the Majorana operators,
\begin{equation}
    \mathcal{U}_Q(\gamma_\mu) = U_Q \gamma_\mu U_Q^\dagger = \sum_{\nu \in [2n]} Q_{\nu\mu} \gamma_\nu
\end{equation}
for each $\mu \in [2n]$. The quantum circuits implementing these transformations take $\O(n^2)$ gates in $\O(n)$ depth~\cite{jiang2018quantum,oszmaniec2022fermion}. While this scaling is necessary in general by parameter counting, constant-factor savings can substantially improve performance in practice, especially on noisy quantum computers.

To this end, we introduce a new compilation algorithm for fermionic Gaussian unitaries, given an arbitrary $Q \in \Orth(2n)$. Our circuit design improves the parallelization of gates compared to prior art~\cite{jiang2018quantum,oszmaniec2022fermion}. The key idea is to observe that two Majorana modes essentially correspond to one qubit under the Jordan--Wigner transformation~\cite{jordanwigner}. Thus, the optimal approach to compiling $U_Q$ into single- and two-qubit gates involves decomposing the matrix $Q$ into elementary blocks of $4 \times 4$ transformations, rather than the $2 \times 2$ Givens rotations utilized in prior designs.

The details of this scheme are provided in \cref{sec:FGU_circuit_design} and implemented in code at our open-source repository~\cite{gitrepo}. We also demonstrate the improvements in circuit size in Figure~\ref{fig:circuit_design_comparison} in the Appendix, with respect to a gate set native to superconducting platforms. From these results we numerically infer roughly $1/3$ reduction in depth and $1/2$ reduction in gate count over prior designs. We make use of this improved design in our numerical simulations, in particular those of Section~\ref{subsec:fermion_qvm}.

\subsection{Spin-adapted matchgate shadows}\label{subsec:spin-adapt_summary}

Systems of spinful fermions often obey a spin symmetry, which allows for compressed block-diagonal representations according to the spin sectors. Such techniques are referred to as symmetry adaptation. We introduce such an adaptation of the matchgate shadows protocol wherein the random distribution is restricted to block-diagonal orthogonal transformations,
\begin{equation}
    Q = \begin{pmatrix}
    Q_\uparrow & 0\\
    0 & Q_\downarrow
    \end{pmatrix} \in \Orth(n) \oplus \Orth(n).
\end{equation}
We call this protocol \emph{spin-adapted} matchgate shadows. This restricted group remains informationally complete over operators which respect the spin sectors, thus sufficing for learning properties in systems with this symmetry. In fact, we show that the shadow norms for $k$-fermion operators under the spin-adapted protocol scale identically as in the unadapted setting. The main advantage of spin adaptation is that the block-diagonal transformation $Q = Q_\uparrow \oplus Q_\downarrow$ can be implemented as $U_{Q_\uparrow} \otimes P_{\downarrow}^s U_{Q_\downarrow}$, where $P_{\downarrow} = Z^{\otimes n/2}$ is the parity operator on the spin-down sector and $s = \delta_{-1, \det Q_{\uparrow}}$. This tensor-product unitary requires roughly half the number of gates and circuit depth compared to implementing a dense element of $\Orth(2n)$. We prove the necessary details in \cref{sec:spin_adaptation} and implement this modified protocol in our numerical experiments when applicable.

\section{Theory of symmetry-adjusted classical shadows}\label{sec:main_theory}

Here we describe the theory behind the symmetry-adjusted classical shadows estimator. This approach uses known symmetry information about the ideal, noiseless state $\rho$ that we wish to prepare (but are only able to produce a noisy version of). In this section we describe the idea for an arbitrary multiplicity-free group $G$;~in Section~\ref{sec:applications_fermions} and \ref{sec:applications_qubits}, we will provide concrete applications to the efficient estimation of local fermionic and qubit observables, respectively.

Suppose $ \rho $ is a quantum state obeying a known symmetry, corresponding to a collection of operators $S_\lambda \in V_\lambda$ for which the values $ s_\lambda \coloneqq \vip{S_\lambda}{\rho} $ are known \emph{a priori}. For example, suppose the system has a conserved quantity with operator $S$. Then we construct the $S_\lambda$ operators using the projectors $\Pi_\lambda$:
\begin{equation}
    \kket{S_\lambda} = \Pi_\lambda \kket{S},
\end{equation}
assuming that $S$ spans multiple irreps. By construction, $S_\lambda$ is an eigenoperator of both $\mathcal{M}$ and $\widetilde{\mathcal{M}}$:
\begin{equation}\label{eq:symmetry_eigenop}
\begin{split}
    \mathcal{M}\kket{S_\lambda} &= f_\lambda \kket{S_\lambda},\\
    \widetilde{\mathcal{M}}\kket{S_\lambda} &= \widetilde{f}_\lambda \kket{S_\lambda}.
\end{split}
\end{equation}
If one is interested in only a subset $R'
\subseteq R_G$ of the irreps, then it suffices to only know those symmetries $S_\lambda$ for which $\lambda \in R'$.

Because the ideal values of $ s_\lambda $ and $ f_\lambda $ are already known, we can use the estimated noisy expectation value of $ S_\lambda $ to build an estimate for $\widetilde{f}_\lambda$. We start with the standard postprocessing of classical shadows:~applying $\mathcal{M}^{-1}$ to the measurement outcomes of the noisy quantum experiments produces, in expectation, the effective state
\begin{equation}\label{eq:noisy_rho}
    \kket{\widetilde{\rho}} \coloneqq \mathcal{M}^{-1}\widetilde{\mathcal{M}} \kket{\rho} = \E_{g \sim G, b \sim \widetilde{\mathcal{U}}_g \kket{\rho}} \mathcal{M}^{-1} \mathcal{U}_{g}^\dagger \kket{b},
\end{equation}
which clearly differs from $\kket{\rho}$ when $\widetilde{\mathcal{M}} \neq \mathcal{M}$. Nonetheless, we can use this noisy data to estimate the value of $\vip{S_\lambda}{\widetilde{\rho}}$, which is equal to
\begin{equation}\label{eq:noisy_symmetry}
    \vip{S_\lambda}{\widetilde{\rho}} = \frac{\widetilde{f}_\lambda}{f_\lambda} s_\lambda
\end{equation}
by Eq.~\eqref{eq:symmetry_eigenop}. In fact, this relation applies to any $O \in V_\lambda$:
\begin{equation}\label{eq:EM_estimator_recovery}
    \vip{O}{\widetilde{\rho}} = \frac{\widetilde{f}_\lambda}{f_\lambda} \vip{O}{\rho}.
\end{equation}
Hence while we use Eq.~\eqref{eq:noisy_symmetry} to learn $\widetilde{f}_\lambda$ from the symmetry $S_\lambda$, this is in turn applicable to all other operators within the same irrep. This leads to the recovery of the ideal expectation values as
\begin{equation}\label{eq:corrected_observable}
    \vip{O}{\rho} = \frac{\vip{O}{\widetilde{\rho}}}{\vip{S_\lambda}{\widetilde{\rho}} / s_\lambda}.
\end{equation}

Having established the theory in expectation, we now analyze the implementation in practice. Let $T$ be the number of classical-shadow snapshots, $\kket{\hat{\rho}_\ell} = \mathcal{M}^{-1} \mathcal{U}_{g_\ell}^\dagger \kket{b_\ell}$ for $\ell = 1, \ldots, T$, obtained by sampling the noisy quantum computer. Recall that these snapshots converge to $\widetilde{\rho}$ rather than $\rho$. From their empirical average, $\hat{\rho}(T) = (1/T) \sum_{\ell=1}^T \hat{\rho}_{\ell}$, we can estimate the lefthand side of Eq.~\eqref{eq:noisy_symmetry} as
\begin{equation}
    \hat{s}_\lambda(T) \coloneqq \vip{S_\lambda}{\hat{\rho}(T)}.
\end{equation}
This in turn provides an estimate for $\widetilde{f}_\lambda$,
\begin{equation}\label{eq:f_hat_est}
    \hat{f}_\lambda(T) \coloneqq f_\lambda \frac{\hat{s}_\lambda(T)}{s_\lambda}.
\end{equation}
This can be understood as a generalization of $\mathrm{NoiseEst}_G(\lambda, g, b)$ from Eq.~\eqref{eq:noiseest}, making the replacements $D_\lambda \to S_\lambda$ and $\op{0^n}{0^n} \to \rho$. Indeed, one can view the calibration state $\ket{0^n}$ as obeying the symmetries given by its stabilizer group.

Consider the estimation of observables $O_1, \ldots, O_L$ with symmetry-adjusted classcial shadows. Suppose without loss of generality that each $O_j \in V_\lambda$ for some $\lambda \in R'$.\footnote{If an observable is supported on multiple subspaces, then we can write it as a linear combination of basis elements across those subspaces.} From the same noisy classical shadow $\hat{\rho}(T)$, we also have estimates for their noisy expectation values:~$\E\vip{O_j}{\hat{\rho}(T)} = \vip{O_j}{\widetilde{\rho}_j}$. Then, following Eq.~\eqref{eq:corrected_observable} we can directly construct error-mitigated estimators as
\begin{equation}\label{eq:EM_estimator}
    \hat{o}_j^{\mathrm{EM}}(T) \coloneqq \frac{\hat{o}_j(T)}{\hat{s}_\lambda(T) / s_\lambda},
\end{equation}
which converges to $\vip{O_j}{\rho}$ in the $T \to \infty$ limit (if Assumptions~\ref{assumption_1} hold). Because $\E[X/Y] \neq \E[X]/\E[Y]$ (for nontrivial random variables $X$ and $Y$), Eq.~\eqref{eq:EM_estimator} describes a biased estimator. In the following theorem, we quantify this bias by bounding the total prediction error of $\hat{o}_j^{\mathrm{EM}}(T)$. This in turn bounds the number of symmetry-adjusted classical-shadow samples $T$ required.

\begin{theorem}\label{thm:main_theorem}
Fix accuracy and confidence parameters $\epsilon, \delta \in (0, 1)$. Let $O_1, \ldots, O_L$ be a collection of observables, each supported on an irrep of $\mathcal{U} : G \to \U(\mathcal{L}(\mathcal{H}))$ as $O_j \in V_\lambda$ for $\lambda \in R' \subseteq R_G$. Let $S_\lambda \in V_\lambda$ be a symmetry operator for each $\lambda \in R'$, for which the ideal values $s_\lambda = \tr(S_\lambda \rho)$ of the target state $\rho$ are known \emph{a priori}. Suppose that each noisy unitary satisfies Assumptions~\ref{assumption_1}, $\widetilde{\mathcal{U}}_g = \mathcal{E} \mathcal{U}_g$, and define the quantities
\begin{align}
    F_{Z,R'}(\mathcal{E}) &\coloneqq \min_{\lambda \in R'} \frac{\tr(\mathcal{E} \mathcal{M}_Z \Pi_\lambda)}{\tr(\mathcal{M}_Z \Pi_\lambda)},\\
    \sigma^2 &\coloneqq \max_{1 \leq j \leq L, \lambda \in R'}\l\{ \V[\hat{o}_j], \V\l[\frac{\hat{s}_\lambda}{s_\lambda} \r] \r\}.
\end{align}
Then, a (noisy) classical shadow $\hat{\rho}(T)$ of size
\begin{equation}
    T = \O\l( \frac{\log((L + |R'|)/\delta)}{F_{Z,R'}(\mathcal{E})^2 \epsilon^2} \sigma^2 \r)
\end{equation}
can be used to construct error-mitigated estimates
\begin{equation}
    \hat{o}_j^{\mathrm{EM}}(T) \coloneqq \frac{\tr(O_j \hat{\rho}(T))}{\tr(S_\lambda \hat{\rho}(T)) / s_\lambda}
\end{equation}
which obey
\begin{equation}
    |\hat{o}_j^{\mathrm{EM}}(T) - \tr(O_j\rho)| \leq (\|O_j\|_\infty + 1) \epsilon + \O(\|O_j\|_\infty \epsilon^2)
\end{equation}
for all $1 \leq j \leq L$, with success probability at least $1 - \delta$.
\end{theorem}

The proof of this statement is provided in \cref{sec:error_analysis}. Note that $\| \cdot \|_\infty$ denotes the spectral norm. We phrase this result in terms of variances, rather than the state-independent shadow norm, because knowledge about $\rho$ (namely, its symmetries) can potentially provide tighter bounds. Note that the variance is with respect to the effective noisy state $\widetilde{\rho}$, which was defined in Eq.~\eqref{eq:noisy_rho}.

We now make a few remarks on this result. First, the symmetry operators appear in the sample complexity, normalized by the value of the symmetry sector as $S_\lambda / s_\lambda$. Therefore we expect that for typical applications, the variance $\V[\hat{s}_\lambda / s_\lambda] \leq \sns{S_\lambda / s_\lambda}{\mathrm{shadow}}^2$ will be comparable to the baseline variance of estimation, $\V[\hat{o}_j] \leq \max_{j'} \sns{O_{j'}}{\mathrm{shadow}}^2$. Additionally, the number of irreps considered is typically $|R'| \ll L$ (for instance, in the concrete examples of Section~\ref{sec:applications_fermions} and \ref{sec:applications_qubits}, $|R'|$ is a constant). Thus, we expect that the inclusion of symmetry operators incurs negligible overheads for most applications.

Instead, the primary overhead arises from the fact that error-mitigated estimation necessarily comes at the cost of larger overall variances~\cite{takagi2022fundamental,takagi2022universal,tsubouchi2022universal,quek2022exponentially}. The quantity
\begin{equation}
    F_{Z,R'}(\mathcal{E}) = \min_{\lambda \in R'} \frac{\tr(\mathcal{E} \mathcal{M}_Z \Pi_\lambda)}{\tr(\mathcal{M}_Z \Pi_\lambda)}
\end{equation}
characterizes an effective noise strength, and it can be seen as a generalization of the average $Z$-basis fidelity of $\mathcal{E}$,
\begin{equation}
    F_Z(\mathcal{E}) = \frac{\tr(\mathcal{E} \mathcal{M}_Z)}{\tr(\mathcal{M}_Z)} = \frac{1}{2^n} \sum_{b \in \{0, 1\}^n} \vev{b}{\mathcal{E}}{b},
\end{equation}
which appears in prior works on noise-robust classical shadows~\cite{chen2021robust,koh2022classical}. In contrast to $F_Z(\mathcal{E})$, the quantity $F_{Z,R'}(\mathcal{E})$ is a more fine-grained characterization of the noise channel, averaged within the relevant subspaces $V_\lambda$. Similar to prior results~\cite{karalekas2020quantum,chen2021robust,koh2022classical,van2022model,arrasmith2023development}, the sampling overhead of our error-mitigated estimates also depends inverse quadratically on this noise fidelity.

Finally, the error bound we obtain is $\O(\|O_j\|_\infty \epsilon)$ when $\epsilon < 1$. Note that $\|O_j\|_\infty = 1$ for Pauli and Majorana operators. Our result also features error terms of order $\O(\|O_j\|_\infty \epsilon^2)$, which reflect the biased nature of $\hat{o}_j^{\mathrm{EM}}(T)$. Nonetheless, this bias vanishes as $\epsilon^2 \sim 1/T$, so that for sufficiently large $T$ the prediction error is dominated by the standard shot-noise scaling of $\epsilon \sim 1/\sqrt{T}$.

\section{Application to fermionic (matchgate) shadows}\label{sec:applications_fermions}

The first application of symmetry-adjusted classical shadows that we consider is the estimation of local fermionic observables. This is achieved efficiently by fermionic classical shadows~\cite{zhao2021fermionic}, wherein the group $G$ corresponds fermionic Gaussian unitaries (also referred to as matchgate shadows~\cite{wan2023matchgate}). We will consider a commonly encountered symmetry in fermionic systems:~fixed particle number. However, it will be clear how the general idea can apply to other symmetries, such as spin. We begin with a review of matchgate shadows.

\subsection{Background on matchgate shadows}\label{subsec:matchgate_bkgd}

Let $a_p^\dagger, a_p$ be creation and annihilation operators for a system of $n$ fermionic modes, $p \in [n]$. The associated Majorana operators are
\begin{equation}
    \gamma_{2p} = a_p + a_p^\dagger, \quad \gamma_{2p + 1} = -\i(a_p - a_p^\dagger).
\end{equation}
Under the Jordan--Wigner transformation~\cite{jordanwigner}, these are mapped to Pauli operators as
\begin{equation}\label{eq:jw_majorana}
    \gamma_{2p} = \l(\prod_{q<p} Z_q\r) X_p, \quad \gamma_{2p + 1} = \l(\prod_{q<p} Z_q\r) Y_p.
\end{equation}
Recall from Eq.~\eqref{eq:majorana_def} that all $d^2$ basis operators are generated by taking arbitrary products:
\begin{equation}
    \Gamma_{\bm{\mu}} = (-\i)^{\binom{m}{2}} \gamma_{\mu_1} \cdots \gamma_{\mu_m},
\end{equation}
where $\bm{\mu} = (\mu_1, \ldots, \mu_m) \subseteq [2n]$. By convention, we order $\mu_1 < \cdots < \mu_m$. We can group all the $m$-degree Majorana indices by defining the set
\begin{equation}
    \comb{2n}{m} \coloneqq \{ \bm{\mu} \subseteq [2n] \mid |\bm{\mu}| = m \}.
\end{equation}
Physical fermionic observables have even degree $m = 2k$. An important subset of such operators comprises those which are diagonal in the standard basis, corresponding to the index set
\begin{equation}
    \diags{2n}{2k} \coloneqq \{ (2p_1, 2p_1 + 1, \ldots, 2p_k, 2p_k + 1) \mid \bm{p} \in \comb{n}{k} \}.
\end{equation}
Using Eq.~\eqref{eq:jw_majorana}, each $\bm{\tau} \in \diags{2n}{2k}$ corresponds to the Pauli-$Z$ operator $\Gamma_{\bm{\tau}} = Z_{p_1} \cdots Z_{p_k}$ under the Jordan--Wigner mapping.

The group of fermionic Gaussian unitaries is the image of the homomorphism $U : \Orth(2n) \to \U(d)$ whose adjoint action obeys
\begin{equation}\label{eq:matchgate_def}
    U_Q \gamma_\mu U_Q^\dagger = \sum_{\nu \in [2n]} Q_{\nu\mu} \gamma_\nu, \quad Q \in \Orth(2n).
\end{equation}
These unitaries are equivalent to (generalized) matchgate circuits~\cite{helsen2022matchgate} and constitute a class of classically simulatable circuits~\cite{valiant2001quantum,knill2001fermionic,terhal2002classical,bravyi2004lagrangian,divincenzo2005fermionic,jozsa2008matchgates}. Fermionic (matchgate) shadows then randomize over certain subgroups 
$G \subseteq \Orth(2n)$ of these Gaussian unitaries. The measurement channel takes the form
\begin{equation}\label{eq:matchgate_channel}
    \mathcal{M} = \sum_{k=0}^n f_{2k} \Pi_{2k},
\end{equation}
where the eigenvalues are
\begin{equation}
    f_{2k} = \l. \binom{n}{k} \middle/ \binom{2n}{2k} \r.
\end{equation}
and each irrep is the image of
\begin{equation}\label{eq:matchgate_projectors}
    \Pi_{2k} = \frac{1}{d} \sum_{\bm{\mu} \in \comb{2n}{2k}} \vop{\Gamma_{\bm{\mu}}}{\Gamma_{\bm{\mu}}}.
\end{equation}
While $\mathcal{U}$ carries $2n + 1$ unique irreps (each labeled by a Majorana degree $m$)~\cite{claes2021character,helsen2022matchgate}, only the $n + 1$ irreps $\lambda = 2k$ have nonvanishing $f_\lambda$~\cite{zhao2021fermionic,wan2023matchgate}. Therefore $\mathcal{M}^{-1}$ is formally the pseudoinverse restricted to those subspaces. Finally, the shadow norm of $k$-body Majorana operators is~\cite{zhao2021fermionic}
\begin{equation}\label{eq:FGU_shadow_norm}
    \sn{\Gamma_{\bm{\mu}}}^2 = f_{2k}^{-1} = \O(n^k).
\end{equation}
Variance expressions for arbitrary observables can be found in Refs.~\cite{wan2023matchgate,ogorman2022fermionic}. For the postprocessing of $T$ shadows into estimates of all $k$-body Majorana observables, we describe an algorithm in \cref{subsec:matchgate_computation} which runs in time $\O(n^k T)$.

We now comment on the choice of $G \subseteq \Orth(2n)$. Fermionic classical shadows were introduced in Ref.~\cite{zhao2021fermionic}, which initially considered the intersection of proper matchgate circuits [the special orthogonal group $\SO(2n)$] with $n$-qubit Clifford unitaries $\Cl(n)$. The result is the group of all $2n \times 2n$ signed permutation matrices with determinant $1$, denoted by $\B^{+}(2n) \subset \SO(2n)$. They also showed that its unsigned subgroup, $\Alt(2n) \subset \B^{+}(2n)$, possesses the same irrep structure~\cite[Supplemental Material, Theorem~11]{zhao2021fermionic}. While the full, continuous group $\SO(2n)$ has not yet been analyzed for classical shadows, it was studied for character randomized benchmarking~\cite{helsen2019new} in Ref.~\cite[Section~VI]{claes2021character}, wherein they demonstrated the presence of multiplicities. These multiplicities can be avoided by enlarging to the generalized matchgate group, i.e., all of $\Orth(2n)$~\cite[Lemma~3]{helsen2022matchgate}. Ref.~\cite{wan2023matchgate} applied these generalized matchgates to fermionic classical shadows, and in particular they prove that the Clifford intersection in this setting (now yielding the subgroup $\B(2n) \subset \Orth(2n)$ of signed permutation matrices with either determinant $\pm 1$) is a $3$-design for $\Orth(2n)$. This implies that $\B(2n)$ is also multiplicity-free.

Due to the variety of options, for the rest of this paper we assume matchgate shadows under any $G$ with the desired irreps. We note that Ref.~\cite{ogorman2022fermionic} introduced a smaller subset of $\B(2n)$ based on perfect matchings, which has the same channel $\mathcal{M}$ and variances;~however its connection to representation theory was not discussed.

\subsection{Utilizing particle-number symmetry}\label{subsec:fermion_symmetry}

Suppose the ideal state we wish to prepare lies in the $\eta$-particle sector of $\mathcal{H}$. This is a $\U(1)$ symmetry generated by the fermion-number operator, $N = \sum_{p\in[n]} a_p^\dagger a_p$. In particular, powers of $N$ obey
\begin{equation}\label{eq:fermion_number}
    \tr(N^k \rho) = \eta^k,
\end{equation}
which provides us a collection of conserved quantities with which to perform symmetry adjustment. Recall from Eq.~\eqref{eq:matchgate_projectors} that $\Pi_{m}$ projects onto the irrep
\begin{equation}
    V_m = \spn\{ \Gamma_{\bm{\mu}} \mid \bm{\mu} \in \comb{2n}{m} \}.
\end{equation}
Then, projecting $N^k$ onto $V_{2k}$ yields the symmetry operators $S_{2k}$, and solving the resulting linear system of equations recovers the ideal values for $s_{2k} = \tr(S_{2k} \rho)$. For ease of exposition we will consider only $k = 1, 2$, but one may generalize to higher $k$ using these ideas.

Concretely, we start with the fact that $a_p^\dagger a_p = (\I - \Gamma_{(2p, 2p+1)})/2$, and $\Gamma_{(2p, 2p+1)} \Gamma_{(2q, 2q+1)} = \Gamma_{(2p, 2p+1, 2q, 2q + 1)}$ for $p < q$. Then, expanding $N$ and $N^2$ into a linear combination of Majorana operators, one finds
\begin{align}
    S_2 &= \Pi_2(N) = -\frac{1}{2} \sum_{\bm{\mu} \in \diags{2n}{2}} \Gamma_{\bm{\mu}},\\
    S_4 &= \Pi_4(N^2) = \frac{1}{2} \sum_{\bm{\mu} \in \diags{2n}{4}} \Gamma_{\bm{\mu}}.
\end{align}
Using Eq.~\eqref{eq:fermion_number} and the relations between $S_2$ and $S_4$ to $N$ and $N^2$ (for example, $N = n\I/2 + S_2$), we arrive at:
\begin{align}
    s_2 &= \tr\l( S_2 \rho \r) = \eta - \frac{n}{2}, \label{eq:s2_val}\\
    s_4 &= \tr\l( S_4 \rho \r) = \frac{1}{2} \binom{n}{2} - \eta(n - \eta). \label{eq:s4_val}
\end{align}

For the sampling cost incurred by these symmetry operators, we argue that the typical shadow norms of these symmetries are $\sn{S_{2k}/s_{2k}}^2 = \O(n^k)$, which is the same as the base estimation. To see this, consider a triangle inequality on the shadow norm:
\begin{equation}
\begin{split}
    \sn{S_{2k}} &\leq \frac{1}{2} \sum_{\bm{\mu} \in \diags{2n}{2k}} \sn{\Gamma_{\bm{\mu}}}\\
    &= \frac{1}{2} \binom{n}{k} \sqrt{\l. \binom{2n}{2k} \middle/ \binom{n}{k} \r.}\\
    &= \O( n^{3k/2} ).
\end{split}
\end{equation}
Thus $\sn{S_{2}}^2 = \O(n^{3})$ and $\sn{S_{4}}^2 = \O(n^{6})$. Next, we need to examine how $s_{2k}^2$ scales with system size. Assuming that $s_2, s_4 \neq 0$ and that the number of electrons is $\eta = \O(n)$, then from Eqs.~\eqref{eq:s2_val} and \eqref{eq:s4_val} we see that $s_2^2 = \Theta(n^2)$ and $s_4^2 = \Theta(n^4)$. Thus
\begin{align}
    \sn{S_2/s_2}^2 &= \O(n),\\
    \sn{S_4/s_4}^2 &= \O(n^2).
\end{align}

\subsubsection{Avoiding division by zero}\label{subsec:ancilla_trick}

One potential obstruction to symmetry adjustment is when some $s_{2k} = 0$. This can occur whenever the particle number takes a specific value:
\begin{align}
     s_2 = 0 &\text{ if } \eta = \frac{n}{2}, \label{eq:s2_zero}\\
     s_4 = 0 &\text{ if } \eta = \frac{n \pm \sqrt{n}}{2}. \label{eq:s4_zero}
\end{align}
Equation~\eqref{eq:s2_zero} occurs at half filling, which is fairly common. On the other hand, Eq.~\eqref{eq:s4_zero} occurs only when the number of modes $n$ is a perfect square and the number of particles $\eta$ is one of two specific values, so it is less likely to occur. Nonetheless, there is a straightforward way to circumvent both possibilities by introducing a single ancilla qubit.

To do so, append an additional fermion mode initialized in the unoccupied state $\ket{0}$, so that the ideal state is now the $(n + 1)$-mode state $\rho' = \rho \otimes \op{0}{0}$. Given that $\rho$ has $\eta$ particles on $n$ modes, $\rho'$ is an $\eta$-particle state on $n + 1$ modes. The new symmetry operators on the $(n + 1)$-mode Hilbert space are
\begin{align}
    S'_2 &= -\frac{1}{2} \sum_{\bm{\mu} \in \diags{2(n + 1)}{2}} \Gamma_{\bm{\mu}},\\
    S'_4 &= \frac{1}{2} \sum_{\bm{\mu} \in \diags{2(n + 1)}{4}} \Gamma_{\bm{\mu}},
\end{align}
which have ideal values
\begin{align}
    s'_2 &= \tr\l( S_2 \rho' \r) = \eta - \frac{n + 1}{2},\\
    s'_4 &= \tr\l( S_4 \rho' \r) = \frac{1}{2} \binom{n + 1}{2} - \eta(n + 1 - \eta).
\end{align}
It is straightforward to check that, if either condition Eq.~\eqref{eq:s2_zero} or Eq.~\eqref{eq:s4_zero} holds, then $s'_2$ and $s'_4$ are always nonzero for $n > 1$.

Under the Jordan--Wigner mapping, this modification is easily achieved by initializing a single ancilla qubit in $\ket{0}$. Recall that the terms in the symmetries $S_2, S_4$ are the diagonal operators $\Gamma_{(2p,2p+1)} = Z_p$ and $\Gamma_{(2p,2p+1,2q,2q+1)} = Z_p Z_q$. Note also that the ancilla qubit is acted on only during the random unitary $U_Q$ (where now $Q$ has dimension $2n + 2$) and otherwise does not interact with the $n$ system qubits.

\section{Application to qubit (Pauli) shadows}\label{sec:applications_qubits}

Now we turn to the application for local observable estimation in systems of spin-$1/2$ particles (qubits). Random Pauli measurements are efficient for this task;~however, for compatibility with the global $\U(1)$ symmetry considered in this work, we must slightly modify the protocol to accommodate its irreps. We begin with a review of the standard Pauli shadows protocol, followed by our modification.

\subsection{Background on standard Pauli shadows}\label{subsec:pauli_shadow_background}

The local Clifford group $\Cl(1)^{\otimes n}$ is implemented by uniformly drawing a single-qubit Clifford gate for each qubit independently. It has $2^n$ irreducible representations, corresponding to all $k$-qubit subsystems $I \subseteq [n]$, where $|I| = k \in \{0, 1, \ldots, n\}$~\cite{gambetta2012characterization}. Twirling $\mathcal{M}_Z$ by this group yields
\begin{equation}
    \mathcal{M} = \sum_{I \subseteq [n]} f_I \Pi_I,
\end{equation}
where $f_I = 3^{-|I|}$ and $\Pi_I$ projects onto the subspace of operators which act nontrivially on precisely the subsystem $I$. The squared shadow norm for $k$-local Pauli operators $P$ is~\cite{huang2020predicting}
\begin{equation}\label{eq:pauli_shadow_norm}
    \sn{P}^2 = 3^k.
\end{equation}
A more general variance bound was derived in Ref.~\cite{paini2021estimating}:~a simple loose bound of their result can be stated as $\V[\hat{o}] \leq 3^k R \| O \|_{\infty}^2$, where $O$ is an arbitrary $k$-local traceless observable and $R$ is the number of terms in its Pauli decomposition. However, they argue that a tighter expression, essentially $3^k \| O \|_{\infty}^2$, is typically a good approximation to the variance.

\subsection{Subsystem symmetrization of Pauli shadows}\label{subsec:ss-pauli}

The irreps of $\Cl(1)^{\otimes n}$ are difficult to reconcile with commonly encountered symmetries. For example, consider a conserved total magnetization $M = \sum_{i \in [n]} Z_i$. In terms of qubits, this is equivalent to the different Hamming-weight sectors. Each term $Z_i$ lies in a different irrep $I = \{i\}$, so $M$ spans multiple irreps rather than having a single conserved quantity per irrep.

To remedy this conflict, we introduce what we call subsystem-symmetrized Pauli shadows, which randomizes over a group whose irreps are labeled only by the qubit locality $k$, rather than any specific subsystem $I$ of $k$ qubits. (This is analogous to how the matchgate irreps depend only on fermionic locality, due to the inherent antisymmetry of fermions.) We formalize the group as follows.

\begin{definition}
\label{def:ss-pauli}
    The subsystem-symmetrized local Clifford group is defined as $\SymCl{n} \coloneqq \Sym(n) \times \Cl(1)^{\otimes n}$, where $\Sym(n)$ is the symmetric group and $\Cl(1)$ is the single-qubit Clifford group. Its unitary action on $\mathcal{H}$ is given by
    \begin{equation}
    U_{(\pi, C)} = S_\pi C,
    \end{equation}
    where $C = \bigotimes_{i \in [n]} C_i \in \Cl(1)^{\otimes n}$ and $\pi \in \Sym(n)$ is represented by a permutation of the $n$ qubits:
    \begin{equation}
    S_\pi \ket{b_0} \cdots \ket{b_{n-1}} = \ket{b_{\pi^{-1}(0)}} \cdots \ket{b_{\pi^{-1}(n-1)}}
    \end{equation}
    for all $b_i \in \{0, 1\}$, $i \in [n]$.
\end{definition}

The unitaries $S_\pi$ can be implemented with $\O(n^2)$ gates and depth $\O(n)$, for example by constructing a parallelized network of nearest-neighbor $\SWAP$ gates according to an odd--even sorting algorithm~\cite{habermann1972parallel} applied to $\pi$. Representing $\pi$ as an array of the permuted elements of $[n]$, the sorting algorithm returns a sequence of adjacent transpositions $i \leftrightarrow i + 1$ which maps $\pi$ to $(0, 1, \ldots, n-1)$. This sequence therefore implements $\pi^{-1}$ as desired. Each such transposition then maps to a $\SWAP_{i,i+1}$ gate to construct the quantum circuit. For the postprocessing of $T$ shadows into $k$-local Pauli estimates, we review in \cref{subsec:pauli_computation} the algorithm which runs in time $\O(n^k T)$.

We prove the relevant properties of subsystem-symmetrized Pauli shadows in \cref{sec:pauli_shadows_appendix}, namely its irreps and the shadow norm of local observables. We summarize the results here:~each irrep is the space of all $k$-local operators,
\begin{equation}
    V_k = \spn(\mathcal{B}_k) \text{ where } \mathcal{B}_k \coloneqq \{ P \in \Pauli(n) : |P| = k \},
\end{equation}
for each $k \in \{0, 1, \ldots, n\}$. Hence the (noisy) measurement channel is
\begin{equation}
    \widetilde{\mathcal{M}} = \sum_{k=0}^n \widetilde{f}_k \Pi_k,
\end{equation}
where $\widetilde{f}_k = \tr(\mathcal{M}_Z \mathcal{E} \Pi_k) / (3^k \binom{n}{k})$ and
\begin{equation}
    \Pi_k = \frac{1}{d} \sum_{P \in \mathcal{B}_k} \vop{P}{P}.
\end{equation}
When $\mathcal{E}$ is the identity channel, we recover $f_k = 3^{-k}$. Also in the absence of noise, the variance formulas are exactly the same as in standard Pauli shadows.\footnote{In fact, all $t$-fold twirls on $\mathcal{M}_Z$ coincide between the symmetrized and unsymmetrized groups.}

\subsection{Utilizing total magnetization symmetry}\label{subsec:qubit_symmetry}

We take the $\U(1)$ symmetry generated by a total magnetization $M = \sum_{i \in [n]} Z_i$. Suppose the ideal state has a known value of $m = \tr(M \rho)$ (equivalently, $\rho$ lives in a sector of fixed Hamming weight $(n - m)/2$). The symmetries projected into the irreps of $\SymCl{n}$ are then
\begin{align}
    S_1 &= \Pi_1(M) = \sum_{i \in [n]} Z_i,\\
    S_2 &= \Pi_2(M^2) = 2 \sum_{i < j} Z_i Z_j,
\end{align}
whose ideal values are
\begin{align}
    s_1 &= m,\\
    s_2 &= m^2 - n.
\end{align}
As in the fermionic setting, we encounter issues if $s_1$ or $s_2$ vanish (i.e., $m = 0$ or $m = \pm\sqrt{n}$, respectively). In this case, we can perform the same ancilla trick, appending a qubit in $\ket{0}$ and modifying the conserved quantities to
\begin{align}
    s'_1 &= m + 1,\\
    s'_2 &= (m + 1)^2 - (n + 1).
\end{align}

The variances of the symmetry operators are
\begin{align}
    \V[\hat{s}_1/s_1] &= \O(n),\\
    \V[\hat{s}_2/s_2] &= \O(1),
\end{align}
whenever the ideal state lives in a symmetry sector of constant $m = \Theta(1)$. We show this in \cref{subsec:ss-pauli_variance}, along with general $m$-dependent expressions in Eqs.~\eqref{eq:s1_noisy_var} and \eqref{eq:s2_noisy_var}. This $n$-dependent variance bound reflects the fact that the symmetries are extensive properties. While local Pauli operators have variances bounded by a constant, we point out that many local observables of interest are linear combinations of an extensive number of Pauli terms. As such, their shadow norms typically grow with system size as well (recall the discussion at the end of Section~\ref{subsec:pauli_shadow_background}). 

\section{Numerical experiments}\label{sec:numerics}

We now demonstrate the error-mitigation capabilities of symmetry-adjusted classical shadows through numerical simulations. We focus on the task of estimating one- and two-body observables in both fermion and qubit systems which obey the global $\U(1)$ symmetries described in Sections~\ref{subsec:fermion_symmetry} and \ref{subsec:qubit_symmetry}.

For each type of system, we first present results when the noise models obey Assumptions~\ref{assumption_1} (readout errors). We demonstrate the successful mitigation at varying sample sizes, noise rates, and system sizes, confirming the correctness of our theory.

Next, we investigate how symmetry adjustment performs under a more comprehensive noise model based on superconducting-qubit platforms. These simulations were performed using the Quantum Virtual Machine (QVM) within the Cirq open-source software package~\cite{cirq,isakov2021simulations}. It uses existing hardware data on a native gate set (single-qubit rotations and two-qubit $\sqrt{\iSWAP}$ gates on a square lattice) to mimic the realistic performance of a noisy quantum computer. We use the calibration data provided of Google's 23-qubit Rainbow processor based on the Sycamore architecture, which was used in quantum experiments simulating quantum chemistry and strongly correlated materials~\cite{arute2020hartree,arute2020observation}. The noise model consists of depolarizing channels, two-qubit coherent errors, single-qubit idling noise, and readout errors. Error rates vary across the chip;~on the $2 \times 4$ grid that we simulated, the average single- and two-qubit Pauli error rates are ${\sim}0.15\%$ and ${\sim}1.5\%$, respectively. A precise description of the noise model can be found in \cref{subsec:noise_model_details}.

Throughout, we use the following conventions for figures. Noiseless data (blue squares) correspond to simulations of an ideal quantum computer, which experiences no noise channel and only exhibits the fundamental sampling error. Unmitigated data (black X's) are simulations of classical shadows on a noisy quantum computer, using standard postprocessing routines. The mitigated estimates (red diamonds) are instead postprocessed as symmetry-adjusted classical shadows, as described in Section~\ref{sec:main_theory}. In some experiments, we also compare against robust shadow estimation~\cite{chen2021robust} (RShadow, green crosses), which involves simulating the calibration protocol on $\ket{0^n}$ under the same noise model. Finally, the true values (teal curves) are the ground truth, against which we determine the prediction error.

Uncertainty bars represent one standard deviation of the combined sampling and postprocessing, computed by empirical bootstrapping~\cite{efron1992bootstrap}. To ease the computational load, we slightly modify the procedure by batching samples;~see \cref{subsec:bootstrap_error_bars} for details.

\subsection{Fermionic systems}

Our first set of numerical experiments consider the application to matchgate shadows to learn and mitigate noise in one- and two-body fermionic observables.

\subsubsection{Readout noise}\label{subsec:fermion_readout_error}

First, we consider the reconstruction of the fermionic two-body reduced density matrix ($2$-RDM) from matchgate shadows. The $2$-RDM elements of a state $\rho$ are given by
\begin{equation}
    {}^2 D_{rs}^{pq} = \tr(a_p^\dagger a_q^\dagger a_s a_r \rho), \quad p, q, r, s \in [n].
\end{equation}
In general, knowledge of the $k$-RDM allows one to calculate any $k$-body observable of the system. By anticommutation relations, there are only $\binom{n}{2}^2$ unique matrix elements, corresponding to the indices $p < q$ and $r < s$. We therefore represent ${}^2 D$ as an $\binom{n}{2} \times \binom{n}{2}$ Hermitian matrix, flattening along those index pairs. Estimates ${}^2 \hat{D}_{rs}^{pq}(T) = \tr(a_p^\dagger a_q^\dagger a_s a_r \hat{\rho}(T))$ are computed from $T$ matchgate-shadow samples. Here, our figure of merit for the prediction error is the spectral-norm difference between the reconstructed and the numerically exact $2$-RDMs, $\epsilon = \| {}^2 \hat{D} - {}^2 D \|_\infty$.

\begin{figure}
\centering
\includegraphics[scale=0.5]{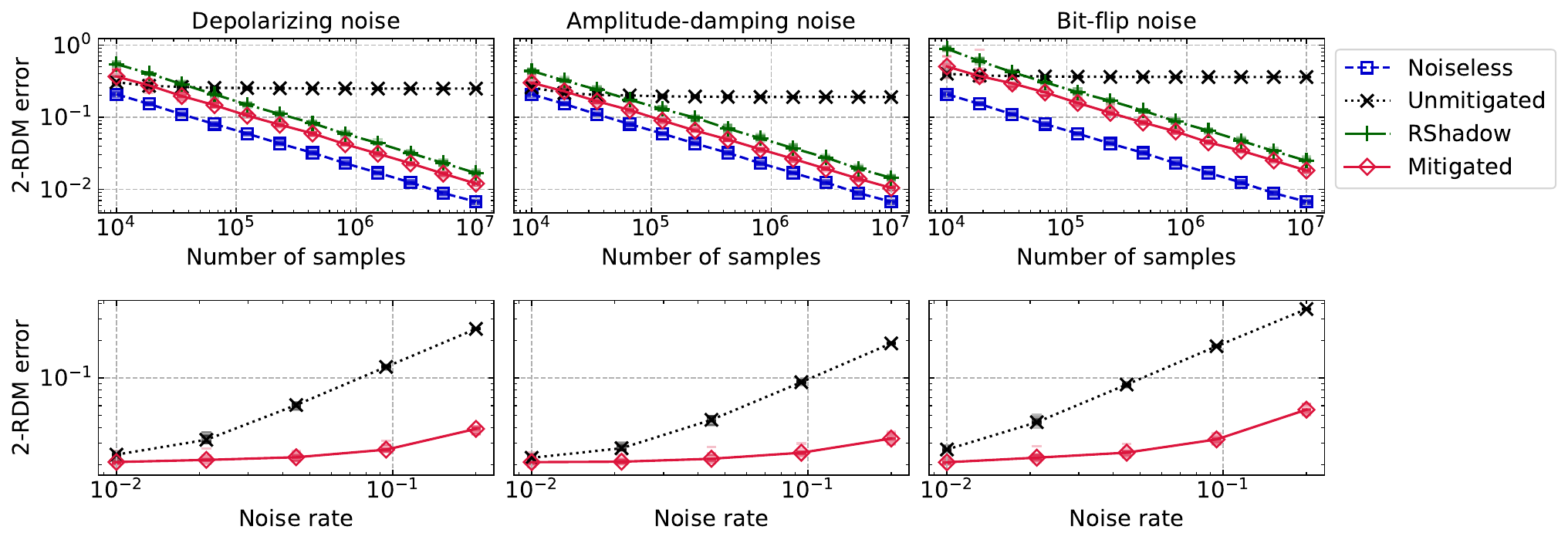}
\caption[Estimation error of the fermionic $2$-RDM reconstructed from matchgate shadows.]{Estimation error of the fermionic $2$-RDM reconstructed from matchgate shadows. Error is quantified by the spectral-norm difference between the estimated and true $2$-RDM. We simulate the protocol on a collection of 20 random Slater determinants on $n = 8$ modes with $\eta = 2$ fermions;~faint dashes are results for individual states, markers indicate the median error across the 20 states. (Top) Scaling of estimation error with the total number of samples $T$, fixing the noise rate to $p = 0.2$ for all noise models. (Bottom) Scaling of the estimation error with the noise rate $p$, fixing the total number of samples to $T = 10^6$.}
\label{fig:random_slater}
\end{figure}

We demonstrate $2$-RDM reconstruction on an ensemble of 20 random Slater determinants (noninteracting-fermion states with particle number $\eta$). An $\eta$-fermion Slater determinant is specified by the first $\eta$ columns of an $n \times n$ unitary matrix, so we generate the random states by uniformly drawing elements of $\U(n)$. This $n \times n$ representation is then lifted to the $2n \times 2n$ fermionic Gaussian representation, which allows us to apply the random matchgate transformations $Q \in \B(2n)$ efficiently. This simulates the action of $\rho \mapsto U_Q \rho U_Q^\dagger$. The measurement of this rotated state is then simulated using the algorithm of Ref.~\cite[Section~5.1]{bravyi2012classical}. Finally, to simulate the readout noise we implement the effective noise channel on the sampled bit strings offline.

While the $2$-RDM of free-fermion states can be computed from the 1-RDM using Wick's theorem, we do not employ any such tricks here. (We use Slater determinants simply to facilitate fast classical simulation.) We also do not use any additional error-mitigation strategies, such as RDM positivity constraints~\cite{rubin2018application}, that could in principle be applied in tandem.

The results are presented in Figure~\ref{fig:random_slater}. We consider a small system size, $n = 8$ and $\eta = 2$, and simulate three types of single-qubit noise channels before readout:~depolarizing, amplitude damping, and bit flip. The noise rate $p$ represents the probability of such an error occurring, independently on each qubit (defined in \cref{subsec:readout_errors}). In the top row, we show how the prediction error varies with the total number of samples $T$. As expected, the noiseless estimates (corresponding to $p = 0$) converge as $\sim T^{-1/2}$, which is the standard shot-limited behavior. Then, setting $p = 0.2$, we see how the unmitigated data experiences an error floor beyond which taking additional samples does not improve the accuracy. On the other hand, the mitigated results clearly bypass this error floor and recover the shot-noise scaling with $T$, thus validating the theory of symmetry-adjusted classical shadows. Compared to the noiseless simulations, our mitigated data exhibit a constant factor increase in the sampling cost, corresponding to the overhead of $\O(F_{Z,R'}^{-2})$ appearing in Theorem~\ref{thm:main_theorem}.

For these experiments, we also compare to the performance of robust shadow estimation (RShadow) by Chen \emph{et al.}~\cite{chen2021robust}, which requires simulating the calibration procedure on $\ket{0^n}$. For a fair comparison, we allocate $T/2$ samples to the calibration step and $T/2$ samples to the estimation step, so that the total number of samples is the same. While Chen \emph{et al.}~\cite{chen2021robust} did not originally consider matchgate shadows, from our generalization in Eq.~\eqref{eq:noiseest} we can construct $\mathrm{NoiseEst}_{\B(2n)}$ by taking $D_{\lambda} = S_{2k}$, which obeys $\ev{0^n}{S_2}{0^n} = -n/2$ and $\ev{0^n}{S_4}{0^n} = n(n-1)/4$. The single-shot estimator is then
\begin{equation}\label{eq:fermion_RSE_estimator}
    \mathrm{NoiseEst}_{\B(2n)}(2k, Q, b) = \frac{\ev{b}{U_Q S_{2k} U_Q^\dagger}{b}}{\ev{0^n}{S_{2k}}{0^n}}.
\end{equation}

As expected, RShadow behaves similarly to symmetry-adjusted classical shadows in this scenario wherein the noise obeys Assumptions~\ref{assumption_1}. However, even here we observe the advantage of our approach in terms of the number of samples. We attribute the additional overhead of RShadow to its calibration procedure, which symmetry adjustment avoids.

In the bottom row of Figure~\ref{fig:random_slater}, we simulate the same collection of random Slater determinants, but now varying the noise rate $p$ at a fixed sample size $T = 10^6$. While the unmitigated errors quickly grow with increasing noise rate, the mitigated estimates remain under control. Note that the errors of the mitigation protocol still grow modestly because we have fixed the number of samples;~in order to achieve a constant prediction error, one would need to scale $T$ proportional to $F_{Z,R'}^{-2}$ (which is $p$-dependent). Our key takeaway is that the combination of both rows of plots indicates the ability to handle a range of noise channels and error rates. Indeed, the growing errors seen in the bottom row can be suppressed by simply taking more samples, which is what the top row demonstrates.

\begin{figure}
\centering
\includegraphics[scale=0.5]{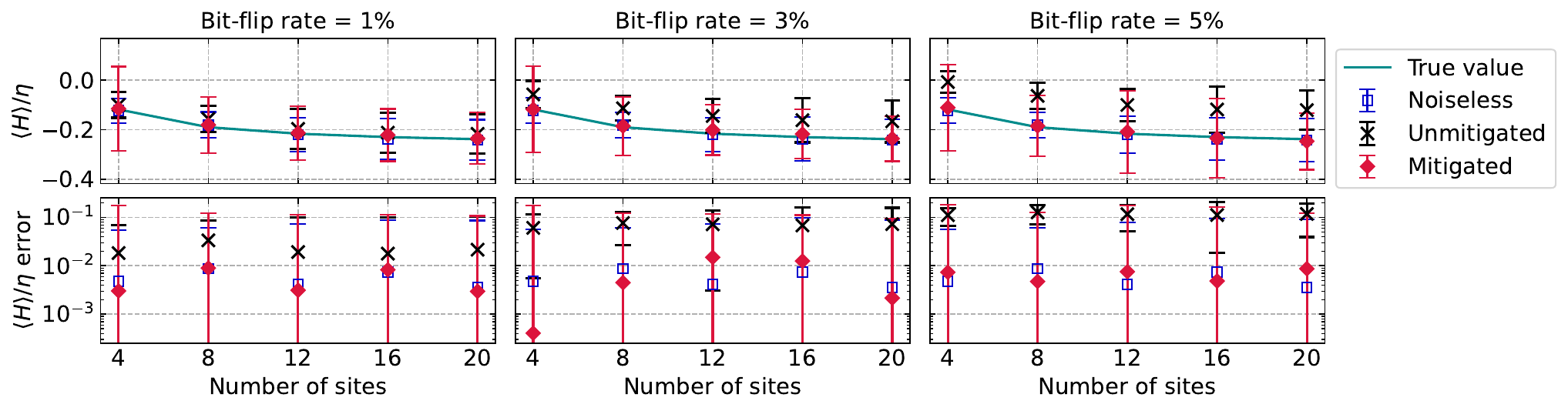}
\caption[Estimation of the energy per electron of a 1D spinful Fermi--Hubbard model, measured using spin-adapted matchgate shadows.]{Estimation of the energy per electron, $\langle H \rangle / \eta$, of a 1D spinful Fermi--Hubbard model at interaction strength $U/t = 4$ on $4 \leq L \leq 20$ sites, measured using spin-adapted matchgate shadows with $T = 2 \times 10^6$ samples. We simulate the protocol on the ground state of the noninteracting component $J$ of the Hamiltonian, Eq.~\eqref{eq:hopping_term}, at half filling in each spin sector, which is a Slater determinant with $\eta_\sigma = L/2$ electrons per sector. These simulations involve $n' = 2L + 2$ noisy qubits experiencing readout bit-flip errors with probabilities $1\%$, $3\%$, and $5\%$. The additional qubit per spin sector is prepared in $\ket{0}$ to avoid division by zero (see Section~\ref{subsec:ancilla_trick}).}
\label{fig:fh_energy}
\end{figure}

Next, we consider the simulation of a 1D spinful Fermi--Hubbard chain of $L = n/2$ sites (for a total of $n$ fermionic modes/qubits). Under open boundary conditions, the Hamiltonian for this model is
\begin{equation}
    H = J + V,
\end{equation}
where
\begin{align}
    J &= -t \sum_{i \in [L-1]} \sum_{\sigma \in \{\uparrow, \downarrow\}} a_{i,\sigma}^\dagger a_{i+1,\sigma} + \mathrm{h.c.}, \label{eq:hopping_term}\\
    V &= U \sum_{i \in [L]} N_{i,\uparrow} N_{i, \downarrow}, \label{eq:interaction_term}
\end{align}
are the hopping and interaction terms, respectively. The creation operators $a_{i,\sigma}^\dagger$ produce an electron at site $i$ with spin $\sigma$, and $N_{i,\sigma} = a_{i,\sigma}^\dagger a_{i,\sigma}$ is the associated occupation-number operator. We set units such that the hopping strength is $t = 1$.

For the target state, we use the ground state of the noninteracting term $J$, which is also a Slater determinant. This allows us to use the same simulation techniques as before to efficiently simulate up to $20$ sites. The number of electrons in each spin sector is $\eta_\sigma = L/2$, for a total of $\eta = \eta_{\uparrow} + \eta_{\downarrow} = n/2$ electrons. Thus the system is at half filling, which requires the use of ancilla qubits to avoid division by zero (described in Section~\ref{subsec:ancilla_trick}). In fact, we simulate $n' = n + 2$ qubits because we append an ancilla qubit to each spin sector. This because we employ spin-adapted matchgate shadows, described in Section~\ref{subsec:spin-adapt_summary} and \cref{sec:spin_adaptation}. This modification essentially treats each spin sector independently in terms of the randomized measurements, and so each sector itself is at half filling.

The Fermi--Hubbard results are shown in Figure~\ref{fig:fh_energy}. We consider the estimation of energy per electron, $\langle H \rangle / \eta$. We set the interaction strength to $U/t = 4$ and the noise model to single-qubit bit-flip errors, with probabilities $p \in \{0.01, 0.03, 0.05\}$. The energy per electron (top) and absolute estimation error (bottom) are plotted as the system size grows, keeping the number of samples fixed at $T = 2 \times 10^6$. Again, these results serve to validate our theory, showing the correctness of symmetry adjustment at larger system sizes (recall that number of electrons scales with the number of sites at half filling). This also demonstrates the use of spin-adapted matchgate shadows and the successful use of ancillas to avoid division by zero in $\hat{o}_j^{\mathrm{EM}}$.

\subsubsection{QVM noise model}\label{subsec:fermion_qvm}

Now we turn to the gate-level noise model simulated through the QVM~\cite{cirq,isakov2021simulations}. This model strongly violates Assumptions~\ref{assumption_1}, reflecting the fact that the state-preparation circuit $U_{\mathrm{prep}}$ is typically the dominant source of errors.

As our testbed fermionic system, we again consider the 1D spinful Fermi--Hubbard chain with open boundary conditions and interaction strength $U/t = 4$. Rather than the static problem, here we simulate a Trotterized time evolution of the Hamiltonian, as the number of Trotter steps provides a systematic way to increase the circuit depth (and hence the cumulative amount of noise). Note that because we are focusing on the mitigation of noisy quantum circuits, the ground truth corresponds to the noiseless Trotter circuit with a finite step size (i.e., we are not interested in the exact non-Trotterized dynamics).

We closely follow the setup of the experiment performed in Ref.~\cite{arute2020observation} (which was in fact performed on a Sycamore processor that our noise model is based on), using code made available by the authors at Ref.~\cite{recirq}. Because simulating the full noisy circuit is exponentially expensive, we restrict to a four-site instance ($n = 2L = 8$). The initial state is the ground state in the $\eta_{\uparrow}, \eta_{\downarrow} = 1$ sector of the noninteracting Hamiltonian
\begin{equation}
\begin{split}
    H_0 = J + \sum_{i \in [L]} \sum_{\sigma \in \{\uparrow, \downarrow\}} \varepsilon_{i,\sigma} N_{i, \sigma},
\end{split}
\end{equation}
where $J$ is the hopping term defined in Eq.~\eqref{eq:hopping_term} and we set the on-site potentials to have a Gaussian form, $\varepsilon_{i,\sigma} = -\lambda_\sigma e^{-\frac{1}{2} (i + 1 - c)^2 / s^2}$. This generates a Slater determinant whose charge density
\begin{equation}
    \varrho_i = \langle N_{i,\uparrow} + N_{i,\downarrow} \rangle.
\end{equation}
has a Gaussian profile, centered around $c$ with width $s$ and magnitude $\lambda_\sigma$. We set the parameters to $c = L/2 + 1/2 = 2.5$, $s = 7/3$, and $\lambda_\sigma = 4\delta_{\sigma,\uparrow}$. This initial state is prepared by the appropriate single-particle basis rotations (a subset of fermionic Gaussian unitaries)~\cite{wecker2015solving,kivlichan2018quantum,jiang2018quantum} on the state $\ket{1000}$ within each spin sector. Denote this unitary by $U(H_0)$. The system is then evolved by Trotterized dynamics according to $H$, with $R \in \{0, 1, \ldots, 5\}$ steps of size $\delta t = 0.2$. Let $J_{\mathrm{even}}$ (resp., $J_{\mathrm{odd}}$) be the terms in $J$ with $i$ even (resp., odd), and similarly for $V_{\mathrm{even}}, V_{\mathrm{odd}}$. One Trotter step is ordered as
\begin{equation}
    U_{\mathrm{Trot}} = e^{-\i J_{\mathrm{odd}} \delta t} e^{-\i V_{\mathrm{odd}} \delta t} e^{-\i V_{\mathrm{even}} \delta t} e^{-\i J_{\mathrm{even}} \delta t},
\end{equation}
which is then compiled into the native gate set. The full state-preparation circuit is then
\begin{equation}
    U_{\mathrm{prep}}(R) = U_{\mathrm{Trot}}^R U(H_0) X_{0,\downarrow} X_{0,\uparrow},
\end{equation}
where $X_{0,\sigma}$ places a spin-$\sigma$ electron on the first site from the vacuum (i.e., prepares $\ket{1000}$ in each spin sector). Note that $R = 0$ corresponds to only preparing the initial Slater determinant. Further details on the construction of these circuits are available in Refs.~\cite{arute2020observation,recirq}.

One final detail of Ref.~\cite{arute2020observation} that we follow is their method of qubit assignment averaging (QAA). This technique is employed as a means of ameliorating inhomogeneities in error rates across the quantum device. QAA works by identifying a collection of different assignments for the physical qubit labels and uniformly averaging over them (note that the Jordan--Wigner convention is kept fixed). For example, one may vary qubit assignments by selecting a different portion of the chip, or rotating/flipping the layout. Here, we fix a $2 \times 4$ grid of qubits and perform QAA over four different orderings of those eight qubits;~see \cref{subsec:qaa} for the specific assignments chosen.

\begin{figure}
\centering
\includegraphics[scale=0.5]{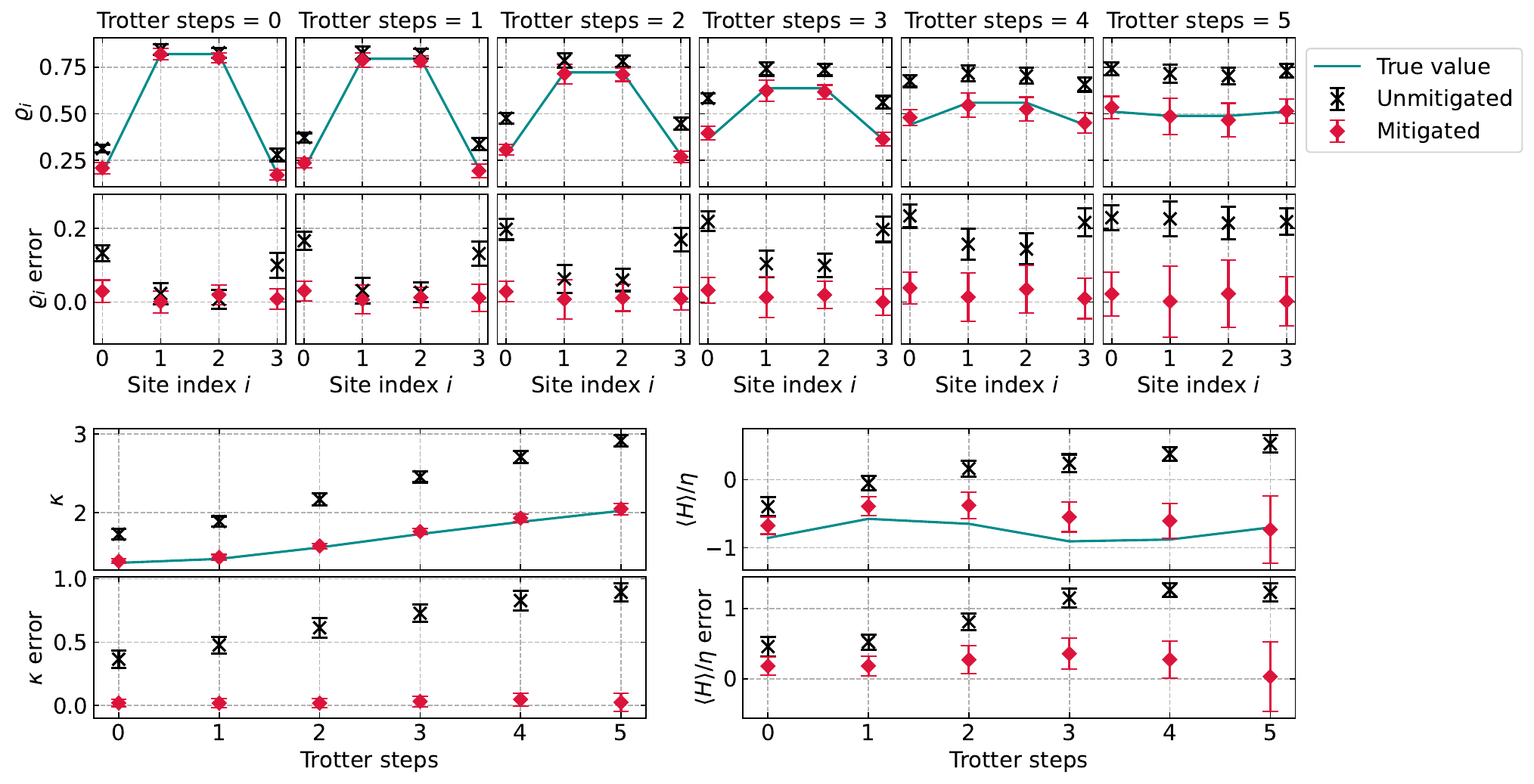}
\caption[Prediction of local properties in the four-site 1D Fermi--Hubbard model undergoing Trotterized time evolution.]{Prediction of local properties in the four-site 1D Fermi--Hubbard model ($U/t = 4$), undergoing Trotterized time evolution (step size $\delta t = 0.2$). The noise model is that of the Google Sycamore Rainbow processor~\cite{arute2020hartree,arute2020observation}, implemented within the Cirq QVM~\cite{isakov2021simulations,cirq}. The size of the noisy circuit grows systematically with the number of Trotter steps. We use spin-adapted matchgate shadows, taking $T = 9.6 \times 10^5$ samples. (Top) Charge density $\varrho_i$ at each site $i \in [L]$. (Bottom left) Charge spread $\kappa$. (Bottom right) Energy per electron $\langle H \rangle / \eta$.}
\label{fig:fh_qvm}
\end{figure}

For each target state $U_{\mathrm{prep}}(R)\ket{0^n}$, we collect $T = 9.6 \times 10^5$ spin-adapted matchgate shadow samples. In Figure~\ref{fig:fh_qvm}, we plot the Trotterized time evolution of charge density throughout the chain, as well as the charge spread
\begin{equation}
    \kappa = \sum_{i \in [L]} \l| i - (L - 1)/2 \r| \varrho_i,
\end{equation}
which quantifies how the density spreads away from the center of the chain. These quantities are only one-body observables, so as an example two-body observable we also plot the energy per electron, $\langle H \rangle / \eta$.

Because Assumptions~\ref{assumption_1} no longer hold, we no longer have the guarantees of Theorem~\ref{thm:main_theorem} and we do not observe an arbitrary amount of error mitigation. We see that as the circuit size grows, so too do the prediction error and uncertainty. This behavior is a reflection of the noise assumptions being increasingly violated. Nonetheless, our results still show a substantial amount of noise reduction, and overall we maintain the qualitative features of the dynamics compared to the unmitigated protocol.

\subsection{Qubit systems}\label{subsec:pauli_numerics}

Next, we study the application of symmetry-adjusted classical shadows to subsystem-symmetrized Pauli shadows, to predict one- and two-body qubit observables in the presence of noise.

\subsubsection{Readout noise}\label{subsec:qubit_readout_error}

For our first demonstration, we simulate random matrix product states (MPS) with maximum bond dimension $\chi \leq n$, lying in the $m = 0$ symmetry sector of $M$. We use the definition of a random MPS from Refs.~\cite{garnerone2010typicality,garnerone2010statistical}. Numerically, we implement all MPS calculations using the open-source software ITensor~\cite{fishman2022itensor}, which can guarantee the correct symmetry sector using efficient tensor-network representations. Within such representations, it is straightforward to apply random local Clifford gates and $\SWAP$ gates, and to sample measurements in the computational basis.

\begin{figure}
\includegraphics[scale=0.5]{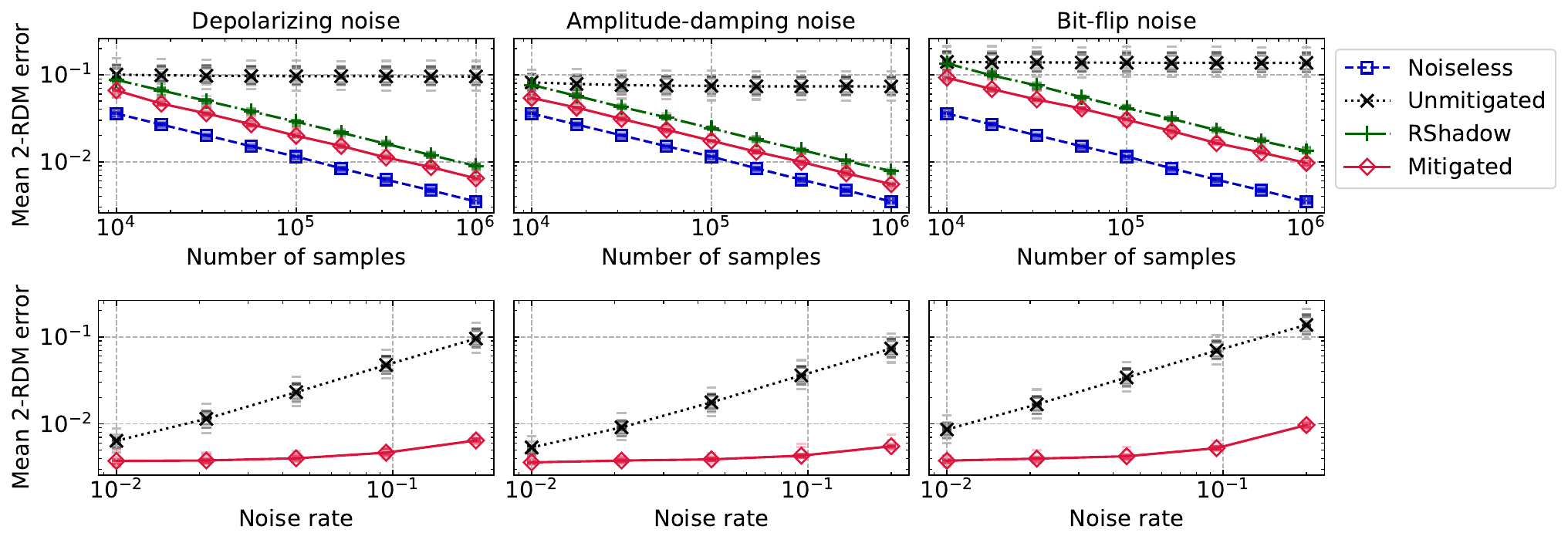}
\caption[Average estimation error of the $2$-RDM over all two-qubit subsystems, reconstructed from (subsystem-symmetrized) Pauli shadows.]{Average estimation error of the $2$-RDM over all two-qubit subsystems, reconstructed from (subsystem-symmetrized) Pauli shadows. Error is quantified by the spectral-norm difference between the estimated and true $2$-RDM for each subsystem, taking the mean over all $\binom{n}{2}$ subsystems. We simulate the protocol on a collection of 20 random eight-qubit matrix product states, with maximum bond dimension $\chi \leq n$ and fixed $Z$ magnetization, $m = 0$. Because $m = 0$ implies $s_1 = 0$, we simulate a system of nine qubits with the ancilla prepared in $\ket{0}$. (Top) Scaling of estimation error with the total number of samples $T$, fixing the noise rate to $p = 0.2$ for all noise models. (Bottom) Scaling of the estimation error with the noise rate $p$, fixing the total number of samples to $T = 10^6$.}
\label{fig:random_mps}
\end{figure}

Unlike fermions, qubits are not symmetrized, so their $2$-RDMs
\begin{equation}
    {}^2 D_{ij} = \tr_{[n] \setminus \{ i, j \}} \rho
\end{equation}
generally differ between different two-qubit subsystems. Our accuracy metric here is therefore the mean $2$-RDM error over all pairs of qubits:
\begin{equation}
    \epsilon = \frac{1}{\binom{n}{2}} \sum_{i<j} \| {{}^2 \hat{D}_{ij}(T)} - {{}^2 D_{ij}} \|_\infty.
\end{equation}
From subsystem-symmetrized Pauli shadows $\hat{\rho}(T)$ of size $T$, we reconstruct the qubit $2$-RDMs by estimating all one- and two-local Pauli expectation values and forming the $4 \times 4$ matrices
\begin{equation}
    {}^2 \hat{D}_{ij}(T) = \frac{1}{4} \sum_{W, W' \in \Pauli(1)} \tr\l( W_i W'_j \hat{\rho}(T) \r) W \otimes W'.
\end{equation}

The results are shown in Figure~\ref{fig:random_mps}. Our conclusions here are entirely parallel to those of Figure~\ref{fig:random_slater}, and we refer the reader to its corresponding discussion. We note here that this simple demonstration also validates the subsystem-symmetrized Pauli shadows protocol and our use of the ancilla trick for Pauli shadows (recall that the random MPS have vanishing symmetry value, $m = s_1 = 0$).

Our next set of numerical experiments are performed on the ground state of an antiferromagnetic XXZ Heisenberg chain with open boundary conditions:
\begin{equation}
    H = J \sum_{i \in [n-1]} \l( X_i X_{i+1} + Y_i Y_{i+1} + \Delta Z_i Z_{i+1} \r).
\end{equation}
Throughout, we set units such that $J = 1$ and consider an anisotropy of $\Delta = 1.5$. This Hamiltonian has the $\U(1)$ symmetry described in Section~\ref{subsec:qubit_symmetry}, and in particular the ground state obeys $m = 0$ (assuming the number of spins $n$ is even). We find the ground state via the density-matrix renormalization group (DMRG) algorithm~\cite{white1992density}, represented as an MPS;~therefore we apply the same classical simulation algorithms as before. Although $m = 0$ implies a vanishing conserved quantity for the one-body subspace, $s_1 = 0$, we do not employ the ancilla technique for these simulations because we will only be interested in strictly two-body observables (for which $s_2 = m^2 - n \neq 0$).

\begin{figure}
\centering
\includegraphics[scale=0.5]{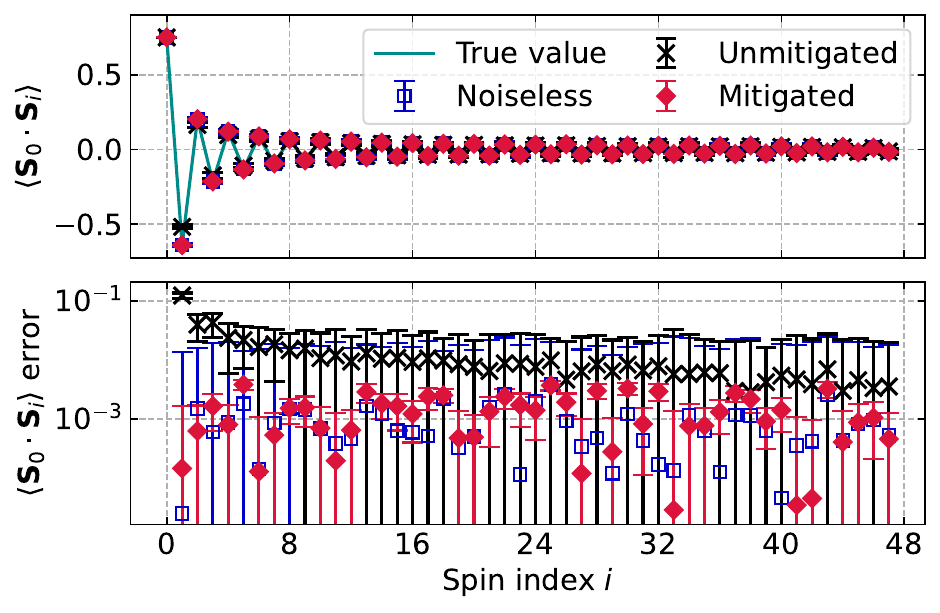}
\caption[Estimation of spin--spin correlations $\langle \bm{S}_0 \cdot \bm{S}_i \rangle$ in the ground state of a $48$-spin XXZ chain.]{Estimation of spin--spin correlations $\langle \bm{S}_0 \cdot \bm{S}_i \rangle$ in the ground state of a $48$-spin XXZ chain ($m = 0$), fixing the first spin. We average over $T = 10^6$ samples collected by subsystem-symmetrized Pauli shadows. The noise model is single-qubit readout errors with bit-flip probability $p = 5\%$.}
\label{fig:xxz_correlation}
\end{figure}

As a first demonstration, in Figure~\ref{fig:xxz_correlation} we plot the mitigation of spin--spin correlation functions in a chain of length $n = 48$, fixing one of the spins to the end of the chain. The noise model is set to a single-qubit bit-flip channel with flip rate $p = 0.05$. The correlation between spins $i$ and $j$ is defined as the expectation value of the operator $\bm{S}_i \cdot \bm{S}_j$, where
\begin{equation}
    \bm{S}_i = \frac{1}{2} \begin{pmatrix}
    X_i \\ Y_i \\ Z_i
    \end{pmatrix}.
\end{equation}

\begin{figure}
\centering
\includegraphics[scale=0.5]{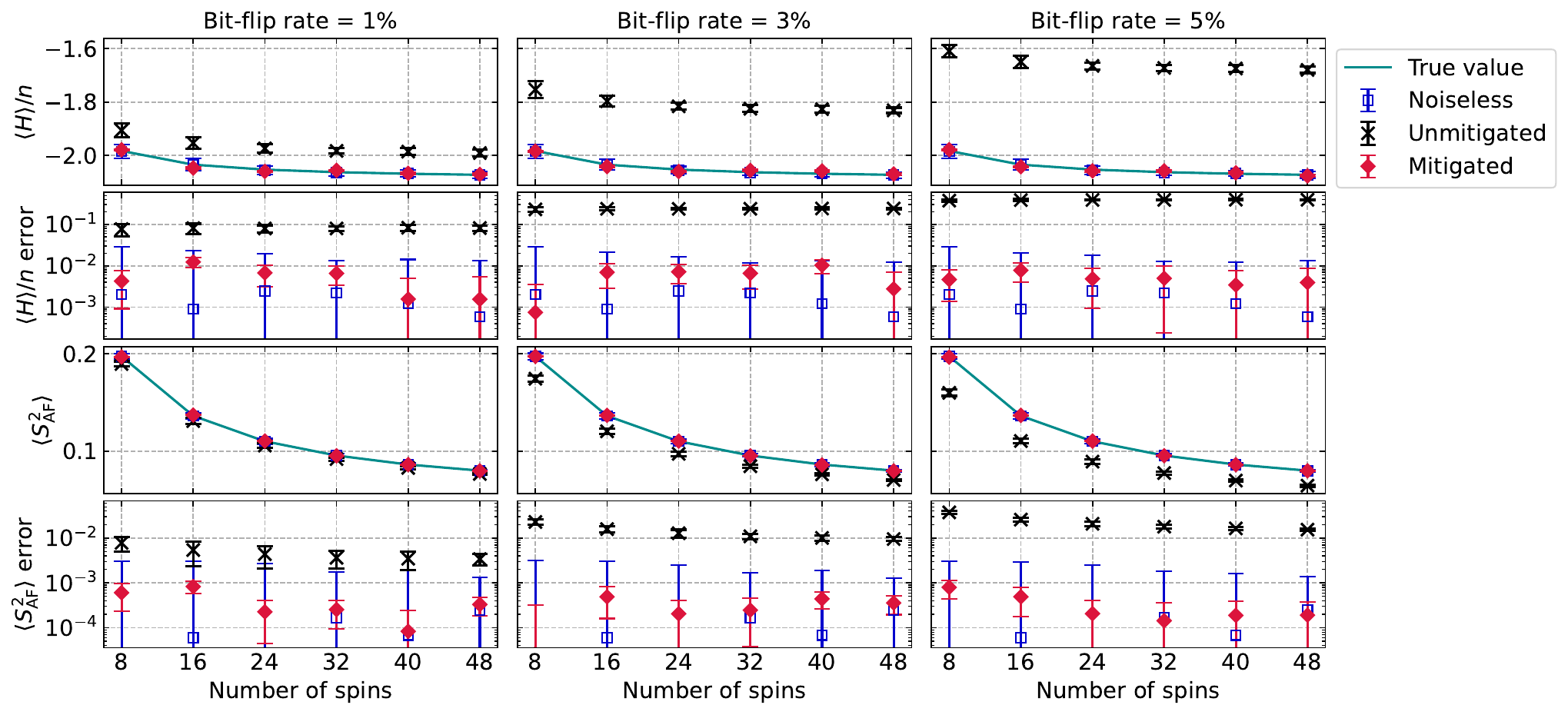}
\caption[Estimation of energy per particle and the N\'{e}el order parameter in the 1D XXZ ground state.]{Estimation of energy per particle $\langle H \rangle / n$ (top) and the N\'{e}el order parameter $\langle S_{\mathrm{AF}}^2 \rangle$ (bottom) in the 1D XXZ ground state ($m = 0$), from $T = 10^6$ subsystem-symmetrized Pauli shadows. The anisotropy of the model is $\Delta = 1.5$. The noise model is single-qubit bit flip with probabilities $p = 1\%$, $3\%$, and $5\%$.}
\label{fig:xxz_S2_and_energy}
\end{figure}

Then in Figure~\ref{fig:xxz_S2_and_energy} we show the mitigation of macroscopic observables at different system sizes and bit-flip rates. The top two rows of plots show the estimation of energy per spin $\langle H \rangle / n$, while the bottom two rows show the estimation of a N\'{e}el order parameter,
\begin{equation}
    \langle S_{\mathrm{AF}}^2 \rangle = \frac{1}{n^2} \sum_{i,j \in [n]} (-1)^{i+j} \langle \bm{S}_i \cdot \bm{S}_j \rangle,
\end{equation}
which quantifies antiferromagnetic correlations throughout the chain. For these experiments, the number of samples taken is $T = 10^6$. Overall, we draw conclusions parallel to those of Figure~\ref{fig:fh_energy}. Namely, the results validate our theory for a range of observables, noise rates, and system sizes.

\subsubsection{QVM noise model}\label{subsec:qubit_qvm}

We now turn to simulations using the QVM noise model, taking the same XXZ Heisenberg spin chain ($\Delta = 1.5$ and $n = 8$) as our testbed system. Similar to our numerical experiments with the Fermi--Hubbard model, we simulate Trotter circuits of the XXZ model starting from a product state within the symmetry sector of $m = 0$. Again, we will only be interested in strictly two-local observables so we do not employ the ancilla trick here either.

Our initial state is a N\'{e}el-ordered product state, $\ket{01010101} = \prod_{j \text{ odd}} X_j \ket{0^n}$. Defining $H_{\mathrm{even}}$ and $H_{\mathrm{odd}}$ as the terms in $H$ with $i$ even and odd, respectively, a single Trotter step is given by
\begin{equation}
    U_{\mathrm{Trot}} = e^{-\i H_{\mathrm{odd}} \delta t} e^{-\i H_{\mathrm{even}} \delta t},
\end{equation}
where we take the step size to be $\delta t = 0.2$. Hence, the full state-preparation circuit for $R$ steps is
\begin{equation}
    U_{\mathrm{prep}}(R) = U_{\mathrm{Trot}}^R \prod_{j \text{ odd}} X_j,
\end{equation}
which is then compiled into the native gate set. For each $R$, we collect $T = 4.8 \times 10^5$ samples using subsystem-symmetrized Pauli shadows. Because the initial state is a simple basis state, we only display results for $R \in \{1, \ldots, 5\}$ for these studies. In line with our Fermi--Hubbard simulations on the QVM, we perform QAA 
here as well, averaging over twelve different assignments of the same $2 \times 4$ qubits;~see \cref{subsec:qaa} for details.

\begin{figure}
\centering
\includegraphics[scale=0.5]{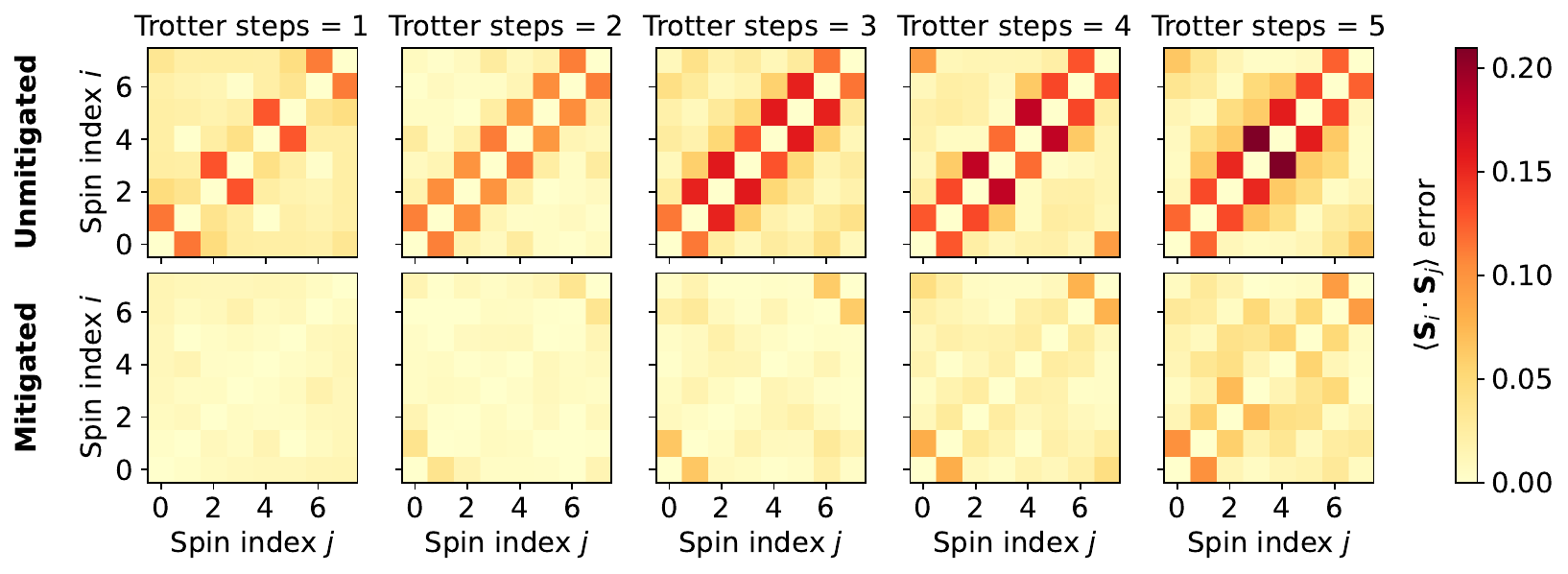}\\
\includegraphics[scale=0.5]{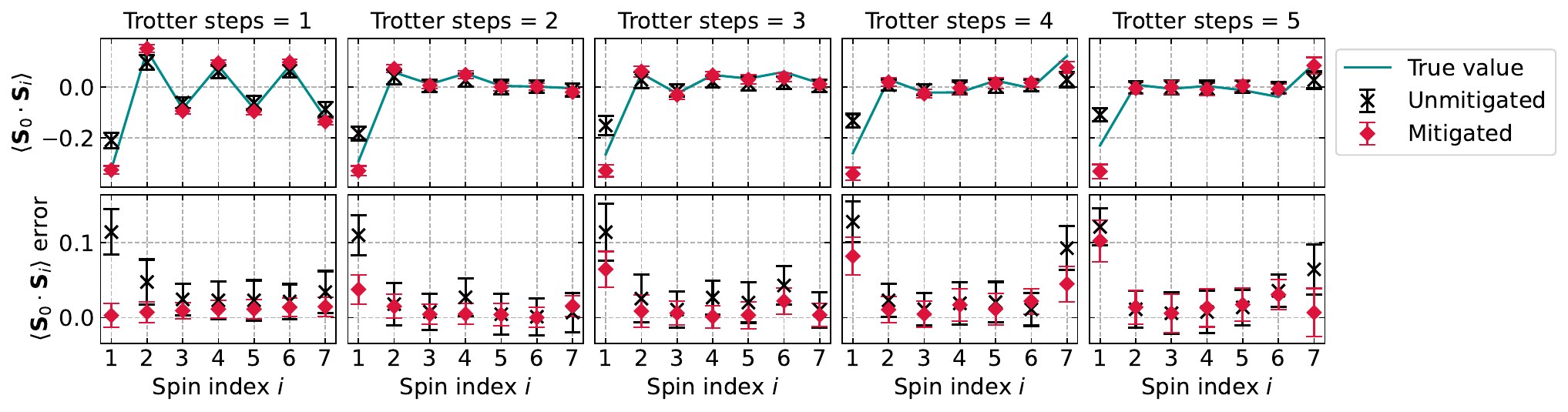}
\caption[Prediction of spin--spin correlations between qubits in the XXZ chain undergoing Trotterized time evolution.]{Prediction of spin--spin correlations between qubits in the XXZ chain ($\Delta = 1.5$), undergoing Trotterized time evolution ($\delta t = 0.2$). The noise model is the superconducting-hardware model implemented within the QVM. We take $T = 4.8 \times 10^5$ subsystem-symmetrized Pauli shadows to estimate the observables. (Top two) Heatmaps of estimation errors for $\langle \bm{S}_i \cdot \bm{S}_j \rangle$, under unmitigated versus our symmetry-enabled mitigated postprocessing. (Bottom two) Plots of the $\langle \bm{S}_0 \cdot \bm{S}_i \rangle$ correlation functions, to show further details of a particular section of the heatmaps.}
\label{fig:xxz_qvm_corr}
\end{figure}

First, we compute the spin--spin correlations for all qubit pairs $(i, j)$ throughout the chain. We plot the prediction errors of these correlation functions in Figure~\ref{fig:xxz_qvm_corr}, with the unmitigated data in the first row and mitigated data in the second row. We observe that, while the shallower circuits are well handled by symmetry-adjusted classical shadows, the mitigation power diminishes as the circuit grows deeper. To examine this effect closer, we plot in the bottom two rows of Figure~\ref{fig:xxz_qvm_corr} the correlation functions between the first spin and the rest of the chain. We see that the $\langle \bm{S}_0 \cdot \bm{S}_{1} \rangle$ errors are particularly dominant due to the magnitude of its true value. Although the absolute error is only marginally improved, we interpret the qualitative behavior as being more faithfully recovered compared to the unmitigated data.

\begin{figure}
\centering
\includegraphics[scale=0.5]{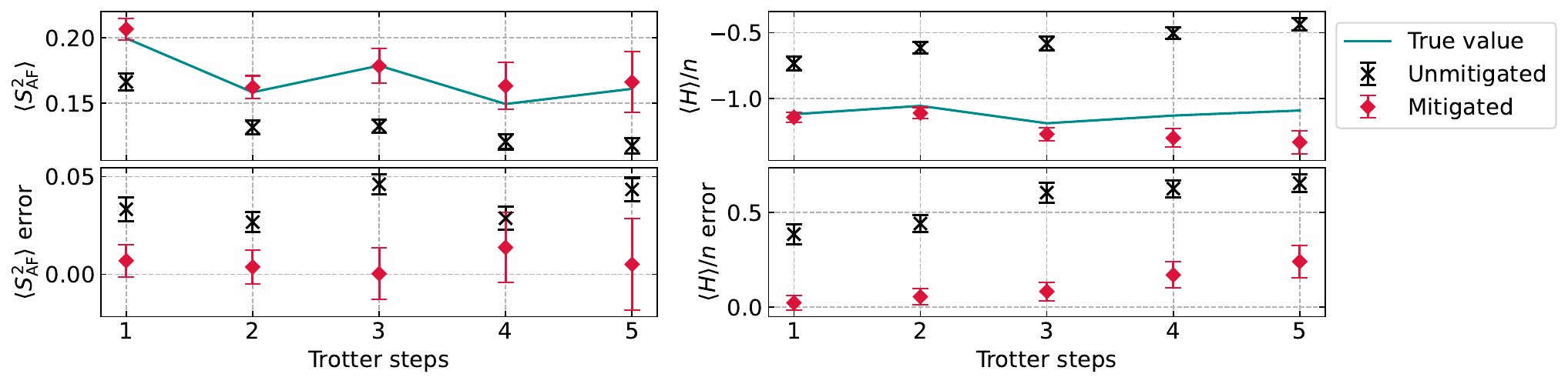}
\caption[Prediction of macroscopic observables in the Trotterized XXZ model under the QVM noise model simulations.]{Prediction of macroscopic observables in the Trotterized XXZ model under the QVM noise model simulations, with $T = 4.8 \times 10^5$ subsystem-symmetrized Pauli shadows. (Left) N\'{e}el order parameter $\langle S_{\mathrm{AF}}^2 \rangle$. (Right) Energy per spin $\langle H \rangle / n$.}
\label{fig:xxz_qvm_props}
\end{figure}

Next, we consider macroscopic observables in Figure~\ref{fig:xxz_qvm_props}, the N\'{e}el order parameter $\langle S_{\mathrm{AF}}^2 \rangle$ and energy per spin $\langle H \rangle / n$. Again we see general trends similar to the other QVM simulations:~the mitigated results are in closer qualitative agreement with the true values than the unmitigated data, at the cost of larger uncertainty bars, and without arbitrary amounts of error mitigation. Symmetry adjustment consistently reduces the absolute error compared to the unmitigated data, although we note that some of the energy estimates are still a few standard deviations away from the true value.

\section{Discussion}\label{sec:discussion}

In this chapter, we have introduced symmetry-adjusted classical shadows, a QEM protocol applicable to quantum systems with known symmetries. Our approach builds on the highly successful classical-shadow tomography~\cite{huang2020predicting,paini2021estimating}, modifying the classically computed linear-inversion step according to symmetry information in the presence of noise. Because our strategy is performed in postprocessing on the noisy measurement data, it allows for straightforward combinations with other QEM strategies. As opposed to prior related works~\cite{karalekas2020quantum,chen2021robust,koh2022classical,van2022model,arrasmith2023development}, the main advantage of our approach is the use of the entire noisy circuit, thereby bypassing the need for calibration experiments and accounting for errors in state preparation. Meanwhile, in contrast with other symmetry-based strategies~\cite{bonet2018low,mcardle2019error,cai2021quantum,jnane2023quantum}, we require no additional quantum resources, utilize finer-grained symmetry information, and can easily take advantage of a wider range of symmetries (e.g., particle number as opposed to only parity conservation).

Overall, our findings reveal that as a low-cost scheme, symmetry-adjusted classical shadows by itself is already potent for practical error mitigation. Our analytical results guarantee the accuracy of prediction under readout noise assumptions. Even when these assumptions are violated in practice, we expect these results to still provide intuition regarding the mitigation behavior. Indeed, this expectation is validated by our numerical experiments with superconducting-qubit noise models on the Cirq QVM~\cite{cirq,isakov2021simulations}. From these simulations, we have observed substantial quantitative improvement when the cumulative circuit noise is sufficiently weak, and qualitative improvements across all experiments performed.

Along the way, we have developed a number of ancillary results that may also be of independent interest. Of note are (1)~the subsystem-symmetrized Pauli shadows, which uniformly symmetrizes the irreps of the local Clifford group among subsystems;~(2)~a new circuit compilation scheme for fermionic Gaussian unitaries, which treats Majorana modes on a more natural footing to improve two-qubit gate parallelization;~and (3)~symmetry-adapted matchgate shadows, which uses block-diagonal transformations within spin sectors to reduce the size of the random matchgate circuits. We expect that these techniques will find broader applicability in quantum simulation beyond the scope of this work.

A number of pertinent open questions and future directions remain. For simplicity of the protocol, and because of the examples that we focused on, we restricted attention to multiplicity-free groups. However, tools to generalize to non-multiplicity-free groups already exist, and in the context of character randomized benchmarking~\cite{helsen2019new} such an extension has been developed successfully~\cite{claes2021character}. It would therefore be useful to extend our ideas similarly, and investigate what effect (if any) multiplicities have on symmetry-adjusted classical shadows.

Regarding the protocols considered, we have focused on local observable estimation in systems with global $\U(1)$ symmetry. However, it is worth noting that the $n$-qubit Clifford group possesses only one nontrivial irrep, making it essentially compatible with any symmetry. Because its shadow norm is exponentially large for local observables, it is an unfavorable choice for typical quantum-simulation applications. One wonders whether this desirable universality of its irrep can nonetheless be harnessed, analogous to our construction of subsystem-symmetrized Pauli shadows. We posit that global $\SU(2)$/$\Cl(1)$ control~\cite{van2022hardware}, or single-fermion $\U(n)$ basis rotations~\cite{low2022classical}, would be particularly promising groups to investigate. Alternatively, one may consider different classes of symmetries, such as local (rather than global) symmetries.

One key advantage of symmetry adjustment is its flexibility, allowing for easy integration with other error-mitigation strategies. Investigating this interplay is a clear target for future work. Particularly valuable would be other techniques to massage the circuit noise into approximately satisfying Assumptions~\ref{assumption_1}, for instance by randomized compiling~\cite{wallman2016noise}. From our usage of QAA~\cite{arute2020observation} in the numerical experiments, we have already shown heuristically that the mere choice of qubit assignments appears to have such an effect.

Indeed, the reliance on such assumptions for rigorous guarantees may be viewed as a limitation of this work. While our numerical results are encouraging, it behooves one to seek a more comprehensive error analysis applicable to a wider range of noise models. For example, while gate-dependent errors are particularly detrimental to our method, they have been closely studied in the context of randomized benchmarking~\cite{proctor2017randomized,wallman2018randomized,carignan2018randomized,merkel2021randomized}. The tools developed therein may be valuable to this setting as well. Establishing a better understanding here may also inspire extensions to surpass the limitations of the current theory. We leave such goals to future work.

\emph{Note added.}---After the manuscript for this work appeared on the arXiv preprint server~\cite{zhao2023group}, two related works~\cite{wu2023error,brieger2023stability} subsequently appeared. The former develops a calibration estimator equivalent to our Eq.~\eqref{eq:fermion_RSE_estimator}, while the latter studies gate dependence analytically (as opposed to our numerical study). Thus the formulation and analyses of symmetry-adjusted classical shadows remain original to our manuscript.


\newpage
\section*{Appendix}

\section{Error analysis}\label{sec:error_analysis}

Here we provide the proof for Theorem~\ref{thm:main_theorem} from the main text, restated below for convenience.

\begin{reptheorem}{thm:main_theorem}[Restated from main text]
Fix accuracy and confidence parameters $\epsilon, \delta \in (0, 1)$. Let $O_1, \ldots, O_L$ be a collection of observables, each supported on an irrep of $\mathcal{U} : G \to \U(\mathcal{L}(\mathcal{H}))$ as $O_j \in V_\lambda$ for $\lambda \in R' \subseteq R_G$. Let $S_\lambda \in V_\lambda$ be a symmetry operator for each $\lambda \in R'$, for which the ideal values $s_\lambda = \tr(S_\lambda \rho)$ of the target state $\rho$ are known \emph{a priori}. Suppose that each noisy unitary satisfies Assumptions~\ref{assumption_1}, $\widetilde{\mathcal{U}}_g = \mathcal{E} \mathcal{U}_g$, and define the quantities
\begin{align}
    F_{Z,R'}(\mathcal{E}) &\coloneqq \min_{\lambda \in R'} \frac{\tr(\mathcal{E} \mathcal{M}_Z \Pi_\lambda)}{\tr(\mathcal{M}_Z \Pi_\lambda)},\\
    \sigma^2 &\coloneqq \max_{1 \leq j \leq L, \lambda \in R'}\l\{ \V[\hat{o}_j], \V\l[\frac{\hat{s}_\lambda}{s_\lambda} \r] \r\}.
\end{align}
Then, a (noisy) classical shadow $\hat{\rho}(T)$ of size
\begin{equation}
    T = \O\l( \frac{\log((L + |R'|)/\delta)}{F_{Z,R'}(\mathcal{E})^2 \epsilon^2} \sigma^2 \r)
\end{equation}
can be used to construct error-mitigated estimates
\begin{equation}
    \hat{o}_j^{\mathrm{EM}}(T) \coloneqq \frac{\tr(O_j \hat{\rho}(T))}{\tr(S_\lambda \hat{\rho}(T)) / s_\lambda}
\end{equation}
which obey
\begin{equation}
    |\hat{o}_j^{\mathrm{EM}}(T) - \tr(O_j\rho)| \leq (\|O_j\|_\infty + 1) \epsilon + \O(\|O_j\|_\infty \epsilon^2)
\end{equation}
for all $1 \leq j \leq L$, with success probability at least $1 - \delta$.
\end{reptheorem}

\begin{proof}
Let $ \hat{\rho}_1, \ldots, \hat{\rho}_T $ be the $T$ noisy classical shadows. Construct the mean of these snapshots,
\begin{equation}
    \hat{\rho}(T) = \frac{1}{T} \sum_{\ell=1}^T \hat{\rho}_\ell.
\end{equation}
(It is straightforward to replace this by a median-of-means estimator if necessary.) In expectation we have $\E[\hat{\rho}(T)] = \widetilde{\rho}$, where the effective noisy state can be described as
\begin{equation}
    \widetilde{\rho} = \mathcal{M}^{-1} \l(\widetilde{\mathcal{M}} (\rho)\r).
\end{equation}
Let $O \in V_\lambda$, with symmetry $s_\lambda = \tr(S_\lambda \rho)$ in the same irrep. Define estimates of the noisy expectation values using $\hat{\rho}(T)$:
\begin{align}
    \bar{X} &= \tr(O \hat{\rho}(T))\\
    \bar{Y} &= \tr\l(\frac{S_\lambda}{s_\lambda} \hat{\rho}(T)\r).
\end{align}
In expectation, these random variables obey $\E[\bar{X}] = \tr(O \widetilde{\rho})$ and $\E[\bar{Y}] = \tr(S_\lambda \widetilde{\rho}) / s_\lambda = \widetilde{f}_\lambda / f_\lambda$. Therefore as established from the main text, we have
\begin{equation}
	r(\E[\bar{X}], \E[\bar{Y}]) = \frac{\tr(O \widetilde{\rho})}{\tr(S_\lambda \widetilde{\rho}) / s_\lambda} = \tr(O \rho),
\end{equation}
where we have defined the function $r(x, y) \coloneqq x/y$. From a finite number of samples, however, we can only construct $\hat{o}^{\mathrm{EM}}(T) \coloneqq r(\bar{X}, \bar{Y})$, which is generally a biased estimator since $ \E[r(\bar{X}, \bar{Y})] \neq r(\E[\bar{X}], \E[\bar{Y}]) $.

To quantify the estimation error, we employ Taylor's remainder theorem:~expanding $r(x, y)$ to first order about a point $ (x_0, y_0) $, we have
\begin{equation}\label{eq:taylor_thm}
\begin{split}
	r(x, y) &= r(x_0, y_0) + \partial_x r(x_0, y_0) (x - x_0) + \partial_y r(x_0, y_0) (y - y_0) + h_1(x, y),
\end{split}
\end{equation}
where the remainder term is
\begin{equation}\label{eq:taylor_remainder}
\begin{split}
    h_1(x, y) &= \frac{1}{2!} \l[ \partial_x^2 r(a, b) (x - x_0)^2 + \partial_y^2 r(a, b) (y - y_0)^2 + 2 \partial_{xy} r(a, b) (x - x_0) (y - y_0) \r]
\end{split}
\end{equation}
for some points $a \in [\min(x, x_0), \max(x, x_0)]$ and $b \in [\min(y, y_0), \max(y, y_0)]$. The relevant partial derivatives of $r(x, y)$ are enumerated below:
\begin{align}
    \partial_x r(x,y) &= 1/y,\\
    \partial_y r(x,y) &= -x/y^2,\\
    \partial_x^2 r(x,y) &= 0,\\
    \partial_{xy} r(x,y) &= -1/y^2,\\
    \partial_y^2 r(x,y) &= 2x/y^3.
\end{align}

Suppose $T$ is large enough such that (with high probability) the estimation error of all \emph{noisy} observables are uniformly bounded by some $\tilde{\epsilon} \in (0, 1)$:
\begin{align}
    \l|\bar{X} - \E[\bar{X}]\r| &\leq \tilde{\epsilon}, \label{eq:X_eps}\\
    |\bar{Y} - \E[\bar{Y}]| &\leq \tilde{\epsilon}. \label{eq:Y_eps}
\end{align}
This is achieved by standard classical shadow arguments, which we will elaborate on later. For now, assuming these error bounds hold, we rearrange Eq.~\eqref{eq:taylor_thm}, set $(x, y) = (\bar{X}, \bar{Y})$ and $(x_0, y_0) = (\E[\bar{X}], \E[\bar{Y}])$, and apply a triangle inequality to obtain
\begin{equation}\label{eq:1st_error_bound}
\begin{split}
    \l| r(\bar{X}, \bar{Y}) - r(\E[\bar{X}], \E[\bar{Y}]) \r| &\leq \frac{1}{|{\E[\bar{Y}]}|} \tilde{\epsilon} + \frac{|{\E[\bar{X}]}|}{\E[\bar{Y}]^2} \tilde{\epsilon}\\
    &\quad + |h_1(\bar{X}, \bar{Y})|.
\end{split}
\end{equation}

To proceed with this error bound, we make the following observations. First, note that
\begin{equation}
    \E[\bar{Y}] = \frac{\widetilde{f}_\lambda}{f_\lambda} = \frac{\tr(\mathcal{M}_Z \mathcal{E} \Pi_\lambda)}{\tr(\mathcal{M}_Z \Pi_\lambda)} \in [0, 1],
\end{equation}
which we will denote by $\xi_\lambda$. We assume that that noise channel $\mathcal{E}$ is such that $\xi_\lambda > 0$, as otherwise the quantity $r(\E[\bar{X}], \E[\bar{Y}])$ diverges. Next, because $\E[\bar{X}]/\E[\bar{Y}] = \tr(O \rho)$, we have the bound
\begin{equation}
    \l| \frac{\E[\bar{X}]}{\E[\bar{Y}]} \r| \leq \| O \|_\infty.
\end{equation}
Thus Eq.~\eqref{eq:1st_error_bound} becomes
\begin{equation}\label{eq:2nd_error_bound}
\begin{split}
    \l| r(\bar{X}, \bar{Y}) - r(\E[\bar{X}], \E[\bar{Y}]) \r| &\leq \frac{1}{\xi_\lambda} \tilde{\epsilon} + \frac{\|O\|_\infty}{\xi_\lambda} \tilde{\epsilon}\\
    &\quad + |h_1(\bar{X}, \bar{Y})|.
\end{split}
\end{equation}

We can bound the remainder term $|h_1(\bar{X}, \bar{Y})|$ as follows. Applying a triangle inequality to Eq.~\eqref{eq:taylor_remainder} yields
\begin{equation}
    | h_1(\bar{X}, \bar{Y}) | \leq \frac{|a|}{|b|^3} \tilde{\epsilon}^2 + \frac{1}{b^2} \tilde{\epsilon}^2.
\end{equation}
Taylor's remainder theorem tells us that the value of $a$ (resp., $b$) lies between $\bar{X}$ and $\E[\bar{X}]$ (resp., $\bar{Y}$ and $\E[\bar{Y}]$), which we know are at most $\tilde{\epsilon}$ apart. We can therefore bound
\begin{equation}
\begin{split}
    |a| &\leq \max\{ |{\E[\bar{X}]}|, |\bar{X}| \}\\
    &\leq \max\{ |{\E[\bar{X}]}|, |{\E[\bar{X}]} + \tilde{\epsilon}|, |{\E[\bar{X}]} - \tilde{\epsilon}| \}\\
    &\leq |{\E[\bar{X}]}| + \tilde{\epsilon}\\
    &\leq \xi_\lambda \|O\|_\infty + \tilde{\epsilon}.
\end{split}
\end{equation}
Similarly for $b$, using the fact that $\E[\bar{Y}] > 0$,
\begin{equation}
\begin{split}
    |b| &\geq \min\{ |{\E[\bar{Y}]}|, |\bar{Y}| \}\\
    &\geq \min\{ \E[\bar{Y}], \E[\bar{Y}] + \tilde{\epsilon}, |{\E[\bar{Y}]} - \tilde{\epsilon}| \}\\
    &= \min\{ \xi_\lambda, |\xi_\lambda - \tilde{\epsilon}| \}.
\end{split}
\end{equation}
If $\tilde{\epsilon} < \xi_\lambda$, then $|b| \geq \xi_\lambda - \tilde{\epsilon} > 0$ always holds. We will see later that this condition is always justified;~for now, we will just suppose that this lower bound on $|b|$ holds. Then the remainder obeys
\begin{equation}\label{eq:remainder_bound}
    | h_1(\bar{X}, \bar{Y}) | \leq \frac{1}{(\xi_\lambda - \tilde{\epsilon})^2} \l( \frac{\xi_\lambda \|O\|_\infty + \tilde{\epsilon}}{\xi_\lambda - \tilde{\epsilon}} + 1 \r) \tilde{\epsilon}^2.
\end{equation}

Combining Eqs.~\eqref{eq:2nd_error_bound} and \eqref{eq:remainder_bound}, we arrive at
\begin{align}
    \l| \hat{o}^{\mathrm{EM}}(T) - \tr(O \rho) \r| &\leq \frac{1}{\xi_\lambda} (\|O\|_\infty + 1) \tilde{\epsilon}\\
    &\quad + \frac{1}{(\xi_\lambda - \tilde{\epsilon})^2} \l( \frac{\xi_\lambda \|O\|_\infty  + \tilde{\epsilon}}{\xi_\lambda - \tilde{\epsilon}} + 1 \r) \tilde{\epsilon}^2. \notag
\end{align}
In order to bound this error by $\O(\|O\|_\infty\epsilon)$ for some desired $\epsilon \in (0, 1)$, we can choose $\tilde{\epsilon} = \xi_\lambda \epsilon$, yielding
\begin{align}\label{eq:error_bound_full}
    \l| \hat{o}^{\mathrm{EM}}(T) - \tr(O \rho) \r| &\leq (\|O\|_\infty + 1) \epsilon\\
    &\quad + \frac{1}{(1 - \epsilon)^2} \l( \frac{\|O\|_\infty  + \epsilon}{1 - \epsilon} + 1 \r) \epsilon^2. \notag
\end{align}
Thus, by demanding $\epsilon < 1$ we ensure that the required technical condition $\tilde{\epsilon} < \xi_\lambda$ is met. Now we need to verify that the remainder term is bounded by $\O(\|O\|_\infty\epsilon^2)$, so that the $\O(\|O\|_\infty \epsilon)$ term dominates asymptotically as $\epsilon \to 0$. Indeed, as long as $\epsilon$ is bounded away from 1 then
\begin{equation}
    \frac{1}{(1 - \epsilon)^2} \l( \frac{\|O\|_\infty  + \epsilon}{1 - \epsilon} + 1 \r) = \O(\|O\|_\infty).
\end{equation}

Finally, we analyze the sample complexity required to achieve the error bound of Eq.~\eqref{eq:error_bound_full}. In Eqs.~\eqref{eq:X_eps} and \eqref{eq:Y_eps} we required that the number of samples $T$ be such that the shot noise of $\bar{X}$ and $\bar{Y}$ are at most $\tilde{\epsilon}$. These random variables correspond to the observables $\{O_j\}_{j=1}^L \cup \{S_\lambda/s_\lambda\}_{\lambda \in R'}$. Standard classical-shadows theory informs us that
\begin{equation}
    T = \O\l( \frac{\log((L + |R'|)/\delta)}{\tilde{\epsilon}^2} \max_{\substack{1 \leq j \leq L\\\lambda \in R'}}\l\{ \V[\hat{o}_j], \V\l[\frac{\hat{s}_\lambda}{s_\lambda} \r] \r\} \r)
    \end{equation}
suffices to accomplish this task (with probability at least $1 - \delta$)~\cite{huang2020predicting}. Then, setting $\tilde{\epsilon} = \min_{\lambda \in R'} \xi_\lambda \epsilon$ ensures that $\tilde{\epsilon}$ is small enough for Eq.~\eqref{eq:error_bound_full} to apply to all target observables.
\end{proof}

\section{Subsystem-symmetrized Pauli shadows}\label{sec:pauli_shadows_appendix}

Here we prove the properties of the subsystem-symmetrized Pauli shadows introduced in Section~\ref{subsec:ss-pauli}. In \cref{subsec:ss-pauli_irreps} we identify the irreps, and in \cref{subsec:ss-pauli_variance} we bound the variance of observables under this protocol, particularly the symmetry operators $S_k$ obtained from $M = \sum_{i \in [n]} Z_i$.

\subsection{Irreducible representations}\label{subsec:ss-pauli_irreps}

Recall that the subsystem-symmetrized local Clifford group is the direct product
\begin{equation}
    \SymCl{n} \coloneqq \Sym(n) \times \Cl(1)^{\otimes n},
\end{equation}
where $\pi \in \Sym(n)$ acts on $(\C^2)^{\otimes n}$ as
\begin{equation}
    S_\pi \ket{b} = \ket{\pi^{-1}(b)}.
\end{equation}
For shorthand, we write $\pi(b)$ for the $n$-bit string $b_{\pi(0)} \cdots b_{\pi(n-1)}$. It is clear that the adjoint representation $\mathcal{U}_{(\pi, C)}$ block diagonalizes into subspaces spanned by $k$-local Pauli operators:
\begin{equation}
    V_k \coloneqq \spn\{ P \in \Pauli(n) : |P| = k \}.
\end{equation}
This can be seen from the fact that neither single-qubit nor $\SWAP$ gates can change the operator locality;~however, $\SWAP$ gates \emph{can} map between equally sized subsystems on which the operator nontrivially acts. What remains is to show that each of these subspaces is irreducible.

First, we define the twirling map.

\begin{definition}\label{def:twirl}
    Let $\phi : G \to \U(V)$ be a unitary representation of a compact group $G$ on a vector space $V$, and let $\Phi : G \to \U(\mathcal{L}(V))$ be its adjoint action, i.e., $\Phi_g(\cdot) = \phi_g (\cdot) \phi_g^\dagger$. The $t$-fold twirl by $\Phi$ is defined as
    \begin{equation}
    \mathcal{T}_{t,\Phi} \coloneqq \E_{g \sim G} \Phi_g^{\otimes t},
    \end{equation}
    which is a linear map on $\mathcal{L}(V)^{\otimes t}$.
\end{definition}

Twirls have a number of convenient properties, mostly arising from the fact that $\Phi$ is a group homomorphism. For example, they are $G$-invariant from the left and right:
\begin{equation}
    \Phi_h^{\otimes t} \circ \mathcal{T}_{t,\Phi} = \mathcal{T}_{t,\Phi} = \mathcal{T}_{t,\Phi} \circ \Phi_h^{\otimes t}
\end{equation}
for all $h \in G$. This furthermore implies that they are in fact projectors:
\begin{equation}
    \mathcal{T}_{t,\Phi}^2 = \mathcal{T}_{t,\Phi}.
\end{equation}
The study of twirls also allows us to determine the irreducible representations of a group. This can be seen by the following well-known result for multiplicity-free groups, which for completeness we provide a self-contained proof of at the end of this subsection.

\begin{proposition}\label{prop:twirl_irreps}
    Let $G$, $V$, $\phi$, and $\Phi$ be as in Definition~\ref{def:twirl}. For any $X \in \mathcal{L}(V)$, the $1$-twirl of $X$ by $\Phi$ takes the form
    \begin{equation}
    \mathcal{T}_{1,\Phi}(X) = \sum_{\lambda \in R_G} \frac{\tr(X \Pi_\lambda)}{\tr(\Pi_\lambda)} \Pi_\lambda
    \end{equation}
    if and only if $\phi$ decomposes irreducibly as $V = \bigoplus_{\lambda \in R_G} V_\lambda$, where $\Pi_\lambda$ is the orthogonal projector onto $V_\lambda$.
\end{proposition}

Our strategy for determining the irreps of $G = \SymCl{n}$ is therefore to directly compute $\mathcal{T}_{1,\Phi}$, from which we can infer the irreps from its block-diagonal structure. To use Proposition~\ref{prop:twirl_irreps}, we will take $\phi$ as the unitary channel $\mathcal{U}$, so that $V = \mathcal{L}(\mathcal{H})$ and $\Phi(\cdot) = \mathcal{U}(\cdot)\mathcal{U}^\dagger$ (note that this is a superchannel). For technical reasons, it will be easier to first compute $\mathcal{T}_{2, \mathcal{U}}$, from which the desired twirl $\mathcal{T}_{1,\Phi}$ can be evaluated. The relation between these two twirls is given by the following lemma.

\begin{lemma}\label{lem:1twirl_to_2twirl}
    Let $\mathcal{U} : G \to \U(\mathcal{L}(\mathcal{H}))$ be a unitary representation and $\Phi : G \to \U(\mathcal{L}(\mathcal{L}(\mathcal{H})))$ its adjoint representation, i.e., $\Phi_g(\mathcal{A}) = \mathcal{U}_g^\dagger \mathcal{A} \mathcal{U}_g$ for any superoperator $\mathcal{A}$. The $1$-twirl by $\Phi$ can be computed from the $2$-twirl by $\mathcal{U}$ as
    \begin{equation}\label{eq:1-twirl_from_2-twirl}
    \mathcal{T}_{1,\Phi}(\mathcal{A})(X) = \tr_2\l[ \mathcal{T}_{2,\mathcal{U}} (\mathcal{A}) (\I \otimes X) \r],
    \end{equation}
    for all $\mathcal{A} \in \mathcal{L}(\mathcal{L}(\mathcal{H}))$ and $X \in \mathcal{L}(\mathcal{H})$. Here, the domain of $\mathcal{T}_{2,\mathcal{U}}$ is understood with respect to the isomorphism $\mathcal{L}(\mathcal{L}(\mathcal{H})) \cong \mathcal{L}(\mathcal{H})^{\otimes 2}$, given by
    \begin{equation}\label{eq:2-twirl_isomorphism}
    \vop{A}{B} \cong A \otimes B^\dagger.
    \end{equation}
\end{lemma}

\begin{proof}
    Write $\mathcal{A} = \sum_{i,j \in [d^2]} \mathcal{A}_{ij} \vop{B_i}{B_j}$, where $\mathcal{A}_{ij} \in \C$ and $\{B_i\}_{i \in [d^2]}$ is an orthonormal operator basis. By a direct calculation:
    \begin{align}
    \mathcal{T}_{1,\Phi}(\mathcal{A})(X) &= \E_{g \sim G} \Phi_g(\mathcal{A})(X) \notag\\
    &= \E_{g \sim G} \mathcal{U}_g^\dagger \mathcal{A} \mathcal{U}_g(X) = \E_{g \sim G} \mathcal{U}_g^\dagger \mathcal{A}(U_g X U_g^\dagger) \notag\\
    &= \E_{g \sim G} \mathcal{U}_g^\dagger \sum_{i,j \in [d^2]} \mathcal{A}_{ij} \kket{B_i} \vip{B_j}{U_g X U_g^\dagger} \notag\\
    &= \E_{g \sim G} \sum_{i,j \in [d^2]} \mathcal{A}_{ij} U_g^\dagger B_i U_g \tr(B_j^\dagger U_g X U_g^\dagger) \notag\\
    &= \E_{g \sim G} \tr_2\l[ \sum_{i,j \in [d^2]} \mathcal{A}_{ij} ( U_g^\dagger B_i U_g )   \otimes ( U_g^\dagger B_j^\dagger U_g X ) \r] \notag\\
    &= \tr_2\l[ \sum_{i,j \in [d^2]} \mathcal{A}_{ij} \mathcal{T}_{2,\mathcal{U}}(B_i \otimes B_j^\dagger)(\I \otimes X) \r] \notag\\
    &= \tr_2\l[ \mathcal{T}_{2,\mathcal{U}} \l( \sum_{i,j \in [d^2]} \mathcal{A}_{ij} \vop{B_i}{B_j} \r) (\I \otimes X) \r] \notag\\
    &= \tr_2\l[ \mathcal{T}_{2,\mathcal{U}} (\mathcal{A}) (\I \otimes X) \r].
    \end{align}
\end{proof}

Before we can compute $\mathcal{T}_{2, \mathcal{U}}$ for the subsystem-symmetrized local Clifford group, we will need a small result about the group orbit of a $k$-local Pauli operator $P$ under the action of $\mathcal{U}$. The orbit is defined as
\begin{equation}
    G \cdot P \coloneqq \{ \mathcal{U}_g(P) \mid g \in G \}.
\end{equation}
This will help us determine how the twirl acts on Pauli operators, which as an basis is used to compute the matrix elements of $\mathcal{T}_{2, \mathcal{U}}$. To this end, we define an orthonormal basis of $k$-local Pauli operators,
\begin{equation}
    \mathcal{B}_k \coloneqq \{ P \in \Pauli(n) / \sqrt{d} : |P| = k \},
\end{equation}
which contains $|\mathcal{B}_k| = 3^k \binom{n}{k}$ elements.

\begin{lemma}\label{lem:sspauli_orbit}
   Let $G = \SymCl{n}$. The orbit $G \cdot P$ of any $P \in \mathcal{B}_k$ is equal to $\pm \mathcal{B}_k$, i.e., the set of all signed $k$-local Pauli operators.
\end{lemma}

\begin{proof}
    Let the nontrivial support of $P$ be $I \subseteq [n]$, $|I| = k$. For each (normalized) Pauli matrix acting on subsystem $I$, its orbit by all single-qubit Clifford gates is $\pm \{X, Y, Z\} / \sqrt{2}$. Meanwhile, the trivial factors $\I/\sqrt{2}$ acting on $[n] \setminus I$ are invariant to any unitary transformation. Therefore $\Cl(1)^{\otimes n} \cdot P$ is the set of all normalized Pauli operators acting nontrivially only on the qubits in $I$ (with both signs $\pm 1$).
    
    Then, conjugation by $S_\pi$ for arbitrary $\pi \in \Sym(n)$ permutes the $k$ nontrivial factors of $P$ among the $n$ qubits. The orbit over all permutations yields all possible $\binom{n}{k}$ supports. Taking the direct product of both these Clifford- and symmetric-group actions therefore yields all $k$-local Pauli operators, with prefactors $\pm 1$.
\end{proof}

We are now ready to compute the $2$-fold twirl by $\mathcal{U}$. We comment that the high-level proof structure of this lemma is inspired by that of Ref.~\cite[Section~IV A 1]{wan2023matchgate}.

\begin{lemma}\label{lem:2-twirl_pauli}
    Let $\mathcal{U} : G \to \U(\mathcal{L}(\mathcal{H}))$ be the unitary representation of $G = \SymCl{n}$, defined by $\mathcal{U}_{(\pi, C)}(\rho) = S_\pi C \rho C^\dagger S_\pi^\dagger$. Its $2$-fold twirl is the projector
    \begin{equation}\label{eq:2-twirl-sspauli}
    \mathcal{T}_{2,\mathcal{U}} = \sum_{k=0}^n \vop{\Sigma_k^{(2)}}{\Sigma_k^{(2)}},
    \end{equation}
    where $\kket{\Sigma_k^{(2)}} \in \mathcal{L}(\mathcal{H})^{\otimes 2}$ is defined as
    \begin{equation}\label{eq:2-twirl-basis}
    \kket{\Sigma_k^{(2)}} = \frac{1}{\sqrt{3^k \binom{n}{k}}} \sum_{P \in \mathcal{B}_k} \kket{P} \kket{P}.
    \end{equation}
\end{lemma}

\begin{proof}
    First, we will establish that for any two basis Pauli operators $P \neq Q$, we have $\mathcal{T}_{2,\mathcal{U}} \kket{P}\kket{Q} = 0$. Thus we only need to consider basis elements of $\mathcal{L}(\mathcal{H})^{\otimes 2}$ of the form $\kket{P}\kket{P}$. Next, we will show that $\mathcal{T}_{2,\mathcal{U}} \kket{P}\kket{P} = \mathcal{T}_{2,\mathcal{U}} \kket{P'}\kket{P'}$ whenever $|P| = |P'|$. Finally, using these two properties we can derive Eq.~\eqref{eq:2-twirl-sspauli}.

    Fix the basis of Pauli operators such that $P, Q \in \bigcup_{0 \leq k \leq n} \mathcal{B}_k$. If $P \neq Q$, then there exists at least one qubit $i \in [n]$ on which $P$ and $Q$ act as a different Pauli matrix. Hence there always exists some $W \in \Pauli(1)$ which anticommutes with one and commutes with the other, e.g., $\mathcal{W}_i \kket{P} = W_i P W_i^\dagger = -P$ and $\mathcal{W}_i \kket{Q} = Q$. Note that $\mathcal{W}_i$ is equal to $\mathcal{U}_{(e, W_i)}$ where $e \in \Sym(n)$ is the identity permutation. Thus using the property that $\mathcal{T}_{2,\mathcal{U}} \circ \mathcal{W}_i^{\otimes 2} = \mathcal{T}_{2,\mathcal{U}}$, we have
    \begin{equation}\label{eq:PQ_vanish}
    \begin{split}
    \mathcal{T}_{2,\mathcal{U}} \kket{P}\kket{Q} &= \mathcal{T}_{2,\mathcal{U}} \mathcal{W}_i^{\otimes 2} \kket{P}\kket{Q}\\
    &= -\mathcal{T}_{2,\mathcal{U}} \kket{P}\kket{Q},
    \end{split}
    \end{equation}
    implying that $\mathcal{T}_{2,\mathcal{U}} \kket{P}\kket{Q} = 0$.

    Now let $P, P'$ be $k$-local Pauli operators for any $k$. If they act nontrivially on different subsets $I, I' \subseteq [n]$ of qubits, then let $\pi \in \Sym(n)$ be a permutation that maps $I$ to $I'$. Given this permutation, if they act as different Pauli matrices on their new shared support $I'$, then furthermore let $C_i \in \Cl(1)$ for $i \in I'$ be Clifford gates that map each one to the other. Writing $C = \bigotimes_{i \in I'} C_i \otimes \I^{\otimes(n-k)}$, this transformation acts as $\mathcal{U}_{(\pi, C)}^{\otimes 2} \kket{P}\kket{P} = \kket{P'}\kket{P'}$, which implies that
    \begin{equation}\label{eq:PP_same}
    \begin{split}
    \mathcal{T}_{2,\mathcal{U}} \kket{P}\kket{P} &= \mathcal{T}_{2,\mathcal{U}} \mathcal{U}_{(\pi, C)}^{\otimes 2} \kket{P}\kket{P}\\
    &= \mathcal{T}_{2,\mathcal{U}} \kket{P'}\kket{P'}.
    \end{split}
    \end{equation}

    We are now ready to derive Eq.~\eqref{eq:2-twirl-sspauli}. As established by Eq.~\eqref{eq:PQ_vanish}, we only need to expand the 2-fold twirl in the basis of $\kket{P}\kket{P}$:
    \begin{equation}
    \begin{split}
    \mathcal{T}_{2,\mathcal{U}} &= \sum_{P, P' \in \Pauli(n)} \bbra{P}\bbra{P} \mathcal{T}_{2,\mathcal{U}} \kket{P'} \kket{P'} \kket{P} \vop{P}{P'} \bbra{P'}\\
    &= \sum_{k=0}^n \sum_{P, P' \in \mathcal{B}_k} \bbra{P}\bbra{P} \mathcal{T}_{2,\mathcal{U}} \kket{P'} \kket{P'} \kket{P} \vop{P}{P'} \bbra{P'},
    \end{split}
    \end{equation}
    where the second simplification is due to the fact that $\mathcal{U}_{(\pi, C)}$ preserves Pauli locality, hence $\bbra{P}\bbra{P} \mathcal{T}_{2,\mathcal{U}} \kket{P'} \kket{P'} = 0$ whenever $|P| \neq |P'|$. Now we invoke Eq.~\eqref{eq:PP_same}, which implies that $\bbra{P}\bbra{P} \mathcal{T}_{2,\mathcal{U}} \kket{P'} \kket{P'} = c_k'$ for all $P, P' \in \mathcal{B}_k$ (i.e., the matrix element does not depend on the particular choice of $P, P'$). Hence
    \begin{equation}
    \begin{split}
    \mathcal{T}_{2,\mathcal{U}} &= \sum_{k=0}^n c_k' \sum_{P, P' \in \mathcal{B}_k} \kket{P}\kket{P} \bbra{P'}\bbra{P'}\\
    &= \sum_{k=0}^n c_k \vop{\Sigma_k^{(2)}}{\Sigma_k^{(2)}},
    \end{split}
    \end{equation}
    where we have rescaled $c_k = c_k' 3^k \binom{n}{k}$ to account for the normalization of $\kket{\Sigma_k^{(2)}}$.

    Finally, we show that all $c_k = 1$ by proving that $\mathcal{T}_{2,\mathcal{U}} \kket{\Sigma_k^{(2)}} = \kket{\Sigma_k^{(2)}}$.  Expand the expression:
    \begin{equation}\label{eq:T2_basis}
    \mathcal{T}_{2,\mathcal{U}} \kket{\Sigma_k^{(2)}} = \frac{1}{|G|} \sum_{g \in G} \frac{1}{\sqrt{|\mathcal{B}_k|}} \sum_{P \in \mathcal{B}_k} \mathcal{U}_{g}\kket{P} \otimes \mathcal{U}_{g}\kket{P}.
    \end{equation}
    We first compute the average over the group for some fixed $P$. By Lemma~\ref{lem:sspauli_orbit}, we know that the orbit $G \cdot P = \{ \mathcal{U}_g\kket{P} \mid g \in G \} = \pm\mathcal{B}_k$. Thus
    \begin{equation}
    \sum_{g \in G} \mathcal{U}_{g}\kket{P} \otimes \mathcal{U}_{g}\kket{P} = 2 \frac{|G|}{|G \cdot P|} \sum_{Q \in \mathcal{B}_k} (\pm 1)^2 \kket{Q} \kket{Q},
    \end{equation}
    where the factor of 2 is due to the fact that for each $Q \in \mathcal{B}_k$, both $\pm Q \in G \cdot P$, and the factor of $|G|/|G \cdot P|$ takes care of double counting when summing over all elements of $G$. Noting that $|G \cdot P| = 2|\mathcal{B}_k|$, we can plug this result into Eq.~\eqref{eq:T2_basis} to find that
    \begin{equation}
    \begin{split}
    \mathcal{T}_{2,\mathcal{U}} \kket{\Sigma_k^{(2)}} &= \frac{1}{|\mathcal{B}_k|^{3/2}} \sum_{P \in \mathcal{B}_k} \sum_{Q \in \mathcal{B}_k} \kket{Q}\kket{Q}\\
    &= \frac{1}{\sqrt{|\mathcal{B}_k|}} \sum_{Q \in \mathcal{B}_k} \kket{Q}\kket{Q}\\
    &= \kket{\Sigma_k^{(2)}},
    \end{split}
    \end{equation}
    as desired.
\end{proof}

We are now ready to prove the main result of this section:~the irreps of $\SymCl{n}$ are labeled by the Pauli weights $k \in \{0, 1, \ldots, n\}$. The proof structure is as follows:~from the expression for $\mathcal{T}_{2, \mathcal{U}}$ from Lemma~\ref{lem:2-twirl_pauli}, we can compute $\mathcal{T}_{1, \Phi}$ by using Lemma~\ref{lem:1twirl_to_2twirl}. Then by examining $\mathcal{T}_{1, \Phi}$, we use Proposition~\ref{prop:twirl_irreps} to infer the irreps.

\begin{theorem}
    The representation $\mathcal{U} : \SymCl{n} \to \U(\mathcal{L}(\mathcal{H}))$, defined by $\mathcal{U}_{(\pi, C)}(\rho) = S_\pi C \rho C^\dagger S_\pi^\dagger$, decomposes into the irreps
    \begin{equation}
    V_k = \spn(\mathcal{B}_k), \quad k \in \{0, 1, \ldots, n\}.
    \end{equation}
\end{theorem}

\begin{proof}
    From Lemma~\ref{lem:2-twirl_pauli}, we have
    \begin{equation}
    \mathcal{T}_{2,\mathcal{U}} = \sum_{k=0}^n \vop{\Sigma_k^{(2)}}{\Sigma_k^{(2)}},
    \end{equation}
    where $\kket{\Sigma_k^{(2)}}$ is defined in Eq.~\eqref{eq:2-twirl-basis}. Using Lemma~\ref{lem:1twirl_to_2twirl}, we compute $\mathcal{T}_{1,\Phi}(\mathcal{A})$ by evaluating $\mathcal{T}_{2,\mathcal{U}}(\mathcal{A})$ for arbitrary superoperators $\mathcal{A}$. Let us express $\mathcal{A}$ in the Pauli basis:
    \begin{equation}
    \mathcal{A} = \sum_{k,\ell=0}^n \sum_{P \in \mathcal{B}_k} \sum_{Q \in \mathcal{B}_\ell} \mathcal{A}_{PQ} \vop{P}{Q}.
    \end{equation}
    Recall from Eq.~\eqref{eq:2-twirl_isomorphism} that in order to evaluate $\mathcal{T}_{2,\mathcal{U}}(\mathcal{A})$, we need $\mathcal{T}_{2,\mathcal{U}}(Q \otimes P)$ for every $P, Q$. But because $\mathcal{T}_{2,\mathcal{U}}$ projects onto symmetrized basis elements $P \otimes P$, we only have to consider the case where $P = Q$:
    \begin{align}
    \mathcal{T}_{2,\mathcal{U}}\kket{P \otimes P} &= \kket{\Sigma_k^{(2)}} \vip{\Sigma_k^{(2)}}{P \otimes P} \notag\\
    &= \kket{\Sigma_k^{(2)}} \frac{1}{\sqrt{|\mathcal{B}_k|}} \sum_{P' \in \mathcal{B}_k} \vip{P'}{P} \vip{P'}{P} \notag\\
    &= \frac{1}{\sqrt{|\mathcal{B}_k|}} \kket{\Sigma_k^{(2)}},
    \end{align}
    where $|\mathcal{B}_k| = 3^k \binom{n}{k}$.

    Inserting this result into Eq.~\eqref{eq:1-twirl_from_2-twirl} yields
    \begin{align}
    \mathcal{T}_{1,\Phi}(\mathcal{A})(X) &= \tr_1\l[ \mathcal{T}_{2,\mathcal{U}}(\mathcal{A}) (X \otimes \I) \r]\\
    &= \tr_1\l[ \sum_{k=0}^n \sum_{P \in \mathcal{B}_k} \mathcal{A}_{PP} \frac{1}{\sqrt{|\mathcal{B}_k|}} \Sigma_k^{(2)} (X \otimes \I) \r] \notag\\
    &= \tr_1\l[ \sum_{k=0}^n \sum_{P \in \mathcal{B}_k} \mathcal{A}_{PP} \frac{1}{\sqrt{|\mathcal{B}_k|}} \r. \notag\\
    &\quad \times \l. \frac{1}{\sqrt{|\mathcal{B}_k|}} \sum_{P' \in \mathcal{B}_k} (P' \otimes P') (X \otimes \I) \r] \notag\\
    &= \sum_{k=0}^n \frac{\sum_{P \in \mathcal{B}_k} \mathcal{A}_{PP}}{|\mathcal{B}_k|} \sum_{P' \in \mathcal{B}_k} \tr(P' X) P'. \notag
    \end{align}
    We make a number of observations here. First, note that $\sum_{P \in \mathcal{B}_k} \mathcal{A}_{PP} = \tr(\mathcal{A} \Pi_k)$, where $\Pi_k = \sum_{P \in \mathcal{B}_k} \vop{P}{P}$.  Also, $|\mathcal{B}_k| = \tr(\Pi_k)$. Finally, the sum over $P'$ can be represented as
    \begin{equation}
    \begin{split}
    \sum_{P' \in \mathcal{B}_k} \tr(P' X) P' &= \sum_{P' \in \mathcal{B}_k} \kket{P'} \vip{P'}{X}\\
    &= \Pi_k \kket{X}.
    \end{split}
    \end{equation}
    Because this holds for all $X \in \mathcal{L}(\mathcal{H})$, we can say that
    \begin{equation}
    \mathcal{T}_{1,\Phi}(\mathcal{A}) = \sum_{k=0}^n \frac{\tr(\mathcal{A} \Pi_k)}{\tr(\Pi_k)} \Pi_k.
    \end{equation}
    By Proposition~\ref{prop:twirl_irreps}, we know that the twirl has this expression if and only if the irreducible subspaces of $\mathcal{U}$ are $V_k = \spn(\mathcal{B}_k)$.
\end{proof}

Finally, we close this subsection with the deferred proof of the well-known result Proposition~\ref{prop:twirl_irreps}, for completeness.

\begin{proof}[Proof (of Proposition~\ref{prop:twirl_irreps})]
    For the forward direction, suppose $\phi = \bigoplus_{\lambda \in R_G} \phi^{(\lambda)}$ where each $\phi^{(\lambda)} : G \to \U(V_\lambda)$ is irreducible. (This is guaranteed by Maschke's theorem, and generalizes to the Peter--Weyl theorem for compact groups~\cite{fulton2004representation}.) Because $\mathcal{T}_{1,\Phi}(X)$ commutes with all $\phi_g$, they are simultaneously block diagonal, so $\mathcal{T}_{1,\Phi}(X) = \bigoplus_{\lambda \in R_G} \mathcal{T}_{1,\Phi^{(\lambda)}}(X)$ where $\Phi^{(\lambda)}(\cdot) = \phi^{(\lambda)}(\cdot)\phi^{(\lambda) \dagger}$. Because $\phi^{(\lambda)}$ is irreducible, by Schur's lemma $\mathcal{T}_{1,\Phi^{(\lambda)}}(X)$ must be a multiple of the identity on $V_\lambda$. Therefore
    \begin{equation}
    \mathcal{T}_{1,\Phi}(X) = \bigoplus_{\lambda \in R_G} c_\lambda(X) \I_{V_\lambda} = \sum_{\lambda \in R_G} c_\lambda(X) \Pi_\lambda.
    \end{equation}
    From the orthogonality of projectors $\Pi_\lambda \Pi_{\lambda'} = \delta_{\lambda \lambda'} \Pi_\lambda$, the scalar $c_\lambda(X)$ is determined by
    \begin{equation}
    \begin{split}
    c_\lambda(X) \tr(\Pi_\lambda) &= \tr(\Pi_\lambda \mathcal{T}_{1,\Phi}(X))\\
    &= \E_{g \sim G} \tr\l( \phi_g^\dagger \Pi_\lambda \phi_g X \r)\\
    &= \E_{g \sim G} \tr(\Pi_\lambda X) = \tr(\Pi_\lambda X).
    \end{split}
    \end{equation}

    For the reverse direction, suppose the twirl takes the form
    \begin{equation}
    \mathcal{T}_{1,\Phi}(X) = \bigoplus_{\lambda} \frac{\tr(X \Pi_\lambda)}{\tr(\Pi_\lambda)} \I_{V_\lambda},
    \end{equation}
    where we denote each block by $\mathcal{T}_{1,\Phi^{(\lambda)}}(X)$. Again because $\mathcal{T}_{1,\Phi}(X)$ and $\phi_g$ commute, the matrix $\phi_g$ is block diagonal in the subspaces $V_\lambda$ for all $g \in G$. We need to show that each block $\phi^{(\lambda)}$ is irreducible.

    Recall that $\phi^{(\lambda)}$ is irreducible if the only subspaces $W \subseteq V_\lambda$ for which $\phi^{(\lambda)}_G(W) \subseteq W$ are $W = \{0\}$ or $W = V_\lambda$. Indeed, let $W \subseteq V_\lambda$ be a subspace such that for any $\ket{v} \in W$ and $g \in G$, $\phi^{(\lambda)}_g\ket{v} \in W$. Suppose there exists a vector $\ket{x} \in V_\lambda$ that is orthogonal to $W$ and set $X = \op{x}{x}$. Then
    \begin{equation}\label{eq:project_W_1}
    \mathcal{T}_{1,\Phi^{(\lambda)}}(\op{x}{x}) \ket{v} = \E_{g \sim G} \phi^{(\lambda)}_g \ket{x} \ev{x}{(\phi^{(\lambda)}_g)^\dagger}{v} = 0
    \end{equation}
    because all $(\phi^{(\lambda)}_g)^\dagger\ket{v} \in W$. However, from $\mathcal{T}_{1,\Phi^{(\lambda)}}(X) = c_\lambda(X) \I_{V_\lambda}$ we see that also
    \begin{equation}\label{eq:project_W_2}
    \mathcal{T}_{1,\Phi}(\op{x}{x})_\lambda \ket{v} = \frac{\ip{x}{x}}{\dim V_\lambda} \ket{v}.
    \end{equation}
    Supposing $\ket{x} \neq 0$, we see that $\ket{v} = 0$ is the only possible element of $W$ to satisfy Eqs.~\eqref{eq:project_W_1} and \eqref{eq:project_W_2} simultaneously. Hence $W = \{0\}$. Otherwise, $\ket{x} = 0$ is the only element of $V_\lambda$ orthogonal to $W$, implying that there is in fact no nontrivial subspace orthogonal to $W$. Thus $W = V_\lambda$ in this case.
\end{proof}

\subsection{Variance of symmetry operators}\label{subsec:ss-pauli_variance}

In this section, we analyze the variance associated with the symmetry operators,
\begin{align}
    \frac{S_1}{s_1} &= \frac{1}{m} \sum_{i \in [n]} Z_i,\\
    \frac{S_2}{s_2} &= \frac{2}{m^2 - n} \sum_{i < j} Z_i Z_j.
\end{align}
Because our error analysis of symmetry-adjusted classical shadows (see \cref{sec:error_analysis}) bootstraps from the variance of unmitigated estimation, we only need to compute quantities related to the noiseless protocol. In this case, the subsystem-symmetrized local Clifford group yields the same channel and variances as the standard local Clifford group because the random permutations have no effect on the twirling on computational basis states.

Specifically, using the two- and three-fold twirls we can express
\begin{equation}
    \mathcal{M}(\rho) = \tr_1\l[ \sum_{b \in \{0,1\}^n} \mathcal{T}_{2,G}\l( \op{b}{b}^{\otimes 2} \r) (\rho \otimes \I) \r]
\end{equation}
and
\begin{align}
    \V_\rho[\hat{o}] &= \tr\l[ \sum_{b \in \{0,1\}^n} \mathcal{T}_{3,G}\l( \op{b}{b}^{\otimes 3} \r) \l( \rho \otimes \mathcal{M}^{-1}(O)^{\otimes 2} \r) \r] \notag\\
    &\quad - \tr(O \rho)^2,
\end{align}
where the $t$-fold twirl by $U : G \to \U(\mathcal{H})$ is defined as $\mathcal{T}_{t,G} \coloneqq \E_{g \sim G} \mathcal{U}_g^{\otimes t}$. These expressions are the same whether we take $G = \SymCl{n}$ or $\Cl(1)^{\otimes n}$, due to the following equivalence:
\begin{equation}
\begin{split}
    \sum_{b \in \{0,1\}^n} \mathcal{T}_{t,\SymCl{n}}\l( \op{b}{b}^{\otimes t} \r) &= \sum_{b \in \{0,1\}^n} \E_{(\pi, C) \sim \SymCl{n}} \l[ \l( C^\dagger S_\pi^\dagger \op{b}{b} S_\pi C \r{)^{\otimes t}} \r]\\
    &= \E_{C \sim \Cl(1)^{\otimes n}} \l[ (C^\dagger)^{\otimes t} \E_{\pi \sim \Sym(n)} \l[ \sum_{b \in \{0,1\}^n} \op{\pi(b)}{\pi(b)}^{\otimes t} \r] C^{\otimes t} \r]\\
    &= \E_{C \sim \Cl(1)^{\otimes n}} \l[ (C^\dagger)^{\otimes t} \sum_{b \in \{0,1\}^n} \op{b}{b}^{\otimes t} C^{\otimes t} \r]\\
    &= \sum_{b \in \{0,1\}^n} \mathcal{T}_{t,\Cl(1)^{\otimes n}}\l( \op{b}{b}^{\otimes t} \r).
\end{split}
\end{equation}
The third equality follows due to the fact that permutations are bijections, hence each $\sum_{b \in \{0,1\}^n} \op{\pi(b)}{\pi(b)}^{\otimes t}$ is just a reordering of the terms in $\sum_{b \in \{0,1\}^n} \op{b}{b}^{\otimes t}$.

As an immediate consequence, we see that the variance of observables under subsystem symmetrization are exactly the same as with standard Pauli shadows. For the rest of this section, we will explicitly compute the variance of $S_k$ using known Haar-averaging formulas over the Clifford group~\cite[Eqs.~(S35) and (S36)]{huang2020predicting}:
\begin{align}
    \E_{U \sim \Cl(1)} U^\dagger \op{x}{x} U \ev{x}{U A U^\dagger}{x} &= \frac{A + \tr(A) \I}{6}\\
    \E_{U \sim \Cl(1)} U^\dagger \op{x}{x} U \ev{x}{U B_0 U^\dagger}{x} \ev{x}{U C_0 U^\dagger}{x} &= \frac{\tr(B_0 C_0) \I + B_0 C_0 + C_0 B_0}{24},
\end{align}
for all unit vectors $\ket{x} \in \C^{2}$ and Hermitian matrices $A, B_0, C_0 \in \C^{2 \times 2}$, with $\tr B_0  = \tr C_0  = 0$. The extension to $\Cl(1)^{\otimes n}$ follows by linearity and statistical independence. We first apply these formulas to $S_1$ to compute its variance. Writing $C = \bigotimes_{i \in [n]} C_i$ and $\ket{b} = \bigotimes_{i \in [n]} \ket{b_i}$, we have
\begin{equation}\label{eq:Es1^2}
\begin{split}
    \E[\hat{s}_1^2] &= \tr\l( \rho \sum_{b \in \{0, 1\}^n} \E_{C \sim \Cl(1)^{\otimes n}} C^\dagger \op{b}{b} C \ev{b}{C \mathcal{M}^{-1}(S_1) C^\dagger}{b}^2 \r)\\
    &= \tr\l( \rho \sum_{b \in \{0, 1\}^n} \E_{C \sim \Cl(1)^{\otimes n}} C^\dagger \op{b}{b} C \times 3^2 \sum_{i, j \in [n]} \ev{b}{C Z_i C^\dagger}{b} \ev{b}{C Z_j C^\dagger}{b} \r)\\
    &= 9 \tr\l( \rho \sum_{i \in [n]} \sum_{b_i \in \{0, 1\}} \E_{C_i \sim \Cl(1)} C_i^\dagger \op{b_i}{b_i} C_i \ev{b_i}{C_i Z_i C_i^\dagger}{b_i}^2 \r)\\
    &\quad + 9 \times 2 \tr\l( \rho \sum_{i < j} \sum_{b_i, b_j \in \{0, 1\}} \E_{C_i, C_j \sim \Cl(1)} C_i^\dagger \op{b_i}{b_i} C_i \otimes C_j^\dagger \op{b_j}{b_j} C_j \ev{b_i}{C_i Z_i C_i^\dagger}{b_i} \ev{b_j}{C_j Z_j C_j^\dagger}{b_j} \r)\\
    &= 9 \tr\l( \rho \sum_{i \in [n]} \frac{\I}{3} \r) + 18 \tr\l( \rho \sum_{i < j} \frac{Z_i}{3} \frac{Z_j}{3} \r)\\
    &= 3n + \tr(S_2 \rho).
\end{split}
\end{equation}
Thus the variance is
\begin{equation}
\begin{split}
    \V_\rho\l[\frac{\hat{s}_1}{s_1}\r] &= \frac{1}{s_1^2} \l( \E[\hat{s}_1^2] - \E[\hat{s}_1]^2 \r)\\
    &= \frac{1}{m^2} \l( 3n + \tr(S_2 \rho) - \tr(S_1 \rho)^2 \r).
\end{split}
\end{equation}
If $\rho$ is the ideal state with symmetries $\tr(S_1 \rho) = m$ and $\tr(S_2 \rho) = m^2 - n$, then $\V_\rho[\hat{s}_1/s_1] = 2n/m^2$. On the other hand, if we make the noisy replacement $\rho \to \widetilde{\rho} = \mathcal{M}^{-1} \widetilde{\mathcal{M}}(\rho)$, then we can obtain a bound
\begin{equation}\label{eq:s1_noisy_var}
\begin{split}
    \V_{\widetilde{\rho}}\l[ \frac{\hat{s}_1}{s_1} \r] &= \frac{1}{m^2} \l( 3n + F_{Z,2}(m^2 - n) - (F_{Z,1} m)^2 \r)\\
    &\leq \frac{2n}{m^2} + 1.
\end{split}
\end{equation}

Next we compute the variance of estimating $S_2$. Analogous to the calculation presented in Eq.~\eqref{eq:Es1^2}, we expand $\ev{b}{C \mathcal{M}^{-1}(S_2) C^\dagger}{b}^2$ and group terms based on the overlapping of indices:
\begin{equation}\label{eq:S_2_squared}
\begin{split}
    \ev{b}{C \mathcal{M}^{-1}(S_2) C^\dagger}{b}^2 &= (3^2 \times 2)^2 \sum_{i < j} \sum_{k < l} \ev{b}{C Z_i Z_j C^\dagger}{b} \ev{b}{C Z_k Z_l C^\dagger}{b}\\
    &= (3^2 \times 2)^2 \l( \sum_{(i = k) < (j = l)} + \sum_{\substack{(i = k) < j, l \\ j \neq l}} + \sum_{k < (i = l) < j} \r.\\
    &\quad \l. + \sum_{i < (j = k) < l} + \sum_{\substack{i, k < (j = l) \\ i \neq k}} + \sum_{\substack{i < j; k < l \\ k \neq i \neq l; k \neq j \neq l}} \r) \ev{b}{C Z_i Z_j C^\dagger}{b} \ev{b}{C Z_k Z_l C^\dagger}{b}.
\end{split}
\end{equation}
We now go through each summation and evaluate the expectations:
\begin{equation}
\begin{split}
    &\quad \sum_{(i = k) < (j = l)} \sum_{b_i, b_j \in \{0, 1\}} \E_{C_i, C_j \sim \Cl(1)} C_i^\dagger \op{b_i}{b_i} C_i \otimes C_j^\dagger \op{b_j}{b_j} C_j \ev{b_i}{C_i Z_i C_i^\dagger}{b_i}^2 \ev{b_j}{C_j Z_j C_j^\dagger}{b_j}^2\\
    &= \sum_{i < j} \frac{\I}{3^2} = \frac{1}{3^2} \binom{n}{2} \I,
\end{split}
\end{equation}
\begin{equation}\label{eq:i=k<j,l}
\begin{split}
    &\quad \sum_{\substack{(i = k) < j, l \\ j \neq l}} \sum_{b_i, b_j, b_l \in \{0, 1\}} \E_{C_i, C_j, C_l \sim \Cl(1)} C_i^\dagger \op{b_i}{b_i} C_i \otimes C_j^\dagger \op{b_j}{b_j} C_j \otimes C_l^\dagger \op{b_l}{b_l} C_l\\
    &\qquad\qquad\qquad\qquad\qquad\qquad\qquad\quad \times \ev{b_i}{C_i Z_i C_i^\dagger}{b_i}^2 \ev{b_j}{C_j Z_j C_j^\dagger}{b_j} \ev{b_l}{C_l Z_l C_l^\dagger}{b_l}\\
    &= 2\sum_{i < j < l} \frac{\I}{3} \frac{Z_j}{3} \frac{Z_l}{3} = \frac{2}{3^3} \sum_{i <
    j < l} Z_j Z_l,
\end{split}
\end{equation}
\begin{equation}
\begin{split}
    &\quad \sum_{k < (i = l) < j} \sum_{b_i, b_j, b_k \in \{0, 1\}} \E_{C_i, C_j, C_k \sim \Cl(1)} C_i^\dagger \op{b_i}{b_i} C_i \otimes C_j^\dagger \op{b_j}{b_j} C_j \otimes C_k^\dagger \op{b_k}{b_k} C_k\\
    &\qquad\qquad\qquad\qquad\qquad\qquad\qquad\quad \times \ev{b_i}{C_i Z_i C_i^\dagger}{b_i}^2 \ev{b_j}{C_j Z_j C_j^\dagger}{b_j} \ev{b_k}{C_k Z_k C_k^\dagger}{b_k}\\
    &= \frac{1}{3^3} \sum_{k < i < j} Z_k Z_j,
\end{split}
\end{equation}
\begin{equation}
\begin{split}
    &\quad \sum_{i < (j = k) < l} \sum_{b_i, b_j, b_l \in \{0, 1\}} \E_{C_i, C_j, C_l \sim \Cl(1)} C_i^\dagger \op{b_i}{b_i} C_i \otimes C_j^\dagger \op{b_j}{b_j} C_j \otimes C_l^\dagger \op{b_l}{b_l} C_l\\
    &\qquad\qquad\qquad\qquad\qquad\qquad\qquad\quad \times \ev{b_i}{C_i Z_i C_i^\dagger}{b_i} \ev{b_j}{C_j Z_j C_j^\dagger}{b_j}^2 \ev{b_l}{C_l Z_l C_l^\dagger}{b_l}\\
    &= \frac{1}{3^3} \sum_{i < j < l} Z_i Z_l,
\end{split}
\end{equation}
\begin{equation}\label{eq:i,k<j=l}
\begin{split}
    &\quad \sum_{\substack{i, k < (j = l) \\ i \neq k}} \sum_{b_i, b_j, b_k \in \{0, 1\}} \E_{C_i, C_j, C_k \sim \Cl(1)} C_i^\dagger \op{b_i}{b_i} C_i \otimes C_j^\dagger \op{b_j}{b_j} C_j \otimes C_k^\dagger \op{b_k}{b_k} C_k\\
    &\qquad\qquad\qquad\qquad\qquad\qquad\qquad\quad \times \ev{b_i}{C_i Z_i C_i^\dagger}{b_i} \ev{b_j}{C_j Z_j C_j^\dagger}{b_j}^2 \ev{b_k}{C_k Z_k C_k^\dagger}{b_k}\\
    &= \frac{2}{3^3} \sum_{i < k < j} Z_i Z_k,
\end{split}
\end{equation}
\begin{equation}
\begin{split}
    \sum_{\substack{i < j; k < l \\ k \neq i \neq l; k \neq j \neq l}} \bigotimes_{q \in \{i, j, k, l\}} \sum_{b_q \in \{0, 1\}} \E_{C_q \sim \Cl(1)} C_q^\dagger \op{b_q}{b_q} C_q \ev{b_q}{C_q Z_q C_q^\dagger}{b_q} &= \frac{1}{3^4} \sum_{\substack{i < j; k < l \\ k \neq i \neq l; k \neq j \neq l}} Z_i Z_j Z_k Z_l\\
    &= \frac{6}{3^4} \sum_{i < j < k < l} Z_i Z_j Z_k Z_l.
\end{split}
\end{equation}
Eqs.~\eqref{eq:i=k<j,l} to \eqref{eq:i,k<j=l} can be combined by relabeling the indices and recognizing that the resulting three-index summation has $3\binom{n}{3} = (n - 2)\binom{n}{2}$ terms of the form $Z_p Z_q$ (symmetric across the index pairs $p, q \in \{i, j, k\}$ with $p \neq q$). Hence there are only $\binom{n}{2}$ unique terms, all of which are repeated $n - 2$ times:
\begin{equation}
\begin{split}
    \frac{2}{3^3} \sum_{i < j < k} ( Z_i Z_j + Z_i Z_k + Z_j Z_k ) &= \frac{2}{3^3} (n - 2) \sum_{i < j} Z_i Z_j.
\end{split}
\end{equation}
Combining these expressions, we obtain
\begin{equation}\label{eq:S2_2nd_moment}
    \sum_{b \in \{0, 1\}^n} \E_{C \sim \Cl(1)^{\otimes n}} C^\dagger \op{b}{b} C \ev{b}{C \mathcal{M}^{-1}(S_2) C^\dagger}{b}^2 = 36 \binom{n}{2} \I + 24 (n - 2) \sum_{i < j} Z_i Z_j + 24 \sum_{i < j < k < l} Z_i Z_j Z_k Z_l.
\end{equation}

Due to the presence of the four-body term, we will need the conserved quantity associated with $M^4$:
\begin{equation}\label{eq:M4}
\begin{split}
    M^4 &= (M^2)^2 = \l( n \I + 2 \sum_{i < j} Z_i Z_j \r{)^2}\\
    &= n^2 \I + 4n \sum_{i < j} Z_i Z_j + 4 \sum_{i < j} \sum_{k < l} Z_i Z_j Z_k Z_l.
\end{split}
\end{equation}
The four-index sum here can be grouped as we did in Eq.~\eqref{eq:S_2_squared}, and the conditions can be simplified as before. Along with the fact that $Z_i^2 = \I$, a straightforward calculation reveals
\begin{equation}
    \sum_{i < j} \sum_{k < l} Z_i Z_j Z_k Z_l = \binom{n}{2} \I + 2 (n - 2) \sum_{i < j} Z_i Z_j + 6 \sum_{i < j < k < l} Z_i Z_j Z_k Z_l.
\end{equation}
Plugging this into Eq.~\eqref{eq:M4}, combined with $\tr(M^4 \rho) = m^4$ and $\tr(S_2 \rho) = m^2 - n$, we arrive at
\begin{equation}
    \tr(S_4 \rho) = 24 \tr\l( \rho \sum_{i < j < k < l} Z_i Z_j Z_k Z_l \r) = m^4 - n^2 - 4\binom{n}{2} - (6n - 8)(m^2 - n),
\end{equation}
where $S_4 \coloneqq \Pi_4(M^4) = 4! \sum_{i < j < k < l} Z_i Z_j Z_k Z_l$. Finally, applying this result to Eqs.~\eqref{eq:S2_2nd_moment}, we can compute the variance with respect to an ideal state lying in the symmetry sector:
\begin{equation}
\begin{split}
    \V_\rho\l[\frac{\hat{s}_2}{s_2}\r] &= \frac{1}{(m^2 - n)^2}\l[ 36\binom{n}{2} + 12(n - 2) \tr(S_2 \rho) + \tr(S_4 \rho) - \tr(S_2 \rho)^2 \r]\\
    &= \frac{1}{(m^2 - n)^2} \l[ 32\binom{n}{2} + (6n - 16)(m^2 - n) + m^4 - n^2 \r] - 1.
\end{split}
\end{equation}
Again making the replacement $\rho \to \widetilde{\rho}$, we instead have the following bound:
\begin{equation}\label{eq:s2_noisy_var}
\begin{split}
    \V_{\widetilde{\rho}}\l[ \frac{\hat{s}_2}{s_2} \r] &= \frac{1}{(m^2 - n)^2}\l[ 36\binom{n}{2} + 12(n - 2) F_{Z,2} \tr(S_2 \rho) + F_{Z,4} \tr(S_4 \rho) \r] - F_{Z,2}^2\\
    &\leq \frac{1}{(m^2 - n)^2} \l[ 18n(n - 1) + 12(n - 2)m^2 + m^4 +6n^2 +8m^2 \r]\\
    &= \frac{1}{(m^2 - n)^2} \l[ 6n(4n - 3) + 12nm^2 + m^2(m^2 - 16) \r].
\end{split}
\end{equation}

If $m = \Theta(1)$, which is the case in our numerical experiments of antiferromagnetic spin systems in Section~\ref{subsec:pauli_numerics}, then $\V_{\widetilde{\rho}}[ \hat{s}_2/s_2 ] = \O(1)$. Recall from Eq.~\eqref{eq:s1_noisy_var} that $\V_{\widetilde{\rho}}[ \hat{s}_1/s_1 ] = \O(n/m^2)$. Therefore the variance overhead of estimating these symmetry operators is, asymptotically,
\begin{equation}
    \max\l\{ \V_{\widetilde{\rho}}\l[ \frac{\hat{s}_1}{s_1}\r], \V_{\widetilde{\rho}}\l[ \frac{\hat{s}_2}{s_2} \r] \r\} = \O(n)
\end{equation}
when $\tr(M \rho) = m = \Theta(1)$.

Systems wherein $m$ depends on $n$ will require a case-by-case analysis, which we leave to the reader. As a pathological example, consider two different functions which are both $\Theta(\sqrt{n})$:~if $m = \alpha\sqrt{n}$ for some constant $\alpha \neq 1$, then $\V_{\widetilde{\rho}}[ \hat{s}_2/s_2 ] = \O(1)$. However, if instead $m = \sqrt{n + c}$ for some constant $c \neq 0$, then $\V_{\widetilde{\rho}}[ \hat{s}_2/s_2 ] = \O(n^2)$. In both cases, $\V_{\widetilde{\rho}}[ \hat{s}_1/s_1 ] = \O(1)$. Thus the specific form of $m = f(n)$ can drastically affect the asymptotic bounds here.

\section{Spin-adapted fermionic shadows}\label{sec:spin_adaptation}

It is well known that number-conserving fermion basis rotations which preserve spin symmetries can be block diagonalized according to the spin sectors, leading to savings in both classical and quantum resources. Here we show how to leverage spin symmetries for the broader class of fermionic Gaussian transformations, and in particular we construct a spin-adapted matchgate shadows protocol. As such, this scheme will be informationally complete only over spin-conserving observables.

Let $n_\uparrow$ and $n_\downarrow$ be the number of spin-up and spin-down fermionic modes, respectively. The total number of modes is $n = n_\uparrow + n_\downarrow$, and we order the labels such that all spin-up modes come first. Gaussian transformations which do not mix between different spin types are block diagonal,
\begin{equation}\label{eq:spin_preserving_transformation}
    Q = \begin{pmatrix}
    Q_\uparrow & 0\\
    0 & Q_\downarrow
    \end{pmatrix} \in \Orth(2n),
\end{equation}
where $Q_\sigma \in \Orth(2n_\sigma) \eqqcolon G_\sigma$ for each $\sigma \in \{\uparrow, \downarrow\}$. Let $G_{\mathrm{spin}} \coloneqq G_\uparrow \times G_\downarrow$ be the spin-adapted group, i.e., the set of all elements of the form of Eq.~\eqref{eq:spin_preserving_transformation}. In order to calculate properties of this ensemble for classical shadows, we shall use the fact that $U_Q$ is nearly equivalent to the tensor product $U_{Q_\uparrow} \otimes U_{Q_\downarrow}$, up to a factor that depends on the determinant of $Q_\uparrow$. More precisely, we have the following.

\begin{lemma}\label{lem:spin_tensor_product}
    For all block-diagonal $Q$ of the form of Eq.~\eqref{eq:spin_preserving_transformation}, the unitary can be written as
    \begin{equation}\label{eq:spin_preserving_tensor_product}
    U_Q = U_{Q_\uparrow} \otimes (P_\downarrow^{s(Q_\uparrow)} U_{Q_\downarrow})
    \end{equation}
    where
    \begin{equation}
    s(Q_\uparrow) = \begin{cases}
    1 & \text{if } \det(Q_\uparrow) = -1,\\
    0 & \text{else},
    \end{cases}
    \end{equation}
    and $P_\sigma = (-i)^{n_\sigma} \prod_{\mu=0}^{2n_\sigma - 1} c_{\mu,\sigma}$ is the parity operator on the $\sigma$-spin sector.
\end{lemma}

For technical reasons, we have introduced two new sets of Majorana operators $\{ c_{\mu,\sigma} \mid \mu \in [2n_\sigma] \}$ on each $n_\sigma$-mode Hilbert space, in order to talk about the different spin sectors in terms of the standard tensor product. Because the tensor product does not respect the antisymmetry of fermions, these new Majorana operators are related to the usual Majorana operators (acting on the full $n$-mode Hilbert space) via
\begin{align}
    \gamma_{\mu,\uparrow} &= c_{\mu,\uparrow} \otimes \I,\\
    \gamma_{\mu,\downarrow} &= P_\uparrow \otimes c_{\mu,\downarrow}.
\end{align}
Lemma~\ref{lem:spin_tensor_product} therefore addresses this technicality of maintaining the anticommutation relations when expressing $U_Q$ as a tensor product.

\begin{proof}[Proof (of Lemma~\ref{lem:spin_tensor_product})]
    It is clear that conjugation by $U_{Q_\sigma}$ transforms each $c_{\mu,\sigma}$ as desired. What we need to ensure is that the Majorana operators $\gamma_{\mu,\sigma}$ on the full Hilbert space transform properly. Indeed, Eq.~\eqref{eq:spin_preserving_tensor_product} performs the desired transformation;~for the spin-up sector, we simply have
    \begin{equation}
    U_Q \gamma_{\mu,\uparrow} U_Q^\dagger = U_{Q_\uparrow} c_{\mu,\uparrow} U_{Q_\uparrow}^\dagger \otimes \I.
    \end{equation}
    
    For the spin-down sector, we will make use of following commutation relations:
    \begin{align}
    P_\sigma U_{Q_\sigma} &= \det(Q_\sigma) U_{Q_\sigma} P_\sigma\\
    P_\sigma c_{\mu,\sigma} &= -c_{\mu,\sigma} P_\sigma.
    \end{align}
    Consider the case $\det(Q_\uparrow) = 1$. Then $s(Q_\uparrow) = 0$, and so $U_Q = U_{Q_\uparrow} \otimes U_{Q_\downarrow}$:
    \begin{equation}
    \begin{split}
    U_Q \gamma_{\mu,\downarrow} U_Q^\dagger &= U_{Q_\uparrow} P_{\uparrow} U_{Q_\uparrow}^\dagger \otimes U_{Q_\downarrow} c_{\mu,\downarrow} U_{Q_\downarrow}^\dagger\\
    &= P_\uparrow \otimes U_{Q_\downarrow} c_{\mu,\downarrow} U_{Q_\downarrow}^\dagger,
    \end{split}
    \end{equation}
    as desired. If instead $\det(Q_\uparrow) = -1$, then $U_Q = U_{Q_\uparrow} \otimes (P_\downarrow U_{Q_\downarrow})$ and so
    \begin{align}
    U_Q \gamma_{\mu,\downarrow} U_Q^\dagger &= U_{Q_\uparrow} P_{\uparrow} U_{Q_\uparrow}^\dagger \otimes P_\downarrow U_{Q_\downarrow} c_{\mu,\downarrow} U_{Q_\downarrow}^\dagger P_\downarrow^\dagger \notag\\
    &= (-P_\uparrow) \otimes (-\det(Q_\downarrow)^2 P_\downarrow P_\downarrow^\dagger U_{Q_\downarrow} c_{\mu,\downarrow} U_{Q_\downarrow}^\dagger)\notag\\
    &= P_\uparrow \otimes U_{Q_\downarrow} c_{\mu,\downarrow} U_{Q_\downarrow}^\dagger,
    \end{align}
    where we have used the fact that $P_\downarrow P_\downarrow^\dagger = P_\downarrow^2 = \I$.
\end{proof}

In the context of classical shadows, the appearance of $P_\downarrow$ is inconsequential because it merely acts as a phasing operator which appears directly before measurement. To see this, first observe that $\mathcal{M}_Z = \mathcal{M}_{Z,\uparrow} \otimes \mathcal{M}_{Z, \downarrow}$, where $\mathcal{M}_{Z, \sigma} = \sum_{b \in \{0, 1\}^{n_\sigma}} \vop{b}{b}$. Using the fact that $\mathcal{P}_\downarrow \kket{b} = P_\downarrow \op{b}{b} P_\downarrow^\dagger = \op{b}{b} = \kket{b}$, we have that that shadow channel of the spin-adapted ensemble is
\begin{align}
    \mathcal{M}_{G_\mathrm{spin}} &= \E_{Q_\uparrow, Q_\downarrow}\l[ \mathcal{U}_{Q_\uparrow}^\dagger \otimes (\mathcal{U}_{Q_\downarrow}^\dagger \mathcal{P}_\downarrow^{s(Q_\uparrow)}) (\mathcal{M}_{Z,\uparrow} \otimes \mathcal{M}_{Z, \downarrow}) \r. \notag\\
    &\qquad \times \l. \mathcal{U}_{Q_\uparrow} \otimes (\mathcal{P}_\downarrow^{s(Q_\uparrow)} \mathcal{U}_{Q_\downarrow}) \r] \notag\\
    &= \E_{Q_\uparrow, Q_\downarrow}\l[ (\mathcal{U}_{Q_\uparrow}^\dagger \mathcal{M}_{Z,\uparrow} \mathcal{U}_{Q_\uparrow}) \otimes (\mathcal{U}_{Q_\downarrow}^\dagger \mathcal{M}_{Z,\downarrow} \mathcal{U}_{Q_\downarrow}) \r] \notag\\
    &= \mathcal{M}_\uparrow \otimes \mathcal{M}_\downarrow.
\end{align}
Therefore the spin-adapted matchgate shadows behaves as two independent instances on each spin sector. The estimators and variance bounds also follow straightforwardly;~first, the shadow channel is
\begin{equation}
\begin{split}
    \mathcal{M}_{G_\mathrm{spin}} &= \mathcal{M}_\uparrow \otimes \mathcal{M}_\downarrow\\
    &= \sum_{j=0}^{n_\uparrow} \sum_{\ell=0}^{n_\downarrow} f_{n_\uparrow,j} f_{n_\downarrow,\ell} \Pi_j \otimes \Pi_\ell,
\end{split}
\end{equation}
where for ease of notation in this section, we define
\begin{equation}
    f_{n_\sigma, j} \coloneqq \l. \binom{n_\sigma}{j} \middle/ \binom{2n_\sigma}{2j} \r..
\end{equation}

For the variance, we use the property that the shadow norm of a tensor-product distribution is the product of shadow norms on each subsystem. This can be seen from the fact that shadow norm of an operator $A$ is the spectral norm of a related operator $A^G \coloneqq \E_{U \sim G} \sum_{b \in \{0, 1\}^n} U^\dagger \op{b}{b} U \ev{b}{U \mathcal{M}_G^{-1}(A) U^\dagger}{b}^2$:
\begin{equation}
\begin{split}
    \sns{A}{G}^2 &= \max_{\text{states } \rho} \tr(\rho A^G)\\
    &= \| A^G \|_\infty,
\end{split}
\end{equation}
which holds because $A^G$ is positive semidefinite. For clarity, in this section we use the notation $\sns{\cdot}{G}$ for the shadow norm associated with the group $G$. Thus for any shadow channel formed as a tensor product $\mathcal{M}_{G_1 \oplus G_2} = \mathcal{M}_{G_1} \otimes \mathcal{M}_{G_2}$, we have
\begin{equation}
\begin{split}
    \sns{A_1 \otimes A_2}{G_1 \oplus G_2}^2 &= \max_{\text{states } \rho} \tr\l( \rho \l( A_1^{G_1} \otimes A_2^{G_2} \r) \r)\\
    &= \| A_1^{G_1} \otimes A_2^{G_2} \|_\infty\\
    &= \| A_1^{G_1} \|_\infty \| A_2^{G_2} \|_\infty\\
    &= \sns{A_1}{G_1}^2 \sns{A_2}{G_2}^2.
\end{split}
\end{equation}
(This argument generalizes to multiple tensor products.) Within the context of our spin-adapted ensemble, this implies that any spin-respecting Majorana operator
\begin{equation}
\begin{split}
    \Gamma_{\bm{p}, \bm{q}} &= (-i)^{j + \ell} \gamma_{p_1,\uparrow} \cdots \gamma_{p_{2j},\uparrow} \gamma_{q_1,\downarrow} \cdots \gamma_{q_{2\ell},\downarrow}\\
    &= (-i)^j c_{p_1,\uparrow} \cdots c_{p_{2j},\uparrow} \otimes (-i)^{\ell} c_{q_1,\downarrow} \cdots c_{q_{2\ell},\downarrow},
\end{split}
\end{equation}
has a squared shadow norm of
\begin{equation}
    \sns{\Gamma_{\bm{p}, \bm{q}}}{G_\mathrm{spin}}^2 = f_{n_\uparrow,j}^{-1} f_{n_\downarrow,\ell}^{-1}.
\end{equation}
Thus, for Majorana operators of constant degree $2(j + \ell) \leq 2k$, the variance scales as $\O(n_\uparrow^j n_\downarrow^\ell) = \O(n^k)$, just as in the unadapted setting. Note that spin-respecting here means that the operator factorizes into an even-degree Majorana operator on each spin sector.

The advantage of this ensemble is that the required circuit depth and gate count are roughly halved, since we only need to implement two independent matchgate circuits on $n_\sigma$ qubits each. Furthermore, one can also check that the shadow norm constant factors in the spin-adapted setting are also slightly smaller (for example, for $k = 2$ and $n_\sigma = n/2$, the ratio of spin-adapted to unadapted shadow norms is asymptotically $\lim_{n\to\infty} f_{n/2,1}^{-2}/f_{n,2}^{-1} = 3/4$).

\section{Improved compilation of fermionic Gaussian unitaries}\label{sec:FGU_circuit_design}

In this section we describe a new scheme for compiling the matchgate circuit $U_Q$ for arbitrary $Q \in \Orth(2n)$, under the Jordan--Wigner mapping. This approach improves upon the circuit depth of prior art~\cite{jiang2018quantum} by optimizing the parallelization of nearest-neighbor single- and two-qubit gates. We accomplish this by modifying previously established ideas to better respect the mapping of Majorana modes to qubits. Our improved design is implemented in code at our open-source repository ({\small\url{https://github.com/zhao-andrew/symmetry-adjusted-classical-shadows}})~\cite{gitrepo}.

\subsection{A previous circuit design}\label{subsec:prior_circuit_design}

First we will review a prior circuit design to encode the action of $U_Q$ into a sequence of single- and two-qubit gates, from which it will become clear where there is room for improved parallelization. While the precise scheme that we describe here has not previously appeared in the literature, the high-level ideas follow from a combination of already developed results~\cite{reck1994experimental,wecker2015solving,kivlichan2018quantum,jiang2018quantum,clements2016optimal,oszmaniec2022fermion}.

Recall that our convention for the Jordan--Wigner mapping is
\begin{align}
    \gamma_{2p} &= Z_0 \cdots Z_{p-1} X_p,\\
    \gamma_{2p + 1} &= Z_0 \cdots Z_{p-1} Y_p
\end{align}
for $p \in [n]$, and our convention for the Gaussian transformation is
\begin{equation}\label{eq:gaussian_transformation_def}
    U_Q \gamma_\mu U_Q^\dagger = \sum_{\nu \in [2n]} Q_{\nu\mu} \gamma_\nu.
\end{equation}
It is straightforward to check that $\mathcal{U} : \Orth(2n) \to \U(\mathcal{L}(\mathcal{H}))$ is a group homomorphism:~$\mathcal{U}_Q \mathcal{U}_{Q'} = \mathcal{U}_{QQ'}$ for any $Q, Q' \in \Orth(2n)$. From this property, a circuit for arbitrary $U_Q$ can be constructed by a QR decomposition of $Q$. Such a decomposition yields a sequence of nearest-neighbor Givens rotations, which we then map to single- and adjacent two-qubit gates.

One possible QR decomposition is
\begin{equation}
    Q = G_1 \cdots G_L D,
\end{equation}
where each $G_j$ is a Givens rotation among adjacent rows and columns, and $D$ is the upper-right triangular matrix from the QR decomposition. Because $Q$ is an orthogonal matrix, $D$ is guaranteed to be a diagonal matrix with $\pm 1$ entries along the diagonal. This is equivalent to the Reck \emph{et al}.~\cite{reck1994experimental} design, and the number of Givens rotations is $L = \O(n^2)$ in depth $\O(n)$. By the homomorphism property of $\mathcal{U}$, this matrix decomposition yields a sequence of circuit elements that implements the desired unitary:
\begin{equation}
    \mathcal{U}_Q = \mathcal{U}_{G_1 \cdots G_L D} = \mathcal{U}_{G_1} \cdots \mathcal{U}_{G_L} \mathcal{U}_D.
\end{equation}
Alternatively, the Clements \emph{et al}.~\cite{clements2016optimal} design computes a decomposition of the form\footnote{The use of the Clements \emph{et al}.~\cite{clements2016optimal} design was first pointed out in Ref.~\cite{huggins2021efficient} by Dominic Berry, in the context of number-preserving matchgate circuits.}
\begin{equation}
    Q = G_{R + 1} \cdots G_{R + L} D G_{R} \cdots G_{1}.
\end{equation}
The total number of Givens rotations here is the same, $R + L = \O(n^2)$. However, by utilizing rotations that act from both left and right, it optimizes parallelization to reduce the depth by a constant factor (roughly $1/2$).

Refs.~\cite{wecker2015solving,kivlichan2018quantum,jiang2018quantum} showed how to convert these Givens rotations into number-preserving quantum gates;~here we seek to generalize to fermionic Gaussian unitaries which do not necessarily conserve particle number. While Ref.~\cite{jiang2018quantum} also considered this scenario, they maintained the representation of Givens rotations as number-preserving gates. Their circuit design breaks particle-number symmetry by interspersing particle--hole transformations throughout the decomposition.

Instead, we will use a representation that inherently features non-number-preserving rotations. Suppose that the Givens rotation $G_j$ acts nontrivially on the axes $(\mu, \mu + 1)$ as
\begin{equation}
    G_j = \begin{pmatrix}
    1 & \cdots & 0 & 0 & \cdots & 0\\
    \vdots & \ddots & \vdots & \vdots &  & \vdots\\
    0 & \cdots & \cos\theta_j & -\sin\theta_j & \cdots & 0\\
    0 & \cdots & \sin\theta_j & \cos\theta_j & \cdots & 0\\
    \vdots &  & \vdots & \vdots & \ddots & \vdots\\
    0 & \cdots & 0 & 0 & \cdots & 1
    \end{pmatrix} \in \SO(2n).
\end{equation}
The quantum gate which achieves this transformation is a single- or two-qubit Pauli rotation, given by
\begin{equation}\label{eq:givens_rotation_gates}
\begin{split}
    U_{G_j} &= e^{-(\theta_j/2) \gamma_{\mu} \gamma_{\mu + 1}}\\
    &= \begin{cases}
    e^{-\i(\theta_j/2) Z_p} & \text{if $\mu = 2p$ is even},\\
    e^{-\i(\theta_j/2) X_p X_{p + 1}} & \text{if $\mu = 2p + 1$ is odd}.
    \end{cases}
\end{split}
\end{equation}
Indeed, one may check that $\mathcal{U}_{G_j}(\gamma_p) = \sum_{q\in[2n]} [G_j]_{qp} \gamma_q$, as desired.

To implement the diagonal matrix $D$ of signs, we require a different scheme. In particular, we can construct $U_D$ as a single layer of Pauli gates. Consider the $2 \times 2$ block along the diagonal
\begin{equation}
	D^{(p)} = \begin{pmatrix}
		D_{2p} & 0 \\ 0 & D_{2p+1}
	\end{pmatrix},
\end{equation}
which describes the transformation
\begin{align}
    \gamma_{2p} &\mapsto D_{2p} \gamma_{2p},\\
    \gamma_{2p+1} &\mapsto D_{2p+1} \gamma_{2p+1}.
\end{align}
If $ D_{2p} = D_{2p+1} = 1 $, then clearly no operations are required. If instead $ D_{2p} = D_{2p+1} = -1 $, then conjugation by $ Z_p $ applies the desired signs on $\gamma_{2p}$ and $\gamma_{2p+1}$ while leaving all other Majorana operators invariant.

The remaining cases, $ D_{2p} = -D_{2p+1} $, can be handled as follows. First, suppose $ D_{2p} = 1 $. We wish to find the gates which perform the transformation
\begin{align}
	\gamma_{2p} &\mapsto \gamma_{2p},\\
	\gamma_{2p+1} &\mapsto -\gamma_{2p+1},
\end{align}
while leaving all other Majorana operators invariant. We can almost accomplish this with $ X_p $, since it will map $ X_p $ to itself and $ Y_p $ to $ -Y_p $. It also commutes with all Majorana operators $ \gamma_{2q}, \gamma_{2q+1} $ for $ q < p $. However, for $ q > p $ this will accrue unwanted signs:
\begin{align}
    X_p \gamma_{2q} X_p &= X_p (Z_0 \cdots Z_p \cdots Z_{q-1} X_q) X_p \notag\\
    &= -Z_0 \cdots Z_p \cdots Z_{q-1} X_q \notag\\
    &= -\gamma_{2q},\\
	X_p \gamma_{2q+1} X_p &= X_p (Z_0 \cdots Z_p \cdots Z_{q-1} Y_q) X_p \notag\\
    &= -Z_0 \cdots Z_p \cdots Z_{q-1} Y_q \notag\\
    &= -\gamma_{2q+1}.
\end{align}
To correct these signs, we introduce a Pauli-$ Z $ string running in the opposite direction of the Jordan--Wigner convention. That is, define
\begin{equation}
	P_{p}^{(X)} \coloneqq X_p Z_{p+1} \cdots Z_{n-1}.
\end{equation}
This unitary has the correct action on $ \gamma_{2p},\gamma_{2p+1} $ and continues to commute with the Majorana operators $ \gamma_{2q}, \gamma_{2q+1} $ with $q < p$. For $q > p$, however, we now have
\begin{align}
	P_{p}^{(X)} \gamma_{2q} P_{p}^{(X)} &= (X_p Z_{p+1} \cdots Z_{n-1}) (Z_0 \cdots Z_p \cdots Z_{q-1} X_q) (X_p Z_{p+1} \cdots Z_{n-1}) \notag\\
	&= Z_0 \cdots (-Z_p) \cdots Z_{q-1} (-X_q) \notag\\
    &= \gamma_{2q},\\
    P_{p}^{(X)} \gamma_{2q+1} P_{p}^{(X)} &= (X_p Z_{p+1} \cdots Z_{n-1}) (Z_0 \cdots Z_p \cdots Z_{q-1} Y_q) (X_p Z_{p+1} \cdots Z_{n-1}) \notag\\
    &= Z_0 \cdots (-Z_p) \cdots Z_{q-1} (-Y_q) \notag\\
    &= \gamma_{2q+1}.
\end{align}
Thus $ P_p^{(X)} $ implements the desired transformation by $ D^{(p)} = \mathrm{diag}(1,-1) $. For $ D^{(p)} = \mathrm{diag}(-1,1) $, we simply replace $ P_p^{(X)} $ by an analogously defined $ P_p^{(Y)} $. This causes the sign of $ \gamma_{2p} $, rather than $ \gamma_{2p+1} $, to flip, while retaining all other properties.

Altogether, we determine these transformations for all $ 2 \times 2 $ diagonal blocks of $D$, resulting in $n$ Pauli strings of the form
\begin{equation}
    W_p = \begin{cases}
    \I & \text{if } D^{(p)} = \mathrm{diag}(1, 1),\\
     Z_p & \text{if } D^{(p)} = \mathrm{diag}(-1, -1),\\
     P_p^{(X)} & \text{if } D^{(p)} = \mathrm{diag}(1, -1),\\
     P_p^{(Y)} & \text{if } D^{(p)} = \mathrm{diag}(-1, 1).
    \end{cases}
\end{equation}
The overall transformation is then simply the product of these Pauli strings, which can be concatenated into a single layer of Pauli gates:
\begin{equation}\label{eq:pauli_gates_D}
    U_D = 
    \prod_{p \in [n]} W_p.
\end{equation}
Note that the order of this product does not matter, since Pauli gates commute up to an unobservable global phase.

\subsection{Discussion on suboptimality}

Now we observe that, depending on the parity of $\mu \in \{0, \ldots, 2n-2\}$, $U_{G_j}$ is either a single- or two-qubit gate. However, the decomposition of $Q$ described above is implicitly optimized under the assumption that only two-qubit gates are present:~each Givens rotation acts on two axes at a time, and it is assumed that this corresponds to physically acting on two wires at a time. This results in underutilized space in the quantum circuit whenever a single-qubit $Z$ rotation occurs, as it leaves a qubit wire needlessly idle. This is true for both the Reck \emph{et al.}~\cite{reck1994experimental} and Clements \emph{et al.}~\cite{clements2016optimal} designs. Ultimately, this suboptimality is due to the fact that $Q$ is a $2n \times 2n$ matrix, so there is a two-to-one correspondence between axes and qubits:~the rows/columns labeled by $(2p, 2p + 1)$ correspond to two Majorana operators, both of which are in turn associated with a single qubit $p$. Note that this discrepancy is not present in circuit designs for the class of number-conserving rotations~\cite{wecker2015solving,kivlichan2018quantum,jiang2018quantum}, which are instead more compactly represented by an $n \times n$ unitary matrix already.

\subsection{Circuit design with improved parallelization}

Now we introduce a circuit design which explicitly accounts for this two-to-one correspondence. The basic idea is to generalize the notion of Givens rotations, which act on a two-dimensional subspace to zero out a single matrix element, to a four-dimensional orthogonal transformation which zeroes out blocks of $2 \times 2$ at a time. Each $4 \times 4$ orthogonal transformation acts on the axes $(2p, 2p + 1, 2p + 2, 2p + 3)$, which corresponds to qubits $p$ and $p + 1$. By performing this process according to the scheme of Clements \emph{et al.}~\cite{clements2016optimal} (but now treating each $2 \times 2$ block of $Q$ as a ``single'' element), we obtain a decomposition wherein the optimal parallelization of the scheme is fully preserved in terms of interactions between nearest-neighbor qubits. Finally, each $4 \times 4$ orthogonal transformation is ultimately decomposed into six rotations of the form of Eq.~\eqref{eq:givens_rotation_gates} and a layer of Pauli gates, achieved by the standard decomposition that we described in \cref{subsec:prior_circuit_design} (i.e., by bootstrapping off the prior scheme within blocks of $2n = 4$). Note that in principle one may instead implement the $4 \times 4$ orthogonal transformations using any gate set of one's choice, rather than $XX$ and $Z$ rotations.

We now describe the algorithm in detail. First we compute a decomposition analogous to the Clements \emph{et al.}~\cite{clements2016optimal} design, 
\begin{equation}\label{eq:Q_block_decomposition}
    Q = B_{R + 1} \cdots B_{R + L} D G B_{R} \cdots B_{1},
\end{equation}
but instead of Givens rotations, each $B_k$ acts nontrivially on a $4 \times 4$ block. (Note that there is a single $2 \times 2$ Givens rotation $G$ as well, which serves to zero out a final matrix element that we will elaborate on later.) We accomplish this by treating $Q$ as an $n \times n$ matrix of $2 \times 2$ blocks,
\begin{equation}\label{eq:left_mult_zero}
    Q_{\bm{p},\bm{q}} = \begin{pmatrix}
    Q_{2p,2q} & Q_{2p,2q+1}\\
    Q_{2p+1,2q} & Q_{2p+1,2q+1}
    \end{pmatrix},
\end{equation}
for each $p, q \in [n]$. Just as Givens rotations are chosen to zero a specific matrix element, each $B_k$ acts to zero out a particular $2 \times 2$ block $Q_{\bm{p},\bm{q}}$.

Suppose we want to find a $B_{R+i}$ which acts from the left ($i = 1, \ldots, L$) to zero out the block $Q_{\bm{p},\bm{q}}$. Then we perform a QR decomposition on the $4 \times 2$ submatrix which includes the target block and the block directly above it:
\begin{equation}
    \l(\begin{array}{c}
    Q_{\bm{p}-1,\bm{q}}\\
    \hline
    Q_{\bm{p},\bm{q}}
    \end{array}\r) = B_{R+i}' \l(\begin{array}{c c}
    * & *\\
    0 & *\\
    \hline
    0 & 0\\
    0 & 0
    \end{array}\r),
\end{equation}
hence zeroing out the lower block as desired. Here, $B_{R+i}' \in \Orth(4)$ is computed from the QR decomposition, and so the orthogonal matrix $B_{R+i} \in \Orth(2n)$ appearing in Eq.~\eqref{eq:Q_block_decomposition} is defined as $B_{R+i}'$ along the axes $(2p-2, 2p-1, 2p, 2p + 1)$ and the identity elsewhere.

Similarly, if we want a $B_{j}$ which acts from the right ($j = 1, \ldots, R$), then we consider instead a $2 \times 4$ submatrix with the target block on the left:
\begin{equation}\label{eq:2x4_block}
    \l(\begin{array}{c | c}
    Q_{\bm{p},\bm{q}} & Q_{\bm{p},\bm{q}+1}
    \end{array}\r).
\end{equation}
This can be zeroed out by performing an LQ decomposition (which is essentially just the transpose of the QR decomposition). For notation in this section, let tildes denote the flipping of rows in a matrix, for example
\begin{equation}
    M = \begin{pmatrix}
    M_{11} & M_{12}\\
    M_{21} & M_{22}\\
    M_{31} & M_{32}\\
    M_{41} & M_{42}
    \end{pmatrix} \mapsto \tilde{M} = \begin{pmatrix}
    M_{41} & M_{42}\\
    M_{31} & M_{32}\\
    M_{21} & M_{22}\\
    M_{11} & M_{12}
    \end{pmatrix}.
\end{equation}
Then performing an LQ decomposition on the row-flipped version of Eq.~\eqref{eq:2x4_block}, we have
\begin{equation}\label{eq:2x4_block_transpose_qr}
\begin{split}
    \l(\begin{array}{c | c}
    \tilde{Q}_{\bm{p},\bm{q}} & \tilde{Q}_{\bm{p},\bm{q}+1}
    \end{array}\r) &= \l(\begin{array}{c c | c c}
    * & 0 & 0 & 0\\
    * & * & 0 & 0
    \end{array}\r) B_{j}'\\
    &= \l(\begin{array}{c c | c c}
    0 & 0 & 0 & *\\
    0 & 0 & * & *
    \end{array}\r) \tilde{B}_{j}'.
\end{split}
\end{equation}
Flipping the rows back to normal on the lefthand side, we get
\begin{equation}
    \l(\begin{array}{c | c}
    Q_{\bm{p},\bm{q}} & Q_{\bm{p},\bm{q}+1}
    \end{array}\r) = \l(\begin{array}{c c | c c}
    0 & 0 & * & *\\
    0 & 0 & 0 & *
    \end{array}\r) \tilde{B}_{j}'
\end{equation}
as desired. Then we define $B_{j} \in \Orth(2n)$ acting as $\tilde{B}_{j}'$ on the axes $(2q, 2q+1, 2q+2, 2q+3)$ and trivially elsewhere.

Now we address the need for the sole Givens rotation $G$ appearing in Eq.~\eqref{eq:Q_block_decomposition}. As the zeroing-out procedure described above progresses, the nonzero blocks get ``pushed'' towards the diagonal until the final matrix is ($2 \times 2$)-block diagonal. These nonzero blocks must be triangular because they are produced by QR/LQ decompositions;~but since $Q$ is orthogonal, this implies that the final triangular blocks along the diagonal must be diagonal themselves. The exception to this is either the leftmost or rightmost block, depending on whether $n$ is even or odd. This is because the decomposition procedure inevitably leaves one of those blocks untouched, so it was never made triangular/diagonal.

This can be visualized as follows:~if $n$ is odd, then we have
\begin{align}
    Q &\to \l(\begin{array}{c c | c c | c c}
    * & * & * & * & * & *\\
    * & * & * & * & * & *\\
    \hline
    * & * & * & * & * & *\\
    * & * & * & * & * & *\\
    \hline
    \bm{0} & \bm{0} & * & * & * & *\\
    \bm{0} & \bm{0} & \bm{0} & * & * & *
    \end{array}\r)
    \to
    \l(\begin{array}{c c | c c | c c}
    * & * & * & * & * & *\\
    \bm{0} & * & * & * & * & *\\
    \hline
    \bm{0} & \bm{0} & * & * & * & *\\
    \bm{0} & \bm{0} & * & * & * & *\\
    \hline
    0 & 0 & * & * & * & *\\
    0 & 0 & 0 & * & * & *
    \end{array}\r) \notag\\
    &\to
    \l(\begin{array}{c c | c c | c c}
    * & * & * & * & * & *\\
    0 & * & * & * & * & *\\
    \hline
    0 & 0 & * & * & * & *\\
    0 & 0 & \bm{0} & * & * & *\\
    \hline
    0 & 0 & \bm{0} & \bm{0} & * & *\\
    0 & 0 & 0 & \bm{0} & * & *
    \end{array}\r) =
    \l(\begin{array}{c c | c c | c c}
    \pm 1 & 0 & 0 & 0 & 0 & 0\\
    0 & \pm 1 & 0 & 0 & 0 & 0\\
    \hline
    0 & 0 & \pm 1 & 0 & 0 & 0\\
    0 & 0 & 0 & \pm 1 & 0 & 0\\
    \hline
    0 & 0 & 0 & 0 & * & *\\
    0 & 0 & 0 & 0 & * & *
    \end{array}\r).
\end{align}
We use boldface to clarify which matrix elements are newly zeroed at each step. The condition that $Q$ is an orthogonal matrix implies the final equality. It also enforces the remaining $2 \times 2$ block to be orthogonal, so that we can diagonalize it by computing the appropriate Givens rotation acting on axes $(2n-2, 2n-1)$. This elucidates the appearance of $G$ in Eq.~\eqref{eq:Q_block_decomposition}. On the other hand, if $n$ is even, then the top-left block remains instead:
\begin{align}
    \hspace*{-3cm} Q &\to \l(\begin{array}{c c | c c | c c | c c}
    * & * & * & * & * & * & * & *\\
    * & * & * & * & * & * & * & *\\
    \hline
    * & * & * & * & * & * & * & *\\
    * & * & * & * & * & * & * & *\\
    \hline
    * & * & * & * & * & * & * & *\\
    * & * & * & * & * & * & * & *\\
    \hline
    \bm{0} & \bm{0} & * & * & * & * & * & *\\
    \bm{0} & \bm{0} & \bm{0} & * & * & * & * & *
    \end{array}\r)
    \to
    \l(\begin{array}{c c | c c | c c | c c}
    * & * & * & * & * & * & * & *\\
    * & * & * & * & * & * & * & *\\
    \hline
    * & * & * & * & * & * & * & *\\
    \bm{0} & * & * & * & * & * & * & *\\
    \hline
    \bm{0} & \bm{0} & * & * & * & * & * & *\\
    \bm{0} & \bm{0} & * & * & * & * & * & *\\
    \hline
    0 & 0 & * & * & * & * & * & *\\
    0 & 0 & 0 & * & * & * & * & *
    \end{array}\r)
    \to
    \l(\begin{array}{c c | c c | c c | c c}
    * & * & * & * & * & * & * & *\\
    * & * & * & * & * & * & * & *\\
    \hline
    * & * & * & * & * & * & * & *\\
    0 & * & * & * & * & * & * & *\\
    \hline
    0 & 0 & * & * & * & * & * & *\\
    0 & 0 & \bm{0} & * & * & * & * & *\\
    \hline
    0 & 0 & \bm{0} & \bm{0} & * & * & * & *\\
    0 & 0 & 0 & \bm{0} & * & * & * & *
    \end{array}\r) \notag\\
    &\to
    \l(\begin{array}{c c | c c | c c | c c}
    * & * & * & * & * & * & * & *\\
    * & * & * & * & * & * & * & *\\
    \hline
    * & * & * & * & * & * & * & *\\
    0 & * & * & * & * & * & * & *\\
    \hline
    0 & 0 & * & * & * & * & * & *\\
    0 & 0 & 0 & * & * & * & * & *\\
    \hline
    0 & 0 & 0 & 0 & \bm{0} & \bm{0} & * & *\\
    0 & 0 & 0 & 0 & \bm{0} & \bm{0} & \bm{0} & *
    \end{array}\r)
    \to
    \l(\begin{array}{c c | c c | c c | c c}
    * & * & * & * & * & * & * & *\\
    * & * & * & * & * & * & * & *\\
    \hline
    * & * & * & * & * & * & * & *\\
    0 & * & * & * & * & * & * & *\\
    \hline
    0 & 0 & \bm{0} & \bm{0} & * & * & * & *\\
    0 & 0 & 0 & \bm{0} & \bm{0} & * & * & *\\
    \hline
    0 & 0 & 0 & 0 & 0 & 0 & * & *\\
    0 & 0 & 0 & 0 & 0 & 0 & 0 & *
    \end{array}\r) \notag\\
    &\to
    \l(\begin{array}{c c | c c | c c | c c}
    * & * & * & * & * & * & * & *\\
    * & * & * & * & * & * & * & *\\
    \hline
    \bm{0} & \bm{0} & * & * & * & * & * & *\\
    0 & \bm{0} & \bm{0} & * & * & * & * & *\\
    \hline
    0 & 0 & 0 & 0 & * & * & * & *\\
    0 & 0 & 0 & 0 & 0 & * & * & *\\
    \hline
    0 & 0 & 0 & 0 & 0 & 0 & * & *\\
    0 & 0 & 0 & 0 & 0 & 0 & 0 & *
    \end{array}\r)
    = \l(\begin{array}{c c | c c | c c | c c}
    * & * & 0 & 0 & 0 & 0 & 0 & 0\\
    * & * & 0 & 0 & 0 & 0 & 0 & 0\\
    \hline
    0 & 0 & \pm 1 & 0 & 0 & 0 & 0 & 0\\
    0 & 0 & 0 & \pm 1 & 0 & 0 & 0 & 0\\
    \hline
    0 & 0 & 0 & 0 & \pm 1 & 0 & 0 & 0\\
    0 & 0 & 0 & 0 & 0 & \pm 1 & 0 & 0\\
    \hline
    0 & 0 & 0 & 0 & 0 & 0 & \pm 1 & 0\\
    0 & 0 & 0 & 0 & 0 & 0 & 0 & \pm 1
    \end{array}\r).
\end{align}
In this case, $G$ needs to act on axes $(0, 1)$.

Thus we have obtained the decomposition of Eq.~\eqref{eq:Q_block_decomposition} as desired. The implementation of each component then follows from bootstrapping the prior techniques:~the diagonal matrix $D$ becomes a layer of Pauli gates, described by Eq.~\eqref{eq:pauli_gates_D};~and the four-dimensional orthogonal transformations $B_j$ are further decomposed into Givens rotations, described in Section~\ref{subsec:prior_circuit_design} (wherein $n = 2$).

\begin{figure}
\centering
\includegraphics[scale=0.5]{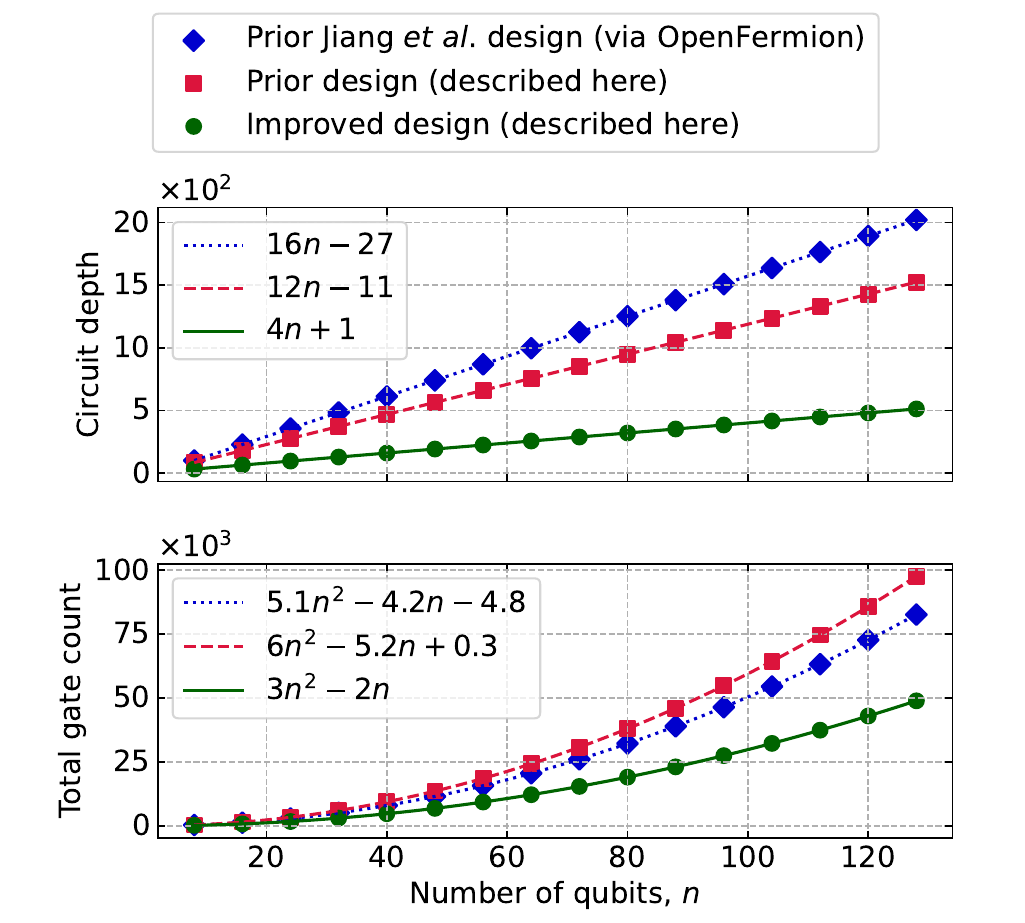}
\caption[Resource comparison of our improved fermionic Gaussian circuit design (green) versus prior designs.]{Resource comparison of our improved fermionic Gaussian circuit design (green) versus prior designs:~that of Jiang \emph{et al.}~\cite{jiang2018quantum} (blue), implemented in OpenFermion~\cite{openfermion}, and the naive scheme described in \cref{subsec:prior_circuit_design} (red). We compare both circuit depth (top) and gate count (bottom). All circuits were optimized to a superconducting-qubit gate set. Linear and quadratic fits are made, demonstrating a roughly $3\times$ and $2\times$ savings, respectively.}
\label{fig:circuit_design_comparison}
\end{figure}

In Figure~\ref{fig:circuit_design_comparison} we demonstrate the circuit depth and gate count of this new design. For each $n$, we run our algorithm on a randomly generated element of $\Orth(2n)$. We further compile the circuits to a gate set native to superconducting-qubit platforms, consisting of arbitrary single-qubit rotations and nearest-neighbor two-qubit $\sqrt{\iSWAP}$ gates (described in \cref{sec:additional_numerics}). For comparison, we also compile the same unitary according to the old design described in \cref{subsec:prior_circuit_design} and the algorithm of Jiang \emph{et al.}~\cite{jiang2018quantum}, which is implemented within the open-source library OpenFermion~\cite{openfermion}. This latter design also uses a Givens-rotation decomposition, but rather than the $\Orth(2n)$ Majorana representation it employs particle--hole transformations on the ladder operators to incorporate non-particle-conserving operations. We also make polynomial fits, demonstrating the resource savings of our design. We infer asymptotic reductions in the circuit depth and gate count by about $1/3$ and $1/2$, respectively. Especially for near-term quantum computers, such savings provide significant improvements to overall performance.

\section{Classical shadows postprocessing details}

In this section we provide details for the classical postprocesisng of local observable estimators from classical shadows. We include this for a self-contained and explicit presentation, and also to address the modified shadows protocols (subsystem symmetrization and spin adaptation) introduced in this paper. These algorithms are implemented at our open-source repository ({\small\url{https://github.com/zhao-andrew/symmetry-adjusted-classical-shadows}})~\cite{gitrepo}.

\subsection{Matchgate shadows}\label{subsec:matchgate_computation}

For any orthogonal matrix $Q \in \Orth(2n)$, Ref.~\cite{wan2023matchgate} derived formulas involving the multiplication of $2n \times 2n$ matrices and the computation of Pfaffians of $2k \times 2k$ submatrices for estimating $k$-body Majorana observables. However when restricting $Q \in \B(2n)$, there exists a significantly cheaper method that does not involve such numerical linear algebra routines. This algorithm was implicitly described in Ref.~\cite{zhao2021fermionic}, but not explicitly outlined. We do so here;~for $k = \O(1)$, it runs in time $\O(n^k T)$ to return estimates for all $2j$-degree Majorana operators, $1 \leq j \leq k$, from $T$ samples. Note that the number of operators is $\O(n^{2k})$, so our approach has significant savings over a naive iteration. Furthermore, it largely involves integer storage and manipulations rather than floating-point operations.

Any $k$-body fermionic observable can be decomposed into a linear combination of polynomially many $(\leq k)$-body Majorana operators. Thus it suffices to consider $\Gamma_{\bm{\mu}}$, for all $\bm{\mu} \in \bigcup_{j \leq k} \comb{2n}{2j}$. Each matchgate-shadow sample $\hat{\rho}_{Q,b}$ is classically stored as $(Q, b)$, where $b \in \{0, 1\}^n$ and $Q$ is represented as an array $\pi$ of the permuted elements of $[2n]$ along with signs $s \in \{-1, +1\}^{2n}$. Specifically, the matrix elements of $Q \in \B(2n)$ are related to $(s, \pi)$ by $Q_{\mu\nu} = s_\mu \delta_{\pi(\mu), \nu}$.

The estimator for $\tr(\Gamma_{\bm{\mu}} \rho)$ can be written as $\tr(\Gamma_{\bm{\mu}} \hat{\rho}_{Q,b}) = f_{2j}^{-1} \ev{b}{U_Q \Gamma_{\bm{\mu}} U_Q^\dagger}{b}$, where $U_Q \Gamma_{\bm{\mu}} U_Q^\dagger$ can be expanded in terms of subdeterminants of $Q$ according to Ref.~\cite[Appendix~A]{chapman2018classical}. However, a simplified derivation is possible here by using the fact that $Q$ implements a signed permutation:
\begin{equation}
\begin{split}
    U_Q \gamma_\mu U_Q^\dagger &= \sum_{\nu \in [2n]} Q_{\nu\mu} \gamma_\nu\\
    &= \sum_{\nu \in [2n]} s_\nu \delta_{\pi(\nu),\mu} \gamma_\nu\\
    &= s_{\pi^{-1}(\mu)} \gamma_{\pi^{-1}(\mu)}.
\end{split}
\end{equation}
Hence for operators of degree $2j$,
\begin{equation}
\begin{split}
    U_Q \Gamma_{\bm{\mu}} U_Q^\dagger &= (-\i)^j U_Q \gamma_{\mu_1} \cdots \gamma_{\mu_{2j}} U_Q^\dagger\\
    &= (-\i)^j s_{\pi^{-1}(\mu_1)} \cdots s_{\pi^{-1}(\mu_{2j})}\\
    &\quad \times \gamma_{\pi^{-1}(\mu_1)} \cdots \gamma_{\pi^{-1}(\mu_{2j})}.
\end{split}
\end{equation}
We would like to retain the ordering of indices when working with the multidegree Majorana operators;~therefore we introduce a further a permutation as $\tilde{\pi}^{-1}(\mu_i)$, which is defined to satisfy $\tilde{\pi}^{-1}(\mu_1) < \cdots < \tilde{\pi}^{-1}(\mu_{2j})$. This incurs another sign factor $(-1)^p$ where $p \in \{0, 1\}$ is the parity of the permutation which sends $\pi^{-1}(\bm{\mu}) \mapsto \tilde{\pi}^{-1}(\bm{\mu})$. Collecting all signs as $\sgn_Q(\bm{\mu}) = (-1)^p s_{\pi^{-1}(\mu_1)} \cdots s_{\pi^{-1}(\mu_{2j})}$, we arrive at
\begin{equation}\label{eq:matchgate_shadow_estimator_permutation}
    \tr(\Gamma_{\bm{\mu}} \hat{\rho}_{Q,b}) = f_{2j}^{-1} \sgn_Q(\bm{\mu}) \ev{b}{\Gamma_{\tilde{\pi}^{-1}(\bm{\mu})}}{b}.
\end{equation}
The matrix element $\ev{b}{\Gamma_{\tilde{\pi}^{-1}(\bm{\mu})}}{b}$ is nonzero if and only if $\tilde{\pi}^{-1}(\bm{\mu}) \in \diags{2n}{2j}$, from which its value of $\pm 1$ is straightforward to determine (e.g., by mapping to Pauli-$Z$ operators). In total, evaluating Eq.~\eqref{eq:matchgate_shadow_estimator_permutation} takes time $\O(n^2 + j^2 + j)$, corresponding respectively to the inversion of $\pi \in \Sym(2n)$, the calculation of $\tilde{\pi}^{-1}(\bm{\mu})$ and its parity on $2j$ indices, and evaluating the product of $2j + 1$ signs and $\ev{b}{\Gamma_{\tilde{\pi}^{-1}(\bm{\mu})}}{b}$, the latter requiring only checking $2j$ indices and $j$ bits of $b$. Assuming $j \leq k = \O(1)$, this implies a computational complexity of $\O(n^2)$ per operator per sample.

To compute this estimator for all $\bm{\mu} \in \comb{2n}{2j}$, a naive approach iterates through each $\bm{\mu}$, of which there are $\binom{2n}{2j} = \O(n^{2j})$ many. Repeating this for each of the $T$ samples would therefore cost $\O(T (n^{2j} + n^2)) = \O(T n^{2j})$ time. Noting that $T = \Ot(n^{j} \epsilon^{-2})$ suffices for $\epsilon$-accurate estimation,\footnote{The notation $\Ot(\cdot)$ suppresses polylogarithmic factors in the complexity.} the total complexity of $\Ot(n^{3j} \epsilon^{-2}) \leq \Ot(n^{3k} \epsilon^{-2})$ would be unacceptably large.

We can speed up the computation over all operators per sample to $\O(n^j)$ by using the fact that many $\ev{b}{\Gamma_{\tilde{\pi}^{-1}(\bm{\mu})}}{b}$ vanish. That is, rather than compute $\tilde{\pi}^{-1}(\bm{\mu})$ for all $\bm{\mu} \in \comb{2n}{2j}$ and checking whether each is an element of $\diags{2n}{2j}$, we work backwards by looping over all target elements $\bm{\tau} \in \diags{2n}{2j}$ and computing $\pi(\bm{\tau})$ to find its preimage. As before, let $\tilde{\pi}(\bm{\tau}) \in \comb{2n}{2j}$ be the reordering of $\pi(\bm{\tau})$ with associated sign $(-1)^p$. Then for each $\bm{\tau} \in \diags{2n}{2j}$, we compute the estimator for $\Gamma_{\tilde{\pi}(\bm{\tau})}$,
\begin{equation}\label{eq:reversed_majorana_estimate}
    \tr(\Gamma_{\tilde{\pi}(\bm{\tau})} \hat{\rho}_{Q,b}) = f_{2j}^{-1} \sgn_Q(\pi(\bm{\tau})) \ev{b}{\Gamma_{\bm{\tau}}}{b},
\end{equation}
where the cumulative sign is $\sgn_Q(\pi(\bm{\tau})) = (-1)^p s_{\tau_1} \cdots s_{\tau_{2j}}$. All other Majorana operators not in the preimage are implicitly assigned an estimate of $0$. Hence we only iterate over the $\binom{n}{j} = \O(n^j)$ elements of $\diags{2n}{2j}$, with each evaluation of Eq.~\eqref{eq:reversed_majorana_estimate} taking $\O(j^2) = \O(1)$ time. Note that this approach also avoids the need to find the inverse permutation $\pi^{-1}$.

Performing this procedure over all $T$ samples results in a time complexity of $\O(n^j T) \leq \O(n^k T)$, running over all $j \in \{1, \ldots, k\}$. We can also include an additive $\O(n^{2k})$ cost to preallocate storage for $\bigcup_{j \leq k} \comb{2n}{2j}$. While not strictly necessary, this is convenient in practice, and besides when $T = \Ot(n^k \epsilon^{-2})$ the total complexity is $\Ot(n^{2k} \epsilon^{-2})$ whether or not we preallocate memory.

For the spin-adapted shadows, because the protocol factorizes across the spin sectors, we perform this algorithm on each sector independently. The estimator for operators of the form $\Gamma_{\bm{\mu}} \otimes \Gamma_{\bm{\nu}}$ is then the product of the independent estimates. Note that if either $|\bm{\mu}|$ or $|\bm{\nu}|$ are odd, then the estimator always vanishes;~this reflects the fact that the spin-adapted ensemble is not informationally complete over such operators.

\subsection{Pauli shadows}\label{subsec:pauli_computation}

Because single-qubit measurements factorize, we consider each qubit $i \in [n]$ independently. Given the random Clifford $C_i \in \Cl(1)$ and measurement outcome $b_i \in \{0, 1\}$, the estimator for $\sigma^{(i)} \in \{\I, X, Y, Z\}$ is~\cite{huang2020predicting}
\begin{equation}\label{eq:shadow_qubit_est}
    \tr(\sigma^{(i)} \hat{\rho}_{C_i, b_i}) = 3 \ev{b_i}{C_i \sigma^{(i)} C_i^\dagger}{b_i} - \tr(\sigma^{(i)}).
\end{equation}
Each Pauli-shadow sample is stored as $(W_i, b_i)$, where $W_i = C_i^\dagger Z C_i \in \pm\{X, Y, Z\}$. Evaluating Eq.~\eqref{eq:shadow_qubit_est} reduces to checking the conditions
\begin{equation}\label{eq:shadow_qubit_cases}
    \tr(\sigma^{(i)} \hat{\rho}_{C_i, b_i}) = \begin{cases}
    \pm 3 \ev{b_i}{Z}{b_i} & \text{if } \sigma^{(i)} = \pm W_i,\\
    1 & \text{if } \sigma^{(i)} = \I,\\
    0 & \text{else}.
    \end{cases}
\end{equation}
The product over $i \in [n]$ then estimates $P = \bigotimes_{i \in [n]} \sigma^{(i)}$ for the full $n$-qubit system. This suffices to estimate any $k$-local observable, which can be decomposed into a linear combination of polynomially many $(\leq k)$-local Pauli operators.

The total time complexity of estimating all $k$-local Pauli operators with $T$ snapshots is $\O(n^k T)$. The algorithm is as follows. For each $W = (W_0, \ldots, W_{n-1})$, we take, for each $j \leq k$, all $\binom{n}{j}$ combinations $W_{i_1}, \ldots, W_{i_j}$ and compute Eq.~\eqref{eq:shadow_qubit_cases} for each $\sigma^{(i)} = \pm W_i$. We assign the result as an estimate for the $j$-local operator $P = W_{i_1} \otimes \cdots \otimes W_{i_j} \otimes \I^{\otimes (n-j)}$, and implicitly assign $0$ to all other Pauli operators. Note that there are a total of $\sum_{j \leq k} 3^j \binom{n}{j} = \O(n^k)$ local Pauli operators, so preallocating storage here is asymptotically negligible.

For the subsystem-symmetrized protocol, the $n$-qubit estimator now takes the form
\begin{equation}
    \tr(P \hat{\rho}_{(\pi, C), b}) = 3^{|P|} \ev{b}{S_\pi C P C^\dagger S_\pi^\dagger}{b}.
\end{equation}
Using the fact that $S_\pi^\dagger \ket{b} = \bigotimes_{i \in [n]} \ket{\pi(b_i)}$, we can simply apply the standard scheme described above, but with the replacement $b_i \to \pi(b_i)$. For each sample this is only an additive $\O(n)$ cost.

\section{Additional details on numerical experiments}\label{sec:additional_numerics}

\subsection{Readout noise models}\label{subsec:readout_errors}

In Section~\ref{subsec:fermion_readout_error} and \ref{subsec:qubit_readout_error}, we demonstrated our mitigation strategy under single-qubit readout errors. The noise channels occur immediately before measurement and are implemented probabilistically:~independently and identically (i.i.d.) on each qubit per circuit repetition. We consider depolarizing, amplitude-damping, and bit-flip errors occurring with probability $p$, which are respectively
\begin{align}
    \mathcal{E}_{\mathrm{dep}}(\rho) &= (1 - p) \rho + p \frac{\I}{2},\\
    \mathcal{E}_{\mathrm{AD}}(\rho) &= E_0 \rho E_0^\dagger + E_1 \rho E_1^\dagger,\\
    E_0 =& \begin{pmatrix}
    1 & 0\\
    0 & \sqrt{1 - p}
    \end{pmatrix}, E_1 = \begin{pmatrix}
    0 & \sqrt{p}\\
    0 & 0
    \end{pmatrix} \notag\\
    \mathcal{E}_{\mathrm{BF}}(\rho) &= (1 - p) \rho + p X \rho X.
\end{align}
These models obey Assumptions~\ref{assumption_1}, although we comment that more complicated noise channels can also satisfy the assumptions, such as non-i.i.d.~errors, correlated multiqubit errors, and even coherent gate errors~\cite{chen2021robust}.

\subsection{QVM gate set and noise model}\label{subsec:noise_model_details}

\begin{figure}
\centering
\includegraphics[width=\textwidth]{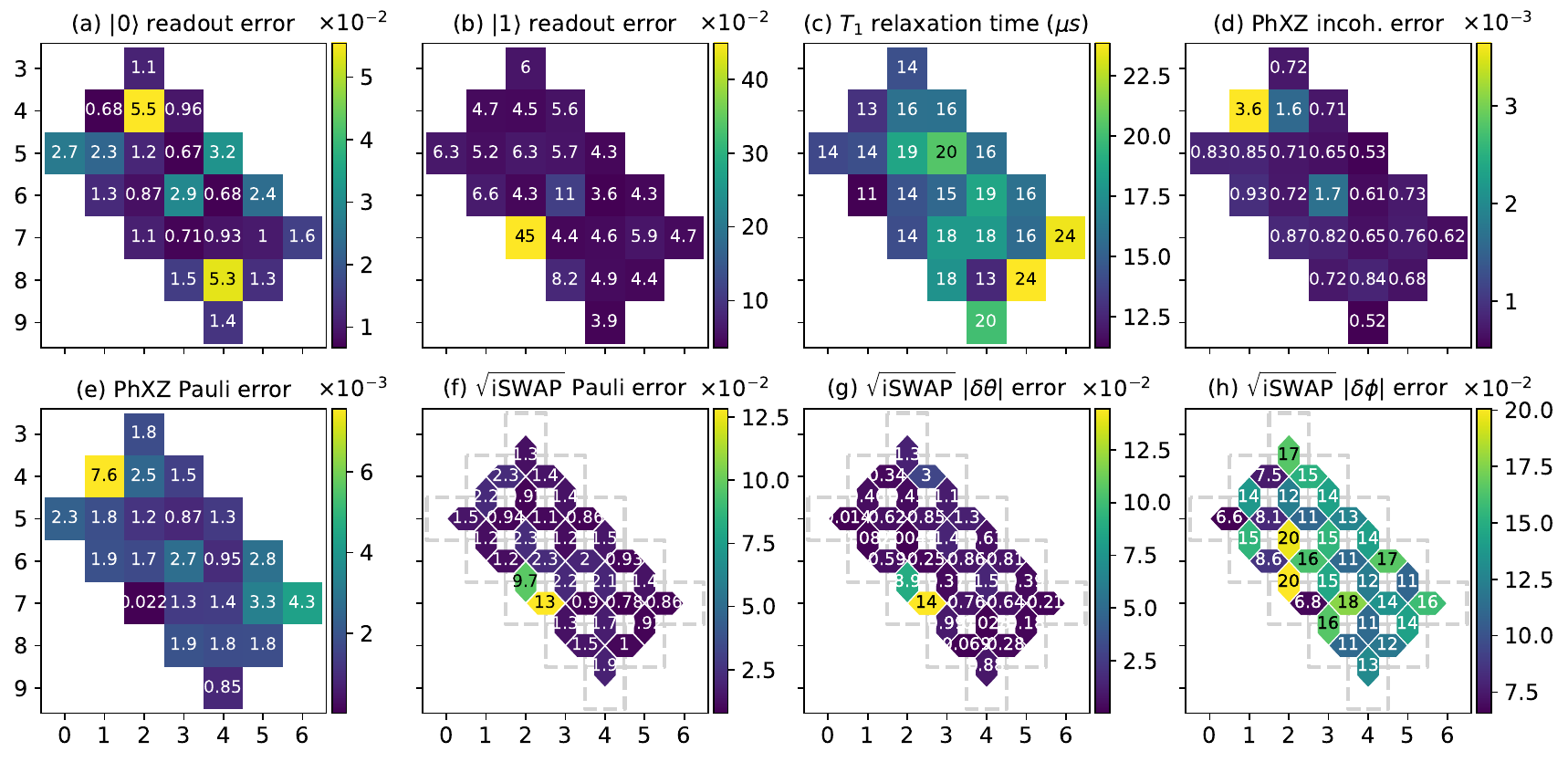}
\caption[Chip layout, connectivity, and error rates of the Google Sycamore Rainbow processor, as simulated by the QVM.]{Chip layout, connectivity, and error rates of the Google Sycamore Rainbow processor, as simulated by the QVM. Our eight-qubit numerical experiments use the $2 \times 4$ grid spanning from qubit $(5, 1)$ to qubit $(6, 4)$. (a), (b) Readout error probabilities, measured in parallel. (c) Characteristic $T_1$ qubit decay times, measured in isolation. (d) Incoherent component of single-qubit gate errors, measured in isolation from RB and purity benchmarking data. Used in conjunction with $T_1$ to infer the $T_2$ dephasing time. (e) Total Pauli error of single-qubit gates, measured in isolation via RB. (f) Total Pauli error of two-qubit $\sqrt{\iSWAP}$ gates, measured in parallel via XEB. (g), (h) Coherent errors in two-qubit gates $\sqrt{\iSWAP} = \fSim(\theta = -\frac{\pi}{4}, \phi = 0)$, measured in parallel from XEB data. We display the magnitudes $|\delta\theta|, |\delta\phi|$ for visualization purposes;~full calibration data (including signs) is available at the Cirq open-source repository~\cite{cirq}.}
\label{fig:calibration_data}
\end{figure}

The noise model we implement on the Cirq Quantum Virtual Machine is based on the Google Sycamore processor ``Rainbow,'' a 2D grid of 23 superconducting qubits. We use the calibration data obtained from November 16, 2021, which can be found in the Cirq open-source repository~\cite{cirq}. The native gate set that we compile our circuits to include single-qubit rotations in the form of phased XZ gates,
\begin{equation}
\begin{split}
    &\PhXZ(x, z, a)\\
    &= \begin{pmatrix}
    e^{\i \frac{\pi x}{2}} \cos\l(\frac{\pi x}{2}\r) & -\i e^{\i \pi(\frac{x}{2} - a)} \sin\l(\frac{\pi x}{2}\r)\\
    -\i e^{\i \pi(\frac{x}{2} + a + z)} \sin\l(\frac{\pi x}{2}\r) & e^{\i \pi(\frac{x}{2} + z)} \cos\l(\frac{\pi x}{2}\r)
    \end{pmatrix}\\
    &= Z^z Z^a X^x Z^{-a}.
\end{split}
\end{equation}
This describes a rotation by $\pi x$ about an axis determined by the parameter $a$ within the $xy$ plane, followed by a phasing of $\pi z$. The native two-qubit gates that we use are
\begin{equation}
\begin{split}
    \sqrt{\iSWAP} &= \begin{pmatrix}
    1 & 0 & 0 & 0\\
    0 & \frac{1}{\sqrt{2}} & \frac{\i}{\sqrt{2}} & 0\\
    0 & \frac{\i}{\sqrt{2}} & \frac{1}{\sqrt{2}} & 0\\
    0 & 0 & 0 & 1
    \end{pmatrix}\\
    &= e^{\i \frac{\pi}{4} (X \otimes X + Y \otimes Y) / 2},
\end{split}
\end{equation}
constrained to the nearest-neighbor connectivity of the chip.

The QVM noise model that we simulate is not fully comprehensive of all types of errors occurring in an actual device, however it captures the most dominant error sources in the superconducting platform~\cite{isakov2021simulations}. It consists of four categories:
\begin{enumerate}
    \item Readout errors are modeled as asymmetric bit-flip channels on each qubit. The asymmetry reflects the fact that the probability of a $\ket{1}$ outcome being erroneously measured as $\ket{0}$ is generally higher than misreading a $\ket{0}$ outcome. Although the errors are modeled as single-qubit channels, the calibration data is taken from parallel experiments, to potentially account for effects such as readout crosstalk and other unintended interactions between qubits.

    \item Decay ($T_1$) and dephasing ($T_2$) errors occur whenever a qubit idles during a moment (layer) of a circuit. Both $T_1$ and $T_2$ relaxations are incorporated into a single channel,
    \begin{equation}
    \mathcal{E}_{\mathrm{idle}}(\rho) = \begin{pmatrix}
    1 - \rho_{11} e^{-t/T_1} & \rho_{01} e^{-t/T_2}\\
    \rho_{10} e^{-t/T_2} & \rho_{11} e^{-t/T_1}
    \end{pmatrix}.
    \end{equation}
    The decay time $T_1$ is characterized by a simple experiment that prepares $\ket{1}$ and measures the survival probability as a function of $t$. This experiment is performed in isolation, i.e., one qubit at a time while all other qubits on the chip idle.
    
    The $T_2$ time is determined from the equation
    \begin{equation}
    \frac{1}{T_2} = \frac{1}{2T_1} + \frac{1}{T_\phi},
    \end{equation}
    where $1/T_\phi$ is the pure dephasing rate that can in principle be measured by Ramsey interferometry. For simplicity, however, this noise model instead approximates $T_\phi$ from the total single-qubit incoherent error $\epsilon_{\mathrm{inc}}$, which is determined by purity benchmarking~\cite{wallman2015estimating,feng2016estimating} performed in isolation. To leading order, $T_\phi$ is approximated using the relation
    \begin{equation}
    \epsilon_{\mathrm{inc}} = \frac{t}{3T_1} + \frac{t}{3T_\phi} + \O(t^2).
    \end{equation}
    The time $t$ which appears in the model channel $\mathcal{E}_{\mathrm{idle}}$ is the longest gate duration occurring within that moment:~$\PhXZ$ gates have a duration of $25$~ns, while $\sqrt{\iSWAP}$ gates take $32$~ns.
    
    \item Single-qubit gate errors are modeled as depolarizing channels occurring after each gate. The depolarizing rate is set to match the total single-qubit Pauli error, which is measured from the device via randomized benchmarking (RB)~\cite{magesan2011scalable,magesan2012characterizing} in isolation.
    
    \item Two-qubit gate errors are modeled with both coherent and incoherent components. The coherent contribution uses the fact that $\sqrt{\iSWAP}$ is an instance of the general fermionic simulation ($\fSim$) gate,
    \begin{equation}
    \fSim(\theta, \phi) = \begin{pmatrix}
    1 & 0 & 0 & 0\\
    0 & \cos\theta & -\i\sin\theta & 0\\
    0 & -\i\sin\theta & \cos\theta & 0\\
    0 & 0 & 0 & e^{-\i\phi}
    \end{pmatrix},
    \end{equation}
    which is a native, tunable interaction on the superconducting platform. The $\sqrt{\iSWAP}$ gate is the instance $(\theta, \phi) = (-\frac{\pi}{4}, 0)$. Coherent errors are thus modeled as an overrotation by $(\delta\theta, \delta\phi)$, which are determined for each pair of connected qubits by fitting to cross-entropy benchmarking (XEB) data using random cycles of gates across the chip~\cite{boixo2018characterizing,neill2018blueprint,arute2019quantum}.
    
    After the coherent overrotation, an incoherent error follows, modeled as a two-qubit depolarizing channel. The depolarizing rate $r_{\mathrm{dep}}^{(i, j)}$ for each pair $(i, j)$ of connected qubits is inferred as follows:~from the total XEB Pauli error $r_{\mathrm{XEB}}^{(i, j)}$, we subtract off the single-qubit incoherent error rates $r_{\mathrm{inc}}^{(i)}, r_{\mathrm{inc}}^{(j)}$ (determined from RB), as well as the average entangling error rate $r_{\mathrm{ent}}^{(i, j)}$, which are calculated using the coherent errors $\delta\theta, \delta\phi$. The model's two-qubit depolarizing rate is then set to account for the remaining amount of error:
    \begin{equation}
    r_{\mathrm{dep}}^{(i, j)} = r_{\mathrm{XEB}}^{(i, j)} - r_{\mathrm{inc}}^{(i)} - r_{\mathrm{inc}}^{(j)} - r_{\mathrm{ent}}^{(i, j)}.
    \end{equation}
    Due to the nature of XEB, both two-qubit error sources are characterized by parallel experimental data.
\end{enumerate}
Further details of the noise model, its numerical implementation, and the calibration-data acquisition are described in Ref.~\cite{isakov2021simulations}, as well as in the Cirq repository~\cite{cirq}. For completeness, in Figure~\ref{fig:calibration_data} we display a series of plots which show the chip connectivity and numerical values of the calibration data used for the various errors described above.

\subsection{Compiling circuits to the native gate set}

Single-qubit rotations are compiled into $\PhXZ$ gates according to an Euler-angle decomposition. Two-qubit unitaries are compiled into at most three $\sqrt{\iSWAP}$ gates (interleaved with single-qubit rotations) by a KAK decomposition, although most two-qubit unitaries (79\% with respect to the Haar measure) can be implemented with just two $\sqrt{\iSWAP}$ gates~\cite{huang2023quantum}. After compiling the entire circuit into this gate set, single-qubit rotations are concatenated into a single $\PhXZ$ gate whenever possible. All operations besides readout are pushed as early into the circuit as possible.

One exception we make is in the random permutation circuits $S_\pi$ appearing in the group $\SymCl{n}$ (for subsystem-symmetrized Pauli shadows). First, we decompose $\pi$ into an parallelized network of adjacent transpositions using an odd--even sorting algorithm~\cite{habermann1972parallel}. Each transposition $i \leftrightarrow j$ corresponds to a $\SWAP$ gate between qubits $i$ and $j$. However, rather than compile $\mathrm{SWAP}$ to the gate set directly (which would require three $\sqrt{\iSWAP}$ gates and four layers of $\PhXZ^{\otimes 2}$ gates), we instead implement the unitary
\begin{equation}
    \iSWAP = \sqrt{\iSWAP} \times \sqrt{\iSWAP},
\end{equation}
which uses only two $\sqrt{\iSWAP}$ gates and no single-qubit gates. The $\iSWAP$ gate differs from $\SWAP$ only by a phasing of $\i$ on the basis states $\ket{01}$ and $\ket{10}$. Such a replacement is valid because $S_\pi$ occurs only at the end of the circuit, immediately before readout. Thus while this phasing is technically unwanted, it has no observable effect on the measurement outcomes.

Finally, we note that the Trotter circuits for our Fermi--Hubbard simulations are optimized for the Sycamore architecture according to Ref.~\cite{arute2020observation}, which we follow closely. In particular, open-source code for their implementation can be found in Ref.~\cite{recirq}.

\subsection{Qubit assignment averaging}\label{subsec:qaa}

Our eight-qubit numerical experiments on the QVM utilize the $2 \times 4$ grid spanning from qubits $(5, 1)$ to $(6, 4)$ (see Figure~\ref{fig:calibration_data}). To map these qubits to the simulated degrees of freedom (fermion modes or spin-$1/2$ particles), we employ qubit assignment averaging (QAA), which was introduced in Ref.~\cite{arute2020observation} in order to handle the issue of inhomogeneous error rates across a noisy quantum device. QAA works by identifying $N$ different assignments of the $n$ qubits and allocating $T/N$ of the experimental repetitions to each realization. Properties are estimated by averaging over all $T$ samples as usual. In principle, one can use a combination of shifting, rotating, and flipping the qubits throughout the chip;~for our simulations, we vary qubit assignments within the same fixed $2 \times 4$ grid.

For the Fermi--Hubbard model, we assign a spin sector to each of the parallel $1 \times 4$ qubit chains. We average over $N = 4$ different qubit assignments, defined by setting either the top or bottom chain as the spin-up chain, and ordering the four site labels starting either from the left or the right.

For the XXZ Heisenberg model, the eight-spin chain is embedded into the $2 \times 4$ grid of qubits. Each qubit assignment ($N = 12$) is defined by setting one of six qubits $\in \{(5, 1), (5, 2), (5, 3), (5, 4), (6, 4), (6, 1)\}$ as either the left end (ordered clockwise) or right end (ordered counterclockwise) of the spin chain.

\begin{figure}
\centering
\includegraphics[scale=0.5]{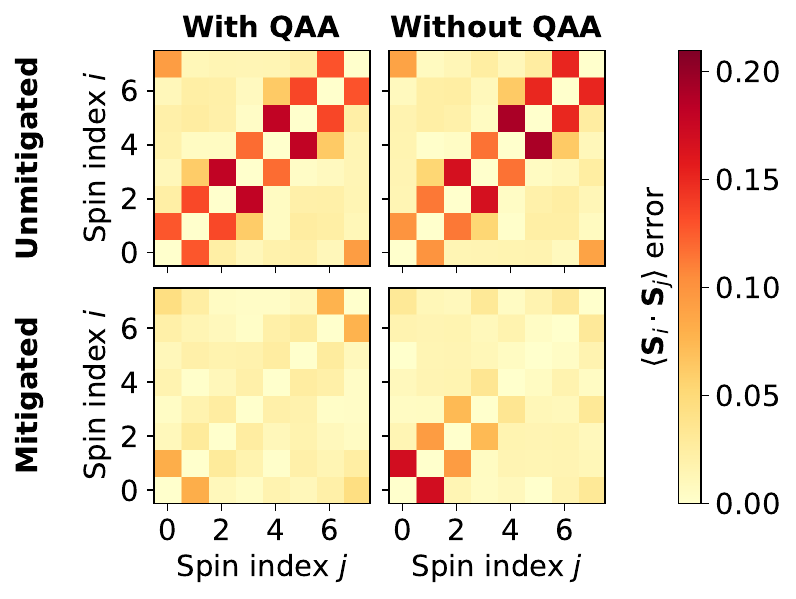}
\caption[Behavior of qubit assignment averaging (QAA), demonstrated with the XXZ spin--spin correlations.]{Behavior of qubit assignment averaging (QAA), demonstrated with the XXZ spin--spin correlations ($R = 4$ Trotter steps) and $T = 4.8 \times 10^5$ subsystem-symmetrized Pauli shadows. Experiments with QAA average over the twelve configurations described in Section~\ref{subsec:qaa}, while experiments without QAA fix the qubit ordering $0 \mapsto (5, 1), \ldots, 7 \mapsto (6, 1)$.}
\label{fig:qaa_comparison}
\end{figure}

While QAA aims to reduce device inhomogeneities, it  cannot lower the total amount of circuit noise. Thus QAA does not necessarily improve prediction accuracy with the unmitigated (standard shadow) estimators. Instead, homogenizing the noise appears to massage it into an effective form which approximately satisfies Assumptions~\ref{assumption_1} better than a single fixed configuration. We substantiate this claim with Figure~\ref{fig:qaa_comparison}, using spin--spin correlations of the XXZ model ($R = 4$ Trotter steps) as a demonstrative example. We see that the unmitigated errors are virtually identical whether or not we perform QAA. On the other hand, the symmetry-adjusted estimates with QAA exhibits a more uniform error profile and overall improved noise suppression. Further investigation into this behavior is left as an open problem.

\subsection{Bootstrapping uncertainty bars}\label{subsec:bootstrap_error_bars}

To estimate uncertainty bars, we employ empirical bootstrapping~\cite{efron1992bootstrap}, modified by batching together samples. First we summarize the original method:~given $T$ classical-shadow snapshots, one resamples that data $T$ times with replacement. Then, averages $\hat{o}_j(T)$ (being either the unmitigated or mitigated estimators) are computed from that resampled data, yielding one bootstrap sample. Repeating this $B$ times and computing the standard deviation among those $B$ bootstrap samples yields the uncertainty bar.

Due to the size $T \sim 10^6$--$10^7$ from our simulations and limitations on classical compute resources, we perform bootstrapping on batches of snapshots. Split the $T$ samples into $K$ batches (each containing $T/K$ samples) and compute $\hat{o}_j^{(k)}(T/K)$ for each batch $k = 1, \ldots, K$. Because these estimates obey $\hat{o}_j(T) = (1/K) \sum_{k=1}^K \hat{o}_j^{(k)}(T/K)$, we resample the $K$ batches (rather than all $T$ shots) to bootstrap uncertainty bars for $\hat{o}_j(T)$. Depending on $T$, we set $K \sim 10^2$--$10^3$, and for all cases we take $B = 200$.

\subsection{Estimating the gate dependence of the QVM noise model}

Here we provide an estimate of how much the QVM noise model violates Assumptions~\ref{assumption_1}. We quantify this by computing a lower bound on the minimal observable error achievable by symmetry-adjusted classical shadows.

Let $U_{\mathrm{prep}}$ be the state-preparation circuit and $U_g$ a random measurement circuit. For Schur's lemma to hold (Assumptions~\ref{assumption_1}), we require that the entire noisy circuit take the form $\mathcal{E} \mathcal{U}_g \mathcal{U}_{\mathrm{prep}}$, where $\mathcal{E}$ is the both time- and $g$-independent. While the noise model that we simulate is indeed time stationary and Markovian, the effective error channel $\mathcal{E} = \mathcal{E}_g$ depends on $g$. (This can be seen, for example, by commuting all the individual gate-level errors throughout $\widetilde{\mathcal{U}}_g$ and $\widetilde{\mathcal{U}}_{\mathrm{prep}}$ to the end of the circuit.)

In order to study this dependence on $g$, consider the decomposition
\begin{equation}
    \mathcal{E}_g = \mathcal{E}_0 + \Delta_g,
\end{equation}
where $\mathcal{E}_0$ is defined to be independent of $g \in G$. Although somewhat of an artificial decomposition, this is always mathematically possible with both $\mathcal{E}_0$ and $\Delta_g$ completely positive;~indeed, a trivial choice is $\mathcal{E}_0 = 0$. Our goal is to find the ``largest'' (in some sense) valid solution for $\mathcal{E}_0$. The remaining contribution $\Delta_g$ will then represent the minimal amount of assumption-violating noise in the model that our rigorous theory currently has no guarantees for.

From the decomposition above, the noisy measurement channel can be written as
\begin{equation}
\begin{split}
    \widetilde{\mathcal{M}} &= \E_{g \sim G} \mathcal{U}_g^\dagger \mathcal{M}_Z \mathcal{E}_g \mathcal{U}_g\\
    &= \widetilde{\mathcal{M}}_0 + \overline{\Delta},
\end{split}
\end{equation}
where
\begin{equation}
\begin{split}
    \widetilde{\mathcal{M}}_0 &= \E_{g \sim G} \mathcal{U}_g^\dagger \mathcal{M}_Z \mathcal{E}_0 \mathcal{U}_g\\
    &= \sum_{\lambda \in R_G} \widetilde{f}_\lambda(\mathcal{E}_0) \Pi_\lambda
\end{split} 
\end{equation}
is diagonal in the irreps of $G$, while the form of $\overline{\Delta} \coloneqq \E_{g \sim G} \mathcal{U}_g^\dagger \mathcal{M}_Z \Delta_g \mathcal{U}_g$ is unknown.

Applying $\mathcal{M}^{-1}$ and taking expectation values for the observables $\{O_j\}_{j=1}^L$ yields (assuming each $O_j \in V_\lambda$)
\begin{align}
    \vev{O_j}{\mathcal{M}^{-1} \widetilde{\mathcal{M}}}{\rho} &= \vev{O_j}{\mathcal{M}^{-1} \widetilde{\mathcal{M}}_0}{\rho} + \vev{O_j}{\mathcal{M}^{-1} \overline{\Delta}}{\rho} \notag\\
    &= \frac{\widetilde{f}_\lambda(\mathcal{E}_0)}{f_\lambda} \vip{O_j}{\rho} + \delta_j.
\end{align}
The terms $\delta_j \coloneqq \vev{O_j}{\mathcal{M}^{-1} \overline{\Delta}}{\rho}$ describe the deviation of observable estimates due to violations of the noise assumptions, which is precisely what we wish to quantify. For notation, denote the noisy expectations by $y_j \coloneqq \vev{O_j}{\mathcal{M}^{-1} \widetilde{\mathcal{M}}}{\rho}$ and noiseless expectations by $x_j \coloneqq \vip{O_j}{\rho}$. We collect these quantities into vectors of length $L$ and define the diagonal matrix $A \in \R^{L \times L}$ with eigenvalues $\widetilde{f}_\lambda(\mathcal{E}_0)/f_\lambda$ (in the appropriate positions corresponding to the irreps). This yields in the linear relationship
\begin{equation}
    \bm{\delta} = \bm{y} - A\bm{x}.
\end{equation}
This equation is underconstrained, so we opt for an estimate of $\bm{\delta}$ by bounding its norm from below. Namely, let $\hat{A}$ be a diagonal matrix of free parameters $0 \leq \xi_\lambda \leq 1$, which we optimize by nonnegative least-squares (NNLS) minimization:
\begin{equation}\label{eq:lower_bound_min}
    \| \bm{\delta} \|_2^2 \geq \min_{0 \leq \{ \xi_\lambda \}_{\lambda \in R'} \leq 1} \| \bm{y} - \hat{A}\bm{x} \|_2^2.
\end{equation}
Define $\hat{\bm{\delta}} \coloneqq \bm{y} - \hat{A}\bm{x}$ as the solution to this problem. In this sense, $\hat{\bm{\delta}}$ represents an error floor beyond which our theory for symmetry adjustment cannot mitigate due to inherent violations of Assumptions~\ref{assumption_1}.

\begin{figure}
\centering
\includegraphics[scale=0.5]{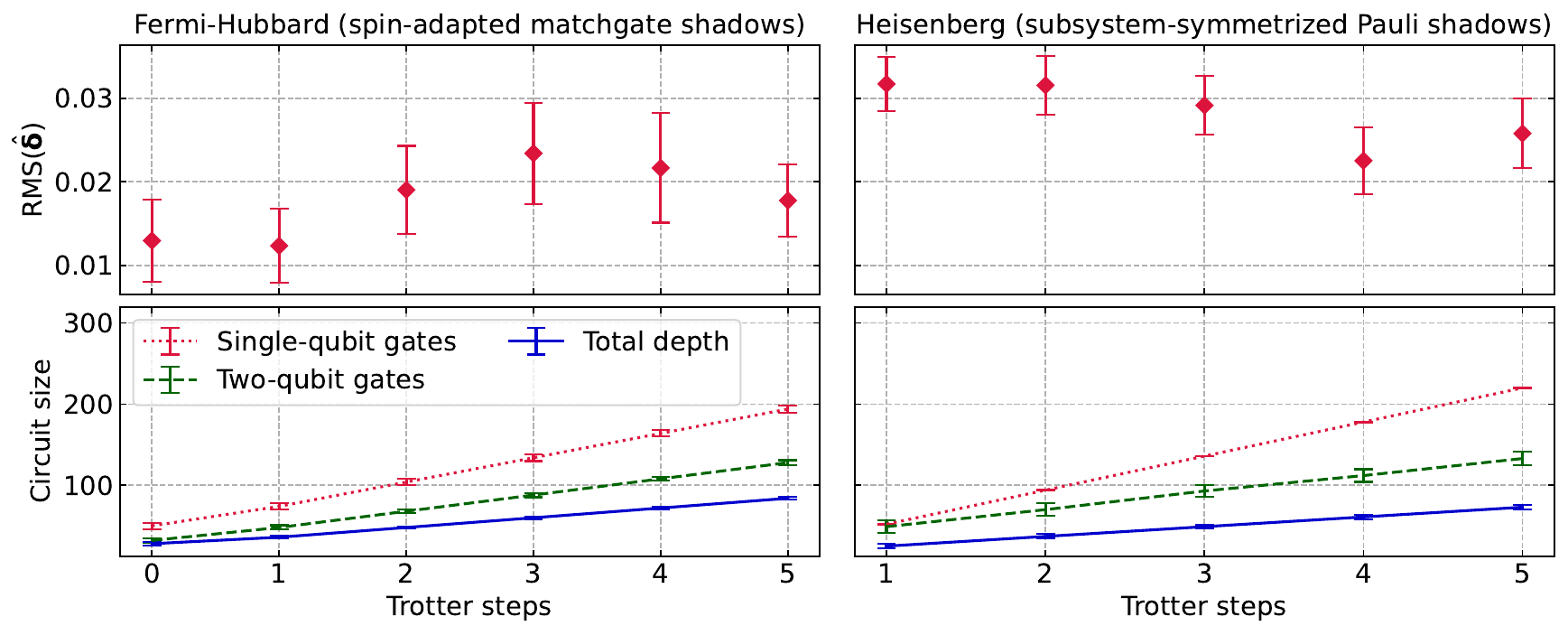}
\caption[Estimated deviation of the QVM noise from a gate-independent model.]{Estimated deviation of the QVM noise from a gate-independent model. (Top) Lower bound on the root-mean-square deviation of one- and two-body Majorana/Pauli expectation values as the number of Trotter steps (noisy circuit depth) grows. The estimated vector of deviations $\hat{\bm{\delta}}$ is described in Eq.~\eqref{eq:lower_bound_min}. Uncertainty bars are calculated by empirical bootstrapping. (Bottom) Metrics for the size of the full circuit (state preparation via Trotterization and the random measurement unitary). We report the number of single-qubit ($\PhXZ$) and two-qubit ($\sqrt{\iSWAP}$) gates used, as well as the overall compiled circuit depth. Uncertainty bars are one standard deviation variations in the random measurement circuits.}
\label{fig:delta_study}
\end{figure}

In Figure~\ref{fig:delta_study} we plot the root mean square of $\hat{\bm{\delta}}$,
\begin{equation}
    \mathrm{RMS}(\hat{\bm{\delta}}) \coloneqq \frac{1}{\sqrt{L}} \| \hat{\bm{\delta}} \|_2,
\end{equation}
which quantifies the average additive error of the estimates. The observables we choose constitute local operators depending on the type of system simulated. For fermions, we consider one- and two-body Majorana operators that respect the spin adaptation. For qubits, we take strictly two-body Pauli operators. Uncertainty bars are bootstrapped as described in Section~\ref{subsec:bootstrap_error_bars}, where each bootstrap sample is obtained from the NNLS solution of the resampled data. We also show data for the Trotter circuit size:~the number of single- and two-qubit gates after compiling to the native gate set, as well as the circuit depth. Uncertainty bars here are given by one standard deviation in the size fluctuations due to the random unitaries $U_g$.

Overall, we assess that there is an error floor on the order of $10^{-2}$ per observable (recall that the observables have unit spectral norm). Interestingly, this lower bound appears roughly independent of circuit size (within uncertainty bars), perhaps indicating a saturation of the $g$-dependent contributions after a certain circuit size. In practice however, we have observed that symmetry-adjusted classical shadows only achieve mitigated errors on the order of $10^{-1}$ at the deepest circuits. We leave a closer analysis of this behavior, and whether this lower bound can actually be achieved, to future work.
\newcommand{\be}{\begin{equation}}
\newcommand{\ee}{\end{equation}}
\newcommand{\ben}{\begin{enumerate}}
\newcommand{\een}{\end{enumerate}}

\chapter{Measurement Reduction in Variational Quantum Algorithms}\label{chap:MR}

\section*{Preface}

This chapter is based on \cite{zhao2020measurement}, coauthored by the author of this dissertation, Andrew Tranter, William M.~Kirby, Shu Fay Ung, Akimasa Miyake, and Peter J.~Love. The material here has been modified in two ways:~first, the presentation has been streamlined to focus on applications to quantum chemistry. Second, a connection to free fermions is underscored in \cref{sec:FGU_from_UP}, which is new to this dissertation.

\section{\label{intro}Introduction}

Quantum simulation is a promising application of future quantum computers~\cite{feynman1982simulating,lloyd1996universal,abrams1997simulation,abrams1999quantum}. Applications in materials science, chemistry, and high-energy physics offer the prospect of significant advantages for simulation of quantum systems~\cite{wu2002polynomial,aspuru2005simulated,preskill1}. Calculations on quantum computers that would challenge the classical state of the art require large-scale, error-corrected quantum computers~\cite{babbush2018encoding}. However, quantum hardware is entering the noisy intermediate-scale quantum (NISQ) era~\cite{preskill2018quantum}, in which the machines are still too small to implement error correction but are already too large to simulate classically~\cite{arute2019quantum}. It is natural to ask whether NISQ computers can perform useful tasks in addition to demonstrations of quantum supremacy~\cite{arute2019quantum,boixo2018characterizing,harrow2017quantum}.

The variational quantum eigensolver (VQE) was developed to enable quantum estimation of ground state energies on noisy small-scale quantum computers~\cite{peruzzo2014variational}. VQE was developed as a method for quantum simulation of electronic structure and concurrently as a simulation method for quantum field theory by cavity QED~\cite{barrett2013simulating}. Contemporaneously, the quantum approximate optimization algorithm (QAOA) was developed as a variational approach to approximate solutions of classical optimization problems~\cite{farhi2014quantum}.
VQE has been widely implemented experimentally due to its simplicity and suitability for NISQ devices~\cite{peruzzo2014variational,wang2015quantum,omalley2016scalable,kandala2017hardware,hempel2018quantum,dumitrescu18a}.

VQE consists of preparation of a variational ansatz state by a low-depth parameterized quantum circuit, followed by estimation of the expectation values of the terms in the Hamiltonian, obtained by measuring each separately. This process is repeated until the statistical error on the expectation value of each term is less than some desired precision threshold. Thus, in VQE the long coherent evolutions of phase estimation are replaced by many independent and short coherent evolutions. However, the necessary number of independent measurements may become overwhelmingly large for problem sizes of ${\sim}50$ qubits, which may soon be accessible. Recently, there has been much activity in addressing this measurement problem, via numerous approaches~\cite{babbush2018low,rubin2018application,wang2019accelerated,verteletskyi2020measurement,jena2019pauli,izmaylov2019unitary,yen2020measuring,huggins2021efficient,gokhale2019on3,bonet2020nearly,crawford2019efficient,torlai2020precise}. In the present chapter, we consider the use of extra coherent resources to reduce the number of separate Pauli terms whose expectation values must be estimated. We refer to this process as \emph{term reduction}.  Our methods are closely related to those introduced in~\cite{izmaylov2019unitary,bonet2020nearly}, which we discuss later.

We consider throughout a $k$-local Pauli Hamiltonian on $n$ qubits:
\begin{equation}\label{ham}
H = \sum_{j=1}^m \alpha_j  P_j,   
\end{equation}
where the $m$ terms $ P_j \in \{ I,X,Y,Z \}^{\otimes n} $ are $k$-local Pauli operators, i.e., tensor products of the Pauli matrices and the $2\times2$ identity containing at most $k$ nonidentity tensor factors. This $k$-locality does not refer to any geometrical locality of the layout of the physical qubits. 

The Hamiltonian $H$ for $1\leq k\leq n$ and $1\leq m\leq 4^{n}$ can represent any qubit observable. Interesting cases occur for $k$ a small constant ($2\leq k\leq 4$)~\cite{farhi2014quantum} and for $k$ scaling logarithmically with $n$~\cite{bravyi2002fermionic,seeley2012bravyi}. Jordan--Wigner mappings of fermions to qubits generate Hamiltonians with $k\leq n$, albeit of a restricted form and in which $m$ is still a polynomial in $n$~\cite{somma2002simulating}. Techniques to map interesting physical Hamiltonians to Pauli Hamiltonians show that the Hamiltonian $H$ is expressive enough to represent problems in physics and chemistry ranging from condensed-matter models to molecular electronic structure to quantum field theory. Restricting to \cref{ham} is therefore not a significant limitation on the applicability of our results to the simulation of quantum systems.

Assuming measurements are to be performed in the $z$ basis on individual qubits, to simulate the terms of \cref{ham} it is necessary to map each $ P_j$ to a measurement in the computational basis (given by the tensor product of the $z$ bases for each qubit). If our NISQ device has all-to-all pairwise connectivity (as is the case for ion trap NISQ devices) then we require $k-1$ CNOT gates and up to $k$ single-qubit Clifford operations to reduce our measurement of a $k$-local Pauli operator $ P_j$ to a $z$-basis measurement~\cite{nielsen2010quantum}. If our NISQ computer has only nearest-neighbor connectivity on the line we may require an additional $O(n)$ CNOT gates to swap the qubits into an adjacent set.

Any completely commuting set of Pauli operators $S_C$ may be mapped to a set of Pauli words over $Z$ and the identity by mapping the common eigenbasis of $S_C$ to the computational basis~\cite{nielsen2010quantum}. Previous works have studied this as a method for reducing the number of measurements;~the resulting technique requires an additional $ O(n^2) $ gates, with numerical evidence for an $ O(n) $ measurement count reduction~\cite{gokhale2019on3,yen2020measuring}.
Because the eigenbasis of $S_C$ is a set of stabilizer states (with stabilizers given by elements of $S_C$ up to a sign), this map is a Clifford operation. 
Clifford operators are known to lack transformation contextuality~\cite{love17a}, i.e., they are describable by positive maps on Wigner functions.

Furthermore, Clifford operations map single Pauli operators to single Pauli operators, which means that if we desire to reduce the number of terms in the Pauli Hamiltonian Eq.~\eqref{ham}, our map must possess some non-Clifford structure.
Hence it must in general possess transformation contextuality.

We describe two methods for term reduction based on such transformations. The first technique, unitary partitioning, was previously and independently obtained in~\cite{izmaylov2019unitary,bonet2020nearly}. Our second technique provides a more efficient realization of the required transformations at the cost of some ancilla state preparation using asymmetric qubitization---an extension of the linear combination of unitaries model~\cite{LCU2012}---introduced in~\cite{babbush2019SYK}. We present these two methods in \cref{termreduction}. \cref{sec4} is devoted to analyzing electronic-structure Hamiltonians in depth. We confirm and extend the previous numerical results of~\cite{izmaylov2019unitary} observing that a linear term reduction with respect to the number of orbitals is possible. We prove that this linear reduction can always be achieved. We close the chapter with discussion and directions for future work.

\section{Term reduction for Pauli Hamiltonians}
\label{termreduction}

Given a Hamiltonian of the form \cref{ham}, we wish to reduce the number of distinct expectation values to estimate in a VQE experiment using the coherent operations of the quantum computer.
Suppose that our ansatz $|\psi_A\rangle$ is prepared by a quantum circuit $ U$ from the state $|\psi_0\rangle\equiv\ket{0}^{\otimes n}$ so that
\be
\ket{\psi_A}= U\ket{\psi_0}.
\ee
Then our experiment estimates the expectation values
\be
\langle P_j\rangle =\bra{\psi_0} U^\dagger P_j U\ket{\psi_0}.
\ee

Suppose instead we rewrite our Hamiltonian in terms of a different set of Pauli operators $\{ Q_l\}_{l=1}^{m_c}$ and unitary operations $\{ R_l\}_{l=1}^{m_c}$ as follows:
\be
H =\sum_{j=1}^m \alpha_j P_j = \sum_{l=1}^{m_c}\gamma_l
R_l^\dagger Q_l R_l.
\ee
Such decompositions give the correct variational estimate:
\begin{eqnarray}
\bra{\psi_A}  H\ket{\psi_A}&=&\sum_{j=1}^m \alpha_j\bra{\psi_A} P_j\ket{\psi_A}\\
&=& \sum_{l=1}^{m_c}\gamma_l \bra{\psi_A} R_l^\dagger Q_l R_l \ket{\psi_A}.
\end{eqnarray}
Each term labeled by $l$ is estimated by a separate prepare and measure ansatz which appends a different unitary $R_l$ to the ansatz preparation. The unitary rotations $ R_l$ therefore represent the additional coherent resources required to reduce the number of separate expectations to be obtained.

Unlike the approach of~\cite{izmaylov2019unitary}, we do not estimate the unitary operators $ R_l^\dagger Q_l R_l$ themselves. Instead, we propose to perform a set of $m_c$ experiments in which the coherent operations $ R_l$ are appended to $ U$, so that the expectation values are obtained by measuring $ Q_l$ in the resultant state. In this case, the $ R_l$ may be made as simple or as complex as the coherent resources available after the state preparation circuit allow. Term reduction therefore allows the use of VQE for larger systems by optimally using the increasing amount of coherent resources available in new devices.

\subsection{Unitary partitioning}
\label{unitarypartitioning}

We will apply rotations in the adjoint representation of $\mathfrak{su}(2^n)$ with the goal of reducing the number of Pauli terms in the Hamiltonian.
For classical algorithms the number of such terms is not a relevant variable, as one must represent all the nonzero terms of the Hamiltonian in some way. There are some general constraints on the form of terms arising from a Pauli matrix by an adjoint unitary action. We now consider what resources the $ R_l$ operations require and give constructions that achieve term reduction. These ideas were previously presented in~\cite{izmaylov2019unitary}. 

We may write
\be
R_l^\dagger Q_l R_l =\sum_j \beta_{lj} P_{f(l,j)},
\ee
where $f$ is a relabeling of generalized Pauli matrices.
Any unitary rotation of a generalized Pauli matrix is self-inverse, so $( R_l^\dagger Q_l R_l)^2={\openone}$, which implies
\be
\label{constraints}
\sum_j \beta_{lj}^2 = 1~~{\rm and}~~\sum_{j<k}\beta_{lj}\beta_{lk}\{ P_{f(l,j)}, P_{f(l,k)}\}=0.
\ee
The first constraint can be satisfied for any subset of terms by scaling the coefficients $\beta_{lj}$ by appropriately defining $\gamma_l$. The second constraint is the defining property of subsets of terms which can be combined into a single term by unitary rotation.
For the technique discussed in this section, we divide the terms of the Hamiltonian into sets in which the operators pairwise anticommute; we call such sets \emph{completely anticommuting sets}.
The second constraint in Eq.~\eqref{constraints} is trivially satisfied within each such set.
We then rescale these terms to satisfy the first constraint and seek unitary operators that map each set to a single Pauli operator.

The \emph{compatibility graph} associated to a set of Pauli operators is an undirected graph whose vertices are the operators in the set, and in which a pair of vertices is connected if the associated operators commute. Completely anticommuting sets of Pauli operators are independent sets of the compatibility graph. A partition of the operators into completely anticommuting sets is provided by a coloring of the vertices of the graph  such that no two vertices connected by an edge have the same color. The number of sets is determined by the number of colors. Graph coloring is a well-known $\mathsf{NP}$-complete problem; however, we only require the number of colors to be less than the number of vertices for our method to provide a reduction in the number of terms. A detailed study of the use of various heuristics for graph coloring for the compatibility graphs of Hamiltonians was performed in~\cite{izmaylov2019unitary}.

We now construct the rotation $ R$ that maps a completely anticommuting set to a single Pauli operator by conjugation.
Let $S$ be a set of Pauli operators appearing in the Hamiltonian such that $\{ P_j, P_k\}=0$ $\forall  P_j \neq  P_k\in S$. It will also be useful to define $s=|S|$. The set of terms corresponding to $S$ in the Hamiltonian is then written
\be
H_S = \sum_{ P_j\in S} \beta_j  P_j.
\ee
We will assume for now that the coefficients satisfy
\be
\sum_j\beta_j^2=1.
\ee
We define the following Hermitian, self-inverse operators:
\be
\mathcal{X}_{sk} = \i  P_s P_k, \quad 1\leq k\leq s-1. \label{eqn:xOperator}
\ee
It is straightforward to verify that $ \mathcal{X}_{sk}$ commutes with all $ P_j\in S$ for $j\neq s$, $j\neq k$, and that it anticommutes with  $ P_k$ and $ P_s$. 

We define the adjoint rotation generated by $ \mathcal{X}_{sk}$:
\be
R_{sk}=\exp\left(-\i\frac{\theta_{sk}}{2} \mathcal{X}_{sk}\right), \label{eqn:adjointRotation}
\ee
whose action on the terms in $H_S$ is given by
\be
\begin{split}
	R_{sk}  P_k R_{sk}^\dagger &=\cos\theta_{sk} P_k+\sin\theta_{sk} P_s,\\
	R_{sk}  P_s R_{sk}^\dagger &=-\sin\theta_{sk} P_k+\cos\theta_{sk} P_s.
\end{split}
\ee
That is, $ R_{sk}$ is an adjoint rotation acting in the space spanned by $ P_s$ and $ P_k$.

If we act on $ H_S$ with $ R_{sk}$, we obtain
\be
\begin{split}
	R_{sk}  H_S  R_{sk}^\dagger =&~(\beta_k\cos\theta_{sk} - \beta_s\sin\theta_{sk}) P_k +(\beta_k\sin\theta_{sk} + \beta_s\cos\theta_{sk}) P_s\\
	&+\sum_{ P_j\in S\setminus \{ P_k, P_s\}} \beta_j P_j.
\end{split}
\ee
Choosing $\beta_k\cos\theta_{sk}=\beta_s\sin\theta_{sk}$ therefore gives a rotation of the Hamiltonian with the $P_k$ term removed and with the norm of the term $ P_s$ increased from $\beta_s$ to $\sqrt{\beta_s^2+\beta_k^2}$.
Defining the operator
\be
\label{axis}
R_S =  R_{s(s-1)}(\theta_{s(s-1)})\cdots R_{s2}(\theta_{s2}) R_{s1}(\theta_{s1}),
\ee
where the angles $\theta_{sk}$ satisfy
\begin{equation}
\beta_1\cos\theta_{s1}=\beta_s\sin\theta_{s1},
\end{equation}
and, for $k > 1$,
\be
\beta_k\cos\theta_{sk}=\sqrt{\left(\beta_s^2 + \sum_{j=1}^{k-1}\beta_j^2\right)} \sin\theta_{sk}, \label{eqn:theta}
\ee
therefore gives
\be
\begin{split}
	R_{S}  H_S  R_{S}^\dagger &=  P_s,
\end{split}
\ee
where we used the fact that $\sum_{j=1}^s\beta_j^2=1$.  Care must be taken when choosing $\theta_{sk}$ so as to obtain the positive root.

Our decomposition strategy is therefore the following:
\be
\label{hamiltoniandecomp}
H=\sum_{j=1}^m \alpha_j P_j=\sum_{l=1}^{m_c}\gamma_{l} H_{S_l},
\ee
where
\be
H_{S_l} = \sum_{ P_j \in S_l} \beta_{lj}  P_j
\ee
has support on a set $S_l$ of self-inverse operators for which $\{ P_j, P_k\}=0$ $\forall j\neq k$ and $\sum_{j}\beta_{lj}^2=1$. Each ${H}_{S_l}$ can be obtained from a single Pauli operator by a unitary rotation as in \cref{axis}, so we can rewrite \cref{hamiltoniandecomp} as
\be
\label{rotatedhamiltonian}
H = \sum_{l=1}^{m_c}\gamma_l R_{S_l}^\dagger P_{s_l} R_{S_l},
\ee
where the $ R_{S_l}$ operators are given for each set of pairwise anticommuting operators by \cref{axis}. 

For each $ H_{S}$ we must therefore append to our ansatz preparation the set of $s-1$ operators $ R_{sk}$ (recall that $s=|S|$). For an $l$-local Hamiltonian, each of these requires $O(l)$ CNOT and single-qubit rotations to implement. Hence one exchanges $s$ separate Pauli expectation value estimations for a single expectation value estimation, at the cost of $O(sl)$ additional coherent operations. Note that directly appending these transformations to the ansatz preparation results in a factor of 2 reduction in the required coherent resources as compared to~\cite{izmaylov2019unitary}, where both $R$ and $R^\dagger$ must be implemented as controlled operations. 

The decomposition given above and in~\cite{izmaylov2019unitary} is the most direct implementation of the transformation of the Hamiltonian.  Improvement can be made through the use of ancilla qubits and more coherent resources, as we now show in Section~\ref{rotations}.

\subsection{Low-depth implementation of the rotations}
\label{rotations}

In Section~\ref{unitarypartitioning} and in Ref.~\cite{izmaylov2019unitary}, an ordered sequence of rotations is used to write a completely anticommuting set of Pauli operators as a single term. Here we will show how to use a single rotation to perform the same reduction, and show how to implement this rotation using the methods based on linear combinations of unitaries (LCU)~\cite{LCU2012}.

We define a set of operators $ H_k$ for $1\leq k\leq n$ such that $ H_1= P_1$, $ H_n=\sin\phi_{n-1}  H_{n-1}+ \cos\phi_{n-1}  P_n$. Each $ H_n$ is self-inverse, and we consider rotations of $ H_n$ around an axis that is Hilbert--Schmidt orthogonal to both $ H_{n-1}$ and $ P_n$. 
The operator defining this axis is:
\be\label{axis_b}
\mathcal{X} = \frac{\i}{2}\left[ H_{n-1}, P_n\right]. 
\ee
The operator $ \mathcal{X}$ is self-inverse, anticommutes with $ H_n$, and so $[\mathcal{X}, H_n]=2 \mathcal{X}  H_n$. Furthermore, we may show that
\begin{equation}
\mathcal{X} H_n = \i(-\sin\phi_{n-1} P_n+\cos\phi_{n-1} H_{n-1}).
\end{equation}
The operator $ \mathcal{X}$ generates the rotation 
\begin{equation}\label{req}
R=\exp(-\i\alpha  \mathcal{X}/2)=\cos(\alpha/2){\openone}-\i\sin(\alpha/2)  \mathcal{X}.
\end{equation} 
The adjoint action of $ R$ on $ H_n$ is given by
\be
R  H_n  R^\dagger = \sin(\phi_{n-1}-\alpha) H_{n-1}+\cos(\phi_{n-1}-\alpha) P_n.
\ee
Choosing $\alpha=\phi_{n-1}$ therefore gives $ R  H_n  R^\dagger =  P_n$. This is a simple constructive demonstration that any self-inverse operator supported on a set of pairwise anticommuting operators $S$ can be mapped to a single Pauli operator. (The details of these calculations can be found in \cref{app:comp_X}.)

The terms in the operator $ \mathcal{X}$ all pairwise anticommute, and $ \mathcal{X}$ squares to the identity. This yields the expression for $ R$ given in \cref{req}. As a linear combination of Pauli operators, which are unitary, this naturally suggests implementation of $ R$ using the LCU method~\cite{LCU2012}. These methods can be combined with qubitization and quantum signal processing to reduce the required gate count~\cite{low16a,low17,poulin18a,babbush2018encoding}. However, $ \mathcal{X}$ has coefficients that are $\ell_2$-normalized, whereas the standard LCU methods naturally treat Hamiltonians with $\ell_1$-normalized coefficients. Fortunately, this issue was already addressed in Ref.~\cite{babbush2019SYK}, in which an asymmetric LCU (ALCU) method was introduced. We propose the ALCU method for the implementation of $ R$. Because $ R$ is equivalent to evolution under the Hamiltonian $ \mathcal{X}$, the cost of asymmetric qubitization scales as the square root of the number of terms in $ \mathcal{X}$, and hence the use of this method offers a quadratic speedup in asymptotic scaling compared to the methods of Section~\ref{unitarypartitioning} and Ref.~\cite{izmaylov2019unitary}. 

ALCU requires $O(\log s)$ additional qubits ($s$ being the maximum size of any of the anticommuting sets) and more complex gate operations than the method of Section~\ref{unitarypartitioning} and~\cite{izmaylov2019unitary}. However, the use of these methods in the context of VQE provides a motivation to implement more sophisticated quantum algorithms on NISQ devices. It should be noted that implementation of ALCU for this purpose is much simpler than its use for direct simulation of time evolution under the original Hamiltonian. This is because the number of terms in $ \mathcal{X}$ is only equal to the number of terms in an anticommuting set. As we discuss in detail below, this can be made smaller in order to take advantage of any additional coherent resources available after state preparation.

\subsection{Commuting terms}

Requiring that the sets of terms to be combined anticommute, as in Sections~\ref{unitarypartitioning} and \ref{rotations}, is sufficient but not necessary to perform term reduction. If there is additional structure on the coefficients of the Hamiltonian, the second constraint in \cref{constraints} may be satisfied without the individual terms all vanishing. Here we consider the possibility that for some $l$,
\be
\sum_{j<k}\beta_{lj}\beta_{lk}\{ P_j, P_k\}=0,
\ee
while the individual terms are nonzero (note that we have simplified the labeling of the Pauli terms). Because generalized Pauli matrices have the property that they either commute or anticommute, we can restrict attention to the subset of the operators that commute. We then require that
\be
\sum_{j<k}\beta_{lj}\beta_{lk}\{ P_j, P_k\}=2\sum_{S(l,j,k)} \beta_{lj}\beta_{lk}  P_j P_k=0,
\ee
where $S(l,j,k)$ is the set of indices satisfying $j<k$ and $[ P_j, P_k]=0$. Each term here is nonzero, so the condition must be enforced by cancellation of pairs, i.e., due to relations of the form
\be
\beta_{lj}\beta_{lk}  P_j P_k + \beta_{ls}\beta_{lr}  P_s P_r= 0.
\ee
This can only be true if $|\beta_{lj}\beta_{lk}|=|\beta_{ls}\beta_{lr}|$, and so this possibility of term reduction depends on the details of the coefficients more sensitively than simply requiring all terms to anticommute in a particular subset. 

Supposing that the conditions on pairs of coefficients are satisfied, we also require that
\be
P_j P_k \pm  P_s P_r =0
\ee
(for $\beta_{lj}\beta_{lk}=\pm \beta_{ls}\beta_{lr}$).
Suppose the pairs $(j,k)$ and $(s,r)$ have one operator in common, $j=s$. Then our requirement is $ P_k=\pm  P_r$, meaning that $(j,k)$ and $(s,r)$ are the same pair. Hence the pairs $(j,k)$ and $(s,r)$ must be completely distinct. This implies that $ P_j P_k= P_t$ and $\pm  P_s P_r= P_t$. This is perfectly possible:~for example, if $ P_k=IX$, $ P_j=XI$, $ P_r=ZZ$, and $ P_s=YY$, then $ P_k P_j=XX$ and $ P_r P_s=-XX$. We leave further investigation of this possibility for term reduction to future work.

\subsection{Total measurement cost estimates}

Achieving precision $ \epsilon $ in the estimate of the expectation value $ \langle {H} \rangle $ requires a statistically significant sample of qubit measurements for each Pauli term in $ {H} $. Naively, this requires approximately $ |\alpha_j|^2/\epsilon^2 $ measurements for the $ j $th term, where $ \alpha_j $ is its associated weight. However, it was proposed in~\cite{wecker2015progress}, and formally proven in~\cite{rubin2018application}, that the optimal number of measurements per term is
\be
\label{singletermmeasurements}
M_j = \frac{|\alpha_j|\sigma_j}{\epsilon^2} \left( \sum_{k=1}^m |\alpha_k| \sigma_k \right),
\ee
where $ \sigma_j^2 = \langle {P}_j^2 \rangle - \langle {P}_j \rangle^2 $ is the operator variance of the $ j $th term. Using $ \sigma_j^2 \leq 1 $ for all self-inverse operators, the upper bound for the total number of measurements to estimate the full Hamiltonian is~\cite{rubin2018application}
\be
M = \sum_{j=1}^m M_j = \left( \frac{1}{\epsilon} \sum_{j=1}^m |\alpha_j|\sigma_j \right)^2 \leq \frac{\Lambda^2}{\epsilon^2},
\ee
where $ \Lambda = \sum_{j=1}^m |\alpha_j| $ is the $ \ell_1 $-norm of the Hamiltonian weights.

Using the standard inequalities
\be
\label{lpinequality}
\frac{1}{\sqrt{d}}\|{x}\|_1 \leq \|{x}\|_2 \leq \|{x}\|_1
\ee
for any $ x \in \mathbb{R}^d $, where $ \| \cdot \|_p $ denotes the $ \ell_p $-norm, we may establish bounds for the value of $ \Lambda^2 $ after transforming the Hamiltonian via unitary partitioning. We reuse the notation of \cref{hamiltoniandecomp,rotatedhamiltonian}, so that
\be
{H} = \sum_{j=1}^m \alpha_j P_j
\ee
is the Hamiltonian as given, and
\be
{H} = \sum_{l=1}^{m_c}\gamma_l R_{S_l}^\dagger P_l R_{S_l}
\ee
is its form after unitary partitioning. Note that $ R_{S_l}^\dagger P_l R_{S_l} $ is self-inverse, so the variances remain bounded by 1. Since the coefficients associated with each anticommuting set $ S_l $ must be $ \ell_2 $-normalized, we have
\be
\gamma_l^2 = \sum_{k\in S_l} \alpha_k^2.
\ee
By abuse of notation, here we use $ S_l $ to denote the index set on which its elements are supported.

Let $ \Lambda $ be the $ \ell_1 $-norm of the weights $ \{ \alpha_j \}_{j=1}^m $ as before, and $ \Lambda_c $ be the $ \ell_1 $-norm of $ \{ \gamma_l \}_{l=1}^{m_c} $. Then, using the right-hand inequality of \cref{lpinequality}, we obtain
\be
\label{Lambdaupper}
\begin{split}
	\Lambda_c = \sum_{l=1}^{m_c} |\gamma_l| &= \sum_{l=1}^{m_c} \sqrt{ \sum_{k\in S_l} \alpha_k^2 } \\
	&\leq \sum_{l=1}^{m_c} \sum_{k\in S_l}  |\alpha_k| \\
	&= \sum_{j=1}^{m} |\alpha_j| = \Lambda.
\end{split}
\ee
Thus $ \Lambda_c \leq \Lambda $, and in fact this bound is saturated only if no partitioning is performed at all.

Applying the left-hand inequality of \cref{lpinequality} to the first line of \cref{Lambdaupper} yields
\be
\label{Lambdalower1}
\sum_{l=1}^{m_c} \Bigg( \frac{1}{\sqrt{|S_l|}} \sum_{k\in S_l}  |\alpha_k| \Bigg) \leq \Lambda_c.
\ee
Let $ s_{\mathrm{max}} = \max_{l} |S_l| $ be the size of the largest set in the partition. Then
\be
\label{Lambdalower2}
\frac{1}{\sqrt{s_{\mathrm{max}}}}  \sum_{l=1}^{m_c} \sum_{k\in S_l}  |\alpha_k| = \frac{\Lambda}{\sqrt{s_{\mathrm{max}}}} \leq \Lambda_c.
\ee
Bounding the set sizes by $ s_{\mathrm{max}} $ is fairly tight if they are all roughly equal, which is both desirable (since the gate complexity scales with the set size) and always possible (one may take a large set and simply divide it into smaller ones, which remain fully anticommuting). Roughly speaking, the number of measurements $ M_c $ may be thought of as being lower bounded by $ M/s_{\mathrm{max}} $, although this is not the whole story, since $ \Lambda $ (resp.~$ \Lambda_c $) is itself an upper bound estimate for $ M $ (resp.~$ M_c $). Equation~\eqref{Lambdalower2} gives only an approximate sense for the maximum amount of measurement reduction possible by unitary partitioning when taking into account the statistical repetitions.

It is worth noting that this lower bound is saturated when $ |\alpha_j| = |\alpha_k| $ $ \forall j,k $. In fact, a weaker condition saturates the tighter bound of \cref{Lambdalower1}. There we require only that $ |\alpha_j| = |\alpha_k| $ $ \forall j,k \in S_l $ for each $ l $---that is, the coefficient magnitudes are uniform within each set. Supposing that this approximately holds, and again that all $ |S_l| $ are roughly the same, yields $ \Lambda_c \approx \Lambda/\sqrt{s_{\mathrm{max}}} $.

Thus partitioning with additional constraints respecting these coefficient conditions may result in more measurement reduction, without requiring any additional coherent rotations. The partitioning algorithm would then require significantly more classical computational resources, as this is now a \emph{weighted} graph coloring problem, but in principle these ideas may be implemented straightforwardly. For the analysis in the following section, we focus only on the number of unique Hamiltonian terms before and after partitioning as a rough estimate for the amount of measurement reduction achieved by our method.

\section{Electronic-structure Hamiltonians}
\label{sec4}

Quantum chemistry simulations are expected to be an important use of variational quantum algorithms~\cite{olsonQuantumInformationComputation2017}. The goal is to find the eigenvalues and eigenvectors of the molecular electronic Hamiltonian
\begin{equation}\label{eq:H_e}
H = \sum_{p,q} h_{pq} a_p^\dagger a_q + \frac{1}{2}\sum_{p,q,r,s} h_{pqrs} a_p^\dagger a_q^\dagger a_r a_s,
\end{equation}
where $ a_p^\dagger $ and $ a_p $ are fermionic creation and annihilation operators acting on the space spanned by molecular spin orbitals $ \chi_p $. For computational purposes, this basis set is truncated to the first $ N $ orbitals. The fermionic operators satisfy the canonical anticommutation relations
\begin{equation}\label{eq:fermion_CAR}
\begin{split}
\{ a_p^\dagger,a_q^\dagger \} &= \{ a_p,a_q \} = 0,\\
\{ a_p,a_q^\dagger \} &= \delta_{pq}{\openone}.
\end{split}
\end{equation}
The weights $h_{pq}$ and $h_{pqrs}$ are defined as
\begin{align}
h_{pq} &= \delta_{\sigma_p\sigma_q} \int d^3r\, \chi_p^*(\mathbf{r}) \left( -\frac{\nabla^2}{2} - \sum_I \frac{\zeta_I}{|\mathbf{r}-\mathbf{R}_I|} \right) \chi_q(\mathbf{r}),\\
h_{pqrs} &= \delta_{\sigma_p\sigma_s} \delta_{\sigma_q\sigma_r} \int d^3r_1 d^3r_2\, \frac{\chi_p^*(\mathbf{r}_1)\chi_q^*(\mathbf{r}_2)\chi_r(\mathbf{r}_2)\chi_s(\mathbf{r}_1)}{|\mathbf{r}_1-\mathbf{r}_2|},
\end{align}
where $ \mathbf{r} $ denotes the electronic spatial coordinates, $ \sigma_p \in \{ \uparrow,\downarrow \} $ is the spin value of the $ p $th orbital, and $ \{ \mathbf{R}_I \}_I $ and $ \{ \zeta_I \}_I $ are the molecule's classical nuclear positions and their associated charges, respectively. These spatial integrals can be efficiently pre-computed on a classical computer. For use in a quantum algorithm, the Hamiltonian is then transformed to a weighted sum of Pauli strings using a fermion-to-qubit encoding, such as the Jordan--Wigner~\cite{jordanwigner}, Bravyi--Kitaev~\cite{bravyi2002fermionic,seeley2012bravyi,tranter2015bravyi}, or other similar~\cite{setia17a} mappings. For the former two encodings, the number $n$ of qubits is the same as the number $N$ of molecular spin orbitals.  The expectation value of each Pauli string is measured independently.  The power of this approach stems from the ability to prepare ansatz states that cannot be efficiently constructed on a classical computer;~these are typically derived from a unitary coupled cluster ansatz~\cite{mccleanTheoryVariationalHybrid2016,romeroStrategiesQuantumComputing2018,leeGeneralizedUnitaryCoupled2019}. This allows for efficient computation of high-precision eigenvalues, which has importance when considering calculations that require such precision, such as reaction kinetics and dynamics.

Implementation of this procedure for chemical systems at the desired accuracy is challenging. For chemistry, the required precision is typically considered to be a constant $1$~kcal/mol, or $1.6$~mHa. This level of precision is roughly commensurate with that obtained by experimental techniques in thermochemistry.  Recall from \cref{singletermmeasurements} that the number of independent measurements that must be performed to estimate the expectation value of a single term with weight $ h $ to precision $ \epsilon $ is $O(\Lambda|h|/\epsilon^2)$.  For chemical accuracy, this means that each term requires on the order of hundreds of thousands of independent measurements, each of which requires a separate ansatz preparation stage.  This must be repeated for each step of the variational optimisation, for each of the $O(N^4)$ terms in the molecular Hamiltonian (noting that using the Jordan--Wigner transformation requires up to 16 Pauli strings for each term).  As such, this quantum chemistry problem has recently garnered much interest with regard to reducing VQE measurement costs~\cite{wang2019accelerated,huggins2021efficient,babbush2018low,bonet2020nearly,gokhale2019on3,yen2020measuring,izmaylov2019unitary}. The term reduction strategy discussed in \cref{termreduction} appears a promising way to reduce the overall resources required by utilising available coherent computational resources subsequent to ansatz preparation.

In the absence of restrictions on the length of circuits that can be performed coherently, the term reduction strategy reduces the number of expectation values that must be independently estimated, going from the number of Hamiltonian terms to the number of fully anticommuting sets of terms.  The main task is therefore to partition the Hamiltonian into such sets.  The effectiveness of this term reduction strategy can be quantified by examining the number of fully anticommuting sets for a given Hamiltonian with respect to both the number of orbitals and the total number terms in the unmodified Hamiltonian. In Section~\ref{subsec:linearReductionInTerms}, we show that it is always possible to reduce the number of terms from $ O(N^4) $ to at most $ O(N^3) $ for any electronic-structure Hamiltonian. In Section~\ref{subsec:pauliLevelColouring}, we perform numerical studies using specific molecules and compare the results to our analytic construction. We also consider how the constraint of circuit size affects one's ability to construct such partitions.

\subsection{Majorana operators}
The approach we take here will be agnostic to the choice of qubit encoding. However, in order to partition the terms into completely anticommuting sets, it will be convenient to express them using Majorana operators. This is because they place all the fermionic operators on an equal footing, are Hermitian and unitary, and obey a single anticommutation relation. Here, we briefly review the properties of these operators essential for our analysis. The single-mode Majorana operators are defined from the fermionic modes as
\begin{equation}
\gamma_{2p} = a_p + a_p^\dagger, \quad \gamma_{2p+1} = -i(a_p - a_p^\dagger).
\end{equation}
In this formalism, the anticommutation relations of Eq.~\eqref{eq:fermion_CAR} become
\begin{equation}\label{eq:majorana_CAR}
\{ \gamma_j, \gamma_k \} = 2\delta_{jk}{\openone}.
\end{equation}
These $ 2N $ single-mode operators generate a basis (up to phase factors) for the full algebra of Majorana operators via arbitrary products, i.e.,
\begin{equation}
\gamma_A = \prod_{j \in A} \gamma_j,
\end{equation}
where $ A \subseteq \{ 0,\ldots,2N-1 \} $ is the support of $ \gamma_A $. From Eq.~\eqref{eq:majorana_CAR}, it is straightforward to show that the anticommutator between two arbitrary Majorana operators $ \gamma_A $ and $ \gamma_B $ is determined by their individual supports and their overlap:
\begin{equation}\label{eq:majorana_gen_AR}
\{ \gamma_A, \gamma_B \} = \left[ 1 + (-1)^{|A||B| + |A \cap B|} \right] \gamma_A \gamma_B.
\end{equation}
This relation provides a clear picture of how to construct fully anticommuting sets of fermionic operators. Since the electronic Hamiltonian contains only terms of quadratic and quartic order, we restrict our attention to even-parity products. In this setting, we only need to examine the overlap of the Majorana operators' supports:~if $ |A \cap B| $ is odd (i.e., the two operators share an odd number of single-mode indices), then they anticommute.

\subsection{Linear reduction in terms}
\label{subsec:linearReductionInTerms}
Since there are no spin interaction terms in our Hamiltonian, we can always choose molecular orbital basis functions $ \chi_p $ which are real-valued. With this, it follows that $ h_{pq}, h_{pqrs} \in \mathbb{R} $, and in particular, we have the permutational symmetries
\begin{align}
h_{pq} &= h_{qp}, \label{eq:hpq_sym}\\
h_{pqrs} = h_{sqrp} &= h_{prqs} = h_{srqp}. \label{eq:hpqrs_sym_1}
\end{align}
Furthermore, the canonical anticommutation relations give $ a_p^\dagger a_q^\dagger a_r a_s = a_q^\dagger a_p^\dagger a_s a_r $, which implies that
\begin{equation}\label{eq:hpqrs_sym_2}
h_{pqrs} = h_{qpsr},
\end{equation}
for a total of eight permutational symmetries in the two-body integrals. Using these symmetries and the generalized anticommutation relation, Eq.~\eqref{eq:majorana_gen_AR}, one can rewrite the Hamiltonian using Majorana operators as
\begin{equation}\label{eq:H_e_majorana}
H = \tilde{h} {\openone} + \sum_{p,q} \tilde{h}_{pq} \i\gamma_{2p}\gamma_{2q+1} + \frac{1}{2} \sum_{\substack{p,q,r,s\\p\neq q,r\neq s}} \tilde{h}_{pqrs} \gamma_{2p}\gamma_{2q}\gamma_{2r+1}\gamma_{2s+1}.
\end{equation}
We refer the reader to \cref{sec:H_e_majorana_appendix} for the details of this derivation. The redefined weights $ \tilde{h} $, $ \tilde{h}_{pq} $, and $ \tilde{h}_{pqrs} $ are given in Eq.~\eqref{eq:new_coeff}. For our present analysis, the only relevant detail here is that each term features an equal number of even and odd indices in its support. In principle, any such combination of terms may appear in the Hamiltonian. In this form, it becomes clear that there are up to $ N^2 $ quadratic terms and $ \binom{N}{2}^2 $ quartic terms.

Furthermore, since the single-mode Majorana operators are Hermitian, there is a one-to-one correspondence between Majorana operators and the respective Pauli strings obtained after a fermion-to-qubit transformation (for encodings that preserve the number of orbitals as the number of qubits). For instance, in the Jordan--Wigner encoding, we have
\begin{equation}
\begin{split}
\gamma_{2p} = X_p Z_{p-1} \cdots Z_0, \\
\gamma_{2p+1} = Y_p Z_{p-1} \cdots Z_0.
\end{split}
\end{equation}
Since the single-mode Majorana operators simply become Pauli strings, arbitrary products of them remain single Pauli strings. In contrast, if one were to deal with the fermionic operators directly, a single $ a_p^\dagger a_q^\dagger a_r a_s $ term would generate a linear combination of up to 16 unique Pauli strings. By writing the Hamiltonian in terms of Majorana operators, we have not circumvented this overhead, but rather, we have explicitly incorporated it into our term counting, while remaining encoding agnostic. In particular, many cancellations and simplifications may occur between the transformed terms, yielding the expression given above in Eq.~\eqref{eq:H_e_majorana}. Also note that any anticommuting partition in the Majorana formalism remains valid after a qubit transformation, since the anticommutation relations are preserved.

Recall from Eq.~\eqref{eq:majorana_gen_AR} that we had determined that every pair of terms anticommutes if and only if their supports intersect an odd number of times. This fact, along with the specific form of the terms appearing in Eq.~\eqref{eq:H_e_majorana}, is crucial for showing that it is always possible to partition this Hamiltonian into at most $ O(N^3) $ completely anticommuting sets.

We note that very recent results have made similar findings. In~\cite{bonet2020nearly}, it was observed that at least $ \Omega(N^3) $ sets would be necessary to divide the set of \emph{all} quartic Majorana operators, rather than the specific terms appearing in electronic-structure Hamiltonians. Meanwhile, in~\cite{gokhale2019on3}, an algorithm was presented which partitions electronic-structure terms into $ O(N^3) $ completely \emph{commuting} sets. The analysis presented there specifies the Jordan--Wigner encoding, but does not assume any of the permutational symmetries in the $ h_{pq}, h_{pqrs} $ coefficients.

We now prove our claim by providing an explicit construction of such a partition.

\begin{theorem}\label{thm:cubicPartitionTheorem}
	Let
	\begin{equation}
	\mathcal{M} = \{ \gamma_{2p}\gamma_{2q}\gamma_{2r+1}\gamma_{2s+1} \mid p < q \;{\rm and }\; r < s \}
	\end{equation}
	be the set of all possible quartic Majorana operators appearing in the electronic-structure Hamiltonian. For each triple $ (q,r,s) \in \{ 0,\ldots,N-1 \}^3 $ satisfying $ r < s $, define
	\begin{equation}
	S_{(q,r,s)} = \{ \gamma_{2p}\gamma_{2q}\gamma_{2r+1}\gamma_{2s+1} \mid p < q \}.
	\end{equation}
	These sets $ S_{(q,r,s)} $ are completely anticommuting, and they form a partition of $ \mathcal{M} $. Furthermore, there are $ O(N^3) $ such sets.
\end{theorem}

\begin{proof}
	By construction, all elements of $ S_{(q,r,s)} $ share support on exactly three indices, hence they all pairwise anticommute, per Eq.~\eqref{eq:majorana_gen_AR}. It is also straightforward to see that these sets form an exact cover of $ \mathcal{M} $:
	\begin{align}
	\left| S_{(q,r,s)} \cap S_{(q',r',s')} \right| &= q \, \delta_{qq'} \delta_{rr'} \delta_{ss'}, \label{eq:disjointness}\\
	\bigcup_{\substack{q,r,s\\r<s}} S_{(q,r,s)} &= \mathcal{M}. \label{eq:covering}
	\end{align}
	There are $ \binom{N}{2} $ values that the pair $ (r,s) $ can take and $ N-1 $ values that $ q $ can take ($ q=0 $ yields the empty set, which we ignore). A slight optimization arises from the observation that the union $ S_{(1,r,s)} \cup S_{(2,r,s)} $ remains a completely anticommuting set. Hence there are a total of $ \binom{N}{2}(N-2) = O(N^3) $ such sets.
\end{proof}

We refer the reader to \cref{sec:majorana_proof_details} for further details of the above proof. Although there are only $ O(N^2) $ quadratic terms, hence not affecting the asymptotic scaling of Theorem~\ref{thm:cubicPartitionTheorem}, they can in fact be included in the above construction with no additional overhead. Intuitively, since there are at most $ N^2 $ such operators which need to be placed into $ O(N^3) $ sets, one has a great deal of freedom in how to allocate them. As one example, consider the set
\begin{equation}
T_p = \{ \i\gamma_{2p}\gamma_{2q+1} \mid 0 \leq q \leq N-1 \}
\end{equation}
for some fixed $ p $. Then all the elements of $ T_p $ anticommute with all of some $ S_{(p,r,s)} $, except for those with $ q=r $ or $ q=s $. The new completely anticommuting set then becomes
\begin{equation}\label{eq:Sjkm_quad_1}
S_{(p,r,s)} \cup T_p \setminus \{ \i\gamma_{2p}\gamma_{2r+1}, \i\gamma_{2p}\gamma_{2s+1} \},
\end{equation}
and those two excluded operators can be placed with any other $ S_{(p,r',s')} $, where all of $ r,r',s $, and $ s' $ are different:
\begin{equation}\label{eq:Sjkm_quad_2}
S_{(p,r',s')} \cup \{ \i\gamma_{2p}\gamma_{2r+1}, \i\gamma_{2p}\gamma_{2s+1} \}.
\end{equation}
Since there are $ N $ such sets $ T_p $, this procedure combines all possible $ N^2 $ quadratic operators with only $ 2N $ of the preexisting sets of quartic operators.

We emphasize that the partition presented here is not an optimal solution to the problem. Rather, it demonstrates that even in the worst case one can always achieve term reduction by at least a factor of $ O(N) $. For a practical demonstration, we move to numerical studies of specific molecular Hamiltonians in \cref{subsec:pauliLevelColouring}.

\subsection{Fermionic Gaussian unitaries from partitions}\label{sec:FGU_from_UP}

Before turning to the numerics, we make a brief observation here regarding the form of the term-combining unitaries arising from our analytical partitioning of \cref{thm:cubicPartitionTheorem}. In particular, we will show how they are equivalent to the fermionic Gaussian unitaries discussed in \cref{chap:fermions}, \cref{subsec:FGU_bkgd}. To do so, we shall consider the quartic partitions $S_{(q,r,s)}$ and quadratic partitions $T_p$ separately, rather than combining them as in \cref{eq:Sjkm_quad_1,eq:Sjkm_quad_2}.

We start with the quadratic partitions, $T_p = \{\Gamma_{pq} \coloneqq \i\gamma_{2p}\gamma_{2q+1} \mid 0 \leq q \leq N - 1\}$, corresponding to the terms
\begin{equation}
    H_{T_p} = \sum_{\Gamma_{pq} \in T_p} \tilde{h}_{pq} \Gamma_{pq}.
\end{equation}
We will rotate this linear combination to a single term $\sqrt{\sum_{q=0}^{N-1} |\tilde{h}_{pq}|^2} \, \Gamma_{pp}$, which is diagonal in the computational basis (for instance, under Jordan--Wigner, $\Gamma_{pp} = \i\gamma_{2p}\gamma_{2p+1} = -Z_p$). Recall that the unitary which effects this transformation is $R_{T_p} = \prod_{q \neq p} R_{pq}$, where
\begin{equation}
    R_{pq} = \exp\l( -\i \frac{\theta_{pq}}{2} \mathcal{X}_{pq} \r), \quad \mathcal{X}_{pq} = \i \Gamma_{pp} \Gamma_{pq},
\end{equation}
and the angle $\theta_{pq}$ is chosen appropriately (according on the coefficients $\tilde{h}_{pq}$). The generator further simplifies to
\begin{equation}
    \mathcal{X}_{pq} = \i \Gamma_{pp} \Gamma_{pq} = \i\gamma_{2p+1}\gamma_{2q+1}.
\end{equation}
Hence $R_{pq}$ is generated by a quadratic Majorana operator, which implies that it is a fermionic Gaussian unitary $U_{Q^{(pq)}}$ for some $2N \times 2N$ orthogonal matrix $Q^{(pq)}$. Because it is generated by a single term, this matrix is easy to determine:~let $A^{(pq)} \in \R^{2N \times 2N}$ be the antisymmetric matrix with all zero entries except for $A^{(pq)}_{2p+1,2q+1} = -\theta_{pq} = -A^{(pq)}_{2q+1,2p+1}$. Then $Q^{(pq)} = e^{A^{(pq)}}$, which is a Givens rotation acting nontrivially only on the two-dimensional subspace spanned by the indices $(2p+1, 2q+1)$ as (using matrix-slicing notation)
\begin{equation}
    Q^{(pq)}[(2p+1, 2q+1), (2p+1, 2q+1)] = \begin{pmatrix}
    \cos\theta_{pq} & -\sin\theta_{pq}\\
    \sin\theta_{pq} & \cos\theta_{pq}
    \end{pmatrix}.
\end{equation}
The Gaussian description of the full circuit $R_{T_p} = U_Q$ is therefore given by simply multiplying all of these $N - 1$ Givens rotations together:
\begin{equation}
    Q = \prod_{q \neq p} Q^{(pq)} \in \Orth(2N),
\end{equation}
where the order is the same as in the definition of $R_{T_p}$ (i.e., the order can be arbitrary as long as we are consistent). The quantum circuit for this unitary can be compiled using the algorithm described in \cref{chap:EM}, \cref{sec:FGU_circuit_design}, although we do not pursue this approach for the numerics presented in this chapter.

It is then straightforward to generalize this idea to the quartic partitions $S_{(q,r,s)} = \{\gamma_{2p}\gamma_{2q}\gamma_{2r+1}\gamma_{2s+1} \mid p < q\}$ arise. The generators here are of the form
\begin{equation}
\begin{split}
    \mathcal{X}_{pp'} &= \i \gamma_{2p}\gamma_{2q}\gamma_{2r+1}\gamma_{2s+1} \times \gamma_{2p'}\gamma_{2q}\gamma_{2r+1}\gamma_{2s+1}\\
    &= \i \gamma_{2p} \gamma_{2p'}
\end{split}
\end{equation}
for $p \neq p' < q$. Because this generator is quadratic, $R_{S_{(q,r,s)}}$ is Gaussian as well. The exact same techniques as above can be used to compile this circuit, with only the differences being:~(1) the angles $\theta_{pp'}$ are determined from the coefficients $\tilde{h}_{pqrs}$, (2)~the Givens rotations act on the subspaces $(2p, 2p')$, and (3)~an extra layer of single-qubit Clifford gates must be placed after $R_{S_{(q,r,s)}}$ in order to map the combined term $\gamma_{2p} \gamma_{2q} \gamma_{2r+1} \gamma_{2s+1}$ to a diagonal Pauli operator. This last point follows because $\gamma_{2p} \gamma_{2q} \gamma_{2r+1} \gamma_{2s+1}$ is only diagonal when $\{p, q\} = \{r, s\}$, which generally does not hold. However, the quartic Majorana operator is still a single Pauli operator, so single-qubit Clifford gates suffice to map it to a Pauli-$Z$ operator.

Finally, we point out that unitary partitioning does not generically yield fermionic Gaussian circuits, as it fundamentally only looks at the anticommutative structure between terms. Instead, what we have shown is that our analytic construction of the $O(N^3)$ anticommuting sets furthermore has this Gaussian property.

\subsection{Pauli-level colouring and numerics}\label{subsec:pauliLevelColouring}

The above analysis demonstrates a reduction in difficulty of VQE by considering the number of fully anticommuting sets of terms in the electronic Hamiltonian.  Equivalently, we may consider fully anticommuting sets of terms at the level of Pauli strings, i.e., subsequent to transforming the electronic Hamiltonian with, for example, the Jordan--Wigner or Bravyi--Kitaev mappings.  This approach could hold advantage by allowing the combination of duplicate strings and allowing the combination of anticommuting Pauli subterms between different fermionic terms.  However, once the fermion-to-qubit mapping is applied, the natural symmetries of the spatial molecular orbital integrals are embedded into a complex structure.  Moreover, the anticommutativity structure of the resulting Pauli terms is difficult to predict.  As such, we turn to numerical methods.

The key metric here is the number of fully anticommuting sets in the Pauli Hamiltonian.  As discussed in \cref{unitarypartitioning}, this is equivalent to a colouring of the compatibility graph---the graph composed of nodes corresponding to terms, with edges drawn where terms commute.  Optimal graph colouring is an $\mathsf{NP}$-hard problem~\cite{gareySimplifiedNPcompleteProblems1974a},  but many approximate algorithms exist~\cite{kosowskiClassicalColoringGraphs2004}.  While minimising the number of sets is advantageous for reducing the number of measurements needed, an approximate solution is sufficient, and diminishing returns are obtained from improving the quality of the approximation.

In order to assess whether this strategy is viable for molecular Hamiltonians, we generated colouring schemes for $65$ Hamiltonians (previously used in Refs.~\cite{tranterComparisonBravyiKitaev2018,tranterOrderingTrotterizationImpact2019} and described in \cref{appendix:systems}).  Geometry specifications were obtained from the NIST CCBDB database~\cite{johnsoniiiNISTComputationalChemistry2016}.  Molecular orbital integrals in the Hartree--Fock basis were gathered using the Psi4 package~\cite{psi4} and OpenFermion~\cite{openfermion}.  Our code was then used to generate Jordan--Wigner and Bravyi--Kitaev Hamiltonians, which were divided into anticommuting subsets using the NetworkX Python package~\cite{networkx} and the greedy independent sets strategy~\cite{kosowskiClassicalColoringGraphs2004}.  As our focus was on quantifying whether the term reduction technique is viable, alternative colouring strategies were not considered;~such an analysis was performed in~\cite{izmaylov2019unitary}.  Our colouring strategy here is relatively computationally expensive, limiting our analysis to a maximum of 36 spin orbitals, with only three systems involving 30 or more.  While our code is unoptimised and can likely be improved upon, this does indicate that it would be difficult to extend this approach to larger systems.  The Majorana-based scheme of Section~\ref{subsec:linearReductionInTerms} was also used to partition the Hamiltonians.  In contrast to the greedy colouring strategy, this  does not require extensive classical computational resources.

\begin{figure}
	\centering
	\includegraphics[width=0.5\columnwidth]{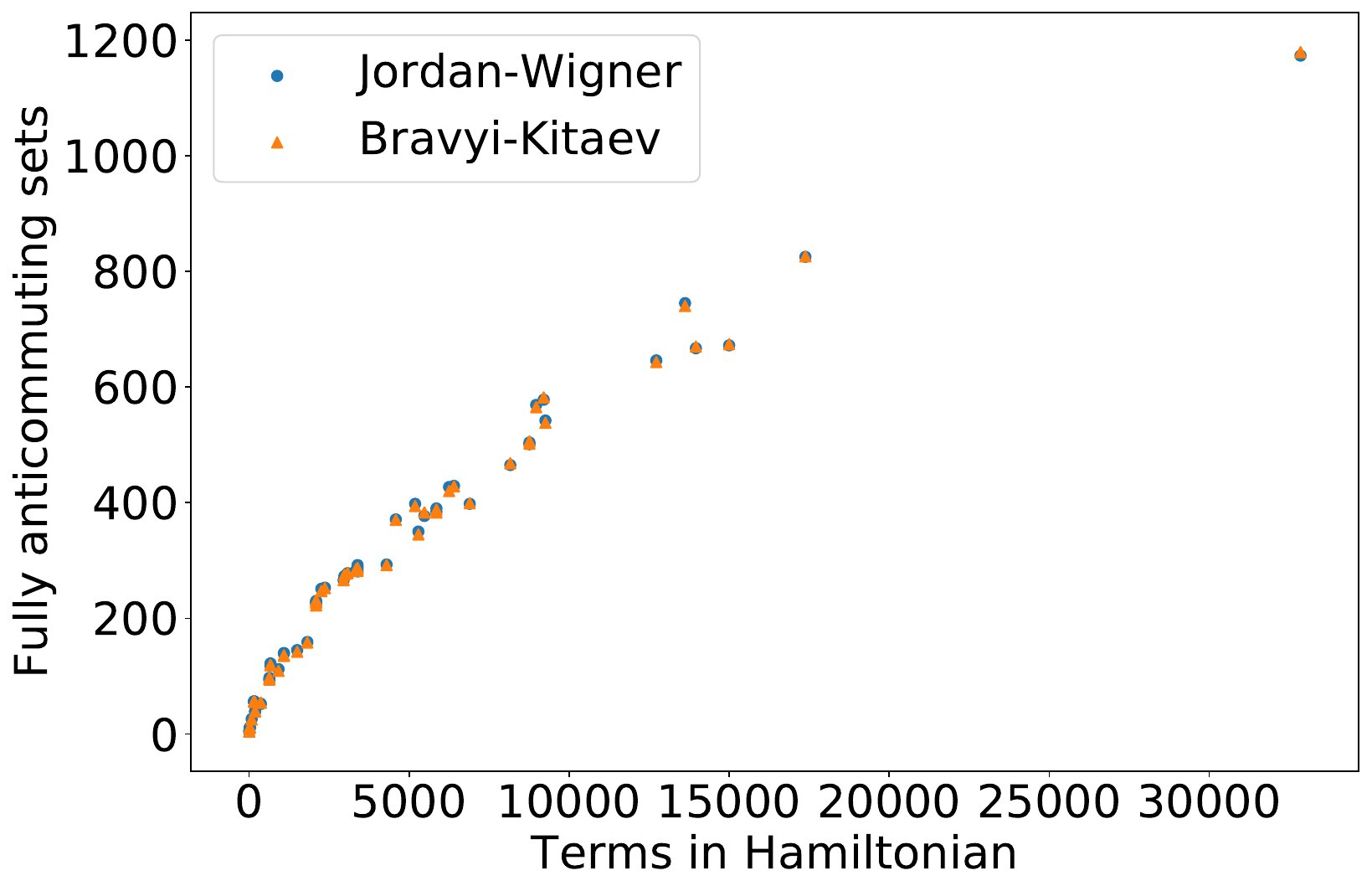}
	\caption[Number of fully anticommuting sets for electronic-structure Hamiltonians versus the number of terms in the full Pauli Hamiltonian.]{Number of fully anticommuting sets for electronic-structure Hamiltonians versus the number of terms in the full Pauli Hamiltonian, using the greedy independent sets strategy.  The number of fully anticommuting sets is at least an order of magnitude less than the number of terms. The Jordan--Wigner and Bravyi--Kitaev mappings perform almost equivalently.}
	\label{fig:numSetsVersusTerms}
\end{figure}

Figure~\ref{fig:numSetsVersusTerms} shows the number of fully anticommuting sets obtained versus the number of terms in the Hamiltonian.  The number of fully anticommuting sets is approximately an order of magnitude less than the number of terms. The choice of Jordan--Wigner and Bravyi--Kitaev mapping does not appear to meaningfully affect the number of fully anticommuting sets found, as the anticommutativity structure is dependent on the underlying molecular Hamiltonian.  Encouragingly, the agreement demonstrated here by Figure~\ref{fig:numSetsVersusTerms} suggests that the greedy independent set strategy is finding close-to-optimal colourings.

\begin{figure}
	\centering
	\includegraphics[width=\textwidth]{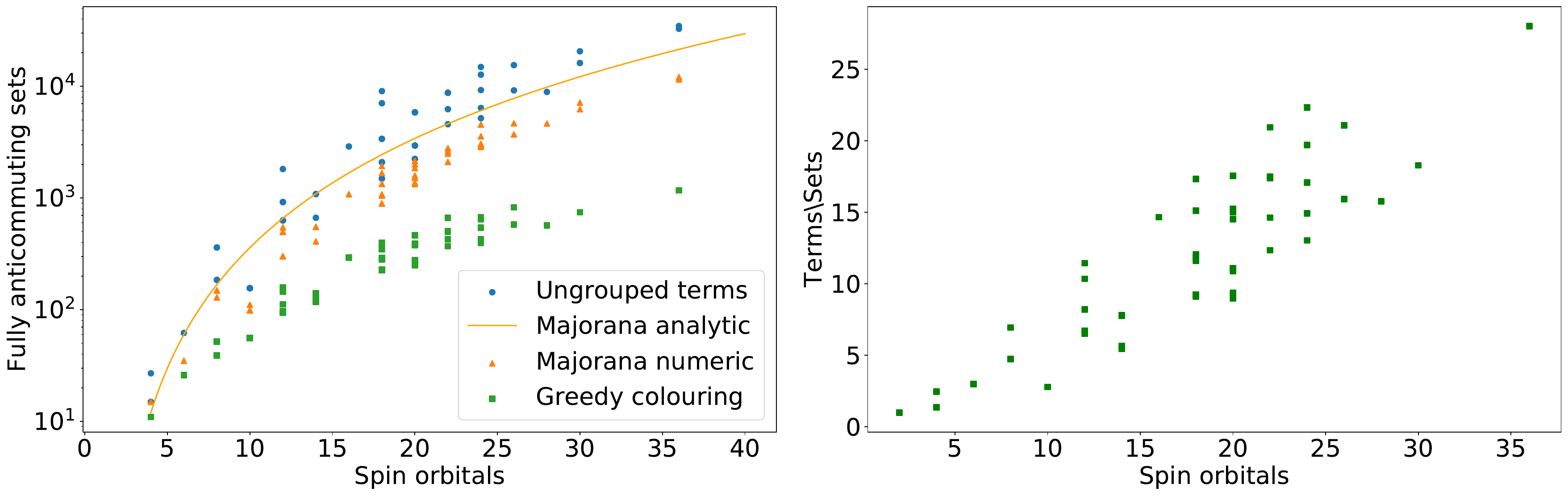}
	\caption[Number of fully anticommuting sets for electronic-structure Hamiltonians versus the number of spin orbitals.]{Number of fully anticommuting sets for electronic-structure Hamiltonians versus the number of spin orbitals, using the Jordan--Wigner mapping.   Left:~Including all partitioning schemes. The ``Majorana analytic'' curve is the $ \binom{N}{2}(N-2) $ upper bound obtained from \cref{thm:cubicPartitionTheorem} for generic Hamiltonians of Eq.~\eqref{eq:H_e_majorana}. The ``Majorana numeric'' data points correspond to the partitions described in \cref{subsec:linearReductionInTerms} without further optimisation. This upper bound is loose, due to sparsity in the molecular Hamiltonians versus the set of all possible terms.  Right:~Ratio of the number of terms to the number of anticommuting sets, for systems with more than 5 spin orbitals.  A roughly linear trend is observed, in agreement with the analytic scaling discussed in \cref{subsec:linearReductionInTerms}.}
	\label{fig:numSets}
\end{figure}

The results for both partitioning schemes against the number of spin orbitals are depicted in Figure~\ref{fig:numSets}.  Both the numerical implementation of the Majorana-based construction and the greedy colouring scheme prove to be consistently effective.  Beyond the smallest Hamiltonians, a roughly linear trend between the number of sets found and the number of Hamiltonian terms is observed, demonstrating that the asymptotic improvement discussed in \cref{subsec:linearReductionInTerms} can be achieved when using numerical approaches to colouring Pauli Hamiltonians.  The numerical Majorana results, and the greedy colouring strategy, consistently outperform the analytic upper bound, as expected.  This may be attributed primarily to the sparsity in the $h_{pq}$ and $h_{pqrs}$ weights, due to geometric molecular symmetries and the locality of the basis functions.  The ratio of the number of terms to the number of sets also appears to increase linearly with the number of spin orbitals (albeit with high variance), in agreement with the scaling properties discussed in \cref{subsec:linearReductionInTerms}.

The greedy colouring scheme yields roughly a factor of $10$ improvement over the numerical Majorana scheme, suggesting that it may be of substantial use in NISQ VQE experiments.  However, it should be emphasised that the substantial classical computing resources required may inhibit its use for systems with more spin orbitals.  The Majorana-based scheme demonstrates the same term reduction scaling, but with substantially reduced classical overhead.

\begin{figure}
	\centering
	\includegraphics[width=\textwidth]{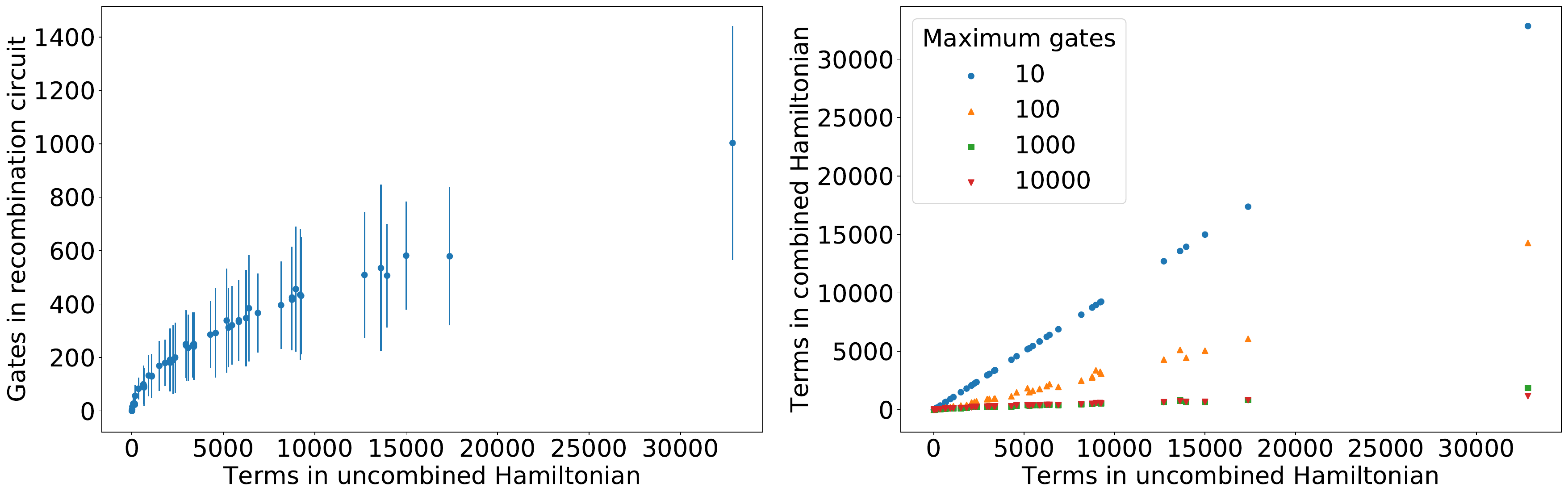}
	\caption[Resource requirements for full and partial term reduction using the greedy algorithm.]{Resource requirements for full and partial term reduction using the greedy algorithm for partitioning. Left:~Average post-ansatz gates required for full term reduction.  Whiskers denote one standard deviation in the length of the circuit required for each anticommuting partition in the Hamiltonian.  The growth in circuit length is dramatically slower than the growth in the number of Hamiltonian terms, but displays high variance between anticommuting sets. Right:~Reduction in the number of required independent expectation values, given restrictions on maximum individual circuit length.  With highly restricted circuit lengths, term recombination is largely impossible.  However, roughly 1000 additional gates at most are sufficient to perform near-maximal term reduction for the molecules considered here (up to 36 spin orbitals), which is in agreement with the figure to the left.}
	\label{fig:circuits}
\end{figure}

Although these results are promising, they do not consider the difficulty of performing the additional coherent operations required for the term recombination procedure. In principle, our analytic construction of anticommuting sets in Section~\ref{subsec:linearReductionInTerms} requires only $ O(N) $ depth circuits under the Jordan--Wigner mapping. This can be shown using well-known gate-compiling techniques~\cite{whitfieldSimulation2011,hastingsImprovingQuantumAlgorithms2015}.  Figure~\ref{fig:circuits} shows that the length of the circuits grows slowly in comparison to the amount of terms in the Hamiltonian.  However, near-term quantum devices are likely to be heavily constrained in the number of operations that can be performed coherently.  As such, it is likely that it will not be possible to combine entire sets of anticommuting terms.  Crucially, however, the term recombination procedure can be applied to \emph{subsets} of the fully anticommuting sets.  Provided the available coherent resources can be quantified prior to execution of the circuits, subsets of terms can be found to maximally use such resources to reduce the overall number of measurements required.  This yields a hardware-dependent tunable parameter---for example, the number of gates that can be implemented coherently subsequent to ansatz preparation---introduced at compile time.  This parameter allows for optimal use of the quantum resources provided by a given hardware option.

In order to assess the implications of varying such a parameter, we generated circuits corresponding to the implementation of the term reduction procedure for each Hamiltonian, introducing a maximum post-ansatz preparation gate count parameter.  For simplicity, these circuits used the standard method of implementing exponentiated Pauli strings given in \cref{unitarypartitioning}, rather than the ALCU circuits of \cref{rotations}.  Where circuits exceeded this length, the corresponding anticommuting set was split in half and new circuits were generated.  This binary splitting process was iterated until sufficiently short circuits were found.  Adjacent self-inverse gates were cancelled, moving through commuting gates where necessary~\cite{hastingsImprovingQuantumAlgorithms2015}.  For verification purposes, we calculated the expectation values with the true ground state of the Hamiltonians predicted by the circuits for systems with less than ten qubits.  As the results presented in Figure~\ref{fig:numSets} suggest that there is little difference between Jordan--Wigner and Bravyi--Kitaev circuits, we consider only Jordan--Wigner circuits.

Figure~\ref{fig:circuits} shows the results of this process.  Using a maximum circuit length of $10\,000$ gates subsequent to ansatz preparation allows all anticommuting sets, in all Hamiltonians, to be combined.  Allowing only $10$ gates removes any possibility of term recombination.  Encouragingly, allowing $100$ gates does not dramatically impede term recombination.  Even for the longest circuit considered, using $100$ gates allows for a reduction in terms by a factor of over 2.  Allowing $1000$ postansatz gates similarly performs as well as full anticommuting set recombination in all systems apart from the bromine atom;~in this instance, the difference between the $1000$- and $10\,000$- gate decompositions is minor. 

Our choice of allowable circuit length here is intended to be illustrative of the practicality of the term recombination procedure.  In a true simulation, the maximum post-ansatz gates parameter should be set to a value that is empirically determined by the ability of the hardware and should not be restricted to an integer power of 10.  Given the relatively low gate counts required for substantial improvement with regard to the number of terms, the results here strongly suggest that this approach is an effective way of reducing the overall runtime of variational quantum algorithms for electronic structure.

\section{Conclusions}

In this chapter we have discussed unitary partitioning---the technique for using anticommuting sets of Hamiltonian terms to reduce the number of measurements needed when performing variational quantum algorithms. The dramatic growth in the number of independent expectation values that must be determined is a key problem in the use of variational quantum algorithms for quantum chemistry in the NISQ era. Applying the technique to electronic-structure Hamiltonians with generic molecular orbital basis sets yielded reduction scaling linearly with the number of qubits. We proved this result in Section~\ref{subsec:linearReductionInTerms} and confirmed its practicality by numerics in Section~\ref{subsec:pauliLevelColouring}.

We report two strategies for partitioning the electronic-structure Hamiltonian into fully anticommuting subsets.  The first of these, based on expressing the fermionic Hamiltonian using Majorana operators, demonstrates the favourable scaling properties, and can be rapidly performed for even large numbers of spin orbitals.  Conversely, using a greedy colouring scheme is relatively expensive with regard to classical computational resources, but demonstrates an order-of-magnitude reduction, even for relatively small systems (less than $30$ qubits). The latter scheme is likely to be useful in NISQ applications where systems are small and greedy solutions can be feasibly computed.  The former yields the same scaling, and is not restricted by the cost of colouring algorithms, but suffers from a constant factor overhead in the number of fully anticommuting sets, compared to the greedy colouring method. The availability of postansatz coherent resources, and the relative difficulty of the classical partitioning step, may determine which scheme is favoured.

Our analysis of circuits for implementing the unitary partitioning procedure indicates that relatively modest additional coherent resources are required, compared to those typically needed for ansatz preparation.  Crucially, this optimisation is tunable, allowing for optimal use of coherent resources by hardware-dependent parameterisation at compile time.  It is also likely that unitary partitioning is compatible with other aspects of VQE optimisation. For instance, while we have remained agnostic to the choice of the parametrised ansatz for this study, the form of the unitaries required to perform term reduction matches those of popular ansatz choices, such as the unitary coupled cluster and related methods~\cite{romeroStrategiesQuantumComputing2018,ryabinkin2018qubit,leeGeneralizedUnitaryCoupled2019}. Thus with proper circuit compilation, one may significantly reduce the effective number of postansatz operations in practice, instead incorporating their rotation angles into the appropriate ansatz parameters. For these reasons, we believe that unitary partitioning could substantially aid in the use of variational quantum algorithms for studying classically intractable systems.


\newpage
\section*{Appendix}

\section{Calculational details}

In this section we give some derivations of the algebraic results used in the text.

\subsection{Computation of $ \mathcal{X}$ for the ALCU method}\label{app:comp_X}

We now derive the results that follow \cref{axis_b}. The operator $ \mathcal{X}$ is given by
\be
\begin{split}
	\mathcal{X} &= \frac{\i}{2}\left[ H_{n-1}, P_n\right]\\
	&=\frac{\i}{2}\sum_{k=1}^{n-1}\beta_k\left[ P_k, P_n\right]\\
	&=\i\sum_{k=1}^{n-1}\beta_k P_k P_n,
\end{split}
\ee
where we wrote $ H_{n-1}=\sum_{k=1}^{n-1}\beta_k P_k$ with $\sum_{k=1}^{n-1}\beta_k^2=1$. Then we can compute:
\be
\begin{split}
	\mathcal{X}^2 &= -\sum_{k=1}^{n-1}\sum_{j=1}^{n-1}\beta_k\beta_j P_k P_n P_j P_n\\
	&= -\sum_{k=1}^{n-1}\beta_k^2 P_k P_n P_k P_n-\sum_{k<j}^{n-1}\beta_k\beta_j\{ P_k P_n, P_j P_n\}\\
	&= -\sum_{k=1}^{n-1}\beta_k^2 P_k P_n P_k P_n\\
	&= \sum_{k=1}^{n-1}\beta_k^2 P_k^2 P_n^2 = {\openone}.
\end{split}
\ee

Now consider the commutator of $ \mathcal{X}$ and $ H_n$. We can use $ \mathcal{X}=\i H_{n-1} P_n$ to write
\begin{equation}
\begin{split}
\mathcal{X}  H_n &= \i H_{n-1} P_n  H_n    \\
&=\i H_{n-1} P_n (\sin\phi_{n-1} H_{n-1}+\cos\phi_{n-1} P_n)\\
&=\i(\sin\phi_{n-1} H_{n-1} P_n  H_{n-1}+\cos\phi_{n-1} H_{n-1} P_n^2).
\end{split}
\end{equation}
Using $\{ H_{n-1}, P_n\}=0$ and $ P_n^2= 1$ we have
\begin{equation}
\begin{split}
\mathcal{X}  H_n &=\i(-\sin\phi_{n-1} P_n +\cos\phi_{n-1} H_{n-1}),
\end{split}
\end{equation}
so that
\begin{equation}
[ \mathcal{X},  H_n] =2\i(-\sin\phi_{n-1} P_n +\cos\phi_{n-1} H_{n-1}).
\end{equation}

This enables us to compute the adjoint action generated by $ \mathcal{X}$ on $H_n$. Using the identity [for any operators $ A$ and $B$, where $ A^2= {\openone}$ so that $e^{-\i(\alpha/2)  A}=\cos(\alpha/2)  {\openone} -\i\sin(\alpha/2)  A$]
\begin{equation}
\begin{split}
e^{-\i(\alpha/2)  A} B e^{\i(\alpha/2)  A} &= \cos^2(\alpha/2)  B +\sin^2(\alpha/2)  A B A +\i\sin(\alpha/2)\cos(\alpha/2) [ A, B],
\end{split}
\end{equation}
we have ($ R = e^{-\i(\alpha/2) \mathcal{X}} $)
\begin{align}
R H_{n} R^\dagger &= \cos^2(\alpha/2)  H_{n} +\sin^2(\alpha/2)  \mathcal{X} H_{n} \mathcal{X} +\i\sin(\alpha/2)\cos(\alpha/2) [ \mathcal{X}, H_{n}] \notag\\
&= (\cos^2\alpha/2 -\sin^2\alpha/2)  H_{n} +(\i/2)2\sin(\alpha/2)\cos(\alpha/2) [ \mathcal{X}, H_{n}] \notag\\
&= \cos\alpha   H_{n} -\sin\alpha(-\sin\phi_{n-1} P_n +\cos\phi_{n-1} H_{n-1}) \notag\\
&= \cos\alpha  (\cos\phi_{n-1} P_n +\sin\phi_{n-1} H_{n-1}) -\sin\alpha(-\sin\phi_{n-1} P_n +\cos\phi_{n-1} H_{n-1}) \notag\\
&= (\cos\alpha\cos\phi_{n-1}+\sin\alpha\sin\phi_{n-1}) P_n +(\cos\alpha\sin\phi_{n-1}-\sin\alpha\cos\phi_{n-1}) H_{n-1} \notag\\ 
&= \cos(\phi_{n-1}-\alpha) P_n +\sin(\phi_{n-1}-\alpha) H_{n-1}.
\end{align}
Choosing $\alpha=\phi_{n-1}$ gives $ R  H_n R^\dagger =  P_n$. Given this role for $ R$, which is generated by $ \mathcal{X}$, we wish to know the commutation relations among the terms of $ \mathcal{X}$. Because $ \mathcal{X}=2\i  P_n H_{n-1}$, the terms of $\mathcal{X}$ have the form $2\i  P_n P_j$ for $j<n$. The commutation relations between any pair of terms are
\begin{equation}
\begin{split}
[ P_n P_j, P_n P_k] &=   P_n P_j P_n P_k- P_n P_k P_n P_j   \\
&=  -( P_n P_n P_j P_k- P_n P_n P_k P_j)   \\
&= -[ P_j, P_k] =2 P_k P_j.
\end{split}
\end{equation}

\subsection{Electronic-structure Hamiltonian using Majorana operators}\label{sec:H_e_majorana_appendix}

Here we derive the form of the Hamiltonian given in Eq.~\eqref{eq:H_e_majorana}. Since the single-mode Majorana operators are linear combinations of the fermionic ladder operators, we have the identities
\begin{equation}\label{eq:fermion_from_majorana}
a_p = \frac{\gamma_{2p} + \i\gamma_{2p+1}}{2}, \quad a_p^\dagger = \frac{ \gamma_{2p} - \i\gamma_{2p+1} }{2}.
\end{equation}
Furthermore, recall the permutational symmetries in the coefficients, given by \cref{eq:hpq_sym,eq:hpqrs_sym_1,eq:hpqrs_sym_2}, and the anticommutation relation for arbitrary Majorana operators, Eq.~\eqref{eq:majorana_gen_AR}. These are the only properties we use, but they allow for considerable simplification to the structure of the Hamiltonian terms. For brevity, we shall make use of such properties freely and often without comment.

First, consider the one-body terms, which are quadratic in fermionic operators. Using Majorana operators, they become
\begin{equation}
\begin{split}
\sum_{p,q} h_{pq} a_p^\dagger a_q = \frac{1}{4} \sum_{p,q} &h_{pq} ( \gamma_{2p}\gamma_{2q} + \gamma_{2p+1}\gamma_{2q+1} + \i\gamma_{2p}\gamma_{2q+1} - \i\gamma_{2p+1}\gamma_{2q} ).
\end{split}
\end{equation}
This expression can be simplified by separating the summation into diagonal and off-diagonal terms, a technique which we employ heavily throughout this derivation. The sum over the $ \gamma_{2p}\gamma_{2q} $ and $ \gamma_{2p+1}\gamma_{2q+1} $ terms simply yields a multiple of the identity:
\begin{equation}
\begin{split}
\sum_{p,q} &h_{pq} ( \gamma_{2p}\gamma_{2q} + \gamma_{2p+1}\gamma_{2q+1} ) \\
&= \sum_{p} h_{pp} \left( \gamma_{2p}^2 + \gamma_{2p+1}^2 \right) + \sum_{\substack{p,q\\p<q}} h_{pq} \left( \{ \gamma_{2p},\gamma_{2q} \} + \{ \gamma_{2p+1},\gamma_{2q+1} \} \right) \\
&= 2 \sum_{p}h_{pp} {\openone}.
\end{split}
\end{equation}
The remaining terms simplify but do not cancel or reduce in order:~by relabeling the indices (another trick which we make frequent use of), we see that $ \sum_{p,q} h_{pq} \i \gamma_{2p}\gamma_{2q+1} = \sum_{p,q} h_{pq} \i\gamma_{2q}\gamma_{2p+1} $, hence
\begin{equation}\label{eq:1b_terms}
\sum_{p,q} h_{pq} a_p^\dagger a_q = \frac{1}{2} \left(\sum_{p} h_{pp} {\openone} + \sum_{p,q} h_{pq} \i\gamma_{2p}\gamma_{2q+1} \right).
\end{equation}

Next, we consider the two-body interaction terms, which feature the quartic order operators. Any such term is written as a linear combination of 16 Majorana operators. To do so, define
\begin{equation}
\Gamma_{pqrs}^{\mathbf{x}} = \i^{|\mathbf{x}|} (-1)^{x_1+x_2} \gamma_{2p+x_1} \gamma_{2q+x_2} \gamma_{2r+x_3} \gamma_{2s+x_4},
\end{equation}
where $ \mathbf{x} = x_1x_2x_3x_4 \in \{ 0,1 \}^4 $ is a binary string encoding the parity of each index and $ |\mathbf{x}| $ is its Hamming weight. Then, from Eq.~\eqref{eq:fermion_from_majorana}, a straightforward algebraic expansion gives the following expression for each two-body term:
\begin{equation}
a_p^\dagger a_q^\dagger a_r a_s = \frac{1}{16} \sum_{\mathbf{x} \in \{ 0,1 \}^4} \Gamma_{pqrs}^{\mathbf{x}}. 
\end{equation}

Consider the set $ B_1 = \{ 0011,1100,0101,1010 \} $. These strings correspond to the quartic Majorana operators appearing in Eq.~\eqref{eq:H_e_majorana}, and as we will see, they are the only such terms which do not vanish. Also, note that since $ a_j^2 = (a_j^\dagger)^2 = 0 $, we impose the trivial constraints in the summations that $ p \neq q $ and $ r \neq s $. Specifying these conditions explicitly will be useful once we relabel the indices. We rewrite these terms as
\begin{equation}
\begin{split}
\sum_{p,q,r,s} h_{pqrs} \Gamma_{pqrs}^{1100} &= - \sum_{\substack{p,q,r,s\\p\neq q;r\neq s}} h_{pqrs} \gamma_{2p+1}\gamma_{2q+1}\gamma_{2r}\gamma_{2s} \\
&= - \sum_{\substack{p,q,r,s\\p\neq q;r\neq s}} h_{pqrs} \gamma_{2r}\gamma_{2s}\gamma_{2p+1}\gamma_{2q+1} \\
&= - \sum_{\substack{p,q,r,s\\p\neq q;r\neq s}} h_{pqrs} \gamma_{2p}\gamma_{2q}\gamma_{2r+1}\gamma_{2s+1},
\end{split}
\end{equation}
and, for $ x,y \in \{0,1\} $ such that $ x \neq y $,
\begin{equation}
\begin{split}
\sum_{p,q,r,s} h_{pqrs} \Gamma_{pqrs}^{xyxy} &= \sum_{\substack{p,q,r,s\\p\neq q;r\neq s}} h_{pqrs} \gamma_{2p+x}\gamma_{2q+y}\gamma_{2r+x}\gamma_{2s+y} \\
&= - \sum_{\substack{p,q,r,s\\p\neq q;r\neq s}} h_{pqrs} \gamma_{2p+x}\gamma_{2r+x}\gamma_{2q+y}\gamma_{2s+y} \\
&= - \sum_{\substack{p,q,r,s\\p\neq r;q\neq s}} h_{pqrs} \gamma_{2p+x}\gamma_{2q+x}\gamma_{2r+y}\gamma_{2s+y}.
\end{split}
\end{equation}
Thus we obtain
\begin{equation}\label{eq:B1}
\sum_{p,q,r,s} h_{pqrs} \left( \sum_{\mathbf{x} \in B_1} \Gamma_{pqrs}^{\mathbf{x}} \right) = -2 \left( \sum_{\substack{p,q,r,s\\p\neq q;r\neq s}} + \sum_{\substack{p,q,r,s\\p\neq r;q\neq s}} \right) h_{pqrs} \gamma_{2p}\gamma_{2q}\gamma_{2r+1}\gamma_{2s+1}.
\end{equation}
Since we would like to completely separate the quadratic terms from the quartic terms, we observe that if $ p=q $ or $ r=s $ in the above expression, then those terms reduce to quadratic order (or the identity, if both equalities hold). The first summation automatically excludes such reduction, so we analyze the second one, again separating the diagonal and off-diagonal summands with respect to each pair $ (p,q) $ and $ (r,s) $:
\begin{align}
\sum_{\substack{p,q,r,s\\p\neq r;q\neq s}} h_{pqrs} \gamma_{2p}\gamma_{2q}\gamma_{2r+1}\gamma_{2s+1} &= \sum_{\substack{p,q,r,s\\p\neq r;q\neq s\\p\neq q;r\neq s}} h_{pqrs} \gamma_{2p}\gamma_{2q}\gamma_{2r+1}\gamma_{2s+1} + \sum_{\substack{p,r\\p\neq r}} h_{pprr} {\openone} \notag \\
&\quad + \sum_{\substack{p,q,r\\p\neq r;q\neq r\\p\neq q}} h_{pqrr} \gamma_{2p}\gamma_{2q} + \!\! \sum_{\substack{p,r,s\\p\neq r;p\neq s\\r\neq s}} h_{pprs} \gamma_{2r+1}\gamma_{2s+1} \notag \\
&= \sum_{\substack{p,q,r,s\\p\neq r;q\neq s\\p\neq q;r\neq s}} h_{pqrs} \gamma_{2p}\gamma_{2q}\gamma_{2r+1}\gamma_{2s+1} + \sum_{\substack{p,r\\p\neq r}} h_{pprr} {\openone} \notag \\
&\quad + \sum_{\substack{p,q,r\\p\neq r;q\neq r\\p<q}} h_{pqrr} \{ \gamma_{2p},\gamma_{2q} \} + \!\!  \sum_{\substack{p,r,s\\p\neq r;p\neq s\\r<s}} h_{pprs} \{ \gamma_{2r+1},\gamma_{2s+1} \} \notag \\
&= \sum_{\substack{p,q,r,s\\p\neq r;q\neq s\\p\neq q;r\neq s}} h_{pqrs} \gamma_{2p}\gamma_{2q}\gamma_{2r+1}\gamma_{2s+1} + \sum_{\substack{p,q\\p\neq q}} h_{ppqq} {\openone}. \label{eq:B1_simp}
\end{align}
So we see that these quadratic terms in fact vanish due to anticommutation.

Now we show that the remaining 12 cases yield the same operators as those already obtained in Eq.~\eqref{eq:1b_terms}. Let $ B_2 = \{ 0000,0110,1001,1111 \} $ and $ x,y \in \{ 0,1 \} $:
\begin{equation}
\begin{split}
\sum_{p,q,r,s} h_{pqrs} \Gamma_{pqrs}^{xyyx} &= \sum_{\substack{p,q,r,s\\p\neq q;r\neq s}} h_{pqrs} \gamma_{2p+x}\gamma_{2q+y}\gamma_{2r+y}\gamma_{2s+x} \\
&= \sum_{\substack{p,q,r,s\\p\neq q;r\neq s\\p\neq s}} h_{pqrs} \gamma_{2p+x}\gamma_{2q+y}\gamma_{2r+y}\gamma_{2s+x} + \sum_{\substack{p,q,r\\p\neq q;r\neq p}} h_{pqrp} \gamma_{2q+y}\gamma_{2r+y}.
\end{split}
\end{equation}
The second sum simplifies to
\begin{align}
\sum_{\substack{p,q,r\\p\neq q;r\neq p}} h_{pqrp} \gamma_{2q+y}\gamma_{2r+y} &= \sum_{\substack{p,q\\p\neq q;r\neq p\\q<r}} h_{pqrp} \{ \gamma_{2q+y},\gamma_{2r+y} \} \notag  + \sum_{\substack{p,q\\p\neq q}} h_{pqqp} {\openone} \notag \\
&= \sum_{\substack{p,q\\p\neq q}} h_{pqqp} {\openone}.
\end{align}
The first sum depends on whether $ x $ and $ y $ are the same or not. If $ x \neq y $, then
\begin{align}
\sum_{\substack{p,q,r,s\\p\neq q;r\neq s\\p\neq s}} &h_{pqrs} \gamma_{2p+x}\gamma_{2q+y}\gamma_{2r+y}\gamma_{2s+x} \notag \\
&= \sum_{\substack{p,q,r,s\\p\neq q;r\neq s\\p<s}} h_{pqrs} ( \gamma_{2p+x}\gamma_{2q+y}\gamma_{2r+y}\gamma_{2s+x} \notag + \gamma_{2s+x}\gamma_{2q+y}\gamma_{2r+y}\gamma_{2p+x} ) \notag\\
&= \sum_{\substack{p,q,r,s\\p\neq q;r\neq s\\p<s}} h_{pqrs} ( \gamma_{2p+x}\gamma_{2q+y}\gamma_{2r+y}\gamma_{2s+x} - \gamma_{2p+x}\gamma_{2q+y}\gamma_{2r+y}\gamma_{2s+x} ) \notag \\
&= 0. \label{eq:quartic_cancellation}
\end{align}
If $ x = y $, we first observe that if $ p \neq r $ and $ q \neq s $, then the sum vanishes, as demonstrated above. Therefore we have the three remaining cases ($ p \neq r $ and $ q = s $, $ p = r $ and $ q \neq s $, and $ p = r $ and $ q = s $):
\begin{align}
\sum_{\substack{p,q,r,s\\p\neq q;r\neq s\\p\neq s}} h_{pqrs} \gamma_{2p+x}\gamma_{2q+x}\gamma_{2r+x}\gamma_{2s+x} &= - \sum_{\substack{p,q\\p\neq q}} h_{pqpq} {\openone} - \sum_{\substack{p,q,r\\p\neq q;r\neq q\\p\neq r}} h_{pqrq} \gamma_{2p+x}\gamma_{2r+x} \notag \\
&\quad - \sum_{\substack{p,q,s\\p\neq q;p\neq s\\q\neq s}} h_{pqps} \gamma_{2q+x}\gamma_{2s+x} \notag \\
&= - \sum_{\substack{p,q\\p\neq q}} h_{pqpq} {\openone} - \sum_{\substack{p,q,r\\p\neq q;r\neq q\\p<r}} h_{pqrq} \{ \gamma_{2p+x},\gamma_{2r+x} \} \notag \\
&\quad - \sum_{\substack{p,q,s\\p\neq q;p\neq s\\q<s}} h_{pqps} \{ \gamma_{2q+x},\gamma_{2s+x} \} \notag \\
&= - \sum_{\substack{p,q\\p\neq q}} h_{pqpq} {\openone}.
\end{align}
Altogether, the terms corresponding to $ B_2 $ are just the identity operator:
\begin{equation}\label{eq:B2}
\sum_{p,q,r,s} h_{pqrs} \left( \sum_{\mathbf{x} \in B_2} \Gamma_{pqrs}^{\mathbf{x}} \right) = \sum_{\substack{p,q\\p\neq q}} \left( 4h_{pqqp} - 2h_{pqpq} \right) {\openone}.
\end{equation}
Let $ B_3 = \{ 0010, 0100, 1011, 1101 \} $. These strings give rise to the same terms, since for $ x \in \{0,1\} $,
\begin{equation}
\begin{split}
\sum_{p,q,r,s} \Gamma_{pqrs}^{x01x} &= \sum_{p,q,r,s} h_{pqrs} \i\gamma_{2p+x}\gamma_{2q}\gamma_{2r+1}\gamma_{2s+x} \\
&= - \sum_{p,q,r,s} h_{pqrs} \i\gamma_{2p+x}\gamma_{2q+1}\gamma_{2r}\gamma_{2s+x} \\
&= \sum_{p,q,r,s} \Gamma_{pqrs}^{x10x}. \label{eq:B3_equiv}
\end{split}
\end{equation}
We simplify the sum using the same type of manipulations as in Eq.~\eqref{eq:quartic_cancellation}:
\begin{align}
\sum_{p,q,r,s} h_{pqrs} \i\gamma_{2p}\gamma_{2q}\gamma_{2r+1}\gamma_{2s} &= \sum_{\substack{p,q,r,s\\p\neq q;r\neq s\\p<s;s\neq q}} h_{pqrs} \i( \gamma_{2p}\gamma_{2q}\gamma_{2r+1}\gamma_{2s} - \gamma_{2p}\gamma_{2q}\gamma_{2r+1}\gamma_{2s} ) \notag \\
&\quad - \sum_{\substack{p,q,r\\p\neq q;r\neq q}} h_{pqrq} \i\gamma_{2p}\gamma_{2r+1} + \sum_{\substack{p,q,r\\p\neq q;r\neq p}} h_{pqrp} \i\gamma_{2q}\gamma_{2r+1} \notag \\
&= \sum_{\substack{p,q,r\\p\neq r;q\neq r}} ( h_{prrq} - h_{prqr} ) \i\gamma_{2p}\gamma_{2q+1}.
\end{align}
Thus we obtain
\begin{equation}\label{eq:B3}
\sum_{p,q,r,s} h_{pqrs} \left( \sum_{\mathbf{x} \in B_3} \Gamma_{pqrs}^{\mathbf{x}} \right) = 4 \sum_{\substack{p,q,r\\p\neq r;q\neq r}} ( h_{prrq} - h_{pqrr} ) \i\gamma_{2p}\gamma_{2q+1}.
\end{equation}
The last set is $ B_4 = \{ 0001,0111,1000,1110 \} $. Again, all four strings correspond to the same terms. We show this by evaluating, for $ w,x,y \in \{0,1\} $ with $ w \neq y $,
\begin{align}
\sum_{p,q,r,s} \Gamma_{pqrs}^{wxxy} &= (-1)^{w+1} \!\! \sum_{p,q,r,s} h_{pqrs} \i\gamma_{2p+w}\gamma_{2q+x}\gamma_{2r+x}\gamma_{2s+y} \notag \\
&= (-1)^{w+1} \!\! \sum_{\substack{p,q,r,s\\p\neq q;r\neq s\\q<r}} h_{pqrs} \i( \gamma_{2p+w}\gamma_{2q+x}\gamma_{2r+x}\gamma_{2s+y} - \gamma_{2p+w}\gamma_{2q+x}\gamma_{2r+x}\gamma_{2s+y} ) \notag \\
&\quad + (-1)^{w+1} \!\! \sum_{\substack{p,q,s\\p\neq q;q\neq s}} h_{pqqs} \i\gamma_{2p+w}\gamma_{2s+y} \notag \\
&= (-1)^{w+1} \!\! \sum_{\substack{p,q,r\\p\neq r;r\neq q}} h_{prrq} \i\gamma_{2p+w}\gamma_{2q+y}.
\end{align}
If we order the Majorana product such that the even index appears first, then the sign of $ (-1)^{w+1} $ cancels with that of swapping $ \gamma_{2p+w} $ with $ \gamma_{2q+y} $, and so we have
\begin{equation}\label{eq:B4}
\sum_{p,q,r,s} h_{pqrs} \left( \sum_{\mathbf{x} \in B_4} \Gamma_{pqrs}^{\mathbf{x}} \right) = 4 \sum_{\substack{p,q,r\\p\neq r;q\neq r}} h_{prrq} \i\gamma_{2p}\gamma_{2q+1}.
\end{equation}
Finally, we collect all the terms from \cref{eq:B1,eq:B2,eq:B3,eq:B4}, along with the slight simplification in Eq.~\eqref{eq:B1_simp}, to write the two-body terms as
\begin{equation}
\begin{split}
\frac{1}{2} \sum_{p,q,r,s} h_{pqrs} a_p^\dagger a_q^\dagger a_r a_s &= \frac{1}{32} \sum_{p,q,r,s} h_{pqrs} \left( \sum_{\mathbf{x} \in \{0,1\}^4} \Gamma_{pqrs}^{\mathbf{x}} \right) \\
&= \frac{1}{8} \sum_{\substack{p,q\\p\neq q}} \left( h_{pqqp} - h_{pqpq} \right) {\openone} + \frac{1}{8} \sum_{\substack{p,q,r\\p\neq r;q\neq r}} ( 2h_{prrq} - h_{pqrr} ) \i\gamma_{2p}\gamma_{2q+1} \\
&\quad - \frac{1}{16} \left( \sum_{\substack{p,q,r,s\\p\neq q;r\neq s}} + \sum_{\substack{p,q,r,s\\p\neq q;r\neq s\\p\neq r;q\neq s}} \right) h_{pqrs} \gamma_{2p}\gamma_{2q}\gamma_{2r+1}\gamma_{2s+1}.
\end{split}
\end{equation}
Including the one-body terms, Eq.~\eqref{eq:1b_terms}, we express the full electronic-structure Hamiltonian in terms of Majorana operators:
\begin{equation}
\begin{split}
H &= \frac{1}{2} \left[\sum_{p} h_{pp} + \frac{1}{4} \sum_{\substack{p,q\\p\neq q}} \left( h_{pqqp} - h_{pqpq} \right) \right] {\openone}  \\
&\quad + \sum_{p,q} \left[ \frac{1}{2} h_{pq} + \!\!\!\! \sum_{\substack{r\\p\neq r;q\neq r}} \!\! \left( \frac{1}{4}h_{prrq} - \frac{1}{8}h_{pqrr} \right) \right] \i\gamma_{2p}\gamma_{2q+1} \\
&\quad - \frac{1}{16} \left( \sum_{\substack{p,q,r,s\\p\neq q;r\neq s}} + \sum_{\substack{p,q,r,s\\p\neq q;r\neq s\\p\neq r;q\neq s}} \right) h_{pqrs} \gamma_{2p}\gamma_{2q}\gamma_{2r+1}\gamma_{2s+1}.
\end{split}
\end{equation}
Defining new coefficients as
\begin{equation}\label{eq:new_coeff}
\begin{split}
\tilde{h} &= \frac{1}{2} \sum_{p} h_{pp} + \frac{1}{8} \sum_{\substack{p,q\\p\neq q}} \left( h_{pqqp} - h_{pqpq} \right) , \\
\tilde{h}_{pq} &= \frac{1}{2} h_{pq} + \sum_{\substack{r\\p\neq r;q\neq r}} \left( \frac{1}{4}h_{prrq} - \frac{1}{8}h_{pqrr} \right), \\
\tilde{h}_{pqrs} &= - \frac{1}{8} [ 1 + (1 - \delta_{pr})(1 - \delta_{qs}) ] h_{pqrs},
\end{split}
\end{equation}
we obtain the Hamiltonian as presented in the main text, Eq.~\eqref{eq:H_e_majorana}.

\section{Proof details for Theorem~\ref{thm:cubicPartitionTheorem}}\label{sec:majorana_proof_details}

\begin{figure}
	\centering
    \includegraphics{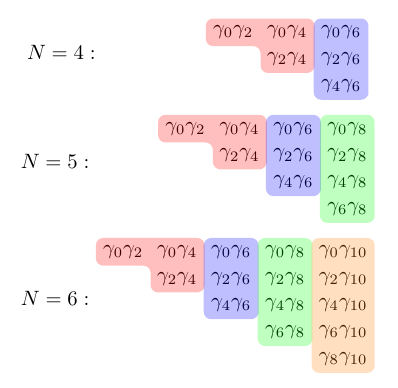}
	\caption[Partitioning of electronic-structure terms.]{Partitioning of electronic-structure terms. Finding an anticommuting partition of the quartic terms can be reduced to finding an anticommuting partition of quadratic terms with exclusively even (equiv.~odd) indices. Each highlighted bin is such an anticommuting set. Excluding the red bin, each set shares one common index $ 2q $ for $ 3 \leq q \leq N-1 $. Although only 3 values of $ N $ are depicted, the induction of this diagram is straightforward for arbitrary $ N $. One thus obtains $ N - 3 $ bins of size $ q $ each and $ 1 $ ``red bin'' of size $ 3 $.}
	\label{fig:quadPartitions}
\end{figure}

To see why \cref{eq:disjointness,eq:covering} hold, we first examine the structure of our anticommuting partition $ \{ S_{(q,r,s)} \} $. Although we have the choice of matching either one or three indices in each term's support, here we only use the condition of three matches. This amounts to matching exactly one even index, since the other two must be odd (or vice versa, by symmetry). In this sense, the problem reduces to finding an anticommuting partition of the set of all quadratic Majorana operators with only even indices in their support. Taking products with the set of all quadratic operators with only odd indices in their support then generates all the relevant quartic operators, $ \mathcal{M} $ (up to phase factors).

One may readily check from the definition of $ S_{(q,r,s)} $ that they do indeed cover $ \mathcal{M} $ and are all pairwise disjoint. However, since we have reduced the problem to considering simply quadratic operators, we may provide a visual argument which clearly demonstrates the partitioning scheme, Figure~\ref{fig:quadPartitions}. Note that for $ N = 2 $, there is only one unique quartic term, and for $ N = 3 $, all the even quadratics already anticommute (i.e., the red bin in the figure). From the figure, we immediately see the disjointness property satisfied, with each set of common index $ 2q $ having size $ q $. The exception, again, is the red bin, which corresponds to the union $ S_{(1,r,s)} \cup S_{(2,r,s)} $ as mentioned in the main text. Hence there are $ N-2 $ anticommuting sets of even-index quadratic operators, and taking products with all $ \binom{N}{2} $ odd-index quadratic operators yields the desired $ O(N^3) $ result.

\section{Electronic-structure systems}
\label{appendix:systems}

Table~\ref{tab:appendixSystems} details the systems from which the electronic-structure Hamiltonians studied in Sec.~\ref{subsec:pauliLevelColouring} were generated.

\begin{table}
	\centering
	\caption[The systems examined in our numerical analysis of unitary partitioning.]{The systems examined in our numerical analysis.  Geometries were obtained from the NIST CCBDB database~\cite{johnsoniiiNISTComputationalChemistry2016}, and molecular orbital integrals in the Hartee--Fock basis obtained from Psi4~\cite{psi4} and OpenFermion~\cite{openfermion}.}\label{tab:appendixSystems}
    \begin{tabular}[t]{lrrlr}
		\toprule
		System & Charge & Multiplicity & Basis & Qubits \\
		\midrule 
		\ce{Ar1} & 0 & 1 & STO-3G & 18 \\
		\ce{B1} & 0 & 2 & STO-3G & 10 \\
		\ce{Be1} & 0 & 1 & STO-3G & 10 \\
		\ce{Br1} & 0 & 2 & STO-3G & 36 \\
		\ce{C1O1} & 0 & 1 & STO-3G & 20 \\
		\ce{C1O2} & 0 & 1 & STO-3G & 30 \\
		\ce{C1} & 0 & 3 & STO-3G & 10 \\
		\ce{Cl1} & 0 & 2 & STO-3G & 18 \\
		\ce{Cl1} & -1 & 1 & STO-3G & 18 \\
		\ce{F1} & 0 & 2 & STO-3G & 10 \\
		\ce{F2} & 0 & 1 & STO-3G & 20 \\
		\ce{H1Cl1} & 0 & 1 & STO-3G & 20 \\
		\ce{H1F1} & 0 & 1 & 3-21G & 22 \\
		\ce{H1F1} & 0 & 1 & STO-3G & 12 \\
		\ce{H1He1} & 0 & 1 & 3-21G & 8 \\
		\ce{H1He1} & 0 & 1 & 6-311G** & 24 \\
		\ce{H1He1} & 0 & 1 & 6-311G & 12 \\
		\ce{H1He1} & 0 & 1 & 6-31G** & 20 \\
		\ce{H1He1} & 0 & 1 & 6-31G & 8 \\
		\ce{H1He1} & 0 & 1 & STO-3G & 4 \\
		\ce{H1Li1O1} & 0 & 1 & STO-3G & 22 \\
		\ce{H1Li1} & 0 & 1 & 3-21G & 22 \\
		\ce{H1Li1} & 0 & 1 & STO-3G & 12 \\
		\ce{H1Na1} & 0 & 1 & STO-3G & 20 \\
		\ce{H1O1} & -1 & 1 & STO-3G & 12 \\
		\ce{H1} & 0 & 2 & STO-3G & 2 \\
		\ce{H2Be1} & 0 & 1 & STO-3G & 14 \\
		\ce{H2C1O1} & 0 & 1 & STO-3G & 24 \\
		\ce{H2C1} & 0 & 3 & 3-21G & 26 \\
		\ce{H2C1} & 0 & 3 & STO-3G & 14 \\
		\ce{H2C1} & 0 & 3 & STO-3G & 14 \\
		\ce{H2C2} & 0 & 1 & STO-3G & 24 \\
		\ce{H2Mg1} & 0 & 1 & STO-3G & 22 \\
		\ce{H2O1} & 0 & 1 & STO-3G & 14 \\
		\ce{H2O2} & 0 & 1 & STO-3G & 24 \\
		\ce{H2S1} & 0 & 1 & STO-3G & 22 \\
		\ce{H2} & 0 & 1 & 3-21G & 8 \\
		\ce{H2} & 0 & 1 & 6-311G** & 24 \\
		\ce{H2} & 0 & 1 & 6-311G & 12 \\
		\ce{H2} & 0 & 1 & 6-31G** & 20 \\
		\ce{H2} & 0 & 1 & 6-31G & 8 \\
		\ce{H2} & 0 & 1 & STO-3G & 4 \\
		\midrule
	\end{tabular}
\end{table}
\null\vfill
\begin{table}[t]
    \centering
    \captionsetup{justification=centering}
    \caption*{\textbf{Table~\ref{tab:appendixSystems}}. (\emph{Continued}.)}
	\begin{tabular}[t]{lrrlr}
        \toprule
		System & Charge & Multiplicity & Basis & Qubits \\
		\midrule 
		\ce{H3N1} & 0 & 1 & STO-3G & 16 \\
		\ce{H3} & 0 & 1 & 3-21G & 12 \\
		\ce{H3} & 1 & 1 & STO-3G & 6 \\
		\ce{H4C1} & 0 & 1 & STO-3G & 18 \\
		\ce{H4C2} & 0 & 1 & STO-3G & 28 \\
		\ce{H4N1} & 1 & 1 & STO-3G & 18 \\
		\ce{He1} & 0 & 1 & STO-3G & 2 \\
		\ce{K1} & 0 & 2 & STO-3G & 26 \\
		\ce{Li1} & 0 & 2 & STO-3G & 10 \\
		\ce{Mg1} & 0 & 1 & STO-3G & 18 \\
		\ce{N1} & 0 & 4 & STO-3G & 10 \\
		\ce{N2} & 0 & 1 & STO-3G & 20 \\
		\ce{Na1} & 0 & 2 & STO-3G & 18 \\
		\ce{Ne1} & 0 & 1 & STO-3G & 10 \\
		\ce{O1} & 0 & 3 & STO-3G & 10 \\
		\ce{O2} & 0 & 1 & STO-3G & 20 \\
		\ce{O2} & 0 & 3 & STO-3G & 20 \\
		\ce{P1} & 0 & 4 & STO-3G & 18 \\
		\ce{S1} & 0 & 3 & STO-3G & 18 \\
		\ce{Si1} & 0 & 3 & STO-3G & 18 \\
		\bottomrule
	\end{tabular}
	\null\hfill
\end{table}
\chapter[Expanding the Reach of Quantum Optimization with Fermionic Embeddings \hfill]{Expanding the Reach of Quantum Optimization with Fermionic Embeddings}\label{chap:QR}

\section*{Preface}

This chapter is based on \cite{zhao2023expanding}, coauthored by the author of this dissertation and Nicholas C.~Rubin.

\section{Introduction}
Finding computational tasks where a quantum computer could have a large speedup is a primary driver for the field of quantum algorithm development. While some examples of quantum advantage are known, such as quantum simulation~\cite{feynman1982simulating,lloyd1996universal}, prime number factoring~\cite{shorfactoring}, and unstructured search~\cite{Grover1996}, generally speaking computational advantages for industrially relevant calculations are scarce. Specifically in the field of optimization, which has attracted a large amount of attention from quantum algorithms researchers due to the ubiquity and relevance of the computational problems, substantial quantum speedups, even on model problems, are difficult to identify.  This difficulty is in part because it is not obvious \emph{a priori} how the unique features of quantum mechanics---e.g., entanglement, unitarity, and interference---can be leveraged towards a computational advantage~\cite{Grover1996,PRXQuantum.2.010103,PRXQuantum.2.030312}.  

In this work we take steps toward understanding how to apply quantum computers to optimization problems by demonstrating that the class of optimization problems involving rotation matrices as decision variables has a natural quantum formulation and efficient embedding. Examples of such problems include the joint alignment of points in Euclidean space by isometries, which has applications within the contexts of structural biology via cryogenic electron microscopy (cryo-EM)~\cite{shkolnisky2012viewing,singer2018mathematics} and NMR spectroscopy~\cite{cucuringu2012eigenvector}, computer vision~\cite{arie2012global,ozyecsil2017survey}, robotics~\cite{rosen2019se,lajoie2019modeling}, and sensor network localization~\cite{cucuringu2012sensor}. The central difficulty in solving these problems is twofold:~first, the set of orthogonal transformations $\Orth(n)$ is nonconvex, making the optimization landscape challenging to navigate in general. Second, the objectives of these problems are quadratic in the decision variables, making them examples of quadratic programming under orthogonality constraints~\cite{nemirovski2007sums}. In this chapter we specifically focus on the problem considered by Bandeira \emph{et al.}~\cite{bandeira2016approximating}, which is a special case of the real little noncommutative Grothendieck (LNCG) problem~\cite{briet2017tight}. While significant progress has been made in classical algorithms development for finding approximate solutions, for example by semidefinite relaxations~\cite{povh2010semidefinite,wang2013exact,naor2014efficient,saunderson2014semidefinite,bandeira2016approximating}, guaranteeing high-quality solutions remains difficult in general. This work therefore provides a quantum formulation of the optimization problem, as a first step in exploring the potential use of a quantum computer to obtain more accurate solutions.

The difficulty of the LNCG problem becomes even more pronounced when restricting the decision variables to the group of rotation matrices $\SO(n)$~\cite{bandeira2017estimation,pumir2021generalized}. One promising approach to resolving this issue is through the convex relaxation of the problem, studied by Saunderson \emph{et al.}~\cite{saunderson2015semidefinite,saunderson2014semidefinite}. They identified that the convex hull of rotation matrices, $\conv\SO(n)$, is precisely the feasible region of a semidefinite program (SDP)~\cite{saunderson2015semidefinite}. Therefore, standard semidefinite relaxations of the quadratic optimization problem can be straightforwardly augmented with this convex-hull description as an additional constraint~\cite{saunderson2014semidefinite}. They prove that when the problem is defined over particular types of graphs, this enhanced SDP is exact, and for more general instances of the problem they numerically demonstrate that it yields significantly higher-quality approximations than the basic SDP. The use of this convex hull has since been explored in related optimization contexts~\cite{matni2014convex,rosen2015convex,saunderson2016convex}. Notably however, the semidefinite description of $\conv\SO(n)$ is exponentially large in $n$. Roughly speaking, this reflects the complexity of linearizing a nonlinear determinant constraint. One such representation is the so-called positive-semidefinite (PSD) lift of $\conv\SO(n)$, which is defined through linear functionals on the trace-1, PSD matrices of size $2^{n-1} \times 2^{n-1}$.

One may immediately recognize this description as the set of density operators on $n - 1$ qubits. In this work we investigate this statement in detail and make a number of connections between the optimization of orthogonal/rotation matrices and the optimization of quantum states, namely fermionic states in second quantization. The upshot is that these connections provide us with a relaxation of the quadratic program into a quantum Hamiltonian problem. Although this relaxation admits solutions (quantum states) which lie outside the feasible space of the original problem, we show that it retains much of the important orthogonal-group structure due to this natural embedding. The notion of quantum relaxations have been previously considered in the context of combinatorial optimization (such as the Max-Cut problem), wherein quantum rounding protocols were proposed to return binary decision variables from the relaxed quantum state~\cite{fuller2021approximate}. In a similar spirit, in this work we consider rounding protocols which return orthogonal/rotation matrices from our quantum relaxation.

Within the broader context of quantum information theory, our work here also provides an alternative perspective to relaxations of quantum Hamiltonian problems. There is a growing interest in classical methods for approximating quantum many-body problems based on SDP relaxations~\cite{brandao2013product, bravyi2019approximation, gharibian_et_al:LIPIcs:2019:11246,anshu_et_al:LIPIcs:2020:12066, parekh_et_al:LIPIcs.ESA.2021.74,parekh2021application,hastings2022optimizing,parekh2022optimal,hastings2022perturbation,king2022improved}. In that context, rounding procedures are more difficult to formulate because the space of quantum states is exponentially large. For instance, the algorithm may only round to a subset of quantum states with efficient classical descriptions such as product states~\cite{brandao2013product, bravyi2019approximation, gharibian_et_al:LIPIcs:2019:11246, parekh_et_al:LIPIcs.ESA.2021.74,parekh2022optimal} or low-entanglement states~\cite{anshu_et_al:LIPIcs:2020:12066,parekh2021application}, effectively restricting the approximation from representing the true ground state. Nonetheless, these algorithms can still obtain meaningful approximation ratios of the optimal energy, indicating that such states can at least capture some qualitative properties of the generically entangled ground state.

Our quantum relaxation can be viewed as working in the opposite direction:~we construct a many-body Hamiltonian where the optimal solution to the underlying classical quadratic program is essentially a product state. Therefore, we propose preparing an approximation to the ground state of the Hamiltonian,\footnote{While the physical problem typically considers the ground-state problem, this work takes the convention of maximizing objectives.} which is then rounded to the nearest product state corresponding to the original classical solution space. This is not unlike quantum approaches to binary optimization such as quantum annealing or the quantum approximate optimization algorithm~\cite{PhysRevLett.101.130504,farhi2014quantum,PhysRevA.101.012320,hauke2020perspectives,PRXQuantum.2.030312}, which explore a state space outside the classical feasible region before projectively measuring, or rounding, the quantum state to binary decision variables. We furthermore provide numerical evidence that the physical qualitative similarity between optimal product and entangled states may translate into quantitative accuracy for the classical optimization problem, in a context beyond discrete combinatorial optimization.

Finally, we remark that Grothendieck-type problems and inequalities have a considerable historical connection to quantum theory. Tsirelson~\cite{tsirelson1987quantum} employed Clifford algebras to reformulate the commutative Grothendieck inequality into a statement about classical XOR games with entanglement. Regev and Vidick~\cite{regev2015quantum} later introduced the notion of quantum XOR games, which they studied through the generalization of such ideas to noncommutative Grothendieck inequalities. The mathematical work of Haagerup and Itoh~\cite{haagerup1995grothendieck} studied Grothendieck-type inequalities as the norms of operators on $C^*$-algebras;~their analysis makes prominent use of canonical anticommutation relation algebras over fermionic Fock spaces. Quadratic programming with orthogonality constraints has also been applied for classical approximation algorithms for quantum many-body problems, for instance by Bravyi \emph{et al.}~\cite{bravyi2019approximation}. Recasting noncommutative Grothendieck problems into a quantum Hamiltonian problem may therefore provide new insights into these connections.

The rest of this chapter is organized as follows:~Section~\ref{sec:problem_statement} provides a formal description of the optimization problem that we study in this chapter and reviews known complexity results of related problems. In Section~\ref{sec:lncg_applications} we describe two well-known applications of the problem:~the group synchronization problem and the generalized orthogonal Procrustes problem. Section~\ref{sec:summary_results} provides a summary of our quantum relaxation which embeds the optimization problem into a Hamiltonian, and two accompanying rounding protocols. In Section~\ref{sec:quantum_formalism} we derive an embedding of orthogonal matrices into quantum states via the Pin and Spin groups. We elaborate on the connection to fermionic theories and provide a quantum perspective on the convex hull of the orthogonal groups. From this embedding, Section~\ref{sec:quantum_relax_section} then establishes the quantum Hamiltonian relaxation of the quadratic optimization problem. Section~\ref{sec:rounding} describes both classical and quantum rounding protocols for relaxations of the problem. Notably, for the classical SDP we derive an approximation ratio for $\SO(n)$ rounding. Finally, in Section~\ref{sec:numerical_experiments} we demonstrate numerical experiments on random instances of the group synchronization problem for $\SO(3)$ on three-regular graphs and report the performance of various classical and quantum rounding protocols. For our simulations of the quantum relaxation, we consider two classes of quantum states:~maximal eigenstates of the Hamiltonian and quasi-adiabatically evolved states. We close in Section~\ref{sec:discussion_quant_relax_ortho} with a discussion on future lines of research.

\section{Problem statement}\label{sec:problem_statement}
In this paper we consider the class of little noncommutative Grothendieck (LNCG) problems over the orthogonal group, as studied previously by Bandeira \emph{et al.}~\cite{bandeira2016approximating}.\footnote{The authors also consider the complex-valued problem over the unitary group, which is outside the scope of this dissertation.} Let $(V, E)$ be an undirected graph with $m = |V|$ vertices and edge set $E$. For integer $n \geq 1$, let $C \in \R^{mn \times mn}$ be a symmetric matrix, which for notation we partition into $n \times n$ blocks as
\begin{equation}
    C = \begin{bmatrix}
    C_{11} & \cdots & C_{1m}\\
    \vdots & \ddots & \vdots\\
    C_{m1} & \cdots & C_{mm}
    \end{bmatrix}.
\end{equation}
The quadratic program we wish to solve is of the form
\begin{equation}\label{eq:LNCG}
    \max_{R_1, \ldots, R_m \in G} \sum_{(u, v) \in E} \langle C_{uv}, R_u R_v^\T \rangle,
\end{equation}
where $G$ is either the orthogonal group
\begin{equation}
    \Orth(n) \coloneqq \{ R \in \R^{n \times n} \mid R^\T R = \I_n \},
\end{equation}
or the special orthogonal group
\begin{equation}
    \SO(n) \coloneqq \{ R \in \Orth(n) \mid \det R = 1 \}
\end{equation}
on $\R^n$. Here, $\langle A, B \rangle = \tr(A^\T B)$ denotes the Frobenius inner product on the space of real matrices and $\I_n$ is the $n \times n$ identity matrix. Note that when $G = \Orth(1) = \{\pm 1\}$, Problem~\eqref{eq:LNCG} reduces to combinatorial optimization of the form
\begin{equation}\label{eq:LCG}
    \max_{x_1, \ldots, x_m \in \{\pm 1\}} \sum_{(u, v) \in E} C_{uv} x_u x_v,
\end{equation}
where now $C \in \R^{m \times m}$. This is sometimes referred to as the commutative instance of the little Grothendieck problem. Problem~\eqref{eq:LNCG} can therefore be viewed as a natural generalization of quadratic binary optimization to 
the noncommutative matrix setting.

We now comment on the known hardness results of these optimization problems. The commutative problem~\eqref{eq:LCG} is already $\mathsf{NP}$-hard in general, as can be seen by the fact that the Max-Cut problem can be expressed in this form. In particular, Khot \emph{et al.}~\cite{khot2007optimal} proved that, assuming the Unique Games conjecture, it is $\mathsf{NP}$-hard to approximate the optimal Max-Cut solution to better than a fraction of $(2/\pi) \min_{\theta \in [0, \pi]} \frac{\theta}{1 - \cos\theta} \approx 0.878$. This value coincides with the approximation ratio achieved by the celebrated Goemans--Williamson (GW) algorithm for rounding the semidefinite relaxation of the problem~\cite{goemans1995improved}. More generally, consider the fully connected graph and let $C \succeq 0$ be arbitrary. Nesterov~\cite{nesterov1998semidefinite} showed that GW rounding guarantees an approximation ratio of $2/\pi \approx 0.636$ in this setting, which Alon and Naor~\cite{alon2004approximating} showed matches the integrality gap of the semidefinite program. Khot and Naor~\cite{khot2009approximate} later demonstrated that this approximation ratio is also Unique-Games-hard to exceed, and finally Bri\"{e}t \emph{et al.}~\cite{briet2017tight} strengthened this result to be unconditionally $\mathsf{NP}$-hard.

For the noncommutative problem~\eqref{eq:LNCG} that we are interested in, less is known about its hardness of approximability. However, it is a subclass of more general optimization problems for which some results are known. The most general instance is the ``big'' noncommutative Grothendieck problem, for which Naor \emph{et al.}~\cite{naor2014efficient} provided a rounding procedure of its semidefinite relaxation. Their algorithm achieves an approximation ratio of at least $1/2\sqrt{2} \approx 0.353$ in the real-valued setting, and $1/2$ in the complex-valued setting (wherein optimization is over the unitary group instead of the orthogonal group). This $1/2$ result was later shown to be tight by Bri\"{e}t \emph{et al.}~\cite{briet2017tight} for both the real- and complex-valued settings;~in fact, they show that this is the $\mathsf{NP}$-hardness threshold of a special case of the problem, called the \emph{little} noncommutative Grothendieck problem.\footnote{See Section 6 of Bri\"{e}t \emph{et al.}~\cite{briet2017tight} for the precise relation between the big and little NCG.} However, the threshold for Problem~\eqref{eq:LNCG}, which is an special case of LNCG, is not known. Algorithmically, Bandeira \emph{et al.}~\cite{bandeira2016approximating} demonstrated constant approximation ratios for Problem~\eqref{eq:LNCG} when $C \succeq 0$ and $G = \Orth(n)$ or $\U(n)$ via an $(n \times n)$-dimensional generalization of GW rounding, along with matching integrality gaps. These approximation ratios exceed $1/2$, indicating that this subclass is quantitatively less difficult than the general instance of the LNCG problem. Although the optimization of rotation matrices is of central importance to many applications, we are unaware of any general approximation ratio guarantees for the $G = \SO(n)$ setting.

\section{Applications of the LNCG problem}\label{sec:lncg_applications}
Before describing our quantum relaxation, here we motivate the practical interest in Problem~\eqref{eq:LNCG} by briefly discussing some applications. Throughout, let $G = \Orth(n)$ or $\SO(n)$ and $(V, E)$ be a graph as before.

\subsection{Group synchronization}\label{sec:group_sync}

The group synchronization problem over orthogonal transformations has applications in a variety of disciplines, including structural biology, robotics, and wireless networking. For example, in structural biology the problem appears as part of the cryogenic electron microscopy (cryo-EM) technique. There, one uses electron microscopy on cryogenically frozen samples of a molecular structure to obtain a collection of noisy images of the structure. The images are noisy due to an inherently low signal-to-noise ratio, and furthermore they feature the structure in different, unknown orientations (represented by rotation matrices). One approach to solving the group synchronization problem yields best-fit estimates for these orientations via least-squares minimization~\cite{boumal2013robust}, from which one can produce a model of the desired 3D structure.\footnote{Note that other loss functions are also considered in the literature, which may not necessarily have a reformulation as Problem~\eqref{eq:LNCG}.} See Ref.~\cite{singer2018mathematics} for a further overview, and Ref.~\cite{ozyecsil2017survey} for a survey of other applications of group synchronization.

The formal problem description is as follows. To each vertex $v \in [m] \coloneqq \{ 1, \ldots, m \}$ we assign an unknown but fixed element $g_v \in G$. An interaction between each pair of vertices connected by an edge $(u, v) \in E$ is modeled as $g_{uv} = g_u g_v^\T$. However, measurements of the interactions are typically corrupted by some form of noise. For instance, one may consider an additive noise model of the form $C_{uv} = g_{uv} + \sigma W_{uv}$, where $\sigma \geq 0$ characterizes the strength of the noise and each $W_{uv} \in \R^{n \times n}$ has independently, normally distributed entries. We would like to recover each $g_v$ given only access to the matrices $C_{uv}$. Therefore, as a proxy to the recovery problem one may cast the solution as the least-squares minimizer
\begin{equation}\label{eq:group_sync_argmin}
    \min_{\mathbf{R} \in G^m} \sum_{(u, v) \in E} \| C_{uv} - R_u R_v^\T \|_F^2,
\end{equation}
where $\| A \|_F = \sqrt{\langle A, A \rangle}$ is the Frobenius norm and we employ the notation $\mathbf{R} \equiv (R_1, \ldots, R_m)$. It is straightforward to see that the minimzer of this problem is equivalent to the maximizer of
\begin{equation}
    \max_{\mathbf{R} \in G^m} \sum_{(u, v) \in E} \langle C_{uv}, R_u R_v^\T \rangle,
\end{equation}
which is precisely in the form of Problem~\eqref{eq:LNCG}.

\subsection{Generalized orthogonal Procrustes problem}

Procrustes analysis has applications in fields such as shape and image recognition, as well as sensory analysis and market research on $n$-dimensional data. In this problem, one has a collection of point clouds, each representing for instance the important features of an image. One wishes to determine how similar these images are to each other collectively. This is achieved by simultaneously fitting each pair of point clouds to each other, allowing for arbitrary orthogonal transformations on each cloud to best align the individual points. We refer the reader to Ref.~\cite{gower2004procrustes} for a comprehensive review.

Consider $m$ sets of $K$ points in $\R^n$, $S_v = \{ x_{v,1}, \ldots, x_{v,K} \} \subset \R^n$ for each $v \in [m]$. We wish to find an orthogonal transformation $R_v \in G$ for each $S_v$ that best aligns all sets of points simultaneously. That is, for each $k \in [K]$ and $u, v \in [m]$ we wish to minimize the Euclidean distance $\| R_u^\T x_{u,k} - R_v^\T x_{v,k} \|_2$. Taking least-squares minimization as our objective, we seek to solve
\begin{equation}\label{eq:procrustes}
    \min_{\mathbf{R} \in G^m} \sum_{u, v \in [m]} \sum_{k \in [K]} \| R_u^\T x_{u,k} - R_v^\T x_{v,k} \|_2^2.
\end{equation}
From the relation between the vector 2-norm and matrix Frobenius norm, Eq.~\eqref{eq:procrustes} can be formulated as
\begin{equation}
    \max_{\mathbf{R} \in G^m} \sum_{u, v \in [m]} \langle C_{uv}, R_u R_v^\T \rangle,
\end{equation}
where each $C_{uv} \in \R^{n \times n}$ is defined as
\begin{equation}
    C_{uv} = \sum_{k \in [K]} x_{u,k} x_{v,k}^\T.
\end{equation}

\section{Summary of results}\label{sec:summary_results}
\begin{figure}
    \centering
    \includegraphics[width=\textwidth]{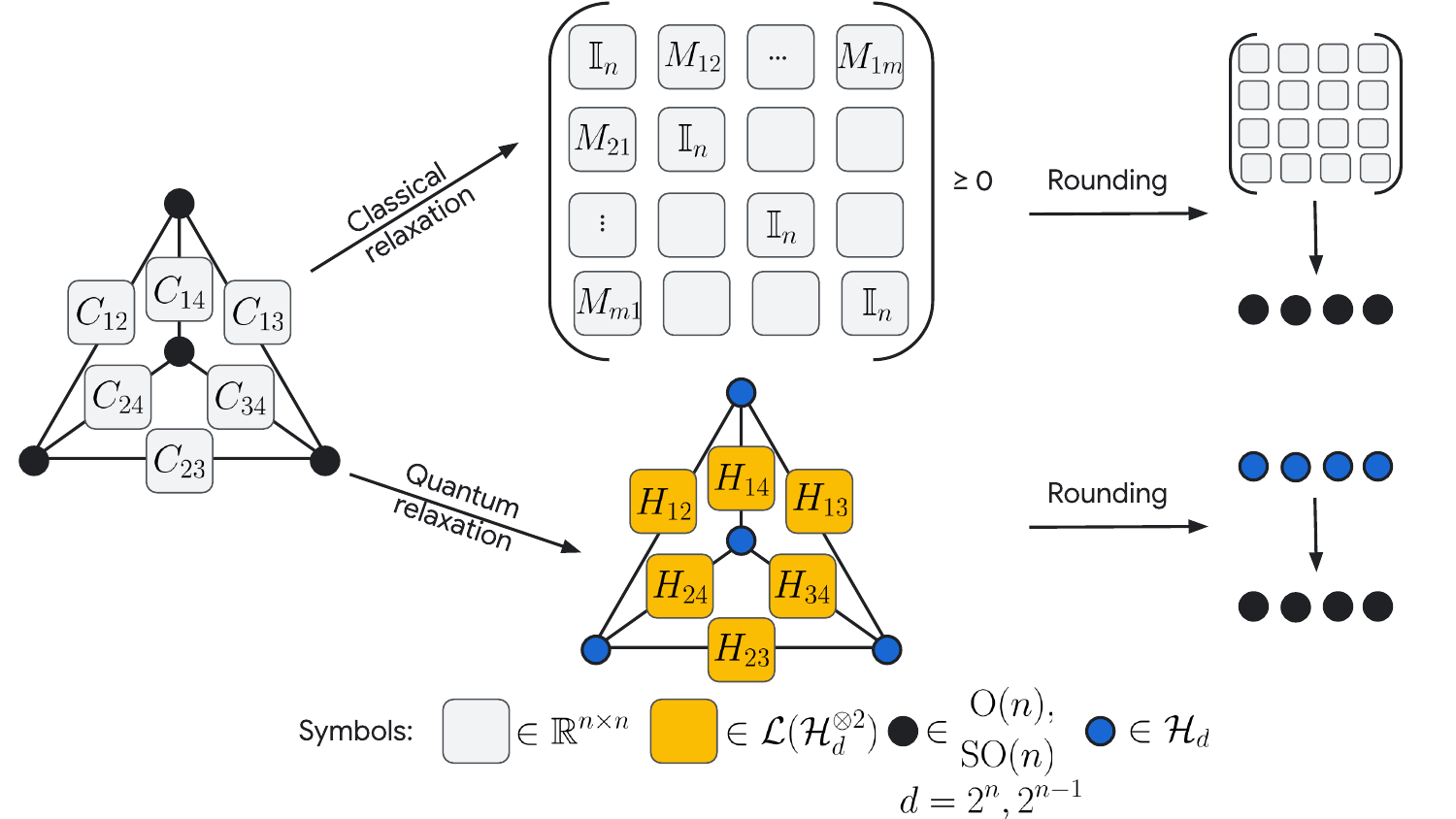}
    \caption[A cartoon description of the quantum and classical encodings of an LNCG problem, followed by classical and quantum rounding.]{A cartoon description of the quantum and classical encodings of an LNCG problem, followed by classical and quantum rounding. (Left) The description of the problem that we consider, which is described by a graph $([m], E)$ and $n \times n$ matrices $C_{uv}$ for each edge $(u, v)$. We wish to assign elements of $\Orth(n)$ or $\SO(n)$ to each vertex such that the quadratic form of Eq.~\eqref{eq:summary_obj} is maximized. (Center top) The description of the standard classical relaxation of the LNCG problem as an $mn \times mn$ PSD matrix $M \succeq 0$, which is optimized using a semidefinite program. (Right top) The classical rounding procedure, which returns a collection of orthogonal matrices from $M$. (Center bottom) A description of our quantum formulation of the LNCG problem as a two-body interacting Hamiltonian. On each vertex we place a $d$-dimensional Hilbert space, and the Hamiltonian corresponds to interaction terms $H_{uv}$ on the edges ($\mathcal{L}(\mathcal{H})$ is the set of linear operators on a Hilbert space $\mathcal{H}$). The classical solution of the LNCG problem lies in a subset of the full Hilbert space containing separable Gaussian states. (Right bottom) Our proposed quantum rounding protocols. One protocol requires knowledge of the two-body reduced density matrices across edges, while the other uses the one-body reduced density matrices on each vertex.}
    \label{fig:quantum_classical_round_cartoon}
\end{figure}

We now provide a high-level overview of the main contributions of this work. We provide summary cartoon in Figure~\ref{fig:quantum_classical_round_cartoon}, depicting the quantum embedding of the problem and the quantum rounding protocols. Let $(V, E)$ be a graph where we label the vertices by $V = [m]$, and denote the objective function of Problem~\eqref{eq:LNCG} by
\begin{equation}\label{eq:summary_obj}
    f(\mathbf{R}) \coloneqq \sum_{(u, v) \in E} \langle C_{uv}, R_u R_v^\T \rangle.
\end{equation}

\subsection{Quantum Hamiltonian relaxation}

First, consider the setting in which $\mathbf{R} = (R_1, \ldots, R_m) \in \Orth(n)^m$. We embed this problem into a Hamiltonian by placing $n$ qubits on each vertex $v \in [m]$, resulting in a total Hilbert space $\mathcal{H}_{2^n}^{\otimes m}$ of $mn$ qubits. Define the $n$-qubit Pauli operators
\begin{equation}
    P_{ij} \coloneqq \begin{cases}
    -X_i Z_{i+1} \cdots Z_{j-1} X_j & i < j,\\
    Z_i & i = j,\\
    -Y_j Z_{j+1} \cdots Z_{i-1} Y_i & i > j,
    \end{cases}
\end{equation}
where $Z_i \coloneqq \I_2^{\otimes (i-1)} \otimes Z \otimes \I_2^{\otimes (n-i)}$ (similarly for $X_i$, $Y_i$). The Hamiltonian
\begin{equation}\label{eq:LNCG_H_summary}
    H \coloneqq \sum_{(u, v) \in E} \sum_{i, j \in [n]} [C_{uv}]_{ij} \sum_{k \in [n]} P_{ik}^{(u)} \otimes P_{jk}^{(v)}
\end{equation}
defines our quantum relaxation of the objective $f$ over $\Orth(n)^m$. The notation $A^{(v)}$ denotes the operator $A$ acting only on the Hilbert space of vertex $v$, and we overload this notation to indicate either the $n$-qubit operator or $mn$-qubit operator acting trivially on the remaining vertices. When the context is clear we typically omit writing the trivial support.

For optimization over $\SO(n)^m$, we consider instead the $(n-1)$-qubit Pauli operators
\begin{align}
    \widetilde{P}_{ij} &\coloneqq \Pi_0 P_{ij} \Pi_0^\T,\\
    \Pi_0 &= \frac{1}{\sqrt{2}} \l( \bra{+} \otimes \I_2^{\otimes (n-1)} + \bra{-} \otimes Z^{\otimes (n-1)} \r),
\end{align}
where $\Pi_0 : \mathcal{H}_{2^n} \to \mathcal{H}_{2^{n-1}}$ represents the projection onto the even-parity subspace of $\mathcal{H}_{2^n}$. The construction of the relaxed Hamiltonian for $\SO(n)$ is then analogous to Eq.~\eqref{eq:LNCG_H_summary}:
\begin{equation}
    \widetilde{H} \coloneqq \sum_{u, v \in E} \sum_{i, j \in [n]} [C_{uv}]_{ij} \sum_{k \in [n]} \widetilde{P}_{ik}^{(u)} \otimes \widetilde{P}_{jk}^{(v)},
\end{equation}
where now the relaxed quantum problem is defined over $m(n-1)$ qubits.

These Hamiltonians serve as relaxations to Problem~\eqref{eq:LNCG} in the following sense. First, we show that for every $R \in \Orth(n)$, there is an $n$-qubit state $\ket{\phi(R)}$ which is the maximum eigenstate of
\begin{equation}
    F(R) = \sum_{i,j \in [n]} R_{ij} P_{ij}.
\end{equation}
In particular, $F(R)$ is a free-fermion Hamiltonian, so $\ket{\phi(R)}$ is a fermionic Gaussian state. If $R \in \SO(n)$, then furthermore $\ket{\phi(R)}$ is an even-parity state, i.e., $\ev{\phi(R)}{Z^{\otimes n}}{\phi(R)} = 1$, so it is only supported on a subspace of dimension $2^{n-1}$ (the image of $\Pi_0$). This correspondence establishes a reformulation of the classical optimization problem as a constrained Hamiltonian problem:
\begin{equation}\label{eq:quantum_reformulation}
    \max_{\mathbf{R} \in G^m} f(\mathbf{R}) = \max_{\substack{\ket{\psi} = \bigotimes_{v \in [m]} \ket{\phi(R_v)} \\ R_v \in G}} \ev{\psi}{H}{\psi}.
\end{equation}
Dropping these constraints on $\ket{\psi}$ implies the inequalities
\begin{align}
    \max_{\mathbf{R} \in \Orth(n)^m} f(\mathbf{R}) &\leq \max_{\rho \in \mathcal{D}(\mathcal{H}_{2^n}^{\otimes m})} \tr(H \rho),\\
    \max_{\mathbf{R} \in \SO(n)^m} f(\mathbf{R}) &\leq \max_{\rho \in \mathcal{D}(\mathcal{H}_{2^{n-1}}^{\otimes m})} \tr(\widetilde{H} \rho),
\end{align}
where $\mathcal{D}(\mathcal{H})$ denotes the set of density operators on a Hilbert space $\mathcal{H}$. This establishes the quantum Hamiltonian relaxation.

\subsection{Quantum rounding}

\begin{algorithm}[t]
\caption{$\conv G$-based rounding of edge marginals.\label{alg:conv_rounding}}
\KwData{Quantum state $\rho \in \mathcal{D}(\mathcal{H}_d^{\otimes m})$ over a graph of $m$ vertices, each with local Hilbert space of dimension $d = 2^n$ if $G = \Orth(n)$, or $d = 2^{n-1}$ if $G = \SO(n)$}
\KwResult{Orthogonal matrices on each vertex, $R_1, \ldots, R_m \in G$}

$\mathcal{M} \gets \I_{mn}$ \;

\For{$u \neq v \in [m]$}{
    \For{$(i, j) \in [n]^2$}{
        \uIf{$G = \Orth(n)$}{
            $[\mathcal{M}_{uv}]_{ij} \gets \frac{1}{n} \tr(\Gamma_{ij}^{(u,v)} \rho)$ \;
        }
        \ElseIf{$G = \SO(n)$}{
            $[\mathcal{M}_{uv}]_{ij} \gets \frac{1}{n} \tr(\widetilde{\Gamma}_{ij}^{(u,v)} \rho)$ \;
        }
    }
}

\For{$v \in [m]$}{
    $R_v \gets \argmin_{Y \in G} \| Y - \mathcal{M}_{1v} \|_F$ \;
}
\end{algorithm}

\begin{algorithm}[t]

\caption{Rounding vertex marginals.\label{alg:vertex_rounding}}
\KwData{Quantum state $\rho \in \mathcal{D}(\mathcal{H}_d^{\otimes m})$ over a graph of $m$ vertices, each with local Hilbert space of dimension $d = 2^n$ if $G = \Orth(n)$, or $d = 2^{n-1}$ if $G = \SO(n)$}
\KwResult{Orthogonal matrices on each vertex, $R_1, \ldots, R_m \in G$}

\For{$v \in [m]$}{
    ${Q}_v \gets 0 \in \R^{n \times n}$ \;
    \For{$(i, j) \in [n]^2$}{
    \uIf{$G = \Orth(n)$}{
        $[{Q}_v]_{ij} \gets \tr(P_{ij}^{(v)} \rho)$ \;
    }
    \ElseIf{$G = \SO(n)$}{
        $[{Q}_v]_{ij} \gets \tr(\widetilde{P}_{ij}^{(v)} \rho)$ \;
    }
    }
}

\For{$v \in [m]$}{
    $R_v \gets \argmin_{Y \in G} \| Y - {Q}_v \|_F$ \;
}
\end{algorithm}

In order to recover orthogonal matrices from a relaxed quantum solution $\rho$, we propose two rounding procedures, summarized in Algorithms~\ref{alg:conv_rounding} and \ref{alg:vertex_rounding}. These rounding procedures operate on local (i.e., single- or two-vertex observables) expectation values of $\rho$ stored in classical memory, which can be efficiently estimated, e.g., by partial state tomography.

Algorithm~\ref{alg:conv_rounding} is inspired by constructing a quantum analogue of the PSD variable appearing in semidefinite relaxations to Problem~\eqref{eq:LNCG}. Consider the $mn \times mn$ matrix of expectation values
\begin{equation}
    \mathcal{M} \coloneqq \begin{bmatrix}
    \I_n & T_{12} & \cdots & T_{1m}\\
    T_{21} & \I_n & \cdots & T_{2m}\\
    \vdots & \vdots & \ddots & \vdots\\
    T_{m1} & T_{m2} & \cdots & \I_n
    \end{bmatrix},
\end{equation}
where the off-diagonal blocks are defined as
\begin{align}
    T_{uv} &\coloneqq \frac{1}{n} \begin{bmatrix}
    \tr(\Gamma_{11}^{(u, v)} \rho) & \cdots & \tr(\Gamma_{1n}^{(u, v)} \rho)\\
    \vdots & \ddots & \vdots\\
    \tr(\Gamma_{n1}^{(u, v)} \rho) & \cdots & \tr(\Gamma_{nn}^{(u, v)} \rho)
    \end{bmatrix} = T_{vu}^\T,\\
    \Gamma_{ij}^{(u, v)} &\coloneqq \sum_{k \in [n]} P_{ik}^{(u)} \otimes P_{jk}^{(v)}
\end{align}
when $G = \Orth(n)$, and we replace the operators $P_{ij}$ with $\widetilde{P}_{ij}$ when $G = \SO(n)$. We show that $\mathcal{M}$ satisfies the following properties for all states $\rho$:
\begin{align}
    &\mathcal{M} \succeq 0,\\
    &\mathcal{M}_{uv} \in \conv G \quad \forall u, v \in [m],
\end{align}
where $\conv G$ is the convex hull of $G$. Thus when $G = \SO(n)$, $\mathcal{M}$ obeys the same constraints as the $\conv\SO(n)$-based semidefinite relaxation proposed by Saunderson \emph{et al.}~\cite{saunderson2014semidefinite}. However, whereas the classical representation of the $\conv\SO(n)$ constraints requires at least matrices of size $2^{n-1} \times 2^{n-1}$ for each edge, our quantum state automatically satisfies these constraints (using only $n - 1$ qubits per vertex).

Algorithm~\ref{alg:vertex_rounding} uses the single-vertex information $\tr(P_{ij}^{(v)} \rho)$ of $\rho$, as opposed to the two-vertex information $\tr(\Gamma_{ij}^{(u, v)} \rho)$. We consider this rounding procedure due to the fact that, if $\rho$ is a pure Gaussian state satisfying the constraint of Eq.~\eqref{eq:quantum_reformulation}, then the matrix of expectation values
\begin{equation}
    Q_v \coloneqq \begin{bmatrix}
    \tr(P_{11}^{(v)} \rho) & \cdots & \tr(P_{1n}^{(v)} \rho)\\
    \vdots & \ddots & \vdots\\
    \tr(P_{n1}^{(v)} \rho) & \cdots & \tr(P_{nn}^{(v)} \rho)
    \end{bmatrix}
\end{equation}
lies in $\Orth(n)$. On the other hand, for arbitrary density matrices we have the relaxation $Q_v \in \conv\Orth(n)$, and again when we replace $P_{ij}^{(v)}$ with $\widetilde{P}_{ij}^{(v)}$ then $Q_v \in \conv\SO(n)$.

Both rounding procedures use the standard projection of the matrices $X \in \conv G$ (e.g., the matrices $T_{uv}$ or $Q_v$ measured from the quantum state) to some $R \in G$ by finding the nearest (special) orthogonal matrix according to Frobenius-norm distance:
\begin{equation}\label{eq:nearest_ortho_summary}
    R = \argmin_{Y \in G} \|X - Y\|_F.
\end{equation}
This can be solved efficiently as a classical postprocessing step, essentially by computing the singular value decomposition of $X = U \Sigma V^\T$. When $G = \Orth(n)$, the solution is $R = UV^\T$. When $G = \SO(n)$, we instead use the so-called special singular value decomposition of $X = U \widetilde{\Sigma} \widetilde{V}^\T$, where $\widetilde{\Sigma} = \Sigma J$ and $\widetilde{V} = VJ$, with $J$ being the diagonal matrix
\begin{equation}\label{eq:summary_J}
    J = \begin{bmatrix}
    \I_{n-1} & 0\\
    0 & \det(UV^\T)
    \end{bmatrix},
\end{equation}
assuming that the singular values $\sigma_i(X)$ are in descending order, $\sigma_1(X) \geq \cdots \geq \sigma_n(X)$. Then the solution to Eq.~\eqref{eq:nearest_ortho_summary} is $R = U \widetilde{V}^\T \in \SO(n)$.

\section{\label{sec:quantum_formalism}Quantum formalism for optimization over orthogonal matrices}
Our key insight into encoding orthogonal matrices into quantum states comes from the construction of the orthogonal group from a Clifford algebra. We review this mathematical construction in Appendix~\ref{sec:clifford} and only discuss the main aspects here. The Clifford algebra $\Cl(n)$ is a $2^n$-dimensional real vector space equipped with an inner product and multiplication operation satisfying the anticommutation relation
\begin{equation}
    e_i e_j + e_j e_i = -2 \delta_{ij} \openone,
\end{equation}
where $e_1, \ldots, e_n$ is an orthonormal basis for $\R^n$ and $\openone$ is the multiplicative identity of the algebra. The orthogonal group is then realized through a quadratic map $Q : \Cl(n) \to \R^{n \times n}$ and the identification of a subgroup $\Pin(n) \subset \Cl(n)$ such that $Q(\Pin(n)) = \Orth(n)$. Notably, the elements of $\Pin(n)$ have unit norm (with respect to the inner product on $\Cl(n)$). The special orthogonal group, meanwhile, is constructed by considering only the even-parity elements of $\Cl(n)$, denoted by $\Cl^0(n)$. The group $\Spin(n) = \Pin(n) \cap \Cl^0(n)$ then yields $Q(\Spin(n)) = \SO(n)$.

Because the Clifford algebra $\Cl(n)$ is a $2^n$-dimensional vector space, we observe that it can be identified with a Hilbert space of $n$ qubits.\footnote{In fact, $n$ rebits suffice since $\Cl(n)$ is a real vector space, but to keep the presentation straightforward we will not make such a distinction.} In this section we explore this connection in detail, showing how to represent orthogonal matrices as quantum states and how the mapping $Q$ acts as a linear functional on those states.

\subsection{Qubit representation of the Clifford algebra}

First we describe the canonical isomorphism between $\Cl(n)$ and $\mathcal{H}_{2^n} \coloneqq (\R^2)^{\otimes n}$ as Hilbert spaces. We denote the standard basis of $\Cl(n)$ by $\{ e_I \coloneqq e_{i_1} \cdots e_{i_k} \mid I = \{i_1, \ldots, i_k\} \subseteq [n] \}$. By convention we assume that the elements of $I$ are ordered as $i_1 < \cdots < i_k$. Each basis element $e_I$ maps onto to a computational basis state $\ket{b}$, where $b = b_1 \cdots b_n \in \{0, 1\}^n$, via the correspondence
\begin{equation}
    e_I \equiv \bigotimes_{i \in [n]} \ket{b_i}, \quad \text{where } b_i = \begin{cases}
1  & \text{if } i \in I,\\
0 & \text{otherwise}.
\end{cases}
\end{equation}
The inner products on both spaces coincide since this associates one orthonormal basis to another. This correspondence also naturally equates the grade $|I|$ of the Clifford algebra with the Hamming weight $|b|$ of the qubits. The notion of parity, $|I| \mod 2 = |b| \mod 2$, is therefore preserved, so $\Cl^0(n)$ corresponds to the subspace of $\mathcal{H}_{2^n}$ with even Hamming weight.

To represent the multiplication of algebra elements in this Hilbert space, we use the fact that left- and right-multiplication are linear automorphisms on $\Cl(n)$, which are denoted by
\begin{equation}
    \lambda_x(y) = xy, \quad \rho_x(y) = yx.
\end{equation}
The action of the algebra can therefore be represented on $\mathcal{H}_{2^n}$ as linear operators. We shall use the matrix representation provided in Ref.~\cite{saunderson2015semidefinite}, as it precisely coincides with the $n$-qubit computational basis described above. Because of linearity, it suffices to specify left- and right-multiplication by the generators $e_i$, which are the operators
\begin{align}
    \lambda_i &\equiv Z^{\otimes (i - 1)} \otimes (-\i Y) \otimes \I_2^{\otimes (n - i)}, \label{eq:lambda}\\
    \rho_i &\equiv \I_2^{\otimes (i - 1)} \otimes (-\i Y) \otimes Z^{\otimes (n - i)}. \label{eq:rho}
\end{align}
It will also be useful to write down the parity automorphism $\alpha(e_I) = (-1)^{|I|} e_I$ under this matrix representation. As the notion of parity is equivalent between $\Cl(n)$ and $\mathcal{H}_{2^n}$, $\alpha$ is simply the $n$-qubit parity operator,
\begin{equation}\label{eq:alpha}
    \alpha \equiv Z^{\otimes n}.
\end{equation}

It will also be useful to represent the subspace $\Cl^0(n)$ explicitly as an $(n-1)$-qubit Hilbert space. This is achieved by the projection from $\Cl(n)$ to $\Cl^0(n)$, expressed in Ref.~\cite{saunderson2015semidefinite} as the $2^{n-1} \times 2^n$ matrix
\begin{equation}\label{eq:projector}
    \Pi_{0} \coloneqq \frac{1}{\sqrt{2}} \l( \bra{+} \otimes \I_2^{\otimes (n - 1)} + \bra{-} \otimes Z^{\otimes (n - 1)} \r).
\end{equation}
It is straightforward to check that $\Pi_0 \ket{b} = 0$ if $|b| \mod 2 = 1$, and that its image is a $2^{n-1}$-dimensional Hilbert space.

\subsection{The quadratic mapping as quantum expectation values}

The quadratic map $Q : \Cl(n) \to \R^{n \times n}$ is defined as
\begin{equation}
    Q(x)(v) \coloneqq \pi_{\R^n}(\alpha(x) v \overline{x}) \quad \forall x \in \Cl(n), v \in \R^n,
\end{equation}
where $\pi_{\R^n}$ is the projector from $\Cl(n)$ to $\R^n$,
\begin{equation}
    \pi_{\R^n}(x) \coloneqq \sum_{i \in [n]} \langle e_i, x \rangle e_i \quad \forall x \in \Cl(n),
\end{equation}
and the conjugation operation $x \mapsto \overline{x}$ is defined as the linear extension of $\overline{e_I} = (-1)^{|I|} e_{i_k} \cdots e_{i_1}$. This map associates Clifford algebra elements with orthogonal matrices via the relations $Q(\Pin(n)) = \Orth(n)$ and $Q(\Spin(n)) = \SO(n)$ (see Appendix~\ref{sec:clifford} for a review of the construction). In the standard basis of $\R^n$, the linear map $Q(x) : \R^n \to \R^n$ has the matrix elements
\begin{equation}
\begin{split}
    [Q(x)]_{ij} &= \langle e_i, Q(x)(e_j) \rangle\\
    &= \langle e_i, \alpha(x) e_j \overline{x} \rangle.
\end{split}
\end{equation}
Using the linear maps $\lambda_i, \rho_j$ of left- and right-multiplication by $e_i, e_j$, as well as the conjugation identity $\langle x, y\overline{z} \rangle = \langle xz, y \rangle$ in the Clifford algebra, these matrix elements of $Q(x)$ can be rearranged as
\begin{equation}\label{eq:Q_matrix_elements}
\begin{split}
    [Q(x)]_{ij} &= \langle e_i, \alpha(x) e_j \overline{x} \rangle\\
    &= \langle e_i x, \alpha(x) e_j \rangle\\
    &= \langle \lambda_i(x), \rho_j(\alpha(x)) \rangle\\
    &= \langle x, \lambda_i^\dagger(\rho_j(\alpha(x))) \rangle.
\end{split}
\end{equation}

We now transfer this expression to the quantum representation developed above. First, define the following $n$-qubit Pauli operators as the composition of the linear maps appearing in Eq.~\eqref{eq:Q_matrix_elements}:
\begin{equation}
    P_{ij} \coloneqq \lambda_i^\dagger \rho_j \alpha = \begin{cases}
    -\I_2^{\otimes (i - 1)} \otimes X \otimes Z^{\otimes (j - i - 1)} \otimes X \otimes \I_2^{\otimes (n - j)} & i < j,\\
    \I_2^{\otimes (i - 1)} \otimes Z \otimes \I_2^{\otimes (n - i)} & i = j,\\
    -\I_2^{\otimes (j - 1)} \otimes Y \otimes Z^{\otimes (i - j - 1)} \otimes Y \otimes \I_2^{\otimes (n - i)} & i > j,
    \end{cases}
\end{equation}
where the expressions in terms of Pauli matrices follow from Eqs.~\eqref{eq:lambda} to \eqref{eq:alpha}. Then we may rewrite Eq.~\eqref{eq:Q_matrix_elements} as
\begin{equation}
    [Q(x)]_{ij} = \langle x | P_{ij} | x \rangle,
\end{equation}
where $\ket{x} \in \mathcal{H}_{2^n}$ is the quantum state identified with $x \in \Cl(n)$. Hence, the matrix elements of $Q(x) \in \R^{n \times n}$ possess the interpretation as expectation values of a collection of $n^2$ Pauli observables $\{ P_{ij} \}_{i, j \in [n]}$. Furthermore, recall that $Q(x) \in \Orth(n)$ if and only if $x \in \Pin(n)$, and $Q(x) \in \SO(n)$ if and only if $x \in \Spin(n)$. Because $\Spin(n) = \Pin(n) \cap \Cl^0(n)$, one can work in the even-parity sector directly by projecting the operators as
\begin{equation}\label{eq:P_ij_son}
    \widetilde{P}_{ij} \coloneqq \Pi_0 P_{ij} \Pi_0^\T.
\end{equation}
These are $(n-1)$-qubit Pauli operators, and we provide explicit expressions in Appendix~\ref{sec:even_subspace}. When necessary, we may specify another map $\widetilde{Q} : \Cl^0(n) \to \R^{n \times n}$,
\begin{equation}
    [\widetilde{Q}(x)]_{ij} \coloneqq \langle x | \widetilde{P}_{ij} | x \rangle,
\end{equation}
for which $\widetilde{Q}(\Spin(n)) = \SO(n)$.

In general, these double covers are only a subset of the unit sphere in $\mathcal{H}_{d}$ ($d = 2^n$ or $2^{n-1}$), so not all quantum states mapped by $Q$ yield orthogonal matrices. In Section~\ref{sec:free-fermion} we characterize the elements of $\Pin(n)$ and $\Spin(n)$ as a class of well-studied quantum states, namely, pure fermionic Gaussian states.

\subsection{\label{sec:free-fermion}Fermionic representation of the construction}

\subsubsection{Notation}

First we establish some notation. A system of $n$ fermionic modes, described by the creation operators $a_1^\dagger, \ldots, a_n^\dagger$, can be equivalently represented by the $2n$ Majorana operators
\begin{align}
    \gamma_i &= a_i + a_i^\dagger,\\
    \widetilde{\gamma}_i &= -\i(a_i - a_i^\dagger),
\end{align}
for all $i \in [n]$. These operators form a representation for the Clifford algebra $\Cl(2n)$, as they satisfy\footnote{Note that we adopt the physicist's convention here, which takes the generators to be Hermitian, as opposed to Eq.~\eqref{eq:anticommutator_cl_n} wherein they square to $-\openone$.}
\begin{align}
    \gamma_i \gamma_j + \gamma_j \gamma_i = \widetilde{\gamma}_i \widetilde{\gamma}_j + \widetilde{\gamma}_j \widetilde{\gamma}_i &= 2 \delta_{ij} \openone,\\
    \gamma_i \widetilde{\gamma}_j + \widetilde{\gamma}_j \gamma_i &= 0.
\end{align}
The Jordan--Wigner mapping allows us to identify this fermionic system with an $n$-qubit system via the relations
\begin{align}
    \gamma_{i} &= Z^{\otimes (i - 1)} \otimes X \otimes \I_2^{\otimes (n - i)}, \label{eq:gamma_X}\\
    \widetilde{\gamma}_i &= Z^{\otimes (i - 1)} \otimes Y \otimes \I_2^{\otimes (n - i)}.\label{eq:gamma_Y}
\end{align}
We will work with the two representations interchangeably.

A central tool for describing noninteracting fermions is the Bogoliubov transformation $\bm{\gamma} \mapsto O \bm{\gamma}$, where $O \in \Orth(2n)$ and
\begin{equation}
    \bm{\gamma} \coloneqq \begin{bmatrix}
    \widetilde{\gamma}_1 & \cdots & \widetilde{\gamma}_n & \gamma_1 & \cdots & \gamma_n
    \end{bmatrix}^\T.
\end{equation}
This transformation is achieved by fermionic Gaussian unitaries, which are equivalent to matchgate circuits on qubits under the Jordan--Wigner mapping~\cite{knill2001fermionic,terhal2002classical,jozsa2008matchgates}. In particular, we will make use of a subgroup of such unitaries corresponding to $\Orth(n) \times \Orth(n) \subset \Orth(2n)$. For any $U, V \in \Orth(n)$, let $\mathcal{U}_{(U, V)}$ be the fermionic Gaussian unitary with the adjoint action
\begin{align}
    \mathcal{U}_{(U, V)} \widetilde{\gamma}_i \mathcal{U}_{(U, V)}^\dagger &= \sum_{j \in [n]} U_{ij} \widetilde{\gamma}_j, \label{eq:gamma_tilde_gaussian}\\
    \mathcal{U}_{(U, V)} \gamma_i \mathcal{U}_{(U, V)}^\dagger &= \sum_{j \in [n]} V_{ij} \gamma_j. \label{eq:gamma_gaussian}
\end{align}
In contrast to arbitrary $\Orth(2n)$ transformations, these unitaries do not mix between the $\gamma$- and $\widetilde{\gamma}$-type Majorana operators.

\subsubsection{\label{sec:linear_free_fermion}Linear optimization as free-fermion models}

Applying the representation of Majorana operators under the Jordan--Wigner transformation, Eqs.~\eqref{eq:gamma_X} and \eqref{eq:gamma_Y},  to the Clifford algebra automorphisms, Eqs.~\eqref{eq:lambda} to \eqref{eq:alpha}, we see that $\lambda_i^\dagger = \i \widetilde{\gamma}_i$ and $\rho_j \alpha = \gamma_j$. Therefore the Pauli operators $P_{ij}$ defining the quadratic map $Q$ are equivalent to fermionic one-body operators,
\begin{equation}
    P_{ij} = \i \widetilde{\gamma}_i \gamma_j.
\end{equation}
Consider now a linear objective function $\ell(X) \coloneqq \langle C, X \rangle$ for some fixed $C \in \R^{n \times n}$, which we wish to optimize over $\Orth(n)$:
\begin{equation}\label{eq:linear_obj}
    \max_{X \in \Orth(n)} \ell(X) = \max_{X \in \Orth(n)} \langle C, X \rangle.
\end{equation}
Because we require $X \in \Orth(n)$, it is equivalent to search over all $x \in \Pin(n)$ through $Q$:
\begin{equation}
    \max_{X \in \Orth(n)} \langle C, X \rangle = \max_{x \in \Pin(n)} \langle C, Q(x) \rangle.
\end{equation}
Writing out the matrix elements explicitly, we see that the objective takes the form
\begin{equation}
\begin{split}
    \ell(X) &= \sum_{i,j \in [n]} C_{ij} [Q(x)]_{ij}\\
    &= \sum_{i,j \in [n]} C_{ij} \langle x | P_{ij} | x \rangle\\
    &= \langle x | F(C) | x \rangle,
\end{split}
\end{equation}
where we have defined the noninteracting fermionic Hamiltonian
\begin{equation}
    F(C) \coloneqq \sum_{i,j \in [n]} C_{ij} P_{ij} = \i \sum_{i,j \in [n]} C_{ij} \widetilde{\gamma}_i \gamma_j.
\end{equation}
The linear optimization problem is therefore equivalent to solving a free-fermion model,
\begin{equation}\label{eq:linear_obj_Pin}
    \max_{X \in \Orth(n)} \langle C, X \rangle = \max_{x \in \Pin(n)} \langle x | F(C) | x \rangle,
\end{equation}
the eigenvectors of which are fermionic Gaussian states. As such, this problem can be solved efficiently by a classical algorithm. In fact, the known classical algorithm for solving the optimization problem is exactly the same as that used for diagonalizing $F(C)$.

We now review the standard method to diagonalize $F(C)$. Consider the singular value decomposition of $C = U \Sigma V^\T$, which is computable in time $\Ord(n^3)$. This decomposition immediately reveals the diagonal form of the Hamiltonian:
\begin{equation}
\begin{split}
    F(C) &= \i \sum_{i, j \in [n]} [U \Sigma V^\T]_{ij} \widetilde{\gamma}_i \gamma_j\\
    &= \i \sum_{k \in [n]} \sigma_k(C) \l( \sum_{i \in [n]} [U^\T]_{ki} \widetilde{\gamma}_i \r) \l( \sum_{j \in [n]} [V^\T]_{kj} \gamma_j \r)\\
    &= {\mathcal{U}}_{(U, V)}^\dagger \l( \sum_{k \in [n]} \sigma_k(C) \i \widetilde{\gamma}_k \gamma_k \r) {\mathcal{U}}_{(U, V)}.
\end{split}
\end{equation}
Because $\i \widetilde{\gamma}_k \gamma_k = Z_k$, it follows that the eigenvectors of $F(C)$ are the fermionic Gaussian states
\begin{equation}
    \ket{\phi_b} = {\mathcal{U}}_{(U, V)}^\dagger \ket{b}, \quad b \in \{0, 1\}^n,
\end{equation}
with eigenvalues
\begin{equation}
    E_b = \sum_{k \in [n]} (-1)^{b_k} \sigma_k(C).
\end{equation}
The maximum energy is $E_{0^n} = \tr \Sigma$ since all singular values are nonnegative. The corresponding eigenstate $\ket{\phi_{0^n}}$ is the maximizer of Eq.~\eqref{eq:linear_obj_Pin}, so it corresponds to an element $\phi_{0^n} \in \Pin(n)$. It is straightforward to see this by recognizing that $[Q(\phi_{0^n})]_{ij} = \ev{\phi_{0^n}}{\i \widetilde{\gamma}_i \gamma_j}{\phi_{0^n}} = [UV^\T]_{ij}$. The fact that $Q(\phi_{0^n}) \in \Orth(n)$ if and only if $\phi_{0^n} \in \Pin(n)$ concludes the argument. 


Indeed, the standard classical algorithm~\cite{schonemann1966generalized} for solving Eq.~\eqref{eq:linear_obj} uses precisely the same decomposition. From the cyclic property of the trace and the fact that $\Orth(n)$ is a group, we have
\begin{equation}
    \max_{X \in \Orth(n)} \langle U \Sigma V^\T, X \rangle = \max_{X' \in \Orth(n)} \langle \Sigma, X' \rangle,
\end{equation}
where we have employed the change of variables $X' \coloneqq U^\T X V$. Again, because $\Sigma$ has only nonnegative entries, $\langle \Sigma, X' \rangle$ achieves its maximum, $\tr \Sigma$, when $X' = \I_n$. This implies that the optimal solution is $X = UV^\T$. Note that this problem is equivalent to minimizing the Frobenius-norm distance, since
\begin{equation}
\begin{split}
    \argmin_{X \in \Orth(n)} \| C - X \|_F^2 &= \argmin_{X \in \Orth(n)} \l( \|C\|_F^2 + \|X\|_F^2 - 2 \langle C, X \rangle \r)\\
    &= \argmax_{X \in \Orth(n)} \, \langle C, X \rangle.
\end{split}
\end{equation}

Now suppose we wish to optimize $\ell$ over $\SO(n)$. In this setting, one instead computes $X = U\widetilde{V}^\T$ from the special singular value decomposition of $C = U \widetilde{\Sigma} \widetilde{V}^\T$. This ensures that $\det(X) = 1$ while maximizing $\ell(X)$, as only the smallest singular value $\sigma_n(C)$ has its sign potentially flipped to guarantee the positive determinant constraint. This sign flip also has a direct analogue within the free-fermion perspective. Recall that the determinant of $Q(x) \in \Orth(n)$ is given by the parity of $x \in \Pin(n)$, or equivalently the parity of the state $\ket{x}$ in the computational basis. Note also that all fermionic states are eigenstates of the parity operator. To optimize over $\SO(n)$, we therefore seek the maximal eigenstate $\ket{\phi_b}$ of $F(C)$ which has even parity. If $\ev{\phi_{0^n}}{Z^{\otimes n}}{\phi_{0^n}} = 1$ then we are done. On the other hand, if $\ev{\phi_{0^n}}{Z^{\otimes n}}{\phi_{0^n}} = -1$ then we need to flip only a single bit in $0^n$ to reach an even-parity state. The smallest change in energy by such a flip is achieved from changing the occupation of the mode corresponding to the smallest singular value of $C$. The resulting eigenstate $\ket{\phi_{0^{n-1}1}}$ is then the even-parity state with the largest energy, $E_{0^{n-1}1} = \tr\Sigma - 2\sigma_n(C)$.

Finally, we point out that all elements of $\Pin(n)$ are free-fermion states. To see this, observe that $C$ is arbitrary. We can therefore construct the family of Hamiltonians $\{F(C) \mid C \in \Orth(n)\}$. Clearly, the maximum $\langle C, X \rangle = n$ within this family is achieved when $X = C$, each of which corresponds to a fermionic Gaussian state $\ket{\phi}$ satisfying $F(C)\ket{\phi} = n\ket{\phi}$ and $Q(\phi) = C$. We note that this argument generalizes the mathematical one presented in Ref.~\cite{saunderson2015semidefinite}, which only considered the eigenvectors lying in $\Spin(n)$.

\subsubsection{\label{sec:mixed_states}Mixed states and the convex hull}

First we review descriptions of the convex hull of orthogonal and rotation matrices, the latter of which was characterized by Saunderson \emph{et al.}~\cite{saunderson2015semidefinite}. The convex hull of $\Orth(n)$ is the set of all matrices with operator norm bounded by 1,
\begin{equation}\label{eq:convOn_singval}
    \conv\Orth(n) = \l\{ X \in \R^{n \times n} \mid \sigma_1(X) \leq 1 \r\}.
\end{equation}
On the other hand, the convex hull of $\SO(n)$ has a more complicated description in terms of special singular values:
\begin{equation}\label{eq:convSOn_singval}
    \conv\SO(n) = \l\{ X \in \R^{n \times n} \mathrel{\Bigg|} \sum_{i \in [n] \setminus I} \widetilde{\sigma}_i(X) - \sum_{i \in I} \widetilde{\sigma}_i(X) \leq n - 2 \quad \forall I \subseteq [n], |I| \text{ odd} \r\}.
\end{equation}
Saunderson \emph{et al.}~\cite{saunderson2015semidefinite} establish that this convex body is a spectrahedron, the feasible region of a semidefinite program. The representation that we will be interested in is called a PSD lift:
\begin{equation}\label{eq:conv-son_psd-lift}
    \conv\SO(n) = \l\{
    \begin{bmatrix}
    \langle \widetilde{P}_{11}, \rho \rangle & \cdots & \langle \widetilde{P}_{1n}, \rho \rangle\\
    \vdots & \ddots & \vdots\\
    \langle \widetilde{P}_{n1}, \rho \rangle & \cdots & \langle \widetilde{P}_{nn}, \rho \rangle
    \end{bmatrix}
    \mathrel{\Bigg|} \rho \succeq 0, \tr \rho = 1 \r\},
\end{equation}
where the $2^{n-1} \times 2^{n-1}$ matrices $\widetilde{P}_{ij}$ are defined in Eq.~\eqref{eq:P_ij_son}.\footnote{Technically, Saunderson \emph{et al.}~\cite{saunderson2015semidefinite} use the definition $\widetilde{P}_{ij} = -\Pi_0 \lambda_i \rho_j \Pi_0^\T$ because they employ the standard adjoint representation, which differs from our use of the twisted adjoint representation which includes the parity automorphism $\alpha$. However since $\alpha(x) = x$ for all $x \in \Cl^0(n)$, both definitions of $\widetilde{P}_{ij}$ coincide.} 

Recall that the density operators on a Hilbert space $\mathcal{H}$ form the convex hull of its pure states:
\begin{equation}
    \mathcal{D}(\mathcal{H}) \coloneqq \conv\{ \op{\psi}{\psi} \mid \ket{\psi} \in \mathcal{H}, \ip{\psi}{\psi} = 1 \} = \{ \rho \in \mathcal{L}(\mathcal{H}) \mid \rho \succeq 0, \tr\rho = 1 \}.
\end{equation}
From Eq.~\eqref{eq:conv-son_psd-lift} one immediately recognizes that the PSD lift of $\conv\SO(n)$ corresponds to $\mathcal{D}(\mathcal{H}_{2^{n-1}})$, where we recognize that $\mathcal{H}_{2^{n-1}} \cong \Cl^0(n)$. Furthermore, the projection of the lift is achieved through the convexification of the map $Q : \Cl(n) \to \R^{n \times n}$, where the fact that $Q$ is quadratic in $\Cl(n)$ translates to being linear in $\mathcal{D}(\Cl(n))$. Specifically, by a slight abuse of notation we shall extend the definition of $Q$ to act on density operators as
\begin{equation}
    Q(\rho) = \sum_{\mu} p_\mu Q(x_\mu), \quad \text{where } \rho = \sum_\mu p_\mu \op{x_\mu}{x_\mu}.
\end{equation}
Then Eq.~\eqref{eq:conv-son_psd-lift} is the statement that $Q(\mathcal{D}(\Cl^0(n))) = \conv\SO(n)$.

In Appendix~\ref{sec:psd_lift_proof} we show that this statement straightforwardly generalizes for $Q(\mathcal{D}(\Cl(n))) = \conv\Orth(n)$. We prove this using the fermionic representation developed in Section~\ref{sec:linear_free_fermion}, and furthermore use these techniques to provide an alternative derivation for the PSD lift of $\conv\SO(n)$. The core of our argument is showing that the singular-value conditions of Eqs.~\eqref{eq:convOn_singval} and \eqref{eq:convSOn_singval} translate into bounds on the largest eigenvalue of corresponding $n$-qubit observables:
\begin{align}
    \sigma_i(X) &= \tr(\i \widetilde{\gamma}_i \gamma_i \rho) \leq 1,\\
    \sum_{i \in [n] \setminus I} \widetilde{\sigma}_i(X) - \sum_{i \in I} \widetilde{\sigma}_i(X) &= \tr\l[ \rho_0 \l( \sum_{i \in [n] \setminus I} \i \widetilde{\gamma}_i \gamma_i - \sum_{i \in I} \i \widetilde{\gamma}_i \gamma_i \r) \r] \leq n - 2,
\end{align}
where $\rho \in \mathcal{D}(\Cl(n))$ and $\rho_0 \in \mathcal{D}(\Cl^0(n))$. The physical interpretation here is that not all pure quantum states map onto to orthogonal or rotation matrices (which is clear from the fact that fermionic Gaussian states are only a subset of quantum states). However, all density operators \emph{do} map onto to their convex hulls, and the distinction between $\conv\Orth(n)$ and $\conv\SO(n)$ can be automatically specified by restricting the support of $\rho$ to the even-parity subspace.


\section{Quantum relaxation for the quadratic problem}\label{sec:quantum_relax_section}
We now arrive at the primary problem of interest in this work, the little noncommutative Grothendieck problem over the (special) orthogonal group. While the linear problem of Eq.~\eqref{eq:linear_obj} can be solved classically in polynomial time, quadratic programs are considerably more difficult. Here, we use the quantum formalism of the Pin and Spin groups developed above to construct a quantum relaxation of this problem. Then in Section~\ref{sec:rounding} we describe rounding procedures to recover a collection of orthogonal matrices from the quantum solution to this relaxation.

Recall the description of the input to Problem~\eqref{eq:LNCG}. Let $(V, E)$ be a graph, and associate to each edge $(u, v) \in E$ a matrix $C_{uv} \in \R^{n \times n}$. We label the vertices as $V = [m]$. We wish to maximize the objective
\begin{equation}
    f(R_1, \ldots, R_m) = \sum_{(u, v) \in E} \langle C_{uv}, R_u R_v^\T \rangle.
\end{equation}
over $(R_1, \ldots, R_m) \in \Orth(n)^m$. First, expand this expression in terms of matrix elements:
\begin{equation}
    \sum_{(u, v) \in E} \langle C_{uv}, R_u R_v^\T \rangle = \sum_{(u, v) \in E} \sum_{i,j \in [n]} [C_{uv}]_{ij} \sum_{k \in [n]} [R_u]_{ik} [R_v^\T]_{kj}.
\end{equation}
From the quadratic mapping $Q : \Cl(n) \to \R^{n \times n}$, we know that for each $R \in G$ there exists some $\phi \in \Pin(n)$ such that $R_{ij} = \langle \phi | P_{ij} | \phi \rangle$. Hence we can express the matrix product as
\begin{equation}
    \begin{split}
    [R_u]_{ik} [R_v^\T]_{kj} &= \langle \phi_u | P_{ik} | \phi_u \rangle \langle \phi_v | P_{jk} | \phi_v \rangle\\
    &= \langle \phi_u \otimes \phi_v | P_{ik} \otimes P_{jk} | \phi_u \otimes \phi_v \rangle,
    \end{split}
\end{equation}
which is now the expectation value of a $2n$-qubit Pauli operator with respect to a product state of two Gaussian states $\ket{\phi_u}$, $\ket{\phi_v}$. To extend this over the entire graph, we define a Hilbert space of $m$ registers of $n$ qubits each. For each edge $(u, v) \in E$ we introduce the Hamiltonian terms
\begin{equation}
    H_{uv} \coloneqq \sum_{i,j \in [n]} [C_{uv}]_{ij} \Gamma_{ij}^{(u, v)},
\end{equation}
where
\begin{equation}
    \Gamma_{ij}^{(u, v)} \coloneqq \l(\sum_{k \in [n]} P_{ik}^{(u)} \otimes P_{jk}^{(v)}\r) \bigotimes_{w \in V \setminus \{u, v\}} \I_{2^n}^{(w)}.
\end{equation}
To simplify notation, we shall omit the trivial support $\bigotimes_{w \in V \setminus \{u, v\}} \I_{2^n}^{(w)}$ when the context is clear.

The problem is now reformulated as optimizing the $mn$-qubit Hamiltonian
\begin{equation}\label{eq:H_LNCG}
    H \coloneqq \sum_{(u, v) \in E} H_{uv} = \sum_{(u, v) \in E} \sum_{i,j \in [n]} [C_{uv}]_{ij} \sum_{k \in [n]} P_{ik}^{(u)} \otimes P_{jk}^{(v)}.
\end{equation}
The exact LNCG problem over $\Orth(n)$ then corresponds to
\begin{equation}\label{eq:quantum_exact}
    \max_{\mathbf{R} \in \Orth(n)^m} f(\mathbf{R}) = \max_{\ket{\psi} \in \mathcal{H}_{2^n}^{\otimes m}} \ev{\psi}{H}{\psi} \quad \text{subject to } \begin{cases}
    \ip{\psi}{\psi} = 1,\\
    \ket{\psi} = \bigotimes_{v \in [m]} \ket{\phi_v},\\
    \ket{\phi_v} = {\mathcal{U}}_{(R_v, \I_n)} \ket{0^n}, & R_v \in \Orth(n) \ \forall v \in [m].
    \end{cases}
\end{equation}
The hardness of this problem is therefore related to finding the optimal separable state for local Hamiltonians, which is $\mathsf{NP}$-hard in general~\cite{gurvits2004classical,ioannou2006computational,gharibian2008strong}. Dropping these constraints on the state provides a relaxation of the problem, since
\begin{equation}
    \max_{\substack{\ket{\psi} \in \mathcal{H}_{2^n}^{\otimes m}, \\ \ip{\psi}{\psi} = 1}} \ev{\psi}{H}{\psi} \geq \max_{\mathbf{R} \in G^m} f(\mathbf{R}).
\end{equation}



We point out here that the Hamiltonian terms $H_{uv}$ can be interpreted as two-body fermionic interactions. Note that there is an important distinction between two-body fermionic operators (Clifford-algebra products of four Majorana operators) and two-body qudit operators (tensor products of two qudit Pauli operators). Recall that $P_{ij} = \i \widetilde{\gamma}_i \gamma_j$ is one-body in the fermionic sense. While the operators $P_{ik}^{(u)} \otimes P_{jk}^{(v)}$ appear to mix both notions, here they in fact coincide. To see this, we consider a global algebra of Majorana operators $\{\gamma_{i + (v-1)n}, \widetilde{\gamma}_{i + (v-1)n} \mid i \in [n], v \in [m] \}$ acting on a Hilbert space of $mn$ fermionic modes. While it is not true that the local single-mode Majorana operators map onto the global single-mode operators, i.e.,
\begin{align}
    \gamma_i^{(v)} \bigotimes_{w \in V \setminus \{v\}} \I_{2^n}^{(w)} \neq \gamma_{i + (v-1)n},
\end{align}
the local two-mode Majorana operators in fact do correspond to global two-mode operators:
\begin{equation}
    \widetilde{\gamma}_i^{(v)} \gamma_j^{(v)} \bigotimes_{w \in V \setminus \{v\}} \I_{2^n}^{(w)} = \widetilde{\gamma}_{i + (v-1)n} \gamma_{j + (v-1)n}.
\end{equation}
Thus, taking the tensor product of two local two-mode Majorana operators on different vertices is equivalent to taking the product of two global two-mode Majorana operators:
\begin{equation}
    \widetilde{\gamma}_i^{(u)} \gamma_j^{(u)} \otimes \widetilde{\gamma}_k^{(v)} \gamma_l^{(v)} \bigotimes_{w \in V \setminus \{u, v\}} \I_{2^n}^{(w)} = \widetilde{\gamma}_{i + (u-1)n} \gamma_{j + (u-1)n} \widetilde{\gamma}_{k + (v-1)n} \gamma_{l + (v-1)n}.
\end{equation}
Therefore Eq.~\eqref{eq:H_LNCG} can be equivalently expressed as a Hamiltonian with two-body fermionic interactions.

Finally, when we wish to optimize over $(R_1, \ldots, R_m) \in \SO(n)^m$, it is straightforward to see that we can simply replace the terms $P_{ij}$ with $\widetilde{P}_{ij}$. Defining
\begin{align}
    \widetilde{\Gamma}_{ij}^{(u, v)} &\coloneqq \l(\sum_{k \in [n]} \widetilde{P}_{ik}^{(u)} \otimes \widetilde{P}_{jk}^{(v)}\r) \bigotimes_{w \in V \setminus \{u, v\}} \I_{2^{n-1}}^{(w)},\\
    \widetilde{H}_{uv} &\coloneqq \sum_{i,j \in [n]} [C_{uv}]_{ij} \widetilde{\Gamma}_{ij}^{(u, v)},
\end{align}
the quantum relaxation for the $\SO(n)$ problem is given by the $m(n-1)$-qubit Hamiltonian
\begin{equation}
    \widetilde{H} \coloneqq \sum_{(u, v) \in E} \widetilde{H}_{uv}.
\end{equation}

\section{\label{sec:rounding}Rounding algorithms}
Optimizing the energy of a local Hamiltonian is a well-studied problem, both from the perspective of quantum and classical algorithms. In this section we will assume that such an algorithm has been used to produce the state $\rho \in \mathcal{D}(\mathcal{H}_d^{\otimes m})$ which (approximately) maximizes the energy $\tr(H \rho)$. We wish to round this state into the feasible space, namely the set of product states of Gaussian states. We do so by rounding the expectation values of $\rho$ appropriately, such that we return some valid approximation $R_1, \ldots, R_m \in G$. In this section we propose two approaches to perform this quantum rounding.

The first uses insight from the fact that our quantum relaxation is equivalent to a classical semidefinite relaxation with additional constraints based on the convex hull of the orthogonal group. This is approach is particularly advantageous when optimizing over $\SO(n)$, as $\conv\SO(n)$ has a matrix representation exponential in $n$ (its PSD lift). To build the semidefinite variable from the quantum state, we require measurements of the expectation values of the two-vertex operators $\Gamma_{ij}^{(u, v)} = \sum_{k \in [n]} P_{ik}^{(u)} \otimes P_{jk}^{(v)}$ for each pair of vertices $(u, v)$. We refer this procedure as \emph{$\conv\SO(n)$-based rounding}.\footnote{This rounding can also be applied to the optimization problem over $\Orth(n)$ as well, but we are particularly interested in the $\conv\SO(n)$ constraints due to their exponentially large classical representation.} Our second rounding protocol uses the expectation values of $P_{ij}^{(v)}$ of each vertex $v$ directly. In this case, rather than expectation values of two-vertex operators as before, we only require the information of single-vertex marginals $\rho_v \coloneqq \tr_{\neg v}(\rho)$. Therefore we call this approach \emph{vertex-marginal rounding}.

If $\rho$ is produced by a deterministic classical algorithm, then the relevant expectation values can be exactly computed (to machine precision). However if the state is produced by a randomized algorithm, or is otherwise prepared by a quantum computer, then we can only estimate the expectation values to within statistical error by some form of sampling. In the quantum setting, this can be achieved either by partial state tomography~\cite{bonet2020nearly,zhao2021fermionic} or a more sophisticated measurement protocol~\cite{huggins2021nearly}.\footnote{For the present discussion we do not consider the effects of finite sampling, although we expect that rounding is fairly robust to such errors since it will always return a solution in the feasible space.} See Appendix~\ref{sec:tomography} for further comments on this quantum measurement aspect. The rounding algorithms then operate entirely as classical postprocessing after estimating the necessary expectation values.

\subsection{Approximation ratio for rounding the classical SDP}

Before describing our quantum rounding protocols, we first review classical relaxations and rounding procedures for Problem~\eqref{eq:LNCG}. The standard semidefinite relaxation can be expressed as the SDP
\begin{equation}\label{eq:LNCG_relaxed}
    \max_{M \in \R^{mn \times mn}} \langle C, M \rangle \quad \text{subject to } \begin{cases}
    M \succeq 0,\\
    M_{vv} = \I_n & \forall v \in [m],
    \end{cases}
\end{equation}
where $C \in \R^{mn \times mn}$ is the matrix with $n \times n$ blocks $C_{uv}$. If an additional nonconvex constraint $\mathrm{rank}(M) = n$ is imposed, then the solution would be exact:
\begin{equation}
    M = \mathbf{R} \mathbf{R}^\T = \begin{bmatrix}
    \I_n & R_1 R_2^\T & \cdots & R_1 R_m^\T\\
    R_2 R_1^\T & \I_n & \cdots & R_2 R_m^\T\\
    \vdots & \vdots & \ddots & \vdots\\
    R_m R_1^\T & R_m R_2^\T & \cdots & \I_n
    \end{bmatrix}.
\end{equation}
Problem~\eqref{eq:LNCG_relaxed} is therefore a relaxation of the original problem. However, the solution $M \in \R^{mn \times mn}$ is still PSD, so it can be decomposed as $M = \mathbf{X} \mathbf{X}^\T$, where
\begin{equation}
    \mathbf{X} = \begin{bmatrix}
    X_1\\
    \vdots\\
    X_m
    \end{bmatrix}, \quad X_v \in \R^{n \times mn}.
\end{equation}
The rounding algorithm of Bandeira \emph{et al.}~\cite{bandeira2016approximating} then computes, for each $v \in [m]$,
\begin{equation}
    O_v = \mathcal{P}(X_v Z) \coloneqq \argmin_{Y \in \Orth(n)} \| Y - X_v Z \|_F,
\end{equation}
where $Z$ is an $mn \times n$ Gaussian random matrix whose entries are drawn i.i.d.~from $\mathcal{N}(0, 1/n)$. When optimizing over $G = \Orth(n)$, this rounded solution guarantees (in expectation) an approximation ratio of
\begin{equation}
    \alpha_{\Orth(n)}^2 = \E\l[ \frac{1}{n} \sum_{i \in [n]} \sigma_i(Z_1) \r{]^2},
\end{equation}
where $Z_1$ is a random $n \times n$ matrix with i.i.d.~entries from $\mathcal{N}(0, 1 / n)$.

In Appendix~\ref{sec:son_approx_ratio} we extend the argument used to obtain this result for the optimization problem over $G = \SO(n)$, and we show a corresponding approximation ratio of
\begin{equation}
    \alpha_{\SO(n)}^2 = \E\l[ \frac{1}{n} \sum_{i \in [n-1]} \sigma_i(Z_1) \r{]^2}
\end{equation}
where the only change to the rounding algorithm is that we project to the nearest $\SO(n)$ element via $\widetilde{\mathcal{P}}$, which is defined as
\begin{equation}
    \widetilde{\mathcal{P}}(X) \coloneqq \argmin_{Y \in \SO(n)} \|Y - X \|_F.
\end{equation}
Note that singular values are nonnegative, and in particular we show that $\E[\sigma_n(Z_1)] > 0$ for all finite $n$. Hence it follows that $\alpha_{\SO(n)}^2 < \alpha_{\Orth(n)}^2$, which provides evidence for the claim that solving for rotations is generally a more difficult problem (see Ref.~\cite[Section~4.3]{pumir2021generalized} for a brief discussion). For small values of $n$, the numerical values of these approximation ratios are (computed using Mathematica):
\begin{align}
    \alpha_{\Orth(2)}^2 \approx 0.6564, &\qquad \alpha_{\SO(2)}^2 \approx 0.3927,\\
    \alpha_{\Orth(3)}^2 \approx 0.6704, &\qquad \alpha_{\SO(3)}^2 \approx 0.5476,\\
    \alpha_{\Orth(4)}^2 \approx 0.6795, &\qquad \alpha_{\SO(4)}^2 \approx 0.6096.
\end{align}
In Appendix~\ref{sec:son_approx_ratio} we provide an integral expression for $\alpha_{\SO(n)}$ which can be evaluated for arbitrary $n$.

For the problem over $\SO(n)$, Saunderson \emph{et al.}~\cite{saunderson2014semidefinite} propose augmenting this SDP by adding the constraints that each block of $M$ lies in $\conv\SO(n)$:
\begin{equation}\label{eq:LNCG_SD_conv}
    \max_{M \in \R^{mn \times mn}} \langle C, M \rangle \quad \text{subject to } \begin{cases}
    M \succeq 0,\\
    M_{vv} = \I_n & \forall v \in [m],\\
    M_{uv} \in \conv\SO(n) & \forall u, v \in [m].
    \end{cases}
\end{equation}
Although they do not prove approximation guarantees for this enhanced SDP, they first show that, if one reintroduces the rank constraint on $M$, then the convex constraint $M_{uv} \in \conv\SO(n)$ in fact suffices to guarantee the much stronger condition $M_{uv} \in \SO(n)$. Then, when dropping the rank constraint (but leaving the $\conv\SO(n)$ constraint) they show that the relaxed problem is still exact over certain types of graphs, such as tree graphs. Finally, they provide numerical evidence that even when the relaxation is not exact, it returns substantially more accurate approximations than the standard SDP~\eqref{eq:LNCG_relaxed}.

\subsection{\label{sec:quantum_gram}Quantum Gram matrix}

Analogous to the classical SDP solution $M$, we can form a matrix $\mathcal{M} \in \R^{mn \times mn}$ from the expectation values of $\rho$ as
\begin{equation}
    \mathcal{M} \coloneqq \begin{bmatrix}
    \I_n & T_{12} & \cdots & T_{1m}\\
    T_{21} & \I_n & \cdots & T_{2m}\\
    \vdots & \vdots & \ddots & \vdots\\
    T_{m1} & T_{m2} & \cdots & \I_n
    \end{bmatrix},
\end{equation}
where
\begin{equation}\label{eq:T_uv}
    T_{uv} \coloneqq \frac{1}{n} \begin{bmatrix}
    \tr(\Gamma_{11}^{(u, v)} \rho) & \cdots & \tr(\Gamma_{1n}^{(u, v)} \rho)\\
    \vdots & \ddots & \vdots\\
    \tr(\Gamma_{n1}^{(u, v)} \rho) & \cdots & \tr(\Gamma_{nn}^{(u, v)} \rho)
    \end{bmatrix}
\end{equation}
and $T_{vu} = T_{uv}^\T$ for $u < v$. Just as $\langle C, M \rangle$ gives the relaxed objective value (up to rescaling and constant shifts), here we have that $\langle C, \mathcal{M} \rangle = \frac{2}{n} \tr(H \rho) + \tr(C)$. In Appendix~\ref{sec:quantum_conv_son} we show that for any quantum state, $\mathcal{M}$ satisfies the following properties:
\begin{equation}
\begin{cases}\label{eq:quantum_M_conditions}
    \mathcal{M} \succeq 0,\\
    \mathcal{M}_{vv} = \I_n & \forall v \in [m],\\
    \mathcal{M}_{uv} \in \conv\Orth(n) & \forall u, v \in [m].
\end{cases}
\end{equation}
Furthermore, we show that when $\rho$ is supported only on the even subspace of each single-vertex Hilbert space (or equivalently, if we replace $\Gamma_{ij}^{(u, v)}$ with $\widetilde{\Gamma}_{ij}^{(u, v)}$ in Eq.~\eqref{eq:T_uv}), then
\begin{equation}\label{eq:conv_son_T}
    \frac{1}{n} \begin{bmatrix}
    \tr(\widetilde{\Gamma}_{11}^{(u, v)} \rho) & \cdots & \tr(\widetilde{\Gamma}_{1n}^{(u, v)} \rho)\\
    \vdots & \ddots & \vdots\\
    \tr(\widetilde{\Gamma}_{n1}^{(u, v)} \rho) & \cdots & \tr(\widetilde{\Gamma}_{nn}^{(u, v)} \rho)
    \end{bmatrix} \in \conv\SO(n) \quad \forall \rho \in \mathcal{D}(\mathcal{H}_{2^{n-1}}^{\otimes m}).
\end{equation}
Therefore when optimizing the relaxed Hamiltonian $\widetilde{H}$ for the $\SO(n)$ setting, we are guaranteed to automatically satisfy the $\conv\SO(n)$ constraints.

\subsection{\label{sec:conv_rounding}$\conv\SO(n)$-based rounding}

Given the construction of the $\mathcal{M}$ from quantum expectation values, we proceed to round the Gram matrix as in the classical SDP with $\conv\SO(n)$ constraints~\cite{saunderson2014semidefinite}. This consists of computing the matrices
\begin{equation}
    R_v = \widetilde{\mathcal{P}}(\mathcal{M}_{1v}),
\end{equation}
where the projection to $\SO(n)$ can be efficiently computed from the special singular value decomposition, i.e.,
\begin{equation}
    \widetilde{\mathcal{P}}(X) = U \widetilde{V}^\T
\end{equation}
(recall Eq.~\eqref{eq:summary_J}). Our choice of rounding using the first $n \times mn$ ``row'' of $\mathcal{M}$ amounts to fixing $R_1 = \I_n$. We note that the same rounding procedure can naturally be applied to the $\Orth(n)$ setting as well, replacing $\widetilde{\mathcal{P}}$ with $\mathcal{P}$.

\subsection{\label{sec:vertex_rounding}Vertex-marginal rounding}

The single-vertex marginals are obtained by tracing out the qudits associated to all but one vertex $v \in [m]$,
\begin{equation}
    \rho_v = \tr_{\neg v}(\rho).
\end{equation}
As $\rho_v \in \mathcal{D}(\mathcal{H}_d^{\otimes m})$, from Section~\ref{sec:mixed_states} we have that $Q(\rho_v) \in \conv G$, where we linearly extend the definition of $Q$ to
\begin{equation}
    Q(\rho_v) \coloneqq \begin{bmatrix}
    \tr(P_{11}^{(v)} \rho) & \cdots & \tr(P_{1n}^{(v)} \rho)\\
    \vdots & \ddots & \vdots\\
    \tr(P_{n1}^{(v)} \rho) & \cdots & \tr(P_{nn}^{(v)} \rho)
    \end{bmatrix}.
\end{equation}
The rounding scheme we propose here then projects $Q(\rho_v)$ to $G$ using either $\mathcal{P}$ or $\widetilde{\mathcal{P}}$:
\begin{equation}
    R_v = \argmin_{Y \in G} \|Y - Q(\rho_v)\|_F.
\end{equation}

We point out that the relaxed Hamiltonian only has two-vertex terms which we seek to maximize. In Appendix~\ref{sec:symmetry_vertex} we show that $H$ commutes with $\mathcal{U}_{(\I_n, V)}^{\otimes m}$ for all $V \in \Orth(n)$, which we further show implies that $H$ may possess eigenstates whose single-vertex marginals obey $Q(\sigma_v) = 0$. This indicates that there may exist eigenstates of $H$ whose single-vertex marginals yield no information, despite the fact that their two-vertex marginals are nontrivial. In our numerical studies, we observe that breaking this symmetry resolves this issue. We accomplish this by including small perturbative one-body terms which correspond to the trace of $Q(\sigma_v)$:
\begin{equation}
    H_1 = \sum_{v \in [m]} \sum_{i \in [n]} P_{ii}^{(v)},
\end{equation}
Note that this trace quantity is importantly invariant with respect to the choice of basis for $\R^n$. We then augment the objective Hamiltonian with $H_1$, defining
\begin{equation}
    H'(\zeta) \coloneqq H + \zeta H_1
\end{equation}
where $\zeta > 0$ is a small regularizing parameter. While this one-body perturbation does not correspond to any terms in the original quadratic objective function, any arbitrarily small $\zeta > 0$ suffices to break the $\Orth(n)$ symmetry. Furthermore, the rounding procedure always guarantees that the solution is projected back into the feasible space $G^m$. When $G = \SO(n)$ we define $\widetilde{H}'(\zeta)$ analogously.

\section{Numerical experiments}\label{sec:numerical_experiments}
To explore the potential of our quantum relaxation and rounding procedures, we performed numerical experiments on randomly generated instances of the group synchronization problem. Because the Hilbert-space dimension grows exponentially in both $m$ and $n$, our classical simulations here are limited to small problem sizes. However, optimizing over rotations in $\R^3$ (requiring only two qubits per vertex) is highly relevant to many practical applications, so here we focus on the problem of $\SO(3)$ group synchronization. For example, this problem appears in the context of cryo-EM as described in Section~\ref{sec:group_sync}. To model the problem, we generated random instances by selecting random three-regular graphs $([m], E)$, uniformly randomly sampling $m$ rotations $g_1, \ldots, g_m \in \SO(n)$, and then constructing $C_{uv} = g_u g_v^\T + \sigma W_{uv}$ for each $(u, v) \in E$, where the Gaussian noise matrix $W_{uv} \in \R^{n \times n}$ has i.i.d.~elements drawn from $\mathcal{N}(0, 1)$ and $\sigma \geq 0$ represents the strength (standard deviation) of this noise.

While the classical $\conv\SO(n)$-based SDP is not guaranteed to find the optimal solution, the problems studied here were selected for such that this enhanced SDP in fact does solve the exact problem. We verify this property by confirming that $\mathrm{rank}(M) = n$ before rounding on each problem instance. In this way we are able to calculate an approximation ratio for the other methods (as it is not clear how to solve for the globally optimal solution in general, even with an exponential-time classical algorithm). The methods compared here include our quantum relaxation with $\conv\SO(n)$-based rounding (denoted CR), vertex-marginal rounding (VR), and the classical SDP (without $\conv\SO(n)$ constraints but using the $\widetilde{\mathcal{P}}$ projection to guarantee that the rounded solutions are elements of $\SO(n)$). When using the vertex-rounding method, we employ $\widetilde{H}'(\zeta)$ as the objective Hamiltonian with $\zeta = 10^{-6}$.

\subsection{Exact eigenvectors}

\begin{figure}
    \includegraphics[width=0.495\textwidth]{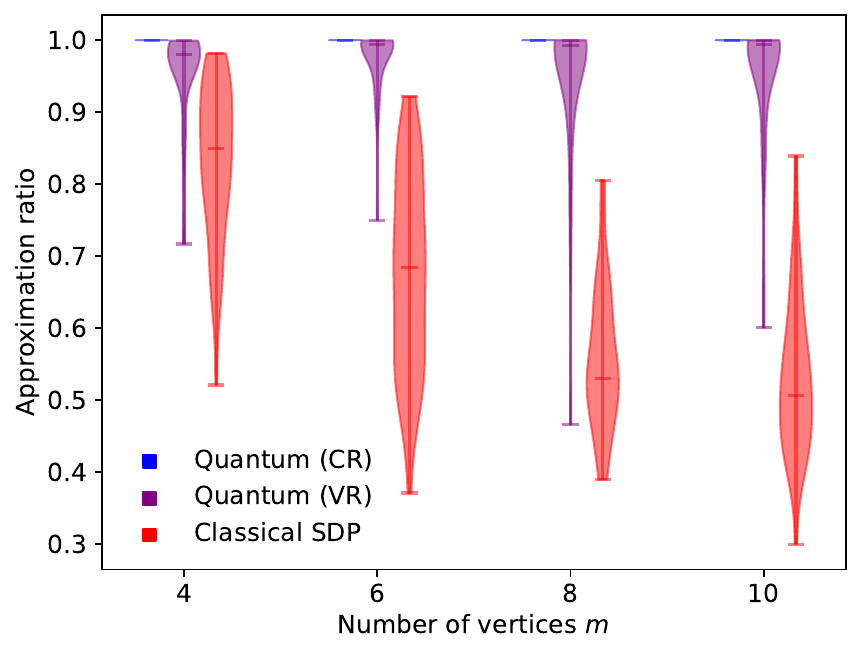}
    \includegraphics[width=0.495\textwidth]{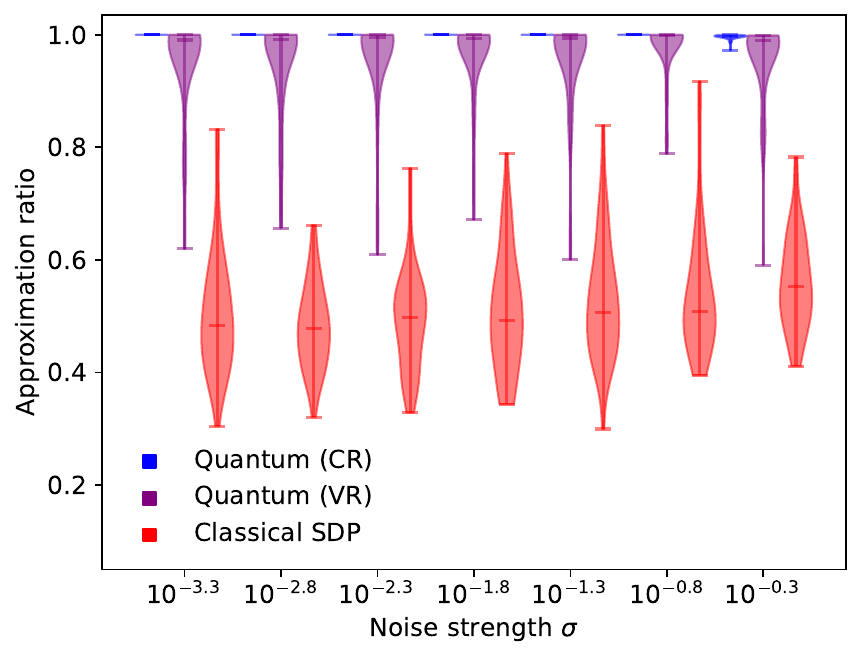}
    \caption[Approximation ratios for solutions obtained from rounding the maximum eigenvector of the relaxed Hamiltonian.]{Approximation ratios for solutions obtained from rounding the maximum eigenvector of the relaxed Hamiltonian $\widetilde{H}$. Violin plots show the distribution of approximation ratios over 50 randomly generated instance, and with the median being indicated by the center marker. CR refers to rounding according to the $\conv\SO(n)$-based scheme (Section~\ref{sec:conv_rounding}), while VR denotes the vertex-marginal rounding scheme (Section~\ref{sec:vertex_rounding}). The classical SDP solution was rounded by the standard randomized algorithm~\cite{bandeira2016approximating}, and we report the best solution over 1000 rounding trials. (Left) Varying the number of vertices $m$ in the graph (random 3-regular graphs). Note that the number of qubits required here is $2m$. (Right) Varying the noise strength parameter $\sigma$ which defines the problem via $C_{uv} = g_u g_v^\T + \sigma W_{uv}$.}
    \label{fig:eigen_group-sync}
\end{figure}

First, we consider the solution obtained by rounding the maximum eigenvector of $\widetilde{H}$. Although the hardness of preparing such a state is equivalent that of the ground-state problem, this nonetheless provides us with a benchmark for the ultimate approximation quality of our quantum relaxation. In Figure~\ref{fig:eigen_group-sync} we plot the approximation ratio of the rounded quantum states and compare to that of the classical SDP on the same problem instances. Each violin plot was constructed from the results of 50 random instances.

The results here demonstrate that, while the approximation quality of the classical SDP quickly falls off with larger graph sizes, our rounded quantum solutions maintain high approximation ratios, at least for the problem sizes probed here. Notably, the $\conv\SO(n)$-based rounding on the quantum state is significantly more powerful and consistent than the vertex-marginal rounding. This feature is not unexpected since, as discussed in Section~\ref{sec:vertex_rounding}, we are maximizing an objective Hamiltonian with only two-body terms, whereas the single-vertex rounding uses strictly one-body expectation values. Furthermore, as demonstrated in previous works~\cite{saunderson2014semidefinite, matni2014convex} the $\conv\SO(n)$ constraints are powerful in practice, and so we expect that the quantum rounding protocol which makes use of this structure enjoys the same advantages.

Meanwhile, when varying the noise parameter $\sigma$, we observe that all methods are fairly consistent. In particular, the $\conv\SO(n)$-based rounding only shows an appreciable decrease in approximation quality when the noise is considerable (note that $\sigma = 0.5 \approx 10^{-0.3}$ is a relatively large amount of noise, since $g_u g_v^\T$ is an orthogonal matrix and therefore has matrix elements bounded in magnitude by 1).

\subsection{\label{sec:asp_numerics}Quasi-adiabatic state preparation}

Because it may be unrealistic to prepare the maximum eigenvector of $\widetilde{H}$, here we consider preparing states using ideas from adiabatic quantum computation~\cite{RevModPhys.90.015002}. Specifically, we wish to demonstrate that states whose relaxed energy may be far from the maximum eigenvalue can still provide high-quality approximations after rounding. If this is the case then we do not need to prepare very close approximations to the maximum eigenstate of $\widetilde{H}$, so the rigorous conditions of adiabatic state preparation may not be required in this context. Hence we consider ``quasi-adiabatic'' state preparation, wherein we explore how time-evolution speeds far from the adiabatic limit may still return high-quality approximations. Our numerical experiments here provide a preliminary investigation into this conjecture.

For simplicity of the demonstration, we consider a linear annealing schedule according to the time-dependent Hamiltonian
\begin{equation}
    H(t) = \l( 1 - \frac{t}{T} \r) H_i + \frac{t}{T} H_f,
\end{equation}
which prepares the state
\begin{equation}
    \ket{\psi(T)} = \mathcal{T}\exp\l( -\i \int_0^T \mathrm{d}t \, H(t) \r) \ket{\psi(0)}
\end{equation}
for some $T > 0$, where $\mathcal{T}$ is the time-ordering operator. The final Hamiltonian $H_f$ is the desired objective LNCG Hamiltonian,
\begin{equation}
    H_f = \widetilde{H}.
\end{equation}
The initial Hamiltonian $H_i$ is the parent Hamiltonian of the initial state, which we choose to be the approximation obtained from the classical SDP, as it can be obtained classically in polynomial time. Let $R_1, \ldots, R_m \in \SO(n)$ be the SDP solution. Our initial state is then the product of Gaussian states
\begin{equation}
    \ket{\psi(0)} = \bigotimes_{v \in [m]} \ket{\phi(R_v)},
\end{equation}
where each $\ket{\phi(R_v)}$ is the maximum eigenvector of the free-fermion Hamiltonian
\begin{equation}
    F(R_v) = \i \sum_{i, j \in [n]} [R_v]_{ij} \widetilde{\gamma}_i \gamma_j.
\end{equation}
Therefore the initial Hamiltonian $H_i$ is a sum of such free-fermion Hamiltonians (here we include the even-subspace projection since we are working with $\SO(n)$):
\begin{equation}
    H_i = \sum_{v \in [m]} \Pi_0 F(R_v) \Pi_0^\T.
\end{equation}
As a Gaussian state, $\ket{\phi(R_v)}$ can be prepared exactly from a quantum circuit of $\Ord(n^2)$ gates~\cite{jiang2018quantum}. Note that since we are working directly in the even subspace of $n - 1$ qubits here, this $n$-qubit circuit must be projected appropriately using $\Pi_0$. We discuss how to perform this circuit recompilation in Appendix~\ref{sec:even_subspace}. We comment that this choice of initial state is that of a mean-field state for non-number-preserving fermionic systems, for instance as obtained from Hartree--Fock--Bogoliubov theory. Suitably, the final Hamiltonian we evolve into is non-number-preserving two-body fermionic Hamiltonian.

\begin{figure}
    \centering
    \includegraphics[width=0.5\textwidth]{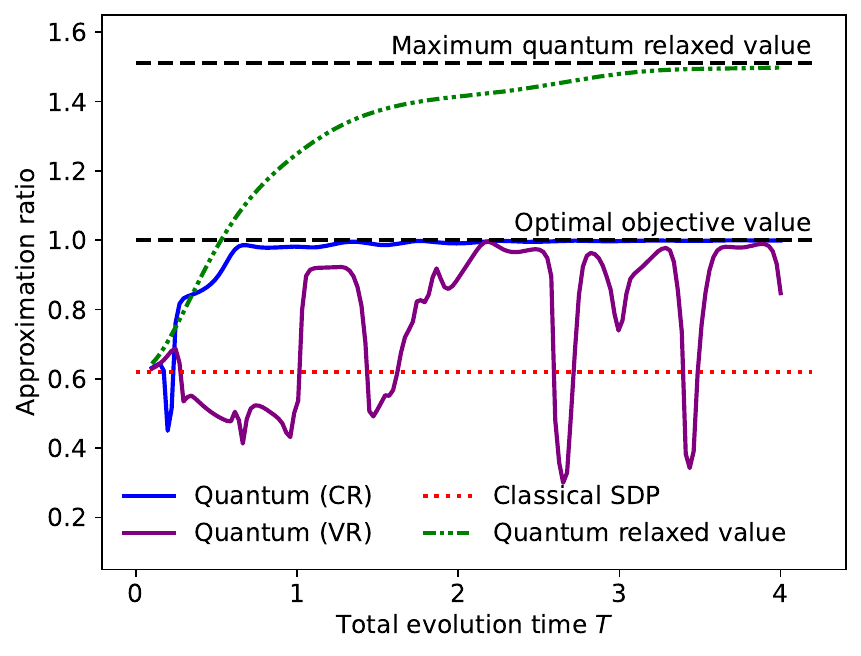}
    \caption[Demonstration of a typical instance of adiabatic state preparation for preparing relaxed quantum solutions.]{Demonstration of a typical instance of adiabatic state preparation for preparing relaxed quantum solutions. The initial state is the product of Gaussian states corresponding to the rounded solution of the classical SDP. As the total evolution time $T$ increases, the evolution becomes more adiabatic, indicated by the convergence of the relaxed value to the maximum eigenvalue (in units of the original problem's optimal value). The rounded solutions of course can never exceed the original problem's optimal value.}
    \label{fig:adiabatic_group-sync}
\end{figure}

In adiabatic state preparation, the total evolution time $T$ controls how close the final state $\ket{\psi(T)}$ is to the maximum eigenstate\footnote{We remind the reader that we are starting in the maximum eigenstate of the initial Hamiltonian, whereas in the physics literature, adiabatic theorems are typically stated in terms of ground states. Of course, the two perspectives are equivalent by simply an overall sign change (note that all Hamiltonians here are traceless).} of the final Hamiltonian $H_f$. One metric of closeness is how the energy of the prepared state, $\ev{\psi(T)}{H_f}{\psi(T)}$, compares to the maximum eigenvalue of $H_f$. On the other hand, as a relaxation, this maximum energy is already larger than the optimal objective value of the original problem. We showcase this in Figure~\ref{fig:adiabatic_group-sync}, using one random problem instance as a demonstrative (typical) example on a graph of $m = 6$ vertices (12 qubits). For each total evolution time point $T$, we computed $\ket{\psi(T)}$ by numerically integrating the time-dependent Schr\"{o}dinger equation, and we plot its relaxed energy as well as its rounded objective values. For large $T$ we approach the maximum eigenstate of $\widetilde{H}$ as expected (thereby also demonstrating that the initial ``mean-field'' state $\ket{\psi(0)}$ has appreciable overlap). Particularly interesting is the behavior for relatively small total evolution times $T$, wherein the energy of $\ket{\psi(T)}$ is far from the maximum eigenenergy. Despite this, the approximation quality after rounding the state using $\mathcal{M}$ is nearly exact around $T \approx 1$. On the other hand, the approximation quality of vertex-marginal rounding is highly inconsistent, which again we attribute to the fact that the single-vertex information is not directly seen by the final Hamiltonian $H_f$.

\begin{figure}
    \centering
    \includegraphics[width=0.495\textwidth]{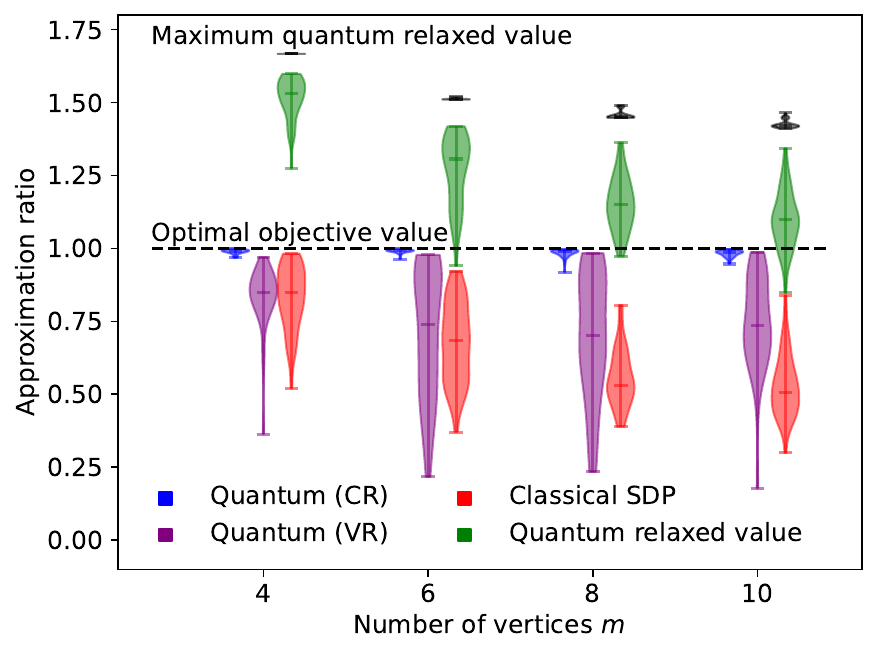}
    \includegraphics[width=0.495\textwidth]{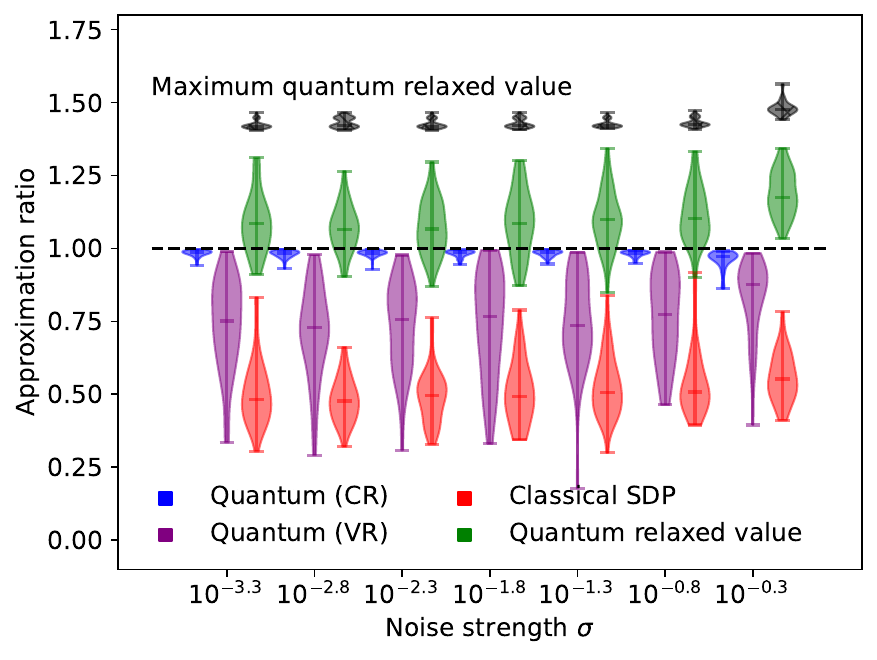}
    \caption[Approximation ratios for solutions obtained from rounding the ``adiabatically'' evolved state with fixed $T$.]{Approximation ratios for solutions obtained from rounding the ``adiabatically'' evolved state $\ket{\psi(T)}$ with fixed $T = 1$ for all $m, \sigma$. Problem instances and visualization is the same as in Figure~\ref{fig:eigen_group-sync}. We also include the maximum eigenvalue of the relaxed Hamiltonian and the energy of the prepared unrounded state, to demonstrate how far $\ket{\psi(T)}$ is from the exact maximum eigenvector. (Left) Varying the number of vertices $m$ in the graph. Note that the number of qubits required here is $2m$. (Right) Varying the noise strength parameter $\sigma$.}
    \label{fig:adiabatic_group-sync_violin}
\end{figure}

Then in Figure~\ref{fig:adiabatic_group-sync_violin} we plot the same 50 problem instances (per graph size/noise level) as in Figure~\ref{fig:eigen_group-sync}, but using the quasi-adibatically prepared state $\ket{\psi(T)}$ where we have fixed $T = 1$ for all graph sizes. The classical SDP results are the same as in Figure~\ref{fig:adiabatic_group-sync_violin}, and for reference we include the energy of the unrounded quantum state and the maximum eigenvalue of the relaxed Hamiltonian (normalized with respect to the optimal objective value). Qualitatively, we observe features similar to those seen in Figure~\ref{fig:adiabatic_group-sync}. Namely, although the annealing schedule is too fast to prepare a close approximation to the maximum eigenstate, the rounded solutions (using the $\conv\SO(n)$-based protocol) consistently have high approximation ratios. Meanwhile, the vertex-rounded solutions are highly inconsistent, which reflects the highly fluctuating behavior seen in Figure~\ref{fig:adiabatic_group-sync}.

\section{Discussion and future work}\label{sec:discussion_quant_relax_ortho}
In this chapter we have developed a quantum relaxation for a quadratic program over orthogonal and rotation matrices, known as an instance of the little noncommutative Grothendieck problem. The embedding of the classical objective is achieved by recognizing an intimate connection between the geometric-algebra construction of the orthogonal group and the structure of quantum mechanics, in particular the formalism of fermions in second quantization. From this perspective, the determinant condition of $\SO(n)$ is succinctly captured by a simple linear property of the state---its parity---and the convex bodies $\conv\Orth(n)$ and $\conv\SO(n)$ (relevant to convex relaxations of optimization over orthogonal matrices) are completely characterized by density operators on $n$ and $n - 1$ qubits, respectively. Recognizing that the reduced state on each vertex therefore corresponds to an element of this convex hull, we proposed vertex-marginal rounding which classically rounds the measured one-body reduced density matrix of each vertex.

We additionally showed that these convex hulls are characterized by density operators on $2n$ and $2(n-1)$ qubits as well, where the linear functionals defining this PSD lift are the Hamiltonian terms appearing in our quantum relaxation. This insight enables our second proposed rounding scheme, $\conv G$-based edge rounding, which is inspired by the fact that the a quantum Gram matrix $\mathcal{M}$ can be constructed from the expectation values of the quantum state which obeys the same properties as the classical SDP of Saunderson~\emph{et al.}~\cite{saunderson2014semidefinite}. Numerically we observe that this approach to quantum rounding is significantly more accurate and consistent than vertex rounding, and it consistently achieves larger approximation ratios than the basic SDP relaxation. However, we are severely limited by the exponential scaling of classically simulating quantum states;~further investigations would be valuable to ascertain the empirical performance of these ideas at larger scales.

The primary goal of this work was to formulate the problem of orthogonal-matrix optimization into a familiar quantum Hamiltonian problem, and to establish the notion of a quantum relaxation for such optimization problems over continuous-valued decision variables. A clear next step is to prove nontrivial approximation ratios from our quantum relaxation. If such approximation ratios exceed known guarantees by classical algorithms, for example on certain types of graphs, then this would potentially provide a quantum advantage for a class of applications not previously considered in the quantum literature. We have proposed one standard, realistically preparable class of states---quasi-adiabatic time evolution---but a variety of energy-optimizing ansatze exist in the literature, especially considering that the constructed Hamiltonian is an interacting-fermion model. From this perspective, it would also be interesting to see if a classical many-body method can produce states which round down to high-quality approximations, even heuristically. Such an approach would constitute a potential example of a quantum-inspired classical algorithm.

From a broader perspective, the quantum formalism described here may also provide new insights into the computational hardness of the classical problem. First, the $\mathsf{NP}$-hard thresholds for Problem~\eqref{eq:LNCG} are not currently known. However, by establishing the classical problem as an instance of Gaussian product state optimization on the many-body Hamiltonian, it may be possible to import tools from quantum computational complexity to study the classical problem. This idea also applies to the more general instances of noncommutative Grothendieck problems,
\begin{equation}\label{eq:NCG}
    \max_{U, V \in \Orth(N)} \sum_{i,j,k,l \in [N]} T_{ijkl} U_{ij} V_{kl},
\end{equation}
where the $N \times N \times N \times N$ tensor $T$ specifies the problem input. It is straightforward to apply our quantum relaxation construction to this problem, yielding a $2N$-qubit Hamiltonian whose terms are of the form $P_{ij} \otimes P_{kl}$. While Bri\"{e}t \emph{et al.}~\cite{briet2017tight} showed that the $\mathsf{NP}$-hardness threshold of approximating this problem is $1/2$, it remains an open problem to construct an algorithm which is guaranteed to achieve this approximation ratio.

Although we have provided new approximation ratios for the instance of Problem~\eqref{eq:LNCG} over $\SO(n)$, it is unclear precisely how much harder the $\SO(n)$ problem is compared to the $\Orth(n)$ problem. The work by Saunderson \emph{et al.}~\cite{saunderson2015semidefinite} establishes a clear distinction between the representation sizes required for $\conv\Orth(n)$ and $\conv\SO(n)$, and this work has connected this structure to properties of quantum states on $n$ qubits. However this does not yet establish a difference of hardness for the corresponding quadratic programs. Again it would be interesting to see if the tools of quantum information theory can be used to further understand this classical problem. For example, one might study the $\mathsf{NP}$-hardness threshold of Problem~\eqref{eq:NCG} where instead $U, V \in \SO(N)$ and leverage the quantum (or equivalently, Clifford-algebraic) representation of $\SO(N)$. In such a setting, the size of the problem is given by a single parameter $N$ and so the exponentially large parametrization of $\conv\SO(N)$ appears to signify a central difficulty of this problem.

We note that it is straightforward to extend our quantum relaxation to the unitary groups $\U(n)$ and $\SU(n)$, essentially by doubling the number of qubits per vertex via the inclusions $\U(n) \subset \Orth(2n)$ and $\SU(n) \subset \SO(2n)$. However this is likely an inefficient embedding, since the $n$-qubit Majorana operators already form a representation of $\Cl(2n)$. It may therefore be possible to encode complex-valued matrices via a complexification of $Q$, using the same amount of quantum space. It is interesting to note that Bri\"{e}t \emph{et al.}~\cite{briet2017tight} in fact utilize  a ``complex extension'' of Clifford algebras when considering Problem~\eqref{eq:NCG} over the unitary group, although the usage is different from ours.


\newpage
\section*{Appendix}

\section{\label{sec:clifford}Clifford algebras and the orthogonal group}
In this appendix we review the key components for constructing the orthogonal and special orthogonal groups from a Clifford algebra. Our presentation of this material broadly follows Refs.~\cite{atiyah1964clifford,saunderson2015semidefinite}.

The Clifford algebra $\Cl(n)$ of $\R^n$ is a $2^n$-dimensional real vector space, equipped with an inner product $\langle \cdot, \cdot \rangle : \Cl(n) \times \Cl(n) \to \R$ and a multiplication operation satisfying the anticommutation relation
\begin{equation}\label{eq:anticommutator_cl_n}
    e_i e_j + e_j e_i = -2 \delta_{ij} \openone,
\end{equation}
where $\{e_1, \ldots, e_n\}$ is an orthonormal basis of $\R^n$ and $\openone$ is the multiplicative identity of the algebra. The basis elements $e_i$ are called the generators of the Clifford algebra, in the sense that they generate all other basis vectors of $\Cl(n)$ as
\begin{equation}
    e_I \coloneqq e_{i_1} \cdots e_{i_k}, \quad I = \{i_1, \ldots, i_k\} \subseteq [n].
\end{equation}
By convention we order the indices $i_1 < \cdots < i_k$, and the empty set corresponds to the identity, $e_\varnothing = \openone$. Taking all subsets $I \subseteq [n]$ and extending the inner product definition from $\R^n$ to $\Cl(n)$, it follows that $\{e_I \mid I \subseteq [n]\}$ is an orthonormal basis with $2^n$ elements. Specifically, we can write any element $x \in \Cl(n)$ as
\begin{equation}
    x = \sum_{I \subseteq [n]} x_I e_I
\end{equation}
with each $x_I \in \R$, and the inner product on $\Cl(n)$ is\footnote{Equipping an inner product to the vector representation of $\Cl(n)$ elements is achieved using the fact that algebra elements square to a multiple of the identity.}
\begin{equation}
    \langle x, y \rangle = \sum_{I \subseteq [n]} x_I y_I,
\end{equation}
where $y = \sum_{I \subseteq [n]} y_I e_I$. Hence $\Cl(n)$ is isomorphic as a Hilbert space to $\R^{2^n}$.

Now we show how to realize the orthogonal group $\Orth(n)$ from this algebra. First observe the inclusion $\R^n = \spn\{e_i \mid i \in [n]\} \subset \Cl(n)$. We shall identify the sphere $S^{n-1} \subset \R^n$ as all $u \in \R^n$ satisfying $\langle u, u \rangle = 1$. We then define the Pin group as all possible products of $S^{n-1}$ elements:
\begin{equation}\label{eq:pin}
    \Pin(n) \coloneqq \{u_1 \cdots u_{k} \mid u_1, \ldots, u_{k} \in S^{n-1}, 0 \leq k \leq n\}.
\end{equation}
It is straightforward to check that this is indeed a group. Each $x \in \Pin(n)$ is also normalized, $\langle x, x \rangle = 1$. In fact, an equivalent definition of this group is all elements $x \in \Cl(n)$ satisfying $x \overline{x} = \openone$, where conjugation $\overline{x}$ is defined from the linear extension of
\begin{equation}
    \overline{e_I} \coloneqq (-1)^{|I|} e_{i_k} \cdots e_{i_1}.
\end{equation}

The Pin group is a double cover of $\Orth(n)$, which can be seen from defining a quadratic map $Q : \Cl(n) \to \R^{n \times n}$. This map arises from the so-called twisted adjoint action, introduced by Atiyah \emph{et al.}~\cite{atiyah1964clifford}:\footnote{Saunderson \emph{et al.}~\cite{saunderson2015semidefinite} consider the standard adjoint action, which is sufficient for describing rotations. However, the ``twist'' due to $\alpha$ is necessary to construct arbitrary orthogonal transformations.}
\begin{equation}
    v \mapsto \alpha(x) v \overline{x}, \quad x, v \in \Cl(n),
\end{equation}
where the linear map $\alpha : \Cl(n) \to \Cl(n)$ is the parity automorphism, defined by linearly extending
\begin{equation}
    \alpha(e_I) \coloneqq (-1)^{|I|} e_I.
\end{equation}
Then for any $x \in \Cl(n)$, the linear map $Q(x) : \R^n \to \R^n$ is defined as
\begin{equation}\label{eq:quad_map_clifford}
    Q(x)(v) \coloneqq \pi_{\R^n}(\alpha(x) v \overline{x}) \quad \forall v \in \R^n,
\end{equation}
where $\pi_{\R^n}$ is the projection from $\Cl(n)$ onto $\R^n$. To show that $Q(\Pin(n)) = \Orth(n)$, it suffices to recognize that, for any $u \in S^{n-1}$, $\alpha(u) v \overline{u} \in \R^n$ is the reflection of the vector $v \in \R^n$ across the hyperplane normal to $u$. To see this, first observe that $uv + vu = -2 \langle u, v \rangle \openone$, which follows from Eq.~\eqref{eq:anticommutator_cl_n} by linearity. Then
\begin{equation}
\begin{split}
    \alpha(u) v \overline{u} &= uvu\\
    &= (-vu - 2 \langle u, v \rangle \openone) u\\
    &= v - 2 \langle u, v \rangle u,
\end{split}
\end{equation}
which is precisely the elementary reflection as claimed. By the Cartan--Dieudonn{\'e} theorem, one can implement any orthogonal transformation on $\R^n$ by composing $k \leq n$ such reflections about arbitrary hyperplanes $u_1, \ldots, u_k$~\cite{gallier2011geometric}. This characterization coincides precisely with the definition of the Pin group provided in Eq.~\eqref{eq:pin}, through the composition of the linear maps $Q(u_1), \ldots, Q(u_k)$ on $\R^n$. Hence for all $x \in \Pin(n)$, $Q(x)$ is an orthogonal transformation on $\R^n$. The double cover property follows from the fact that $Q$ is quadratic in $x$, so $Q(x) = Q(-x)$. 

The special orthogonal group arises from the subgroup $\Spin(n) \subset \Pin(n)$ containing only even-parity Clifford elements. First observe that $\Cl(n)$ is a $\Z_2$-graded algebra:
\begin{equation}
    \Cl(n) = \Cl^{0}(n) \oplus \Cl^{1}(n),
\end{equation}
where
\begin{align}
    \Cl^0(n) &\coloneqq \spn\{e_I \mid |I| \text{ even}\},\\
    \Cl^1(n) &\coloneqq \spn\{e_I \mid |I| \text{ odd}\}.
\end{align}
By a $\Z_2$ grading we mean that for each $x \in \Cl^a(n)$ and $y \in \Cl^b(n)$, their product $xy$ lies in $\Cl^{a + b \mod 2}(n)$. We say that elements in $\Cl^0(n)$ (resp., $\Cl^1(n)$) have even (resp., odd) parity. In particular, this grading implies that $\Cl^{0}(n)$ is a subalgebra, hence its intersection with the Pin group is also a group, which defines
\begin{equation}
    \Spin(n) \coloneqq \Pin(n) \cap \Cl^{0}(n) = \{u_1 \cdots u_{2k} \mid u_1, \ldots, u_{2k} \in S^{n-1}, 0 \leq k \leq \lfloor n/2 \rfloor\}.
\end{equation}
Just as the Pin group double covers $\Orth(n)$, so does the Spin group double cover $\SO(n)$. This is again a consequence of the Cartan--Dieudonn{\'e} theorem, wherein all rotations on $\R^n$ can be decomposed into an even number of (at most $n$) arbitrary reflections.

\section{\label{sec:quantum_conv_son}Convex hull of orthogonal matrices and quantum states}
Recall the following characterizations of the convex hulls:
\begin{align}
\conv\Orth(n) &= \l\{ X \in \R^{n \times n} \mid \sigma_1(X) \leq 1 \r\}\\
\conv\SO(n) &= \l\{ X \in \R^{n \times n} \mathrel{\Bigg|} \sum_{i \in [n] \setminus I} \widetilde{\sigma}_i(X) - \sum_{i \in I} \widetilde{\sigma}_i(X) \leq n - 2 \quad \forall I \subseteq [n], |I| \text{ odd} \r\}, \label{eq:convSOn_app}
\end{align}
where $\{\sigma_i(X)\}_{i \in [n]}$ and $\{\widetilde{\sigma}_i(X\}_{i \in [n]}$ are the singular values and special singular values of $X$ in descending order, respectively. Note that $\sigma_i(X) = \widetilde{\sigma}_i(X)$ for all $i \leq n-1$ and $\sigma_n(X) = \sign(\det(X)) \sigma_n(X)$.

\subsection{\label{sec:psd_lift_proof}PSD lift of $\conv\Orth(n)$ and $\conv\SO(n)$}

In this section we show that $Q(\mathcal{D}(\Cl(n))) = \conv\Orth(n)$ and $Q(\mathcal{D}(\Cl^0(n))) = \conv\SO(n)$, using the quantum formalism described in Section~\ref{sec:quantum_formalism}.

First we show that for all $X \in \conv\Orth(n)$, there exists some $\rho \in \mathcal{D}(\mathcal{H}_{2^n})$ which generates $X$, essentially by the convex extension of $Q$. Every $X \in \conv\Orth(n)$ can be expressed as a convex combination ($\sum_\mu p_\mu = 1$, $p_\mu \geq 0$) of orthogonal matrices $R_\mu \in \Orth(n)$:
\begin{equation}
    X = \sum_{\mu} p_\mu R_\mu.
\end{equation}
For each $R_\mu$ there exists some $x_\mu \in \Pin(n)$ such that $[R_\mu]_{ij} = \ev{x_\mu}{P_{ij}}{x_\mu}$. Therefore the matrix elements of $X$ can be expressed as
\begin{equation}
    X_{ij} = \tr\l( P_{ij} \sum_{\mu} p_\mu \op{x_\mu}{x_\mu} \r) = \tr(P_{ij} \rho),
\end{equation}
where $\rho \coloneqq \sum_{\mu} p_\mu \op{x_\mu}{x_\mu} \in \mathcal{D}(\mathcal{H}_{2^n})$.

Next we show the reverse direction, that for all $\rho \in \mathcal{D}(\mathcal{H}_{2^n})$, the matrix $X \coloneqq [\tr(P_{ij} \rho)]_{i,j \in [n]}$ is an element of $\conv\Orth(n)$. Recall that $X \in \conv\Orth(n)$ if and only if $\sigma_1(X) \leq 1$. Therefore we take the singular value decomposition of $X = U \Sigma V^\T$ and, using $P_{ij} = \i \widetilde{\gamma}_i \gamma_j$, each singular value is equal to
\begin{equation}
\begin{split}
    \sigma_k(X) &= [U^\T X V]_{kk}\\
    &= \sum_{i,j \in [n]} U_{ik} \tr(\i \widetilde{\gamma}_i \gamma_j \rho) V_{jk}\\
    &= \tr(\i \mathcal{U}_{(U, \I_n)}^\dagger \widetilde{\gamma}_k \mathcal{U}_{(U, \I_n)} \mathcal{U}_{(\I_n, V)}^\dagger \gamma_k \mathcal{U}_{(\I_n, V)} \rho)\\
    &= \tr(\i \widetilde{\gamma}_k \gamma_k \rho'),
\end{split}
\end{equation}
where $\rho' \coloneqq \mathcal{U}_{(U, V)} \rho \mathcal{U}_{(U, V)}^\dagger$. Because $\i \widetilde{\gamma}_k \gamma_k$ has eigenvalues $\pm 1$, we see that $\sigma_k(X) \leq 1$ for all $k \in [n]$.

For the restriction to $\conv\SO(n)$, the first argument is essentially the same. One merely replaces $P_{ij}$ with $\widetilde{P}_{ij}$, hence $\sum_\mu p_\mu \op{x_\mu}{x_\mu} \in \mathcal{D}(\mathcal{H}_{2^{n-1}})$. For the reverse direction, we instead employ the special singular value decomposition which yields
\begin{equation}
    \widetilde{\sigma}_k(X) = \tr(\i \widetilde{\gamma}_k \gamma_k \rho'),
\end{equation}
where now $\rho' \coloneqq \mathcal{U}_{(U, \widetilde{V})} \rho \mathcal{U}_{(U, \widetilde{V})}^\dagger \in \mathcal{D}(\mathcal{H}_{2^{n-1}})$. Note that we have not projected to the even subspace this time, as it is more convenient to work in the full $n$-qubit space when handling the Gaussian unitaries. Instead, we will impose the constraint that $\rho$ only has support on the even-parity subspace, so $\tr(Z^{\otimes n} \rho) = 1$. Furthermore, because the special singular value decomposition guarantees that $\det(U)\det(\widetilde{V}) = 1$, $\mathcal{U}_{(U, \widetilde{V})}$ is parity preserving so that $\tr(Z^{\otimes n} \rho') = 1$ as well. Now recall that $X \in \conv\SO(n)$ if and only if
\begin{equation}\label{eq:conv-son_ineq_X}
    \sum_{k \in [n] \setminus I} \widetilde{\sigma}_k(X) - \sum_{k \in I} \widetilde{\sigma}_k(X) \leq n - 2
\end{equation}
for all subsets $I \subseteq [n]$ of odd size. By linearity,
\begin{equation}
    \sum_{k \in [n] \setminus I} \widetilde{\sigma}_k(X) - \sum_{k \in I} \widetilde{\sigma}_k(X) = \tr\l( \rho' \sum_{k \in [n]} (-1)^{z_k} Z_k \r),
\end{equation}
where $z = z_1 \cdots z_n \in \{0, 1\}^n$ is defined as $z_k = 1$ if $k \in I$ and $z_k = 0$ otherwise, and we have used the fact that $\i \widetilde{\gamma}_k \gamma_k = Z_k$. It therefore suffices to examine the spectrum of $A_z \coloneqq \sum_{k \in [n]} (-1)^{z_k} Z_k$:
\begin{equation}\label{eq:conv-son-eigval}
    A_z \ket{b} = \l( \sum_{k \in [n]} (-1)^{[z \oplus b]_k} \r) \ket{b}, \quad b \in \{0, 1\}^n,
\end{equation}
where $\oplus$ denotes addition modulo 2. As we are only interested in the subspace spanned by even-parity states, we restrict attention to the eigenvalues for which $|b| \mod 2 = 0$. Because $|I|$ is odd, so too is $|z|$, hence $|z \oplus b| \mod 2 = 1$. This implies that there must be at least one term in the sum of Eq.~\eqref{eq:conv-son-eigval} which is negative, so it can only take integer values at most $n - 2$. This establishes Eq.~\eqref{eq:conv-son_ineq_X}, hence $X \in \conv\SO(n)$.
\subsection{Relation to $\conv\SO(n)$-based semidefinite relaxation}

Here we provide details for our claim that the relaxed quantum solution obeys the same constraints as the classical SDP which uses the exponentially large representation of $\conv\SO(n)$. Recall that this relaxation can be formulated as
\begin{equation}\label{eq:conv_sdp_app}
    \max_{M \in \R^{mn \times mn}} \sum_{(u, v) \in E} \langle C_{uv}, M_{uv} \rangle \quad \text{subject to } \begin{cases}
    M \succeq 0,\\
    M_{vv} = \I_n & \forall v \in [m],\\
    M_{uv} \in \conv\SO(n) & \forall u, v \in [m].
    \end{cases}
\end{equation}
We will show that the Gram matrix $\mathcal{M}$ constructed from the measurements of a quantum state $\rho$, defined in Section~\ref{sec:quantum_gram} as
\begin{equation}\label{eq:quantum_gram_elements}
    [\mathcal{M}_{uv}]_{ij} = \begin{cases}
    \delta_{ij} & u = v,\\
    \frac{1}{n} \tr(\Gamma_{ij}^{(u, v)} \rho) & u < v,\\
    [M_{vu}]_{ji} & u > v,
    \end{cases}
\end{equation}
obeys the constraints of Eq.~\eqref{eq:conv_sdp_app}. Specifically, when the marginals of $\rho$ on each vertex are even-parity states (recall this is equivalent to replacing $\Gamma_{ij}$ with $\widetilde{\Gamma}_{ij}$), we obtain the $\conv\SO(n)$ condition, whereas when the parity of $\rho$ is not fixed then $M_{uv} \in \conv\Orth(n)$.

First, we show that $\mathcal{M}$ is positive semidefinite for all quantum states.

\begin{lemma}\label{lem:quantum_gram_matrix}
    Let $\mathcal{M} \in \R^{mn \times mn}$ be defined as in Eq.~\eqref{eq:quantum_gram_elements}. For all $\rho \in \mathcal{D}(\mathcal{H}_{2^n}^{\otimes m})$, $\mathcal{M} \succeq 0$.
\end{lemma}

\begin{proof}
We prove the statement by a sum-of-squares argument. To see where the fact of $1/n$ appears in the quantum definition of $\mathcal{M}$ above, we first construct a matrix $\mathcal{M}' \succeq 0$ which turns out to simply be $\mathcal{M}' = n\mathcal{M}$.

For each $k \in [n]$ define the Hermitian operator
\begin{equation}
    A_k = \sum_{v \in V} \sum_{i \in [n]} c_i^{(v)} P_{ik}^{(v)},
\end{equation}
where $c_i^{(v)} \in \R$ are arbitrary coefficients. Consider its square,
\begin{equation}\label{eq:A_k}
\begin{split}
    A_k^2 &= \l( \sum_{v \in V} \sum_{i \in [n]} c_i^{(v)} P_{ik}^{(v)} \r{)^2}\\
    &= \sum_{v \in V} \sum_{i, j \in [n]} c_i^{(v)} c_j^{(v)} P_{ik}^{(v)} P_{jk}^{(v)} + \sum_{\substack{u, v \in V \\ u \neq v}} \sum_{i, j \in [n]} c_i^{(u)} c_j^{(v)} P_{ik}^{(u)} \otimes P_{jk}^{(v)}.
\end{split}
\end{equation}
Note that the terms with $u \neq v$ feature the two-vertex operators as desired, while the diagonal terms of the sum contain products of the Pauli operators acting on the same vertex. Because $P_{ik} = \i \widetilde{\gamma}_i \gamma_k$, the diagonal terms reduce to (suppressing superscripts here)
\begin{equation}
\begin{split}
    \sum_{i, j \in [n]} c_i c_j P_{ik} P_{jk} &= - \sum_{i, j \in [n]} c_i c_j \widetilde{\gamma}_i \gamma_k \widetilde{\gamma}_j \gamma_k\\
    &= \sum_{i, j \in [n]} c_i c_j \widetilde{\gamma}_i \widetilde{\gamma}_j\\
    &= \sum_{i \in [n]} c_i^2 \I_{2^n} + \sum_{1 \leq i < j \leq n} c_i c_j (\widetilde{\gamma}_i \widetilde{\gamma}_j + \widetilde{\gamma}_j \widetilde{\gamma}_i)\\
    &= \l( \sum_{i \in [n]} c_i^2 \r) \I_{2^n}.
\end{split}
\end{equation}
Plugging this result into Eq.~\eqref{eq:A_k} and summing over all $k \in [n]$, we obtain
\begin{equation}
\begin{split}
    \sum_{k \in [n]} A_k^2 &= \l( \sum_{v \in V} \sum_{i, k \in [n]} |c_i^{(v)}|^2 \r) \I_{2^n} + \sum_{\substack{u, v \in V \\ u \neq v}} \sum_{i, j, k \in [n]} c_i^{(u)} c_j^{(v)} P_{ik}^{(u)} \otimes P_{jk}^{(v)}\\
    &= n \langle c, c \rangle \I_{2^n} + \sum_{\substack{u, v \in V \\ u \neq v}} \sum_{i, j \in [n]} c_i^{(u)} c_j^{(v)} \Gamma_{ij}^{(u, v)},
\end{split}
\end{equation}
where we have collected the coefficients $c_i^{(v)}$ into a vector $c \in \R^{mn}$. Similarly, if we arrange the expectation values $\tr(\Gamma_{ij}^{(u, v)} \rho)$ into a matrix $T \in \R^{mn \times mn}$ (where the $n \times n$ blocks on the diagonal are $0$), then the expectation value of the sum-of-squares operator is
\begin{equation}
\begin{split}
    \tr\l( \sum_{k \in [n]} A_k^2 \rho \r) &= n \langle c, c \rangle + \langle c, T c \rangle\\
    &= \langle c, \mathcal{M}' c \rangle
\end{split}
\end{equation}
where we have defined $\mathcal{M}' \coloneqq n\I_{mn} + T$. Because $\sum_{k \in [n]} A_k^2$ is a sum of PSD operators, its expectation value is always nonnegative, hence $\langle c, \mathcal{M}' c \rangle \geq 0$. This inequality holds for all vectors $c \in \R^{mn}$, so $\mathcal{M}' \succeq 0$ and hence $\mathcal{M} = \mathcal{M}'/n \succeq 0$ as claimed.
\end{proof}

The fact that the diagonal blocks $\mathcal{M}_{vv} = \I_n$ holds by definition. Finally, we need to show that each block of $\mathcal{M}$ lies in $\conv(\SO(n))$ when $\rho$ has even parity. A straightforward corollary of this result is that the blocks lie in $\conv\Orth(n)$ when $\rho$ does not have fixed parity. Note that in Section~\ref{sec:mixed_states} we showed that the matrix of expectation values $\tr(P_{ij} \rho_1)$ lies in $\conv\Orth(n)$ for any $n$-qubit density operator $\rho_1$, a straightforward extension of the PSD-lift representation of $\conv\SO(n)$ presented in Ref.~\cite{saunderson2015semidefinite}. Here we instead show that the matrix of expectation values $\frac{1}{n} \tr(\Gamma_{ij} \rho_2)$ for any $2n$-qubit density operator $\rho_2$ also lies in $\conv\Orth(n)$, and is an element of $\conv\SO(n)$ when $\rho_2$ has support only in the even-parity sector.

Because $\I_n \in \conv\SO(n) \subset \conv\Orth(n)$, and because these convex hulls are closed under transposition, all that remains is to prove the statement for the blocks $\mathcal{M}_{uv}$ when $u < v$. We show this by considering all density matrices on the reduced two-vertex Hilbert space.

\begin{lemma}
    Let $\rho_2 \in \mathcal{D}(\mathcal{H}_{2^n}^{\otimes 2})$. Define the $n \times n$ matrix $T$ by
    \begin{equation}
    T_{ij} \coloneqq \tr(\Gamma_{ij} \rho_2) = \tr\l( \sum_{k \in [n]} P_{ik} \otimes P_{jk} \rho_2 \r), \quad i, j \in [n].
    \end{equation}
    Then $T/n \in \conv\Orth(n)$. Furthermore, if $\tr(Z^{\otimes 2n} \rho_2) = 1$, then $M_{uv} \in \conv\SO(n)$.
\end{lemma}

\begin{proof}
    Consider the special singular value decomposition of $T = U \widetilde{\Sigma} \widetilde{V}^\T$. We can express the special singular values as
    \begin{equation}\label{eq:ssv_T}
    \begin{split}
    \widetilde{\sigma}_i(T) &= [U^\T T \widetilde{V}]_{ii}\\
    &= \sum_{j,\ell \in [n]} U_{ji} T_{j\ell} \widetilde{V}_{\ell i}\\
    &= \sum_{j,\ell \in [n]} U_{ji} \widetilde{V}_{\ell i} \sum_{k \in [n]} \tr\l( \i \widetilde{\gamma}_j \gamma_k \otimes \i \widetilde{\gamma}_\ell \gamma_k \rho_2 \r)\\
    &= \sum_{k \in [n]} \tr\l( {\mathcal{U}}_{(U, \I_n)}^\dagger \i \widetilde{\gamma}_i \gamma_k {\mathcal{U}}_{(U, \I_n)} \otimes {\mathcal{U}}_{(\widetilde{V}, \I_n)}^\dagger \i \widetilde{\gamma}_i \gamma_k {\mathcal{U}}_{(\widetilde{V}, \I_n)} \rho_2 \r)\\
    &= \sum_{k \in [n]} \tr\l( \i \widetilde{\gamma}_i \gamma_k \otimes \i \widetilde{\gamma}_i \gamma_k \rho'_2 \r),
    \end{split}
    \end{equation}
    where $\rho'_2 \coloneqq \l({\mathcal{U}}_{(U, \I_n)} \otimes {\mathcal{U}}_{(\widetilde{V}, \I_n)}\r) \rho_2 \l({\mathcal{U}}_{(U, \I_n)} \otimes {\mathcal{U}}_{(\widetilde{V}, \I_n)}\r{)^\dagger}$. The fact that $T/n \in \conv\Orth(n)$ follows immediately from the fact that the spectrum of $\i \widetilde{\gamma}_i \gamma_k \otimes \i \widetilde{\gamma}_i \gamma_k$ is $\{\pm 1\}$:
    \begin{equation}
    \begin{split}
    \sigma_1(T/n) &= \frac{1}{n} \l| \widetilde{\sigma}_1(T) \r|\\
    &= \frac{1}{n} \l| \sum_{k \in [n]} \tr\l( \i \widetilde{\gamma}_1 \gamma_k \otimes \i \widetilde{\gamma}_1 \gamma_k \rho'_2 \r) \r|\\
    &\leq \frac{1}{n} \sum_{k \in [n]} \l| \tr\l( \i \widetilde{\gamma}_1 \gamma_k \otimes \i \widetilde{\gamma}_1 \gamma_k \rho'_2 \r) \r|\\
    &\leq 1.
    \end{split}
    \end{equation}
    
    Now we examine the inclusion in $\conv\SO(n)$. Suppose that $\rho_2$ is an even-parity state. By virtue of the special singular value decomposition, we have that $\det(U\widetilde{V}^\T) = 1$ which implies that the Gaussian unitary ${\mathcal{U}}_{(U, \I_n)} \otimes {\mathcal{U}}_{(\widetilde{V}, \I_n)}$ preserves the parity of $\rho_2$:~$\tr(Z^{\otimes 2n} \rho_2) = \tr(Z^{\otimes 2n} \rho_2') = 1$. For $T/n$ to lie in $\conv\SO(n)$, the following inequality from Eq.~\eqref{eq:convSOn_app} must hold:
    \begin{equation}\label{eq:conv_ineq}
    \sum_{i \in [n] \setminus I} \widetilde{\sigma}_i(T) - \sum_{i \in I} \widetilde{\sigma}_i(T) \leq n(n - 2)
    \end{equation}
    for all subsets $I \subseteq [n]$ of odd size. To show this, first we write the left-hand side in terms of the result derived from Eq.~\eqref{eq:ssv_T}:
    \begin{equation}\label{eq:conv_T}
    \begin{split}
    \sum_{i \in [n] \setminus I} \widetilde{\sigma}_i(T) - \sum_{i \in I} \widetilde{\sigma}_i(T) &= \sum_{i \in [n] \setminus I} \sum_{k \in [n]} \tr\l( \i \widetilde{\gamma}_i \gamma_k \otimes \i \widetilde{\gamma}_i \gamma_k \rho'_2 \r) - \sum_{i \in I} \sum_{k \in [n]} \tr\l( \i \widetilde{\gamma}_i \gamma_k \otimes \i \widetilde{\gamma}_i \gamma_k \rho'_2 \r)\\
    &= \tr\l( \rho_2' \sum_{i,k \in [n]} (-1)^{z_i} \i \widetilde{\gamma}_i \gamma_k \otimes \i \widetilde{\gamma}_i \gamma_k \r),
    \end{split}
    \end{equation}
    where the string $z = z_1 \cdots z_n \in \{0, 1\}^n$ is defined as
    \begin{equation}
    z_i = \begin{cases}
    1 & i \in I,\\
    0 & i \notin I.
    \end{cases}
    \end{equation}
    Note that $|I|$ being odd implies that the Hamming weight of $z$ is also odd. To bound Eq.~\eqref{eq:conv_T} we shall seek a bound on the largest eigenvalue of the Hermitian operator
    \begin{equation}
    \sum_{i,k \in [n]} (-1)^{z_i} \i \widetilde{\gamma}_i \gamma_k \otimes \i \widetilde{\gamma}_i \gamma_k = \l( \sum_{i \in [n]} (-1)^{z_i \oplus 1} \widetilde{\gamma}_i \otimes \widetilde{\gamma}_i \r) \l( \sum_{k \in [n]} \gamma_k \otimes \gamma_k \r) \eqqcolon A_z B
    \end{equation}
    over the space of even-parity states. Here we use $\oplus$ to denote addition modulo 2.
    
    The operator $A_z B$ can in fact be exactly diagonalized in the basis of Bell states. First, observe that $[A_z, B] = 0$, which follows from the fact that $\widetilde{\gamma}_i \gamma_k = -\gamma_k \widetilde{\gamma}_i$, hence $[\widetilde{\gamma}_i \otimes \widetilde{\gamma}_i, \gamma_k \otimes \gamma_k] = 0$ for all $i, k \in [n]$. We can therefore seek their simultaneous eigenvectors, which can be determined from looking at the Jordan--Wigner representation of the Majorana operators:
    \begin{align}
    \widetilde{\gamma}_i \otimes \widetilde{\gamma}_i &= (Z_1 \cdots Z_{i-1} Y_i) \otimes (Z_1 \cdots Z_{i-1} Y_i), \label{eq:tensor_tilde_gamma}\\
    \gamma_k \otimes \gamma_k &= (Z_1 \cdots Z_{k-1} X_k) \otimes (Z_1 \cdots Z_{k-1} X_k). \label{eq:tensor_gamma}
    \end{align}
    These operators are diagonalized by the $2n$-qubit state
    \begin{equation}
    \ket{\beta(x, y)} = \bigotimes_{j=1}^n \ket{\beta(x_j, y_j)},
    \end{equation}
    where $x, y \in \{0, 1\}^n$, and the Bell state $\ket{\beta(x_j, y_j)}$ between the $j$th qubits across the two subsystems is defined as
    \begin{equation}
    \ket{\beta(x_j, y_j)} \coloneqq \frac{\ket{0} \otimes \ket{y_j} + (-1)^{x_j} \ket{1} \otimes \ket{y_j \oplus 1}}{\sqrt{2}}.
    \end{equation}
    Indeed there are $2^{2n}$ such states $\ket{\beta(x,y)}$, so they form an orthonormal basis for the $2n$ qubits. The eigenvalues of Eqs.~\eqref{eq:tensor_tilde_gamma} and \eqref{eq:tensor_gamma} can be determined by a standard computation,
    \begin{align}
    (X_j \otimes X_j) \ket{\beta(x_j, y_j)} &= (-1)^{x_j} \ket{\beta(x_j, y_j)},\\
    (Y_j \otimes Y_j) \ket{\beta(x_j, y_j)} &= (-1)^{x_j \oplus y_j \oplus 1} \ket{\beta(x_j, y_j)},\\
    (Z_j \otimes Z_j) \ket{\beta(x_j, y_j)} &= (-1)^{y_j} \ket{\beta(x_j, y_j)}.\label{eq:ZZ_beta}
    \end{align}
    Taking the appropriate products furnishes the eigenvalues of the $B$ and $A_z$ as
    \begin{align}
    \begin{split}
    B \ket{\beta(x, y)} &= \sum_{k \in [n]} (Z_1 \otimes Z_1) \cdots (Z_{k-1} \otimes Z_{k-1}) (X_k \otimes X_k) \ket{\beta(x,y)}\\
    &= \sum_{k \in [n]} (-1)^{y_1 \oplus \cdots \oplus y_{k-1}} (-1)^{x_k} \ket{\beta(x, y)},
    \end{split}\\
    \begin{split}
    A_z \ket{\beta(x, y)} &= \sum_{i \in [n]} (-1)^{z_i \oplus 1} (Z_1 \otimes Z_1) \cdots (Z_{k-1} \otimes Z_{k-1}) (Y_k \otimes Y_k) \ket{\beta(x,y)} \\
    &= \sum_{i \in [n]} (-1)^{z_i \oplus 1} (-1)^{y_1 \oplus \cdots \oplus y_{i-1}} (-1)^{x_i \oplus y_i \oplus 1} \ket{\beta(x, y)}\\
    &= \sum_{i \in [n]} (-1)^{z_i} (-1)^{y_1 \oplus \cdots \oplus y_{i}} (-1)^{x_i} \ket{\beta(x, y)}
    \end{split}
    \end{align}
    Altogether we arrive at the expression for the eigenvalues of $A_z B$,
    \begin{equation}\label{eq:AzB_eigval}
    \ev{\beta(x, y)}{A_z B}{\beta(x, y)} = \l( \sum_{i \in [n]} (-1)^{z_i} (-1)^{y_1 \oplus \cdots \oplus y_i} (-1)^{x_i} \r) \l( \sum_{k \in [n]} (-1)^{y_1 \oplus \cdots \oplus y_{k-1}} (-1)^{x_k} \r).
    \end{equation}
    We wish to find the largest value this can take over even-parity states. First, observe that the eigenstates $\ket{\beta(x, y)}$ have fixed parity according to
    \begin{equation}
    \ev{\beta(x, y)}{Z^{\otimes 2n}}{\beta(x, y)} = (-1)^{|y|},
    \end{equation}
    which follows from Eq.~\eqref{eq:ZZ_beta}. Hence we shall only consider $y$ to have even Hamming weight. Additionally recall that $z$ has odd Hamming weight, while the Hamming weight of $x$ is unrestricted.
    
    Let us denote the sums in Eq.~\eqref{eq:AzB_eigval} by
    \begin{align}
    a_z(x, y) &\coloneqq \sum_{i \in [n]} (-1)^{z_i} (-1)^{y_1 \oplus \cdots \oplus y_{i-1} \oplus y_i} (-1)^{x_i},\\
    b(x, y) &\coloneqq \sum_{k \in [n]} (-1)^{y_1 \oplus \cdots \oplus y_{k-1}} (-1)^{x_k}.
    \end{align}
    Clearly, they can be at most $n$, and this occurs whenever all the terms in their sum are positive. For $b(x, y)$, this is possible if and only if $x = p(y)$, where $p : \{0, 1\}^n \to \{0, 1\}^n$ stores the parity information of the $(k-1)$-length substring of its input into the $k$th bit of its output:
    \begin{equation}
    [p(y)]_k \coloneqq y_1 \oplus \cdots \oplus y_{k-1}.
    \end{equation}
    For notation clarity we point out that $[p(y)]_1 = 0$ and $[p(y)]_2 = y_1$. The forward direction, $b(p(y), y) = n$, is clear by construction. The reverse direction, that $b(x, y) = n$ implies $x = p(y)$, follows from the bijectivity of modular addition.
    
    Plugging this value of $x = p(y)$ into $a_z(x, y)$ yields
    \begin{equation}\label{eq:sum_i}
    a_z(p(y), y) = \sum_{i \in [n]} (-1)^{z_i} (-1)^{y_i} = \sum_{i \in [n]} (-1)^{[z \oplus y]_i}.
    \end{equation}
    Because $y$ has even and $z$ has odd Hamming weight, their sum $y \oplus z$ must have odd Hamming weight. Therefore at least one term in Eq.~\eqref{eq:sum_i} must be negative, implying that $a_z(p(y), y) \leq n - 2$. It follows that
    \begin{equation}
    \ev{\beta(p(y), y)}{A_z B}{\beta(p(y), y)} = a_z(p(y), y) b(p(y), y) \leq (n-2) n.
    \end{equation}
    We now show that no other assignment of $(x, y)$ can exceed this bound. Recall that $b(x, y) = n$ if and only if $x = p(y)$. Thus any other choice of $x$ necessarily returns a smaller value of $b(x, y)$. Because sums of $\pm 1$ cannot yield $n - 1$, the next largest value would be $b(x,y) = n - 2$. However we can always trivially bound $a_z(x, y) \leq n$ for all $x, y$. This implies that such a choice of $x$ for which $b(x, y) = n - 2$ (whatever it is) also cannot provide a value of $a_z(x, y) b(x, y)$ exceeding $n(n-2)$.
    
    Note that there is another assignment that saturates the upper bound, which is simply considering a global negative sign in front of both products. Specifically, let $x = p(y) \oplus 1^n$. In this case $b(x,y) = -n$ and $a_z(x, y) \geq -(n-2)$, yielding the same bound $a_z(x,y) b(x, y) \leq n(n-2)$.

    To conclude, we use the fact that the maximum eigenvalue of $A_z B$ in the even-parity subspace is $n(n-2)$ for any odd-weight $z$ (equiv., any odd-size $I \subseteq [n]$). This bounds the value of Eq.~\eqref{eq:conv_T} by $n(n-2)$, hence validating the inequality of Eq.~\eqref{eq:conv_ineq}. Thus $T/n \in \conv\SO(n)$ whenever $\rho_2$ has even parity.
\end{proof}

As usual, replacing the operators $\Gamma_{ij}$ with $\widetilde{\Gamma}_{ij}$ is equivalent to enforcing the even-parity constraint. In fact, since $\widetilde{\Gamma}_{ij} = \sum_{k \in [n]} \widetilde{P}_{ik} \otimes \widetilde{P}_{jk}$, the equivalent constraint involves the reduced single-vertex marginals, $\tr(Z^{\otimes n} \otimes \I_{2^n} \rho_2) = \tr(\I_{2^n} \otimes Z^{\otimes n} \rho_2) = 1$, rather than the entire $2n$-qubit Hilbert space. Of course, if both single-vertex parity constraints are satisfied, then the two-vertex constraint automatically follows.

\section{\label{sec:even_subspace}Details for working in the even-parity subspace}
First we provide an expression for $n$-qubit Pauli operators projected to $\Cl^0(n)$. Let
\begin{equation}
    A \coloneqq W_1 \otimes \cdots \otimes W_n, \quad W_i \in \{\I, X, Y, Z\}.
\end{equation}
A straightforward calculation yields the conditional expression:
\begin{equation}\label{eq:projected_paulis}
    \widetilde{A} = \Pi_0 A \Pi_0^\T = \begin{cases}
    0 & \text{if } [A, Z^{\otimes n}] \neq 0\\
    \begin{cases}
    W_2 \otimes \cdots \otimes W_n & \text{if } W_1 = \I, X\\
    \i (W_2 Z) \otimes \cdots \otimes (W_n Z) & \text{if } W_1 = Y\\
    (W_2 Z) \otimes \cdots \otimes (W_n Z) & \text{if } W_1 = Z.
    \end{cases} & \text{if } [A, Z^{\otimes n}] = 0.
    \end{cases}
\end{equation}
Notably, if $A$ does not commute with the parity operator then $\widetilde{A} = 0$.

Now we generalize from the main text, defining the operator
\begin{equation}
    \Pi_k \coloneqq \frac{1}{\sqrt{2}} \l( \bra{+} \otimes \I_2^{\otimes (n-1)} + (-1)^k \bra{-} \otimes Z^{\otimes (n-1)} \r), \quad k \in \{0, 1\},
\end{equation}
which is the projector $\Cl(n) \to \Cl^k(n)$. These operators obey
\begin{align}
    \Pi_k \Pi_k^\T &= \I_{2^{n-1}}, \label{eq:PPT}\\
    \Pi_k^\T \Pi_k &= \frac{\I_{2^n} + (-1)^k Z^{\otimes n}}{2}.\label{eq:PTP}
\end{align}

Given a state $\widetilde{\rho} \in \mathcal{D}(\mathcal{H}_{2^{n-1}})$, the expectation values of $\widetilde{P}_{ij}$ satisfy
\begin{equation}
\begin{split}
    \tr(\widetilde{P}_{ij} \widetilde{\rho}) &= \tr(P_{ij} \Pi_0^\T \widetilde{\rho} \Pi_0)\\
    &= \tr(P_{ij} \rho_0),
\end{split}
\end{equation}
where $\rho_0 \coloneqq \Pi_0^\T \widetilde{\rho} \Pi_0 \in \mathcal{D}(\mathcal{H}_{2^n})$ has only support on even-parity computational basis states. By Eq.~\eqref{eq:PPT} we can ``invert'' this relation in the following sense:~given a state $\rho \in \mathcal{D}(\mathcal{H}_{2^n})$ which only has support on the even subspace, its $(n-1)$-qubit representation is $\Pi_0 \rho \Pi_0^\T \in \mathcal{D}(\mathcal{H}_{2^{n-1}})$.

This translation is useful when using the fermionic interpretation of the $P_{ij}$ but we wish to work directly in the $(n-1)$-qubit subspace. For example, suppose we wish to prepare a product of Gaussian states in a quantum computer (as is done in Section~\ref{sec:asp_numerics} to prepare the initial state of the quasi-adiabatic evolution). It is well known how to compile linear-depth circuits for this task~\cite{jiang2018quantum}, however this is within the standard $n$-qubit representation. Here we show how to translate those circuits into the $(n-1)$-qubit representation under $\Pi_0$. In fact, these techniques apply to any sequence of gates which commute with the parity operator $Z^{\otimes n}$.

Let $\ket{\psi} \coloneqq V \ket{0^n} \in \mathcal{H}_{2^n}$, where the circuit $V$ is constructed from $L$ gates, $V = V_L \cdots V_1$. We can assume without loss of generality that the initial state is the vacuum $\ket{0^n}$ and that all gates $V_\ell$ commute with the parity operator, as otherwise $\ket{\psi}$ would not lie in the even subspace of $\mathcal{H}_{2^n}$.\footnote{While it is possible to have an even number of gates which anticommute with the parity operator, for simplicity we assume that the circuit has been compiled such that each $V_\ell$ preserves parity.} For example, parity-preserving Gaussian unitaries can be decomposed into single- and two-qubit gates of the form $e^{-\i \theta Z_i}$, $e^{-\i \theta X_i X_{i+1}}$. We wish to obtain a circuit description for preparing the state $\Pi_0 \ket{\psi} \in \mathcal{H}_{2^{n-1}}$. By Eq.~\eqref{eq:PTP}, $\Pi_0^\T \Pi_0 \ket{0^n} = \ket{0^n}$, and furthermore $\Pi_0 \ket{0^n} = \ket{0^{n-1}}$ is the $(n-1)$-qubit representation of the vacuum. Therefore
\begin{equation}
\begin{split}
    \Pi_0 \ket{\psi} &= \Pi_0 V \ket{0^n}\\
    &= \Pi_0 V_L \cdots V_1 \Pi_0^\T \Pi_0 \ket{0^n}\\
    &= \Pi_0 V_L \cdots V_1 \Pi_0^\T \ket{0^{n-1}}.
\end{split}
\end{equation}
Because $\Pi_0$ is not unitary ($\Pi_0^\T$ is merely an isometry), we cannot simply insert terms like $\Pi_0^\T \Pi_0$ in between each gate. However, observe that if each $V_\ell$ preserves parity, then they can be block diagonalized into the even and odd subspaces of $\mathcal{H}_{2^n}$,
\begin{equation}\label{eq:V_l_block}
    V_\ell = \begin{bmatrix}
    V_{\ell,0} & 0\\
    0 & V_{\ell,1}
    \end{bmatrix},
\end{equation}
where $V_{\ell,k} \coloneqq \Pi_k V_\ell \Pi_k^\T$ are $2^{n-1}$-dimensional unitary matrices. Thus
\begin{equation}
    V = \begin{bmatrix}
    V_{L,0} \cdots V_{1,0} & 0\\
    0 & V_{L,1} \cdots V_{1,1}
    \end{bmatrix},
\end{equation}
and conjugation by the projector $\Pi_0$ precisely extracts the first block of this matrix:
\begin{equation}
    \Pi_0 V \Pi_0^\T = V_{L,0} \cdots V_{1,0}.
\end{equation}
This sequence of gates is what we wish to implement on the physical $(n-1)$-qubit register. When the gates $V_\ell$ take the form
\begin{equation}
    V_\ell = e^{-\i \theta A_\ell}
\end{equation}
for some $n$-qubit Pauli operator $A_\ell$, then
\begin{equation}\label{eq:V_l0}
\begin{split}
    V_{\ell,0} &= \Pi_0 V_\ell \Pi_0^\T\\
    &= \Pi_0 \l( \I_{2^n} \cos\theta - \i A_\ell \sin\theta \r) \Pi_0^\T\\
    &= \I_{2^{n-1}} \cos\theta - \i \widetilde{A}_\ell \sin\theta\\
    &= e^{-\i \theta \widetilde{A}_\ell},
\end{split}
\end{equation}
where $\widetilde{A}_\ell \coloneqq \Pi_0 A_\ell \Pi_0^\T$. Note that this calculation assumes that $V_\ell$ commutes with parity, hence $[A_\ell, Z^{\otimes n}] = 0$, which guarantees that $\widetilde{A}_\ell$ is unitary and Hermitian and hences furnishes the final line of Eq.~\eqref{eq:V_l0}. (If they did not commute then the decomposition of Eq.~\eqref{eq:V_l_block} would not be valid to begin with.)

\section{Small $n$ examples}
In this appendix we write down the Pauli operators $P_{ij}$ and $\widetilde{P}_{ij}$ for small but relevant values of $n$, to provide the reader with some concerte examples of how to construct the LNCG Hamiltonian.

\subsection{The Ising model from the $\Orth(1)$ setting}

As a warm-up we first demonstrate that the LNCG Hamiltonian reduces to the Ising formulation of the commutative combinatorial optimization problem when considering $G = \Orth(1)$. Each local Hilbert space $\mathcal{H}_d$ with $d = 2$ is simply a qubit, and we only have to consider a single Pauli operator on each local qubit,
\begin{equation}
    P_{11} = \i \widetilde{\gamma}_1 \gamma_1 = Z.
\end{equation}
It then follows that the LNCG interaction terms are
\begin{equation}
    \Gamma_{11}^{(u, v)} = Z^{(u)} \otimes Z^{(v)},
\end{equation}
and so the full Hamiltonian acting on $\mathcal{H}_2^{\otimes |V|}$ is indeed the classical Ising Hamiltonian,
\begin{equation}
    H = \sum_{(u, v) \in E} C_{uv} Z^{(u)} \otimes Z^{(v)},
\end{equation}
with weights $C_{uv} \in \R$. It is also instructive to write down the elements of $\Pin(1)$ as quantum states. The Clifford algebra $\Cl(1)$ is spanned by $e_1$ and $e_\varnothing = \openone$, with the only elements of $S^0$ being $\pm e_1$. Therefore, taking all possible products of elements in $S^0$ (including the empty product), we arrive at
\begin{equation}
    \Pin(1) = \{\pm e_\varnothing, \pm e_1\},
\end{equation}
which corresponds to the qubit computational basis states $\{\ket{0}, \ket{1}\}$ (up to global phases), as expected. Indeed, one sees that the mappings $Q(e_\varnothing) = \ev{0}{Z}{0} = 1$ and $Q(e_1) = \ev{1}{Z}{1} = -1$ fully cover $\Orth(1)$.

\subsection{The projected operators of the $\SO(3)$ setting}

As $\SO(3)$ is arguably the most ubiquitous group for physical applications, for reference we explicitly write down its Pauli operators under the projection to $\Cl^0(3) \cong \mathcal{H}_4$. As seen by the dimension of this Hilbert space, only two qubits per variable (for a total of $2m$ qubits) are required to represent this problem. Using Eq.~\eqref{eq:projected_paulis} we have
\begin{equation}
    \begin{bmatrix}
    \widetilde{P}_{11} & \widetilde{P}_{12} & \widetilde{P}_{13}\\
    \widetilde{P}_{21} & \widetilde{P}_{22} & \widetilde{P}_{23}\\
    \widetilde{P}_{31} & \widetilde{P}_{32} & \widetilde{P}_{33}
    \end{bmatrix} =
    \begin{bmatrix}
    Z_1 Z_2 & -X_1 & -Z_1 X_2\\
    X_1 Z_2 & Z_1 & -X_1 X_2\\
    X_2 & - Y_1 Y_2 & Z_2
    \end{bmatrix}.
\end{equation}

\section{\label{sec:son_approx_ratio}Classical approximation ratio for $\SO(n)$}
Approximation ratios for the rounded solution of the classical semidefinite relaxation of Problem~\eqref{eq:LNCG} were obtained in Ref.~\cite{bandeira2016approximating} for the cases $G = \Orth(n)$ and $\U(n)$. However, no such approximation ratios were derived for the case of $\SO(n)$. Here we adapt the argument of Ref.~\cite{bandeira2016approximating} to this setting, wherein the rounding algorithm performs the special singular value decomposition to guarantee that the rounded solutions have unit determinant. As we will see, this feature results in a approximation ratio for the classical semidefinite program over $\SO(n)$ that is strictly worse than the previously studied $\Orth(n)$ case.

Recall that the semidefinite relaxation of Problem~\eqref{eq:LNCG} can be formulated as
\begin{equation}\label{eq:LNCG_SDP_relax}
    \max_{X_1, \ldots, X_m \in \R^{n \times mn}} \sum_{(u, v) \in E} \langle C_{uv}, X_u X_v^\T \rangle \quad \text{subject to} \quad X_v X_v^\T = \I_n.
\end{equation}
To round the relaxed solution back into the feasible space of orthogonal matrices, Ref.~\cite{bandeira2016approximating} proposes the following randomized algorithm with a guarantee on the approximation ratio.

\begin{theorem}[{\cite[Theorem 4]{bandeira2016approximating}}]\label{thm:O(n)_approx_ratio}
    Let $X_1, \ldots, X_n \in \R^{n \times mn}$ be a solution to Problem~\eqref{eq:LNCG_SDP_relax}. Let $Z$ be an $mn \times n$ Gaussian random matrix whose entries are drawn i.i.d.~from $\mathcal{N}(0, 1/n)$. Compute the orthogonal matrices
    \begin{equation}
        Q_v = \mathcal{P}(X_v Z),
    \end{equation}
    where $\mathcal{P}(X) = \argmin_{Y \in \Orth(n)} \| Y - X \|_F$. The expected value of this approximate solution (averaged over $Z$) obeys
    \begin{equation}
        \E\l[ f(Q_1, \ldots, Q_m) \r] \geq \alpha_{\Orth(n)}^2 \max_{R_1, \ldots, R_m \in \Orth(n)} f(R_1, \ldots, R_m).
    \end{equation}
    The approximation ratio $\alpha_{\Orth(n)}^2$ is defined by the average singular value of random Gaussian $n \times n$ matrices $Z_1 \sim \mathcal{N}(0, \I_n / n)$,
    \begin{equation}
        \alpha_{\Orth(n)} \coloneqq \E\l[ \frac{1}{n} \sum_{i \in [n]} \sigma_i(Z_1) \r],
    \end{equation}
    where $\sigma_i(Z_1)$ is the $i$th singular value of $Z_1$.
\end{theorem}

Our adaptation to the $\SO(n)$ setting simply replaces the rounding operator $\mathcal{P}$ with $\widetilde{\mathcal{P}}(X) \coloneqq \argmin_{Y \in \SO(n)} \| Y - X \|_F$. With this change we obtain an analogous result for optimizing over $\SO(n)$ elements with the same classical semidefinite program:
\begin{theorem}\label{thm:SO(n)_approx_ratio}
    Let $X_1, \ldots, X_n$ and $Z$ be as in Theorem~\ref{thm:O(n)_approx_ratio}. Compute the rotation matrices
    \begin{equation}
        Q_v = \widetilde{\mathcal{P}}(X_v Z),
    \end{equation}
    where $\widetilde{\mathcal{P}}(X) = \argmin_{Y \in \SO(n)} \| Y - X \|_F$. The expected value of this approximate solution (averaged over $Z$) obeys
    \begin{equation}
        \E\l[ f(Q_1, \ldots, Q_m) \r] \geq \alpha_{\SO(n)}^2 \max_{R_1, \ldots, R_m \in \SO(n)} f(R_1, \ldots, R_m).
    \end{equation}
    The approximation ratio $\alpha_{\SO(n)}^2$ is defined by the average $n - 1$ largest singular values of random Gaussian $n \times n$ matrices $Z_1 \sim \mathcal{N}(0, \I_n / n)$,
    \begin{equation}
        \alpha_{\SO(n)} \coloneqq \E\l[ \frac{1}{n} \sum_{i \in [n-1]} \sigma_i(Z_1) \r],
    \end{equation}
    where $\sigma_i(Z_1)$ is the $i$th singular value of $Z_1$, in descending order $\sigma_1(Z_1) \geq \cdots \geq \sigma_n(Z_1) \geq 0$.
\end{theorem}
Because singular values are nonnegative, it is clear that
\begin{equation}
    \alpha_{\Orth(n)} - \alpha_{\SO(n)} = \frac{1}{n} \E\l[ \sigma_n(Z_1) \r] \geq 0.
\end{equation}
In particular we will see that $\E\l[ \sigma_n(Z_1) \r] > 0$ for all finite $n$, so the rounding algorithm guarantees a strictly smaller approximation ratio for the problem over $\SO(n)$ than over $\Orth(n)$.

The proof of Theorem~\ref{thm:O(n)_approx_ratio} requires two lemmas regarding the expected value of random Gaussian matrices under the rounding operator $\mathcal{P}$. Analogously, our proof of Theorem~\ref{thm:SO(n)_approx_ratio} requires a modification of those lemmas when $\mathcal{P}$ is replaced by $\widetilde{\mathcal{P}}$.
\begin{lemma}[{Adapted from \cite[Lemma 5]{bandeira2016approximating}}]\label{lem:proj_arbitrary}
    Let $M, N \in \R^{n \times mn}$ obey $MM^\T = NN^\T = \I_n$. For $Z \in \R^{mn \times n}$ with i.i.d.~entries drawn from $\mathcal{N}(0, n^{-1})$, we have
    \begin{equation}
        \E\l[ \widetilde{\mathcal{P}}(MZ) (NZ)^\T \r] = \E\l[ (MZ) \widetilde{\mathcal{P}}(NZ)^\T \r] = \alpha_{\SO(n)} MN^\T.
    \end{equation}
\end{lemma}
This lemma is proved with the help of the following lemma.
\begin{lemma}[{Adapted from \cite[Lemma 6]{bandeira2016approximating}}]\label{lem:proj_gaussian}
    Let $Z_1 \in \R^{n \times n}$ with i.i.d.~entries drawn from $\mathcal{N}(0, 1/n)$. Then
    \begin{equation}
        \E\l[ \widetilde{\mathcal{P}}(Z_1) Z_1^\T \r] = \E\l[ Z_1 \widetilde{\mathcal{P}}(Z_1)^\T \r] = \alpha_{\SO(n)} \I_n.
    \end{equation}
\end{lemma}
Before we prove these two lemmas, we will use them to prove Theorem~\ref{thm:SO(n)_approx_ratio}. The proof idea here is entirely analogous to the original argument of Theorem~\ref{thm:O(n)_approx_ratio} from Ref.~\cite{bandeira2016approximating}, but with the appropriate replacements of $\mathcal{P}$ by $\widetilde{\mathcal{P}}$. Nonetheless we sketch the proof below for completeness.

\begin{proof}[Proof (of Theorem~\ref{thm:SO(n)_approx_ratio})]
We wish to lower bound the average rounded value
\begin{equation}
    \E[f(Q_1, \ldots, Q_m)] = \E\l[ \sum_{(u,v) \in E} \langle C_{uv}, \widetilde{\mathcal{P}}(X_u Z) \widetilde{\mathcal{P}}(X_v Z)^\T \rangle \r]
\end{equation}
in terms of the relaxed value $\sum_{(u,v) \in E} \langle C_{uv}, X_u X_v^\T \rangle$. Assuming we have such a lower bound with ratio $0 < \alpha^2 \leq 1$, this leads to a chain of inequalities establishing the desired approximation ratio to the original problem:
\begin{equation}\label{eq:approx_ratio_from_relaxation}
\begin{split}
    \E\l[ \sum_{(u,v) \in E} \langle C_{uv}, \widetilde{\mathcal{P}}(X_u Z) \widetilde{\mathcal{P}}(X_v Z)^\T \rangle \r] &\geq \alpha^2 \sum_{(u,v) \in E} \langle C_{uv}, X_u X_v^\T \rangle\\
    &\geq \alpha^2 \max_{R_1, \ldots, R_m \in \Orth(n)} \sum_{(u,v) \in E} \langle C_{uv}, R_u R_v^\T \rangle\\
    &\geq \alpha^2 \max_{R_1, \ldots, R_m \in \SO(n)} \sum_{(u,v) \in E} \langle C_{uv}, R_u R_v^\T \rangle,
\end{split}
\end{equation}
where the second inequality follows from the fact that the relaxation provides an upper bound to the original problem, and the third inequality is a consequence of $\SO(n) \subset \Orth(n)$. The task is then to determine such an $\alpha$ which satisfies the first inequality of Eq.~\eqref{eq:approx_ratio_from_relaxation}. The core argument is a generalization of the Rietz method~\cite{alon2004approximating}, which proceeds by constructing a positive semidefinite matrix $S \in \R^{mn \times mn}$ whose $(u, v)$th block is defined as
\begin{equation}
    S_{uv} \coloneqq \l( X_u Z - \alpha^{-1} \widetilde{\mathcal{P}}(X_u Z) \r) \l( X_v Z - \alpha^{-1} \widetilde{\mathcal{P}}(X_v Z) \r)^\T.
\end{equation}
The expected value of this matrix is
\begin{equation}
\begin{split}
    \E[S_{uv}] &= \E\l[ X_u Z (X_v Z)^\T - \alpha^{-1} \widetilde{\mathcal{P}}(X_u Z) (X_v Z)^\T \r.\\
    &\quad \l. - \alpha^{-1} (X_u Z) \widetilde{\mathcal{P}}(X_v Z)^\T + \alpha^{-2} \widetilde{\mathcal{P}}(X_u Z) \widetilde{\mathcal{P}}(X_v Z)^\T \r]\\
    &= X_u \E\l[ ZZ^\T \r] X_v - \alpha^{-1} \E\l[ \widetilde{\mathcal{P}}(X_u Z) (X_v Z)^\T \r] \\
    &\quad - \alpha^{-1} \E\l[ (X_u Z) \widetilde{\mathcal{P}}(X_v Z)^\T \r] + \alpha^{-2} \E\l[ \widetilde{\mathcal{P}}(X_u Z) \widetilde{\mathcal{P}}(X_v Z)^\T \r].
\end{split}
\end{equation}
Because $ZZ^\T$ is a Wishart matrix with covariance matrix $\I_n / n$, we have $\E[ZZ^\T] = \I_n$. Meanwhile, $\E\l[ \widetilde{\mathcal{P}}(X_u Z) \widetilde{\mathcal{P}}(X_v Z)^\T \r]$ is the quantity we wish to bound. To compute the expected values of the two cross terms, we invoke Lemma~\ref{lem:proj_arbitrary} which holds because $X_u X_u^\T = X_v X_v^\T = \I_n$:
\begin{equation}
    \E\l[ \widetilde{\mathcal{P}}(X_u Z) (X_v Z)^\T \r] = \E\l[ (X_u Z) \widetilde{\mathcal{P}}(X_v Z)^\T \r] = \alpha_{\SO(n)} X_u X_v^\T.
\end{equation}
Thus, setting $\alpha = \alpha_{\SO(n)}$, we obtain
\begin{equation}
\begin{split}
    \E[S_{uv}] &= X_u X_v^\T - X_u X_v^\T - X_u X_v^\T + \alpha_{\SO(n)}^{-2} \E\l[ \widetilde{\mathcal{P}}(X_u Z) \widetilde{\mathcal{P}}(X_v Z)^\T \r]\\
    &= - X_u X_v^\T + \alpha_{\SO(n)}^{-2} \E\l[ \widetilde{\mathcal{P}}(X_u Z) \widetilde{\mathcal{P}}(X_v Z)^\T \r].
\end{split}
\end{equation}
Finally, using the fact that $C, S \succeq 0$, we have that $\langle C, S \rangle \geq 0$ and so $\E\langle C, S \rangle \geq 0$, which implies that
\begin{equation}
    \E\l[ \sum_{(u,v) \in E} \langle C_{uv}, \widetilde{\mathcal{P}}(X_u Z) \widetilde{\mathcal{P}}(X_v Z)^\T \rangle \r] \geq \alpha_{\SO(n)}^2 \sum_{(u,v) \in E} \langle C_{uv}, X_u X_v^\T \rangle.
\end{equation}
Then by Eq.~\eqref{eq:approx_ratio_from_relaxation} the claim follows.
\end{proof}

We now establish the value of
\begin{equation}
    \alpha_{\SO(n)} = \E\l[ \frac{1}{n} \sum_{i \in [n-1]} \sigma_i(Z_1) \r]
\end{equation}
from Lemmas~\ref{lem:proj_arbitrary} and \ref{lem:proj_gaussian}. Because Lemma~\ref{lem:proj_arbitrary} is somewhat technical and the argument is virtually unchanged by replacing $\mathcal{P}$ with $\widetilde{\mathcal{P}}$, we refer the reader to Ref.~\cite{bandeira2016approximating} for proof details. Instead, we simply note that the only part of the proof for Lemma~\ref{lem:proj_arbitrary} which does depend on the change to $\widetilde{\mathcal{P}}$ is the final result, wherein it is established that
\begin{equation}
    \E\l[ (MZ) \widetilde{\mathcal{P}}(NZ)^\T \r] = \E\l[ Z_1 \widetilde{\mathcal{P}}(Z_1)^\T \r] MN^\T,
\end{equation}
where $Z_1 \in \R^{n \times n}$ has entries i.i.d.~from $\mathcal{N}(0, 1 / n)$.
Thus proving Lemma~\ref{lem:proj_gaussian} is the key component in establishing the value of the approximation ratio $\alpha_{\SO(n)}^2$.

\begin{proof}[Proof (of Lemma~\ref{lem:proj_gaussian})]
    Consider the singular value decomposition of $Z_1 = U \Sigma V^\T \in \R^{n \times n}$. Its special singular value decomposition can be written as $Z_1 = U (\Sigma J_{UV^\T}) (V J_{UV^\T})^\T$, where $J_{UV^\T}$ is the $n \times n$ diagonal matrix
\begin{equation}
    J_{UV^\T} \coloneqq \begin{bmatrix}
    I_{n-1} & 0\\
    0 & \det UV^\T
    \end{bmatrix}.
\end{equation}
Note that $J_{UV^\T} = J_U J_V$. Using the fact that the (special) rounding operator returns
\begin{equation}
    \widetilde{\mathcal{P}}(Z_1) = U J_{UV^\T} V^\T,
\end{equation}
we have
\begin{equation}\label{eq:PZZ}
    \widetilde{\mathcal{P}}(Z_1) Z_1^\T = U J_U J_V \Sigma U^\T.
\end{equation}
Because $Z_1$ is a random Gaussian matrix with i.i.d.~entries, its singular values and left- and right-singular vectors are distributed independently~\cite{tulino2004random}. In particular, both $U$ and $V$ are distributed according to the Haar measure on $\Orth(n)$. The expected value of Eq.~\eqref{eq:PZZ} can therefore be split into three independent averages:
\begin{equation}
    \E_{Z_1 \sim \mathcal{N}(0, \I_n / n)} \l[\widetilde{\mathcal{P}}(Z_1) Z_1^\T \r] = \E_{\Sigma \sim D} \E_{U \sim \Orth(n)} \l[ U J_U \E_{V \sim \Orth(n)} \l[ J_V \r] \Sigma U^\T \r].
\end{equation}
(We shall comment on the distribution $D$ of singular values later.) Because $\Orth(n)$ is evenly divided into its unconnected $({+1})$- and $({-1})$-determinant components, the average determinant vanishes:~$\E_{V \sim \Orth(n)}[\det V] = 0$. This leaves us with
\begin{equation}\label{eq:special_haar_avg}
    \E\l[\widetilde{\mathcal{P}}(Z_1) Z_1^\T \r] = \E\l[ U \overline{\Sigma} U^\T \r],
\end{equation}
where
\begin{equation}
    \overline{\Sigma} = \begin{bmatrix}
    \sigma_1(Z_1) & & & \\
    & \ddots & & \\
    & & \sigma_{n-1}(Z_1) & \\
    & & & 0
    \end{bmatrix}.
\end{equation}
The Haar average over $U \sim \Orth(n)$ in Eq.~\eqref{eq:special_haar_avg} is well-known~\cite{collins2006integration} to be proportional to the identity, $\E[U \overline{\Sigma} U^\T] = \lambda \I_n$, and the constant of proportionality can be determined by considering its trace:
\begin{equation}
\begin{split}
    n\lambda = \tr\l( \lambda \I_n \r) &= \tr\l( \E[ U \overline{\Sigma} U^\T ] \r)\\
    &= \E[\tr\overline{\Sigma}]\\
    &= \E\l[ \sum_{i \in [n-1]} \sigma_i(Z_1) \r].
\end{split}
\end{equation}
Hence $\lambda = \alpha_{\SO(n)}$ and so
\begin{equation}
    \E\l[\widetilde{\mathcal{P}}(Z_1) Z_1^\T \r] = \alpha_{\SO(n)} \I_n.
\end{equation}
The corresponding statement for $\E\l[ Z_1 \widetilde{\mathcal{P}}(Z_1)^\T \r]$ follows completely analogously, essentially by interchanging the roles of $U$ and $V$. The entire argument is equivalent because $U$ and $V$ are i.i.d.
\end{proof}

To numerically evaluate $\alpha_{\SO(n)}$ we can use the linearity of expectation,
\begin{equation}
    \alpha_{\SO(n)} = \alpha_{\Orth(n)} - \frac{1}{n} \E\l[ \sigma_n(Z_1) \r].
\end{equation}
The distribution of singular values of random Gaussian matrices can be analyzed from the theory of Wishart matrices. In particular, $Z_1 Z_1^\T = U \Sigma^2 U^\T$ is a Wishart matrix with covariance matrix $\I_n / n$, so the distribution of singular values $\Sigma$ is the square root of the Wishart distribution of eigenvalues. The quantity $\alpha_{\Orth(n)}$ in terms of the marginal distribution $p_n^{\text{(avg)}}(x)$ of Wishart eigenvalues $x \in (0, \infty)$ was studied in Ref.~\cite{bandeira2016approximating}, yielding the expression
\begin{equation}
    \alpha_{\Orth(n)} = \frac{1}{\sqrt{n}} \int_0^\infty p_n^{\text{(avg)}}(x) \sqrt{x} \, dx.
\end{equation}
Note the factor of $n^{-1/2}$, which is introduced because the distribution $p_n^{\text{(avg)}}(x)$ is normalized to have unit variance. An explicit expression of $p_n^{\text{(avg)}}(x)$ can be found in Refs.~\cite[Lemma 21]{bandeira2016approximating} and \cite[Eq.~(16)]{livan2011momets}.

For our newly derived approximation ratio $\alpha_{\SO(n)}$, we need to additionally evaluate the expected smallest singular value of this Wishart distribution. This minimum-eigenvalue distribution was studied in Ref.~\cite{edelman1988eigenvalues}, wherein an analytical expression was derived (again assuming unit variance):
\begin{equation}
    p_n^{\text{(min)}}(x) = \frac{n}{2^{n - 1/2}} \frac{\Gamma(n)}{\Gamma(n/2)} \frac{e^{-xn/2}}{\sqrt{x}} U\l( \frac{n-1}{2}, -\frac{1}{2}, \frac{x}{2} \r).
\end{equation}
Here, $U(a, b, z)$ with $a > 0$ and $b < 1$ is the Tricomi confluent hypergeometric function, the unique solution to the differential equation
\begin{equation}
    z \frac{d^2 U}{dz^2} + (b - z) \frac{dU}{dz} - aU = 0
\end{equation}
with boundary conditions $U(a, b, 0) = \Gamma(1-b)/\Gamma(1+a-b)$ and $\lim_{z \to \infty} U(a, b, z) = 0$. The expression for the average smallest singular value is therefore
\begin{equation}
    \E\l[ \sigma_n(Z_1) \r] = \frac{1}{\sqrt{n}} \int_0^\infty p_n^{\text{(min)}}(x) \sqrt{x} \, dx.
\end{equation}
Altogether, we arrive at the integral expression for
\begin{equation}
    \alpha_{\SO(n)} = \frac{1}{\sqrt{n}} \int_0^\infty \l[ p_n^{\text{(avg)}}(x) - \frac{1}{n} p_n^{\text{(min)}}(x) \r] \sqrt{x} \, dx.
\end{equation}

\section{\label{sec:symmetry_vertex}LNCG Hamiltonian symmetries}
Here we demonstrate the local $\Orth(n)$ symmetry discussed in Section~\ref{sec:vertex_rounding}. Consider an edge term
\begin{equation}
    H_{uv} = \sum_{i,j \in [n]} [C_{uv}]_{ij} \sum_{k \in [n]} P_{ik}^{(u)} \otimes P_{jk}^{(v)}.
\end{equation}
Because $P_{ik} = \i \widetilde{\gamma}_i \gamma_k$ and the sum over $k$ is independent of $C_{uv}$, we can factor out each $\gamma_k^{(u)} \otimes \gamma_k^{(v)}$ and rewrite the Hamiltonian term as as
\begin{equation}
    H_{uv} = -\sum_{i,j \in [n]} [C_{uv}]_{ij} \l( \widetilde{\gamma}_i^{(u)} \otimes \widetilde{\gamma}_j^{(v)} \r) \l( \sum_{k \in [n]} \gamma_k^{(u)} \otimes \gamma_k^{(v)} \r).
\end{equation}
The operator $\sum_{k \in [n]} \gamma_k^{(u)} \otimes \gamma_k^{(v)}$ is invariant to any orthogonal transformation $V \in \Orth(n)$ which acts identically on both vertices:
\begin{equation}
\begin{split}
    \mathcal{U}_{(\I_n, V)}^{\otimes 2} \l( \sum_{k \in [n]} \gamma_k^{(u)} \otimes \gamma_k^{(v)} \r) (\mathcal{U}_{(\I_n, V)}^{\otimes 2})^\dagger &= \sum_{k \in [n]} \l( \sum_{\ell \in [n]} V_{k\ell} \gamma_\ell^{(u)} \r) \otimes \l( \sum_{\ell' \in [n]} V_{k\ell'} \gamma_{\ell'}^{(v)} \r)\\
    &= \sum_{\ell,\ell' \in [n]} \l( \sum_{k \in [n]} [V^\T]_{\ell k} V_{k\ell'} \r) \gamma_\ell^{(u)} \otimes \gamma_{\ell'}^{(v)}\\
    &= \sum_{\ell,\ell' \in [n]} \delta_{\ell\ell'} \gamma_\ell^{(u)} \otimes \gamma_{\ell'}^{(v)}\\
    &= \sum_{\ell \in [n]} \gamma_\ell^{(u)} \otimes \gamma_\ell^{(v)}.
\end{split}
\end{equation}
Because $\mathcal{U}_{(\I_n, V)}$ acts trivially on all $\widetilde{\gamma}_i$, it follows that
\begin{equation}
    \mathcal{U}_{(\I_n, V)}^{\otimes 2} H_{uv} (\mathcal{U}_{(\I_n, V)}^{\otimes 2})^\dagger = H_{uv}
\end{equation}
for each $(u, v)$. Finally, this symmetry can be straightforwardly extended to all $m$ vertices:
\begin{equation}\label{eq:O(n)_sym}
    \mathcal{U}_{(\I_n, V)}^{\otimes m} H (\mathcal{U}_{(\I_n, V)}^{\otimes m})^\dagger = H.
\end{equation}

Now we investigate some consequences of this continuous symmetry. The following lemma is particularly important, as it necessitates the use of the one-body perturbation $\zeta H_1$ to break this symmetry when preparing of eigenstates of $H$.

\begin{lemma}
Let $\ket{\psi}$ be a nondegenerate eigenstate of $H$. Then for each single-vertex marginal $\sigma_v \coloneqq \tr_{\neg v} \op{\psi}{\psi}$, $v \in [m]$, we have
\begin{equation}
    Q(\sigma_v) = 0.
\end{equation}
\end{lemma}

\begin{proof}
Consider the expansion of its density matrix $\op{\psi}{\psi}$ in the Majorana operator basis, up to the relevant one-body expectation values:
\begin{equation}
    \op{\psi}{\psi} = \frac{1}{d^m} \l( \openone^{\otimes m} + \sum_{v \in [m]} \sum_{i,j \in [n]} [Q(\sigma_v)]_{ij} \i \widetilde{\gamma}_i^{(v)} \gamma_j^{(v)} + \cdots \r),
\end{equation}
where we recall that $[Q(\sigma_v)]_{ij} = \ev{\psi}{\i \widetilde{\gamma}_i^{(v)} \gamma_j^{(v)}}{\psi}$. Due to the symmetry [Eq.~\eqref{eq:O(n)_sym}], for every $V \in \Orth(n)$ the state $\ket{\psi(V)} = {\mathcal{U}}_{(\I_n, V)}^{\otimes m} \ket{\psi}$ is also an eigenvector of $H$ with the same eigenvalue. The one-body expectation values of $\ket{\psi}$ are therefore transformed as
\begin{equation}
\begin{split}
    {\mathcal{U}}_{(\I_n, V)}^{\otimes m} \l( \sum_{v \in [m]} \sum_{i,j \in [n]} [Q(\sigma_v)]_{ij} \i \widetilde{\gamma}_i^{(v)} \gamma_j^{(v)} \r) ({\mathcal{U}}_{(\I_n, V)}^{\otimes m})^\dagger &= \sum_{v \in [m]} \sum_{i,j \in [n]} [Q(\sigma_v)]_{ij} \i \sum_{i' \in [n]} V_{ii'} \widetilde{\gamma}_{i'}^{(v)} \gamma_{j}^{(v)}\\
    &= \sum_{v \in [m]} \sum_{i',j \in [n]} [V^\T Q(\sigma_v)]_{i'j} \i \widetilde{\gamma}_{i'}^{(v)} \gamma_{j}^{(v)}.
\end{split}
\end{equation}
Now suppose that $\ket{\psi}$ is nondegenerate. Then we have that $\op{\psi}{\psi} = \op{\psi(V)}{\psi(V)}$ for all $V \in \Orth(n)$, and in particular we can take the Haar integral over $\Orth(n)$ of this identity:
\begin{equation}
\begin{split}
    \int_{\Orth(n)} d\mu(V) \op{\psi(V)}{\psi(V)} = \int_{\Orth(n)} d\mu(V) \op{\psi}{\psi} = \op{\psi}{\psi},
\end{split}
\end{equation}
where $\mu$ is the normalized Haar measure satisfying $\mu(\Orth(n)) = 1$. Because the Haar integral over linear functions vanishes, i.e., $\int_{\Orth(n)} d\mu(V) \, V_{ij} = 0$~\cite{collins2006integration}, it follows that
\begin{equation}
    \int_{\Orth(n)} d\mu(V) \, V^\T Q(\sigma_v) = 0.
\end{equation}
Furthermore, because $\i \widetilde{\gamma}_{i}^{(v)} \gamma_{j}^{(v)}$ are linearly independent (as elements of an operator basis), the equality $\op{\psi}{\psi} = \int_{\Orth(n)} d\mu(V) \op{\psi(V)}{\psi(V)}$ implies that
\begin{equation}
    Q(\sigma_v) = \int_{\Orth(n)} d\mu(V) \, V^\T Q(\sigma_v) = 0
\end{equation}
for all $v \in [m]$.
\end{proof}

\section{\label{sec:pin_circuits}The Pin group from quantum circuits}
In the main text we showed that each $x \in \Pin(n)$ corresponds to the eigenstates of a family of free-fermion Hamiltonians, which are (pure) fermionic Gaussian states. Here we provide an alternative perspective of this correspondence, using quantum circuits which prepare such states.

Recall that every $x \in \Pin(n)$ can be written as
\begin{equation}
    x = u_1 \cdots u_k
\end{equation}
for some $k \leq n$, where we may expand each $u_j \in S^{n-1}$ in the standard basis as
\begin{equation}
    u_j = \sum_{i \in [n]} v_i^{(j)} e_i
\end{equation}
for some unit vector $v^{(j)} \in \R^n$. The product of these unit vectors can be expressed using the right-multiplication operator $\rho_{u_j}$ acting on the identity element,
\begin{equation}
\begin{split}
    x &= e_\varnothing x\\
    &= e_\varnothing u_1 \cdots u_k\\
    &= (\rho_{u_k} \cdots \rho_{u_1})(e_\varnothing)\\
\end{split}
\end{equation}
On the other hand, consider the so-called Clifford loader~\cite{kerenidis2022quantum}, a circuit primitive defined (in our notation) as
\begin{equation}
    \Gamma(v) = \sum_{i \in [n]} v_i \gamma_i
\end{equation}
for any unit vector $v \in \R^n$. It is straightforward to check that this operator is Hermitian and unitary, and Ref.~\cite{kerenidis2022quantum} provides an explicit circuit constructions based on two-qubit Givens rotation primitives.\footnote{Givens rotations themselves are representations of fermionic Gaussian transformations acting on two modes at a time.} Using the relation $\gamma_i = \rho_i \alpha$ and acting this circuit on the vacuum state $\ket{0^n} \equiv \ket{e_\varnothing}$, we see that the state
\begin{equation}
\begin{split}
    \ket{x} &= \Gamma(v^{(k)}) \cdots \Gamma(v^{(1)}) \ket{0^n}\\
    &= (\rho_{u_k} \alpha \cdots \rho_{u_1} \alpha)\ket{e_\varnothing}\\
    &= (-1)^{\binom{k}{2}} (\rho_{u_k} \cdots \rho_{u_1})\ket{e_\varnothing}
\end{split}
\end{equation}
indeed is equivalent to $x \in \Pin(n)$, up to a global sign (recall that $\alpha^2 = \openone$ and $\alpha\ket{e_\varnothing} = \ket{e_\varnothing}$). In other words, the Clifford loader is precisely the quantum-circuit representation of generators of the $\Pin$ group.

It is worth noting that in Ref.~\cite{kerenidis2022quantum} they construct ``subspace states'' from this composition of Clifford loaders. In the language of fermions, subspace states are Slater determinants:~free-fermion states with fixed particle number. Preparing Slater determinants in this fashion requires that the unit vectors $v^{(1)}, \ldots, v^{(k)}$ be linearly independent (and thus, without loss of generality, they can be made orthonormal while preserving the subspace that they span, hence the alternative name). However, the definition of the Pin group demands all possible unit vectors in such products, not just those which are linearly independent. Indeed, one can see that if the state $\ket{x}$ is a Slater determinant, then its trace is an integer, as
\begin{equation}
    \tr[Q(x)] = \langle x | \sum_{i \in [n]} \i \widetilde{\gamma}_i \gamma_i | x \rangle = \langle x | (n \I_{2^n} - 2 N) | x \rangle = n - 2k \in \{-n, \ldots, n\},
\end{equation}
where $N = \sum_{i \in [n]} a_i^\dagger a_i$ is the total number operator. Clearly not all orthogonal matrices have integer trace, so Slater determinants are insufficient to cover all of $\Pin(n)$. To reach the remaining elements, we note that if the unit vectors are linearly dependent, then one can show that the $\ket{x} = \Gamma(v^{(k)}) \cdots \Gamma(v^{(1)}) \ket{0^n}$ does not have fixed particle number, so $\ev{x}{N}{x}$ is not necessarily an integer.

\section{\label{sec:tomography}Measurement schemes}
In this section we comment on the efficient schemes available for measuring the relevant expectation values. This is important even in the context of a phase-estimation approach, as one needs to obtain the values of the decision variables to perform the rounding procedure.

\subsection{Tomography of edge marginals}

To measure the energy (for variational approaches) or to perform edge rounding, we require the expectation values of the two-body observables
\begin{align}
    \Gamma_{ij}^{(u, v)} &= \sum_{k \in [n]} P_{ik}^{(u)} \otimes P_{jk}^{(v)}, \quad G = \Orth(n),\\
    \widetilde{\Gamma}_{ij}^{(u, v)} &= \sum_{k \in [n]} \widetilde{P}_{ik}^{(u)} \otimes \widetilde{P}_{jk}^{(v)}, \quad G = \SO(n),
\end{align}
for each $(u, v) \in E$ and $i, j \in [n]$. When considering $G = \Orth(n)$, because each $P_{ik}^{(u)} \otimes P_{jk}^{(v)}$ is a fermionic two-body operator, we can straightforwardly apply the partial tomography schemes developed for local fermionic systems, such as Majorana swap networks~\cite{bonet2020nearly} or classical shadows~\cite{zhao2021fermionic}. In either case, the measurement circuits required are fermionic Gaussian unitaries and the sample complexity is $\Ord(N^2/\epsilon^2)$, where $N = n|V|$ is the total number of qubits and $\epsilon > 0$ is the desired estimation precision of each expectation value.

\subsection{Tomography of vertex marginals}

The vertex-rounding procedure requires the expectation values of only single-qudit observables $P_{ij}^{(v)}$ or $\widetilde{P}_{ij}^{(v)}$ on each vertex $v \in V$, $i, j \in [n]$. In this case the observables being measured commute across vertices, so it suffices to talk about the tomography of a single vertex, as the same process can be executed in parallel across all vertices. Again, because these operators are fermionic one-body observables, the same fermionic partial tomography technology~\cite{bonet2020nearly,zhao2021fermionic} can be applied here, incurring a sampling cost of $\Ord(n/\epsilon^2)$. In fact, further constant-factor savings can be achieved in the one-body setting by using the measurement scheme introduced in Ref.~\cite{arute2020hartree}. This scheme requires only particle-conserving fermionic Gaussian unitaries, which can be compiled with only half the depth of the more general Gaussian unitaries required of the previous two methods. Note that each operator $P_{ij}$ is of the form of either $XX$ or $YY$ when $|i - j| = 1$, and so they correspond precisely to the observables measured to reconstruct the real part of the fermionic one-body reduced density matrix~\cite{arute2020hartree}.

\subsection{Estimating observables via gradient method}

Ref.~\cite{huggins2021nearly} introduces a quantum algorithm for estimating a large collection of (generically noncommuting) $M$ observables $\{O_j \mid j \in [M]\}$ to precision $\epsilon$ by encoding their expectation values into the gradient of a function. This function is implemented as a quantum circuit which prepares the state of interest and applies $\Ot(\sqrt{M}/\epsilon)$ gates of the form $c\text{-}e^{-\i\theta O_j}$, controlled on $\Ord(M\log(1/\epsilon))$ ancilla qubits. Finally, using the algorithm of Ref.~\cite{gilyen2019optimizing} for gradient estimation, one calls this circuit $\Ot(\sqrt{M}/\epsilon)$ times to estimate the encoded expectation values (the notation $\Ot(\cdot)$ suppresses polylogarithmic factors). Although this approach demands additional qubits and more complicated circuitry, it has the striking advantage of a quadratically improved scaling in the number of state preparations with respect to estimation error $\epsilon$, compared to the refinement of sampling error in tomographic approaches. In our context, we have either $M = n^3|E|$ or $M = n^2|V|$ observables of interest (satisfying a technical requirement of having their spectral norms bounded by $1$), corresponding to the measurement of edge or vertex terms respectively. The gates required are then simply controlled Pauli rotations.
\chapter{Conclusion}\label{chap:conclusion}


The theory of quantum computation has sparked deep and profound questions about the nature of computation and the computation of Nature. For computer scientists, it has forced a reevaluation of what a ``reasonable'' computational model means within the extended Church--Turing thesis~\cite{deutsch1985quantum,bernstein1993quantum}. Physicists, meanwhile, have had to confront with the possibility that their models of Nature might not properly reflect what she is actually capable of~\cite{kempe2006complexity,poulin2011quantum}. 

Fortunately, quantum computers also offer to help us solve some difficult problems that we face in concrete domains. This dissertation has addressed this more down-to-earth affair, focusing on improvements to quantum algorithms for simulating fermions so that we might realize that promise sooner rather than later. As is always the case, we are left with more questions than when we had started. In this chapter, I will opine on some of the most outstanding questions that remain following the results presented within this dissertation.

In \cref{chap:FCS}, we developed an optimal extension of classical shadows for fermionic systems, and in \cref{chap:EM} we studied how such protocols enjoy inherent robustness properties in the presence of quantum symmetries. The motivation for these works is to squeeze out every drop of performance from NISQ machines, from an algorithmic-design perspective. While motivated for use in the NISQ era, I am interested in seeing how these results can find applications more broadly. For example, in one collaboration that I was involved with, we studied the efficiency of classical shadows within the fault-tolerant regime~\cite{babbush2023quantum}. There, we considered simulating the time evolution of interacting electrons, wherein partial-tomography techniques are necessary to extract dynamical properties from the system (as opposed to techniques like phase estimation, which primarily address static properties). What can be said of enhancing classical shadows for fault-tolerant applications and dynamics simulations? Impressive work on the former aspect has already been accomplished by van Apeldoorn \emph{et al.}~\cite{van2023quantum}, showing how shadow tomography can be substantially improved with access to state-preparation circuits. In particular, they achieve Heisenberg-limited estimation of multiple observables, albeit with substantial space complexity. To the latter aspect, Rall~\cite{rall2020quantum} showed how to learn time-correlation functions using post-NISQ resources. Such properties are both essential for studies of quantum many-body physics, yet highly challenging to extract from a quantum computer. How much further can classical shadows/shadow tomography be pushed, to offer practical solutions for these problems?

More generally, the promise of NISQ appears to be wavering at the time of this writing~\cite{obrien2023purification}. Nonetheless, this line of research continues to provide us with interesting questions that we had not even thought to ask before the advent of such machines. For example, a running theme of NISQ algorithms has been to explore how much complexity can be offloaded onto a classical processor before encountering exponential costs. The initial formulation, VQE~\cite{peruzzo2014variational}, supposed that the quantum computer needs only to prepare a low-energy state;~determining the circuit for doing so can be offloaded to a classical processor. However, the hybrid ``quantum--classical quantum Monte Carlo'' (QC-QMC) algorithm introduced by Huggins \emph{et al.}~\cite{huggins2022unbiasing} appears to challenge that notion. Their experimental success is not necessarily attributed to how low energy of a state that their noisy quantum computer produced, but instead how closely its sign structure (relative phases between amplitudes of the wavefunction) matched that of the true ground state. This is because a significant amount of the computational hardness in QMC manifests as the so-called sign problem~\cite{loh1990sign}, whereby quantum negativity obstructs the convergence of Monte Carlo simulations. With knowledge of the sign structure, the QMC algorithm (run entirely classically) can be modified to avoid the sign problem.

This new perspective leads us to ask how we might further ease the job of the NISQ machine in practice. To this end, we need to pinpoint precisely where the hardness manifests within various classical simulation algorithms. While the complexity-theoretic hardness is equivalent in all forms, it appears that some tasks (perhaps like approximating the sign structure) might be robust than others (like directly preparing an approximate ground state). As we move toward larger and more accurate machines, can the hope for noisy quantum information processing be rekindled? For example, suppose one is given a device with below-threshold error rates. Such a machine could produce a handful of perfect logical qubits, an impressive technological feat but still easily classically simulated. On the other hand, using the physical qubits \emph{directly} might grant access to ${\sim} 10^4$ very-low-noise qubits. This regime might not necessarily be ``intermediate scale'' anymore, but nonetheless might be where the hidden promise of NISQ lies.

Finally, in \cref{chap:QR} we investigated a hard classical optimization problem and showed how to naturally embed it into a quantum (indeed, fermionic) problem, a procedure referred to as ``quantum relaxation.''\footnote{To my knowledge, this terminology/concept has only been considered by Fuller \emph{et al.}~\cite{fuller2021approximate} and us~\cite{zhao2023expanding}, although I believe it is a very natural generalization of relaxations from classical approximation theory that merits wider study.} This work took the initial steps of this subject, and we have already mentioned the many potential further directions to pursue at the end of that chapter. Here, I will recap the vision I have for this line of work with regards to approximation theory. For instance, Naor, Regev, and Vidick~\cite{naor2014efficient} constructed an efficient algorithm for the noncommutative Grothendieck problem which achieves an approximation ratio of $1/2$ in the complex case, but only $1/2\sqrt{2}$ in the real case. On the other hand, Bri{\" e}t, Regev, and Saket demonstrated an $\mathsf{NP}$-hard threshold of approximation of $1/2$ for both cases. The real variant is precisely what our fermionic representation addresses;~can it then provide the tools for designing a classical algorithm which improves the $1/2\sqrt{2}$ result to match the hardness threshold?

More broadly, what can be said about the hardness of the $\SO(N)$ problem, which has yet to be properly studied from a complexity-theoretic perspective? Building off the work of Saunderson, Parrilo, and Willsky~\cite{saunderson2014semidefinite}, our results showed how to encode the determinant condition into a linear constraint on an exponentially large vector space. Ideally, this representation could elucidate rigorous statements about the hardness of quadratic $\SO(N)$ optimization. In connection with quantum algorithms, I would ideally like to see approximation thresholds in terms of $\mathsf{QMA}$-hardness, which would establish the limits of approximation even with quantum computers. Unfortunately, this avenue currently seems intractable without an established quantum PCP theorem~\cite{aharonov2013guest}.

\setstretch{\dnormalspacing}
\backmatter

\end{document}